%% file: Smith_Geneva_M_2023June_PhDSoftwareEngineering.tex
\documentclass[11pt]{report}
\usepackage{gscale_thesis_singlespace}
\usepackage{fancyheadings}     
\usepackage{natbib}
\usepackage[nottoc,notlof,notlot,numbib]{tocbibind}

\usepackage{graphicx}
\usepackage{subcaption}
\usepackage[justification=centering]{caption}
\usepackage{array}
\newcolumntype{P}[1]{>{\raggedright\let\newline\\\arraybackslash\hspace{0pt}}m{#1}}
\newcolumntype{C}[1]{>{\centering\arraybackslash}m{#1}}
\usepackage{booktabs}
\usepackage{longtable}
\usepackage{multirow}
\usepackage{lscape}
\usepackage{threeparttable}
\usepackage{threeparttablex}
\usepackage{float}
\usepackage{enumerate}
\usepackage[shortlabels, inline]{enumitem}
\usepackage[shortcuts]{extdash}
\usepackage{amsmath}
\usepackage{amsfonts}
\usepackage{amssymb}
\usepackage{textcomp}
\usepackage{wasysym}
\usepackage{listings}
\usepackage{epigraph}
\usepackage[usenames,dvipsnames,table]{xcolor}
\usepackage{xargs}
\usepackage{afterpage}

\usepackage[framemethod=tikz]{mdframed}
\usetikzlibrary{shadows}

\usepackage[british]{babel}
\usepackage{halloweenmath}
\usepackage{marvosym}
\usepackage{hieroglf}
\usepackage{relsize}
\usepackage[colorinlistoftodos,prependcaption,textsize=tiny]{todonotes}

\usepackage{bbding}

\usepackage[hyphens]{url}
\usepackage[hidelinks]{hyperref}

\usepackage{wrapfig}

\usepackage{epstopdf}

\numberwithin{equation}{section} 

\makeatletter
\def\namedlabel#1#2{\begingroup
    \def\@currentlabel{#2}%
    \label{#1}\endgroup
}
\makeatother

\newcommandx{\unsure}[2][1=]{\todo[inline,linecolor=red,backgroundcolor=red!25,
bordercolor=red,size=\normalsize,#1]{#2}}
\newcommandx{\change}[2][1=]{\todo[inline,linecolor=blue,
backgroundcolor=blue!25,bordercolor=blue,size=\normalsize,#1]{#2}}
\newcommandx{\info}[2][1=]{\todo[inline,linecolor=green,
backgroundcolor=green!25,bordercolor=green,size=\normalsize,#1]{#2}}
\newcommandx{\improvement}[2][1=]{\todo[linecolor=purple,
backgroundcolor=Purple,bordercolor=purple,size=\normalsize,#1]{#2}}
\newcommandx{\citeme}[1][1=]{\todo[linecolor=Goldenrod,
backgroundcolor=Goldenrod!25,bordercolor=Goldenrod,size=\normalsize,#1]{Cite
me!}}

\newcommand{\citepg}[2]{\citeauthor{#1}, \citeyear{#1}, p.~#2}
\newcommand{\citeg}[1]{\citeauthor{#1}, \citeyear{#1}}

\newcommand{\parasep}{
    \begin{center}
        \includegraphics[scale=0.8]{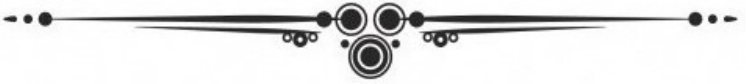}
    \end{center}
}

\newcommand\RStar{{\footnotesize\FiveStarOpen}}
\newcommand\doubleRStar{{\RStar\kern-0.4em\RStar}}
\newcommand\tripleRStar{{\RStar\kern-0.4em\RStar\kern-0.4em\RStar}}

\newcommand{\strong}{\tripleRStar}
\newcommand{\good}{\doubleRStar}
\newcommand{\weak}{\RStar}
\newcommand{\disqualified}{--}

\newcommand{\colourRow}{\rowcolor[rgb]{0.851,0.851,1}}
\newcommand{\colourCell}{\cellcolor[rgb]{0.851,0.851,1}}

\definecolor{CoolPurple}{RGB}{117, 0, 235}
\definecolor{NatureGreen}{RGB}{8, 104, 11}
\definecolor{SeaBlue}{RGB}{6, 97, 122}
\definecolor{Lilac}{RGB}{178, 131, 206}

\newmdenv[%
leftmargin=0.1in, rightmargin=0.1in,
roundcorner=5pt,
shadow=true, shadowsize=2pt,
linecolor=DarkOrchid,
subtitlebelowline=true,subtitleaboveline=true,
subtitlebackgroundcolor=BlueViolet!20,
frametitlerule=false,
frametitlebackgroundcolor=BlueViolet!20,
]{notes}

\newmdenv[%
leftmargin=0.1in, rightmargin=0.1in,
roundcorner=5pt,
shadow=true, shadowsize=2pt,
linecolor=DarkOrchid,
subtitlebelowline=true,subtitleaboveline=true,
subtitlebackgroundcolor=Mulberry!20,
frametitlerule=false,
frametitlebackgroundcolor=Periwinkle!20,
]{highlight}

\newmdenv[%
leftmargin=0.45in, rightmargin=0.45in,
roundcorner=5pt,
shadow=true, shadowsize=2pt,
linecolor=Lilac,
subtitlebelowline=true,subtitleaboveline=true,
subtitlebackgroundcolor=Lilac,
frametitle={\textbf{Key Points}},frametitlerule=false,
frametitlebackgroundcolor=Lilac,
]{keypoints}

\def\signed #1{{\leavevmode\unskip\nobreak\hfil\penalty50\hskip2em
        \hbox{}\nobreak\hfil#1%
        \parfillskip=0pt \finalhyphendemerits=0 \endgraf}}

\newsavebox\mybox
\newenvironment{aquote}[1]
{\savebox\mybox{#1}\begin{quote}}
    {\signed{\usebox\mybox}\end{quote}}

\hypersetup{
    colorlinks=true,
    linktoc=all,
    linkcolor=CoolPurple,
    citecolor=NatureGreen,
    filecolor=WildStrawberry,
    urlcolor=SeaBlue
}

\input{commands}

\begin{document}
\include{definitions}  

\beforepreface         
  \include{layabstr}   
  \include{abstr}      
  \include{dedic}      
  \include{acknowledgements}  
  \referencepages 
  \include{notation}
  
    \newpage
    \input{academicachievementdeclaration}
    \label{NumPrefacePages}

\afterpreface

   \part[Ready Player One]{Ready Player One \\
    \begin{center}
        \begin{minipage}[l]{4.5in}
            \vspace*{5mm}
            \def\epigraphflush{center}
            \setlength{\epigraphwidth}{0.85\linewidth}
            \def\textflush{center}
            \epigraph{\textnormal{Would you like to play a
            game?}}{\textnormal{Joshua, \textit{Wargames}}}
            \textnormal{\normalsize\parbox{\linewidth}{
                    \begin{center}
                        Welcome!
                    \end{center}
                    This part presents relevant background information
                    necessary to understand \progname{}, its purpose, and its
                    location in the broader research field. ``Introduction''
                    (Chapter~\ref{chapter:intro}) introduces the work, briefly
                    motivates it, and describes the research questions it wants
                    to answer. ``Engaging Players with Believable Characters''
                    (Chapter~\ref{chapter:believable}) describes the motivation
                    in further depth, linking player engagement with video
                    games to believable game characters, and ``Meet Emotion
                    (Briefly)'' (Chapter~\ref{chapter:primer}) reviews
                    essential information about affect and emotion necessary
                    for this work. ``On Designing Emotion Engines''
                    (Chapter~\ref{chapter:se-ee-design}) describes how software
                    engineering does and could influence the development of
                    computational systems ``with emotion'', and supporting
                    development methodologies proposed as an outcome of the
                    work on \progname{}. Finally, ``Affective Theories in
                    Computational Models'' (Chapter~\ref{chapter:cmeOverview})
                    surveys existing software systems that model emotion and
                    related phenomena for common design decision trends and
                    theoretical roots in \ref{as}.
                    \begin{center}
                        When you are Ready Player One, Start Your \progname{}!
                    \end{center}
                }
            }
            \parasep
        \end{minipage}
    \end{center}}

   \include{introduction}
    \setcounter{figure}{0}
    \setcounter{equation}{0}
    \setcounter{table}{0}

   \include{believablecharacters}
    \setcounter{figure}{0}
    \setcounter{equation}{0}
    \setcounter{table}{0}

   \include{psychprimer}
    \setcounter{figure}{0}
    \setcounter{equation}{0}
    \setcounter{table}{0}

   \include{SEforEEs}
    \setcounter{figure}{0}
    \setcounter{equation}{0}
    \setcounter{table}{0}

   \include{theoriesInCMEs}
    \setcounter{figure}{0}
    \setcounter{equation}{0}
    \setcounter{table}{0}

   \part[Enter the \progname{}]{Enter the \progname{} \\
       \begin{center}
           \begin{minipage}[l]{4.5in}
               \vspace*{5mm}
               \def\epigraphflush{center}
               \setlength{\epigraphwidth}{0.85\linewidth}
               \def\textflush{center}
               \epigraph{\textnormal{Warning: Game
               Corruption.}}{\textnormal{Main Frame Game Voice,
               \textit{Reboot}}}
                \textnormal{\normalsize\parbox{\linewidth}{
                        \begin{center}
                            Good to see you again!
                        \end{center}
                        This part gives an account of \progname{}'s design and
                        development. ``Start Your \progname{}: Requirements and
                        Scope'' (Chapter~\ref{chapter:reqsAndScope}) reviews
                        \progname{}'s high-level requirements and design scope,
                        then begins describing the process for choosing its
                        underlying emotion theories and models. The process
                        concludes in ``Support Your \progname{}: The
                        Requirements Choose the Theories''
                        (Chapter~\ref{chapter:theoryAnalysis}). Additional
                        examples of this process in ``Interlude: Choosing
                        Theories for Other CMEs''
                        (Chapter~\ref{chapter:choosingExamples}) demonstrate
                        how changes in high-level requirements and design scope
                        can change the outcome of the theory selection process.
                        ``Spec Your \progname{}: Defining the Pieces''
                        (Chapter~\ref{chapter:equations}) describes
                        \progname{}'s models and their documentation method,
                        followed by ``Build Your \progname{}: Some Assembly
                        Required'' (Chapter~\ref{chapter:designImplement}) that
                        describes its architecture design, along with its
                        documentation method, and relevant implementation
                        details. And last, but never least, ``Gather Your
                        Tools: Defining Acceptance Test Case Templates''
                        (Chapter~\ref{chapter:testcasedefinition}) and
                        ``Inspect Your \progname{}: Extending Acceptance Test
                        Case Templates'' (Chapter~\ref{chapter:testcaseEMgine})
                        details the steps taken to build acceptance test cases
                        that borrow elements from \progname{}'s models, yet
                        exist independently of \progname{} itself.
                        \begin{center}
                            And now, the main event: Enter the \progname{}!
                        \end{center}
                    }
                }
               \parasep
           \end{minipage}
   \end{center}}

   \include{highleveldesign}
    \setcounter{figure}{0}
    \setcounter{equation}{0}
    \setcounter{table}{0}

   \include{theory2reqs}
    \setcounter{figure}{0}
    \setcounter{equation}{0}
    \setcounter{table}{0}

   \include{choosingtheoriesexamples}
    \setcounter{figure}{0}
    \setcounter{equation}{0}
    \setcounter{table}{0}

   \include{appraisalequations}
    \setcounter{figure}{0}
    \setcounter{equation}{0}
    \setcounter{table}{0}

   \include{designAndImplement}
    \setcounter{figure}{0}
    \setcounter{equation}{0}
    \setcounter{table}{0}

   \include{validation}
    \setcounter{figure}{0}
    \setcounter{equation}{0}
    \setcounter{table}{0}

   \include{validationExtend}
    \setcounter{figure}{0}
    \setcounter{equation}{0}
    \setcounter{table}{0}

      \part[Continue?]{Continue? \\
          \begin{center}
              \begin{minipage}[l]{4.5in}
                  \vspace*{5mm}
                  \def\epigraphflush{center}
                  \setlength{\epigraphwidth}{0.85\linewidth}
                  \def\textflush{center}
                  \epigraph{\textnormal{The future is our
                  time.}}{\textnormal{Agent Smith, \textit{The Matrix}}}
                  \textnormal{\normalsize\parbox{\linewidth}{
                          \begin{center}
                              You made it!
                          \end{center}
                          This part reviews the motivation and work on
                          \progname{}. ``Looking Up at the Sky From Down the
                          Rabbit Hole'' (Chapter~\ref{chapter:conclusion})
                          revisits the work as a whole, reflects on its
                          limitations, and suggests avenues for future work.
                          The bibliography and supplementary material follow
                          (Appendices~\ref{appendix:survey},
                          \ref{chapter:affect}, \ref{chapter:reqsTheoryNotes},
                          and \ref{appendixEmotionClassification}).
                          \begin{center}
                              Would you like to Continue?
                          \end{center}
                      }
                  }
                  \parasep
              \end{minipage}
      \end{center}}

   \include{conclusion}
    \setcounter{figure}{0}
    \setcounter{equation}{0}
    \setcounter{table}{0}

\bibliographystyle{ACM-Reference-Format}
\bibliography{../BibFiles/references_Geneva,
../BibFiles/references_SEPerspective, ../BibFiles/references,
../BibFiles/references_psych, ../BibFiles/references_media,
../BibFiles/references_definitions, ../BibFiles/references_gamedesign,
../BibFiles/references_documentation, ../BibFiles/references_test}

\def\epigraphflush{center}
\setlength{\epigraphwidth}{0.85\textwidth}
\def\textflush{center}
\epigraph{End of Line.}{Master Control Program, \textit{TRON}}
\parasep

\begin{appendix}
    \include{appendix_CMESurveySupp}
     \setcounter{figure}{0}
     \setcounter{equation}{0}
     \setcounter{table}{0}

    \include{affect_and_emotion}
     \setcounter{figure}{0}
     \setcounter{equation}{0}
     \setcounter{table}{0}

    \include{appendix_theory2reqsNotes}
     \setcounter{figure}{0}
     \setcounter{equation}{0}
     \setcounter{table}{0}

    \include{appendix_emotionClassifications}
     \setcounter{figure}{0}
     \setcounter{equation}{0}
     \setcounter{table}{0}
\end{appendix}

\clearpage


\vspace*{\fill}
\begin{figure}[!h]
    \centering
    \includegraphics[width=0.3\linewidth]{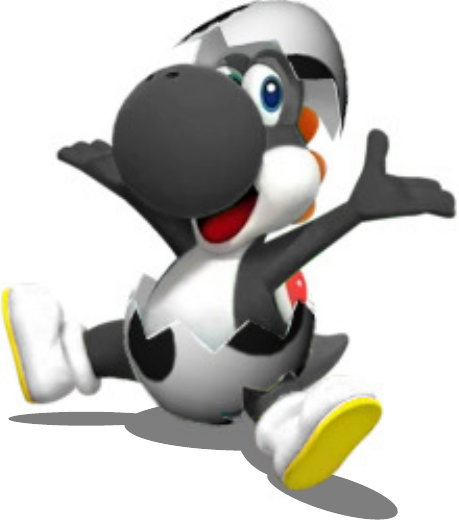}
    \caption*{\scriptsize Black Yoshi \\\textcopyright{}1997 Nintendo}
\end{figure}
\parasep
\vspace*{\fill}

\label{NumDocumentPages}

\end{document}

%% file: commands.tex
\newcommand{\progname}{EMgine} 

\newcommand{\timetype}{\mathbb{T}}
\newcommand{\deltatimetype}{\mathbb{T}_{\Delta}}

\newcommand{\emotionintensitytype}{\mathbb{I}}
\newcommand{\responsestrength}{\mathbb{I}_{\Delta}}
\newcommand{\emotionstatetype}{\mathbb{ES}}
\newcommand{\emotionstatedecaytype}{\mathbb{ES}_\lambda}
\newcommand{\emotiontype}{\mathbb{E}}
\newcommand{\emotionkindstype}{\mathbb{K}}

\newcommand{\emotiondecaytype}{\mathbb{I_{\lambda}}}

\newcommand{\goaltype}{\mathbb{G}}
\newcommand{\plantype}{\mathbb{P}}
\newcommand{\padpoint}{P_{\left(P,A,D\right)}}

\newcommand{\worldtype}{\mathbb{W}}
\newcommand{\worldstatetype}{\mathbb{S}}
\newcommand{\worldstatechangetype}{\mathbb{S}_{\Delta}}
\newcommand{\statedistancetype}{\mathbb{D}}
\newcommand{\statedistancechangetype}{\mathbb{D}_{\Delta}}

\newcommand{\socialattachmenttype}{\mathbb{SA}}

\newcommand{\attentiontype}{\mathbb{AT}_x}

\newcommand{\True}{\mathit{True}}
\newcommand{\False}{\mathit{False}}
\newcommand{\defEq}{\text{ } \mathlarger{\mathlarger\circeq} \text{ }}

\newcommand{\mFear}{\mathtt{Fear}}
\newcommand{\mAnger}{\mathtt{Anger}}
\newcommand{\mSadness}{\mathtt{Sadness}}
\newcommand{\mJoy}{\mathtt{Joy}}
\newcommand{\mInterest}{\mathtt{Interest}}
\newcommand{\mSurprise}{\mathtt{Surprise}}
\newcommand{\mDisgust}{\mathtt{Disgust}}
\newcommand{\mTrust}{\mathtt{Acceptance}}

%% file: definitions.tex
\title{Start Your EM(otion En)gine: Towards Computational Models of Emotion for
Improving the Believability of Video Game Non-Player Characters}
\halftitle{Start Your EM(otion En)gine}

\author{Geneva M. Smith}
\shortauthor{G. M. Smith} 

\dept{Computing \& Software} 

\field{Software Engineering} 

\prevdegreeone{M.A.Sc. (Software Engineering), \\McMaster University, Hamilton,
ON, Canada}
\prevdegreetwo{M.A.Sc. (Software Engineering)}

\submitdate{June 2023}
\copyrightyear{2023}

\principaladviser{Dr. Jacques Carette} 

%% file: layabstr.tex
\vspace*{\fill}
\begin{center}
    \begin{minipage}[l]{4.8in}
        \vspace*{-50mm}
        \prefacesection{Lay Abstract} 

        Video games can deeply engage players using characters that appear to
        have emotionally-driven behaviours. One way that developers encode and
        carry knowledge between projects is by creating development tools,
        allowing them to focus on how they use that knowledge and create new
        knowledge. 

        \vspace{\baselineskip}

        This work draws from software engineering to propose three methods for
        creating development tools for game characters ``with emotion'': a
        process for analyzing academic emotion literature so that the tool's
        functions are plausible with respect to real-life emotion; a process
        for translating academic emotion literature into mathematical notation;
        and a process for creating tests to evaluate these kinds of development
        tools using narrative characters. The development of an example tool for
        creating game characters ``with emotion'', \progname{}, demonstrates
        these methods and serves as an example of good development practices.
    \end{minipage}
\end{center}
\vspace*{\fill}

%% file: abstr.tex
\vspace*{\fill}
\begin{center}
    \begin{minipage}[l]{4.8in}
        \vspace*{-40mm}
        \prefacesection{Abstract} 

        Believable Non-Player Characters (NPCs) help motivate player
        engagement with narrative-driven games. An important aspect of
        believable characters is their contextually-relevant reactions to
        changing situations, which emotion often drives in humans. Therefore,
        giving NPCs ``emotion'' should enhance their believability. For
        adoption in industry, it is important to create processes for
        developing tools to build NPCs ``with emotion'' that fit with current
        development practices. 

        \vspace{\baselineskip}

        Psychological validity---the grounding in affective science---is a
        necessary quality for plausible emotion-driven NPC behaviours.
        Computational Models of Emotion (CMEs) are one solution because they use
        at least one affective theory/model in their design. However, CME
        development tends to be insufficiently documented such that its
        processes seem unsystematic and poorly defined. This makes it difficult
        to reuse a CME's components, extend or scale them, or compare it to
        other CMEs. 

        \vspace{\baselineskip}

        This work draws from software engineering to propose three methods for
        acknowledging and limiting subjectivity in CME development to improve
        their reusability, maintainability, and verifiability:
        \begin{itemize}[noitemsep]

            \item A systematic, document analysis-based methodology for
            choosing a CME's underlying affective theories/models using its
            high-level design goals and design scope, which critically
            influence a CME's functional requirements;

            \item An approach for transforming natural language descriptions of
            affective theories into a type-based formal model using an
            intermediate, second natural language description refining the
            original descriptions and showing where and what assumptions
            informed the formalization; and

            \item A literary character analysis-based methodology for
            developing acceptance test cases with known believable characters
            from professionally-crafted stories that do not rely on specific CME
            designs.

        \end{itemize}

        Development of \progname{}, a game development CME for generating NPC
        emotions, shows these methods in practice. 
    \end{minipage}
\end{center}
\vspace*{\fill}

%% file: dedic.tex
\vspace*{\fill}
\begin{center}
    \textsl{For Oma \\ Finally! A {\scriptsize non-medical} doctor in the
    family!}

    \begin{figure}[h]
        \centering
        \includegraphics[width=0.2\textwidth]{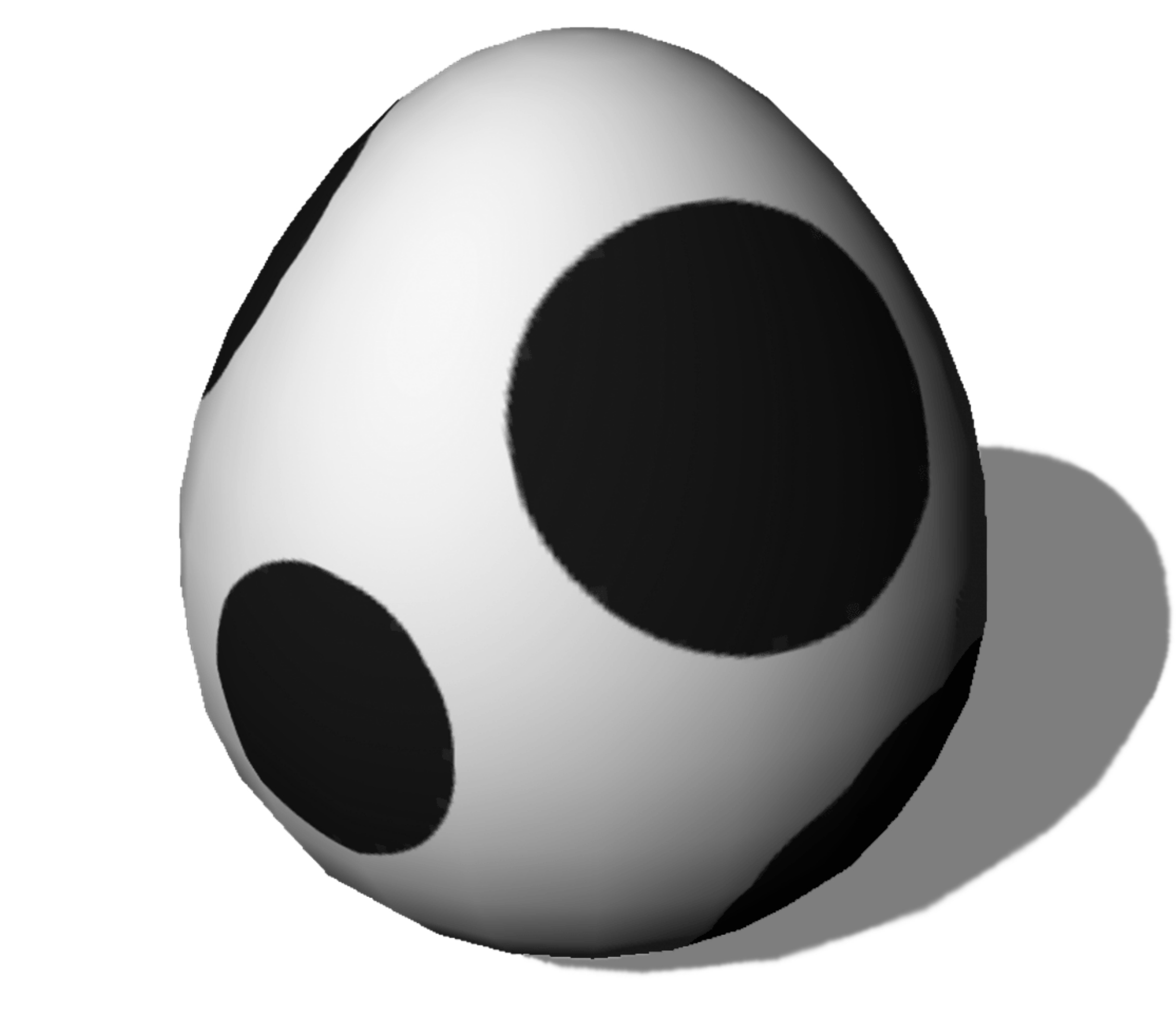}
        \caption*{\scriptsize Black Yoshi Egg \\\textcopyright{}1997 Nintendo}
    \end{figure}
\end{center}
\vspace*{\fill}

%% file: acknowledgements.tex
\vspace*{\fill}
\begin{center}
    \begin{minipage}[l]{\linewidth}
        \vspace*{-35mm}
\prefacesection{Acknowledgements}

Over the highest mountains and through the deepest dungeons, this thesis has
been a journey. It would not have been possible---or much fun---if I had to do
it alone. I have many thanks to bestow on my ragtag party of knights, mages,
archers, machinists, and foebreakers.

Thank you Dr. Carette for seeing this through to the end (and arriving
precisely when you meant to.). We had our ups and downs, but many hours and
several rewrites later \progname{} is something we can both be proud of. Thank
you Drs. Wassyng and Geiskkovitch for your feedback on horrible drafts and
half-baked ideas, and Dr. Smith for also reviewing some of the design documents
despite your busy schedule. Thank you Dr. Nacke for your suggestion to use
flowcharts and assuring me that I'm doing Ph.D. worthy work. Thank you Dr.
Prada for reviewing and suggesting ways to improvement this manuscript. Many,
many thanks to Leeanne Romane for being Librarian Extraordinaire, reminding me
how much fun research can be, and fielding my \textit{interesting} questions
about research questions. I do not know who you are, but if you were one of the
librarians responding to my constant stream of research material request: you
are amazing and I could not have done this without you. Thanks also to
Catherine Maybrey and Dr. Bandler for helping me find the lighthearted fun in
this work too.

Thank you to my guild of counsellors who were always supportive (even if you
had no idea what I was talking about). Thanks Mom for always refilling my
ration supply, making sure my bottomless teapot was indeed bottomless, and
making sure that I still know that the outside world exists. Thanks Dad for
always being ready with bear hugs, a canned goods supply, and reading my
published papers (even when you don't always understand them). Thank you Arnie
for helping me learn my guzz-in-tahs, keeping my ``spirits'' up, and
beta-reading my abstracts (I still think you only did half a bag of Nibs worth
of work though). Thank you Dr. M. Bear for helping me review drafts and always
being ready to lend a sympathetic ear. And so many thank yous to all the rest
of my born and adopted family from here to Northern Ontario to the East Coast
to Texas and everywhere betwixt and between. Your encouraging words have been
invaluable.

And of course, thank you to those adventurers who were down in the muck with
me, hacking our way through the bog. To George, Krystien, and Nicky for the
Zoom chats and escape rooms. To Omar, for cheerfully handling my panicked
emails about reviewer comments and revising manuscripts.
\begin{wrapfigure}{r}{0.25\linewidth}
    \centering
    \includegraphics[width=0.98\linewidth]{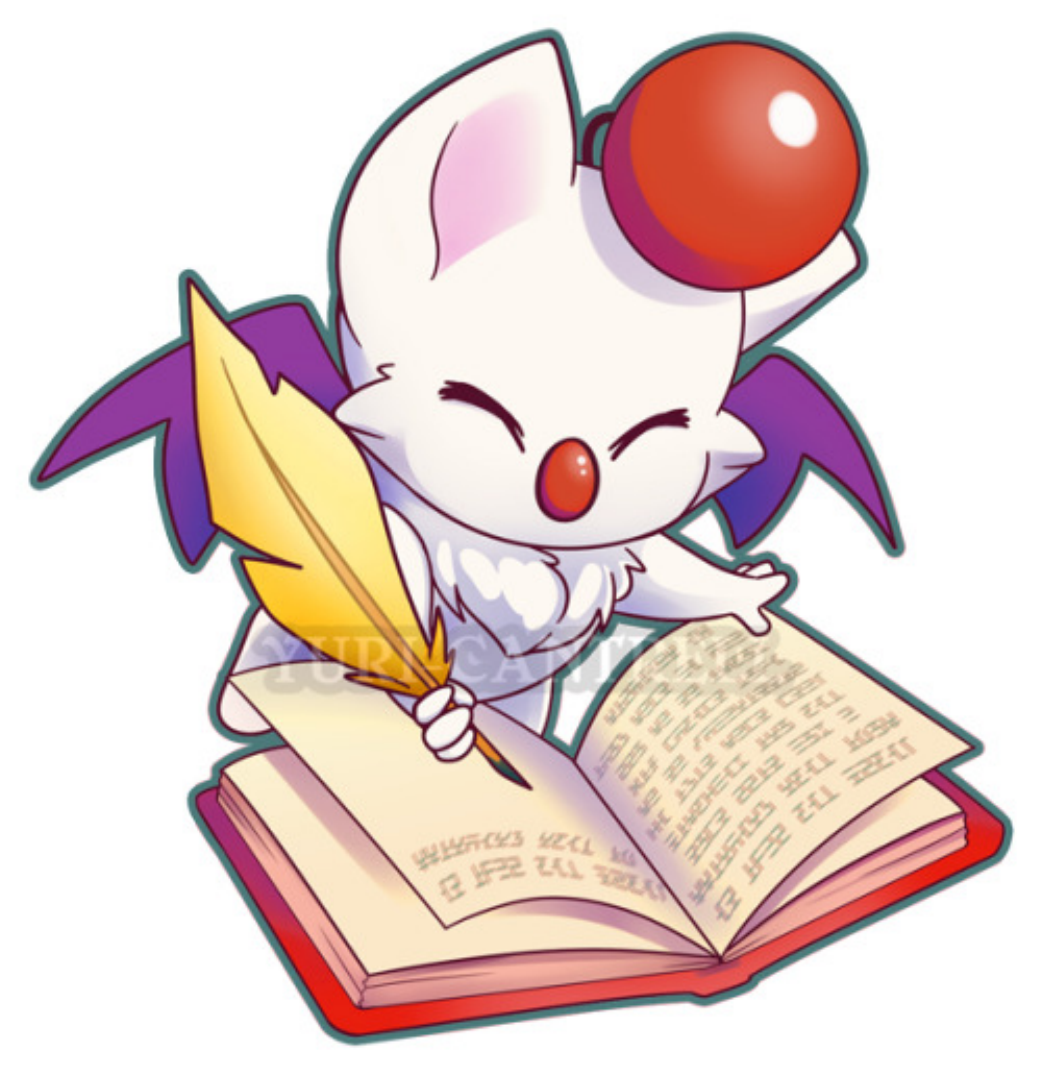}
    \caption*{\scriptsize Moogle \\\textcopyright{}1990 Square Enix}
\end{wrapfigure}
To Sam and Bea, who I don't talk to nearly enough but I am always so happy when
we do. To G-ScalE Classic: Adrian, Adele, Gabe, Noel, and Pav. Let's be
honest---despite what Shakari and Seanua thought, sometimes the only thing
keeping me going was the absolute chaos of our D \& D campaigns (I definitely
blame them for my dice hoard). Extra blame and thanks to Sasha for supplying
advice, encouragement, and many kinds of tea. To Neo G-ScalE, Xin and Vansh,
for being willing to tag along on our wild ideas; and many, many thanks to
Brendan for your enthusiasm, optimism, and showing me that supervising isn't so
bad. Thank you to Storm, Nami, Bella, Salem, and Xena for your antics,
headbutts, and many, MANY comments. Finally, thank you to Devin for coming
with me on this wild and unpredictable journey.

\end{minipage}
\end{center}
\vspace*{\fill}

%% file: notation.tex
\prefacesection{Notation, Definitions, and Abbreviations}

\section*{Notation}
\begin{description}[font=\rmfamily\bfseries, leftmargin=4.3cm, style=nextline]

    \item [$\mathbb{B}$] The set of Boolean values $\True$ and $\False$

    \item [$\neg x$] NOT x

    \item [$x \land y$] x AND y

    \item [$x \lor y$] x OR y

    \item [$\forall x$] Universal quantifier

    \item [$\exists x$] Existential quantifier

    \item [$\mathbb{N}$] The natural numbers, zero inclusive

    \item [$\mathbb{Z}$] The integers

    \item [$\mathbb{R}$] The reals

    \item [${\left[x,y\right]}$] A closed interval bounded by $x$ and $y$,
    inclusive

    \item [${\left(x,y\right)}$] An open interval bounded by $x$ and $y$,
    exclusive

    \item[$|x|$] Absolute value of $x$

    \item[$\lceil x \rceil$] Ceiling function

    \item[$\mathlarger\sum_{i=x}^{y} s_i$] Summation of $s_i$ from $i = x$ to
    $y$ inclusive

    \item[$x \cdot y$] Multiply $x$ and $y$

    \item [$P(B|A)$] Conditional probability of $B$ given $A$

    \item [$\left\{x_0, x_1, ..., x_n\right\}$] A set of $n$ elements

    \item [$\left\{X\right\}$] A set of elements of type $X$

    \item [$X \subset Y $] $X$ is a proper/strict subset of $Y$

    \item [$X \subseteq Y $] $X$ is a subset of $Y$

    \item [$A \times B$] The Cartesian product of $A$ and $B$

    \item [$\left(x_0, x_1, ..., x_n\right)$] A sequence of $n$ elements

    \item [$x : X$] A variable $x$ of type $X$

    \item [$x : X \rightarrow y : Y$] A function mapping $x :  X$ to $y : Y$

    \item [$x \oplus y$] Shorthand for $\mathtt{apply}(x, y)$, where
    $\mathtt{apply}$ is a function that changes $x$ by $y$

    \item [$X^?$] An Option/Maybe type that returns a value of type $X$ or
    Nothing/None

    \item [$\left< e_1, ..., e_n \right>$] An enumeration with $n$ labels

    \item [$\left\{ l_1 = v_1, ..., l_n = v_n \right\}$] A record with $n$
    fields each with a label $l$ and value $v$ such that no two sets of $l$
    and $v$ must have the same type

    \item [$\left\{ r \text{ \normalfont with } x = y \right\}$] Update a
    record $r$ that has the label $x$ with the value $y$

    \item [$x \defEq y $] $x$ ``is defined by'' $y$

\end{description}

\section*{Definitions}
\begin{description}[font=\rmfamily\bfseries, leftmargin=4.3cm, style=nextline]

    \item [Affective Computing\namedlabel{ac}{Affective Computing}] Computing
    that relates to, arises from, or deliberately influences emotion and other
    affective phenomena~\citep{oxfordAffectiveComputing}. It combines
    engineering and computer science with disciplines like psychology,
    cognitive science, neuroscience, sociology, linguistics, education,
    medicine, psychophysiology, value-centred design, and ethics.

    \item [Affective Science\namedlabel{as}{Affective Science}] An
    interdisciplinary field of study devoted to all aspects of affect and
    emotion~\citep{oxfordAffectiveSciences}. It draws from biology, psychology,
    economics, political science, law, psychiatry, neuroscience, education,
    sociology, ethology, literature, linguistics, history, and anthropology.

    \item [Antecedent\namedlabel{antecedent}{antecedent}] A real or imagined
    event or stimulus that an organism perceives as important to its physical,
    social, or personal well-being~\citep{oxfordAntecedents}. It often elicits,
    signals, or sets occasion for a particular behaviour or response.

    \item [Arousal\namedlabel{arousal}{Arousal}] A short-term state of
    excitement or energy expenditure~\citep{oxfordArousal}.

    \item [Autonomous Agent] Artificial entities that interact with their
    environment with a relatively high level of independence to make and
    execute their own decisions driven by their relation to and perception of
    their internal state and the external environment~\citep{oxfordAutonomous}.

    \item [Circumplex\namedlabel{circumplex}{Circumplex}] A circular depiction
    of the similarities between variables shown by their distance from each
    other on the circle~\citep{oxfordCircumplex}.

    \item [Coping\namedlabel{coping}{coping}] Cognitive and behavioural
    strategies to manage the demands of a taxing or stressful
    situation~\citep{oxfordCoping}.

    \item[Valence\namedlabel{valence}{Valence}] The anticipated satisfaction of
    goal attainment or event outcome~\citep{oxfordValence}.

\end{description}

\section*{Abbreviations}
\begin{description}[font=\rmfamily\bfseries, leftmargin=4.3cm, style=nextline]

    \item [AI] Artificial Intelligence

    \item [CME] Computational Model of Emotion

    \item [DDD] Document Driven Design

    \item [FPS] First Person Shooter

    \item [HCI] Human-Computer Interaction

    \item [IDE] Integrated Development Environment

    \item [MG] Module Guide

    \item [MIS] Module Interface Specification

    \item [NPC] Non-Player Character

    \item [OS] Operating System

    \item [PX] Player Experience

    \item [SRS] Software Requirements Specification

    \item [UX] User Experience

    \item [V-A] \ref{valence}-\ref{arousal} (as affective dimensions)

\end{description}

%% file: academicachievementdeclaration.tex
\begin{center}
    \begin{minipage}[l]{4.5in}
        \prefacesection{Declaration of Academic Achievement}

        I, Geneva M. Smith, declare this thesis to be my own work. I am the
        sole author of this document.

        \vspace{\baselineskip}

        I published parts of this work as journal articles in the \textit{IEEE
        Transactions on Affective Computing} and \textit{Eludamos: Journal for
        Computer Game Culture} and a presentation for the IDEA workshop at
        AAMAS 2023. Reviewers are considering parts of this work for
        publication as a journal article in \textit{Entertainment Computing}.
        Chapter~\ref{sec:RQs} has details about these publications.

        \vspace{\baselineskip}

        I reprinted published content in this thesis with permission. To the
        best of my knowledge, the content of this document does not infringe on
        anyone's copyright.

        \vspace{\baselineskip}

        Dr. Jacques Carette provided valuable support and guidance at all
        research stages where I was the sole contributor. While I was the
        primary contributor, Dr. Carette made significant contributions to:
        model development in Chapters~\ref{chapter:equations},
        \ref{sec:formalATC}, and \ref{chapter:testcaseEMgine}; module
        decomposition in Chapter~\ref{sec:moduleDecomp}; and reviews of
        \progname{}'s source code.
    \end{minipage}
\end{center}

%% file: introduction.tex
\chapter{Introduction}\label{chapter:intro}
\def\epigraphflush{center}
\setlength{\epigraphwidth}{0.75\textwidth}
\def\textflush{center}
\epigraph{Let's get this party started!}{Claptrap, \textit{Borderlands: The
Pre-Sequel}}

\begin{figure}[!b]
    \begin{center}
        \subfloat[An annoyed Fran subtley warns Vaan that he is asking an
        intrusive question in Square Enix's \textit{Final Fantasy XII: The
        Zodiac Age}~\citep{ffxii}]{
            \includegraphics[width=0.495\linewidth]{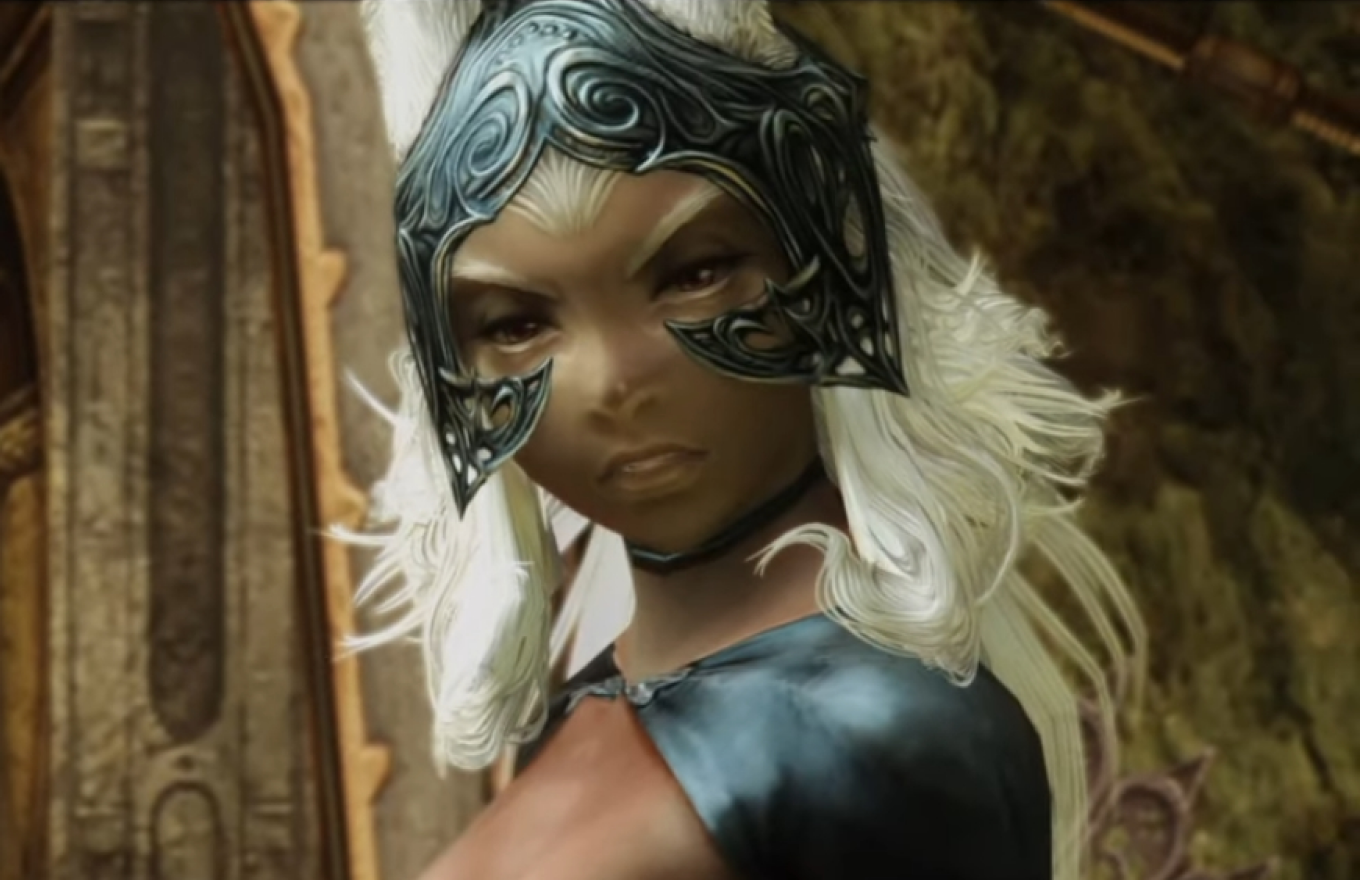}
        }
        \subfloat[Espeon and Deerling celebrate a reunion in Nintendo's
        \textit{New Pok\'emon Snap}~\citep{pokemonsnap}]{
            \includegraphics[width=0.45\linewidth]{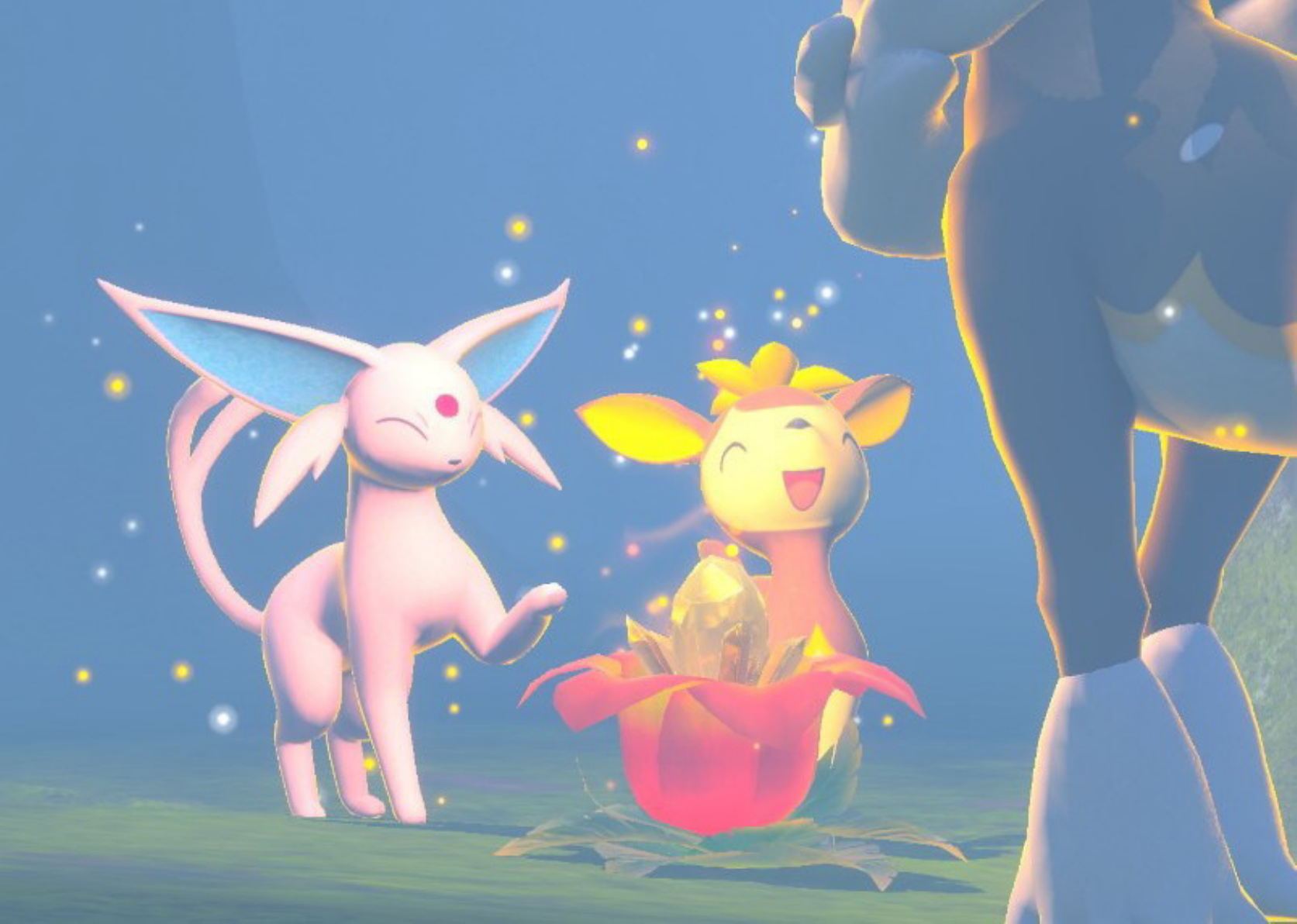}
        }

        \caption{Examples of Believable Game Characters}
        \label{fig:believablegamechars}
    \end{center}
\end{figure}

In both literature and film, richly layered characters make significant
narrative contributions and increase the audience's enjoyment and emotional
attachment to the story world's events and characters. However, video games
might be a more influential medium because the direct interaction with game
characters can make it feel ``more real'' to players
(\citepg{rusch2009mechanisms}{2}; \citepg{rusch2008emotional}{28}). The
increasing complexity of game narratives is a prominent and generally
well-received evolution in video game technology and
design~\citep[p.~117]{kuo2017from}. Many players fondly remember favourite game
characters and tend to talk about how they---the player---feel about them as if
they are real. This is due, in part, to the \textit{believability} of those
characters shown through their interactions with the world. One element of
believability is their emotional behaviours
(Figure~\ref{fig:believablegamechars}), which help players empathize with these
characters. This suggests that game developers can leverage character emotion
to increase the impact they have on players.

Development approaches in industry emphasize the fast-paced and iterative
nature of game development~\citep{mckenzie2019software}. This environment
requires reliable tools and processes which a systematic software engineering
approach can help create. The proprietary nature of the video game industry
means that \textit{postmortem} reports are often the sole source of information
on the development process~\citep[p.~553]{politowski2020dataset}. However, even
this limited information reveals that some problems relate to inadequate or
absent tools (\citepg{ullmann2022video}{13};
\citepg{politowski2020dataset}{556}) despite tools being a common form of code
reuse in game development~\citep[p.~4]{murphy2014cowboys}. The
\textit{postmortems} also reveal that familiar technologies, prototyping, and
testing-related factors contribute to successful
projects~\citep[p.~20--21]{ullmann2022what}. This suggests that methods for
creating game development tools aimed at ``emotional'' NPC creation are ideal
because it allows developers to build their own tools and evaluate other ones
to see if they are a good fit for the intended task.

\section{Research Questions and Contributions}\label{sec:RQs}
Prior research on believable computer agents focused on developing specific
tools such as GAMYGDALA~\citep{popescu2014gamygdala},
Em/Oz~\citep{reilly1996believable}, and The Soul~\citep{bidarra2010growing}.
Each tool has proven successful in their domains and beyond, but their
development process is unclear. Consequently, it is difficult for others to
build on them to create other tools to suit their needs. This is untenable in
game development where time is a scarce resource and building new tools from
scratch is expensive. Therefore, assuming that game developers aim to engage
players with ``emotional'' NPCs (Chapter~\ref{chapter:believable}) and prefer
to have tool creation \textit{methods} over any single tool itself, a research
question that follows is:
\begin{quote}
    \centering
    \textbf{RQ0} \textit{What software engineering-based methods and/or
    techniques can aid the creation of game development tools for believable
    characters with emotion?}
\end{quote}

\noindent An exploration of what ``emotion'' means in \ref{as} and psychology
(Chapter~\ref{chapter:primer}), then Computational Models of Emotion (CMEs) and
how developers apply software engineering practices to their development and
testing (Chapter~\ref{chapter:se-ee-design}, which
\citet{osuna2020seperspective} critically influenced) led to the creation of
\progname{}, a domain-specific CME for generating NPC emotions. Initially,
there was a naive assumption that \progname{}'s development would merely follow
the software design stages---requirements analysis, design, implementation,
verification, and validation. During development, questions concerning how to
choose theories and/or models for CMEs (Section~\ref{sec:rq1}), how to build
acceptance test cases to see if it produced expected emotions
(Section~\ref{sec:rq2}), and how \progname{} could be of practical use to the
CME development community (Section~\ref{sec:rq3}) quickly proved this
assumption false.

\subsection{Choosing Theories and/or Models for CMEs}\label{sec:rq1}
By definition, a CME takes at least one emotion theory and/or model to base its
design and testing on that aligns with its requirements. Often, considering
additional domain knowledge helps clarify user needs. This connection moves
theory selection into \textit{requirements analysis} because it influences
model specifications, ultimately defining what to build. For \progname{},
several theories that other CMEs use appeared promising. However, those CMEs
have only a partially satisfactory answer to \textit{why} they chose those
theories. It effectively made the choice seem arbitrary and lacking
justification. A shallow decision of theories for \progname{} based on what
``looked promising'' led to significant modelling challenges that ultimately
became untenable, raising the question:
\begin{quote}
    \centering
    \textbf{RQ1} \textit{How can user-oriented software requirements and domain
    knowledge inform the selection of emotion theories and/or models for CMEs
    built as game development tools?}
\end{quote}

\paragraph{Work \& Contributions} A systematic survey of 67 CMEs that generate
emotion and whose design rely on at least one emotion theory revealed that each
one tended to serve a particular purpose, implying that it strongly supports
certain kinds of CME requirements (Chapter~\ref{chapter:cmeOverview}). This
prompted the design of a document analysis-based methodology for choosing
emotion theories for CMEs using their high-level requirements and target
domain (Chapter~\ref{sec:reqProcess}). \progname{} served as a test to evaluate
the methodology's viability (Chapters~\ref{chapter:reqsAndScope} and
\ref{chapter:theoryAnalysis}) and a series of sketches about choosing theories
for other types of CMEs shows the methodology's potential
(Chapter~\ref{chapter:choosingExamples}).

\paragraph{Publications} The systematic survey of CMEs appears in the
\textit{IEEE Transactions on Affective Computing} as a survey
article (\href{https://doi.org/10.1109/TAFFC.2022.3197456}{What Lies
Beneath---A Survey of Affective Theory Use in Computational Models of Emotion},
\citet{smith2021what}). A description of the methodology with illustrative
examples is under review for publication as a research article in
\textit{Entertainment Computing}, currently available as a preprint
(\href{https://doi.org/10.2139/ssrn.4327741}{Start Your EMgine---A Methodology
for Choosing Emotion Theories for Computational Models of Emotion},
\citet{smith2023start}).

\subsection{Building Acceptance Test Cases for CMEs}\label{sec:rq2}
One of the challenges in CME design is translating natural language theories
into formal models for implementation. Often, testing with information defined
independently of how CMEs work is the only way to know if its models are
``correct''. However, simply using empirical observations of emotions is
unproductive because \textit{believable} is not necessarily \textit{realistic}.
Often, story characters have exaggerated behaviours that many would claim
unrealistic if they saw them in real life. If a CME is for creating believable
characters, the question becomes:
\begin{quote}
    \centering
    \textbf{RQ2} \textit{How can existing narratives inform the development of
    test cases for evaluating CMEs built as game development tools?}
\end{quote}

\paragraph{Work \& Contributions} This led to the development of a character
analysis-based methodology for developing acceptance test cases
(Chapter~\ref{sec:atcProcess}), illustrated with examples
(Chapters~\ref{chapter:testcasedefinition} and \ref{chapter:testcaseEMgine}).
\progname{} uses these as part of its validation test plan
(Section~\ref{sec:rq3}).

\paragraph{Publications} A presentation at the Interdisciplinary Design of
Emotion Sensitive Agents (IDEA) workshop at AAMAS 2023 described the
methodology with an illustrative example demonstrating the specification of a
test case template and extension for \progname{}
(\href{https://en.uit.no/project/idea/accepted_papers}{Building Test Cases for
Video Game-Focused Computational Models of Emotion}\footnote{Originally titled
``Inspect Your EMgine---Building Acceptance Test Cases from Narratives for
Entertainment-Focused Computational Models of Emotion'', shortened to stay in
the page limit.}, \citet{smith2023building}). Pre-\progname{} work on test case
development appears in a research article, which the workshop submission and
thesis expands on, in \textit{Eludamos: Journal for Computer Game Culture}
(\href{https://doi.org/10.7557/23.6175}{Design Foundations for Emotional Game
Characters}, \citet{smith2019design}).

\subsection{\progname{} as a Basis for Future Research and
Development}\label{sec:rq3}
Ultimately, \progname{} demonstrates the value of a software engineering
approach to the creation of game development tools. This includes embodying
software qualities such as reusability and replicability to encourage others to
use and expand on \progname{}'s design. Some qualities are especially desirable
for researchers, allowing them to independently verify \progname{}'s, and other
CMEs', abilities now and in the future~\citep{benureau2018rerun}. Moving
towards this goal means asking:
\begin{quote}
    \centering
    \textbf{RQ3} \textit{What steps can be taken during the development of
    domain-specific CMEs to improve their reusability and replicability?}
\end{quote}

\paragraph{Work \& Contributions} To this end, a significant effort went into
documenting \progname{}'s development process. Descriptions of \progname{}'s
underlying models trace the translation from: informal and natural language
concepts based on two affective theories (Plutchik's psycho\-/evolutionary
synthesis and Oatley \& Johnson-Laird's Communicative Theory of Emotions), one
affective model (Mehrabian's PAD Space), and other sources from \ref{as} and
existing CMEs; to a second natural language description refining the first with
\progname{}'s assumptions about the concept, using specific types of data; and
finally into a type-based formal model (Chapter~\ref{chapter:equations}).

\progname{}'s architecture design documents the rationale for choosing a
component-based style and the organization of models into modules with known
issues resulting from the decomposition
(Chapter~\ref{chapter:designImplement}). There is also documentation about
\progname{}'s implementation, including why it uses C\# and the development
environment configuration.

\paragraph{Publications} Open-source versions of \progname{}'s documentation,
implementation, and test documents are available on GitHub
(\href{https://github.com/GenevaS/EMgine}{https://github.com/GenevaS/EMgine}).
There are currently no peer-reviewed publications of this work.

\section{A Word About the Cultural Dependence of the Language of
Emotion}\label{sec:language}
\progname{}'s design heavily relies on emotion terms from an English-speaking,
North American lexicon. It is unwise to proceed without acknowledging the role
of language and culture in our understanding of emotions. Some emotion theories
refer to emotions that people mainly recognize in English, which can cause
issues when using \progname{} in other cultures and languages~\citep[p.~8,
10]{ortony2021all}. Even among English-speaking theorists, there is little
consistency between definitions of ``emotion'' in the
literature~\citep[p.~80]{robert1980emotion}. Some languages lack an equivalent
term for ``emotion''~\citep[p.~3]{wierzbicka1999emotions}. How is it possible
to model something that appears to be fundamental to the human experience but
does not exist in everyone's vocabulary?

Based on the lexical sedimentation hypothesis, researchers propose that
everyday languages have captured useful information about emotions based on the
importance of emotion in human social interactions and
literature~\citep[p.~8]{scherer2013measuring}. Evidence strongly suggests that
there are, albeit limited, dimensional representations and categorical clusters
of emotion terms common across different languages and cultures~\citep[p.~37,
40, 43]{fontaine2013dimensional}. They are not incompatible, as researchers
have replicated the clusters in dimensional space and found that they remain
systematically differentiated.

However, the specific \textit{word} someone uses to describe their internal
state does not exist reliably across languages---often with no direct
equivalent~\citep[p.~62]{ogarkova2013folk}. A word is only a representation of
an emotion and ``...surely we should not be so narrow as to insist that it has
to be an English word!''~\citep[p.~199]{ekman2007emotions}. An emotion term
might only be an abbreviation for common scenarios that are noteworthy to
members of the cultural group~\citep[p.~548]{wierzbicka1992defining}. This
aligns with a suggestion that self-reports about emotional experiences reveal
affective properties rather than emotion
categories~\citep[p.~37]{barrett2006emotions}. In this sense, emotions are
social constructions---the specific scenarios and conditions for them---based
on a common need for cognitive management~\citep[p.~119]{oatley1992best} that
often relates to observable body ``symptoms'' of those states. Therefore, it is
imperative that descriptions of \progname{}'s generated emotions use terms such
as ``feel'', ``good'', and ``bad''~\citep[p.~275]{wierzbicka1999emotions} so
that localizing it for different languages and cultures will be easier.

\parasep

%% file: believablecharacters.tex
\chapter{Engaging Players with Believable
Characters}\label{chapter:believable}
\def\epigraphflush{center}
\setlength{\epigraphwidth}{0.75\textwidth}
\def\textflush{center}
\epigraph{I hope the motivation is effective.}{EDI, \textit{Mass Effect 3}}

Like ``good'' books and ``good'' movies, specifying what makes a ``good'' game
is not straightforward. Instead, game designers often aim to create a good
player experience (\citepg{mcallister2015video}{11, 13}). Player experience
(PX) is a complex concept with many moving parts that developers aim to
understand and design by adapting user experience (UX) concepts and methods
(\citepg{bernhaupt2015user}{2, 7}; \citepg{scacchi2015research}{4}). An
important UX concept for games is \textit{engagement}: a quality of the user
experience describing a positive human-computer
interaction~\citep[p.~1094]{o2013examining}\footnote{It differs from related
concepts like flow and immersion, although people do use the terms
interchangeably (\citepg{brockmyer2009development}{624};
\citepg{turner2014figure}{33}; \citepg{glas2015definitions}{944};
\citepg{cairns2016engagement}{81}; \citepg{doherty2018engagement}{99:4}).}.

Player research suggests that an engaging \textit{narrative} is essential to
the gaming experience in non-linear sandbox and multiplayer games in a wide
variety of genres such as action, fighting, role-playing, shooter, simulator,
and survival~\citep[p.~67--68]{carvalho2020framework}---suggesting that most
video games have some form of narrative~\citep[p.~110]{qin2009measuring}. Even
in games that do not traditionally have a strong or complete narrative, like
First Person Shooters (FPSs), the inclusion of one made players more
physiologically aroused, feel more involved in the game, and liked their
experience more than one without a story (\citepg{schneider2004death}{370};
\citepg{kuo2017from}{107}).

Computer-controlled characters---the \textit{Non-Player Characters}
(NPCs)---often drive game narratives and populate the game world, filling
important mechanical and narrative roles (\citepg{jorgensen2010game}{315};
\citepg{lee2015cognitive}{47--48}; \citepg{phan2016development}{1231};
\citepg{warpefelt2017model}{40}; \citepg{harth2017empathy}{2};
\citepg{emmerich2018m}{142}). As with narrative, players have stated that NPCs
help them connect to the game world~\citep[p.~67]{carvalho2020framework} and
can help them identify with their own
character~\citep[p.~278]{rogers2018exploring}. However, creating NPCs who react
``correctly'' is challenging because games with a lot of player agency have a
lot of unpredictable situations (\citepg{reilly1996believable}{2};
\citepg{loyall1997believable}{2}; \citepg{gebhard2003adding}{48};
\citepg{ochs2009simulation}{281}; \citepg{harth2017empathy}{4};
\citepg{bidarra2010growing}{337}; \citepg{carbone2020radically}{465}). The
subjective nature of what \textit{players} see as \textit{believable}
(\citepg{livingstone2006turing}{4}; \citepg{lee2015cognitive}{55};
\citepg{warpefelt2017model}{42}) further exacerbates the problem. Inconsistent
NPC behaviours ``...remind the player that it is just a
game''~\citep[p.~321]{sweetser2004player} or that the NPC is ``broken''
\citep[p.~70]{carvalho2020framework}. Consistent---believable---NPCs on the
other hand help reinforce the game narrative's
believability~\citep[p.~328]{yannakakis2014panorama} and often results in a
higher emotional investment from the player
(\citepg{yannakakis2015emotion}{465}; \citepg{lankes2015evaluating}{116};
\citepg{emmerich2018m}{150}). This can ``...evoke social effects similar to
human co-players''~\cite[p.~143]{emmerich2018m}. In many cases, this is ideal
because it promotes continued interactions with the game.

\section{Games and Player Engagement}
Engaging players is a fundamental goal of games, having a key role in player
satisfaction---``the degree to which the player feels gratified with his or her
experience while playing a video game''~\citep[p.~1220]{phan2016development}.
Players have a disposition towards being, and expect to be, engaged when they
play~\citep[p.~84]{cairns2016engagement}. This could be because playing games
is a voluntary activity done for pleasure (\citepg{poels2007always}{86--87};
\citepg{yannakakis2015emotion}{459}) in which they are an active
participant~\citep[p.~94]{mayra2007fundamental}. The inherent disposition
players have towards engagement shifts the game designer's focus from
\textit{why} players engage with a particular game to \textit{how} they become
engaged~\citep[p.~85]{cairns2016engagement}.

\citet[p.~406--407]{schonau2012sure} propose six general causes of player
engagement: intellectual, physical, sensory, social, narrative, and
emotional. Emotional engagement emerges from the player's personal feelings
aroused by an in-game event, character, asset attributes, or another player
which causes them to want to continue playing. Many have argued that to deepen
engagement, a game \textit{must} affect the player's emotions, both positive
and negative (\citepg{brown2004grounded}{1299}; \citeg{emotioneering};
\citepg{jennett2008measuring}{657}; \citepg{hudlicka2008affective}{5};
\citepg{yannakakis2015emotion}{465}; \citepg{bernhaupt2015user}{3};
\citepg{takatalo2015understanding}{89}; \citepg{de2015conceptual}{3};
\citepg{lankes2015evaluating}{116}; \citepg{zhang2017spatial}{5}). This
suggests that emotional engagement could be the most potent type of engagement.
Players themselves have said that a game must elicit an emotional response from
them for it to deeply engage them~\citep[p.~323]{sweetser2004player} and are
generally open to and actively seek emotional experiences when playing games
(\citepg{yannakakis2015emotion}{460}). Researchers have found games well-suited
for emotion-related studies~\citep[p.~290]{scherer2021towards} further
supporting their ability to elicit player emotions. As well as the control that
emotion has over one's actions and
decision-making~\citep[p.~38]{turner2014figure}, this need for emotional
engagement could be due to the personal value attached to an emotional
experience which can be a powerful motivation to play
(\citepg{ryan2006motivational}{353};
\citepg{takatalo2015understanding}{98--99}).

Emotion-invoking game content can create more engaging experiences that players
perceive as more realistic, natural, and believable than their non-emotional
counterparts~\citep[p.~739]{banos2004immersion}. It also does not depend on
interaction medium \citep[p.~653]{aymerich2010presence} or technology,
following the observation that interacting with, forming attachments to, and
empathizing with game elements elicit player emotions
(\citepg{yannakakis2015emotion}{461}). For example, a puzzle game can
emotionally engage a player with its level of challenge whereas a role-playing
game can emotionally engage a player with its
narrative~\citep[p.~407]{schonau2012sure}. This indicates that it is possible
to engage a player emotionally in any kind of game because it depends less on
what the game is and more on the inclusion of well-crafted game content.

\section{Engaging Player Emotions with Narratives and Characters}
Narratives play a significant role in human cognition and
affect~\citep[p.~362]{schneider2004death}---one does not have to look far to
see the prevalence of stories in human culture. It follows that creating a game
narrative can be an effective way to elicit player emotions, leading to their
emotional engagement with the game (\citepg{qin2009measuring}{128};
\citepg{chilukuri2011algorithm}{292}; \citepg{adams2014fundamentals}{183};
\citepg{yannakakis2015emotion}{462}; \citepg{takatalo2015understanding}{89}).
Narratives for games are also not limited to ``happy'' stories, since players
willingly extend their interactions with uncomfortable game narratives if there
is something they want to do~\citep[p.~228--229]{schoenau2011hooked}. The
interactive nature of games has the potential to engage players more deeply in
a narrative than other mediums because games can give players an active role in
the unfolding story (\citepg{o2008user}{946}; \citepg{qin2009measuring}{111};
\citepg{takatalo2015understanding}{89}; \citepg{kuo2017from}{107--108, 110};
\citepg{carvalho2020framework}{70}). This helps players establish who their
character is (\citepg{sweetser2004player}{323};
\citepg{schneider2004death}{371}; \citepg{calvillo2015assessing}{46}) and make
their role personal (\citepg{ng2014review}{80};
\citepg{takatalo2015understanding}{98}).

An NPC can have a significant impact on a player's emotional investment and
engagement if the player believes that they have an impact on the
NPC~\citep[p.~736]{hall2005achieving}, sometimes to the point where the
player's attachment to them influences their in-game actions
(\citepg{harth2017empathy}{16}; \citepg{bopp2019exploring}{319}). In general,
players become attached to NPCs that they feel responsible for and do not see
as a burden, share personal experiences with, or view as a friend in that they
are loyal, caring, and accommodating~\citep[p.~319]{bopp2019exploring}.
A player's ability to empathize with an NPC nurtures their attachment to them
(\citepg{paiva2005learning}{244}; \citepg{mayra2007fundamental}{101--102};
\citepg{adams2014fundamentals}{157}; \citepg{phan2016development}{1231};
\citepg{broekens2021emotion}{356--357}) as they build their relationship via
interactions over time (\citepg{yannakakis2015emotion}{462};
\citepg{harth2017empathy}{13}). A player might also become emotionally attached
to NPCs that they must work with to complete
tasks~\citep[p.~13]{de2015conceptual}. This appears to be consistent with
strategies to build prosocial behaviours~\citep[p.~85--86]{roseman2001model}.
Industry recognizes the power of this connection between player and NPC and
have been designing their games to encourage this ``character
experience''~\citep{walktall}. Prerequisites of this experience include their
\textit{functionality} in the game and their \textit{believability} so that
player can become comfortable with the NPCs.

\section{What Makes a Believable Character?}\label{sec:believable}
A believable character, central in artistic mediums like literature and film,
``...allows the audience to suspend their disbelief and...provides a convincing
portrayal of the personality they expect or come to expect [from the
character]''~\citep[p.~1]{loyall1997believable}. Disney animators have
described how they give this ``illusion of life'' to their characters to ensure
that their actions are understood by the audience. This includes building a
conceptual model of their internal processes and state---even when no such
processes are taking place~\citep{thomas1981illusion}. Believability is not
limited to ``smart'' or ``normal'' characters because it depends on the
situational context and the character's personality
(\citepg{reilly1996believable}{10--12}; \citepg{loyall1997believable}{3--4};
\citepg{lisetti2015and}{95}). What ``believable'' means also depends on the
application domain---the expectations in entertainment differ from those in
soft skills training~\citep{ortony2002making}. In short: for an NPC to be
believable, it must behave reasonably within the context of its game world.
Generally, NPCs are believable when they (\citepg{loyall1997believable}{15--26};
\citepg{lankoski2007gameplay}{417}; \citepg{warpefelt2013analyzing}{10}):
\clearpage\begin{itemize}
    \item Appear to be self-motivated,

    \item Are aware of what is happening around them, and

    \item React in ways appropriate for their surrounding context while
    adhering to their personality.
\end{itemize}

A character's \textit{emotion} is one element that makes them believable
(\citepg{loyall1997believable}{19}; \citeauthor{gard2000building},
\citeyear{gard2000building}; \citepg{paiva2005learning}{237};
\citepg{lankoski2007gameplay}{417}; \citepg{warpefelt2013analyzing}{4};
\citepg{de2015beyond}{116}; \citepg{lisetti2015and}{95};
\citepg{emmerich2018m}{145}). Characters with emotion address the core features
of believability because they convey a character's goals and desires
(\textit{self-motivated}) by showing their \textit{awareness} of,
\textit{responsiveness} to, and care (\textit{personality}-driven) for their
surroundings (\citepg{bates1994}{124}; \citepg{reilly1996believable}{12};
\citepg{broekens2021emotion}{356}). It follows that one way to improve an NPC's
believability is to have them react emotionally to their
surroundings.

Emotional behaviours are not necessary for all NPC types, but their importance
does increase as their context becomes more complicated and their narrative
importance grows (\citepg{warpefelt2017model}{49};
\citepg{emmerich2018m}{143}). For example, players expect to have a stronger
emotional and social bond with their constant NPC companion than they do an
unnamed merchant~\citep[p.~229]{isbister2006nonplayer}. If they are necessary
for a game, the game's interactivity makes emotional NPCs difficult to realize
because it is impossible to plan the NPCs' behaviours for every potential game
scenario. Instead, one can generate emotions and/or emotion-driven behaviours
as the NPCs' surroundings change.

\section{Summary}
A fundamental goal of game design is to create good player experiences,
regardless of the game's scope or genre, where player engagement is one
element. Playing games is a voluntary activity, predisposing players to
engagement. However, players do not automatically become engaged with every
game that they play---the question is not \textit{why}, but \textit{how} to
engage players. Challenges, physical movements, sensory aspects, social
interactions with other players, and narratives all have the potential to
engage players. While personal preferences influence what a player finds
engaging, it is a common sentiment that the game must engage them emotionally
to have a lasting impact. A game narrative's NPCs can have a significant impact
on the player's emotional investment and engagement if they affect the player's
decisions and actions. This requires the NPCs to be believable, convincing the
player of their ``realness''.

Believable characters convince the player of their personality and display the
``illusion of life''. An NPC's emotions are a key factor in this,
communicating that they are self-motivated, self-aware, and care about what
happens around and to them. Ultimately, deciding if, when, and how to use
emotional NPCs is left to the designer.

\clearpage
\vspace*{\fill}
\begin{keypoints}
    \begin{itemize}
        \item The goal of game design is to create good player experiences
        \item Engagement is an aspect of the player experience and players are
        predisposed to being engaged
        \item Evidence suggests that a game must emotionally engage a player to
        have a long-lasting effect on them
        \item Game narratives and their characters can be an effective way to
        emotionally engage a player
        \item The Non-Player Characters (NPCs) of a game's narrative can
        significantly impact a player's emotional engagement if they are
        believable
        \item Believable characters with emotion can show the player that they
        are self-motivated, self-aware, and care about what happens around and
        to them
        \item It is the game designer's decision to determine if their game
        needs a narrative, any characters, and if their NPCs require emotional
        behaviour
    \end{itemize}
\end{keypoints}

\parasep
\vspace*{\fill}

%% file: psychprimer.tex
\chapter{Meet Emotion (Briefly)}\label{chapter:primer}
\def\epigraphflush{center}
\setlength{\epigraphwidth}{0.75\textwidth}
\def\textflush{center}
\epigraph{I wish you'd just tell me rather than trying to engage my enthusiasm
because I haven't got one.}{Marvin the Paranoid Android, \textit{The
Hitchhiker's Guide to the Galaxy}}

Before giving emotion to Non-Player Characters (NPCs), it is prudent to
understand what that means. This is not as simple as it seems. Like immersion,
no one truly knows what emotions are~\citep[p.~9]{ortony2021all}. Are they
explanations for why people behave in certain ways or are they classification
schemes that people impose on their perception of the
world~\citep[p.~46]{barrett2006emotions}? There might not even be a ``real''
meaning to the term ``emotion''~\citep[p.~75]{dorner2003mathematics}. People
typically find it difficult to descriptively articulate
them~\citep[p.~53]{clore2000cognition}, likely because it is not always clear
what an emotion state is~\citep[p.~9]{ortony2021all}. It might not even refer
to a uniform entity~\citep[p.~22]{hudlicka2014habits}. People often use it to
reference a broad range of mental states (\citepg{de2015conceptual}{2};
\citepg{sloman2005architectural}{208}), and have trouble distinguishing between
bodily sensations, cognitive states, and affective states
(\citepg{feldman1995variations}{815}; \citepg{oatley1992best}{75}). This is not
an issue for daily use but it is for scientific
study~\citep[p.~9]{ortony2021all}. ``Emotion'' might be best described as a
fuzzy set of definitions (\citepg{russell1980circumplex}{1165};
\citepg{sloman2005architectural}{209, 211}), which inevitably leads to
borderline cases defying classification~\citep[p.~611]{smith1990emotion}.
This fundamental, unanswered question on the nature of emotion makes specifying
a computational model difficult. To begin, it is useful to know what emotion
is, how it differs from other types of affect, and what its potential functions
are~\citep[p.~10]{scherer2010emotion}. This helps clarify what to
model~\citep[p.~297]{hudlicka2014computational}, directing an affective
theories analysis to determine which ones best fit a CME's requirements
(\citepg{lisetti2015and}{99}; \citepg{osuna2020seperspective}{4}).

\section[The Form of Emotion]{The Form of
Emotion\footnote{\normalfont{\footnotesize\textcopyright{}} 2022 IEEE.
Reprinted with permission from \citeauthor{smith2021what}
(\citeyear{smith2021what}, p.~1793).}}\label{sec:affectiveDefs}
While there is no agreed-on, precise definition of emotion, researchers agree
on a fuzzy working definition and typical examples of emotions, which are
sufficient for meaningful comparisons between theories without favouring any
single one~\citep[p.~248--249]{reisenzein2013computational}. An
\textbf{emotion} is a short-term affective state representing the coordinated
physiological and behavioural response of the brain and body to events that an
organism perceives as relevant (\citepg{jeon2017emotions}{4};
\citepg{frijda1986emotions}{249}; \citepg{scherer2000psychological}{138--139};
\citepg{broekens2021emotion}{349}; \citepg{smith2001toward}{121}).

Some researchers hypothesize that each emotion has a \textit{signature}---a
coordinated response pattern it typically causes in an organism
(\citepg{hudlicka2019modeling}{133}; \citepg{scherer2001appraisalB}{108})
including: behavioural and expressional characteristics; somatic and
neurophysiological factors that prepare the body for action; cognitive and
interpretive evaluations that give rise to the emotion; and experiential and
subjective qualities unique to the individual. Elements of the signature can be
innate or learned~\cite[p.~6]{carlson1992psychology}. Emotions are also
characterized by their: high intensity relative to other types of affect (e.g.
personality, mood); tendency to come and go quickly; association with a
specific triggering event, object, or person; and clear cognitive contents
(\citepg{jeon2017emotions}{4}; \citepg{scherer2000psychological}{139--140};
\citepg{broekens2021emotion}{350}). These attributes distinguish emotion from
other types of affect. However, there is general agreement that each affect type
interacts with emotion, influencing individual experiences:
\begin{itemize}
    \item \textbf{``Affect''} is a general term for any body-linked state that
    influences the mind~\citep{oxfordAffect}. A general affective state is
    typically weaker than an emotion state~\citep[p.~5]{jeon2017emotions}. Since
    this is a nebulous concept, the scope is on the more specific
    \textbf{``core affect''} (\citepg{vastfjall2002measurement}{20, 27};
    \citepg{russell2009emotion}{1264--1266}; \citepg{scarantino2009core}{948}),
    defined as ``...a state of pleasure or displeasure with some degree of
    \ref{arousal}''~\citep[p.~170]{barrett2009affect}.

    \item \textbf{Feelings} are conscious mental representations and
    interpretations of an emotional response that follow emotions
    evolutionarily and experientially (\citepg{jeon2017emotions}{4};
    \citepg{oxfordFeelings}{184}; \citepg{scherer2000psychological}{139}), but
    are not critical for understanding emotional
    behaviours~\citep[p.~40]{fellous2004human} as they are personal reflections
    on affective states~\citep[p.~251--252]{frijda1986emotions}. Feelings are
    ill-defined from a modelling
    perspective~\citep[p.~133]{hudlicka2019modeling}, and rarely appear
    computationally.

    \item \textbf{Moods} are enduring, less intense, and more diffuse states
    than emotions (\citepg{oxfordMood}{258}; \citepg{jeon2017emotions}{4};
    \citepg{hudlicka2019modeling}{133}; \citepg{scherer2000psychological}{140};
    \citepg{broekens2021emotion}{351}) with no focused
    object~\citep[p.~54]{clore2000cognition}. Their presence is typically
    unclear to the experiencing individual and often have a more prolonged
    influence on an individual's cognition and behaviours.

    \item \textbf{Attitudes, Opinions, Sentiments, and Relations} are enduring
    emotional dispositions towards objects and people, formed over repeated
    exposures and appraisals of the same
    stimulus~\citep[p.~351]{broekens2021emotion}. These help structure an
    individual's relationships, which can influence their knowledge of and
    plans concerning that stimulus~\citep[p.~81]{oatley2000sentiments}.

    \item \textbf{Personality} is a set of stable affective traits
    (\citepg{oxfordPersonality}{304}; \citepg{hudlicka2014computational}{300};
    \citepg{jeon2017emotions}{5}; \citepg{scherer2000psychological}{141},
    \citepg{broekens2021emotion}{351}) that influence affective processes.
\end{itemize}

\section{The Function of Emotion}
Historically, people viewed emotionality and rationality as mutually exclusive
concepts (\citepg{damasio2002second}{12}; \citepg{desousa1987rationality}{1}).
Psychologists have now come to view affect as an integral element in a healthy
cognitive system that developed evolutionarily, likely a result of co-evolution
with perceptual, cognitive, and motor
abilities~\citep[p.~40]{fellous2004human}, whose purpose might be to improve
adaptive action beyond what information can achieve alone, uniting the
informational, attentional, and motivational effects of emotion. Emotion
provides essential functions including homeostasis regulation, reproductive
and survival behaviours, and adaptive behaviours in complex and uncertain
environments~\citep[p.~301]{hudlicka2014computational}. Affect and emotion are
now assumed to have many potential roles, including:
\begin{itemize}
    \item Rapid resource mobilization and allocation
    (\citepg{izard1977human}{108}; \citepg{hudlicka2014computational}{303})

    \item Goal management~\citep[p.~133]{hudlicka2019modeling}

    \item Decision-making via goal-directed processes
    (\citepg{fridja1994emotions}{118}; \citepg{lewis2005getting}{215};
    \citepg{reisenzein2013computational}{251}), likely directed by the
    underlying appraisal dimensions~\citep[p.~485]{lerner2000beyond}

    \item Influencing judgments and risk assessments
    (\citepg{izard1977human}{109}; \citepg{lerner2000beyond}{485};
    \citepg{storbeck2007interdependence}{1226--1227};
    \citepg{hudlicka2019modeling}{133})

    \item Attention (\citepg{izard1977human}{108};
    \citepg{lewis2005getting}{215--216};
    \citepg{storbeck2007interdependence}{1226};
    \citepg{hudlicka2019modeling}{133})

    \item Memory (\citepg{storbeck2007interdependence}{1226};
    \citepg{hudlicka2019modeling}{133})

    \item Learning and information acquisition
    (\citepg{hudlicka2019modeling}{133})

    \item Multi-level interpersonal communication and regulation that is simple
    but highly impactful~\citep[p.~41]{fellous2004human}
\end{itemize}

Clearly affect and emotion are part of a functioning system. But what about
emotions that do not make sense? Emotions indicate dispositions and
sensitivities to certain events and a process of relevance signalling for
deliberative actions. In this view, dysfunctional responses might indicate a
functional system~ that has overtaxed
resources~\citep[p.~121]{fridja1994emotions}, is unusually sensitive, has
ineffective \ref{coping} strategies, and/or conflicting responses which results
in undesirable and non-functional side effects such as a reduced resource pool
for the duration of the emotional
episode~\citep[p.~251--252]{reisenzein2013computational}. Mood could also cause
dysfunctional responses by amplifying a low intensity
emotion~\citep[p.~28]{siemer2007appraisals}, the intensity of the current
emotion which could impact unrelated memories and
processes~\citep[p.~476--477]{lerner2000beyond}, or to maladaptive patterns in
the emotion-cognition-action patterns (\citepg{izard2000motivational}{254};
\citepg{clore2000cognition}{49}). Distinguishing between the concept of emotion
and its individual elements means that assuming the adaptive function of
emotions does not imply that all emotions serve that function.

\section{Summary}
Although there is no agreement on what emotions even are, a working definition
and the collection of agreed-on examples is sufficient for meaningful affective
research. This should also be sufficient for creating believable NPCs with
emotions. Emotion differs from other types of affect in its duration,
intensity, and focus:
\begin{itemize}
    \item Emotion is shorter than moods, attitudes, and personality but longer
    than affect

    \item Emotion is more intense than affect, moods, and attitudes, whereas
    personality is typically not associated with intensity in this way

    \item Emotion relates to a specific event, person, or object at a given
    point in time, whereas the associations of attitudes develop over time, and
    affect and mood have no focus at all

    \item Feelings are reflections on these states, implying that mechanisms
    outside the affective system trigger them
\end{itemize}

Emotions also affect different aspects of behaviour, including goal management,
memory functionality, and attention. Knowing this makes it clearer what a
Computational Model of Emotion (CME) might need in its design.

\clearpage
\vspace*{\fill}
\begin{keypoints}
    \begin{itemize}

        \item An emotion is a short-term affective state representing the
        physiological and behavioural response of the brain and body to
        perceived opportunities and threats

        \item Emotion is part of a healthy cognitive system, likely developed
        during evolution to serve adaptive responses

        \item Emotions are a feature of an adaptive system, but some emotions
        can be maladaptive

    \end{itemize}
\end{keypoints}

\parasep
\vspace*{\fill}

%% file: SEforEEs.tex
\chapter{On Designing Emotion Engines}\label{chapter:se-ee-design}
\def\epigraphflush{center}
\setlength{\epigraphwidth}{0.75\textwidth}
\def\textflush{center}
\epigraph{As you see, there's no escape and resistance is futile!}{Mingy Jongo,
\textit{Banjo-Tooie}}

A crucial ingredient in the success of human-like or believable behaviours in
Non-Player Characters (NPCs) is their \textit{plausibility}, meaning that they
make sense to players~\citep[p.~216--217]{broekens2016emotional}. This is
directly influenced by the \textit{psychological validity} of those behaviours.
This means that it must ground itself in \ref{as}, the interdisciplinary study
of affective phenomena, related processes, and its influencing
factors~\cite[p.~xiii]{davidson2003handbook}. Many systems for creating
emotional game characters acknowledge this, building on existing psychological
theories~\citep[p.~462]{yannakakis2015emotion}.

Just as believability does not equal intelligence, plausibility does not
necessarily mean ``normal'' or appropriate
behaviours~\citep[p.~217]{broekens2016emotional}---it also encompasses
exaggerated or intentionally broken ones. \ref{as} supports this because it
includes models of undesirable and abnormal behaviours. This provides
opportunities to design unbalanced or mentally ill characters as a game's
design requires. Basing a computational design in affective science also helps
identify relevant empirical data and model validation methods, enable
communication with other researchers, and increase the design's reusability
potential~\citep[p.~305]{hudlicka2014computational}. This has resulted in a
class of software systems, called Computational Models of Emotion (CMEs), that
are influenced by affective science (Section~\ref{sec:CMEs}). An exploration of
a typical CME design process reveals unique development steps
(Section~\ref{sec:se}) and any applied software engineering practices.

\section[Computational Models of Emotion]{Computational Models of
Emotion\footnote{\normalfont Content up to and including the definition of CME
{\footnotesize\textcopyright{}} 2022 IEEE. Reprinted with permission from
\citet[p.~1793]{smith2021what}.}}\label{sec:CMEs}
\textit{\ref{ac}} introduces emotion as a concern in programs so that they may
recognize and respond to human users more intelligently~\citep[p.~3,
50]{picard1997affective}. There are three main affective computing tasks
(\citepg{scherer2010emotion}{4}; \citepg{fathalla2020emotional}{2}) that enable
this human-centred approach:
\begin{itemize}
    \item \textit{Emotion Recognition}, to capture user information like speech
    and gesture to infer the user's current affective state,

    \item \textit{Emotion Generation}, to produce an affective state given the
    current program and environment state, and

    \item \textit{Emotion Effects on Behaviour}, to change a program's
    behaviour (e.g. facial expressions, gestures, or movements) given its
    affective state.
\end{itemize}

Infrequently, the list includes another task: \textit{Emotion Effects on
Cognitive Processes} or \textit{Cognitive Consequences of
Emotions}~\citep[p.~100]{lisetti2015and}.

A \textit{Computational Model of Emotion (CME)} is a software system that is
influenced by emotion research, embodying at least one emotion theory as the
basis for its stimuli evaluation, emotion elicitation, and emotional behaviour
generation mechanisms~\citep[p.~2, 14]{osuna2020seperspective}. This
theoretical foundation helps define a CME's mechanisms, components, phases, and
architecture which software engineering techniques and methods can
implement~\citep[p.~139]{osuna2021toward}.

Broadly, there are two types of CME with different foci in both requirements
and validation: research-oriented and domain-specific/applied
(\citepg{wehrle1995potential}{600--601};
\citepg{hudlicka2019modeling}{130--131}; \citepg{osuna2020seperspective}{4--6}).
\begin{itemize}

    \item \textit{Research-oriented} systems emulate structures, processes, and
    mechanisms with the goal of understanding their design and structure in
    biological agents. CME designers extract requirements directly from
    affective theories and/or models---informing what mechanisms it must have,
    their order, and other aspects that are characteristic of that
    theory/model---to test hypotheses about affective phenomena and their
    eliciting mechanisms and processes. A research-oriented system is valid if
    it corresponds to the modelled phenomenon in both structure and function
    (i.e. validates a hypothesis about what structures produce a phenomenon).
    In software engineering terms, research-oriented systems are white-box
    models because they must have explainable behaviours and mechanisms for
    their outputs to enable validation. Examples include
    ACRES~\citep{frijda1987can}, EMA~\citep{gratch2004domain}, and
    ELSA~\citep{meuleman2015computational}.

    \item \textit{Domain-specific} or \textit{applied} models aim to produce
    specific aspects of affective phenomena and do not care about the specific
    structures, processes, and mechanisms behind them. Designers extract
    requirements from the qualities and behaviours that their CME should
    produce. This often relates to the CME's intended application domain (e.g.
    video games, conversational agents for health). Domain-specific systems are
    valid if it meets the designer and end-user's performance criteria, which
    can vary between domains (e.g. criteria for believability, effective user
    interactions). In software engineering terms, these systems are black-box
    models because it does not matter how they produce outputs if they have the
    desired effects. Examples include GAMYGDALA~\citep{popescu2014gamygdala},
    Em~\citep{reilly1996believable}, and APF~\citep{klinkert2021artificial}.

\end{itemize}

Building game development tools for believable NPCs ``with emotion'' requires
\textit{domain-specific} CMEs because its ability to engage players determines
its validity rather than how closely it resembles true affective phenomena. In
general, domain-specific CMEs have fewer design constraints than
research-oriented systems because they are not strict models of affective
phenomena~\citep[p.~233]{sloman2005architectural}. This affords CME designers
freedom to choose any combination of theories and/or models they wish, and make
the assumptions and design decisions necessary to realize it as a computational
model that existing research might not be able to support---an unavoidable task
when working with informally defined theories
(\citepg{marsella2010computational}{21, 23};
\citepg{hudlicka2019modeling}{130}). Domain-specific CMEs are also likely to be
significantly less complex than research-oriented ones because their
``realism'' is proportional to the complexity of their models
(\citepg{rosis2003from}{83}; \citepg{osuna2021toward}{139}). ``Realism'' might
even be detrimental to a domain-specific CME~\citep{reilly1996believable} (see
Chapter~\ref{realvsefficient} for a discussion). For example, a game developer
might exaggerate their NPCs' behaviours---which is at odds with
``realism''---to improve their believability for player interactions and,
consequently, player engagement. Taken together, this implies that
domain-specific CMEs are unlikely to be exactly alike, even if they target the
same domain.

\section{A Software Engineering Approach to CME Development}\label{sec:se}
CME development typically follows a general set of
steps~\citep[p.~2]{osuna2020seperspective}:
\begin{enumerate}
    \item Choose one or more emotion theories/models as the design foundation
    and translating them into formal languages
    \item Implement the theories/models and other software artifacts necessary
    to meet design requirements while addressing missing information in the
    theories/models
    \item If applicable, integrate the CME into a larger cognitive agent
    architecture
    \item Test that the CME produces the expected behaviours
\end{enumerate}

However, many CME designs appear to be \textit{ad hoc}---they do not
systematically apply design methods or techniques, nor follow a well-defined
development sequence (\citepg{marsella2010computational}{21};
\citepg{osuna2020seperspective}{14}). In some cases, they do not systematically
apply emotion theories/models~\citep[p.~291]{scherer2021towards}. This makes it
difficult to reuse CME components (\citepg{reisenzein2013computational}{261}),
compare different CMEs, and extend or scale them
(\citepg{osuna2020seperspective}{14--15}).

As a software system, CMEs would benefit from a disciplined software
engineering approach \citep{ghezzifundamentals2003}. It offers systematic
development processes and proven design tools which would help address issues
around CME comparisons, and support desirable software qualities such as:
reusability; modularity, which influences understandability; flexibility and
scalability, which influence maintainability; and
interoperability~\citep[p.~60]{osuna2023towards}. This would also encourage the
creation of multi-disciplinary teams for CMEs with a broader design focus
(\citepg{hudlicka2014habits}{21}; \citepg{osuna2021toward}{141}). The software
design process---requirements analysis, design, implementation, and
verification and validation---can define CME development stages to show
potential areas for improvement and increase the likelihood of creating a
well-designed system.

\subsection{Requirements Analysis}\label{sec:reqProcess}
A CME's requirements restrict the number, type, and nature of its
components~\citep[p.~441, 449]{rodriguez2015computational}, essentially
defining the design process boundaries. CME development tends to use a two-stage
requirement gathering process:
\begin{enumerate}
    \item A high-level requirement specification---including stakeholder and
    user needs---sets the system's goals, identifying which affective theories
    and/or models could achieve them (\citepg{lisetti2015and}{99};
    \citepg{osuna2020seperspective}{4}; \citepg{scherer2021towards}{281}); then

    \item Identifying which of those theories/models fit the system's goals
    best by systematically comparing them and analyzing existing
    CMEs~\citep[p.~20]{hudlicka2014habits} to generate functional and
    non-functional requirements.
\end{enumerate}

Many requirements-related pitfalls concern the scope of the \ref{as}
literature, which has diverse perspectives and lacks common terms to discuss
them (\citepg{marsella2010computational}{21};
\citepg{osuna2020seperspective}{15}). Affective theories/models often have
abstract, natural language descriptions that leave them unsystematically and
informally defined (\citepg{marsella2015appraisal}{54};
\citepg{jones2011fuzzy}{657}). This makes it difficult to identify which
theories/models suit a CME's requirements without making, typically
subjective~\citep[p.~449--450]{rodriguez2015computational}, assumptions about
unspecified behaviours~\citep[p.~15]{elliott1989ective}. These problems are so
prevalent that a group of psychologists and computer scientists called for the
deconstruction of emotion theories into basic assumptions to translate them
into a common system, language, or
architecture~\citep[p.~261]{reisenzein2013computational}. This also means that
there is ``...a lack of guidelines to determine which theory of emotion should
be used to ensure a successful development of a
CME''~\cite[p.~15]{osuna2020seperspective}.

Another common pitfall---which the informal nature of affective theory/model
descriptions are at least partially to blame---is the tendency to use
high-level descriptions for the final version of a CME's requirements
(\citepg{osuna2020seperspective}{15}), leaving questions as to how the
requirements realize design goals and influence subsequent development stages.
Developers should write the requirements at a level that establishes a context
for finding potential opportunities to use software design patterns that promote
desirable software qualities~\citep[p.~143--144]{osuna2021toward}.

\subsubsection{Choosing a Domain-Specific CME's Foundations}
A systematic method for deciding which theories to use helps minimize the
pitfall consequences by making design decisions traceable. For domain-specific
CMEs, the choice of theories/models follows from the desired system behaviours
and qualities. Expanding on the described two-stage requirement gathering
process, I propose a four-stage methodology to serve as a guideline for
examining and choosing affective theories/models for domain-specific CMEs:
\begin{enumerate}

    \item Using the CME's high-level requirements/design goals and other domain
    knowledge (e.g. game type, target player interactions, NPC embodiment),
    define the CME's design scope

    \item Using the CME's design scope and high-level requirements/design
    goals, identify broad groups of affective theories/models that could serve
    those needs

    \item Using the CME's high-level requirements/design goals, examine each
    theory/model within those groups to see ``how well'' they satisfy those
    requirements by:
    \begin{enumerate}
        \item Examining each theory/model with respect to the high-level
        requirement/design goal and recording pertinent information

        \item Using the recorded information to assign each theory/model a
        \textit{score} representing their relative suitability for that
        requirement/goal

        \item Tallying those scores to evaluate a theory's overall
        ``suitability'' for satisfying the high-level requirements/design goals
    \end{enumerate}

    \item Choose a theory/model or set of theories/models to use for the CME's
    design using the ``suitability'' scores, potentially influenced by factors
    such as domain knowledge, ease of formalization (requires experimentation),
    existing CME designs, and/or personal preference (see a proposed process in
    Chapter~\ref{sec:choosetheories})

\end{enumerate}

This methodology relies on \textit{document analysis}, which developers
commonly use for CME requirements
analysis~\citep[p.~4]{osuna2020seperspective}. Each stage aligns with a
component of document analysis~\citep[p.~32]{bowen2009document}:
\begin{itemize}

    \item The analysis context is defined by the high-level requirements and
    derived design scope,

    \item Thematic analysis of affective theories/models identifies the broad
    theory groups

    \item Content analysis organizes theories/models within those groups into
    levels of requirement ``satisfaction'' by identifying pertinent aspects of
    theories/models directly from the affective science literature, and

    \item Drawing recommendations/conclusions from the gathered data is the
    process of choosing a theory or set of theories for the CME

\end{itemize}

This methodology is specific to domain-specific CMEs---which focus on
\textit{what} they must do rather than \textit{how} they must do it---because
their development begins by defining desirable qualities and tasks it should
have \textit{before} choosing affective theories and/or models. It offers a
structured approach to establish and justify a CME's foundations and encourages
the documentation of assumptions and choices necessary for replicability and
reuse (see Chapters~\ref{chapter:reqsAndScope}, \ref{chapter:theoryAnalysis},
and \ref{chapter:choosingExamples} for example applications of this
methodology).

Based on one's understanding of the requirements and \ref{as} literature, this
methodology has many potential outcomes that might all be useful. One should
also not expect to create a strictly ``correct'' implementation of those
theories and/or models. Even the OCC theory, created with computational
tractability in mind, lacks a standard implementation~\citep[p.~218,
229]{occ2022}. It is also impossible to model, or even to choose, theories
without running into the subjectivity pitfall. Despite this, deeply ingraining
the primary literature in this methodology is beneficial for: helping CME
developers identify assumptions and decisions in existing CMEs that are not
part of the affective theory/model, informing decisions to reuse, build on, or
emulate aspects of that CME; to gain familiarity with the domain to more easily
identify aspects of a theory/model that can ``be flexible'' while remaining
faithful to its intention; and to establish psychological validity, which
directly influences the plausibility---and subsequent believability---of the
CME's outputs.

\subsection{Design}
Software development often divides design efforts into the high-level
architecture and the low-level modules~\citep[p.~8,
16]{osuna2020seperspective}. CME development appears to use these design stages,
but it is difficult to determine if there was a guiding methodology or if it
was mostly \textit{ad hoc}. CME designs tend to have an architecture to show
connections between modules---though the approach can differ widely---and do
not always document module interfaces or internal data structures~\citep[p.~11,
14]{osuna2020seperspective}. This contributes to the difficulty in comparing
different systems.

The inability to compare CMEs that use similar underlying theories/models
largely disappears when they have components with identical modularization
points~\citep[p.~26, 31, 38]{marsella2010computational}. If done correctly,
deciding which modules to include and how to connect them contain most of a
CME's design differences. This kind of modularization encourages reusability
and allows independent, empirical assessments of design decisions using
software metrics and quality attributes. Recent work towards a CME-focused
software architecture with software design patterns for each
component~\citep{osuna2021toward}, a reference
architecture~\citep{osuna2023towards}, and a framework for supporting affective
and cognitive component communication~\citep{osuna2022interoperable} is
beginning to address these issues.

\subsection{Implementation}
There is generally little to no documented information about a CME's
implementation process, making it difficult to replicate them~\citep[p.~13,
16]{osuna2020seperspective}. CMEs sometimes report their---often
object-oriented---implementation language, but not the followed practices. It
also is uncommon to find open-source versions of CMEs, though some do exist
(e.g. \citet{wasabiGit}, \citet{derekGit}, \citet{gebhardGit},
\citet{mmSource}, \citet{broekensGit}, \citet{fatimaModularSource},
\citet{fatimaGit}, \citet{orientSource}, and \citet{fearnotSource}). These
issues concern the broader scientific computing community, where there are
acknowledged characteristics for improving the reproducibility and
replicability of code~\citep{benureau2018rerun}.

\subsection{Verification and Validation}\label{sec:atcProcess}
There are no known standards or benchmarks for verifying or validating CMEs
(\citepg{osuna2020seperspective}{16}; \citepg{hudlicka2014habits}{21}), nor is
there evidence of robustness testing~\citep[p.~13]{osuna2020seperspective} or
methods for comparing CMEs~\citep[p.~373--374]{broekens2021emotion}. Other
researchers also often have difficultly replicating a CME's validation because
its testers did not formally report the
process~\citep[p.~63]{osuna2023towards}. Testers often have difficultly
adapting verification techniques to CMEs because it is usually a question of
how realistic or convincing the resulting affective behaviours are rather than
their technical functionality~\citep[p.~13--14, 16]{osuna2020seperspective}.

Test cases, usually called example scenarios or simulations, are a useful
technique for validation. They have been widely used to verify that CMEs meet
their requirements as defined from the underlying emotion theories. However,
developers run tests in specific environments and under specific conditions and
there is no standard framework for designing or running the tests. Should a
framework exist, it would not be possible to use the same one for both
research-oriented and domain-specific CMEs due to differing validation
criteria (Section~\ref{sec:CMEs}). However, CMEs of the same domain should be
able to use the same test cases and would also be an avenue for comparison. A
design framework would also allow for parallel test case creation, making test
suite development more objective, verifiable, reusable,
and---consequently---build confidence in the soundness of the test suite. A
design process would also make test suite development feasible. It is unclear
how many test cases are necessary for evaluating a CME for generating
believable emotion, but one designer claims that they analyzed approximately
600 scenarios for a model with twenty-six
emotions~\citep[p.~21--22]{elliott1998hunting}---an average of 23 per emotion.

It is not enough to test a CME's implementation (i.e. satisfies its technical
specification) because that cannot determine if it \textit{behaves} as expected
(i.e.  satisfies its external requirements). This requires acceptance tests
derived from data and/or behaviours specified independently of specific
theories, models, and/or CMEs. They must be reproducible and specific enough
for implementation, and to build confidence that the test cases are reasonable
for CME validation.

\subsubsection{Building Acceptance Test Cases for Evaluating Emotion
Believability}
CMEs generating or portraying aspects of believable agent emotions must focus on
what makes emotion \textit{believable}, not how it functions in biological
beings. Storytellers---such as novelists, playwrights, and actors---are an
excellent source for such tests because they know how to express emotion
believably in their characters (\citepg{reilly1996believable}{10};
\citepg{oatley1992best}{123}). Building test cases from stories with characters
is possible when testers know~\citep[p.~123]{smith2019design}:
\begin{enumerate}
    \item A character's narrative design (goals, motivation, current state,
    etc.),
    \item Aspects of the current world state relevant to that character, and
    \item That character's emotional reaction to the world state.
\end{enumerate}

The ``expected output'' of an acceptance test case is a character's emotional
reaction to a situation, phrased using known behavioural and expressive
characteristics of emotion kinds/categories or affective dimensions. The
character's narrative design and the current world state are inputs---the
factors causing the character's emotional reaction. These are less clear and
one must infer them from narrative elements. This inference step makes a
methodology important for replicability due to the inherent subjectivity of
character and story interpretation. Specifically, the methodology must guide
the development of subjective interpretations from an objective investigation
of a character, like a detective at a crime scene~\citep[p.~14,
20]{kusch2016literary}, to systematically identify and organize salient aspects
of a character to support deductions about them. I propose a five-stage
methodology for building acceptance test cases from stories:
\begin{enumerate}
    \item Using the CME's target domain, identify a source medium (e.g.
    literature, film, theatre) to gather information from

    \item Using the source medium and the CME's expected emotion kinds, build
    \textit{profiles} for each emotion using knowledge of how storytellers
    encode them in their medium and---to build in some psychological
    validity---information from \ref{as}

    \item From an instance of the source medium, choose a character to analyze
    and identify data collection ``trigger points'' (e.g. changes in a
    character's emotion):
    \begin{enumerate}

        \item Using the ``profiles'', identify the emotion and record elements
        of the ``profile'' that apply to the character in that moment

        \item Record elements of the scene that might have contributed to the
        emotion's elicitation (i.e. ``transient'' knowledge)

    \end{enumerate}

    \item At the end of data collection, organize the information and infer
    ``persistent'' knowledge about the character, deducible from observations
    such as the character's tendencies to act (e.g. always greeting a certain
    entity when they appear) and patterns of elements across scenes (e.g. the
    character is only calm when they have a particular item)

    \item Translate natural language descriptions into formal statements (e.g.
    ``close to death'' could become ``$\mathtt{health} \leq 5\
    \mathtt{units}$''), recording how statements from the character analysis
    map to mathematical representations
\end{enumerate}

This methodology relies on \textit{character studies/analyses}, a literary
analysis tool for examining a character's external aspects (e.g. physical
description, relationships/social status, actions, dialogue) to deduce their
internal ones (e.g. personality, motivations, emotions)~\citep[p.~22, 154, 158,
188--189]{hebert2022introduction}. Many aspects of literary works
also apply to theatre. In the broadest sense, a character is an actor in a
performance (medium) who delivers their lines (dialogue) following stage
directions (storyteller-planned actions). Therefore, source mediums do not have
to be strictly literary ones. This process offers a structured method for
structuring test case information and helps build confidence in the test cases'
validity (see Chapters~\ref{chapter:testcasedefinition} and
\ref{chapter:testcaseEMgine} for a demonstration of this methodology).

\subsubsection{Test Case Input Types}\label{sec:synthtests}
Recalling that believable characters must appear self-motivated, aware of what
is happening around them, and react appropriately in the context while adhering
to their personality (Chapter~\ref{sec:believable}), the ``data'' that
contributes to a character's emotion state can be split into two groups:
\begin{enumerate}
    \item Local data that changes between scenarios (i.e. aware of what is
    happening around them, react appropriately in the context), and
    \item Global data that does not change or changes very slowly (i.e.
    self-motivated, adhering to their personality)
\end{enumerate}
\noindent where the latter improves the coherence of the character's
behaviours~\citep[p.~189--190, 203]{ortony2002making}. Consequently, test case
inputs are either ``transient'' (i.e. local) knowledge about \textit{what is
happening to} a character and ``persistent'' (i.e. global) knowledge
\textit{about} them.

\paragraph{``Transient'' Knowledge} Emotion is a short-term state related to
events (Chapter~\ref{sec:affectiveDefs}). Knowing how a story event changes the
``world state'' is necessary to understand how the event affects a character.
As the ``world'' evolves independently, emotion evaluation happens concurrently
with each event that is significant to one or more characters.

Audiences build conceptual models of a character's internal state from their
visible~\textit{actions} \citep{thomas1981illusion}. Therefore, collecting the
following ``transient'' knowledge relies on a careful examination of story
events and their impact on the characters:
\begin{itemize}
    \item The character's action(s) and dialogue,
    \item The character's physical state (e.g. injuries), and
    \item If other characters and/or entities (e.g. the environment) are
    present/related to the character's action(s):
    \begin{itemize}
        \item The character's relation to them,
        \item Their action(s) and dialogue (actual or the character's
        assumption of them), and
        \item Their physical state.
    \end{itemize}
\end{itemize}

\paragraph{``Persistent'' Knowledge} To understand what events a character
deems \textit{relevant} (Chapter~\ref{sec:affectiveDefs}), they must possess
some static---or very slowly changing---attributes such as personality and
goals. These help explain a character's motivation and their world perception,
which is ``persistent'' knowledge because it is tied to the character rather
than the ``world''. ``Persistent'' information is usually implicit and must be
inferred from multiple sets of ``transient'' knowledge. Therefore, this
process is easier when the character appears frequently in the narrative (i.e.
main characters).

A character's important actions are the ones that they deem \textit{useful},
interpreted as actions the character does while attempting to obtain or
preserve a \textit{desirable} (to themselves) ``world state''. \textit{How} a
character performs those actions is also important because it illustrates how
they perceive the world. From this, it is possible to deduce the following
``persistent'' knowledge about a character:
\begin{itemize}
    \item Goal(s)/motivations, ranked by relative priority to the character,
    \item Personality traits, and
    \item Principles and preferences.
\end{itemize}

\subsection{Implications for CME Development}\label{sec:implications}
A common thread linking these CME development issues is a lack of
documentation, which developers must create alongside the design process---not
after~\citep[p.~949]{parnas1994precise}:
\begin{itemize}
    \item Requirements analysis must be more rigorous, capturing assumptions
    and design decisions, and showing a clear path from the CME's high-level
    design goals to specific functional and non-functional requirements for
    easy reference during development~\citep[p.~253, 255]{parnas1986rational}

    \item CME architecture and module development should localize each system
    function to one module or a related family of
    modules~\citep[p.~260--261]{parnas1985modular} to encourage desirable
    software qualities like reusability, verifiability, and
    maintainability~\citep[p.~108]{SmithAndLai2005}

    \item Implementations---even systematic
    ones~\citep[p.~948]{parnas1994precise}---and associated documentation must
    be made available so that others can reproduce, test, and compare existing
    systems

    \item Verification and validation must go beyond test cases to build
    confidence in the CME's
    capabilities~\citep[p.~270]{ghezzifundamentals2003}, which documentation
    can aid by showing how developers verified it and how the CME realizes its
    design goals, and revealing additional elements to test such as
    performance~\citep[p.~252]{parnas1986rational}
\end{itemize}

\section{Summary}
The success of believable NPCs with ``emotion'' hinges on the plausibility of
their behaviours. Psychological validity influences their plausibility,
requiring their grounding in \ref{as}. By definition, a CME---a software system
that represents some aspect of affective processing based on one or more
affective theories/models---meets this need. Historically, CME development has
typically been informal and/or poorly documented. Drawing from the systematic
and disciplined field of software engineering can alleviate these issues. To
forward this effort, I have proposed:
\begin{itemize}
    \item A document analysis-based methodology for choosing a domain-specific
    CME's underlying affective theories/models based on their high-level
    requirements/design goals and domain knowledge

    \item A character analysis/study-based methodology for deriving acceptance
    test cases from characters in professionally-crafted stories
\end{itemize}

These methodologies target domain-specific CMEs because a tool for enhancing
NPC believability via ``emotional'' behaviours requires a domain-specific CME
rather than a research-oriented one, which might work against ``believability''.
Its goals should focus on system behaviour as it relates to NPC believability
rather than specific affective phenomena. The need for psychological validity
still mandates that the design's foundation use at least one affective theory
and/or model.

\clearpage
\vspace*{\fill}
\begin{keypoints}
    \begin{itemize}

        \item The plausibility of Non-Player Character (NPC) behaviour is
        essential to their success as ``emotional'' agents and can include
        exaggerated, undesirable, and/or abnormal behaviours

        \item Plausibility is directly influenced by psychological
        validity---the grounding of the behaviours in \ref{as}

        \item \ref{ac} introduces emotion as a concern in programs to improve
        their interactions with human users

        \item A Computational Model of Emotion (CME) is a software system that
        at least one affective theory/model inspires and represents at least
        one part of affective processing

        \item Research-oriented CMEs test hypotheses about affective phenomena
        and their underlying mechanisms, whereas domain-specific CMEs aim to
        mimic affective phenomena without worrying about the true nature of the
        mechanisms

        \item CME development appears to be \textit{ad hoc}, and would benefit
        from a software engineering approach and more rigorous documentation
        practices

        \item The proposed methodology for choosing theories during
        requirements analysis for domain-specific CMEs uses document analysis
        for systematically analyzing emotion theories using the CME's
        high-level requirements, derived design scope, and other relevant
        domain knowledge

        \item The proposed methodology for building acceptance test cases for
        domain-specific CMEs uses character analyses/studies for systematically
        collecting data from stories and translating them into formal,
        implementable statements

        \item Recent research on software design patterns, a reference
        architecture, and a communication framework is promising for improving
        the rigour of CME architecture and module design

        \item Existing guidelines in the scientific computing community for
        improving the reproducibility and replicability of code are also
        applicable to CME implementation, which developers tend to under-report

        \item Improving the documentation of CME development would make each
        development stage more rigorous, make it possible to begin comparing
        CMEs, and promote desirable software qualities such as reusability,
        verifiability, and maintainability

    \end{itemize}
\end{keypoints}

\parasep
\vspace*{\fill}

%% file: theoriesInCMEs.tex
\chapter[Affective Theories in Computational Models]{Affective Theories in
Computational Models\footnote{\normalfont{\footnotesize\textcopyright{}} 2022
IEEE. Reprinted with permission from
\citet{smith2021what}. Consequently, this chapter refers to \citet{occ} rather
than \citet{occ2022}.}}\label{chapter:cmeOverview}
\def\epigraphflush{center}
\setlength{\epigraphwidth}{0.75\textwidth}
\def\textflush{center}
\epigraph{Scanning...scanning...}{Cyborg, \textit{Zack Snyder's Justice League}}

Choosing emotion theories for CME creation is difficult because each theory
typically focuses on a subset of emotion process stages and has their own
assumptions on how different components integrate and how to differentiate
emotions~\cite[pp.~10--11]{scherer2010emotion}. Given the large number of
emotion theories available (we've seen at least $27$), trying to understand
them all is unrealistic. By first focusing on families of theories, grouped by
core assumptions or focus~\cite[pp.~11, 20]{scherer2010emotion}, one can
identify a subset of theories that might satisfy a CME's requirements,
including their level of empirical validation and how they might be used
together. We choose to focus on emotion generation and some aspects of emotion
effects on behaviour because the relevant literature is vast.

This survey explores 67 CMEs that are stand-alone applications (e.g. GAMYGDALA
(\ref{gamygdala})) or part of a broader system (e.g. in Kismet (\ref{kismet})).
Its aim is to give an overview of some affective theories that appear in CME
designs and the reasons for that choice\footnote{See
\citet[pp.~370--372]{broekens2021emotion} for some historical context too.}.
Seeing affective theories in context has two advantages. First, CMEs translate
theories into concrete computational representations, thus dispelling the
fuzziness of the theories' natural language presentations. The second and
greater advantage is that a CME targeted at a specific application domain will
illustrate the underlying theory's strengths and how it could be mechanized. In
practice, designers often combine theories---sometimes implicitly---to achieve
the desired CME functionality because single theories do not address all
aspects of emotion or the available empirical
data~\cite[pp.~10]{hudlicka2014habits}. The role assigned to a theory in a CME
could be an indicator of its strengths.

Section~\ref{sec:surveyscope} reviews the survey's scope and methods, then
Section~\ref{sec:classify} organizes CMEs into categories by the creator's
original intent to help give context for their design decisions. Next,
Section~\ref{sec:survey} presents how CMEs use theories for emotion
representation, elicitation, and expression. In Section~\ref{sec:results} we
examine the theories that appear in CMEs at least five times
(Table~\ref{tab:theoryOverview}) to synthesize commonalities and strengths. We
also note theory combinations. Section~\ref{sec:discussion} explores other
information that could be useful for designing CMEs. We use
abbreviations---some of them our own---throughout the survey to increase the
legibility of the text.

\begin{table}[!tb]
    \renewcommand{\arraystretch}{1.2}
    \centering
    \caption{Overview of the Main Theories Used in Surveyed CMEs
    \textcopyright{} 2022 IEEE}
    \label{tab:theoryOverview}
    \small
    \begin{tabular}{P{0.3\linewidth}P{0.12\linewidth}P{0.48\linewidth}}
        \toprule
        \textbf{Theory} & \textbf{Abbr.} & \textbf{References} \\ \midrule

        \colourRow\textbf{Izard} & Iz. & \cite{izard1993stability,
        izard1993four, izard2000motivational} \\

        \textbf{Ekman} & Ek. & \cite{ekman2007emotions, facs} \\

        \colourRow\textbf{Plutchik} & Plu. & \cite{robert1980emotion,
        plutchik1984emotions} \\

        \textbf{\ref{valence} \& \ref{arousal}} & V-A & \cite{oxfordValence},
        \citep{oxfordArousal}, \citep{wundt1912introduction} \\

        \colourRow\textbf{Pleasure-Arousal-Dominance Space} & PAD &
        \cite{mehrabian1996pleasure} \\

        \textbf{Frijda} & Frj. & \cite{frijda1986emotions, frijda2001appraisal}
        \\

        \colourRow\textbf{Lazarus} & Laz. & \cite{lazarus1991emotion} \\

        \textbf{Scherer} & Sch. & \cite{scherer2001appraisalB} \\

        \colourRow\textbf{Roseman} & Ros. & \cite{roseman1996appraisal,
        roseman2011emotional, roseman2018functions} \\

        \textbf{Ortony, Clore, \& Collins} & OCC & \cite{occ, ortony2002making,
            ortony2005affect} \\

        \colourRow\textbf{Smith \& Kirby} & S \& K & \cite{smith1996toward,
        smith2001toward} \\

        \textbf{Oatley \& Johnson-Laird} & O \& JL & \cite{oatley1987towards,
            johnson1992basic, oatley1992best, oatley2000sentiments} \\

        \colourRow\textbf{Sloman} & Slo. & \cite{sloman2005architectural} \\

        \textbf{Damasio} & Dam. & \cite{damasio2005descartes} \\

        \colourRow\textbf{LeDoux} & LD & \cite{ledoux1996emotional} \\

        \midrule\bottomrule
    \end{tabular}
\end{table}

\section{Survey Scope and Methods}\label{sec:surveyscope}
We only include CMEs that generate emotion due to our focus. Our search
protocol follows the PRISMA-S guidelines~\citep{prismas}\footnote{See
Appendix~\ref{sec:searchprotocol} for the full protocol.}. Fifteen systems are
direct iterations of prior designs\footnote{See Appendix~\ref{sec:genes} for
CME ``genealogy''.}. Prior systems are not surveyed unless they are
sufficiently different (i.e. use different emotion theories, have differing
designer intents) to warrant exploration. Prior systems that are not
psychologically grounded (e.g. based on physical brain structures, empirical
data) are also omitted, though mentioned when important ideas are borrowed from
them.

We found 166 CMEs accompanied by a published description. We removed one
because its bibliographic data was uncertain. Our selection protocol is
partially based on citations, which take time to accumulate, so recent papers
(2020 and later) were examined by hand for scope fit. Two of seven did. Of
those from 2019 or earlier, 73 had strictly more than our threshold of 1.5
Citations per Year (C/Y). We included all CMEs with C/Y $> 2.5$ in the survey,
and an additional handful chosen subjectively as they seemed to bring something
interesting to the discussion. We made an exception for ELSA (\ref{elsa}) with
$0.43$ C/Y due to its unique implementation of Sch., and for Scherer's
involvement in its creation (see Section~\ref{sec:psychologists}). We survey 67
CMEs in total.

\section{Classifying CMEs}\label{sec:classify}
We group CMEs by application domain according to the \textit{creator's
documented intent} (Table~\ref{tab:domainOverview}) because this would have
guided their selection of emotion theories. These categories are not
exclusive---someone could use a CME successfully in a different domain.

\begin{table}[!b]
    \renewcommand{\arraystretch}{1.2}
    \centering
    \caption{Documented Application Domains \textcopyright{} 2022 IEEE}
    \label{tab:domainOverview}
    \small
    \begin{tabular}{P{0.26\linewidth}P{0.67\linewidth}}
        \toprule
        \textbf{Domain} & \textbf{Systems} \\ \midrule

        \colourRow\textbf{Multi-Purpose} &
        \begin{enumerate*}[series=systemList]
            \item\label{ar} Affective Reasoner (AffectR),

            \item\label{cathexis} Cathexis,

            \item\label{emMod} Emotion Model (EmMod),

            \item\label{flame} FLAME,

            \item\label{scream} SCREAM,

            \item\label{mamid} MAMID,

            \item\label{tabasco} TABASCO,

            \item\label{wasabi} WASABI,

            \item\label{maggie} Maggie,

            \item\label{akr} AKR Scheme,

            \item\label{gvh} General Virtual Human (GVH),

            \item\label{parlee} ParleE,

            \item\label{impmeb} Interdependent Model of Personality,
            Motivations, Emotion, and Mood (IM-PMEB),

            \item\label{genia3} GenIA$^3$,

            \item\label{infra} InFra,

            \item\label{fatimaM} FAtiMA Modular (FAtiMA-M),

            \item\label{hybridc} Hybrid Model of Emotion-Eliciting Conditions
            (HybridC),

            \item\label{gema} GEmA
        \end{enumerate*} \\

        \textbf{Natural Language Processing} &
        \begin{enumerate*}[resume=systemList]
            \item\label{som} SOM
        \end{enumerate*} \\

        \colourRow\textbf{Cognitive Architecture} &
        \begin{enumerate*}[resume=systemList]
            \item\label{soar} Soar,

            \item\label{lida} LIDA,

            \item\label{clarion} CLARION
        \end{enumerate*}  \\

        \textbf{Scientific Research} & \begin{enumerate*}[resume=systemList]
            \item\label{acres} ACRES,

            \item\label{ema} EMA,

            \item\label{will} Will,

            \item\label{elsa} ELSA,

            \item\label{gamae} GAMA-E

        \end{enumerate*} \\

        \colourRow\textbf{Military and Emergency Training}
        & \begin{enumerate*}[resume=systemList]
            \item\label{emile} \'Emile,

            \item\label{emotion} EMOTION,

            \item\label{humdpme} HumDPM-E,

            \item\label{jbdiemo} JBdiEmo,

            \item\label{dett} DETT,

            \item\label{epbdi} EP-BDI,

            \item\label{microcrowd} MicroCrowd

        \end{enumerate*} \\

        \textbf{Soft Skills Training} & \begin{enumerate*}[resume=systemList]
            \item\label{puppet} Puppet,

            \item\label{cbi} CBI,

            \item\label{fatima} FAtiMA,

            \item\label{tardis} TARDIS,

            \item\label{puma} PUMAGOTCHI

        \end{enumerate*} \\

        \colourRow\textbf{Virtual Social Agents} &
        \begin{enumerate*}[resume=systemList]
            \item\label{greta} Greta,

            \item\label{alma} ALMA,

            \item\label{eva} Eva,

            \item\label{ppad} PPAD-Algorithm (PPAD-Algo),

            \item\label{peedy} Peedy the Parrot,

            \item\label{erdams} ERDAMS,

            \item\label{teatime} TEATIME,

            \item\label{mmt} Mobile Medical Tutor (MMT),

            \item\label{presence} Presence

        \end{enumerate*} \\

        \textbf{Social Robots} & \begin{enumerate*}[resume=systemList]
            \item\label{pomdp} Partially Observable Markov Decision Process for
            Cognitive Appraisal (POMDP-CA),

            \item\label{iphonoid} iPhonoid,

            \item\label{eegs} Ethical Emotion Generation System (EEGS),

            \item\label{pwe} Plutchik's Wheel of Emotions Inspired (PWE-I),

            \item\label{kismet} Kismet,

            \item\label{robo} Roboceptionist (R-Cept),

            \item\label{grace} GRACE,

            \item\label{tame} TAME

        \end{enumerate*} \\

        \colourRow\textbf{Art and Entertainment} &
        \begin{enumerate*}[resume=systemList]
            \item\label{aee} Artificial Emotion Engine\texttrademark~(AEE),

            \item\label{feelme} FeelMe,

            \item\label{socio} Socioemotional State (SocioEmo),

            \item\label{soul} The Soul,

            \item\label{gamygdala} GAMYGDALA,

            \item\label{mobsim} Mob Simulation (MobSim),

            \item\label{apf} Artificial Psychosocial Network (APF),

            \item\label{mexica} MEXICA,

            \item\label{npe} Narrative Planning with Emotions (NPE),

            \item\label{em} Em/Oz,

            \item\label{s3a}\label{sysCount} S3A

        \end{enumerate*} \\

        \midrule\bottomrule
    \end{tabular}
\end{table}

\begin{itemize}
    \item \textit{Multi-Purpose} CMEs (Systems \ref{ar}--\ref{gema}) are not
    limited to one domain. These systems: explicitly list multiple,
    sufficiently different potential uses (\citeg{hudlicka2019modeling};
    \citepg{elliott1989ective}{3--6}; \citeg{el2000flame};
    \citeg{salichs2012new}); name a general type of CME environment
    (\citepg{becker2008wasabi}{10}; \citeg{velasquez1998robots};
    \citeg{ushida1998emotion}; \citeg{prendinger2004mpml};
    \citeg{petta2002role}; \citeg{lisetti2002can};
    \citeg{kshirsagar2002multilayer}; \citeg{duy2004creating};
    \citeg{shvo2019interdependent}; \citeg{castellanos2018computational};
    \citepg{jain2019modeling}{60}; \citeg{kazemifard2011design}); and allow
    users to integrate their own implementations of emotion
    theories~\citep{alfonso2017toward, dias2014fatima}.

    \item \textit{Natural Language Processing} CMEs (Systems \ref{som}) read,
    decipher, comprehend, and analyze human language, focusing on affective
    content~\citep{yanaru1997emotion}. \clearpage

    \item \textit{Cognitive Architectures} (Systems \ref{soar}--\ref{clarion})
    implement theories concerned with the components of the mind and
    interactions between them~\citep{soarIntro, lida, sun2016emotion}.

    \item \textit{Scientific Research} CMEs (Systems \ref{acres}--\ref{gamae})
    explore aspects of affect or affective system design. They are typically
    stricter about the system's behaviours, as they aim to test an affective
    theory~\citep{frijda1987can, meuleman2015computational} or replicate
    observed affective phenomena~\citep{moffat1997personality, gratch2004domain,
        bourgais2017enhancing}.

    \item \textit{Military and Emergency Training} CMEs (Systems
    \ref{emile}--\ref{microcrowd}) help train personnel for
    emotionally\-/charged scenarios in consequence-free
    environments~\citep{gratch2000emile, el2004modelling,
    aydt2011computational, korecko2016jadex}, or run simulations where emotion
    is a factor~\citep{parunak2006model, zoumpoulaki2010multi,
    lhommet2011never}.

    \item CMEs for \textit{Soft Skills Training} (Systems
    \ref{puppet}--\ref{puma}) help train life skills that can be difficult to
    hone with traditional techniques, such as emotional
    intelligence~\citep[pp.~153]{andre2000integrating}, problem solving under
    pressure~\citep{marsella2000interactive}, empathy~\citep{dias2005feeling},
    interview skills~\citep{jones2013tardis}, healthy eating habits, and
    responsibility for pets~\citep{laureano2012design}.

    \item \textit{Virtual Social Agents} with CMEs (Systems
    \ref{greta}--\ref{presence}) have a virtual embodiment, interacting with
    users in a conversational capacity. They focus on:
    believability~\citep{rosis2003from, gebhard2005alma, kasap2009making,
    zhang2016modeling}; improving interface usability~\citep{ball2000emotion,
    ochs2012formal, yacoubi2018teatime, alepis2011automatic}; or
    both~\citep[pp.~158]{andre2000integrating}.

    \item CMEs for \textit{Social Robots} (Systems \ref{pomdp}--\ref{tame})
    are different from virtual assistants because of a robot's physical
    embodiment~\citep[p.~120]{breazeal2003emotion}. These CMEs aim to humanize
    robots and improve human-robot interactions by adding a social dimension to
    them (\citeg{kim2010computational}; \citeg{masuyama2018personality};
    \citeg{ojha2016ethically}; \citepg{qi2019building}{209})---sometimes over
    extended time frames (\citepg{breazeal2003emotion}{122--124};
    \citeg{kirby2010affective})---and to provide
    companionship~\citep{dang2009experimentation, moshkina2011tame}.

    \item CMEs for \textit{Art and Entertainment} (Systems
    \ref{aee}--\ref{s3a}) are often used for improving agent
    believability, changing the focus from strict adherence to psychological
    validity to interesting and entertaining behaviours. However, agent
    behaviours must remain plausible to be
    effective~\citep[p.~216--217]{broekens2016emotional}. There are CMEs for:
    developer tools~\citep{wilson2000artificial, broekens2004scalable,
        ochs2009simulation, bidarra2010growing, popescu2014gamygdala,
        durupinar2016psychological, klinkert2021artificial}; narrative
    planning~\citep{y2007employing, shirvani2020formalization}; and agent
    architectures (\citepg{reilly1996believable}{31};
    \citeg{martinho2000emotions}).
\end{itemize}

\section{Survey}\label{sec:survey}
We document the following tasks performed by CMEs:
\begin{itemize}

    \item \textit{Emotion Representation} (Table~\ref{tab:reprOverview}): CMEs
    might use a theory to specify what kinds of emotion it supports. Although
    several CMEs tend to use the same theory to both represent and elicit
    emotion, these are examined separately because differences might indicate
    other aspects of the theory relevant to CME design.

    \item \textit{Emotion Elicitation} (Table~\ref{tab:elicitOverview}): Since
    there are emotion theories that do not, or vaguely, describe the
    \textit{process} of emotion generation, this use is separated from emotion
    representation to clarify the difference.

    \item \textit{Emotion Expression} (Table~\ref{tab:expressionOverview}): We
    examine affective theories selected for expression separately because they
    are distinct tasks. CMEs that do both generation and expression might use
    separate theories for each task or the same combination of theories for
    both.

\end{itemize}

Eight CMEs (i.e. EmMod (\ref{emMod}), WASABI (\ref{wasabi}), FAtiMA-M
(\ref{fatimaM}), HybridC (\ref{hybridc}), CLARION (\ref{clarion}), Greta
(\ref{greta}), Presence (\ref{presence}), GRACE (\ref{grace})) do not implement
one or more theories, but instead use them as design guides. These also reveal
decision rationale, so we make note of this. When a CME can be programmed with
a user's choice of theories (i.e. GenIA$^3$ (\ref{genia3}), InFra
(\ref{infra}), FAtiMA-M (\ref{fatimaM}), FeelMe (\ref{feelme})), we examine its
default implementation.

\afterpage{\clearpage
    \begin{landscape}
        {\topskip0pt
        \vspace*{\fill}
        \begin{table}[!tbh]
            \renewcommand{\arraystretch}{1.2}
            \centering
            \caption{Theories Used for Emotion Representation \textcopyright{}
            2022 IEEE}
            \label{tab:reprOverview}
            \small
            \begin{tabular}{@{}clccccccccccccccc@{}}
                \toprule

                \multicolumn{1}{c}{} & \multicolumn{1}{c}{} & \textbf{Iz.} &
                \textbf{Ek.} & \textbf{Plu.} & \textbf{V-A} & \textbf{PAD} &
                \textbf{Frj.} & \textbf{Laz.} & \textbf{Sch.} & \textbf{Ros.} &
                \textbf{OCC} & \textbf{S\&K} & \textbf{O\&JL} & \textbf{Slo.} &
                \textbf{Dam.} & \textbf{LD} \\
                \midrule

                \colourRow\ref{ar} &
                \textbf{AffectR}$\dagger$ & -- & -- & -- & -- & -- & -- & -- &
                -- & -- & \textbf{r} &--&--&--&--&-- \\

                \ref{cathexis}& \textbf{Cathexis} & \textbf{R} & \textbf{R} &
                --& -- & -- &-- &-- & -- & -- & -- &--  & \textbf{R} &--& -- &
                -- \\

                \colourRow\ref{emMod}& \textbf{EmMod} & -- &
                \textbf{R} & --& -- &--  &-- &-- &-- & -- & -- & -- &
                \textbf{R} &--&-- &--  \\

                \ref{flame}& \textbf{FLAME}$\dagger$ & --& --& -- & -- & --& --
                &-- & -- & \textbf{r} & \textbf{r} &-- & --& --& --& -- \\

                \colourRow\ref{scream}&
                \textbf{SCREAM}$\dagger$ & -- & --& -- & --&  -- & --& -- &  --
                & -- & \textbf{r} &--&--&--&-- &-- \\

                \ref{mamid}& \textbf{MAMID}$\dagger$ & -- & -- & -- &
                \textbf{r} & -- & -- & -- & \textbf{r} & -- & -- & \textbf{r} &
                -- & -- & -- & -- \\

                \colourRow\ref{tabasco} &
                \textbf{TABASCO}$\dagger$ & -- & -- & -- & -- & --& -- & -- &
                -- & -- & --& \textbf{r} & --&  --&--&-- \\

                \ref{wasabi}&\textbf{WASABI} & -- & \textbf{R} & \textbf{r} &
                -- & \textbf{R} & -- & -- &--  &-- & \textbf{r} & -- & --&
                \textbf{r} & \textbf{R} & \textbf{R} \\

                \colourRow\ref{maggie} & \textbf{Maggie} &
                -- & -- & -- & --&-- & -- & -- & -- & -- & \textbf{r}$^1$ & --&
                --&--  &--&-- \\

                \ref{akr}& \textbf{AKR} & -- & \textbf{r} & --& --& --&
                \textbf{r} & -- & \textbf{r} & \textbf{r} & \textbf{r} & -- &
                -- &--&--&-- \\

                \colourRow\ref{gvh}& \textbf{GVH} & -- &
                \textbf{R} & -- &-- &-- & -- & -- &-- & -- & \textbf{r}
                &--&--&--&--&-- \\

                \ref{parlee}& \textbf{ParleE}$\dagger$ & -- & --& -- & -- & --
                & -- & -- & -- & \textbf{R} (\ref{flame}) & \textbf{R} & -- &
                -- & --  & -- & -- \\

                \colourRow\ref{impmeb}& \textbf{IM-PMEB} &
                -- &-- & -- &-- & \textbf{r} (\ref{alma}) & --&-- &-- & -- &
                \textbf{r} &--&--&--&--&-- \\

                \ref{genia3} & \textbf{GenIA\textsuperscript{3}} & -- & -- & --
                & -- & \textbf{r} (\ref{alma}) & -- & -- & -- & -- & \textbf{r}
                (\ref{ema}, \ref{alma}) & -- & -- & -- & -- & -- \\

                \colourRow\ref{infra}& \textbf{InFra} & -- &
                --& \textbf{r} & --& --& -- &-- & -- &-- & -- &--&--&--&--&-- \\

                \ref{fatimaM}& \textbf{FAtiMA-M}$\dagger$ & -- &-- &  --& --
                &-- & -- & -- & --&-- & \textbf{R} (\ref{fatima})
                &--&--&--&--&-- \\

                \colourRow\ref{hybridc}& \textbf{HybridC} &
                \textbf{R} & \textbf{R} & \textbf{R} & -- & --& -- &-- & -- &
                -- & -- &--& r & -- & -- & -- \\

                \ref{gema}& \textbf{GEmA}$\dagger$ & -- & -- & -- & -- & --& --
                &-- & -- & -- & \textbf{R} &--& -- & -- & -- & -- \\

                \midrule

                \colourRow\ref{som}& \textbf{SOM} & -- & --&
                \textbf{R} &-- & -- & -- & --& --& -- &-- &--&--&--&--&-- \\

                \midrule

                \ref{soar}& \textbf{Soar}$\dagger$ &-- & -- & -- & --& --& --&
                -- & \textbf{r}$^2$ & -- & -- & --& --&-- &--&-- \\

                \colourRow\ref{lida}& \textbf{LIDA}$\dagger$
                & -- &--&-- & -- & --& -- & --& \textbf{r}$^2$ & -- &--
                &--&--&--&--&-- \\

                \ref{clarion}& \textbf{CLARION}$\dagger$ & -- &--&-- &
                -- &-- & -- &-- & \textbf{r} & -- &-- &--&--&--&--&-- \\

                \colourRow\ref{acres}&
                \textbf{ACRES}$\dagger$ & -- & -- &--  & --& --& \textbf{R} &
                -- & -- & -- & -- & --& -- &--&--&--  \\

                \ref{ema}& \textbf{EMA}$\dagger$ &-- & -- & -- &-- & -- &-- &
                -- &-- &--  & \textbf{r} (\ref{ar}) & -- & -- &--&--&-- \\

                \colourRow\ref{will}& \textbf{Will}$\dagger$
                & -- & --& -- & -- & -- & \textbf{R} &-- &-- & -- &--
                &--&--&--&--&-- \\

                \midrule\bottomrule
                \multicolumn{17}{r}{\footnotesize\textit{Continued on next
                page}}
            \end{tabular}
        \end{table}
        \vspace*{\fill}}

        \clearpage

        {\topskip0pt
        \vspace*{\fill}
        \addtocounter{table}{-1}
        \captionsetup{list=no}
        \begin{table}[!tbh]
            \renewcommand{\arraystretch}{1.2}
            \centering
            \caption{(\textit{Continued.}) Theories Used for Emotion
            Representation \textcopyright{} 2022 IEEE}
            \small
            \begin{tabular}{@{}clccccccccccccccc@{}}
                \toprule

                \multicolumn{1}{c}{} & \multicolumn{1}{c}{} &
                \textbf{Iz.} & \textbf{Ek.} & \textbf{Plu.} &
                \textbf{V-A} & \textbf{PAD} & \textbf{Frj.} &
                \textbf{Laz.} & \textbf{Sch.} & \textbf{Ros.} &
                \textbf{OCC} & \textbf{S\&K} & \textbf{O\&JL} &
                \textbf{Slo.} & \textbf{Dam.} & \textbf{LD} \\
                \midrule

                \colourRow\ref{elsa}& \textbf{ELSA}$\dagger$
                & -- & -- & -- & -- & -- & -- & -- & \textbf{R} & -- & -- & --
                & -- & -- & -- & -- \\

                \ref{gamae}& \textbf{GAMA-E} & -- & -- & -- & -- & -- & -- & --
                & -- & -- & \textbf{r} & -- & -- & -- & -- & -- \\

                \midrule

                \colourRow\ref{emile}&
                \textbf{\'Emile}$\dagger$ & -- & --& -- & -- &-- & -- & -- & --
                & -- & \textbf{r} (\ref{ar}) & -- & -- &--& -- &-- \\

                \ref{emotion}& {\small\textbf{EMOTION}} & -- &-- & --& --& --&
                --& --& --& --& \textbf{r} (\ref{gvh})$^3$ &-- & --& --&--&--
                \\

                \colourRow\ref{humdpme}&
                {\renewcommand{\arraystretch}{1}
                \begin{tabular}[x]{@{}l@{}}\textbf{\small Hum-} \\
                \textbf{\small DPM-E}$\dagger$ \end{tabular}} & -- & --
                & -- & -- & -- & -- & -- & \textbf{r} & -- & -- & -- &
                -- & -- & -- & -- \\

                \ref{jbdiemo}& \textbf{JBdiEmo}$\dagger$ & -- &-- & -- & -- &--
                & --&-- & --& -- & \textbf{r}$^4$ & -- &-- &--& --&-- \\

                \colourRow\ref{dett}& \textbf{DETT}$\dagger$
                & -- & -- & -- & -- & -- & -- & -- & -- & -- & \textbf{r} & --
                & -- & -- & -- & -- \\

                \ref{epbdi}& \textbf{EP-BDI} & -- & -- & -- & -- & -- & -- & --
                &-- & -- & \textbf{r}$^1$ & -- & -- & -- & -- & -- \\

                \colourRow\ref{microcrowd}&
                {\renewcommand{\arraystretch}{1}
                \begin{tabular}[x]{@{}l@{}}\textbf{\small Micro-} \\
                \textbf{\small Crowd}$\dagger$ \end{tabular}} & -- & -- & -- &
                -- & -- & -- & -- & -- & -- & \textbf{r} & -- & -- & -- & -- &
                -- \\

                \midrule

                \ref{puppet}& \textbf{Puppet} & -- & \textbf{R} & -- & -- &--
                &-- & -- &-- & -- & \textbf{R} &-- & --& --&--& -- \\

                \colourRow\ref{cbi}& \textbf{CBI}$\dagger$
                &-- &-- & --& --& --& --& \textbf{R} & -- & -- &-- &--&--&--
                &--&-- \\

                \ref{fatima}& \textbf{FAtiMA}$\dagger$ & -- & -- & -- & -- & --
                & -- & -- & -- & -- & \textbf{r} (\ref{ema}, \ref{s3a}) & --
                & -- & -- & -- & -- \\

                \colourRow\ref{tardis}& \textbf{TARDIS} &--
                &-- & -- &-- & \textbf{r} (\ref{alma}, \ref{socio}) & -- & -- &
                -- &-- & \textbf{r} & -- &-- &--& --&-- \\

                \ref{puma}& {\renewcommand{\arraystretch}{1}
                \begin{tabular}[x]{@{}l@{}}\textbf{\small PUMA-} \\
                \textbf{\small GOTCHI}$\dagger$ \end{tabular}} &-- &-- & -- &--
                & -- & -- & -- & -- &-- & \textbf{r} & -- &-- &--& --&-- \\

                \midrule

                \colourRow\ref{greta}& \textbf{Greta} & -- &
                \textbf{r} & -- & -- & -- & -- & -- & -- & -- & \textbf{r} & --
                & -- & -- & -- & -- \\

                \ref{alma}& \textbf{ALMA} & -- & --& -- & -- & \textbf{R} &--
                & --& -- &  --& \textbf{r} & -- & --&--&--&--  \\

                \colourRow\ref{eva}& \textbf{Eva} & -- & --
                & -- & --& \textbf{r} (\ref{alma}) & -- & -- & -- &-- &
                \textbf{r}$^{1}$ &--&--&--&--&-- \\

                \ref{ppad}& {\footnotesize\textbf{PPAD-Algo}} & -- & -- & --
                &-- & \textbf{R} & -- & -- & -- &-- & \textbf{r} (\ref{alma})
                &--&--&--&-- &-- \\

                \colourRow\ref{peedy}& \textbf{Peedy} & -- &
                -- & -- & \textbf{R} & -- &--  & -- & -- & -- & -- &--&--&--
                &--&-- \\

                \ref{erdams}& \textbf{ERDAMS} & -- & -- & -- & -- & -- & -- &
                -- & -- & -- & \textbf{R} &--&--&--&--&-- \\

                \colourRow\ref{teatime}& \textbf{TEATIME} &
                -- & -- & -- & --& -- & -- & -- & -- & \textbf{r} & --
                &--&--&--&--&-- \\

                \ref{mmt}& \textbf{MMT}$\dagger$ & -- & -- & -- & -- & -- & --
                & -- & -- & --& \textbf{r} &--&--& -- & -- &-- \\

                \colourRow\ref{presence}& \textbf{Presence}
                & -- & -- & -- & -- & -- & -- & -- & -- & --& \textbf{R}
                &--&--& -- & -- &-- \\

                \midrule\bottomrule
                \multicolumn{17}{r}{\footnotesize\textit{Continued on
                next page}}
            \end{tabular}
        \end{table}
        \vspace*{\fill}}

        \clearpage

        {\topskip0pt
        \vspace*{\fill}
        \addtocounter{table}{-1}
        \captionsetup{list=no}
            \begin{table}[!tbh]
                \renewcommand{\arraystretch}{1.2}
                \centering
                \caption{(\textit{Continued.}) Theories Used for Emotion
                    Representation \textcopyright{} 2022 IEEE}
                \small
                \begin{threeparttable}
                    \begin{tabular}{@{}clccccccccccccccc@{}}
                        \toprule

                        \multicolumn{1}{c}{} & \multicolumn{1}{c}{} &
                        \textbf{Iz.} & \textbf{Ek.} & \textbf{Plu.} &
                        \textbf{V-A} & \textbf{PAD} & \textbf{Frj.} &
                        \textbf{Laz.} & \textbf{Sch.} & \textbf{Ros.} &
                        \textbf{OCC} & \textbf{S\&K} & \textbf{O\&JL} &
                        \textbf{Slo.} & \textbf{Dam.} & \textbf{LD} \\
                        \midrule

                        \colourRow\ref{pomdp}&
                        \textbf{POMDP-CA}$\dagger$ & -- & -- & -- & -- &--
                        & -- &-- &-- & \textbf{R} & -- & -- &-- & -- & -- & --
                        \\

                        \ref{iphonoid}& \textbf{iPhonoid}
                        & -- & -- & -- & -- &
                        \textbf{r} & -- & -- & --  &-- &--&--&--&--&--&-- \\

                        \colourRow\ref{eegs}&
                        \textbf{EEGS}$\dagger$
                        & -- & -- & -- & -- & -- & -- & -- & -- & -- &
                        \textbf{r} & --
                        & -- & -- & -- & -- \\

                        \ref{pwe}& \textbf{PWE-I} & -- & -- & \textbf{R} & -- &
                        -- & --
                        & -- & --  &-- &--&--&--&--&--&-- \\

                        \colourRow\ref{kismet}&
                        \textbf{Kismet} &
                        \textbf{R} & \textbf{R} & \textbf{R} & \textbf{R$^{5}$}
                        & --&
                        -- &-- & --& -- & -- & -- & -- & -- & -- & -- \\

                        \ref{robo}& \textbf{R-Cept} & -- & \textbf{r} & -- & --
                        &-- &
                        -- & --& -- & -- & -- & -- &-- & -- & -- & -- \\

                        \colourRow\ref{grace}&
                        \textbf{GRACE}$\dagger$ & -- & -- & -- & --& --& -- &
                        -- &
                        \textbf{r} & -- & \textbf{r} & -- &-- &--&--& -- \\

                        \ref{tame}& \textbf{TAME} & -- & \textbf{R} & -- & -- &
                        -- & --
                        &-- &-- &--  & -- & -- & --& -- & -- & -- \\

                        \midrule

                        \colourRow\ref{aee}& \textbf{AEE}
                        &-- &
                        \textbf{r} &-- & --& -- & --& -- & -- &-- & -- & --
                        &--&--&--&-- \\

                        \ref{feelme}& \textbf{FeelMe} &-- & -- &-- & --&
                        \textbf{R} &
                        --& -- & -- &-- & -- & -- &--&--&--&-- \\

                        \colourRow\ref{socio}&
                        \textbf{SocioEmo} & -- & -- & -- & -- & \textbf{r}
                        (\ref{alma}) & -- & -- &  -- & -- & \textbf{R$^1$}
                        (\ref{parlee}) & -- & -- &-- &--& -- \\

                        \ref{soul} & \textbf{The Soul} & -- & -- & -- & -- &
                        \textbf{R} & -- & -- & -- &--& \textbf{r} (\ref{alma})
                        &--&--&--&--&-- \\

                        \colourRow\ref{gamygdala}&
                        \textbf{GAMYGDALA} & --& -- & -- & -- & \textbf{R} &--
                        &-- & -- & -- & \textbf{R} & -- & --& -- &--&-- \\

                        \ref{mobsim}& \textbf{MobSim} &-- & -- &-- & --&
                        \textbf{R} & --& -- & -- &-- & \textbf{R} (\ref{alma})
                        & -- &--&--&--&-- \\

                        \colourRow\ref{apf}& \textbf{APF} &
                        -- & -- & -- & -- & -- & -- & -- &  -- & -- &
                        \textbf{r$^4$} (\ref{socio}) & -- & -- &-- &--& -- \\

                        \ref{mexica}& \textbf{MEXICA} & -- & -- & -- & -- & --
                        & -- & -- &  -- & -- & \textbf{r} & -- & -- &-- &--& --
                        \\

                        \colourRow\ref{npe}& \textbf{NPE} &
                        -- & -- & -- & -- & -- & -- & -- &  -- & -- &
                        \textbf{r} & -- & -- &-- &--& -- \\

                        \ref{em}& \textbf{Em/Oz} & --& -- & -- &-- & --& -- &
                        -- & -- & --& \textbf{R} &--&--&--&--&-- \\

                        \colourRow\ref{s3a} &
                        \textbf{S3A}$\dagger$ & -- &-- & -- & -- & -- & -- & --
                        & -- & -- & \textbf{r} & -- & -- &--&--&-- \\

                        \midrule\bottomrule
                    \end{tabular}
                    \begin{tablenotes}

                        \footnotesize
                        \vspace*{2mm}

                        \item \textbf{R}: \textit{Reasons for choosing the
                        theory are clear;} \textbf{r}: \textit{Reasons are
                        unclear;} (\#): \textit{System borrowed from/is
                        influenced by System \#}

                        \item [$\dagger$] \textit{Used as a consequence of
                        theories chosen for emotion elicitation
                        (Section~\ref{sec:elicit}).}

                        \item [1] \textit{Based on
                        \citep[p.~193]{ortony2002making}, a simplified model
                        developed by Ortony for believable ``artifacts''
                        (\citepg{kasap2009making}{23};
                        \citepg{salichs2012new}{62};
                        \citepg{ochs2009simulation}{285}).}

                        \item [2] \textit{Not yet implemented
                        (\citepg{soarEmotion}{271}; \citepg{lida}{26}).}

                        \item [3] \textit{Builds on
                        \citep[p.~2]{egges2004generic}, a successor of GVH
                        (\ref{gvh}).}

                        \item [4] \textit{Uses the OCCr model
                        (\citepg{korecko2016jadex}{195, 197};
                        \citepg{klinkert2021artificial}{702}) a
                        reinterpretation of the OCC model that aims to clarify
                        the model's logical structure and address
                        ambiguities~\citep[p.~1]{steunebrink2009occ}.}

                        \item [5] \textit{Also uses a \texttt{stance} dimension
                        to measure the approachability of a
                        stimulus~\citep[p.~133]{breazeal2003emotion}.}

                    \end{tablenotes}
                \end{threeparttable}%
            \end{table}
        \vspace*{\fill}}
    \end{landscape}
    \captionsetup{list=yes}
}

\afterpage{\clearpage
    \begin{landscape}
        {\topskip0pt
            \vspace*{\fill}
            \begin{table}[!tbh]
                \renewcommand{\arraystretch}{1.2}
                \centering
                \caption{Theories Used for Emotion Elicitation \textcopyright{}
                2022 IEEE}
                \label{tab:elicitOverview}
                \small
                \begin{tabular}{@{}clccccccccccccc@{}}
                    \toprule

                    \multicolumn{1}{c}{} & \multicolumn{1}{c}{} & \textbf{Iz.} &
                    \textbf{V-A} & \textbf{PAD} & \textbf{Frj.} & \textbf{Laz.}
                    & \textbf{Sch.} & \textbf{Ros.} &\textbf{OCC} &
                    \textbf{S\&K} & \textbf{O\&JL} & \textbf{Slo.} &
                    \textbf{Dam.} & \textbf{LD} \\
                    \midrule

                    \colourRow\ref{ar}& \textbf{AffectR} &
                    -- & -- & --& -- & -- &-- &  -- & \textbf{e}
                    &--&--&--&--&-- \\

                    \ref{cathexis}& \textbf{Cathexis} & e & -- & -- &-- &-- &
                    -- & -- & -- &--  & -- &--& \textbf{E} & \textbf{E} \\

                    \colourRow\ref{emMod}& \textbf{EmMod} &
                    -- & -- &--  &-- &-- &-- & -- & \textbf{e} & -- & -- &--&
                    \textbf{e}  &--  \\

                    \ref{flame}& \textbf{FLAME} & -- & -- & --& -- &-- & -- &
                    \textbf{e} & \textbf{e} &-- & --& --& --& \textbf{e} \\

                    \colourRow\ref{scream}& \textbf{SCREAM}
                    & -- & --& -- & --& -- &  -- & -- & \textbf{e} (\ref{em})
                    &--&--&--&-- &-- \\

                    \ref{mamid}& \textbf{MAMID} & -- & \textbf{e} & -- & -- &
                    -- & \textbf{e} & -- & -- & \textbf{e} & -- & -- & -- & --
                    \\

                    \colourRow\ref{tabasco} &
                    \textbf{TABASCO} & -- & -- & --& \textbf{e} & -- &
                    \textbf{e} & -- & --& \textbf{e} & --& --&--&-- \\

                    \ref{wasabi}&\textbf{WASABI} & -- &--  & -- & -- & -- &--
                    &-- & \textbf{e} & -- & --& \textbf{e} & \textbf{E} &
                    \textbf{E} \\

                    \colourRow\ref{maggie} & \textbf{Maggie}
                    & -- & --&-- & -- & \textbf{E} & -- & -- & \textbf{E$^1$} &
                    --& --&--  &--&-- \\

                    \ref{akr}& \textbf{AKR}$\ddagger$ & -- & -- & -- &
                    \textbf{e} & -- & \textbf{e} & \textbf{e} & \textbf{e} & --
                    & -- & -- & -- & -- \\

                    \colourRow\ref{gvh}& \textbf{GVH} & --
                    &-- &-- & -- & -- &-- & -- & \textbf{e} &--&--&--&--&-- \\

                    \ref{parlee}& \textbf{ParleE} & -- & -- & -- & -- & -- & --
                    & \textbf{E} & \textbf{E}  (\ref{flame}, \ref{em}) & -- &
                    \textbf{e} & -- & -- & -- \\

                    \colourRow\ref{impmeb}& \textbf{IM-PMEB}
                    & -- &-- & --& --&-- &-- & -- & \textbf{e} &--&--&--&--&--
                    \\

                    \ref{genia3}& \textbf{GenIA\textsuperscript{3}} & -- &-- &
                    --&  --&-- &-- & -- & \textbf{e} (\ref{ema}, \ref{alma},
                    \ref{erdams}) &--&--&--&--&-- \\

                    \colourRow\ref{infra}& \textbf{InFra} &
                    -- & --& --& -- &-- & \textbf{e} &-- & \textbf{e}
                    (\ref{flame}) &--&--&--&--&-- \\

                    \ref{fatimaM}& \textbf{FAtiMA-M} & -- & -- &-- & -- &
                    \textbf{e} (\ref{fatima}) & \textbf{E} &-- & \textbf{E}
                    (\ref{fatima}) &--&--&--&--&-- \\

                    \colourRow\ref{hybridc}&
                    \textbf{HybridC} & -- & -- & --& -- &-- & \textbf{E} &
                    \textbf{E} & \textbf{E} &--& -- &--&--&-- \\

                    \ref{gema}& \textbf{GEmA} & -- & -- & --& -- &-- & -- & --
                    & \textbf{E} &--& -- & -- & -- & -- \\

                    \midrule

                    \colourRow\ref{soar}& \textbf{Soar} &--
                    & --& --& --& -- & \textbf{e}$^2$ & -- & -- & --& --&--
                    &--&-- \\

                    \ref{lida}& \textbf{LIDA} & -- & -- & --& -- & --&
                    \textbf{e}$^2$ & -- &-- &--&--&--&--&-- \\

                    \colourRow\ref{clarion}&
                    \textbf{CLARION} & -- & -- &-- & -- &-- & \textbf{e} & --
                    &-- &--&--&--&--&-- \\

                    \midrule

                    \ref{acres}& \textbf{ACRES} & -- & --& --& \textbf{E} & --
                    & -- & -- & -- & --& -- &--&--&--  \\

                    \colourRow\ref{ema}& \textbf{EMA} &--
                    &-- & -- &-- & \textbf{e} &-- &--  & \textbf{e} (\ref{ar},
                    \ref{emile}) & -- & -- &--&--&-- \\

                    \ref{will}& \textbf{Will} & -- & -- & -- & \textbf{E} &--
                    &-- & -- &-- &--&--&--&--&-- \\

                    \colourRow\ref{elsa}& \textbf{ELSA} & --
                    & -- & -- & -- &-- & \textbf{E} & -- &-- &--&--&--&--&-- \\

                    \midrule\bottomrule
                    \multicolumn{15}{r}{\footnotesize\textit{Continued on next
                            page}}
                \end{tabular}
            \end{table}
            \vspace*{\fill}}

        \clearpage

        {\topskip0pt
            \vspace*{\fill}
            \addtocounter{table}{-1}
            \captionsetup{list=no}
            \begin{table}[!tbh]
                \renewcommand{\arraystretch}{1.2}
                \centering
                \caption{(\textit{Continued.}) Theories Used for Emotion
                Elicitation \textcopyright{} 2022 IEEE}
                \small
                \begin{tabular}{@{}clccccccccccccc@{}}
                    \toprule

                    \multicolumn{1}{c}{} & \multicolumn{1}{c}{} & \textbf{Iz.} &
                    \textbf{V-A} & \textbf{PAD} & \textbf{Frj.} & \textbf{Laz.}
                    & \textbf{Sch.} & \textbf{Ros.} &\textbf{OCC} &
                    \textbf{S\&K} & \textbf{O\&JL} & \textbf{Slo.} &
                    \textbf{Dam.} & \textbf{LD} \\
                    \midrule

                    \colourRow\ref{gamae}& \textbf{GAMA-E} &
                    -- & -- & -- & -- &-- & -- & -- & \textbf{e}
                    &--&--&--&--&-- \\

                    \midrule

                    \ref{emile}& \textbf{\'Emile} & -- & -- &-- & -- &
                    \textbf{E}$^2$ & -- & -- & \textbf{e} (\ref{ar}, \ref{em})
                    & -- & \textbf{e} & \textbf{e} & -- &-- \\

                    \colourRow\ref{emotion}&
                    \textbf{EMOTION} & -- & --& --& --& --& --& --& \textbf{e}
                    (\ref{gvh})$^3$ &-- & --& --&--&--  \\

                    \ref{humdpme}&
                    \textbf{HumDPM-E} & -- & -- & -- & -- & -- & \textbf{e} &
                    -- & -- & -- & -- & -- & -- & -- \\

                    \colourRow\ref{jbdiemo}&
                    \textbf{JBdiEmo} & -- &-- &-- & --&-- & --& -- &
                    \textbf{e}$^4$ & -- &-- &--& --&-- \\

                    \ref{dett}& \textbf{DETT} &-- & -- & --& -- & --& -- &-- &
                    \textbf{e} &--&--&--&-- &-- \\

                    \colourRow\ref{epbdi}& \textbf{EP-BDI}
                    &-- & -- & --& -- & --& -- &-- & \textbf{e}$^1$
                    &--&--&--&-- &-- \\

                    \ref{microcrowd}& \textbf{MicroCrowd} & -- & -- & -- & -- &
                    -- & -- & -- & \textbf{E} & -- & -- & -- & -- & -- \\

                    \midrule

                    \colourRow\ref{puppet}& \textbf{Puppet}
                    & -- & -- &-- &-- & -- &-- & -- & \textbf{E} &-- & --&
                    --&--& -- \\

                    \ref{cbi}& \textbf{CBI} &-- & --& --& --& \textbf{E} &--
                    &-- &-- &--&--&-- &--&-- \\

                    \colourRow\ref{fatima}& \textbf{FAtiMA}
                    &-- & \textbf{e} &-- &--  & \textbf{e} (\ref{ema}) & -- &--
                    & \textbf{e} (\ref{ema}, \ref{s3a}) & -- & -- &--&--&-- \\

                    \ref{tardis}& \textbf{TARDIS} &-- &-- &-- &-- & -- & -- &--
                    & \textbf{e} & -- &-- &--& --&-- \\

                    \colourRow\ref{puma}&
                    \textbf{PUMAGOTCHI} &-- &-- &-- &--  & -- & -- &-- &
                    \textbf{e} & -- &-- &--& --&-- \\

                    \midrule

                    \ref{greta}& \textbf{Greta}$\ddagger$ & -- & -- & -- & -- &
                    -- & -- & -- & \textbf{e} & -- & \textbf{e} & -- & -- & --
                    \\

                    \colourRow\ref{alma}& \textbf{ALMA} & --
                    & -- &-- &--  & --& -- & --& \textbf{e} & -- & --&--&--&--
                    \\

                    \ref{eva}& \textbf{Eva} & -- &
                    --& -- & -- & -- & -- &-- & \textbf{e}$^{1}$
                    &--&--&--&--&-- \\

                    \colourRow\ref{ppad}& \textbf{PPAD-Algo}
                    & -- &-- & \textbf{E} & -- & -- & -- &-- & \textbf{e}
                    (\ref{alma}) &--&--&--&-- &-- \\

                    \ref{erdams}& \textbf{ERDAMS} & -- & -- & -- & -- & -- &
                    \textbf{E} & \textbf{e} & \textbf{E} (\ref{ar}, \ref{em})
                    &--&--&--&--&-- \\

                    \colourRow\ref{teatime}&
                    \textbf{TEATIME}$\ddagger$ & -- & -- & -- &\textbf{E} & --
                    & -- & \textbf{E} & -- & -- & -- & -- & -- & -- \\

                    \ref{mmt}& \textbf{MMT} & -- & -- & -- & -- & -- &-- & --&
                    \textbf{e} &--&--& -- & -- &-- \\

                    \colourRow\ref{presence}&
                    \textbf{Presence} & -- & -- & -- & \textbf{e} & -- & -- &
                    --& \textbf{E} &--&--& \textbf{e} & \textbf{e} &-- \\

                    \midrule

                    \ref{pomdp}& \textbf{POMDP-CA} & -- & -- &-- & -- &-- &-- &
                    \textbf{E} & -- & -- &-- & -- & -- & -- \\

                    \colourRow\ref{iphonoid}&
                    \textbf{iPhonoid} & -- & -- &-- & -- & --&-- & -- &
                    \textbf{e} & -- & -- & -- &  -- & -- \\

                    \ref{eegs}& \textbf{EEGS} & -- &-- &--  &-- & -- &
                    \textbf{e} & -- & \textbf{e} & \textbf{E} &--& --&--&-- \\

                    \colourRow\ref{kismet}& \textbf{Kismet}
                    & -- & -- & --& -- &-- & --& -- & -- & --&-- & --&
                    \textbf{E} & -- \\

                    \midrule\bottomrule
                    \multicolumn{15}{r}{\footnotesize\textit{Continued on
                            next page}}
                \end{tabular}
            \end{table}
            \vspace*{\fill}}

        \clearpage

        {\topskip0pt
            \vspace*{\fill}
            \addtocounter{table}{-1}
            \captionsetup{list=no}
            \begin{table}[!tbh]
                \renewcommand{\arraystretch}{1.2}
                \centering
                \caption{(\textit{Continued.}) Theories Used for Emotion
                    Elicitation \textcopyright{} 2022 IEEE}
                \small
                \begin{threeparttable}
                    \begin{tabular}{@{}clccccccccccccc@{}}
                        \toprule

                        \multicolumn{1}{c}{} & \multicolumn{1}{c}{} &
                        \textbf{Iz.} & \textbf{V-A} & \textbf{PAD} &
                        \textbf{Frj.} & \textbf{Laz.} & \textbf{Sch.} &
                        \textbf{Ros.} &\textbf{OCC} & \textbf{S\&K} &
                        \textbf{O\&JL} & \textbf{Slo.} & \textbf{Dam.} &
                        \textbf{LD} \\
                        \midrule

                        \colourRow\ref{grace}&
                        \textbf{GRACE} &
                        -- & --& --& -- & -- & \textbf{e} & -- & \textbf{e} &
                        --
                        &-- &--&--& -- \\

                        \midrule

                        \ref{feelme}& \textbf{FeelMe} &-- & --& -- & --& -- &
                        \textbf{e} &-- & -- & \textbf{e} &--&--&--&-- \\

                        \colourRow\ref{socio}&
                        \textbf{SocioEmo} & -- & -- & -- & -- & --&  -- & -- &
                        \textbf{E} (\ref{emile}, \ref{em}) & -- & -- &-- &--&--
                        \\

                        \ref{soul} & \textbf{The Soul}$\ddagger$ & -- & -- & --
                        & -- & -- & -- & -- & \textbf{e} (\ref{alma}) & -- & --
                        &
                        -- & -- & -- \\

                        \colourRow\ref{gamygdala}&
                        \textbf{GAMYGDALA} & -- & -- & -- &-- &--& -- & -- &
                        \textbf{E} (\ref{em}) & -- & --& -- &--&-- \\

                        \ref{mobsim}& \textbf{MobSim} &-- &--& -- & --& -- & --
                        &-- & \textbf{E}$^5$ & -- &--&--&--&-- \\

                        \colourRow\ref{apf}& \textbf{APF} &
                        -- & -- & -- & -- & -- &  -- & -- & \textbf{e$^4$}
                        (\ref{socio}) & -- & -- &-- &--& -- \\

                        \ref{mexica}& \textbf{MEXICA} & -- & -- & -- & -- & --
                        & -- & -- & \textbf{e} & -- & -- & -- & -- & -- \\

                        \colourRow\ref{npe}& \textbf{NPE} &
                        -- & -- & -- & -- & -- & -- & -- & \textbf{e} & -- & --
                        & -- & -- & -- \\

                        \ref{em}& \textbf{Em/Oz} & --&-- & --& -- & -- & -- &
                        --& \textbf{E} &--&--&--&--&-- \\

                        \colourRow\ref{s3a} & \textbf{S3A} &
                        -- & -- &-- & \textbf{E} & -- & -- & -- & \textbf{e} &
                        -- & -- &--&--&-- \\

                        \midrule\bottomrule
                    \end{tabular}
                    \begin{tablenotes}

                        \footnotesize
                        \vspace*{2mm}

                        \item \textbf{E}: \textit{Reasons for choosing the
                        theory are clear;} \textbf{e}: \textit{Reasons are
                        unclear;} (\#): \textit{System borrowed from/is
                        influenced by System \#}

                        \item [$\ddagger$] \textit{Used as a consequence of
                        theories chosen for emotion representation
                        (Section~\ref{sec:repr}).}

                        \item [1] \textit{Based on
                        \citep[p.~193]{ortony2002making}, a simplified model
                        developed by Ortony for believable ``artifacts''
                        (\citepg{kasap2009making}{23};
                        \citepg{salichs2012new}{62};
                        \citepg{ochs2009simulation}{285}).}

                        \item [2] \textit{Not yet implemented
                        (\citepg{soarEmotion}{271}; \citepg{lida}{26};
                        \citepg{gratch2000emile}{331}).}

                        \item [3] \textit{Builds on
                        \citep[p.~2]{egges2004generic}, a successor of GVH
                        (\ref{gvh}).}

                        \item [4] \textit{Uses the OCCr model~\citep[p.~195,
                        197]{korecko2016jadex} a reinterpretation of the OCC
                        model that aims to clarify the model's logical
                        structure and address
                        ambiguities~\citep[p.~1]{steunebrink2009occ}.}

                        \item [5] \textit{Process derived from
                        \citep{bartneck2002integrating}.}

                    \end{tablenotes}
                \end{threeparttable}%
            \end{table}
            \vspace*{\fill}}
    \end{landscape}
    \captionsetup{list=yes}
}

\afterpage{\clearpage
    \begin{landscape}
        {\topskip0pt
            \vspace*{\fill}
            \begin{table}[!tbh]
                \renewcommand{\arraystretch}{1.2}
                \centering
                \caption{Theories Used for Emotion Expression \textcopyright{}
                2022 IEEE}
                \label{tab:expressionOverview}
                \small
                \begin{tabular}{@{}clccccccccccccc@{}}
                    \toprule
                    &  & \textbf{Iz.} & \textbf{Ek.} & \textbf{Plu.} &
                    \textbf{V-A} & \textbf{PAD} & \textbf{Frj.} & \textbf{Laz.}
                    & \textbf{Ros.} & \textbf{OCC} & \textbf{S\&K} &
                    \textbf{Slo.}  & \textbf{Dam.} & \textbf{LD} \\
                    \midrule

                    \colourRow\ref{ar}& \textbf{AffectR}
                    &--& --& --& -- & -- & --& --&--& \textbf{x}$^{1}$ & --&--
                    &-- & --\\

                    \ref{cathexis}& \textbf{Cathexis} & -- & -- & -- & -- & --
                    & -- & -- & -- & -- &--& --& \textbf{x} &--\\

                    \colourRow\ref{emMod}& \textbf{EmMod} &
                    -- & \textbf{x} & -- & -- & -- & -- & -- &--& -- & -- & --&
                    \textbf{x} &--\\

                    \ref{scream}& \textbf{SCREAM} & -- & \textbf{x} &-- &  --&
                    -- & -- & -- &-- & \textbf{x}$^{2}$ & --&-- &--& --\\

                    \colourRow\ref{tabasco} &
                    \textbf{TABASCO} & --& --& --& -- & --& \textbf{X} &
                    \textbf{x} &--&-- &--&-- &--& --\\

                    \ref{wasabi}& \textbf{WASABI} & -- & \textbf{X} & --& --&
                    -- & --& -- & --&--&-- &--& --&--\\

                    \colourRow\ref{akr}& \textbf{AKR} & -- &
                    \textbf{x} & --& --& --& \textbf{x} & -- & -- & -- & --&
                    --& --&--\\

                    \ref{gvh}& \textbf{GVH} & -- & \textbf{X} &-- &--  & -- &
                    -- & -- &-- &--&-- &--& --&--\\

                    \colourRow\ref{parlee}& \textbf{ParleE}
                    & \textbf{X} & \textbf{X} &-- & -- &-- & --& -- &-- &--&--
                    &--& --&--\\

                    \ref{genia3}& \textbf{GenIA$^3$} &--  & -- &-- & -- & -- &
                    -- & \textbf{x} (\ref{ema}) &-- & \textbf{x}$^{2}$ &--&--
                    &--& --\\

                    \colourRow\ref{infra}& \textbf{InFra}
                    &--  & -- &-- & -- & -- & -- & -- & -- &--&-- &--& --&
                    \textbf{x} \\

                    \midrule

                    \ref{soar}& \textbf{Soar} &-- & --& --& \textbf{X} & -- &
                    -- & -- & --&--&-- &--& --&--\\

                    \colourRow\ref{clarion}&
                    \textbf{CLARION} &-- & --& --& --& -- & -- & \textbf{x} &--
                    &--&--&-- &--& --\\

                    \midrule

                    \ref{acres}& \textbf{ACRES} &-- &-- & --& --&--  &
                    \textbf{X} & -- &-- &-- &-- &--& --&--\\

                    \colourRow\ref{ema}& \textbf{EMA} &-- &
                    -- & --& --& -- & \textbf{x} & \textbf{x} &-- &--&
                    \textbf{x} &--&--& --\\

                    \ref{will}& \textbf{Will} &-- & -- & --& --& -- &
                    \textbf{X} & -- &-- & -- &-- &--& --&--\\

                    \midrule

                    \colourRow\ref{emile}& \textbf{\'Emile}
                    & --& --&-- &-- & -- &-- &-- & --& --&--& --& \textbf{x}
                    &--\\

                    \midrule

                    \ref{puppet}& \textbf{Puppet} & -- & \textbf{X} & --& --&
                    --& --& --& -- &-- &-- &--& --&--\\

                    \colourRow\ref{cbi}& \textbf{CBI}
                    &-- &-- &-- &-- &--  & -- & \textbf{X} & --&-- &-- &--&
                    --&--\\

                    \ref{fatima} & \textbf{FAtiMA} &-- &-- & --& --& --&--
                    & \textbf{x} (\ref{cbi}) & --& --&--& \textbf{x} &--
                    &--\\

                    \midrule

                    \colourRow\ref{greta}&
                    \textbf{Greta} & -- & \textbf{X} & -- & -- & -- & -- &
                    -- &-- &--&-- &--& --&--\\

                    \ref{teatime}& \textbf{TEATIME} & -- & -- & -- & -- &
                    -- & -- & -- & \textbf{X} &--&-- &--& -- & --\\

                    \midrule\bottomrule
                    \multicolumn{14}{r}{\footnotesize\textit{Continued on next
                            page}}
                \end{tabular}
            \end{table}
            \vspace*{\fill}}

        \clearpage

        {\topskip0pt
            \vspace*{\fill}
            \addtocounter{table}{-1}
            \captionsetup{list=no}
            \begin{table}[!tbh]
                \renewcommand{\arraystretch}{1.2}
                \centering
                \caption{(\textit{Continued.}) Theories Used for Emotion
                Expression \textcopyright{} 2022 IEEE}
                \small
                \begin{threeparttable}
                    \begin{tabular}{@{}clccccccccccccc@{}}
                        \toprule
                        &  & \textbf{Iz.} & \textbf{Ek.} & \textbf{Plu.} &
                        \textbf{V-A} & \textbf{PAD} & \textbf{Frj.} &
                        \textbf{Laz.} & \textbf{Ros.} & \textbf{OCC} &
                        \textbf{S\&K} & \textbf{Slo.}  & \textbf{Dam.} &
                        \textbf{LD} \\
                        \midrule

                        \colourRow\ref{presence}&
                        \textbf{Presence} & -- &-- &-- & \textbf{X} & -- & -- &
                        -- &--& \textbf{X} & -- &-- &--& --\\

                        \midrule

                        \ref{iphonoid}& \textbf{iPhonoid} & -- & \textbf{x} &
                        -- & -- &-- & -- & -- & --&--&-- &--& --&--\\

                        \colourRow\ref{kismet}&
                        \textbf{Kismet} & \textbf{X} & \textbf{x} & \textbf{X}
                        & \textbf{X$^{3}$} &-- & -- & -- & --&--&-- &--& --&--\\

                        \ref{grace}& \textbf{GRACE} &-- &-- &-- &-- & -- & -- &
                        \textbf{x} &-- &--&-- &--& --&--\\

                        \midrule

                        \colourRow\ref{soul}& \textbf{The
                        Soul} & -- & \textbf{X} & -- &-- & \textbf{X} & -- & --
                        &-- &--&-- &--& --&--\\

                        \ref{em}& \textbf{Em/Oz} & --&-- &-- & --& --&-- &--&
                        --& \textbf{X$^{1}$} &--&-- &--& --\\

                        \colourRow\ref{s3a} & \textbf{S3A}
                        &--  &-- &-- & --&-- & -- & --&--& \textbf{X}
                        (\ref{em}) &--&-- &--& --\\

                        \midrule\bottomrule
                    \end{tabular}
                    \begin{tablenotes}

                        \footnotesize
                        \vspace*{2mm}

                        \item \textbf{X}: \textit{Reasons for choosing the
                        theory are clear;} \textbf{x}: \textit{Reasons are
                        unclear;} (\#): \textit{System borrowed from/is
                        influenced by System \#}

                        \item [1] \textit{Based on an unpublished work which
                        AffectR describes~\citep[p.~50]{elliott1989ective} and
                        Em/Oz duplicates and expands
                        on~\cite[pp.~104]{reilly1996believable}.}

                        \item [2] \textit{Based on \citep[p.~193,
                        198]{ortony2002making}, a simplified model developed by
                        Ortony for believable ``artifacts''
                        (\citepg{prendinger2004mpml}{234};
                            \citepg{alfonso2017toward}{5:5}).}

                        \item [3] \textit{Also uses a \texttt{stance} dimension
                        to measure the approachability of a
                        stimulus~\citep[p.~133, 140]{breazeal2003emotion}.}

                    \end{tablenotes}
                \end{threeparttable}%
            \end{table}
            \vspace*{\fill}}
    \end{landscape}
    \captionsetup{list=yes}
}

\subsection{Emotion Representation}\label{sec:repr}
Twenty-eight CMEs appear to use the same theory to represent \textit{and}
elicit emotion, with the decision driven by elicitation requirements (marked
with a $\dagger$ in Table~\ref{tab:reprOverview}). The others make
representation choices independently of elicitation or appear to start with a
representation and build an elicitation process from it (see
Section~\ref{sec:elicit}).

Four CMEs reference Plu. for emotion representation because of its ability to
``create'' new emotions as combinations of its emotion categories (i.e. InFra
(\ref{infra})~\citep[p.~35]{castellanos2018computational}\footnote{\label{foot:infra}
    Inferred from InFra's (\ref{infra}) design
    goals~\citep[p.~27]{castellanos2018computational}.}, HybridC
(\ref{hybridc}) alongside Iz.~\citep[p.~63]{jain2019modeling}, SOM
(\ref{som})~\citep[p.~217--218]{yanaru1997emotion}, and PWE-I (\ref{pwe})).
PWE-I also uses Plu.'s emotion structure which can be implemented as a 2D
space, affording emotion dynamics and interactions while also using its emotion
categories~\citep[p.~210--211]{qi2019building}.

Kismet (\ref{kismet}) uses a dimensional space that includes V-A to combine
disparate information sources and unify the emotion elicitation process,
internal representations, and facial expression generation~\citep[p.~133, 148,
151]{breazeal2003emotion}. However, it also found that a third dimension,
\textit{stance}, was necessary to prevent accidental activation of emotions that
are similar in the simpler 2D space~\citep[p.~139--140]{breazeal2003emotion}.
PAD, a 3D space, appears in eleven CMEs for emotion representation as a common
space to define elicitation and expression mechanisms, as well as their
interactions. FeelMe (\ref{feelme}) uses PAD because ``[it] argues that any
emotion can be expressed in terms of values on these three dimensions, and
provides extensive evidence for this claim...makes his three dimensions
suitable for a computational approach. Second, since the PAD scales are
validated for both emotional-states and traits, they provide a useful basis for
a computational framework that consistently integrates states and
traits...provides an extensive list of emotional labels for points in the PAD
space''~\citep[p.~212]{broekens2004scalable}. The Soul (\ref{soul}) uses PAD
because it is ``[a] simple yet powerful model for representing emotional
reactions...'', ``... is able to represent a broad range of emotions. It can be
compared to creating a whole spectrum of colours using only red, green and
blue'', and ``...it uses only three axes, which furthermore are almost
orthogonal to each other, as we are used to, for example, in 3D
space''~\citep[p.~338--339]{bidarra2010growing}. WASABI (\ref{wasabi}) uses PAD
because it felt that ``...three dimensions are necessary and sufficient to
capture the main elements of an emotion's connotative meaning---at least in
case of simpler emotions such as primary or basic
ones''~\citep[p.~58]{becker2008wasabi}. A dimensional space also affords
numerical measurements and calculations so that emotions and other types of
affect can influence each other and another view of the emotion
state~\citep[p.~89, 97]{becker2008wasabi}.

Eleven of the CMEs using PAD pair it with OCC for emotion representation.
GAMYGDALA (\ref{gamygdala}) starts with OCC because it is ``...a well-known and
accepted theory of emotions, it is a componential model of emotion that fits
the needs of a computational framework, components are generic enough to allow
for a wide set of emotions, it accounts for both internal emotions and social
relationships which in games are quite important, and most importantly many
computational models have been built on it'', combining it with PAD because it
``...complements the OCC model...''~\citep[p.~33, 37]{popescu2014gamygdala}.
Eight CMEs reference ALMA (\ref{alma}) for the combination of OCC and
PAD (\citepg{shvo2019interdependent}{68};
\citepg{alfonso2017toward}{5:17--5:18}; \citepg{jones2013tardis}{5};
\citepg{kasap2009making}{24}; \citepg{zhang2016modeling}{216--217, 224};
\citepg{ochs2009simulation}{289}; \citepg{bidarra2010growing}{340};
\citepg{durupinar2016psychological}{2146--2148}). ALMA maps OCC emotion
categories to points in PAD space to afford interactions with other types of
affect~\citep[p.~31]{gebhard2005alma}, and MobSim (\ref{mobsim}) found that
using PAD as an intermediary representation between elicitation and expression
prevents ``erratic behaviours'' due to rapid changes in emotion
intensity~\citep[p.~2151--2152]{durupinar2016psychological}. APF (\ref{apf})
does \textit{not} reference PAD to accompany its use of OCC, but \textit{does}
create a dimensional space using Multidimensional Scaling
(MDS)~\citep[p.~698--699]{klinkert2021artificial}.

Representing emotions with OCC \textit{does} appear to be connected to how CMEs
elicit emotions, perhaps by limiting which categories a CME includes due to the
needs of the domain (i.e. Maggie (\ref{maggie})~\citep[p.~62]{salichs2012new},
GAMA-E (\ref{gamae})~\citep[p.~94]{bourgais2017enhancing}, EMOTION
(\ref{emotion})~\citep[p.~2]{el2004modelling}, TARDIS
(\ref{tardis})~\citep[p.~2]{jones2013tardis}, MMT
(\ref{mmt})~\citep[p.~9844]{alepis2011automatic}, GRACE
(\ref{grace})~\citep[p.~137--139]{dang2009experimentation}, NPE
(\ref{npe})~\citep[p.~118]{shirvani2020formalization}), but there might be
other reasons too. For example, Puppet (\ref{puppet}) and Presence
(\ref{presence}) use OCC because it is ``...readily amenable to the intentional
stance, and so ideally suited to the task of creating concrete
representations/models of...emotions with which to enhance the illusion of
believability in computer characters.''~\citep[p.~151]{andre2000integrating}.
WASABI (\ref{wasabi}) requires an emotion representation that depends on
cognition (``secondary emotions'') because it ``...affords a more complex
interconnection of the agent's emotion dynamics and its cognitive reasoning
abilities''~\citep[p.~93]{becker2008wasabi}. It uses OCC for this, choosing a
subset of emotion categories that rely on past events and future
expectations~\citep[p.~87, 100]{becker2008wasabi}. WASABI makes a clear
distinction between cognitive and non-cognitive-dependent emotions, choosing
simpler emotion representations from other theories, even though they are
defined in OCC: \textit{Fear} as proposed by LD due to its work on animal brain
studies~\citep[p.~47, 87]{becker2008wasabi}; and Plu. to define \textit{Anger}
as a reactive response tendency~\citep[p.~85]{becker2008wasabi}.

CMEs also use OCC to represent emotion because it: distinguishes emotions about
the self and about others (``empathetic emotions'') necessary for some
conversational agents like Greta (\ref{greta})~\citep[p.~94,
111]{rosis2003from}\footnote{Inferred from Greta's
    example~\citep[p.~84]{rosis2003from}.}, Eva (\ref{eva})~\citep[p.~21,
23]{kasap2009making}, and ERDAMS (\ref{erdams})~\citep[p.~412]{ochs2012formal},
social simulations like GAMA-E
(\ref{gamae})~\citep[p.~94]{bourgais2017enhancing}, EP-BDI
(\ref{epbdi})~\citep[p.~425--426]{zoumpoulaki2010multi}, and MicroCrowd
(\ref{microcrowd})~\citep[p.~91]{lhommet2011never}, and narrative planners
like MEXICA (\ref{mexica})~\citep[p.~90]{y2007employing} and NPE
(\ref{npe})~\citep[p.~118]{shirvani2020formalization}; represents emotion as
both categories and classes of triggering conditions (i.e. PUMAGOTCHI
(\ref{puma})~\citep[p.~64]{laureano2012design}, SocioEmo
(\ref{socio})~\citep[p.~282, 285]{ochs2009simulation}); and its hierarchical
organization of emotion categories (APF
(\ref{apf})~\citep[p.~703]{klinkert2021artificial}, Em/Oz
(\ref{em})~\citep[p.~73]{reilly1996believable}).

Six CMEs that aim to express emotions via facial expressions (see
Section~\ref{sec:express}) chose an emotion representation to ensure a smooth
connection between them. Ek. emotion categories appear for this, as in TAME
(\ref{tame}) ``...in part because these basic emotions have universal,
well-defined facial expressions, are straightforwardly elicited, and would be
expected, perhaps subconsciously, on a humanoid's face, as appearance does
affect expectations''~\citep[p.~211]{moshkina2011tame} and AEE
(\ref{aee})~\cite{wilson2000artificial}. It is possible to generate expressions
from affective dimensions, but these might be more difficult than distinct
categories such as those provided by Ek.~\citep[p.~324]{kirby2010affective}. GVH
(\ref{gvh}) uses the OCC categories to define emotions, but reorganizes and
expands them into the six categories defined by Ek. which ``...enables us to
handle relatively less number of emotional states still retaining completeness
necessary for expressive
conversation''~\citep[p.~108--109]{kshirsagar2002multilayer}. Puppet
(\ref{puppet}) chose a subset of OCC emotions to match Ek. facial expressions
due to evidence of the associated emotions' universality and distinctive facial
expressions which children can recognize~\citep[p.~155]{andre2000integrating},
whereas Greta (\ref{greta}) also cites Ek. and OCC for their representation but
adds facial expressions to match the possible emotion representation
states~\citep[p.~91]{rosis2003from}. Representations based on categories from
Ek., Iz., Plu., and/or O \& JL have also been cited for reasons such as:
evidence of universality and facial
expressions~\citep[p.~71]{velasquez1998robots}; ``...they are easy to explain
and understand''~\citep[p.~65]{ushida1998emotion}; their association with
evolutionary, cross-species, and social
functions~\citep[p.~129]{breazeal2003emotion}; and their connection to emotions
that are hard-wired and do not require cognitive processing (``primary
emotions'' in WASABI (\ref{wasabi})~\citep[p.~84, 100]{becker2008wasabi},
HybridC (\ref{hybridc})~\citep[p.~63--64]{jain2019modeling}).

Three CMEs appear to choose their emotion representation before their
elicitation methods because they align with the CMEs' goals. Peedy
(\ref{peedy}) represents emotion with V-A because it ``...corresponds more
directly to the universal responses...that people have to the events that
affect them''~\citep[p.~199--200]{ball2000emotion}. TEATIME (\ref{teatime})
aims to strongly connect emotion to speech acts, stating that ``...emotions
cannot be reduced to a label or a vector: these are only a description of the
state of the individual'', and therefore focuses on ``...action
tendency...defined as the will to establish, modify, or maintain a particular
relationship between the person and a stimulus'' as defined by
Frj.~\citep[p.~144, 145--150]{yacoubi2018teatime}. This led it to draw from
both Frj. and Ros., which emphasize action tendencies in their theories, to
represent emotion. In the case of AKR (\ref{akr}), part of its goal is to
define a taxonomy of emotion and other types of affect~\citep[p.~594, 596--597,
599--600, 606]{lisetti2002can}. This lead it to pull from a range of emotion
theories to represent emotion: Ek., presumably to connect to facial
expressions; Sch., Ros., and OCC for appraisal variables, although Ros. is
presumably for representing \textit{Surprise}; and Frj. for action tendencies.

\subsection{Emotion Elicitation}\label{sec:elicit}
Three CMEs appear to choose theories for emotion elicitation based on their
choice for representation (marked with a $\ddagger$ in
Table~\ref{tab:elicitOverview}) and six others elicit emotion independently of
a theory with methods such as affine mapping and fuzzy inference mechanisms
(SOM (\ref{som})~\citep[p.~219, 244]{yanaru1997emotion}), Bayesian Networks
(Peedy (\ref{peedy})~\citep[p.~204]{ball2000emotion}), hard-coded values
(R-Cept (\ref{robo})~\citep[p.~324]{kirby2010affective}), and/or signal
processing-based approaches (TAME
(\ref{tame})~\citep[p.~211]{moshkina2011tame}, PWE-I
(\ref{pwe})~\citep[p.~211--213]{qi2019building}, AEE
(\ref{aee})~\citep{wilson2000artificial}). The rest ground elicitation methods
directly in emotion theories. Methods can be broadly grouped into cognitive and
non-cognitive elicitation and a CME need not be limited to one type.

None of the CMEs implement non-cognitive elicitation alone, instead realizing
it as a mechanism or process that complements cognitive elicitation. One CME,
Kismet (\ref{kismet}), uses Dam. alone to create a ``mixed'' elicitation
system~\citep[p.~133--134]{breazeal2003emotion}. Six CMEs use multiple,
coexisting theories for this purpose. TABASCO (\ref{tabasco}) references Sch.
and S \& K to create a multi-layer appraisal system which has different
appraisal mechanisms for different types of information~\citep[p.~265--266,
268--269]{petta2002role}. It also applies this to a Frj.-based monitor which
ensures that actions influence appraisals. The five remaining CMEs reference at
least one of Slo., Dam., and LD to define a ``mixed'' elicitation system.
Presence (\ref{presence}) differentiates cognitive and non-cognitive emotion
processes using Slo. and Dam. so that it aligns with recent \ref{ac}
research~\citep[p.~160--161]{andre2000integrating}. It implements non-cognitive
emotions using heuristics and combines Frj.'s process with OCC in a BDI model
for cognitive emotion elicitation. FLAME (\ref{flame}) uses LD for learning
non-cognitive, conditioned behaviour and Ros. and OCC for cognitive
appraisal~\citep[p.~227--228, 237--238]{el2000flame}. Cathexis (\ref{cathexis})
also uses LD for non-cognitive behaviour, this time combined with Dam. for
cognitive, memory-driven emotion
elicitation~\citep[p.~71--72]{velasquez1998robots}. It references Iz. to
differentiate between cognitive and non-cognitive emotion elicitors.

WASABI (\ref{wasabi}) references both OCC and Dam. for the division of its
cognitive layer into a reactive and reasoning layer to differentiate between
cognitive and non-cognitive elicitation~\citep[p.~50, 54, 84, 87, 90--92,
97--98, 102]{becker2008wasabi}. WASABI drew assumptions about Dam.'s connection
between memories and cognitive elicitation such that it could be formalized. In
a separate emotion module, it uses Slo. to define a dynamics system that
accepts valenced pulses as inputs and creates an ``alarm'' signal that is
translated into PAD as a primary emotion encoded by \textit{pleasure} and
\textit{arousal} values. WASABI uses OCC for cognitive emotion elicitation
because it requires high-level reasoning, but manually codes their intensity
values~\citep[p.~95, 100]{becker2008wasabi}.

The other 51 CMEs choose to focus exclusively on cognitive elicitation. Sch.
appears in the three cognitive architectures (\citepg{soarEmotion}{272--273,
277--278}; \citepg{sun2016emotion}{9, 11}), chosen---at least in part---for its
focus on the cognitive contents of emotion~\citep[p.~26--27]{lida}. ELSA
(\ref{elsa}) chose Sch. due to its dynamic systems view and focus on emotion as
emergent phenomena of time-dependent, componential changes rather than events
with specific labels~\citep[p.~99--102, 143]{meuleman2015computational}.
HumDPM-E (\ref{humdpme}) uses Sch. to generate emotion patterns such that each
possible emotion type is assigned a value~\citep[p.~74,
76]{aydt2011computational}. This allows HumDPM-E to define different agents
based on their ``susceptibility'' to different emotions. MAMID (\ref{mamid})
uses Sch., in combination with S \& K, for its domain independent appraisal
variables, multiple levels of resolution, multi-stage appraisals,
and---potentially---because they account for some effects of emotion on
cognition~\citep[p.~134, 136]{hudlicka2019modeling}. MAMID also draws from V-A
for part of its emotion intensity specification and to serve as another
perspective on the emotion state. FeelMe (\ref{feelme}) is also based on a
combination of Sch. and S \& K to enable a scalable design with modular
components~\citep[p.~210--211, 213]{broekens2004scalable}. It shows that this
structure can be combined with dimensions from other theories, such as PAD.
Although it also draws from Sch. for emotion elicitation, EEGS (\ref{eegs})
creates a parallel appraisal process as in S \& K: ``The rationale behind this
is that human brain is multi-processing and several evaluations occur
simultaneously. This is why EEGS uses multi-threading approach to represent the
true mechanism of emotion generation that occurs in
humans''~\citep[p.~236]{ojha2016ethically}. FAtiMA-M (\ref{fatimaM}), although
it implements OCC as its default, generalized its design requirements so that
it could represent Sch., ``...one of the most complex Appraisal
Theories''~\citep[p.~45]{dias2014fatima}.

Four CMEs choose theories because they provide functionality central to their
design, such as CBI's (\ref{cbi}) use of Laz. for its integration of coping in
appraisal~\citep[p.~301--302, 306]{marsella2000interactive}. ACRES
(\ref{acres}) and Will (\ref{will}) chose their underlying
theory---Frj.---because their aim is to implement that theory as a
computational system (\citepg{frijda1987can}{247};
\citepg{moffat1997personality}{138, 151--152}). POMDP-CA (\ref{pomdp}) uses
Ros., not for its functionality, but because ``[i]t has concrete definitions of
criteria of cognitive appraisal and a structure that is amenable to
computational implementation''~\citep[p.~268]{kim2010computational}.

Twenty-six CMEs use OCC alone to define rules and/or conditions for emotion
elicitation, both independently of an
architecture (\citepg{prendinger2004mpml}{230};
\citepg{kshirsagar2002multilayer}{109, 112};
\citepg{shvo2019interdependent}{68}; \citeg{kazemifard2011design};
\citepg{el2004modelling}{4--5}; \citepg{jones2013tardis}{4};
\citepg{laureano2012design}{63}; \citepg{gebhard2005alma}{33};
\citepg{kasap2009making}{23}; \citepg{alepis2011automatic}{9841};
\citepg{masuyama2018personality}{217, 220};
\citepg{popescu2014gamygdala}{37--39}; \citepg{klinkert2021artificial}{698,
702}; \citepg{y2007employing}{90}; \citepg{shirvani2020formalization}{117,
121}) or integrated into a Belief-Desire-Intention (BDI) design
(\citepg{alfonso2017toward}{5:2, 5:4--5:5, 5:12, 5:17--5:18};
\citepg{bourgais2017enhancing}{92}; \citepg{parunak2006model}{993--994};
\citepg{zoumpoulaki2010multi}{424}; \citepg{lhommet2011never}{90};
\citepg{andre2000integrating}{153--154}; \citepg{rosis2003from}{88, 94--95,
97}; \citepg{ochs2012formal}{417}). It is not always clear why CMEs use OCC,
but at least four reference its computational tractability
(\citepg{duy2004creating}{135--136}; \citepg{jain2019modeling}{66};
\citepg{lhommet2011never}{89}) and/or prevalence in \ref{ac}
\citep[p.~37]{popescu2014gamygdala}. Em/Oz (\ref{em}) is explicit in its
reasoning, stating that it chose OCC because it was ``...designed to be
implemented computationally...reasonably simple to understand...'', and because
Em/Oz's users ``...will not have much formal  psychology
training...''~\citep[p.~28, 52--54, 59--60]{reilly1996believable}. The emphasis
on computational tractability and intuitiveness motivated other versions of OCC
(e.g. \citet{ortony2002making}) which appear in CMEs
(\citepg{salichs2012new}{62}; \citepg{korecko2016jadex}{195, 197}). MobSim
(\ref{mobsim}) claims that OCC allows one to ``...formally define the rules
that determine an agent's evaluation of its surrounding events and
relationships with other agents, [providing] a suitable basis for crowd
simulation applications'' and uses \citet{bartneck2002integrating} to aid in
its mechanization of OCC~\citep[p.~2149--2150]{durupinar2016psychological}.
SocioEmo (\ref{socio}) uses the OCC version in \citet{ortony2002making},
partially because ``[g]ame developers are usually not specialists of AI
[Artificial Intelligence] or cognitive psychology. This guided us toward models
which are relatively simple to use''~\citep[p.~282, 285]{ochs2009simulation}.
GEmA (\ref{gema}) uses OCC for ``...events and actions assessment [because] it
includes comprehensive local and global variables to compute intensity of
emotions and methods for [assessing] events and
actions.''~\citep[p.~2642]{kazemifard2011design}. AffectR (\ref{ar}) is less
clear, but its focus on \textit{reasoning} about an agent's emotion might be
the motivation~\citep[p.~27, 30]{elliott1989ective}. Both AffectR and Em/Oz
have influenced later CMEs, such as ParleE
(\ref{parlee})~\citep[p.~117--125]{duy2004creating}, EMA
(\ref{ema})~\citep[p.~282--283, 285]{gratch2004domain}, \'Emile
(\ref{emile})~\citep[p.~326--329]{gratch2000emile}, and ERDAMS
(\ref{erdams})~\citep[p.~421--422]{ochs2012formal}.

Fifteen CMEs also use OCC for emotion elicitation but combine it with other
theories for their unique strengths, such as:
\begin{itemize}
    \item Emotion intensity functions based on a PAD vector space (PPAD-Algo
    (\ref{ppad})~\citep[p.~217, 223--224]{zhang2016modeling}), single
    dimensions like \textit{arousal} (FAtiMA
    (\ref{fatima})~\citep[p.~131]{dias2005feeling}), and explicit plan
    representations in Slo. and/or O \& JL (ParleE
    (\ref{parlee})~\citep[p.~117--125]{duy2004creating}, \'Emile
    (\ref{emile})~\citep[p.~328]{gratch2000emile})

    \item Ros. for defining eliciting conditions for \textit{Surprise} (ParleE
    (\ref{parlee})~\citep[p.~118--119, 135--136]{duy2004creating}, HybridC
    (\ref{hybridc})~\citep[p.~66]{jain2019modeling}) or \textit{Anger} (ERDAMS
    (\ref{erdams}~\citep[p.~417]{ochs2012formal}))

    \item Appraisal variables from Sch. (InFra (\ref{infra})~\citep[p.~30,
    32]{castellanos2018computational}, HybridC
    (\ref{hybridc})~\citep[p.~66]{jain2019modeling}, ERDAMS
    (\ref{erdams})~\citep[p.~416]{ochs2012formal}, EEGS
    (\ref{eegs})~\citep[p.~214--216]{ojha2018essence}, GRACE
    (\ref{grace})~\citep[p.~137--138]{dang2009experimentation})

    \item Elicitation process from Frj. because it ``...complements the OCC
    model'' (S3A (\ref{s3a})~\citep[p.~48]{martinho2000emotions})

    \item Laz. process (FAtiMA-M (\ref{fatimaM})~\citep[p.~44,
    46--48]{dias2014fatima}, EMA (\ref{ema})~\citep[p.~272]{gratch2004domain}),
    coping (\'Emile (\ref{emile})~\citep[p.~331]{gratch2000emile}, FAtiMA
    (\ref{fatima})~\citep[p.~130--134]{dias2005feeling}), or emotion themes
    (Maggie (\ref{maggie})~\citep[p.~60]{salichs2012new}) which are integrated
    into emotion elicitation

    \item Dam. to define a deliberative architecture layer that relies on
    cognition (EmMod (\ref{emMod})~\citep[p.~63]{ushida1998emotion})

    \item O \& JL to frame cognition as a knowledge transformation process to
    drive cognitive appraisals (Greta
    (\ref{greta})~\citep[p.~94]{rosis2003from})
\end{itemize}

\subsection{Emotion Expression}\label{sec:express}
Twenty-nine emotion-generating CMEs also specify how the emotion state is
expressed. Two CMEs draw from emotion theories to define an interface between
internal emotion states and external behaviour systems (e.g. InFra
(\ref{infra}) uses LD to define an emotion-to-expression
interface~\citep[p.~27]{castellanos2018computational}) and Presence
(\ref{presence}) uses OCC emotion types and V-A values to annotate actions such
as speech and body gesture
generation~\citep[p.~161--162]{andre2000integrating}). One CME relates their
potential emotions to the functions they serve (Kismet (\ref{kismet})
references Iz. and Plu. for this~\citep[p.~129]{breazeal2003emotion}), but
eleven reference action tendencies---``...readiness for different actions
having the same intent'' and that ``...account for behaviour
flexibility''~\citep[p.~70--71]{frijda1986emotions}.

CMEs use four emotion theories to define action tendencies (e.g. Laz. in FAtiMA
(\ref{fatima})~\citep[p.~131, 134]{dias2005feeling}, Ros. in TEATIME
(\ref{teatime})~\citep[p.~149--150]{yacoubi2018teatime}), the most commonly
referenced ones being Frj. and OCC. AKR is unclear in its choice to use Frj.
for this~\citep[p.~596--597]{lisetti2002can}, whereas TABASCO
(\ref{tabasco})---which uses an underlying system that is ``very close to the
functionality'' of Frj.---compares the action tendencies to ``flexible
programs'' that allow behaviour variations and can be influenced by feedback
processes~\citep[p.~267]{petta2002role}. ACRES (\ref{acres}) is an
implementation of Frj.~\citep[p.~247]{frijda1987can}. With Will as ACRES's
successor~\citep[p.~138, 146]{moffat1997personality}, Frj.'s use in these two
CMEs is unsurprising.

Two CMEs use a ``simplified'' version of OCC (\citepg{prendinger2004mpml}{234};
\citepg{alfonso2017toward}{5:5--5:6}) that includes a hierarchy of response
tendencies grouped by type~\citep{ortony2002making}. The hierarchy might be a
simplification of unpublished work intended for the full theory, which AffectR
(\ref{ar}) and Em/Oz (\ref{em})---which S3A (\ref{s3a}) builds
on~\citep[p.~52--53]{martinho2000emotions}---incorporate
(\citepg{elliott1989ective}{50--53}; \citepg{reilly1996believable}{86, 100,
104}). The hierarchy elements are not uniquely associated with OCC emotion
categories, so the hierarchy can be implemented to allow the categorization of
display mechanisms---encouraging modular development---and assign the same
behaviour to different tendencies, affording more control over emotional
displays.

CMEs targeting specific domains typically specify what types of behaviours
their CMEs produce. At least nine systems intended for face-to-face
interactions with people use facial expressions to convey emotion. Ek. is often
referenced for this. For example, GVH (\ref{gvh}) is concerned with the facial
representation of virtual humans and uses Ek.'s facial expression specification
because they are ``...recognized as universal by many facial expression and
emotion researchers''~\citep[p.~108--109]{kshirsagar2002multilayer}. Puppet
(\ref{puppet}) chose Ek. due to evidence of the associated emotions'
universality and distinctive facial expressions that children can
recognize~\citep[p.~155]{andre2000integrating}. ParleE (\ref{parlee}) cites the
universality of Ek. and Iz.'s given facial expressions, building a generation
system on FACS~\citep[p.~142--143, 146]{duy2004creating}, which documents
facial muscles with respect to expressions~\citep{facs}. Although unclear,
several other CMEs also seem to cite Ek. for its work on facial expressions
(\citepg{becker2008wasabi}{84, 100}; \citepg{ushida1998emotion}{66};
\citepg{lisetti2002can}{596--597}\footnote{This decision is inferred.};
\citepg{masuyama2017application}{740}), potentially in connection to
FACS~\citep[p.~88--89, 91]{rosis2003from}. Kismet (\ref{kismet}) and The Soul
(\ref{soul}) use Ek. to define points in dimensional models so that facial
expressions can be procedurally generated (\citepg{breazeal2003emotion}{140,
143}; \citepg{bidarra2010growing}{338, 340--343}). SCREAM (\ref{scream})
references Ek.'s rules for when emotions are outwardly displayed given social
and interaction contexts (``display rules'') to regulate when their CME can
show their emotions~\citep[p.~231--232]{prendinger2004mpml}.

Specific effects of emotions on behaviour can also refer to the effects that
emotions have on other processes within or directly connected to the CME.
Cathexis (\ref{cathexis}), EmMod (\ref{emMod}), and \'Emile (\ref{emile})
reference Dam. to specify how emotion influences decision-making and
planning (\citepg{velasquez1998robots}{72}; \citepg{ushida1998emotion}{63};
\citepg{gratch2000emile}{330}). EMA (\ref{ema}) draws from Frj. and S \& K to
define attentional focus necessary for coping~\citep[p.~286,
297]{gratch2004domain}.

CMEs tend to use Laz. when coping itself is central to the CME's purpose (CBI
(\ref{cbi})~\citep[p.~302, 306]{marsella2000interactive}) whose design has been
adopted by others (FAtiMA (\ref{fatima})~\citep[p.~131, 134]{dias2005feeling}),
and because it can be implemented with a planner when viewed as a ``planful
process'' (TABASCO (\ref{tabasco})~\citep[p.~267]{petta2002role}). EMA
(\ref{ema}) and GRACE (\ref{grace}) are unclear in their reasons for choosing
Laz. for coping (\citepg{gratch2004domain}{272, 278};
\citepg{dang2009experimentation}{136, 138}). CLARION (\ref{clarion}) uses Laz.
for coping so that emotions can influence decision-making, goal management, and
regulatory processes~\citep[p.~10, 12]{sun2016emotion}. GenIA$^3$
(\ref{genia3}) is more modest in its use of Laz.-based coping, allowing it to
return to a previous emotion state and/or modify the agent's
beliefs~\citep[p.~5:5--5:6]{alfonso2017toward}. Emotions can also influence:
learning, such as in Soar's (\ref{soar}) use of V-A to define reward signals
for reinforcement learning because the dimensions can be unified with appraisal
theories~\citep[p.~279--280]{soarEmotion}; and emotion-driven plan selection
such as the use of Slo. in FAtiMA
(\ref{fatima})~\citep[p.~134]{dias2005feeling}.

\section{Observations from the Survey}\label{sec:results}
Surveying CMEs and the affective theories they use brought out some
commonalities. We discuss some \textit{use trends} for each theory and
\textit{psychologist influences}.

\subsection{Use Trends}\label{sec:uses}
The CMEs use a variety of theories for different purposes
(Table~\ref{tab:theoryCounts}). It is not always clear why CMEs use particular
theories. However, there are clear trends in \textit{how} CMEs use affective
theories. Even in CMEs without a documented choice rationale, these uses align
with different aspects of the theories. This is indicative of their strengths,
which tend to be similar within each perspective.

\begin{table}[!tb]
    \renewcommand{\arraystretch}{1.3}
    \centering
    \caption{Number of Uses of Emotion Theories \textcopyright{} 2022 IEEE}
    \label{tab:theoryCounts}
    \small
    \begin{tabular}{lccc|c}
        \toprule
        & \textbf{\begin{tabular}[c]{@{}c@{}}Emotion \\
                Representation\end{tabular}} &
        \textbf{\begin{tabular}[c]{@{}c@{}}Emotion \\
                Elicitation\end{tabular}} &
        \textbf{\begin{tabular}[c]{@{}c@{}}Emotion
                \\ Expression\end{tabular}} & \textbf{Total}\\
        \hline

        \colourRow\textbf{OCC} & 42 & 46 & 6 & 94 \\

        \textbf{Ek.} & 12 & -- & 11 & 23 \\

        \colourRow\textbf{Sch.} & 8 & 15 & -- & 23 \\

        \textbf{PAD} & 13 & 1 & 1 & 15 \\

        \colourRow\textbf{Frj.} & 3 & 7 & 5 & 15 \\

        \textbf{Laz.} & 1 & 6 & 7 & 14 \\

        \colourRow\textbf{Ros.} & 5 & 7 & 1 & 13 \\

        \textbf{Dam.} & 1 & 5 & 3 & 9 \\

        \colourRow\textbf{V-A} & 3 & 2 & 3 & 8 \\

        \textbf{Plu.} & 6 & -- & 1 & 7 \\

        \colourRow\textbf{S \& K} & 2 & 4 & 1 & 7 \\

        \textbf{Iz.} & 3 & 1 & 2 & 6 \\

        \colourRow\textbf{O \& JL} & 2 & 3 & -- & 5 \\

        \textbf{Slo.} & 1 & 3 & 1 & 5 \\

        \colourRow\textbf{LD} & 1 & 3 & 1 & 5 \\

        \hline\bottomrule
    \end{tabular}%
\end{table}

We use the broad categories of~\citet{lisetti2015and}---\textit{discrete},
\textit{dimensional}, \textit{appraisal}, and \textit{neurophysiologic}---to
organize emotion theories. Other ways are available, such as
\citet[p.~280]{scherer2021towards}.

\subsubsection{Discrete Theories}
These appear when emotion ``types'' must be clearly distinguished. This reflects
a strength of discrete theories, which build a small set of emotion categories
that are theorized to have evolved via natural
selection~\citep[p.~305]{hudlicka2014computational}. The discrete perspective is
associated with the most empirical evidence of observed emotion effects to
emotions~\citep[p.~10]{hudlicka2014habits}. However, discrete theories do not
give many details on emotion generation processes, so they are often combined
with another theory or used in hand coded designs (e.g. R-Cept
(\ref{robo})~\citep[p.~324]{kirby2010affective}). This is due to a core
assumption that emotions are innate, hard-wired features with dedicated neural
circuitry which circumvents cognitive
processing~\citep[p.~250]{reisenzein2013computational}. There are differences
in the definition of ``primary'' emotions but these are not mutually
exclusive~\citep[p.~2--3]{ortony2021all}. However, they do change which emotions
are considered ``basic''. Ek. and Iz. are part of the ``biologically basic''
view, which tend to focus on facial expressions as indicators of primality,
whereas Plu. is part of the ``elemental'' view that seeks emotions that cannot
be defined with other emotions (i.e. ``mixtures'' of other
emotions). Still, identifying and labelling emotion categories helps delimit
them, making it easier to talk about them both formally and
informally (\citepg{broekens2021emotion}{353};
\citepg{scherer2021towards}{286}).

\paragraph{Izard (Iz.)}
Although it does not tend to appear by itself, CMEs use Iz. to define facial
expressions along side Ek. (e.g. ParleE (\ref{parlee})) and ``mixed'' emotions
and the functional role of emotions with Plu. (e.g. HybridC (\ref{hybridc}) and
Kismet (\ref{kismet})). This is likely because Iz. shares some of the
same assumptions with them~\citep[p.~64--65, 83, 85--92, 97]{izard1977human}.
However, there could be untapped potential in Iz., such as its differentiation
between cognitive and non-cognitive emotion elicitors (e.g. Cathexis
(\ref{cathexis})).

\paragraph{Ekman (Ek.)}
This theory is common in CMEs that express emotions via facial
expressions (Section~\ref{sec:express}), and often use Ek.'s emotion categories
to ensure a one-to-one mapping from internal state to facial configuration.
This aligns with Ek.'s focus~\citep[p.~1]{ekman2007emotions} and the
resulting FACS~\citep{facs}, which breaks the face down into individual muscles
and shows how they can combine into expressions. This makes Ek. a strong
candidate, potentially ``...the de facto standard for analysis and description
of facial expressions, and serves as the foundation of...the synthesis of
emotion expressions in virtual agents and
robots.''~\citep[p.~4]{hudlicka2014habits}.

Ek. could be combined with: Iz. (e.g. ParleE (\ref{parlee})), which shares
similar views~\citep[p.~3]{ekman2007emotions} and also has a system for
identifying facial expressions~\citep{izard1979maximally}; and O \& JL (e.g.
Cathexis (\ref{cathexis})) as there is deliberate overlap in their ``primary''
emotion categories~\citep[p.~209, 217]{johnson1992basic}.

\paragraph{Plutchik (Plu.)}
Plu. appears most often when CMEs want to represent ``mixed'' emotions as
combinations of emotion categories, which allows a CME to add ``more'' emotion
types. This is unsurprising, as Plu. ``...has one of the better developed
theories of emotion mixes''~\citep[p.~113]{ledoux1996emotional} and experiments
have shown that laypeople tend to agree on the components of emotion
``mixtures''~\citep[p.~204--205]{plutchik1984emotions}. Plu. identifies its
``primary'' emotions from evidence of a finite set of adaptive behaviours that
aim to maintain internal homeostasis by acting on the
environment~\citep[p.~203, 215]{plutchik1984emotions}. This effectively
connects behaviours to action tendencies~\citep[p.~72]{frijda1986emotions} and
motivations~\citep[p.~13]{scherer2010emotion}, which can help specify an
emotion's function (e.g. WASABI (\ref{wasabi}), Kismet (\ref{kismet})).

Using self-reports on the meanings of emotion words, Plu. arranges its emotion
categories on a circumplex~\citep[p.~204]{plutchik1984emotions}. This affords
the use of arbitrarily chosen axes because they are only reference
points~\citep[p.~13]{plutchik1997circumplex}, which can serve as affective
dimensions. The result is a 3D colour space analogy---with intensity as the
third dimension---that is familiar to computer
scientists~\citep[p.~21]{becker2008wasabi}. There is also evidence that the
circumplex can act as a common space for different types of
affect~\citep[p.~30--31]{plutchik1997circumplex}, which can help visualize
affective dynamics using Plu.'s colour analogy (e.g. PWE-I (\ref{pwe})).

\subsubsection{Dimensional Theories}
These appear when CMEs need a simple and effective emotion model, as another
perspective of emotion categories, and/or as a common space for modelling
different affective phenomena and their interactions. The dimensional
perspective's strength lies in its description of affect in a simple
way---usually two or three dimensions~\citep[p.~97]{lisetti2015and}---where any
affective phenomena, including emotions~\citep[p.~9]{hudlicka2014habits}, can
be mapped. However, the dimensions can lose information about an emotion state
if it has a higher information resolution than their dimensions can represent
(\citepg{broekens2021emotion}{353}; \citepg{schaap2008towards}{172}). This
might not be appropriate for all CMEs. Dimensional theories focus on what kind
of mental states emotions are, how to construct them, and how they fit into a
general taxonomy of mental states~\citep[p.~250]{reisenzein2013computational}.
Consequently, they say little about how to generate emotions and what their
effects are~\citep[p.~10]{hudlicka2014habits}, making them unsuitable for
defining a complete computational
model~\citep[p.~250]{reisenzein2013computational}.

\paragraph{Valence-Arousal (V-A)}
The \textit{valence} and \textit{arousal} dimensions are the two most widely
agreed on affective dimensions~\citep[p.~168]{picard1997affective} and are
common in dimensional theories~\citep[p.~280]{scherer2021towards}. They form a
simple model that captures most affective phenomena, including aspects of
emotion, in a numerical form that is computationally efficient and can be used
in emotion intensity functions (e.g. MAMID (\ref{mamid})), as inputs to other
CME processes (e.g. Soar (\ref{soar})), and to coordinate emotion expression
modalities (e.g. Presence (\ref{presence})) and/or generation (e.g. Kismet
(\ref{kismet})).

The ability to represent different kinds of affective information can make V-A
useful for combining disparate information sources and external behaviour
systems with a single representation, helping them work in concert so that the
CME ``...not only does the right thing, but also at the right time and in the
right manner''~\citep[p.~151]{breazeal2003emotion}. However, two dimensions
might not be enough to distinguish between every emotion a CME might
need~\citep[p.~139--140]{breazeal2003emotion}. This implies that V-A is only
ideal for CMEs with a set of emotions that are conceptually easy to distinguish
both as internal representations and external expressions.

\paragraph{Pleasure-Arousal-Dominance Space (PAD)}
PAD is similar to V-A. Its \textit{pleasure} dimension fills the same role as
\textit{valence}, and \textit{arousal} is shared by both theories. The third
dimension, \textit{dominance}, distinguishes emotions such as \textit{Anger}
and \textit{Fear} that are otherwise indistinguishable (i.e. have similar
\textit{valence} and \textit{arousal} values) by quantifying how much control
one believes they have (i.e. one tends to feel that they have low control when
experiencing  \textit{Fear}, and high control in
\textit{Anger})~\citep[p.~263--264]{mehrabian1996pleasure}. The empirical
nature and ability to map emotions to three continuous dimensions might make PAD
easy to understand using parallels to RGB colour
space~\citep[p.~339]{bidarra2010growing} and ``...suitable for a computational
approach''~\citep[p.~212]{broekens2004scalable}.

As with V-A, CMEs often choose PAD to specify a simple model for representing
emotion and its interactions with other types of affect (e.g. CMEs \ref{alma},
\ref{iphonoid}, \ref{feelme}) that can also be used in numerical-based
functions such as emotion intensity (e.g. PPAD-Algo (\ref{ppad})), affective
dynamics (e.g. WASABI (\ref{wasabi})), facial expression generation (e.g. The
Soul (\ref{soul})) and behaviour mediation (e.g. MobSim (\ref{mobsim})), or as
an alternate view of emotion categories (e.g. WASABI (\ref{wasabi}), GAMYGDALA
(\ref{gamygdala})). PAD's pervasiveness in CMEs suggests its usefulness for
creating a unified space for multiple types of affect and interfacing between
theories. Caution is required as soundness depends on how rigorously concepts
are matched.

\subsubsection{Appraisal Theories}
CMEs often use these theories (\citepg{lisetti2015and}{97};
\citepg{marsella2015appraisal}{55}). This might be due to the theories' ability
to comprehensively represent the complexity of emotion processes, receiving
consistent empirical support for their hypothesized
mechanisms~\citep[p.~281--282]{scherer2021towards}. However, they are based on
cognition and CMEs seeking to use these theories must be able to account for
it~\citep[p.~354]{broekens2021emotion}.

Appraisal theories emphasize distinct components of emotion, including
appraisal dimensions or variables~\citep[p.~97]{lisetti2015and}. Analyzing
stimuli for meaning and consequences with respect to an individual produce
values for these variables (\citepg{reisenzein2013computational}{250};
\citepg{smith1985patterns}{819}), regardless of process
sophistication~\citep[p.~273]{gratch2004domain} and independent of biological
processes~\citep[p.~559]{arbib2004emotions}. Appraisals are continuous and
change with the situation, the individual's behaviours, and their attempts to
appraise the situation differently~\citep[p.~28]{siemer2007appraisals}. This
can account for the personal, transactional, and temporal character of emotion
with respect to a changing environment, applied coping strategies, and
continuous appraisals. This makes appraisal theories of particular interest for
decision-making, action selection, facial animations, and
personality~\citep[p.~274]{gratch2004domain}. However, there is little
empirical data associating individual appraisal variables to expressive
behaviours or behavioural choices~\citep[p.~10]{hudlicka2014habits}.

\paragraph{Frijda (Frj.)}
CMEs tend to use Frj. to explicitly connect emotions to action tendencies (e.g.
TEATIME (\ref{teatime})) and define an action-driven appraisal process (e.g.
CMEs \ref{tabasco}, \ref{acres}, \ref{will}). This aligns with Frj.'s proposal
that emotions---outputs of a continuous information processing system---are
changes in action readiness~\citep[p.~453, 466]{frijda1986emotions}. ``Action
readiness'' refers to motivational states which are associated with goals
rather than actions or behaviours~\citep[p.~143]{frijda2001appraisal}. Frj.'s
description of action tendencies appears to transfer to designs that do not
implement its appraisal process (e.g. AKR (\ref{akr}), Presence
(\ref{presence})), since many of the identified action tendencies are
associated with an emotion label~\citep[p.~87--90]{frijda1986emotions}.

The conceptualization of emotion elicitation as an information processing
system is a useful analogy and can provide the necessary mechanization
framework for structure-oriented theories like Ros. (e.g. TEATIME
(\ref{teatime})) and OCC (e.g. S3A (\ref{s3a})). It is also possible to
abstract and apply different elements independent of the broader theory, such
as implicit appraisal checks (e.g. TABASCO (\ref{tabasco})), information
filtering (e.g. S3A (\ref{s3a})), and mechanisms whose behaviour
changes with the system state (e.g. EMA (\ref{ema}), TAME (\ref{tame})).

\paragraph{Lazarus (Laz.)}
CMEs tend to use Laz. to specify coping behaviour, a deliberative process
whereby the individual can suppress action tendencies and choose other
strategies to influence the current situation~\citep[p.~628]{smith1990emotion}.
The appearance of Laz. in this context is unsurprising, as coping plays a
critical role in the theory~\citep[p.~39--40]{lazarus1991emotion}. Coping can
be incorporated directly into the appraisal process as an influencing factor
(e.g. CMEs \ref{genia3}, \ref{ema}, \ref{cbi}) and to plan agent behaviours
(e.g. CMEs \ref{tabasco}, \ref{clarion}, \ref{emile}, \ref{grace}). FAtiMA
(\ref{fatima}) successfully paired Laz. with a separately defined component for
quick, reactionary behaviours. The coping models in EMA (\ref{ema}), \'Emile
(\ref{emile}), and CBI (\ref{cbi}) have been particularly influential for other
CMEs (\citepg{marsella2004expressive}{353}; \citeg{traum2005fight}).

Laz. also describes a reappraisal process to explain the continuous and
responsive nature of the emotion system~\citep[p.~134]{lazarus1991emotion}.
This is directly tied to coping which can affect changes in an individual's
interpretation of the environment. This concept has also appeared alone in CMEs
that reprocess information after deliberative processes like coping (e.g.
FAtiMA-M (\ref{fatimaM}), FAtiMA (\ref{fatima})), which could result in
different emotions compared to purely reactive systems.

Another feature of Laz. is its connection between relational themes and
emotions~\citep[p.~122]{lazarus1991emotion} which treats appraisal as a
comprehensive unit rather than a set of individual dimensions. This emulates
discrete categories, allowing CMEs to treat each emotion separately (e.g.
Maggie (\ref{maggie})).

\paragraph{Scherer (Sch.)}
Sch. tends to appear where CMEs need multi-level and/or multi-stage appraisals
(e.g. CMEs \ref{mamid}, \ref{tabasco}, \ref{feelme}), allowing them to use
different appraisal mechanisms and/or sources of variable complexity together.
These features are inherent in Sch.~\citep[p.~99, 103]{scherer2001appraisalB}.
Notably, the list of CMEs that use Sch. include the cognitive architectures
(e.g. CMEs \ref{soar}--\ref{clarion}). This is likely because Sch. ``...is the
most elaborate appraisal theory, [and] doesn't necessarily make it the most
suitable starting point for an affective computing
researcher''~\citep[p.~58]{marsella2015appraisal}. FAtiMA-M (\ref{fatimaM})
explicitly mentioned this complexity in their requirements to ensure that it
could support Sch. if desired. ELSA (\ref{elsa}) calls itself a neural network
(NN), which makes it difficult to
understand~\citep[p.~143--144]{meuleman2015computational}, but aligns with
NN-based illustrations of connections and activation patterns in
Sch.~\citep[p.~105]{scherer2001appraisalB}.

CMEs can simplify Sch. by only using its appraisal variables (e.g. AKR
(\ref{akr}))---sometimes combining them with variables from other theories like
OCC (e.g. CMEs \ref{infra}, \ref{hybridc}, \ref{erdams}, \ref{eegs})---or take
inspiration from its process model to connect emotion generation to other
subsystems (e.g. GRACE (\ref{grace})). HumDPM-E (\ref{humdpme}) cleverly
leverages Sch.'s ``modal'' emotions~\citep[p.~113]{scherer2001appraisalB},
allowing it to produce and store different emotions simultaneously. This
suggests that some CMEs can comfortably use pieces of Sch. independent of the
complete theory.

\paragraph{Roseman (Ros.)}
CMEs commonly use Ros. to define \textit{Surprise} as an emotion because they
use other theories---usually Sch. and/or OCC---that do not explicitly define it
(e.g. \ref{akr}, \ref{parlee}, \ref{hybridc}). These unions appear to be sound.
OCC agrees with Ros. that \textit{unexpectedness} elicits
\textit{Surprise}~\citep[p.~32]{occ}, and Sch.'s \textit{suddenness} variable
in the novelty check appears to do a comparable
evaluation~\citep[p.~95]{scherer2001appraisalB}. \textit{Anger} is an emotion
that Ros. shares with OCC, but it limits its scope to events caused by other
agents, which a CME might find more helpful (e.g. ERDAMS (\ref{erdams})). Ros.
can also help define action tendencies and map emotions to them (e.g. TEATIME
(\ref{teatime}), combined with Frj.).

Choosing Ros. is partially driven by its computational tractability. POMDP-CA
(\ref{pomdp}) cites this, also noting that---of the two theories identified in
\citet{picard1997affective} for cognitive appraisal---Ros. systematically built
a model between appraisal variables and emotions from empirical
studies~\citep[p.~267--268]{roseman1996appraisal} which makes it more
plausible~\citep[p.~265]{kim2010computational}. The larger issue is that Ros.
does not specify an emotion generation process. CMEs have compensated for this
by using Markov Models (e.g. POMP-CA
(\ref{pomdp})~\citep[p.~267]{kim2010computational}), fuzzy logic (e.g. FLAME
(\ref{flame})~\citep[p.~227--228]{el2000flame}), and combining Ros. with
process-based theories like Sch. (e.g. HybridC (\ref{hybridc})) and Frj. (e.g.
TEATIME (\ref{teatime})).

\paragraph{Ortony, Clore, and Collins (OCC)}
This is the most used (\citepg{bourgais2017enhancing}{91};
\citepg{dang2009experimentation}{136}; \citepg{lim2012creating}{292}) and
widely accepted theory in affective computing~\citep[p.~1]{el2004modelling}
despite cautioning that ``...we view each emotion specification, or
characterization, as a \textit{proposal} rather than as an empirically
established fact.''~\citep[p.~87--88]{occ} and not being as popular in
psychology~\citep[p.~278]{gratch2004domain}.

The widespread use of OCC is partially due to its hierarchical emotion
structure and event-driven eliciting conditions~\citep[p.~18--19]{occ} (e.g.
CMEs \ref{scream}, \ref{hybridc}, \ref{gema}, \ref{epbdi}, \ref{microcrowd},
\ref{puma}, \ref{mmt}, \ref{gamygdala}--\ref{mexica}) which feels familiar to
computer scientists (\citepg{becker2008wasabi}{44};
\citepg{marsella2015appraisal}{57}) and is more amenable to computation than
other theories (\citepg{picard1997affective}{195};
\citepg{broekens2021emotion}{362}; \citepg{hudlicka2014habits}{8};
\citepg{occ}{2, 181--182}). There has been significant strides towards
refining~\citep{bartneck2002integrating}, formalizing
(\citeg{steunebrink2009occ}; \citeg{steunebrink2012formal}), and re-framing OCC
for applications like agent believability~\citep{ortony2002making}. A further
benefit of OCC's comparison to a computational approach is that it can be
easier to understand without a background in psychology
(\citepg{ochs2009simulation}{282}; \citepg{reilly1996believable}{28}). This
might make it more ``clear and
convincing''~\citep[p.~741]{masuyama2017application} than other appraisal
theories.

OCC's structure also shows which variables contribute to an emotion's
intensity, proposing that it is evaluated with a weighted
function~\citep[p.~69, 82]{occ}. Unfortunately, it does not propose what those
weights should be, nor the function's nature. CMEs have compensated by
designing a separate tool for empirically deriving intensity
parameters (e.g. ALMA (\ref{alma})~\citep[p.~209]{kipp2011designing}, The Soul
(\ref{soul})), translating OCC emotion categories to a dimensional space (e.g.
PPAD-Algo (\ref{ppad})), defining their own functions or values from OCC
variables with no clear empirical basis (e.g. CMEs \ref{flame}, \ref{wasabi},
\ref{maggie}, \ref{gema}, \ref{dett}, \ref{eva}, \ref{erdams}, \ref{mmt},
\ref{iphonoid}, \ref{eegs}, \ref{gamygdala}, \ref{mexica}, \ref{em},
\ref{s3a}), or not concerning themselves with intensity at all (e.g. CMEs
\ref{ar}, \ref{gamae}, \ref{puma}).

Strictly speaking, the weighted function used by these CMEs is not an intensity
function. OCC proposes that a weighted combination of the variables leading to
an emotion category along the hierarchy is an \textit{emotion potential}---a
higher potential means a higher chance of experiencing that kind of
emotion~\citep[p.~81--82]{occ}. The \textit{difference} between an emotion
threshold and this value is its intensity, which MMT (\ref{mmt})
incorporates~\citep[p.~9844]{alepis2011automatic}. CMEs have also used this
difference modulate to simulate other types of affect
(\citepg{becker2008wasabi}{92--93}; \citepg{ushida1998emotion}{65--66};
\citepg{duy2004creating}{119}; \citepg{kazemifard2011design}{2645};
\citepg{dias2014fatima}{48}; \citepg{dias2005feeling}{131};
\citepg{rosis2003from}{103, 109--110}; \citepg{martinho2000emotions}{51}).

Ironically, OCC's authors believe that computers \textit{cannot} have emotion
but it is still useful to reason about them: ``...we do not consider it
possible for computers to experience anything until and unless they are
conscious. Our suspicion is that machines are simply not the kinds of things
that can be conscious...There are many AI endeavours in which the ability to
understand and reason about emotions or aspects of emotions could be
important''~\citep[p.~182]{occ}. AffectR (\ref{ar}) adheres to this when
reasoning about another agent's actions~\citep[p.~27]{elliott1989ective}. One
could also view narrative planners (e.g. MEXICA (\ref{mexica}), NPE
(\ref{npe})) as an exercise in reasoning about character emotions. However, OCC
can be applied to emotion generation as
well~\citep[p.~195]{picard1997affective}, also shown by AffectR, because the
process of reasoning about emotions could be understood as reasoning about the
emotional significance of an event to the
agent~\citep[p.~230]{prendinger2004mpml}. The focus on reasoning makes OCC
amenable to an intentional stance, which enhances agent
believability~\citep[p.~151--152]{andre2000integrating}, because users can
``see'' the agent's thought processes.

The ``fortunes of others'' emotions (e.g. \textit{Happy-For}) might be unique
to OCC which rely on evaluations of how \textit{someone else} feels. These are
critical for empathetic agents (e.g. Greta (\ref{greta}), ERDAMS
(\ref{erdams})) and agents that model relationships (e.g. CMEs \ref{gamae},
\ref{epbdi}, \ref{microcrowd}, \ref{eva}, \ref{socio}, \ref{gamygdala},
\ref{apf}--\ref{npe}). However, OCC omits \textit{Surprise}---which is
important for some CMEs---because they believe that it is not inherently
positive or negative~\citep[p.~32]{occ}. Instead, they categorize it as a
cognitive state tied to a global \textit{unexpectedness} variable. CMEs that
need \textit{Surprise} draw from Ros. (e.g. ParleE (\ref{parlee}), HybridC
(\ref{hybridc})) because it shares this hypothesis and explicitly defines
\textit{Surprise} as an emotion~\citep[p.~269]{roseman1996appraisal}.

A shortcoming of OCC is a lack of emotion elicitation processes. This is a
deliberate omission because OCC views it as a general cognitive psychology
problem, but stresses the role of cognition in such processes~\citep[p.~2]{occ}.
CMEs have realized OCC in plan-based systems (e.g. CMEs \ref{parlee},
\ref{ema}, \ref{emile}, \ref{fatima}, \ref{mexica}, \ref{npe}), which are a
step towards explainable behaviours. They provide context for elicited
emotions~\citep[p.~328]{gratch2000emile}, aligning with the OCC's focus on
reasoning about emotions~\citep[p.~182]{occ}. Another approach, supported by
\citet{ortony2005affect}, is to integrate OCC in a biologically-inspired
approach (e.g. Maggie (\ref{maggie}), IM-PMEB (\ref{impmeb})) or architecture
(e.g. EmMod (\ref{emMod}), WASABI (\ref{wasabi})) due to OCC's reliance on
cognition. CME commonly use a BDI-inspired system or architecture to account
for cognitive activities (e.g. CMEs \ref{genia3}, \ref{gamae},
\ref{jbdiemo}--\ref{puppet}, \ref{erdams}), but this can make the CME difficult
to modify if it is integrated too deeply into the host
architecture~\citep[p.~111--112]{rosis2003from}. Other approaches include
combining OCC with process-oriented theories like Frj. (e.g. S3A (\ref{s3a}))
or Sch. (e.g. HybridC (\ref{hybridc}), GRACE (\ref{grace})), other resources
such as \citet{picard1997affective} (e.g. AKR (\ref{akr})), and fuzzy
logic (e.g. CMEs \ref{flame}, \ref{infra}, \ref{tardis}, \ref{puma}). Many CMEs
set their emotion model between input and output modules to mediate their
interactions (e.g. CMEs \ref{gvh}, \ref{emotion}, \ref{fatima}, \ref{alma},
\ref{eva}, \ref{iphonoid}, \ref{socio}, \ref{apf}). If the goal is not to
create ``correct'' behaviours, this strategy is sufficient if it meets the
CME's other design goals~\citep[p.~44--45]{reilly1996believable}.

\paragraph{Smith \& Kirby (S \& K)}
This theory only seems to appear when CMEs want to integrate multiple, parallel
input sources into one unit for appraisal, which is its distinguishing
feature~\citep[p.~129--130]{smith2001toward}.

S \& K always appears with Sch. to combine appraisal information from sources
on multiple levels of resolution (e.g. CMEs \ref{mamid}, \ref{tabasco},
\ref{eegs}, \ref{feelme}). This might be because Sch. is better
validated~\citep[p.~129]{smith2001toward} and computationally tractable.
Scherer also draws parallels between sequential check registers and S \& K's
integrated appraisal~\citep[p.~105, 120]{scherer2001appraisalB}. Another
possibility for this pairing is a misconception that Sch. is strictly a
sequential appraisal process~\citep[p.~236]{ojha2016ethically} when it is
not~\citep[p.~100, 103]{scherer2001appraisalB}.

An exception is EMA (\ref{ema}), which combines S \& K with Frj. to define an
attention mechanism~\citep[p.~286]{gratch2004domain}. This is likely because
\citet[pp.~149]{frijda2001appraisal} compares its ``blackboard control
structure'' to S \& K's appraisal register. This suggests that CMEs can combine
S \& K with other theories that have some comparable work to the appraisal
register concept.

\paragraph{Oatley \& Johnson-Laird\protect\footnote{\normalfont Although it
does not name appraisal dimensions, O \& JL talk about evaluating events
relevant to plans and goals such that changes in achievement probability induce
emotions~\citep[p.~50]{oatley1992best}. Therefore, it is grouped with the
appraisal theories.} (O \& JL)}
CMEs use O \& JL to define what emotions they support and connect emotion
intensity to changes in computational plans. O \& JL typically have a
supporting role for defining emotions in CMEs with Ek. as the main theory
present for defining CME emotions (e.g. CMEs \ref{cathexis}, \ref{emMod},
\ref{hybridc}). This connection is sound, as O \& JL considered Ek. as evidence
when identifying their set of basic emotions~\citep[p.~57--61]{oatley1992best}.

O \& JL propose that there is no emotion process, arguing that emotions are
states entered at plan junctures, that might include conflicts between
different goals, agents, and resource demands~\citep[p.~22, 24--25,
31--36]{oatley1992best}. CMEs have taken this information to define emotion
intensity in relation to an agent's goals and plans (e.g. ParleE
(\ref{parlee}), \'Emile (\ref{emile})). This also frames cognition as a
knowledge transformation process, which is amenable to computation (e.g. Greta
(\ref{greta})).

Perhaps the most useful element of O \& JL is its focus on the social and
communicative role of emotions~\citep[p.~41--42]{oatley1987towards}. This has
implications for multi-agent applications with affective content because each
agent is an independent module in a larger
system~\citep[p.~178, 181--182]{oatley1992best}. Conversational agents might
also benefit from this view, which casts conversations as a form of mutual
planning. As the field of social affective agents progresses, O \& JL could
come to play a larger role in the field.

\subsubsection{Neurophysiologic Theories}
Biological neural circuitry and brain structures inspire the
\textit{neurophysiologic} theories of affect, which offer a grounded view of
how emotion systems might be organized and connected to the
body~\citep[p.~98--99]{lisetti2015and}. They tend to appear when a CME wants to
distinguish between reactive, non-cognitive and deliberative, cognitive emotion
processes. All three theories claim mechanisms for fast, ``stupid'' reactions
and slower, deliberative plans that people collectively call
``emotions'' (\citepg{sloman2005architectural}{230};
\citepg{damasio2005descartes}{133}; \citepg{ledoux1996emotional}{161--165}).

\paragraph{Sloman\protect\footnote{\normalfont Since Sloman views the brain as
an information processing system~\citep[p.~206--207]{sloman2005architectural},
it is grouped with the neurophysiologic theories.} (Slo.)}
Slo. conceptualizes emotion as a product of a central information-processing
system, distinguishing between types of emotion based on their architectural
requirements~\citep[p.~204, 211]{sloman2005architectural}. CMEs use this
distinction to specify elicitation mechanisms with varying performance
requirements (e.g. WASABI (\ref{wasabi}), Presence (\ref{presence})). The
distinction also makes it possible to specify individual aspects of a CME such
as goal importance for emotion intensity functions (e.g. \'Emile (\ref{emile}))
and emotion-driven plan selection (e.g. FAtiMA (\ref{fatima})).

WASABI (\ref{wasabi}) explicitly models aspects of Slo. for emotion
elicitation using signal impulses that ``disturb'' its homeostatic
state~\citep[p.~90]{becker2008wasabi}. Slo. views these ``disturbances'' as a
kind of emotion~\citep[p.~230]{sloman2005architectural} which could be useful
for CMEs that do not have deliberative processes. When deliberative processes
are needed, Slo. might be particularly amenable to BDI-based CMEs because it
explicitly references ``beliefs'', ``desires'', and ``intentions'' as
architectural features~\citep[p.~208]{sloman2005architectural}.

\paragraph{Damasio (Dam.)}
Dam. proposes two emotion types: innate, evolution-based primary emotions and
learned, cognition-driven secondary emotions that trigger the primary
system~\citep[p.~131--139]{damasio2005descartes}. CMEs use this to motivate
multiple, coexisting emotion elicitation processes (e.g. \ref{cathexis},
\ref{wasabi}, \ref{presence}).

Two of Dam.'s features have proven useful for CMEs. One is emotion's influence
on decision-making~\citep[p.~126, 128]{damasio2005descartes} which can drive
the design of CME behaviour (e.g. Cathexis (\ref{cathexis}), \'Emile
(\ref{emile})) and/or the design of connections between emotion elicitation and
cognitive processes (e.g. EmMod (\ref{emMod})). Directly related to
decision-making, the second feature is the Somatic Marker Hypothesis (SMH)
which describes how secondary emotions are learned and connected to the primary
emotion system~\citep[p.~137, 145, 174]{damasio2005descartes}. CMEs have used
the SMH as-described to elicit emotions from memories via learned associations
between stimuli and emotions (e.g. Cathexis (\ref{cathexis})) and as a clever
way to mark different types of inputs with common information to coordinate
further functions (e.g. Kismet (\ref{kismet})).

Damasio posits that CMEs \textit{cannot} use this theory because of the
biological connection between the mind and
body~\citep[p.~249--250]{damasio2005descartes}, suggesting that Dam. cannot be
implemented in agents without a physical body. However, there is a version of
SMH that bypasses the body~\citep[p.~155--158]{damasio2005descartes} which
WASABI (\ref{wasabi}) uses successfully in a virtual agent~\citep[p.~50,
56]{becker2008wasabi}.

\paragraph{LeDoux (LD)}
LD views emotions as biological functions with different neural systems that
evolution maintained across species~\citep[p.~106--107,
171]{ledoux1996emotional}. It proposes that each emotion has a mechanism
programmed to detect and react to innate stimuli relevant to the system's
function~\citep[p.~134, 143, 161--163, 165, 175--176]{ledoux1996emotional}.
This suggests that some emotions like \textit{Fear} do not necessarily require
higher reasoning to elicit (e.g. WASABI (\ref{wasabi})) and they could be a
direct map to behaviours (e.g. InFra (\ref{infra})). This proposal also sets
the stage for LD's work on emotional conditioning mechanisms---specifically
\textit{Fear}~\citep[p.~127--128]{ledoux1996emotional}---suggests methods for
emotional learning in CMEs (e.g. FLAME (\ref{flame})).

Damasio and LeDoux applaud each other's---mutually
relevant---work (\citepg{damasio2005descartes}{133};
\citepg{ledoux1996emotional}{250, 298}). One focuses on the ``low road'' (i.e.
non-cognitive) and the other on the ``high road'' (i.e. cognitive) which could
explain their co-use or connection in some CMEs (e.g. Cathexis
(\ref{cathexis}), WASABI (\ref{wasabi})).

\subsection{Psychologists Directly Involved in CME
Design}\label{sec:psychologists}
Translating a psychological theory into a CME is difficult because it involves
formalizing informal concepts and documenting hidden
assumptions~\citep[p.~22--23]{marsella2010computational}. CMEs designed with
the participation of the theory's creator stand out as being ``truest'' to the
theory.

Frijda supervised the development of both ACRES (\ref{acres}) and the Will
architecture (\ref{will}). ACRES is designed as a partial test of its
functionality, treating the theory as a design specification~\citep[p.~237,
247]{frijda1987can}. Will---the spiritual successor of ACRES---proposes a
reasonable extension of Frj.: emotion, moods, sentiments, and personality are
related by focus and duration~\citep[p.~135--136, 138]{moffat1997personality}.
This is convenient for CMEs as it shows that these affective types can share
the same underlying structure. Scherer directly influenced the design of ELSA
(\ref{elsa}) which is particularly relevant as its purpose is to show that
Sch.---which takes an information systems view on emotion
processes~\citep[p.~103]{scherer2001appraisalB}---can be implemented and used
as a research tool~\citep[p.~142--143]{meuleman2015computational}. Ortony
provided direct supervision for AffectR
(\ref{ar})~\citep[p.~iv]{elliott1989ective} which presumably makes it the most
faithful account of OCC emotion generation processes and action tendency
hierarchy.

In other cases, theory creators acted as consultants to CME designers
(\citepg{becker2008wasabi}{vii}; \citepg{petta2002role}{281};
\citepg{gratch2004domain}{303}; \citepg{gratch2000emile}{332};
\citepg{marsella2009ema}{89}) and/or drew from other CMEs that were developed
under that creator's guidance~\citep[p.~325]{gratch2000emile}. Caution must be
used in evaluating their faithfulness to the theories, as it is usually not
documented what parts relied on consultation and which did not.

\section{Discussion}\label{sec:discussion}
In our examination of these CMEs, we also found design decisions and trade-offs
relevant to \textit{implementing theories}, \textit{how CMEs could combine
theories from different perspectives}, \textit{CME realism versus efficiency},
and \textit{other sources of design influence}.

\subsection{Implementing Theories}
Implementing an affective theory is challenging. Some theories---Frj., Sch.,
OCC, O \& JL, and Slo.---were explicitly designed to be computationally
tractable (\citepg{frijda1987can}{247}; \citepg{scherer2021towards}{279};
\citepg{occ}{181}; \citepg{oatley1987towards}{30};
\citepg{sloman1987motives}{231}) while others---like Dam. and LD---argue that
their theories \textit{cannot} be computationally realized
(\citepg{damasio2005descartes}{249--250}; \citepg{ledoux1996emotional}{41,
176}). Regardless, they have been implemented. Nonetheless, how accurately a
CME adheres to a theory and/or observed emotion phenomenon tends to be directly
proportional to how complex the CME is.

Neurophysiologic theories might be more plausible than appraisal
theories~\citep[p.~72--73]{velasquez1998robots} and better align with current
findings~\citep[p.~160]{andre2000integrating}. However, they require modelling
parts of the brain and body, which this is neither feasible nor desirable for
many CMEs. Furthermore, the resulting system will not necessarily be accurate
due to gaps in our understanding of anatomical structures and functions
(although complete accuracy might not be useful to
anyone~\citep[p.~60]{marsella2015appraisal}).

Appraisal theories might be best suited for CMEs as they touch on all
components and phases of emotion  processing~\citep[p.~13]{scherer2010emotion}.
They are also relatively easy to implement as they are often
rule-based~\citep[p.~225]{picard1997affective} and built on information
processing analogies~\citep[p.~59]{marsella2015appraisal}. While some have
integrated neurophysiologic aspects, this increases their complexity. For
example, empirical test of Sch. have been relatively successful in predicting
different patterns in emotion processes~\citep[p.~93, 103,
117--118]{scherer2001appraisalB} but it is very complex and involves
implementations of components like the Autonomic Nervous System (ANS) and
memory while allowing for multiple levels of information processing. This might
be why Sch. is favoured by cognitive architectures and research CMEs like
CLARION (\ref{clarion}) and MAMID (\ref{mamid}), whose assumptions closely
follow Sch.'s (\citepg{hudlicka2019modeling}{136}; \citepg{sun2016emotion}{6}).
These systems purposefully sacrifice computational efficiency for accuracy
since their aim is to study emotion phenomena. This complexity also makes them
are to explain and debug~\citep[p.~143--144]{meuleman2015computational}.

\subsection{How CMEs Could Combine Perspectives}
Some theories are easily combined as they share a perspective based on coherent
assumptions. For example, Izard, Ekman, and Plutchik agree on the function of
at least four primary emotions---\textit{Joy/Happiness}, \textit{Sadness},
\textit{Anger}, and \textit{Fear}---and their ability to interact to produce
what people recognize as other, more complex, emotions
(\citepg{izard2000motivational}{254, 258--259}; \citepg{ekman2007emotions}{69};
\citepg{plutchik1984emotions}{200, 204--205}). Similar overlaps exist in the
appraisal theories' evaluation dimensions and how they label distinct
combinations. The dimensional theories, V-A and PAD, are also obviously
compatible---one could directly layer V-A over the P-A plane. By staying within
one perspective, a CME design can use the individual strengths of each theory
with little worry of conflicting assumptions or views.

Combining theories from \textit{different} perspectives poses a more complex
challenge, but often necessary to address all aspects of affect needed in the
design. For example, OCC is frequently combined with Ek.---which focuses on
automatic, hard-wired appraisals rather than
evaluations~\citep[p.~51]{ekman1999basic}---to produce facial expressions from
cognitively-evaluated events (\citepg{kshirsagar2002multilayer}{109};
\citepg{jain2019modeling}{66}; \citepg{andre2000integrating}{155};
\citepg{masuyama2017application}{740--741}). This connection is presumably due
to the OCC's association of characteristically similar ``linguistic tokens''
with each emotion~\citep[p.~1--2, 87--88]{occ}. By finding similar words, one
can fit the discrete theories' emotions into the OCC structure. However, this
relies on subjective interpretations, and even the given lists lack empirical
validation~\citep[p.~172--176]{occ}. More pressingly, emotions of the same name
might represent different concepts. \textit{Fear} and \textit{Anger} in OCC, as
with many appraisal theories, are complex emotions requiring flexible,
cognitive evaluations~\citep[p.~85, 87]{becker2008wasabi} but the same emotions
in discrete theories are simpler, triggered by inflexible hard-wired systems.
While they might be expressed with the same physiological changes, behaviours,
and expressions, their eliciting mechanisms are not of the same
kind~\citep[p.~15--16]{scherer2010emotion}. Whether or not this distinction is
important for a CME, it should still be addressed as it affects how accurately
the theories are modelled. Similar considerations must be made when attempting
to align the dimensional theories with the dimensions of appraisal theories and
locating discrete emotions in dimensional space.

These conceptual mismatches does not mean that there are ``correct'' and
``incorrect'' theories or that they are incompatible, especially considering
how they overlap and converge on the role of emotion
(\citepg{scherer2010emotion}{10--11, 14--15}; \citepg{broekens2021emotion}{352,
354--355}; \citepg{scherer2021towards}{281}). Rather, they are different
views---perspectives---of a complete system, each focusing on different aspects
of emotions~\citep[p.~259]{frijda1986emotions}. Emotion systems seem to rely on
both fast, primary and deliberative, secondary
emotions~\citep[p.~70]{picard1997affective}. This idea of two emotion types is
present in some affective theories, such as \citet[p.~74]{izard1993four},
\citet[p.~51]{ekman1999basic}, \citet[p.~155]{frijda2001appraisal},
\citet[p.~102]{scherer2001appraisalB}, and \citet[p.~93]{smith2000consequences},
and is also supported by empirical investigations of the
brain (\citepg{damasio2005descartes}{136--139};
\citepg{ledoux1996emotional}{177--178}). Some CMEs have explicitly modelled
these two ``pathways'', including \citet[p.~98]{becker2008wasabi},
\citet[p.~73]{velasquez1998robots}, \citet[p.~63]{ushida1998emotion}, and
\citet[p.~160]{andre2000integrating}. Correspondingly, one way that each
perspective could be assigned roles in CME designs to address different aspects
of emotion generation is:
\begin{itemize}
    \item \textit{Neurophysiologic theories} provide guidelines for how to
    unite disparate emotion processing pathways into a coherent system,

    \item \textit{Discrete theories} drive the creation of a limited set of
    fast, hard-wired~\citep[p.~366]{broekens2021emotion} reactions to specific
    stimuli (``primary emotions'', ``low road''),

    \item \textit{Appraisal theories} drive the deliberative, slower systems
    for emotion elicitation that require planning and/or reasoning (``secondary
    emotions'', ``high road'') that can account for language and sociocultural
    factors, allowing for a broader range of identifiable emotions, and

    \item \textit{Dimensional theories} provide a common space for merging the
    outcomes of each emotion pathway in the spirit of appraisal
    registers (\citepg{scherer2001appraisalB}{105};
    \citepg{smith2000consequences}{93}) while allowing other types of affect to
    interact with emotion.
\end{itemize}

\subsection{CME Realism versus Computational Efficiency}\label{realvsefficient}
For CMEs that focus on agent believability, enforcing realism and rational
intelligence can be detrimental to their
goals~\citep[p.~11]{reilly1996believable}. These systems typically interact
with users in (soft) real-time, so efficiency is more important than accuracy.
Being able to test and debug models is also important. For example, dimensional
theories are arguably the simplest to implement efficiently. However, they also
have the lowest affective resolution. Nevertheless they are considered
``universal'' and individual emotion ``points'' can be labelled as
needed~\citep[p.~12, 15]{scherer2010emotion}. However, since they do not define
emotion generation, the designer must determine how much of it they want to
implement.

Efficient implementation of \textit{believable} but not necessarily sound
emotion generation can be done in myriad ways: with
metadata (\citepg{prendinger2004mpml}{236}; \citepg{gebhard2005alma}{33});
concepts like Bayesian Networks (\citepg{kshirsagar2002multilayer}{108};
\citepg{ball2000emotion}{204}), Markov Models (\citepg{lisetti2002can}{603};
\citepg{duy2004creating}{115}), and fuzzy logic (\citepg{el2000flame}{229};
\citepg{castellanos2018computational}{29--30}; \citepg{jones2013tardis}{3, 7};
\citepg{laureano2012design}{68}); and AI techniques like behaviour trees and
goal-oriented action planning~\citep[p.~698]{klinkert2021artificial}. Other
theories---typically appraisal---are also conscripted, but might not be
modelled in full (\citepg{popescu2014gamygdala}{36};
\citepg{reilly1996believable}{52}). CMEs have also improved their efficiency by
considering their target domain's limitations, which might require fewer
emotion categories (\citepg{salichs2012new}{58};
\citepg{kshirsagar2002multilayer}{109}; \citepg{el2004modelling}{2}), and
appraisal variables (\citepg{yacoubi2018teatime}{150};
\citepg{shirvani2020formalization}{118}), while others are able to scale as
needed (\citepg{prendinger2004mpml}{239};
\citepg{castellanos2018computational}{27}; \citepg{dias2014fatima}{44};
\citepg{moshkina2011tame}{217}; \citeg{broekens2004scalable};
\citepg{popescu2014gamygdala}{35}). As these have all found some success in
achieving their goals, this further emphasizes that accuracy is not always
necessary. This opens up the design space to create a CME that behaves ``well
enough'' for its intended tasks.

\subsection{Other Sources of Design Influence}
Several systems strengthen or extend their chosen theoretical foundations by
supporting it with other comparable or complementary theories
(\citepg{lisetti2002can}{594, 599--600, 606};
\citepg{castellanos2018computational}{27}; \citepg{jain2019modeling}{63--64};
\citepg{yanaru1997emotion}{218, 247}; \citepg{rosis2003from}{91};
\citepg{breazeal2003emotion}{129}; \citeg{wilson2000artificial}). Some cite
additional work to support perceived short-comings in a foundational
theory~\citep[p.~223, 233--234, 239]{el2000flame} or formally define
concepts\footnote{POMDP-CA (\ref{pomdp}), which uses
\citet{weaver1948probability} to define \textit{unexpectedness}, similar to
\textit{suddenness} in Sch.~\citep[p.~95]{scherer2001appraisalB}. This is
necessary to appraise \textit{Surprise} in both
Ros.~\citep[p.~267]{roseman1996appraisal} and
OCC~\citep[p.~126]{occ}.}~\citep[p.~269]{kim2010computational}. Yet other CMEs
use additional sources to connect emotions with other system components, such as
social variables (\citepg{bourgais2017enhancing}{93};
\citepg{kasap2009making}{22}; \citepg{ochs2009simulation}{288};
\citepg{klinkert2021artificial}{698}) and emotion contagion
(\citepg{aydt2011computational}{77}; \citepg{lhommet2011never}{91};
\citepg{durupinar2016psychological}{2151}).

Emotion theories do not address all aspects of emotion generation, such as
emotion intensity and cognitive organization. Other sources of information
are needed. Three stand out: the work of Picard~\citep{picard1997affective},
Minsky's theory~\citep{minsky1988society}, and empirical data.

A pioneer of \ref{ac}, Picard offers a computer science-friendly view of
emotion, proposing models and ideas for CMEs that often guide the selection of
their underlying emotion theories. AKR (\ref{akr}) references them to justify
its use of Markov Models for emotion dynamics~\citep[p.~603]{lisetti2002can}.
IM-PMEB (\ref{impmeb}), FAtiMA (\ref{fatima}), and SocioEmo (\ref{socio})
reference Picard to define an emotion intensity decay function
(\citepg{shvo2019interdependent}{68}; \citepg{dias2005feeling}{130--131};
\citepg{ochs2009simulation}{289}). Presence (\ref{presence}) cites Picard for
their separation of primary and secondary emotion processing channels,
motivating its use of Slo. and Dam.~\citep[p.~160]{andre2000integrating}. Greta
(\ref{greta}) cites Picard's ``tub of water'' metaphor, comparable to Plu., for
addressing coexistent emotions in its design
considerations~\citep[p.~99]{rosis2003from}. TAME (\ref{tame}) uses Picard for
defining emotion dynamics as a system response~\citep[p.~211]{moshkina2011tame}.

Minsky offers a model of human intelligence amenable to AI. Since many emotion
theories---especially appraisal theories---rely on cognition, Minsky's
\textit{Society of Mind} presents a way to model it. O \& JL explicitly draw
parallels to it~\citep[p.~32, 39]{oatley1987towards}. EmMod (\ref{emMod}) cites
Minsky as the main inspiration for its architecture, producing complex
behaviours via the interactions of many, simple
units~\citep[p.~63]{ushida1998emotion}. Cathexis (\ref{cathexis}) compares its
models of secondary emotions to Minsky \textit{k}-lines, connecting primary
emotions to encountered stimuli~\citep[p.~73]{velasquez1998robots}.

Empirical data, as the best source for replicating observable phenomena,
has been used for: defining degrees of emotion
positivity and negativity~\citep[p.~4]{ojha2017emotional} and emotion
effects~\citep[p.~136]{hudlicka2019modeling}; deriving emotion intensity
functions~\citep[p.~419]{ochs2012formal}; quantifying the relationship between
emotion intensity, desires, and expectations (\citepg{el2000flame}{232};
\citepg{duy2004creating}{125}); and gesture
models~\citep[p.~302]{marsella2000interactive}. Some systems have moved to
purely data-driven approaches (e.g. \citet{ojha2019integrating},
\citet{bai2021enhancing}).

\section[Summary]{Summary\protect\footnote{\normalfont ``Conclusion'' in
\citet{smith2021what}.}}
We have examined how CMEs from different application domains use emotion
theories for emotion generation (i.e. for emotion representation and
elicitation) and expression. We found that each type of emotion theory filled a
similar role regardless of the domain: \textit{discrete} theories define which
emotions a CME can represent and express, and how it does so;
\textit{dimensional} theories can provide a simple and powerful representation
that describes emotion numerically and with respect to other types of affect;
\textit{appraisal} theories chiefly drive the elicitation process; and
\textit{neurophysiologic} theories unite the reactionary and deliberative
emotion views, tying them to measurable body states.

These roles can be complementary. Appraisal theories seem the best starting
point for emotion generation CMEs, as they explicitly describe emotion
processes and are relatively easy to implement. Discrete theories can improve
a CME's comprehensibility, and dimensional theories are useful for their
quantitative representation of emotions (and other types of affect) in a common
space. The neurophysiologic theories can contribute to emotion process
definitions, but tends to increase a CME's overall complexity---they should be
used with care. Generally, the less realistic that generated emotions need to
be, the more efficient a CME can be.

There are even more theories (e.g. see \citep[p.~280--281]{scherer2021towards}),
models, and data to draw from for CME designs. For example, if one is
considering V-A then they might consider \citet{russell1980circumplex} too due
to their similarities. Lastly, future CMEs might get inspiration from unlikely
places (e.g. \citet{nallaperuma2011eme}). This survey aims to be a resource for
creating new CME designs, providing a practical view of \textit{some} emotion
theories and existing CMEs to borrow from and build on.

\clearpage
\vspace*{\fill}
\begin{keypoints}
    \begin{itemize}
        \item It is common for CME designs to combine theories that have
        overlapping concepts or when a chosen theory does not meet all of the
        CME's design requirements

        \item Discrete theories---such as Izard, Ekman, and Plutchik---are
        commonly used to define what emotion kinds a CME can generate

        \item Dimensional theories like V-A and PAD Space appear when a CME
        wants to represent different types of affect in a common space or to
        provide an alternate view of emotions

        \item Appraisal theories---Frijda, Lazarus, Scherer, Roseman, OCC,
        Smith \& Kirby, and Oatley \& Johnson-Laird---appear in CMEs to define
        variables that map an evaluation of the world to emotion kinds and to
        specify the emotion generation process

        \item Neurophysiological inspired theories such as Sloman, Damasio, and
        LeDoux appear to distinguish between innate and deliberative emotion
        generation or when emotions must influence planning

        \item Nearly all of the surveyed CMEs used at least one appraisal theory

    \end{itemize}
\end{keypoints}

\parasep
\vspace*{\fill}

%% file: highleveldesign.tex
\chapter{Start Your \progname{}: Requirements and
Scope}\label{chapter:reqsAndScope}
\def\epigraphflush{center}
\setlength{\epigraphwidth}{0.85\textwidth}
\def\textflush{center}
\epigraph{Everything that follows, is a result of what you see here.}{Dr.
Lanning's Hologram, \textit{I, Robot (2004)}}

Here, the design of \progname{} begins in earnest. \progname{} is a
Computational Model of Emotion (CME) for emotion generation that strives to
create a more engaging player experience via believable Non-Player Characters
(NPCs). Based on the proposed requirements analysis process
(Chapter~\ref{sec:reqProcess}), the first step is defining high-level
requirements (Section~\ref{sec:userReqs}). These guide the definition of
\progname{}'s scope (Section~\ref{sec:scope}) and which broad groups of emotion
theories could support it (Section~\ref{sec:perspective_candidates}). This
ensures that \progname{} builds on theories that serve its larger design goals,
embedding support for them at the core of its design.

The theories and models examined are the same as those found in the survey of
existing CMEs (Chapter~\ref{chapter:cmeOverview}) because it already organizes
emotion theories by perspective---discrete, dimensional, appraisal, and
neurophysiological---based on their general agreement about some aspects of
emotion such its representation (e.g. kinds, dimensions).

While one need not be an expert in \ref{as} or psychology to reproduce this
example, it does require some understanding of the material to be able to
recognize, interpret, and synthesize pertinent information to draw conclusions
from. This might involve reading additional material in both affective science
(e.g. \citet{oxfordcompanion}) and computing  (e.g. \citet{picard1997affective})
to contextualize terms and concepts. It might also involve---as is common in
document analysis~\citep[p.~32]{bowen2009document}---iteratively skimming,
reading, and interpreting the literature until enough information is available
to draw conclusions (i.e. choose theories) from.

\section{Requirements and the Game Designer}\label{sec:userReqs}
While CME designers do take emotion theories and the needs of the application
domain into consideration, they rarely consider future
users~\citep[p.~15]{osuna2020seperspective}. In particular, game
designers/developers attempting to create emotional NPCs have specific needs
that CME designers must integrate early in the design
process~\citep[p.~162--163]{ghezzifundamentals2003} to improve the chances of a
CME's adoption.

Although game developers do see the potential player experience (PX)
improvements of emotional NPCs (\citeg{walktall};
\citepg{yannakakis2014panorama}{328}), some believe that there is no player
demand~\citep[p.~218]{broekens2016emotional} and doubt that there would be
sufficient return on investment. One should also consider how a CME might
introduce unnecessary game design restrictions. For example, relying on a
specific agent architecture~\citep[p.~218]{broekens2016emotional} and
generating a specific set of emotion kinds (e.g. \textit{Joy},
\textit{Sadness}) could be problematic, as they might not be compatible with
many games in both technical and design capacities. By extension, when and how
to implement emotion in NPCs should also be the developer's
decision~\cite[p.~8:13]{mascarenhas2022fatima}. There also appears to be a link
between the ``authoring experience'' and the adoption rate of potential
solutions~\citep[p.~5]{guimaraes2022fatima}. These concerns prompt several
requirements that a game development-oriented CME must have to have a chance of
being adopted, collected here into two broad groups---\textit{flexibility}
requirements (RF) and \textit{ease-of-use} requirements
(RE)\footnote{\progname{}'s software requirements specification (SRS) documents
these as nonfunctional requirements. The full specification is at
\href{https://github.com/GenevaS/EMgine/blob/main/docs/SRS/EMgine_SRS.pdf}{https://github.com/GenevaS/EMgine/blob/main/docs/SRS/EMgine\_SRS.pdf}.}.
\vspace*{\fill}
\begin{highlight}[frametitle=Flexibility High-Level Requirements]
    \textbf{Flexibility} is about making \progname{} adaptable so that it can
    meet game designer needs~\citep[p.~30]{reilly1996believable}. The aim is
    for \progname{} to be applicable to a range of game designs while avoiding
    making decisions for the game developer. These requirements include:
    \begin{enumerate}[label=RF\arabic*]
        \item\label{flexArch} Independence from an agent architecture so that
        designers can choose how to integrate \progname{} into their game
        (\citepg{loyall1997believable}{25--26};
        \citepg{rodriguez2015computational}{443};
        \citepg{broekens2016emotional}{218})

        \item\label{flexTasks} Allowing the game designer to choose which of
        \progname{}'s tasks to use, as well as when and how to use
        them (\citepg{mascarenhas2022fatima}{8:13};
        \citepg{guimaraes2022fatima}{20})

        \item\label{flexCustom} Allowing the customization or redefinition of
        \progname{}'s preexisting configuration parameters
        (\citepg{reilly1996believable}{30}; \citepg{guimaraes2022fatima}{20})
        such as the definition of time and emotion decay rates

        \item\label{flexNew} Allowing designers to integrate new components
        into \progname{} that influence or are influenced by emotion
        (\citepg{rodriguez2015computational}{450};
        \citepg{castellanos2019mechanism}{353}), such as mood, personality,
        motivations, culture, gender, and physical state

        \item\label{flexEm} Allowing designers to choose which kinds of emotion
        \progname{} produces (i.e. which emotions an NPC can
        have)~\citep[p.~331]{hudlicka2014computational}, (e.g. \textit{Anger},
        \textit{Joy})

        \item\label{flexOut} Allowing designers to specify how to use
        \progname{}'s outputs to accommodate different ways of expressing
        emotion and/or using emotion as an influence to external
        systems~\citep[p.~86]{loyall1997believable}

        \item\label{flexComplex} Allowing designers to use \progname{} with
        NPCs of different complexities~\citep[p.~220]{broekens2016emotional},
        e.g. a \textit{Pac-man} ghost~\citep{pacman} and a \textit{Skyrim}
        citizen~\citep{skyrim} might not require the same type and/or quantity
        of information to evaluate if they are experiencing an emotion

        \item\label{flexScale} Being resource/time efficient and scalable to
        minimize \progname{}'s impact on overall game performance
        \citep[p.~42]{popescu2014gamygdala}
    \end{enumerate}
\end{highlight}
\vspace*{\fill}
\clearpage
\begin{highlight}[frametitle=Ease-Of-Use High-Level Requirements]
    \textbf{Ease-of-use} concerns the usability of \progname{} and showing how
    it supports game development. These requirements include:
    \begin{enumerate}[label=RE\arabic*]
        \item\label{easeHide} Hiding the complexity of emotion generation so
        that game designers do not have to be knowledgeable of affective
        science or computing to use \progname{}
        (\citepg{reilly1996believable}{28};
        \citepg{broekens2016emotional}{220}; \citepg{guimaraes2022fatima}{5})

        \item\label{easeAPI} Providing a clear and understandable Application
        Programming Interface (API) or similar that shows how to use the
        different aspects of \progname{}~\citep[p.~218]{broekens2016emotional}

        \item\label{easeAuthor} Minimizing authorial burden as game developers
        add NPCs to their game~\citep[p.~5]{guimaraes2022fatima}

        \item\label{easeTrace} Allowing \progname{}'s outputs to be traceable
        and understandable (\citepg{loyall1997believable}{86};
        \citepg{guimaraes2022fatima}{5, 19--20})---critical for testing---by
        providing ways to view the range, intensity, and causes of emotion per
        NPC, per NPC group, and per game world
        area~\citep[p.~219--220]{broekens2016emotional}

        \item\label{easeAuto} Giving developers the option to automate the
        storing and decaying of \progname{}'s internal emotion
        state~\citep[p.~86]{loyall1997believable}

        \item\label{easePX} Showing that \progname{} improves PX, since a
        subpar design could be a detriment to the overall game and would not be
        useful for game development

        \item\label{easeNovel} Providing examples as to how \progname{} can
        create novel game experiences~\citep[p.~221]{broekens2016emotional}
    \end{enumerate}
\end{highlight}

Note that this is not a static list---it is intended to change as more
information emerges. These requirements are also theory-agnostic, so any
affective theory/model that supports them is a reasonable choice. They might
also be applicable to other game entities that are not NPCs because they
describe what the CME must allow game developers to do, not what they must
create.

\section{Defining \progname{}'s Scope}\label{sec:scope}
A CME's design scope (i.e. analysis context) describes \textit{what} it should
do, which must be coherent with its requirements. Here, \progname{} considers
what general \ref{ac} tasks it must do, what an NPC's embodiment might be, and
what emotion components \progname{} should give NPCs. This helps judge the
suitability of broad groups of emotion theories
(Section~\ref{sec:perspective_candidates}).

While it is tempting to dismiss some theories based on design scope alone, it
is too early to make these decisions. Some theories might not explicitly suit
the design scope, but they might help define an interface between \progname{}
and external modules to support \textit{Allowing the Integration of New
Components} (\ref{flexNew}).

\begin{figure}[!tb]
    \begin{center}
        \includegraphics[width=\linewidth]{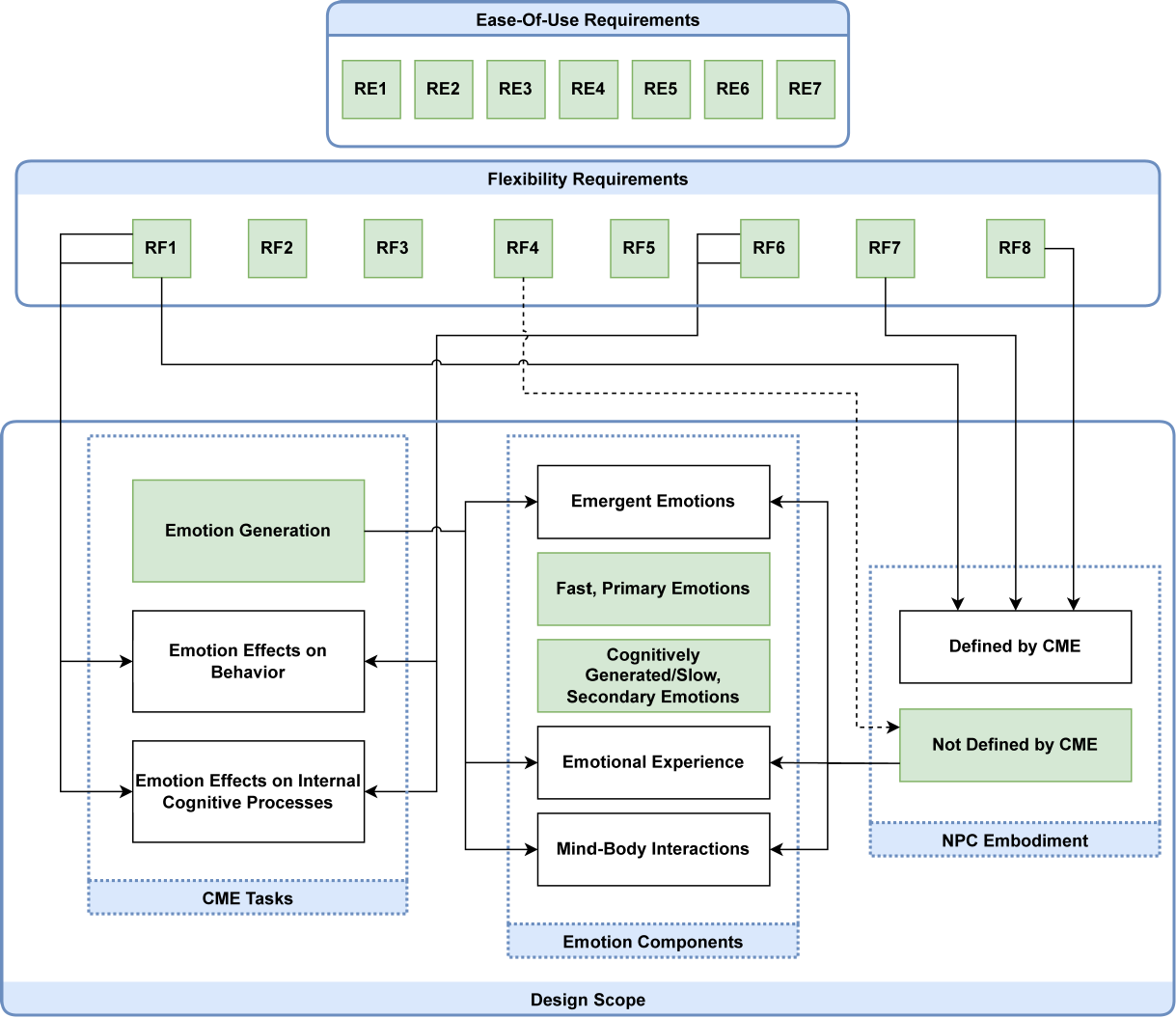}

        \caption[Overview of Influences on \progname{} Design Scope]{Overview
        of Influences on \progname{}'s Design Scope Showing the Connection to
        the High-Level Requirements (Shaded boxes are within scope; Solid
        arrows indicate conflicts with, dotted arrows indicate support for)}
        \label{fig:scopeOverview}
    \end{center}
\end{figure}

Figure~\ref{fig:scopeOverview} illustrates the flexibility requirements'
influence on \progname{}'s design scope. \progname{} automatically considers
all requirements to be in scope, although the ease-of-use requirements did not
contribute to the design at this point. Boxes with outgoing solid arrows point
to boxes they conflict with, and are therefore eliminated from \progname{}'s
scope. Conversely, dotted arrows indicate support for a box/component. Shaded
boxes without incoming solid arrows and are therefore within \progname{}'s
scope.

\subsection{CME Tasks}\label{sec:cmetasks}
\ref{ac} tasks (Chapter~\ref{sec:CMEs}) of relevance to NPCs ``with emotion''
include:
\begin{itemize}
    \item \textit{Emotion Generation} to produce an emotion state given the
    current program and environment state

    \item \textit{Emotion Effects on Behaviour} to change a NPC's
    ``observable'' behaviour (e.g. facial expressions, gestures, or movements)
    given an emotion state, and

    \item \textit{Emotion Effects on Internal Cognitive Processes} or
    \textit{Cognitive Consequences of Emotions} to change a NPC's internal
    processing behaviours.
\end{itemize}

Since it only produces emotion states, \progname{} should focus on
\textit{emotion generation} alone. In terms of the design scope, this means
that it excludes anything outside of taking inputs for evaluation and
producing an emotion state. \progname{} leaves these tasks to other modules
which might not be available. Consequently, the tasks \textit{Emotion Effects
on Behaviour} and \textit{Emotion Effects on Internal Cognitive Processes} are
out of scope. There are other reasons to exclude these tasks from \progname{}'s
scope: they both conflict with \textit{Allowing Developers to Specify How to
Use Outputs} (\ref{flexOut}), as they decide what to do based on a given
emotion state and even suggesting an association between behaviours and
emotions encodes an assumption about how the game designer should use
them~\citep[p.~57]{marsella2015appraisal}; and it could also conflict with
\textit{Independence from an Agent Architecture} (\ref{flexArch}) depending on
the implemented behaviours and effects.

\subsection{NPC Embodiment}\label{sec:embodiment}
\progname{} should not impose constraints on an NPC's embodiment, which would
make it dependent on what components and/or processes that NPC has. Doing so
could violate several requirements, including \textit{Independence from an
Agent Architecture} (\ref{flexArch}), \textit{Ability to Operate on Different
Levels of NPC Complexity} (\ref{flexComplex}), and---potentially---\textit{Be
Efficient and Scalable} (\ref{flexScale}). Instead, and in support of the
requirement to \textit{Allow the Integration of New Components}
(\ref{flexNew}), it should be simple to integrate \progname{} with external
modules at its input and output points.

\subsection{Emotion Components}\label{sec:emotioncomponents}
Emotions have multiple components that affect different aspects of behaviour and
experience (\citepg{hudlicka2019modeling}{133};
\citepg{scherer2001appraisalB}{92}). Based on evidence from psychology and
neuroscience, there are up to five components in computers ``with
emotions''~\cite[p.~60--70]{picard1997affective}. A CME need not have all of
these components to be effective for its job, ``just like simple animal forms
do not need more than a few primary
emotions...''~\cite[p.~68]{picard1997affective}:
\begin{itemize}
    \item \textit{Emergent Emotions} attributed to the system based on their
    ``observable'' behaviours,

    \item \textit{Fast Primary Emotions}, the hard-wired and potentially
    inaccurate responses to innate knowledge elicited by fundamental mechanisms
    (e.g. instinctual fear of pain),

    \item  \textit{Cognitively Generated/Slow Secondary Emotions}, those
    emotions that require some level of reasoning to elicit (e.g. learned fear
    of public speaking),

    \item \textit{Emotional Experience}, comprised of cognitive awareness of
    emotions being experienced, physi-ological changes, and subjective
    feelings---requiring self-awareness and consciousness to identi-fy---and

    \item \textit{Mind-Body Interactions}, the interactions between emotions and
    other cognitive and non-cognitive system components (e.g. bidirectional
    interactions between emotions and decision-making).
\end{itemize}

\textit{Emergent Emotions} are out of scope because they depend on
\textit{Emotion Effects on Behaviour} (Section~\ref{sec:cmetasks}), which is
itself out of scope. These would fall to game developers as part of
\textit{Allowing Developers to Specify How to Use Outputs} (\ref{flexOut}).
\textit{Emotional Experience} is also out of scope because NPCs do not
necessarily have to reason \textit{about} their emotions to interact with
players. \textit{Mind-Body Interactions} are also out of scope due to their
reliance on external components and processes which are not reliably available
due to the lack of restrictions on NPC embodiment
(Section~\ref{sec:embodiment}).

Both the \textit{fast primary emotions} and \textit{cognitively
generated/slow secondary emotions} are within \progname{}'s scope because they
describe two ways of generating emotion. The \ref{as} literature supports
modelling these as distinct processes (e.g. \cite{damasio2005descartes,
ledoux1996emotional}), and would also integrate some support for the
\textit{Ability to Operate on Different Levels of NPC Complexity}
(\ref{flexComplex}), and---potentially---\textit{Be Efficient and Scalable}
(\ref{flexScale}). Game developers can choose to use both, only primary---best
for smaller games with few emotional stimuli---or only secondary emotions---for
NPCs that have some planning or reasoning abilities---as their game requires.

\section{Examining Perspectives of Emotion}\label{sec:perspective_candidates}
With its design scope defined, \progname{} examines broad groups of emotion
theories (i.e. ``themes'' for thematic analysis) to see if they could support
it and the high-level requirements. For this example, \progname{} defines
groups following the themes/perspectives from the CME survey
(Chapter~\ref{sec:results}) which found that the perspectives frequently appear
in specific roles in CME designs:
\begin{itemize}
    \item Discrete theories\footnote{\citet{rodriguez2015computational} call
    these \textit{hierarchical theories}.} for defining a representation of an
    emotion state with clearly distinguishable categories and consequences of
    that state (e.g. facial expressions)

    \item Dimensional theories for creating simple models of emotion, viewing
    emotion categories from a different perspective, and defining a common space
    for representing different kinds of affect (e.g. emotion, personality,
    mood) and their interactions

    \item Appraisal theories for defining emotion processes and mechanisms

    \item Neurophysiological theories\footnote{Although they rarely appear
    (\citepg{lisetti2015and}{98}; \citepg{rodriguez2015computational}{451}).}
    guide architecture-related decisions that distinguish between \textit{fast,
    primary emotions} and \textit{cognitively generated/slow, secondary
    emotions}
\end{itemize}

\begin{figure}[!tb]
    \begin{center}
        \includegraphics[width=\linewidth]{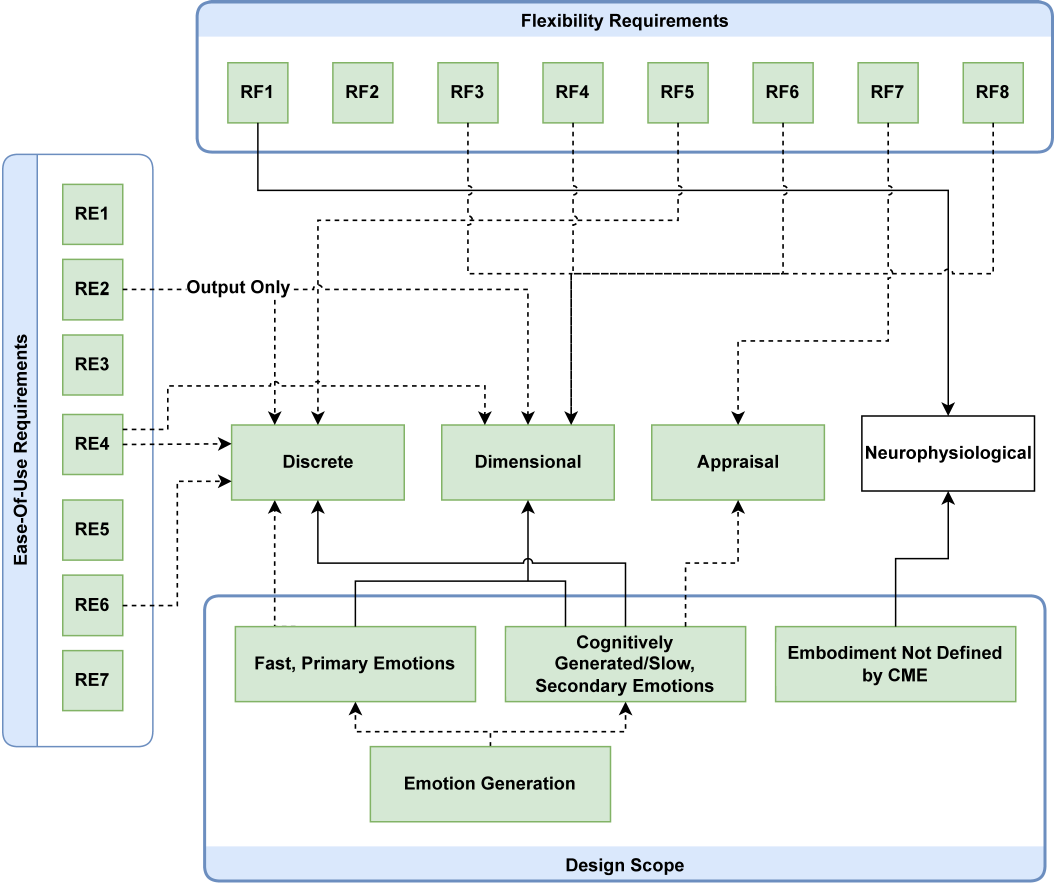}

        \caption[Overview of Influences on \progname{}'s View on Perspectives
        of Emotion]{Overview of Influences on \progname{}'s View on
        Perspectives of Emotion Showing the Connection to the High-Level
        Requirements and Design Scope (Shaded boxes are in scope; Solid arrows
        indicate conflicts with, dotted arrows indicate support for)}
        \label{fig:perspectivesOverview}
    \end{center}
\end{figure}

\progname{} uses these observations as a starting point to determine which ones
warrant an exploration of individual members
(Chapter~\ref{chapter:theoryAnalysis}). Figure~\ref{fig:perspectivesOverview}
shows the conclusions. \progname{} automatically considers all requirements to
be in scope. Boxes with outgoing solid arrows point to boxes they conflict
with, and therefore do not support \progname{}'s scope. Conversely, dotted
arrows indicate support for a box/component. Shaded boxes might be useful for
\progname{}'s design, warranting a closer examination.

\subsection{Discrete Theories}\label{sec:discreteperspective}
One of the core features of the discrete theories is the definition of distinct
emotion kinds, like \textit{Fear} and \textit{Anger}, that are recognizable by
a set of observable features (e.g. facial expression, typical behaviours). This
``...fits with the way we talk about emotion every day...people automatically
and effortlessly perceive emotion in themselves and
others...''~\citep[p.~47--48]{barrett2006emotions}. While true facial
expressions~\citep[p.~46]{barrett2019emotional} and survival-based
behaviours~\citep[p.~177]{barrett2018concepts} are likely more variable and
context-dependant than originally thought, caricatures of emotions are the most
unambiguous depictions of them. Animators use these to amplify the
believability of their characters (\citeg{gard2000building};
\citepg{loyall1997believable}{2};
\citepg{williams2001animator}{315--316})\footnote{The animated movie
\textit{Inside Out} credits Dr. Ekman as a scientific
advisor~\citep{insideout_ekman}.}. NPCs are also animated characters, so they
also benefit from exaggerated expressions~\citep[p.~5]{livingstone2006turing}.
This suggests that players could recognize which emotion an NPC is expressing
based on emotion kinds, which is useful for creating player studies for
\textit{Showing that \progname{} improves PX} (\ref{easePX}).

The simplicity offered by emotion kinds also suggests that the discrete theories
have a greater chance of being understood by game developers, which they can
use to convey the intended internal state of an NPC to
players~\citep[p.~352--353]{broekens2021emotion}. This implies excellent
support for several requirements, including \textit{Allowing developers to
choose which kinds of emotion \progname{} produces} (\ref{flexEm}),
\textit{Providing a clear and understandable API} (\ref{easeAPI}) for outputs,
and \textit{Allowing \progname{}'s outputs to be traceable and understandable}
(\ref{easeTrace}).

The discrete theories propose that innate, hardwired circuits or programs elicit
emotions (\citepg{ortony2021all}{41}; \citepg{scherer2021towards}{280}). While
this is within the scope of \textit{fast, primary emotions}
(Section~\ref{sec:emotioncomponents}), they are unable to define
\textit{cognitively generated/slow, secondary emotions} because there are no
clearly defined processes or mechanisms to do so. This makes discrete theories
unable to satisfy many of \progname{}'s requirements such as \textit{Allowing
the game developer to choose which of \progname{}'s tasks to use}
(\ref{flexTasks}), \textit{Allowing the customization or redefinition of the
\progname{}'s preexisting configuration parameters} (\ref{flexCustom}),
\textit{Hiding the complexity of emotion generation} (\ref{easeHide}),
\textit{Providing a clear and understandable API} (\ref{easeAPI}) for inputs,
\textit{Allowing \progname{}'s outputs to be traceable and understandable}
(\ref{easeTrace}), and \textit{Allowing developers the option to automate the
storing and decaying of \progname{}'s emotion state} (\ref{easeAuto}). However,
they cannot conflict with these requirements either.

Although they cannot satisfy \progname{}'s needs alone, the benefits of
discrete theories might outweigh their lack of emotion elicitation processes.
Therefore, \progname{} examines the discrete theories individually with respect
to each high-level requirement.

\subsection{Dimensional Theories}
These theories can more easily distinguish between different emotions
(\citepg{scherer2010emotion}{12}; \citepg{smith1985patterns}{813}) using a
small number of continuous dimensions. The dimensional theories view emotion as
an individual's interpretation of their current ``core
affect''~\citep[p.~353]{broekens2021emotion} (i.e. elementary affective
feelings). While any point in dimensional space is part of ``core
affect''~\citep[p.~97]{lisetti2015and}, it is possible for individuals to
verbally label points representing their subjective feeling of an
emotion~\citep[p.~12]{scherer2010emotion}. This creates an effect of
``plotting'' emotion kinds (e.g. as in discrete theories) as points in the
space, implying that the dimensional theories could also support
\textit{Providing a clear and understandable API} (\ref{easeAPI}) for outputs
and \textit{Allowing \progname{}'s outputs to be traceable and understandable}
(\ref{easeTrace}). However, emotions defined by dimensions alone might be more
difficult to label with everyday language~\citep[p.~213]{johnson1992basic}
implying that they are harder to understand intuitively than discrete emotion
kinds without supplemental qualitative information.

``Core affect'' can also relate other psychological constructs such as
personality and thought \citep[p.~353]{broekens2021emotion}, making dimensional
theories ideal for creating seamless interaction dynamics, \textit{Allowing
developers to integrate new components into \progname{} that influence or are
influenced by emotion} (\ref{flexNew}). The nature of dimensions also affords a
more granular way to define emotion-driven responses, affording more ways for
\textit{Allowing developers to specify how to use \progname{}'s outputs}
(\ref{flexOut}).

The numerical nature of dimensional theories could also support requirements
like \textit{Allowing the customization or redefinition of \progname{}'s
preexisting configuration parameters} (\ref{flexCustom}), and are likely to
\textit{Be efficient and scalable to minimize the overall impact on game
performance} (\ref{flexScale}). As a bonus for CME development, these theories
are convenient to implement~\citep[p.~440]{rodriguez2015computational}.

The dimensional theories ``...say comparatively little...about how the
different emotions are produced, and what useful or other effects they
have...''~\cite[p.~250]{reisenzein2013computational}. This makes them, like the
discrete theories, unable to satisfy or conflict with \progname{}'s ability to
produce \textit{cognitively generated/slow, secondary emotions}. Unlike the
discrete theories, the dimensional theories cannot define \textit{fast, primary
emotions} either.

Given the number of requirements they could support compared to those they
cannot, \progname{} examines the dimensional theories individually with respect
to each high-level requirement. However, like the discrete theories, it is
unlikely that they could support \progname{}'s needs sufficiently on their own.

\subsection{Appraisal Theories}
Appraisal theories propose that emotions arise from evaluations of the
relationship between an individual's well-being and their environment rather
than the environment's objective
qualities~\citep[p.~86]{smith2000consequences}. This naturally accounts for
individual differences since someone can appraise a situation differently than
another~\citep[p.~1353]{smith2009putting}. These theories often conceptualize
more than one emotion processing pathway~\citep[p.~129]{smith2001toward},
allowing for flexible behaviours with different response times. This implies
that appraisal theories do not conflict with the requirement to have both
\textit{fast, primary emotions} and \textit{cognitively generated/slow,
secondary emotions} in \progname{}'s design scope
(Section~\ref{sec:emotioncomponents}). However, these theories mainly focus on
the link between cognitive processing and emotion
elicitation~\citep[p.~354]{broekens2021emotion} so \progname{} should only rely
on them to define \textit{cognitively generated/slow, secondary emotions}. The
existence of multiple processing pathways implies the potential for variable
levels of component complexity in multi-component emotion generation depending
on the information available to the CME. This supports \textit{Allowing
developers to use \progname{} on different levels of NPC complexity}
(\ref{flexComplex}).

Since the appraisal perspective is the only one that addresses the
\textit{cognitively generated/slow, secondary emotions} aspect of the design
scope, \progname{} must incorporate it. This is unsurprising: appraisal
theories appear the most frequently in \ref{ac} (\citepg{lisetti2015and}{97};
\citepg{marsella2015appraisal}{55}) compared to discrete, dimensional, and
neurophysiological theories. They can be relatively simple to realize
computationally and often meet several CME requirements concerning input
evaluation, emotion elicitation, and emotion response
generation~\citep[p.~439]{rodriguez2015computational}. However, \textit{which}
of the appraisal theories \progname{} should use is less clear so it must
examine them more closely.

\subsection{Neurophysiological Theories}
Although they support the inclusion of the \textit{fast, primary emotions} and
\textit{cognitively generated/slow, secondary emotions} components in the
design scope (Section~\ref{sec:emotioncomponents}), processes and mechanisms in
neurophysiological theories rely on the surrounding agent
architecture~\citep[p.~218, 225--226]{sloman1987motives}, learning mechanisms
(e.g. Somatic Marker Hypothesis~\citep[p.~179--180]{damasio2005descartes}),
and/or the body's somatic and visceral responses
(\citepg{damasio2005descartes}{249--250}; \citepg{ledoux1996emotional}{41,
176}). Since the \textit{Mind-Body Interactions} emotion component is beyond
the scope of \progname{} (Section~\ref{sec:emotioncomponents}), and
consequently cannot truly support \textit{Independence from an Agent
Architecture} (\ref{flexArch}), \progname{} should not use theories from the
neurophysiological perspective. There might be variations of these theories
that could work, but the effort necessary to determine this is beyond the
current scope.

\section{Summary}
Without criteria to measure with, choosing affective theories for \progname{}
is ineffectual. To ensure that \progname{} embeds its focus in the design, the
criteria are its high-level requirements and scope. The high-level requirements
capture some of the needs of game designers, categorized as:
\begin{itemize}
    \item Flexibility requirements, such that designers can adapt the CME to a
    variety of games, and
    \item Ease-of-use requirements, to minimize the burden of its use during
    development.
\end{itemize}

\progname{}'s scope is emotion generation alone. This means that it ignores any
tasks concerned with expressing emotion, reasoning about emotion, or defining
interactions between emotions and other components. There are also no
assumptions about NPC embodiment to maximize \progname{}'s flexibility. Within
the emotion generation task, \progname{} should be able to generate fast,
reactive emotions and slow, cognitively driven ones. Designing these
independently allows game designers to control them individually for maximal
flexibility.

Examining the theories broadly as perspectives shows that the
neurophysiological theories are not suitable for satisfying \progname{}'s
requirements. Therefore, \progname{} only further examines theories from the
discrete, dimensional, and appraisal perspectives with respect to high-level
requirements.

\clearpage
\vspace*{\fill}
\begin{keypoints}
    \begin{itemize}

        \item Defining the high-level requirements and scope of \progname{}
        provides a framework for choosing which affective theories to use

        \item One must consider the game designer's needs to increase a CME's
        chance of being adopted in practice

        \item The high-level requirements of a CME for believable NPCs include
        those that make the CME flexible and easy to use and guide the
        selection of affective theories that form the CME's theoretical
        foundation

        \item \progname{}'s scope includes \textit{fast, primary emotions}
        and \textit{slow, secondary emotions} within the \textit{emotion
        generation} task with no assumptions about NPC embodiment

        \item The \textit{discrete}, \textit{dimensional}, and
        \textit{appraisal} theoretical perspectives on emotion are likely to be
        the most helpful for \progname{}'s design, whereas the
        \textit{neurophysiological} theories are not ideal due to conflicts with
        some high-level requirements
    \end{itemize}
\end{keypoints}

\parasep
\vspace*{\fill}

%% file: theory2reqs.tex
\chapter{Support Your \progname{}: The Requirements Choose the Theories}
\label{chapter:theoryAnalysis}
\def\epigraphflush{center}
\setlength{\epigraphwidth}{0.75\textwidth}
\def\textflush{center}
\epigraph{Scan complete.}{Baymax, \textit{Big Hero 6}}

With the broad ``themes'' of emotion theories identified, \progname{} proceeds
to examine theories (Table~\ref{tab:theories}) within those themes to see ``how
well'' they satisfy each high-level requirement individually given
\progname{}'s design scope and domain of NPC behaviour for player engagement
(i.e. content analysis stage in document analysis). It does this by identifying
features of each theory and interpreting their ``ability'' to support a
high-level requirement (Section~\ref{sec:theoryanalysis}). This data guides the
final selection process (i.e. drawing recommendations/conclusions,
Section~\ref{sec:choosetheories}). \textit{This analysis is, in part,
subjective} because a judgment---however well-supported by evidence from the
literature---is made without true objective measures or methods. A different
understanding of the requirements and evolution of \ref{as} could produce
variations in the results.

\begin{table}[!b]
    \renewcommand{\arraystretch}{1.2}
    \centering
    \caption{Theories Analyzed for \progname{}}
    \label{tab:theories}
    \small
    \begin{tabular}{P{0.2\columnwidth}P{0.65\columnwidth}}
        \toprule
        \textbf{Perspective} & \textbf{Theories} \\

        \midrule

        \colourRow Discrete & Ekman \& Friesen (Ek.), Izard
        (Iz.), Plutchik (Plu.) \\

        Dimensional & Valence-Arousal (V-A), PAD Space (PAD) \\

        \colourRow Appraisal & Frijda (Frj.), Lazarus (Laz.),
        Scherer (Sch.), Roseman
        (Ros.), Ortony, Clore, and Collins (OCC), Smith \& Kirby (S \& K),
        Oatley \& Johnson-Laird (O \& JL) \\

        \bottomrule

        & \\[-4mm]

        \multicolumn{2}{l}{\footnotesize{Appendix~\ref{chapter:affect} has
        notes about these theories/models}}

    \end{tabular}
\end{table}

\section{Approaching Theory Analysis}\label{sec:theoryanalysis}
\progname{} divides its high-level requirements into \textit{system-level},
which applies to \progname{} as a whole, and \textit{component-level} for
requirements that only apply to specific pieces of \progname{}
(Table~\ref{tab:req-division-summary}). The analysis assigns categories for
component-level requirements solely to appraisal theories because they concern
process-related elements that discrete and dimensional theories do not address.
Each theory has unique elements which might make it better or worse for
satisfying a particular requirement. The analysis notes these. However, is not
unusual for theories from the same perspective to satisfy a requirement equally
well. In these cases, the requirement examination treats these theories as a
collective unit. Since the perspective-level analysis showed that the
neurophysiological theories are ill-suited for \progname{}, it does not analyze
them in detail.

\begin{table}[!tb]
    \renewcommand{\arraystretch}{1.2}
    \centering
    \caption{Summary of High-Level Requirement Division}
    \label{tab:req-division-summary}
    \small
    \begin{tabular}{@{}c P{0.61\linewidth} c c@{}}
        \toprule
        \multicolumn{2}{c}{} & \textbf{System} & \textbf{Component} \\
        \midrule

        \colourRow \ref{flexArch} & \textit{Independence from
        an Agent Architecture} & \checkmark &  \\

        \ref{flexTasks} & \textit{Choosing Which Tasks to Use} &  & \checkmark
        \\

        \colourRow \ref{flexCustom} & \textit{Customization
        of Existing Task Parameters} &  & \checkmark \\

        \ref{flexNew} & \textit{Allowing the Integration of Components} &
        \checkmark &  \\

        \colourRow \ref{flexEm} & \textit{Allowing Designers to Choose What
        Emotions an NPC can Have} & \checkmark &  \\

        \ref{flexOut} & \textit{Allowing Developers to Specify How to Use CME
        Outputs} & \checkmark &  \\

        \colourRow \ref{flexComplex} & \textit{Ability to
        Operate on Different Levels of NPC Complexity} & \checkmark &  \\

        \ref{flexScale} & \textit{Be Efficient and Scalable} & \checkmark &  \\

        \colourRow \ref{easeHide} & \textit{Hiding the
        Complexity of Emotion Generation} &  & \checkmark \\

        \ref{easeAPI} & \textit{Having a Clear API (Input)} &  & \checkmark \\

        \colourRow \ref{easeAPI} & \textit{Having a Clear API
        (Output)} & \checkmark &  \\

        \ref{easeAuthor} & \textit{Minimizing Authorial Burden} &
        \multicolumn{2}{c}{Excluded from analysis} \\

        \colourRow \ref{easeTrace} & \textit{Traceable CME
        Outputs} &  & \checkmark \\

        \ref{easeAuto} & \textit{Allowing the Automatic Storage and Decay of
        the Emotion State} &  & \checkmark \\

        \colourRow \ref{easePX} & \textit{Showing that
        Emotions Improve the Player Experience} & \checkmark &  \\

        \ref{easeNovel} & \textit{Providing Examples of Novel Game Experiences}
        & \checkmark &  \\

        \hline\bottomrule
    \end{tabular}
\end{table}

\subsection{Scoping Some Requirements}
While most of the high-level requirements do not need additional scoping for
this analysis, some do to better focus on what information to search for.

The analysis separates \textit{Providing a clear and understandable API}
(\ref{easeAPI}) into two---Input and Output---to get a better feel for each
theory's usefulness and to acknowledge that some theories are better for
emotion expression such as Ekman \& Friesen.

The requirement for \textit{Allowing the Integration of New Components}
(\ref{flexNew}) is broad and \progname{} should scope it for its initial
design. New \progname{} components could be non-affective---such as
attention---and affective---like personality---in nature. Integrating
non-affective components should be theory-agnostic because they are in separate
components of mind~\citep{cognitiondef}. From a software engineering
perspective, one could view the mind as a system with distinct, interacting
subsystems. Modular interfaces that control interactions between
\progname{}---the affective subsystem---and components from other subsystems
would support this concept while also supporting \progname{}'s requirements for
\textit{Independence from an Agent Architecture} (\ref{flexArch}) and
\textit{Allowing Developers to Specify How to Use Outputs} (\ref{flexOut}).
Integrating other affective components would depend on \progname{} and its
foundational theories, limited to only those components that its emotions or
other types of affect can represent and connect to.

This requirements analysis focuses on three other types of affect as defined in
Chapter~\ref{sec:affectiveDefs}: ``core'' affect, mood, and personality. Of
these, it prioritizes personality because it is necessary for creating the
consistent and coherent agent behaviours that influence believability
(\citepg{reilly1996believable}{26}; \citepg{loyall1997believable}{19};
\citepg{ortony2002making}{203}).

Finally, the analysis excludes \textit{Minimizing authorial burden}
(\ref{easeAuthor}) because it focuses on helping game developers manage the
creation of an increasing NPC population which is agnostic of the underlying
theories.

\subsection{Making Notes About and Scoring Emotion Theories}\label{sec:analysis}
With these more specific high-level requirements, examining each theory with
guidance from individual requirements produces a set of notes. After reviewing
the notes, each theory has an assigned \textit{score} describing its relative
``suitability'' for that requirement (Table~\ref{tab:scoring}). This step is
somewhat subjective because the evaluations do not have true objective measures
or methods. As an example, notes about the dimensional theories and
\ref{flexEm} are here, while the rest are in
Appendix~\ref{chapter:reqsTheoryNotes}.
Tables~\ref{tab:theory-req-sys-summary-flexibility},
\ref{tab:theory-req-sys-summary-easeofuse},
\ref{tab:theory-req-comp-summary-flexibility}, and
\ref{tab:theory-req-comp-summary-easeofuse} summarize the scores for each
theory and requirement.

\vspace*{\fill}
\begin{table}[!hbt]
    \renewcommand{\arraystretch}{1.2}
    \centering
    \caption{Summary of Scoring Categories}
    \label{tab:scoring}
    \small
    \begin{tabular}{ccP{0.65\columnwidth}}
        \toprule
        \textbf{Score Category} & \textbf{Symbol} & \textbf{Definition} \\

        \midrule

        \colourRow \textit{Strong} & \strong & The theory
        appears to satisfy the requirement in a clear, understandable way and
        is likely to aid in \progname{}'s usability \\

        \textit{Good} & \good & The theory appears to satisfy the requirement
        and is somewhat defined \\

        \colourRow \textit{Weak} & \weak & The theory
        describes ways that \textit{could} satisfy the requirement, but it is
        not fully defined or could make \progname{} harder to use \\

        \textit{Disqualified} & \disqualified & The theory does not seem likely
        to be able to satisfy the requirement, or it violates other
        requirements when it can (including psychological validity) \\

        \bottomrule
    \end{tabular}
\end{table}
\vspace*{\fill}

\clearpage

\vspace*{\fill}

\begin{notes}[frametitle=Dimensional Theories: Allowing Designers to Choose
What Emotions an NPC can Have (\ref{flexEm})]
    Neither V-A or PAD strictly enforce the inclusion of specific emotion types.
    Instead, the use of dimensions allows for an infinite number of affective
    states. While this trivially supports the ability to \textit{Allow
    Designers to Choose What Emotions an NPC can Have} (\ref{flexEm}), the
    dimensional theories might not be practical for \progname{} on their own.
    Instead, point locations representing named emotions guide the addition of
    specific ones. This removes the burden of deciding where an emotion's
    location in dimensional space is from game designers if they do not want to
    do so themselves.

    \begin{itemize}
        \item \textbf{V-A} (\weak)
        \begin{itemize}
            \item Space represented by \ref{valence} and \ref{arousal} only
            represents part of an emotion episode (\citepg{yik2002relating}{90};
            \citepg{roseman2011emotional}{441}; \citepg{lisetti2015and}{97})

            \item Not ideal $\rightarrow$ some emotions, like \textit{Anger} and
            \textit{Fear}, are difficult to differentiate without additional
            information
            \begin{itemize}
                \item Might be some of the most common emotions that a game
                designer will use $\rightarrow$ could be the \textit{only} two
                emotions required in some games (e.g. NPCs in oppositional
                First Person Shooters (FPSs) due to the game's pace and the
                limited time and ways that players interact with them)
            \end{itemize}

            \item Adding new emotions cannot be adequately contained in V-A
        \end{itemize}

        \item \textbf{PAD} (\good)
        \begin{itemize}
            \item Accompanied by a list of 151 emotion
            labels~\citep[p.~42--45]{mehrabian1980basic} identified from
            empirical data $\rightarrow$ notes their average location in PAD
            space and the standard deviation in the data

            \item List is still finite and cannot account for cultural
            differences, might not cover all of the affective states that a game
            designer needs $\rightarrow$ list is long enough that there is a
            reasonable chance that a game designer can find all the affective
            state labels that they require

            \item Prone to interpretation errors, as the designer's definition
            and the definition used to locate points in PAD space might not be
            the same $\rightarrow$ designers can make their own judgments of
            the suitability of a term based on its coordinates, potentially
            violating \textit{Hiding the Complexity of Emotion Generation}
            (\ref{easeHide})
        \end{itemize}
    \end{itemize}
\end{notes}

\vspace*{\fill}

\clearpage
\begin{landscape}
    \vspace*{\fill}
    \begin{table}[!tbh]
        \renewcommand{\arraystretch}{1.3}
        \centering
        \caption[Support for System-Level Flexibility High-Level
        Requirements]{Support for System-Level \textbf{Flexibility} High-Level
            Requirements}
        \label{tab:theory-req-sys-summary-flexibility}
        \footnotesize
        \begin{threeparttable}
            \begin{tabular}{@{}cP{0.3\linewidth}cccccccccccc@{}}

                \toprule
                \multicolumn{2}{c}{} & \textbf{Ek.} & \textbf{Iz.} &
                \textbf{Plu.} & \textbf{V-A} & \textbf{PAD} & \textbf{Frj.} &
                \textbf{Laz.}{\footnotesize\textpmhg{\Hi}} & \textbf{Sch.} &
                \textbf{Ros.} & \textbf{OCC} & \textbf{S\&K} & \textbf{O\&JL} \\
                \midrule

                \colourRow \ref{flexArch} &
                \textit{Independence from an Agent Architecture} &
                {\normalsize\good} & {\normalsize\good} & {\normalsize\good} &
                {\normalsize\good} & {\normalsize\good} & {\normalsize\strong}
                & {\normalsize\good} & \disqualified & {\normalsize\good} &
                {\normalsize\strong} & {\normalsize\strong} &
                {\normalsize\strong} \\

                \ref{flexNew} & \textit{Allowing the Integration of
                Components}{\large\textpmhg{\Hl}} & {\normalsize\weak} &
                {\normalsize\weak}\textsuperscript{\normalsize\Moon} &
                {\normalsize\good}\textsuperscript{\normalsize\Moon} &
                {\normalsize\strong}\textsuperscript{\large\Jupiter} &
                {\normalsize\strong}\textsuperscript{\large\Jupiter} &
                {\normalsize\strong} & {\normalsize\weak} & {\normalsize\weak} &
                {\normalsize\weak} & {\normalsize\strong} & {\normalsize\weak} &
                {\normalsize\strong} \\

                \colourRow \ref{flexEm} & \textit{Allowing Designers to Choose
                What Emotions an NPC can Have} & {\normalsize\weak} &
                \disqualified & {\normalsize\strong} & {\normalsize\weak} &
                {\normalsize\good} & {\normalsize\weak} & {\normalsize\weak} &
                {\normalsize\good} & {\normalsize\good} & {\normalsize\good} &
                {\normalsize\weak} & {\normalsize\strong} \\

                \ref{flexOut} & \textit{Allowing Developers to Specify How to
                Use CME Outputs} & {\normalsize\good} & {\normalsize\good} &
                {\normalsize\good} & {\normalsize\strong} &
                {\normalsize\strong} & {\normalsize\strong} &
                {\normalsize\strong} & {\normalsize\strong} &
                {\normalsize\strong} & {\normalsize\strong} &
                {\normalsize\strong} & {\normalsize\strong} \\

                \colourRow \ref{flexComplex} &
                \textit{Ability to Operate on Different Levels of NPC
                Complexity} & {\normalsize\good} & {\normalsize\good} &
                {\normalsize\good} & {\normalsize\weak} & {\normalsize\weak} &
                {\normalsize\weak} & {\normalsize\strong} &
                {\normalsize\strong} & {\normalsize\good} & {\normalsize\good} &
                {\normalsize\strong} & {\normalsize\good} \\

                \ref{flexScale} & \textit{Be Efficient and Scalable} &
                {\normalsize\good} & {\normalsize\good} & {\normalsize\good} &
                {\normalsize\weak} & {\normalsize\weak} &
                {\normalsize\good}\textsuperscript{\large\Pluto} &
                {\normalsize\strong} & {\normalsize\good} & {\normalsize\weak} &
                {\normalsize\good} & {\normalsize\strong} &
                {\normalsize\strong} \\

                \hline\bottomrule
            \end{tabular}
            \begin{tablenotes}

                \footnotesize
                \vspace*{2mm}

                \item \textit{See Table~\ref{tab:scoring} for score category
                descriptions.}

                \item {\small\textpmhg{\Hi}} \textit{Excludes the
                \textit{Coping Process} because it is part of action
                generation.}

                \item {\Large\textpmhg{\Hl}} \textit{Strictly focusing on
                ``core'' affect, mood, and personality.}

                \item {\normalsize\Moon} \textit{Natively supports integration
                with Personality.}

                \item {\Large\Jupiter} \textit{Personality integration based on
                the Five Factor Model (OCEAN).}

                \item {\Large\Pluto} \textit{Might improve if some factors are
                not necessary for implementation scope.}

            \end{tablenotes}
        \end{threeparttable}%
    \end{table}

    \begin{table}[!tbh]
        \renewcommand{\arraystretch}{1.3}
        \centering
        \caption[Support for System-Level Ease-of-Use High-Level
        Requirements]{Support for System-Level \textbf{Ease-of-Use}
            High-Level Requirements}
        \label{tab:theory-req-sys-summary-easeofuse}
        \footnotesize
        \begin{threeparttable}
            \begin{tabular}{@{}cP{0.33\linewidth}cccccccccccc@{}}

                \toprule
                \multicolumn{2}{c}{} & \textbf{Ek.} & \textbf{Iz.} &
                \textbf{Plu.} & \textbf{V-A} & \textbf{PAD} & \textbf{Frj.} &
                \textbf{Laz.}{\footnotesize\textpmhg{\Hi}} & \textbf{Sch.} &
                \textbf{Ros.} & \textbf{OCC} & \textbf{S\&K} & \textbf{O\&JL} \\
                \midrule

                \colourRow \ref{easeAPI} & \textit{Having a
                Clear API (Output)} & {\normalsize\strong} & {\normalsize\weak}
                & {\normalsize\good} & {\normalsize\weak} & {\normalsize\weak} &
                {\normalsize\good} & {\normalsize\strong} & {\normalsize\weak} &
                {\normalsize\strong} & {\normalsize\good} &
                {\normalsize\strong} & {\normalsize\strong} \\

                \ref{easePX} & \textit{Showing that Emotions Improve the Player
                    Experience} & {\normalsize\good} & {\normalsize\good} &
                {\normalsize\good} & {\normalsize\good} & {\normalsize\good} &
                {\normalsize\good} & {\normalsize\good} & {\normalsize\strong} &
                {\normalsize\good} & {\normalsize\weak} & {\normalsize\good} &
                {\normalsize\good} \\

                \colourRow \ref{easeNovel} & \textit{Providing
                    Examples of Novel Game Experiences} & {\normalsize\weak} &
                {\normalsize\weak} & {\normalsize\weak} & {\normalsize\good} &
                {\normalsize\good} & {\normalsize\good} & {\normalsize\good} &
                {\normalsize\good} & {\normalsize\good} & {\normalsize\good} &
                {\normalsize\good} & {\normalsize\good} \\

                \hline\bottomrule
            \end{tabular}
            \begin{tablenotes}

                \footnotesize
                \vspace*{2mm}

                \item \textit{See Table~\ref{tab:scoring} for score category
                    descriptions.}

                \item {\small\textpmhg{\Hi}} \textit{Excludes the
                \textit{Coping Process} because it is part of action
                generation.}

            \end{tablenotes}
        \end{threeparttable}%
    \end{table}
    \vspace*{\fill}
\end{landscape}

\begin{table}[!tbh]
    \renewcommand{\arraystretch}{1.3}
    \centering
    \caption[Support for Component-Level Flexibility High-Level
    Requirements]{Support for Component-Level \textbf{Flexibility}
    High-Level Requirements}
    \label{tab:theory-req-comp-summary-flexibility}
    \footnotesize
    \begin{threeparttable}
        \begin{tabular}{@{}cP{0.3\linewidth}ccccccc@{}}
            \toprule
            \multicolumn{2}{c}{} & \textbf{Frj.} &
            \textbf{Laz.}{\footnotesize\textpmhg{\Hi}} & \textbf{Sch.} &
            \textbf{Ros.} & \textbf{OCC} & \textbf{S\&K} & \textbf{O\&JL} \\
            \midrule

            \colourRow \ref{flexTasks} & \textit{Choosing
            Which Tasks to Use} & {\normalsize\weak} & {\normalsize\weak} &
            {\normalsize\strong} & \disqualified & {\normalsize\good} &
            {\normalsize\good} & {\normalsize\good} \\

            \ref{flexCustom} & \textit{Customization of Existing Task
            Parameters} & {\normalsize\strong} & \disqualified &
            {\normalsize\good} & \disqualified & {\normalsize\good} &
            {\normalsize\good} & {\normalsize\good} \\

            \hline\bottomrule
        \end{tabular}
        \begin{tablenotes}

        \footnotesize
        \vspace*{2mm}

        \item \textit{See Table~\ref{tab:scoring} for score category
            descriptions.}

        \item {\small\textpmhg{\Hi}} \textit{Excludes the
        \textit{Coping Process} because it is part of action
        generation.}

    \end{tablenotes}
\end{threeparttable}%
\end{table}

\begin{table}[!tbh]
    \renewcommand{\arraystretch}{1.3}
    \centering
    \caption[Support for Component-Level Ease-of-Use High-Level
    Requirements]{Support for Component-Level \textbf{Ease-of-Use}
    High-Level Requirements}
    \label{tab:theory-req-comp-summary-easeofuse}
    \footnotesize
    \begin{threeparttable}
        \begin{tabular}{@{}cP{0.3\linewidth}ccccccc@{}}
            \toprule
            \multicolumn{2}{c}{} & \textbf{Frj.} &
            \textbf{Laz.}{\footnotesize\textpmhg{\Hi}} & \textbf{Sch.} &
            \textbf{Ros.} & \textbf{OCC} & \textbf{S\&K} & \textbf{O\&JL} \\
            \midrule

            \colourRow \ref{easeHide} & \textit{Hiding the
            Complexity of Emotion Generation} & {\normalsize\strong} &
            {\normalsize\strong} & {\normalsize\strong} & {\normalsize\good} &
            {\normalsize\good} & {\normalsize\good} & {\normalsize\good} \\

            \ref{easeAPI} & \textit{Having a Clear API (Input)} &
            {\normalsize\good} & {\normalsize\disqualified} &
            {\normalsize\good} & {\normalsize\good} & {\normalsize\weak} &
            {\normalsize\good} & {\normalsize\strong} \\

            \colourRow \ref{easeTrace} & \textit{Traceable
            CME Outputs} & {\normalsize\strong} & {\normalsize\strong} &
            {\normalsize\weak} & {\normalsize\strong} & {\normalsize\strong} &
            {\normalsize\strong} & {\normalsize\strong} \\

            \ref{easeAuto} & \textit{Allowing the Automatic Storage and Decay
            of the Emotion State} & {\normalsize\good} & {\normalsize\good} &
            {\normalsize\good} & {\normalsize\disqualified} &
            {\normalsize\good} & {\normalsize\good} & {\normalsize\good} \\

            \hline\bottomrule
        \end{tabular}
    \begin{tablenotes}

        \footnotesize
        \vspace*{2mm}

        \item \textit{See Table~\ref{tab:scoring} for score category
            descriptions.}

        \item {\small\textpmhg{\Hi}} \textit{Excludes the
            \textit{Coping Process} because it is part of action
            generation.}

    \end{tablenotes}
\end{threeparttable}%
\end{table}

\section{Theory Selection}\label{sec:choosetheories}
After examining theories from the perspective of a high-level requirement,
\progname{} can compare their scores for their relative suitability (i.e.
drawing recommendations/conclusions). For each requirement, counting
occurrences of the score categories assigned to each theory gives them a
``rank'' relative to the others
(Tables~\ref{tab:theory-req-sys-summary-flexibility},
\ref{tab:theory-req-sys-summary-easeofuse},
\ref{tab:theory-req-comp-summary-flexibility}, and
\ref{tab:theory-req-comp-summary-easeofuse}). This means that choosing theories
that best serve a high-level requirement becomes a min/max problem where the
goal is to maximize the number of \textit{Strong} and \textit{Good} scores while
minimizing the \textit{Weak} and \textit{Disqualified} ones. More sophisticated
analyses are not possible because the score categories are
\textit{nominal/categorical data}---code assignments representing a
``suitability'' attribute---so mathematical operations on them do not yield
meaningful information~\cite[p.~134--135]{mackenzie2012human}.

If there is only one requirement, it is a straightforward selection process
that could result in few theories to pick from. Choosing one of those theories
requires some additional experimentation to see how well one can formalize them
and which ones might be preferable. In practice, it is likely that there are
multiple requirements to satisfy.

Choosing theories when there are multiple requirements means finding a balance
between those theories that strongly satisfy some, but often not all,
requirements. In some cases, it might be necessary to choose a theory that
satisfies a high-priority requirement strongly and a low-priority one weakly.
Referencing the CME's design scope helps with requirement prioritization and
guides the theory selection task.

\subsection{Choosing Only One Theory}\label{sec:singletheory}
If a CME design can use only one theory as its foundation (e.g. limiting design
complexity, target game is relatively simple) the decision process is simple:
pick the theory that most strongly satisfies the most requirements in priority
order.

If this was the case for \progname{}, eliminating the theories that have
\textit{Disqualified} scores in
Tables~\ref{tab:theory-req-sys-summary-flexibility},
\ref{tab:theory-req-sys-summary-easeofuse},
\ref{tab:theory-req-comp-summary-flexibility}, and
\ref{tab:theory-req-comp-summary-easeofuse} leaves Frijda, OCC, Smith \& Kirby,
and Oatley \& Johnson-Laird. Of these, only Smith \& Kirby and Oatley \&
Johnson-Laird have \textit{Good} or \textit{Strong} scores for the prioritized
requirements. Either of these could be good choices, but experimenting with
Oatley \& Johnson-Laird for its suitability should be first because it has only
\textit{Good} and \textit{Strong} scores for all requirements whereas Smith \&
Kirby have a few \textit{Weak} scores for non-prioritized requirements.

\subsection{Choosing Multiple Theories}\label{sec:multireqs}
It would be ideal if there was a single theory to explain affective phenomena.
However, creating one theory of affect would require reconciling narrowly
defined existing theories and their architectural
assumptions~\citep[p.~99]{lisetti2015and}. Instead of trying to create one
unified model of emotion, one can combine theories to address different
modelling needs.

If a CME can use more than one theory, the selection process becomes more
difficult. One tactic would be to immediately eliminate theories that cannot
satisfy a requirement (i.e. have a \textit{Disqualified} score), but this might
eliminate a theory that is well-suited in other aspects. For example,
Tables~\ref{tab:theory-req-sys-summary-flexibility},
\ref{tab:theory-req-sys-summary-easeofuse},
\ref{tab:theory-req-comp-summary-flexibility}, and
\ref{tab:theory-req-comp-summary-easeofuse} (no assigned score implies
\textit{Disqualified}) shows that the discrete and dimensional theories have
considerably more \textit{Disqualified} scores (6 to 7) than the appraisal
theories (0 to 3). This is unsurprising because those requirements relate to
the \textit{process} of emotion generation, which reinforces the discrete and
dimensional theories' inability to satisfy the need for \textit{cognitively
generated/slow, secondary emotions} specified in \progname{}'s design scope
(Chapter~\ref{sec:scope}). However, by choosing theories by this information
alone we might prematurely eliminate theories that we could use to specify some
requirements very well (e.g. PAD Space for integrating multiple types of
affect). Choosing theories based on occurrences of \textit{Strong} scores
creates a similar problem, where a theory might be excellent for a few
requirements but a poor choice for many others. Instead, the \textit{coverage}
achieved by the \textit{set} of chosen theories must satisfy all prioritized
requirements. It also becomes possible to build better support for
non-prioritized requirements due to overall requirements coverage, unlike in
the single-theory example (Section~\ref{sec:singletheory}).

Finding such a set of theories benefits from a systematic decision-making
process where information about the CME's high-level requirement scores and
their derivation justifies each step (Figure~\ref{fig:choosetheoriesFlow}):
\afterpage{
    \clearpage
    \vspace*{\fill}
    \begin{figure}[!ht]
        \centering
        \includegraphics[width=\linewidth]{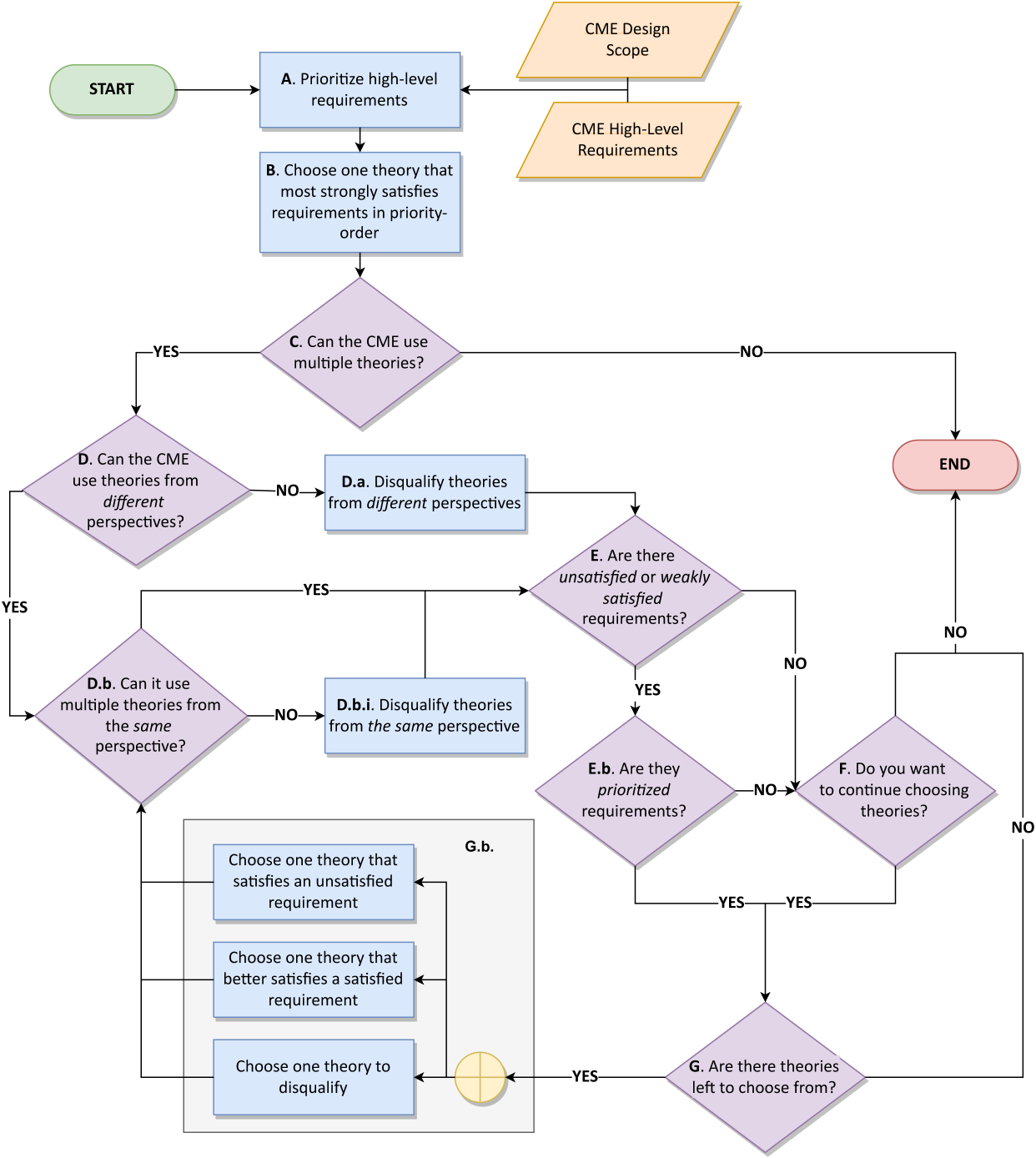}

        \caption{Overview of the Proposed Decision Process for Choosing a CME's
        Emotion Theories/Models}
        \label{fig:choosetheoriesFlow}
    \end{figure}
    \vspace*{\fill}
    \clearpage
}

\begin{enumerate}[label=(\Alph*), ref=\Alph*.]

    \item\label{step:priority} Prioritize (``sort'') the high-level
    requirements using the CME's design scope, then go to \ref{step:pickfirst}

    \item\label{step:pickfirst} Choose one theory or model that most strongly
    satisfies the high-level requirements in priority-order, then go to
    \ref{step:decideMulti}

    \item\label{step:decideMulti} Decide if the CME can use more than one
    theory and/or model in its design.
    \begin{enumerate}[label=(\alph*), ref=\alph*.]

        \item\label{step:decideMultiNO} If \textsc{NO} (e.g. limiting design
        complexity, target application is relatively simple), then go to
        \ref{step:finish}

        \item\label{step:decideMultiYES} If \textsc{YES},  then go to
        \ref{step:decideDiff}

    \end{enumerate}

    \item\label{step:decideDiff} Decide if the CME can use theories and/or
    models from \textit{different} perspectives (e.g. discrete, dimensional,
    appraisal) in its design (see Section~\ref{sec:cancombine} for discussion).
    \begin{enumerate}[label=(\alph*), ref=\alph*.]

        \item\label{step:decideDiffNO} If \textsc{NO}, disqualify all theories
        from the perspectives that the initial theory/model \textit{does not}
        belong to and go to \ref{step:decideSatisfy}

        \item\label{step:decideDiffYES} If \textsc{YES}, decide if the CME can
        use more than one theory and/or model from the \textit{same}
        perspective in its design.
        \begin{enumerate}[label=(\roman*), ref=\roman*.]

            \item\label{step:decideDiffYESNO} If \textsc{NO}, disqualify all
            theories from the perspectives that the initial theory/model
            \textit{does} belong to and go to \ref{step:decideSatisfy}

            \item\label{step:decideDiffYESYES} If \textsc{YES}, then go to
            \ref{step:decideSatisfy}

        \end{enumerate}

    \end{enumerate}

    \item\label{step:decideSatisfy} Determine if there are any unsatisfied or
    weakly satisfied requirements.
    \begin{enumerate}[label=(\alph*), ref=\alph*.]

        \item\label{step:decideSatisfyNO} If \textsc{NO}, then go to
        \ref{step:decideContinue}

        \item\label{step:decideSatisfyYES} If \textsc{YES}, determine if they
        are priority requirements.
        \begin{enumerate}[label=(\roman*), ref=\roman*.]
            \item If \textsc{NO}, then go to \ref{step:decideContinue}

            \item If \textsc{YES}, then go to \ref{step:choiceLeft}

        \end{enumerate}

    \end{enumerate}

    \item\label{step:decideContinue} Decide if the decision-making process
    should continue. Requirement priority, time constraints, and personal
    preference influence this decision.
    \begin{enumerate}[label=(\alph*), ref=\alph*.]

        \item\label{step:decideContinueNO} If \textsc{NO}, then go to
        \ref{step:finish}

        \item\label{step:decideContinueYES} If \textsc{YES}, then go to
        \ref{step:choiceLeft}

    \end{enumerate}

    \item\label{step:choiceLeft} Determine if there are still emotion theories
    and/or models to choose from.
    \begin{enumerate}[label=(\alph*), ref=\alph*.]

        \item\label{step:choiceLeftNO} If \textsc{NO}, then go to
        \ref{step:finish}

        \item\label{step:choiceLeftYES} If \textsc{YES}, do one of:
        \begin{itemize}
            \item Choose a theory/model that satisfies an unsatisfied
            requirement, in priority-order, then go to
            \ref{step:decideDiff}\ref{step:decideDiffYES}

            \item Choose a theory/model that satisfies an already satisfied
            requirement more strongly, in priority-order, then go to
            \ref{step:decideDiff}\ref{step:decideDiffYES}

            \item Disqualify a theory/model, then go to
            \ref{step:decideDiff}\ref{step:decideDiffYES}

        \end{itemize}
    \end{enumerate}

    \item\label{step:finish} End the process.

\end{enumerate}

The decision-making process terminates after satisfying all prioritized
requirements and the process is voluntarily stopped, or when there are no
theories and/or models left to choose from because they have already been
chosen or disqualified. This process also works for designs that can only use
one theory because stpdf \ref{step:priority}, \ref{step:pickfirst}, and
\ref{step:decideMulti} are the same as those described in
Section~\ref{sec:singletheory}.

\subsubsection{Can You Combine Theories From Different Theoretical
    Perspectives?}\label{sec:cancombine}
\begin{figure}[!t]
    \centering
    \includegraphics[width=\linewidth]{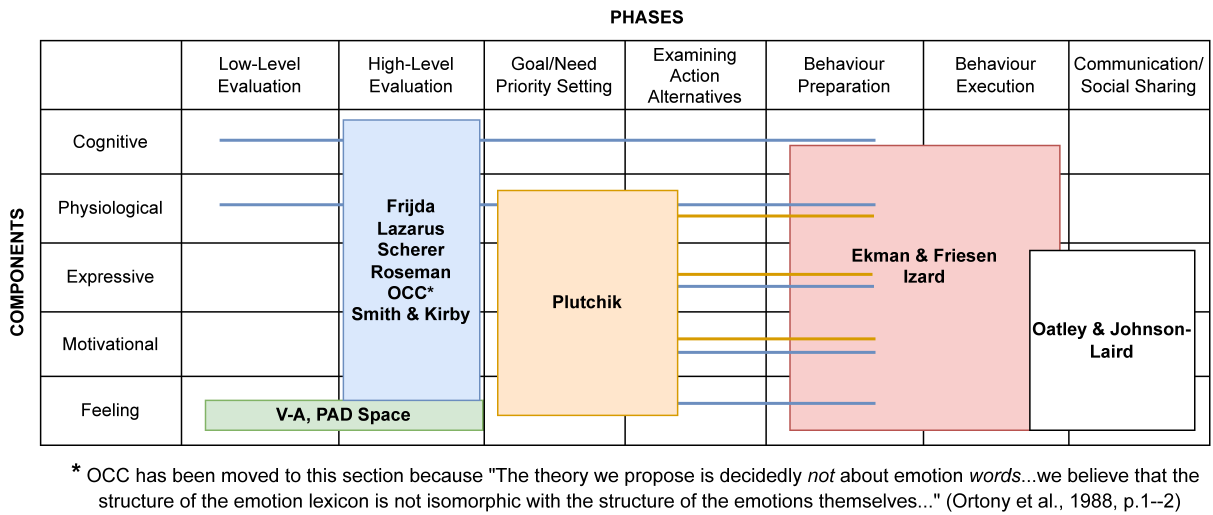}

    \caption[Coverage of Different Perspectives for Emotion Processes and
    Components]{Coverage of Different Perspectives for Emotion Processes and
        Components (Adapted and Modified from
        \citet[p.~11]{scherer2010emotion})}
    \label{fig:theoryCoverage}
\end{figure}

Depending on the CME's intended purpose, it is possible to combine theories
from different perspectives in a single design. The discrete, dimensional, and
appraisal perspectives are complementary\footnote{Appraisal and discrete
    theories appear to be especially
    compatible~\citep[p.~250]{reisenzein2013computational}.}~\citep[p.~354--355]{broekens2021emotion}
and capture key features of emotion phenomena (Figure~\ref{fig:theoryCoverage}).
Theorists from different perspectives generally agree that emotions are part of
a coherent, organized system that largely serves adaptive
functions~\citep[p.~121]{smith2001toward}. People should, instead, view each
perspective as an alternate conceptualization of the same ideas with their own
explanations, scope, and empirical
data~\citep[p.~306--307]{hudlicka2014computational}. Consequently, and as shown
by the perspective-level analysis (Chapter~\ref{sec:perspective_candidates}),
they provide different degrees of support for modelling emotion processes that
designers can combine to address gaps in models. It is not surprising that CME
designs, at least implicitly, combine multiple theories in their
design~\citep[p.~10]{hudlicka2014habits}.

The complementary nature of these theoretical perspectives implies that
choosing theories from different perspectives can support different functions
and provide better support for more requirements than any perspective
individually. However, this increases the CME's internal complexity and also
comes with the risk of lowering a CME's psychological validity if design-time
decisions create conflicting assumptions and/or
models~\citep[p.~99]{lisetti2015and}. A CME's tolerance for this risk depends
on its intended purpose and application domain (see Chapter~\ref{sec:CMEs} for
discussion).

\subsubsection{Using the Decision-Making Process for
\progname{}}\label{sec:emginechoosestheories}
Referring to the theory scores in
Tables~\ref{tab:theory-req-sys-summary-flexibility},
\ref{tab:theory-req-sys-summary-easeofuse},
\ref{tab:theory-req-comp-summary-flexibility}, and
\ref{tab:theory-req-comp-summary-easeofuse} and the information that drove
their derivation (i.e. design scope (Chapter~\ref{sec:scope}), analysis of broad
perspectives (Chapter~\ref{sec:perspective_candidates}), and notes about how
theories could satisfy requirements (Appendix~\ref{chapter:reqsTheoryNotes})),
\progname{} choose three theories for \progname{}---Oatley \& Johnson-Laird,
Plutchik, and PAD Space:
\begin{enumerate}
    \item \textit{Prioritize High-Level Requirements} (Step \ref{step:priority})

    \progname{} prioritizes \ref{flexArch}, \ref{flexOut}, \ref{flexComplex},
    \ref{flexScale}, \ref{easeHide}, \ref{easeAPI}, and \ref{easeTrace} because
    it is unlikely that developers will adopt \progname{} without them. The
    remaining requirements offer more options to tailor it to different game
    designs (\ref{flexTasks}, \ref{flexCustom}, \ref{easeAuto}) and/or could be
    satisfied as an extension later on (\ref{flexNew}, \ref{flexEm},
    \ref{easePX}, \ref{easeNovel}).

    \item \textit{Choose an Initial Theory or Model} (Step \ref{step:pickfirst})

    The first chosen theory must support as many requirements as possible for
    both the system-level and component-level
    (Section~\ref{sec:theoryanalysis}) to maximize coverage. For \progname{},
    this means choosing an \textit{appraisal theory} because discrete and
    dimensional theories cannot support the prioritized component-level
    requirements \ref{easeHide}, \ref{easeAPI} (Input), and \ref{easeTrace}.

    There are three appraisal theories that have \textit{Disqualified} scores
    (Table~\ref{tab:appraisal-theory-req-summary}): Lazarus, Scherer, and
    Roseman. Both Lazarus and Scherer are \textit{Disqualified} for a priority
    requirement (\ref{easeAPI} (Input) and \ref{flexArch} respectively), so
    \progname{} should not choose them. Neither should it choose Roseman
    because it has a \textit{Weak} score for a priority requirement. Roseman
    also has an ill-defined emotion elicitation process compared to the others
    (e.g. Appendix~\ref{sec:notes-appraisal-choose-tasks}) which makes it a
    weak initial choice for \progname{}, an emotion generation-focused CME.

    Of the remaining theories, Oatley \& Johnson-Laird is the most promising:
    it has \textit{Strong} scores for all but one priority requirement
    (\ref{easeHide}) and only \textit{Good} and \textit{Strong} scores for
    non-prioritized ones. Compared to the remaining appraisal
    theories---Frijda, OCC, and Smith \& Kirby---Oatley \& Johnson-Laird has
    the same score as the next strongest candidate for all requirements except
    for \ref{flexCustom} (Frijda), \ref{flexComplex} (Smith \& Kirby), and
    \ref{easeHide} (Frijda). While two of these are priority requirements, it
    is a reasonable trade-off for achieving a minimum score of \textit{Good}
    for \textit{all} requirements. Therefore, \progname{}'s initial theory is
    Oatley \& Johnson-Laird.

    \item \textit{Can \progname{} use more than one theory?} \textsc{YES} (Step
    \ref{step:decideMulti}\ref{step:decideMultiYES})

    \progname{} could use one theory but would be difficult for it to have all
    of the required emotion components (Chapter~\ref{sec:emotioncomponents}),
    as none of the discrete, dimensional, or appraisal perspectives provides
    them all. There is no reason to limit \progname{} to one theory because the
    added complexity would be internal to \progname{} (otherwise it would
    conflict with \ref{easeHide}, a priority requirement), so it should
    consider a combination of them.

    \item\label{emgine:multiTheory} \textit{Can \progname{} use theories from
        different perspectives?} \textsc{YES} (Step
    \ref{step:decideDiff}\ref{step:decideDiffYES})

    From the analysis of broad perspectives
    (Chapter~\ref{sec:perspective_candidates}), the discrete and dimensional
    theories cannot satisfy the need for \textit{cognitively generated/slow,
        secondary emotions} in \progname{}'s design scope
    (Chapter~\ref{sec:emotioncomponents}). This means that it must use at least
    one appraisal theory. It also knows that the discrete theories can best
    address the need for \textit{fast, primary emotions}. Therefore,
    \progname{} should choose at least one appraisal and one discrete theory.
    This approach is sufficient because \progname{} is a domain-specific model
    (see Chapter~\ref{sec:CMEs} for discussion)---it does not need to
    faithfully replicate affective
    processes~\citep[p.~131]{hudlicka2019modeling}.

    \afterpage{
        \begin{table}[!tb]
        \renewcommand{\arraystretch}{1.2}
        \centering
        \caption{Summary of Support for High-Level Requirements by Appraisal
            Theories}
        \label{tab:appraisal-theory-req-summary}
        \footnotesize
        \begin{threeparttable}
            \begin{tabular}{@{}c c P{0.33\linewidth} C{0.05\linewidth}
            C{0.059\linewidth} C{0.059\linewidth} C{0.059\linewidth}
            C{0.05\linewidth} C{0.05\linewidth} C{0.05\linewidth} @{}}
                \toprule
                \multicolumn{3}{c}{} & \textbf{Frj.} & \textbf{OCC} &
                \textbf{S\&K} & \textbf{O\&JL} & \textbf{Laz.} & \textbf{Sch.}
                & \textbf{Ros.} \\
                \midrule

                \colourRow {\scriptsize\Snowflake} &
                \ref{flexArch} & \textit{Independence from an Agent
                Architecture} & {\normalsize\strong} & {\normalsize\strong} &
                {\normalsize\strong} & {\normalsize\strong} &
                {\normalsize\good} & \disqualified & {\normalsize\good} \\

                {\scriptsize\Snowflake} & \ref{flexOut} & \textit{Allowing
                Developers to Specify How to Use CME Outputs} &
                {\normalsize\strong} & {\normalsize\strong} &
                {\normalsize\strong} & {\normalsize\strong} &
                {\normalsize\strong} & {\normalsize\strong} &
                {\normalsize\strong} \\

                \colourRow {\scriptsize\Snowflake} &
                \ref{flexComplex} & \textit{Ability to Operate on Different
                Levels of NPC Complexity} & {\normalsize\weak} &
                {\normalsize\good} & {\normalsize\strong} & {\normalsize\good}
                & {\normalsize\strong} & {\normalsize\strong} &
                {\normalsize\good} \\

                {\scriptsize\Snowflake} & \ref{flexScale} & \textit{Be
                Efficient and Scalable} & {\normalsize\good}&
                {\normalsize\good} & {\normalsize\strong} &
                {\normalsize\strong}  & {\normalsize\strong} &
                {\normalsize\good} & {\normalsize\weak} \\

                \colourRow {\scriptsize\Snowflake} &
                \ref{easeHide} & \textit{Hiding the Complexity of Emotion
                Generation} & {\normalsize\strong} & {\normalsize\good} &
                {\normalsize\good} & {\normalsize\good} & {\normalsize\strong}
                & {\normalsize\strong} & {\normalsize\good} \\

                {\scriptsize\Snowflake} & \ref{easeAPI} & \textit{Having a
                Clear API (Input)} & {\normalsize\good} & {\normalsize\weak} &
                {\normalsize\good} & {\normalsize\strong} &
                {\normalsize\disqualified} & {\normalsize\good} &
                {\normalsize\good} \\

                \colourRow {\scriptsize\Snowflake} &
                \ref{easeAPI} & \textit{Having a Clear API (Output)} &
                {\normalsize\good} & {\normalsize\good} & {\normalsize\strong}
                & {\normalsize\strong} & {\normalsize\strong} &
                {\normalsize\weak} & {\normalsize\strong} \\

                {\scriptsize\Snowflake} & \ref{easeTrace} & \textit{Traceable
                CME Outputs} & {\normalsize\strong} & {\normalsize\strong} &
                {\normalsize\strong} & {\normalsize\strong} &
                {\normalsize\strong} & {\normalsize\weak} &
                {\normalsize\strong} \\

                \colourRow & \ref{flexTasks} &
                \textit{Choosing Which Tasks to Use} & {\normalsize\weak} &
                {\normalsize\good} & {\normalsize\good} & {\normalsize\good} &
                {\normalsize\weak} & {\normalsize\strong} & \disqualified \\

                & \ref{flexCustom} & \textit{Customization of Existing Task
                Parameters} & {\normalsize\strong} & {\normalsize\good} &
                {\normalsize\good} & {\normalsize\good} & \disqualified &
                {\normalsize\good} & \disqualified \\

                \colourRow & \ref{flexNew} & \textit{Allowing
                the Integration of Components} & {\normalsize\strong}  &
                {\normalsize\strong} & {\normalsize\weak} &
                {\normalsize\strong} & {\normalsize\weak} & {\normalsize\weak}
                & {\normalsize\weak}\\

                & \ref{flexEm} & \textit{Choosing NPC Emotions} &
                {\normalsize\weak} & {\normalsize\good} & {\normalsize\weak} &
                {\normalsize\strong} & {\normalsize\weak} & {\normalsize\good}
                & {\normalsize\good} \\

                \colourRow & \ref{easeAuthor} &
                \textit{Minimizing Authorial Burden} &
                \multicolumn{7}{c}{Excluded from analysis} \\

                & \ref{easeAuto} & \textit{Allowing the Automatic Storage and
                Decay of the Emotion State} & {\normalsize\good} &
                {\normalsize\good} & {\normalsize\good} & {\normalsize\good} &
                {\normalsize\good} & {\normalsize\good} &
                {\normalsize\disqualified} \\

                \colourRow & \ref{easePX} & \textit{Showing
                that Emotions Improve the Player Experience} &
                {\normalsize\good} & {\normalsize\weak} & {\normalsize\good}
                & {\normalsize\good} & {\normalsize\good} &
                {\normalsize\strong} & {\normalsize\good} \\

                & \ref{easeNovel} & \textit{Providing Examples of Novel Game
                Experiences} & {\normalsize\good} & {\normalsize\good} &
                {\normalsize\good} & {\normalsize\good} & {\normalsize\good} &
                {\normalsize\good} & {\normalsize\good} \\

                \hline\bottomrule
            \end{tabular}
            \begin{tablenotes}

                \footnotesize
                \vspace*{2mm}

                \item {\scriptsize\Snowflake} \textit{Priority requirement}

            \end{tablenotes}
    \end{threeparttable}%
    \end{table}}

    \item\label{emgine:decideDiff} \textit{Can \progname{} use
    multiple theories from the same perspective?} \textsc{NO} (Step
    \ref{step:decideDiff}\ref{step:decideDiffYES}\ref{step:decideDiffYESNO})

    There does not appear to be a need for more appraisal theories because the
    motivation for using multiple theories is to address different aspects of
    \progname{}'s design scope that only one type of theory cannot satisfy.
    Since Oatley \& Johnson-Laird already satisfy every high-level requirement
    with a minimum score of \textit{Good}, \progname{} removes the remaining
    appraisal theories from the candidate pool.

    \item \textit{Are there any unsatisfied or weakly satisfied requirements?}
    \textsc{NO} (Step \ref{step:decideSatisfy}\ref{step:decideSatisfyNO})

    Every requirement is currently satisfied with a minimum score of
    \textit{Good}.

    \item\label{emgine:continue} \textit{Should \progname{} continue to choose
    theories?} \textsc{YES} (Step
    \ref{step:decideContinue}\ref{step:decideContinueYES})

    The motivation for using theories from different perspectives is to address
    different aspects of \progname{}'s design scope (see step
    \ref{emgine:multiTheory}). Therefore, the decision-making process continues
    because there are other theories to choose from.

    \clearpage\item \textit{Are there still emotion theories and/or models for
    \progname{} to choose from?} \textsc{YES} (Step
    \ref{step:choiceLeft}\ref{step:choiceLeftYES})

    The other appraisal theories are no longer in the candidate pool (see
    step \ref{emgine:decideDiff}), leaving the discrete and dimensional ones.
    Of these, \progname{} likely needs a discrete theory (see step
    \ref{emgine:multiTheory}), so it examines these next. Since Oatley \&
    Johnson-Laird satisfy all requirements, \progname{} is looking for a theory
    to either more strongly satisfy an already satisfied requirement or a
    theory to disqualify.

    Table~\ref{tab:discrete-theory-req-summary} shows that the discrete
    theories' scores vary for only three requirements: \ref{flexNew},
    \ref{flexEm}, and \ref{easeAPI} (Output). Izard is \textit{Disqualified}
    for one requirement that the others are not (\ref{flexEm}) and has no
    \textit{Strong} scores. Therefore, it cannot satisfy any requirement more
    strongly than Oatley \& Johnson-Laird---which has a \textit{minimum} score
    of \textit{Good} for all requirements---and \progname{} should not use it.
    Therefore, it disqualifies Izard from the candidate pool.

    \item \textit{Can \progname{} use multiple theories from the same
    perspective?} \textsc{NO}
    (Step~\ref{step:decideDiff}\ref{step:decideDiffYES}\ref{step:decideDiffYESNO})

    \progname{} has only chosen one theory so far, an appraisal theory, and
    removed all other appraisal theories from the candidate pool. Therefore, it
    continues to the next step.

    \item \textit{Are there any unsatisfied or weakly satisfied requirements?}
    \textsc{NO} (Step~\ref{step:decideSatisfy}\ref{step:decideSatisfyNO})

    Every requirement remains satisfied with a minimum score of \textit{Good}.

    \item \textit{Should we continue to choose theories for \progname{}?}
    \textsc{YES} (Step~\ref{step:decideContinue}\ref{step:decideContinueYES})

    The reasoning is the same as step~\ref{emgine:continue}.

    \item\label{step:discrete} \textit{Are there still emotion theories and/or
    models for \progname{} to choose from?} \textsc{YES}
    (Step~\ref{step:choiceLeft}\ref{step:choiceLeftYES})

    \progname{} has not yet considered all discrete theory candidates, so it
    continues its examination to find one that better satisfies a requirement
    than Oatley \& Johnson-Laird or that it can disqualify.

    Ekman \& Friesen and Plutchik have different scores for three requirements
    (Table~\ref{tab:discrete-theory-req-summary}): \ref{easeAPI} (Output)
    (Ekman \& Friesen \textit{Strong}, Plutchik \textit{Good}), \ref{flexNew}
    (Ekman \& Friesen \textit{Weak}, Plutchik \textit{Good}), and \ref{flexEm}
    (Ekman \& Friesen \textit{Weak}, Plutchik \textit{Strong}). By scores
    alone, neither of these theories satisfy any requirement better than Oatley
    \& Johnson-Laird. Certainly not for \ref{flexNew} which Oatley \&
    Johnson-Laird have a \textit{Strong} score for. However, scrutinizing Ekman
    \& Friesen for \ref{easeAPI} (Output) and Plutchik for \ref{flexEm} shows
    small but significant differences in their support of those requirements.

    \begin{table}[!tb]
        \centering
        \renewcommand{\arraystretch}{1.2}
        \caption{Summary of Support for High-Level Requirements by Discrete
            Theories}
        \label{tab:discrete-theory-req-summary}
        \footnotesize
        \begin{threeparttable}
            \begin{tabular}{@{}c c P{0.58\linewidth} C{0.05\linewidth}
                    C{0.05\linewidth} C{0.05\linewidth}@{}}
                \toprule
                \multicolumn{3}{c}{} & \textbf{Ek.} & \textbf{Plu.} &
                \textbf{Iz.} \\
                \midrule

                \colourRow {\scriptsize\Snowflake} &
                \ref{flexArch} & \textit{Independence from an Agent
                    Architecture} & {\normalsize\good} & {\normalsize\good} &
                {\normalsize\good} \\

                {\scriptsize\Snowflake} & \ref{flexOut} & \textit{Allowing
                    Developers to Specify How to Use CME Outputs} &
                {\normalsize\good} & {\normalsize\good} & {\normalsize\good} \\

                \colourRow {\scriptsize\Snowflake} &
                \ref{flexComplex} & \textit{Ability to Operate on Different
                    Levels of NPC Complexity} & {\normalsize\good} &
                {\normalsize\good} & {\normalsize\good} \\

                {\scriptsize\Snowflake} & \ref{flexScale} & \textit{Be
                    Efficient and Scalable} & {\normalsize\good}&
                {\normalsize\good} & {\normalsize\good} \\

                \colourRow {\scriptsize\Snowflake} &
                \ref{easeHide} & \textit{Hiding the Complexity of Emotion
                    Generation} & \disqualified & \disqualified & \disqualified
                    \\

                {\scriptsize\Snowflake} & \ref{easeAPI} & \textit{Having a
                    Clear API (Input)} & \disqualified & \disqualified &
                \disqualified \\

                \colourRow {\scriptsize\Snowflake} &
                \ref{easeAPI} & \textit{Having a Clear API (Output)} &
                {\normalsize\strong} & {\normalsize\good} & {\normalsize\weak}
                \\

                {\scriptsize\Snowflake} & \ref{easeTrace} & \textit{Traceable
                    CME Outputs} & \disqualified & \disqualified &
                    \disqualified \\

                \colourRow & \ref{flexTasks} &
                \textit{Choosing Which Tasks to Use} & \disqualified &
                \disqualified & \disqualified \\

                & \ref{flexCustom} & \textit{Customization of Existing Task
                    Parameters} & \disqualified & \disqualified & \disqualified
                    \\

                \colourRow & \ref{flexNew} & \textit{Allowing
                    the Integration of Components} & {\normalsize\weak} &
                {\normalsize\good} & {\normalsize\weak} \\

                & \ref{flexEm} & \textit{Choosing NPC Emotions} &
                {\normalsize\weak} & {\normalsize\strong} & \disqualified \\

                \colourRow& \ref{easeAuthor} &
                \textit{Minimizing Authorial Burden} &
                \multicolumn{3}{c}{Excluded from analysis} \\

                & \ref{easeAuto} & \textit{Allowing the Automatic Storage and
                    Decay of the Emotion State} & \disqualified & \disqualified
                    & \disqualified \\

                \colourRow & \ref{easePX} & \textit{Showing
                    that Emotions Improve the Player Experience} &
                {\normalsize\good} & {\normalsize\good} & {\normalsize\good} \\

                & \ref{easeNovel} & \textit{Providing Examples of Novel Game
                    Experiences} & {\normalsize\weak} & {\normalsize\weak} &
                {\normalsize\weak} \\

                \hline\bottomrule
            \end{tabular}
            \begin{tablenotes}

                \footnotesize
                \vspace*{2mm}

                \item {\scriptsize\Snowflake} \textit{Priority requirement}

            \end{tablenotes}
        \end{threeparttable}%
    \end{table}

    For \ref{easeAPI} (Output), \progname{} notes that Ekman \& Friesen wrote
    their publication on facial expressions for the general public
    (Appendix~\ref{sec:notes-discrete-api-output}). However, Oatley \&
    Johnson-Laird's identification of basic emotions are partially based on and
    ``[are] consistent with'' Ekman \& Friesen's findings~\citep[p.~33,
    47]{oatley1987towards}. This implies that \progname{}'s overall design would
    not benefit from using Ekman \& Friesen explicitly because it can leverage
    these strengths implicitly through Oatley \& Johnson-Laird.

    Concerning Plutchik and \ref{flexEm}
    (Appendix~\ref{sec:notes-discrete-choose-emotions}), \progname{} sees that
    it uses a colour wheel analogy to describe the creation of ``new'' emotions
    based on how a layperson understands ``emotion combinations''. This does
    not require additional processes/data sources like Oatley \& Johnson-Laird
    do, which create ``new/complex'' emotions by adding semantic content to
    ``basic'' emotions (Appendix~\ref{sec:notes-appraisal-choose-emotions}).
    This implies that \progname{} would benefit from Plutchik as it could
    support \ref{flexEm} in entities with low complexity as required by
    \ref{flexComplex}. Happily, Plutchik also has a better score distribution
    than Ekman \& Friesen (Seven to five \textit{Good} scores and equal counts
    of \textit{Disqualified} and \textit{Strong} scores). Therefore,
    \progname{} chooses Plutchik to address \textit{fast, primary emotions} in
    its design scope while also improving coverage of \ref{flexEm}.

    \item \textit{Can \progname{} use multiple theories from the same
    perspective?} \textsc{NO} (Step
    \ref{step:decideDiff}\ref{step:decideDiffYES}\ref{step:decideDiffYESNO})

    \progname{} does not appear to need multiple discrete theories for (see
    step~\ref{emgine:multiTheory} for rationale) and Ekman \& Friesen's
    benefits are implicitly captured by Oatley \& Johnson-Laird (see
    Step~\ref{step:discrete} for rationale). Therefore, \progname{} removes all
    remaining discrete theories from the candidate pool.

    \item \textit{Are there any unsatisfied or weakly satisfied requirements?}
    \textsc{NO} (Step \ref{step:decideSatisfy}\ref{step:decideSatisfyNO})

    Every requirement remains satisfied with a minimum score of \textit{Good}.

    \item\label{emjine:dimensional} \textit{Should \progname{} continue to
    choose theories?} \textsc{YES} (Step
    \ref{step:decideContinue}\ref{step:decideContinueYES})

    One could stop here, ignoring the dimensional theories. The coverage of
    Oatley \& Johnson-Laird and Plutchik satisfies all requirements with a
    minimum score of \textit{Good}. However, from the perspective-level
    analysis (Chapter~\ref{sec:perspective_candidates}), \progname{} knows that
    dimensional theories are especially suitable for representing different
    types of affect and their interactions in a common space---which is
    promising for \ref{flexNew}---and afford more control over what emotions
    could ``do'' in an NPC, which is promising for \ref{flexOut}. The
    additional design freedom afforded to game developers for believable NPC
    behaviours suggests that including a dimensional theory could increase
    \progname{}'s overall flexibility. Therefore, \progname{} continues the
    decision-making process.

    \item \textit{Are there still emotion theories and/or models for
    \progname{} to choose from?} \textsc{YES} (Step
    \ref{step:choiceLeft}\ref{step:choiceLeftYES})

    The dimensional theories remain, so \progname{} examines these now. V-A and
    PAD Space have identical scores except for one---\ref{flexEm}---which PAD
    Space scores higher on (Table~\ref{tab:dimensional-theory-req-summary}).
    Therefore, \progname{} considers this model first.

    The scores for PAD Space do not make it obvious if it better satisfies any
    requirement than the coverage achieved with Oatley \& Johnson-Laird and
    Plutchik. However, the rationale in step~\ref{emjine:dimensional} suggests
    that PAD Space could add new ways to address \ref{flexNew} and
    \ref{flexOut}. Therefore, \progname{} chooses to include it to improve its
    coverage of these requirements.

    \afterpage{\begin{table}[!t]
        \renewcommand{\arraystretch}{1.2}
        \centering
        \caption{Summary of Support for High-Level Requirements by Dimensional
            Theories}
        \label{tab:dimensional-theory-req-summary}
        \footnotesize
        \begin{threeparttable}
            \begin{tabular}{@{}c c P{0.58\linewidth} C{0.085\linewidth}
                    C{0.085\linewidth}@{}}
                \toprule
                \multicolumn{3}{c}{} & \textbf{V-A} & \textbf{PAD} \\
                \midrule

                \colourRow {\scriptsize\Snowflake} &
                \ref{flexArch} & \textit{Independence from an Agent
                    Architecture} &
                {\normalsize\good} & {\normalsize\good} \\

                {\scriptsize\Snowflake} & \ref{flexOut} & \textit{Allowing
                    Developers to Specify How to Use CME Outputs} &
                {\normalsize\strong} & {\normalsize\strong} \\

                \colourRow {\scriptsize\Snowflake} &
                \ref{flexComplex} & \textit{Ability to Operate on Different
                Levels of NPC Complexity} & {\normalsize\weak} &
                {\normalsize\weak} \\

                {\scriptsize\Snowflake} & \ref{flexScale} & \textit{Be Efficient
                    and Scalable} & {\normalsize\weak}& {\normalsize\weak} \\

                \colourRow {\scriptsize\Snowflake} &
                \ref{easeHide} & \textit{Hiding the Complexity of Emotion
                    Generation} & \disqualified & \disqualified \\

                {\scriptsize\Snowflake} & \ref{easeAPI} & \textit{Having a Clear
                    API (Input)} & \disqualified & \disqualified \\

                \colourRow {\scriptsize\Snowflake} &
                \ref{easeAPI}
                & \textit{Having a Clear API (Output)} & {\normalsize\weak} &
                {\normalsize\weak} \\

                {\scriptsize\Snowflake} & \ref{easeTrace} & \textit{Traceable
                    CME Outputs} & \disqualified & \disqualified \\

                \colourRow & \ref{flexTasks} &
                \textit{Choosing Which Tasks to Use} & \disqualified &
                \disqualified \\

                & \ref{flexCustom} & \textit{Customization of Existing Task
                    Parameters} & \disqualified & \disqualified \\

                \colourRow & \ref{flexNew} & \textit{Allowing
                    the Integration of Components} & {\normalsize\strong} &
                {\normalsize\strong} \\

                & \ref{flexEm} & \textit{Choosing NPC Emotions} &
                {\normalsize\weak} & {\normalsize\good} \\

                \colourRow& \ref{easeAuthor} &
                \textit{Minimizing Authorial Burden} &
                \multicolumn{2}{c}{Excluded
                    from analysis} \\

                & \ref{easeAuto} & \textit{Allowing the Automatic Storage and
                    Decay of the Emotion State} & \disqualified & \disqualified
                    \\

                \colourRow & \ref{easePX} & \textit{Showing
                    that Emotions Improve the Player Experience} &
                {\normalsize\good} & {\normalsize\good} \\

                & \ref{easeNovel} & \textit{Providing Examples of Novel Game
                    Experiences} & {\normalsize\good} & {\normalsize\good} \\

                \hline\bottomrule
            \end{tabular}
            \begin{tablenotes}

                \footnotesize
                \vspace*{2mm}

                \item {\scriptsize\Snowflake} \textit{Priority requirement}

            \end{tablenotes}
        \end{threeparttable}%
    \end{table}}

    \item \textit{Can \progname{} use multiple theories from the same
    perspective?} \textsc{NO} (Step
    \ref{step:decideDiff}\ref{step:decideDiffYES}\ref{step:decideDiffYESNO})

    There does not appear to be a benefit in including a second dimensional
    theory in \progname{}. Additionally, \progname{} could construct V-A in PAD
    Space due to their overlapping dimensions
    (Table~\ref{tab:affectiveDimensions}), so there is no need to explicitly
    include it. Therefore, \progname{} removes all remaining dimensional
    theories from the candidate pool.

    \afterpage{\begin{table}[!b]
        \renewcommand{\arraystretch}{1.2}
        \centering
        \caption{Comparison of Dimensions in Dimensional Theories}
        \label{tab:affectiveDimensions}
        \begin{tabular}{lcc}
            \toprule
            \textbf{Dimension} &
            {\begin{tabular}[c]{@{}c@{}}\textbf{\ref{valence}-\ref{arousal}} \\
                    (\textbf{V-A})\end{tabular}} &
            {\begin{tabular}[c]{@{}c@{}}\textbf{PAD} \\
                    \textbf{\citep{mehrabian1996pleasure}}\end{tabular}} \\
            \hline

            \colourRow Pleasure/\ref{valence} & \checkmark &
            \checkmark \\

            \ref{arousal} & \checkmark & \checkmark \\

            \colourRow Dominance &  & \checkmark \\
            \hline\bottomrule
        \end{tabular}
    \end{table}}

    \item \textit{Are there any unsatisfied or weakly satisfied requirements?}
    \textsc{NO} (Step \ref{step:decideSatisfy}\ref{step:decideSatisfyNO})

    Every requirement remains satisfied with a minimum score of \textit{Good}.

    \item \textit{Should \progname{} continue to choose theories?} \textsc{YES}
    (Step \ref{step:decideContinue}\ref{step:decideContinueYES})

    Answering this question with \textsc{NO} here does not change the outcome.
    For illustrative purposes, \progname{} chooses \textsc{YES} to show that
    this process terminates when every candidate theory/model has been chosen or
    disqualified.

    \item \textit{Are there still emotion theories and/or models for
    \progname{} to choose from?} \textsc{NO} (Step
    \ref{step:choiceLeft}\ref{step:choiceLeftNO})

    Since \progname{} chose not to use multiple theories from the \textit{same}
    perspective, it disqualified all theories/models except for the chosen
    three: Oatley \& Johnson-Laird, Plutchik, and PAD Space. The decision-making
    process ends.

\end{enumerate}

This execution of the process for \progname{} produced three theories that fully
cover all its high-level requirements: Oatley \& Johnson-Laird because it has
only \textit{Strong} or \textit{Good} scores for \textit{all} requirements;
Plutchik because it provides additional, complementary support for
\ref{flexEm}; and PAD Space to afford more flexibility for \ref{flexNew} and
\ref{flexOut} without detracting from Oatley \& Johnson-Laird and Plutchik.

\section{Summary}
Although somewhat subjective, examining each theory in the discrete,
dimensional, and appraisal perspectives and assigning them a categorical score
reveals which high-level requirements they could satisfy and how well. It is
clear that \progname{} needs at least one appraisal theory because they are the
only ones that can satisfy component-level requirements. The next question is
which and how many appraisal theories \progname{} should use, and if it also
needs theories from the discrete and/or dimensional perspectives.

\progname{} must have an appraisal theory to specify \textit{cognitively
generated/slow, secondary emotions} and a discrete theory for \textit{fast,
primary emotions} to satisfy \progname{}'s design scope. Using a systematic
decision-making process, \progname{} chose three theories to fully cover all
its high-level requirements (Table~\ref{tab:req-coverage}): Oatley \&
Johnson-Laird as its appraisal theory because it has the best requirements
score distribution with only \textit{Strong} or \textit{Good} scores; Plutchik
as its discrete theory because it provides additional, complementary support
for \ref{flexEm}; and PAD Space, a dimensional theory,  because it affords more
flexibility for \ref{flexNew} and \ref{flexOut} without detracting from Oatley
\& Johnson-Laird and Plutchik. \progname{} does not necessarily need to use PAD
Space, so it can be safely removed from the design if it proves difficult to
integrate.

\vspace*{\fill}
\begin{table}[!ht]
    \renewcommand{\arraystretch}{1.2}
    \centering
    \caption{Coverage of \progname{} High-Level Requirements by Chosen Theories}
    \label{tab:req-coverage}
    \footnotesize
    \begin{threeparttable}
        \begin{tabular}{@{}c c P{0.56\linewidth} C{0.07\linewidth}
                C{0.07\linewidth} C{0.07\linewidth}@{}}
            \toprule
            \multicolumn{3}{c}{} & \textbf{O\&JL} & \textbf{Plu.} &
            \textbf{PAD} \\
            \midrule

            \colourRow {\scriptsize\Snowflake} &
            \ref{flexArch} & \textit{Independence from an Agent Architecture} &
            {\normalsize\strong\textpmhg{\HY}} & {\normalsize\good} &
            {\normalsize\good} \\

            {\scriptsize\Snowflake} & \ref{flexOut} & \textit{Allowing
            Developers to Specify How to Use CME Outputs} &
            {\normalsize\strong} & {\normalsize\good} &
            {\normalsize\strong\textpmhg{\HY}} \\

            \colourRow {\scriptsize\Snowflake} &
            \ref{flexComplex} & \textit{Ability to Operate on Different Levels
            of NPC Complexity} & {\normalsize\good\textpmhg{\HY}} &
            {\normalsize\good} & {\normalsize\weak} \\

            {\scriptsize\Snowflake} & \ref{flexScale} & \textit{Be Efficient
            and Scalable} & {\normalsize\strong\textpmhg{\HY}} &
            {\normalsize\good} & {\normalsize\weak} \\

            \colourRow {\scriptsize\Snowflake} &
            \ref{easeHide} & \textit{Hiding the Complexity of Emotion
            Generation} & {\normalsize\good\textpmhg{\HY}} & \disqualified &
            \disqualified \\

            {\scriptsize\Snowflake} & \ref{easeAPI} & \textit{Having a Clear
            API (Input)} & {\normalsize\strong\textpmhg{\HY}} & \disqualified &
            \disqualified \\

            \colourRow {\scriptsize\Snowflake} &
            \ref{easeAPI} & \textit{Having a Clear API (Output)} &
            {\normalsize\strong\textpmhg{\HY}} & {\normalsize\good} &
            {\normalsize\weak} \\

            {\scriptsize\Snowflake} & \ref{easeTrace} & \textit{Traceable CME
            Outputs} & {\normalsize\strong\textpmhg{\HY}} & \disqualified &
            \disqualified \\

            \colourRow & \ref{flexTasks} & \textit{Choosing
            Which Tasks to Use} & {\normalsize\good\textpmhg{\HY}} &
            \disqualified & \disqualified \\

            & \ref{flexCustom} & \textit{Customization of Existing Task
            Parameters} & {\normalsize\good\textpmhg{\HY}} & \disqualified &
            \disqualified \\

            \colourRow & \ref{flexNew} & \textit{Allowing the
            Integration of Components} & {\normalsize\strong} &
            {\normalsize\good} & {\normalsize\strong\textpmhg{\HY}} \\

            & \ref{flexEm} & \textit{Choosing NPC Emotions} &
            {\normalsize\strong} & {\normalsize\strong\textpmhg{\HY}} &
            {\normalsize\good} \\

            \colourRow& \ref{easeAuthor} &
            \textit{Minimizing Authorial Burden} & \multicolumn{3}{c}{Excluded
            from analysis} \\

            & \ref{easeAuto} & \textit{Allowing the Automatic Storage and Decay
            of the Emotion State} & {\normalsize\good\textpmhg{\HY}} &
            \disqualified & \disqualified \\

            \colourRow & \ref{easePX} & \textit{Showing that
            Emotions Improve the Player Experience} & {\normalsize\good} &
            {\normalsize\good} & {\normalsize\good} \\

            & \ref{easeNovel} & \textit{Providing Examples of Novel Game
            Experiences} & {\normalsize\good} & {\normalsize\weak} &
            {\normalsize\good} \\

            \hline\bottomrule
        \end{tabular}
        \begin{tablenotes}

            \footnotesize
            \vspace*{2mm}

            \item {\scriptsize\Snowflake} \textit{Priority requirement}

            \item {\normalsize\textpmhg{\HY}} \textit{Best satisfies
            requirement}

        \end{tablenotes}
    \end{threeparttable}%
\end{table}
\vspace*{\fill}

\clearpage
\vspace*{\fill}
\begin{keypoints}
    \begin{itemize}

        \item \progname{}'s high-level requirements are the evaluation
        perspective that it views theories from the discrete, dimensional, and
        appraisal perspectives from

        \item For each requirement and theory, notes record information from
        the \ref{as} literature relevant to their evaluation

        \item After reviewing those notes, categorical scores show a theory's
        ``suitability'' for satisfying a requirement

        \item Assigned scores are at least partially subjective, which could
        change based on how requirements and theories are understood and
        developments in affective science

        \item Scores form the basis for selecting theories to based
        \progname{}'s design on

        \item A combination of theories is best for \progname{} as there are no
        strict restrictions on its design complexity and it takes advantage of
        each perspective's strengths while mitigating their weaknesses

        \item After examining their scores in
        Tables~\ref{tab:theory-req-sys-summary-flexibility},
        \ref{tab:theory-req-sys-summary-easeofuse},
        \ref{tab:theory-req-comp-summary-flexibility}, and
        \ref{tab:theory-req-comp-summary-easeofuse}, \progname{} uses Oatley \&
        Johnson-Laird (appraisal), Plutchik (discrete), and PAD Space
        (dimensional)

    \end{itemize}
\end{keypoints}

\parasep
\vspace*{\fill}

%% file: choosingtheoriesexamples.tex
\chapter{Interlude: Choosing Theories for Other
CMEs}\label{chapter:choosingExamples}
\def\epigraphflush{center}
\setlength{\epigraphwidth}{0.85\textwidth}
\def\textflush{center}
\epigraph{Come with me if you want to live.}{The Terminator, \textit{Terminator
2: Judgment Day}}

The vision for \progname{} is to afford game developers a way to integrate
emotion in their Non-Player Characters (NPCs) such that they enhance player
engagement. It also aims to give game developers complete freedom to specify
how to use its outputs (\ref{flexOut}) because they know how to best engage
players with ``emotional'' NPCs.

Due to the choice of theories, \progname{} can provide both categorical (e.g.
emotion types) and dimensional (e.g. variables) data, which offers a wider
range of mechanisms for expressing NPC emotions. This is important because
these mechanisms vary with the game---a text-based game does not have the same
ways to express NPC emotions as one with graphics, which itself varies with
resolution (e.g. 2D sprites versus 3D models). This also allows developers to
specify how \progname{} synchronizes with other game elements (e.g. camera,
story manager) that might have their own concept of ``emotion'' to create a
cohesive, contextually-relevant player experience (PX). The methodology and
decision-making process led to three theories---Oatley \& Johnson-Laird,
Plutchik, and PAD Space---that support the vision for \progname{}
(Section~\ref{sec:emginechoosestheories}). This begs the question: is the
methodology useful for choosing theories for Computational Models of Emotion
(CMEs) with different requirements and/or scope?

\section{Sketch of Choosing Theories for a CME that Expresses
    Emotion}\label{sec:sketch}
\begin{figure}[!t]
    \begin{center}
        \subfloat[Orgnar from \textit{The Elder Scrolls V:
            Skyrim}~\citep{skyrim}]{
            \includegraphics[width=0.435\linewidth]{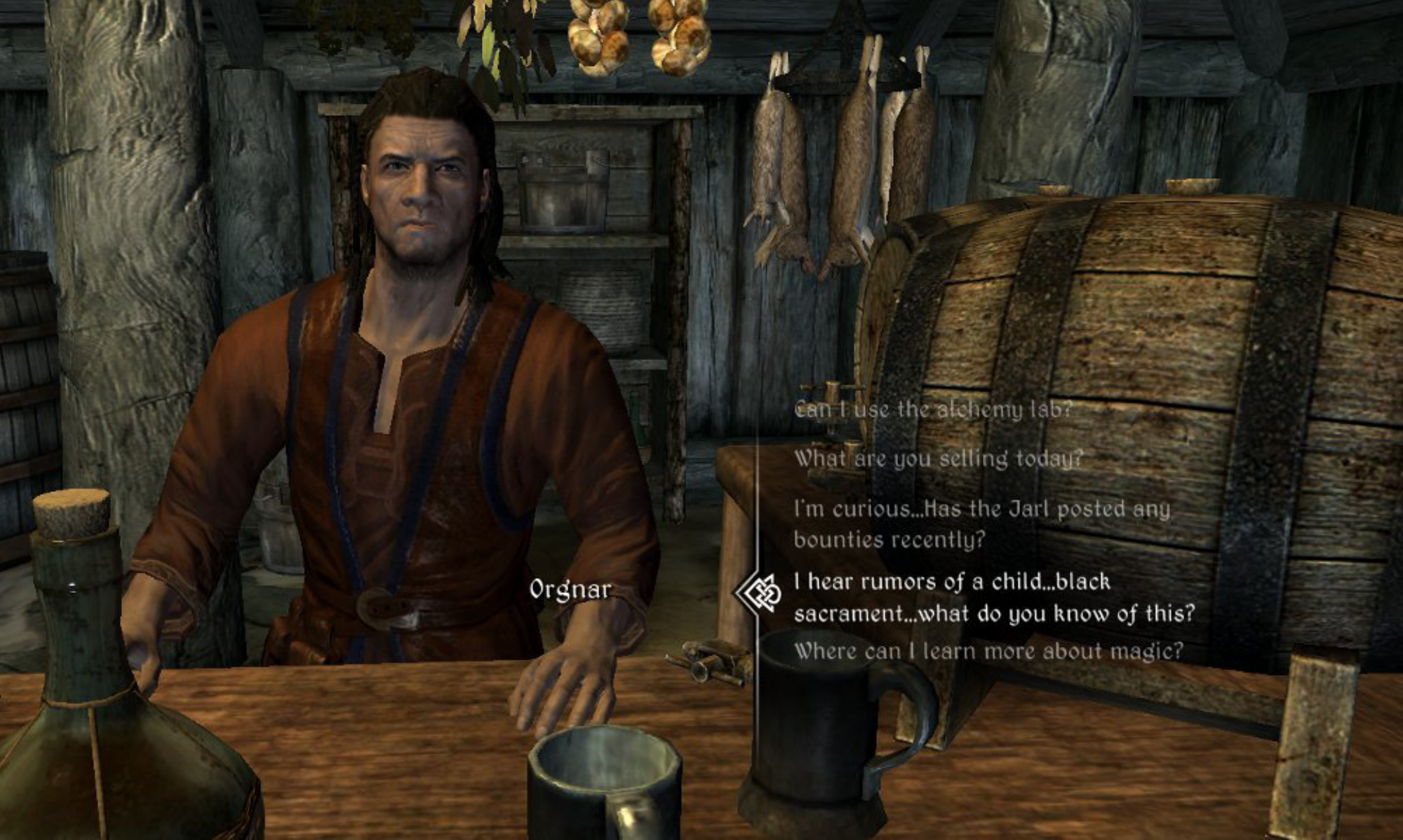}
        }
        \subfloat[Samara from \textit{Mass Effect 2}~\citep{masseffect2}]{
            \includegraphics[width=0.465\linewidth]{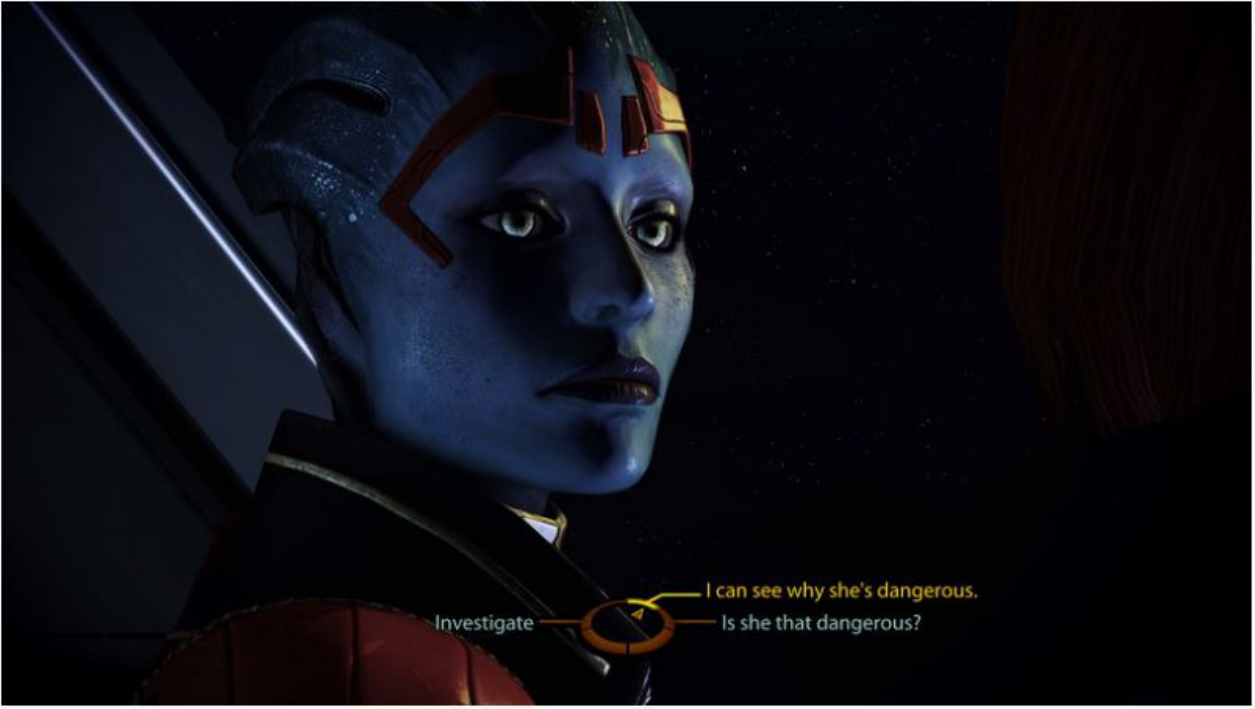}
        }

        \caption{Character dialogue screens where their facial expressions are
        a prominent aspect of the interface and contribute to the interaction}
        \label{fig:dialogue}
    \end{center}
\end{figure}

Another kind of CME that is beneficial for games---and complementary to one
that generates emotion---is one that selects an NPC's facial expression based
on its current emotion. During player-NPC dialogue interactions, the NPC's
facial expression often conveys more information that influences a player's
response because ``the face is one of the most powerful channels of nonverbal
communication''~\citep[p.~377]{torre2011facial}. For example, Orgnar's
expression helps convey how he feels about different conversation topics to add
``flavor'' to the interaction, while Samara's expression shows us her grief for
her estranged daughter which could interest a player in other conversation
options that could lead to new tasks or missions (Figure~\ref{fig:dialogue}).

It is common for Ekman \& Friesen to appear in CMEs that model facial
expressions~\citep[p.~4]{hudlicka2014habits}. This is an expected outcome for
this short example of a facial expression-oriented
CME---\progname/Express---aimed at character dialogue screens where the
character's face is prominently visible:
\begin{enumerate}
    \item Reviewing the high-level requirements (Chapter~\ref{sec:userReqs})
    with respect to generating facial expressions from emotion labels or
    dimensions/appraisal variables, \ref{flexOut} and \ref{easeNovel} no longer
    apply because \progname/Express targets a specific output method with a
    specific purpose (facial expressions in NPC dialogue screens)---there is no
    need to account for emotion expression mechanisms or their influence on CME
    processes. \ref{easeHide} is also no longer applicable because
    \progname/Express gives the designer control over how and when a character
    shows their emotion, which requires some level of understanding of those
    mechanisms/rules. \ref{easeAuto} might still apply if game designers want
    to treat the ``decay'' of facial expressions separately from the emotions
    themselves. The remaining requirements are still applicable to
    \progname/Express.

    \item For the design scope (Chapter~\ref{sec:scope}), \progname/Express
    focuses on \textit{Emotion Effects on Behaviour} because facial expressions
    are an observable (i.e. external) behaviour of an internal emotion state.
    Since it is targeting NPC dialogue screens where their face is visible,
    \progname/Express scopes the NPC's embodiment to those that include a face.
    Finally, \progname/Express chooses to focus on the \textit{emergent
    emotions} emotion component because it is not eliciting emotions (primary
    and secondary emotion components), reasoning about them in connection with
    other aspects of the NPC (emotional experience), nor connecting facial
    expressions into the NPC's internal systems (mind-body interactions).
    Including these components requires assumptions about the NPC's other
    elements and additional aspects of its embodiment that \progname/Express
    does not want to make.

    \item From the examination of the perspectives of emotion
    (Chapter~\ref{sec:perspective_candidates}), \progname/Express chooses to
    delve into discrete theories because they categorize emotion types by their
    observable features that laypeople typically recognize. It also chooses to
    delve into dimensional theories because they can relate different types of
    affect. Their numerical representation might be also useful for generating
    ``in-between'' expressions for more fluid animation transitions.
    \progname/Express will not examine the appraisal and neurophysiological
    theories because of their focus on internal processes and their relative
    complexity.

    \item After analyzing each candidate theory in the discrete and dimensional
    perspectives with respect to the high-level requirements
    (Chapter~\ref{sec:theoryanalysis}), \progname/Express derives the example
    scores shown in Table~\ref{tab:exampleScores-Exp}. Finally, it begins the
    decision-making process (Chapter~\ref{sec:choosetheories}):

    \begin{table}[!b]
        \centering
        \caption{Example Scores for \progname{}/Express, a Facial
        Expression-Oriented CME}
        \label{tab:exampleScores-Exp}
        \begin{tabular}{@{}cclccccc@{}}
            \toprule
            & &  & \multicolumn{3}{c}{\textbf{Discrete}} &
            \multicolumn{2}{c}{\textbf{Dimensional}} \\ \cmidrule(l){4-8}
            & &  & Ekman \& Friesen & Izard & Plutchik & V-A & PAD Space \\
            \midrule
            \multirow{8}{*}{\textbf{Flexibility}} &
            \colourCell{\scriptsize\Snowflake} & \colourCell RF1 &
            \colourCell\good & \colourCell\good & \colourCell\good &
            \colourCell\weak & \colourCell\weak \\

            & & RF2 & \good & \good & \good & \strong & \strong \\

            & \colourCell{\scriptsize\Snowflake} & \colourCell RF3 &
            \colourCell\strong & \colourCell\good & \colourCell\good &
            \colourCell\strong & \colourCell\strong \\

            & & RF4 & \strong & \good & \weak & \strong & \strong \\

            & \colourCell{\scriptsize\Snowflake} & \colourCell RF5 &
            \colourCell\strong & \colourCell\strong & \colourCell\good &
            \colourCell\strong & \colourCell\strong \\

            & & RF6 & \multicolumn{5}{c}{Not applicable} \\

            & \colourCell{\scriptsize\Snowflake} & \colourCell RF7 &
            \colourCell\strong & \colourCell\good & \colourCell\good &
            \colourCell\strong & \colourCell\strong \\

            & {\scriptsize\Snowflake} & RF8 & \good & \good & \good & \strong &
            \strong \\

            \midrule

            \multirow{7}{*}{\textbf{Ease-of-Use}} & \colourCell& \colourCell
            RE1 & \multicolumn{5}{c}{\colourCell Not applicable} \\

            & {\scriptsize\Snowflake} & RE2 & \strong & \strong & \good & \weak
            & \weak \\

            & \colourCell{\scriptsize\Snowflake} & \colourCell RE3 &
            \colourCell\strong & \colourCell\strong & \colourCell\good &
            \colourCell\strong & \colourCell\strong \\

            & {\scriptsize\Snowflake} & RE4 & \strong & \strong & \good & \good
            & \good \\

            & \colourCell& \colourCell RE5 & \colourCell\weak &
            \colourCell\weak & \colourCell\weak & \colourCell\strong &
            \colourCell\strong \\

            & {\scriptsize\Snowflake} & RE6 & \strong & \strong & \weak & \weak
            & \weak \\

            & \colourCell& \colourCell RE7 & \multicolumn{5}{c}{\colourCell Not
            applicable} \\
            \bottomrule

            &  &  &  &  &  &  &  \\[-3mm]

            \multicolumn{8}{r}{{\scriptsize\Snowflake} = \textit{Priority
                    Requirement}} \\
            \multicolumn{8}{r}{\strong = \textit{Strong}, \good =
                \textit{Good}, \weak = \textit{Weak}, -- =
                \textit{Disqualified}}
        \end{tabular}
    \end{table}

    \begin{enumerate}

        \item \progname/Express prioritizes requirements that enforce a game
        developer's ability to tailor an NPC's expressions (perhaps in
        collaboration with their character artist(s)) in their chosen
        development environment easily and to ensure that players can observe
        the results with minimal effort so that it does not interfere with the
        intended message. Therefore, it prioritizes \ref{flexArch},
        \ref{flexCustom}, \ref{flexEm}, \ref{flexComplex}, \ref{flexScale},
        \ref{easeAPI}, \ref{easeAuthor}, \ref{easeTrace}, and \ref{easePX}.

        \item For the initial theory, \progname/Express sees that: V-A and PAD
        Space weakly satisfy priority requirements \ref{flexArch},
        \ref{easeAPI}, and \ref{easePX}; Izard weakly satisfies \ref{flexEm};
        and Plutchik weakly satisfies \ref{easePX}. Since Ekman \& Friesen have
        only \textit{Strong} and \textit{Good} scores for all priority
        requirements, it is the best option for an initial theory. Therefore,
        \progname/Express chooses it.

        This makes sense because Ekman \& Friesen focus on facial expressions
        as an indicator of emotion kinds~\citep{ekman2007emotions} and has a
        well-known system for identifying facial expressions from facial muscle
        movements~\citep{facs}. This makes it ideal for guiding the design of a
        system that controls them. This example also illustrates that relying
        on counts of \textit{Strong} alone is not reliable because Ekman \&
        Friesen, V-A, and PAD Space have equal counts of \textit{Strong} scores.

        \item For argument's sake, \progname/Express assumes that it can use
        multiple theories. They do not have to be from the same perspective,
        nor be the only one chosen from a perspective. Now \progname/Express
        must decide if it should continue the decision-making process or end it.
        Ekman \& Friesen have only one weakly satisfied requirement and it is
        not a priority one (\ref{easeAuto}). Although it could stop the process
        here, \progname/Express chooses to continue because there are only four
        other options to examine and it will be easy to exhaust them.

        \item In this cycle, \progname/Express compares the discrete theories
        and finds that Izard also uses facial expressions as an indicator of
        emotion kinds~\citep{izard1977human} and has a system for identifying
        facial expressions from facial muscle
        movements~\citep{izard1979maximally} like Ekman \& Friesen. Izard also
        appears to identify more expressions than they do. However, Izard is
        unable to better satisfy any requirement than Ekman \& Friesen, so
        \progname/Express disqualifies it.

        Later, \progname/Express might use Izard to justify the configuration
        of expressions for emotion kinds that Ekman \& Friesen do not specify
        (e.g. \textit{Interest}, \textit{Shame}~\citep{izard1977human,
        izard1971face}). Ekman \& Friesen could implicitly support this due to
        their significant overlap and by specifying them as a ``blend'' of
        other expressions~\citep[p.~3, 69]{ekman2007emotions}.

        \item Continuing the selection process, \progname/Express examines
        Plutchik. This theory identifies emotion kinds by their intended
        ``effect'' (e.g. \textit{Fear} has the intended effect of
        ``Protection'')~\citep{plutchik1984emotions} and does not explicitly
        connect them to expression mechanisms. Since \progname/Express is
        focusing on facial expressions specifically, it chooses to disqualify
        Plutchik from the candidate pool.

        \item \progname/Express now considers V-A and PAD Space because they
        can satisfy \ref{flexTasks}, \ref{flexScale} (a prioritized
        requirement), and \ref{easeAuto} better than Ekman \& Friesen. In this
        case, V-A and PAD Space have identical scores for all requirements so
        \progname/Express must look more closely to see which, if either, one
        it chooses to use.

        People tend to judge facial expressions in a circumplex structure along
        the V-A dimensions~\citep{barrett2009affect} and researchers have
        connected these dimensions to individual facial muscle
        movements~\citep{smith_scott_mandler_1997}. PAD Space also makes some
        connections between its dimensions and facial
        expressions~\citep[p.~186]{mehrabian1980basic}, although they are less
        defined than V-A. However, PAD Space offers an additional dimension
        that developers can manipulate to produce more variation in NPC
        expressions while retaining the benefits of the V-A
        dimensions~\citep{bakker2014pleasure}. Therefore, \progname/Express
        chooses PAD Space over V-A.

        \item In the previous step \progname/Express reasoned that PAD Space
        could adequately capture V-A within itself, so it disqualifies V-A. The
        decision-making process now stops because there are no more theories to
        choose from.

    \end{enumerate}

\end{enumerate}

By following the methodology and decision-making process, \progname/Express
first chose Ekman \& Friesen---an expected outcome---and found that PAD Space
could also be useful. Existing facial expression-generating CMEs have combined
these theories successfully~\citep{bidarra2010growing,
breazeal2003emotion}\footnote{The Soul (\ref{soul}) and Kismet (\ref{kismet}),
included in the CME survey
(Chapter~\ref{chapter:cmeOverview})}\textsuperscript{,}\footnote{\citet{breazeal2003emotion}
 uses \textit{stance} rather than \textit{dominance} but it appears to serve a
comparable purpose.}, implying that they are reasonable outcomes.

\clearpage

\section{Sensitivity to Changes in High-Level Requirements and Scope}
The sketch of \progname/Express demonstrates that changing the relevance of
three high-level require-ments---\ref{flexOut}, \ref{easeHide}, and
\ref{easeNovel}---and the CME task from \textit{Emotion Generation} to
\textit{Emotion Effects on Behaviour} causes cascading changes through other
design scope elements, and the focus of the analysis of emotion perspectives
and individual theories. This suggests that the methodology is sensitive to
high-level requirements and design scope changes. Therefore, it is reasonable
to expect that it culminates in any examined theory or group of theories given
a relevant set of requirements and scope. For example, a CME:
\begin{itemize}

    \item That uses experienced emotions to change its behaviours over time
    might use requirements like \progname{}'s---such as \ref{flexTasks},
    \ref{flexEm}, and \ref{easeTrace}---but have an entirely different scope
    (i.e. \textit{Emotion Effects on Internal Cognitive Processes},
    \textit{Mind-Body Interactions}) which could lead to Izard because it
    describes emotion-dependent personality
    development~\citep{izard2000motivational}

    \item That needs a simple emotion model which can relate to observable
    behaviours and account for influencing factors such as personality (e.g.
    \citet{ball2000emotion}) with \textit{valence} and \textit{arousal} as
    internal reward signal for reinforcement learning (e.g.
    \citet[p.~279--280]{soarEmotion}) might need requirements like \textit{Must
    operate in real-time interactions} and \textit{Must only use one affective
    theory/model}---which would likely change the scope as well (e.g.
    \textit{Emotion Effects on Behaviour + Internal Cognitive Processes},
    \textit{Emergent Emotions + Mind-Body Interactions}) could culminate in V-A
    due to its dissociation from specific expressions and action
    tendencies~\citep{broekens2021emotion} and the numerical nature of
    dimensional theories

    \item That acts as/is part of an agent's behaviour control system (e.g.
    \citet{martinho2000emotions, petta2002role}) and has similar requirements as
    \progname{} has a different design scope---\textit{Emotion Effects on
    Internal Cognitive Processes}, favours physical embodiment (e.g. toy robot),
    and \textit{Mind-Body Interactions}---could result in Frijda because its
    continuous information processing nature incorporates control signals and a
    feedback loop~\citep[p.~454]{frijda1986emotions}

    \item That uses emotion to plan social interactions (e.g.
    \citet{marsella2000interactive}) might have the same requirements and scope
    as the CME that chooses Frijda (perhaps with no constraints on agent
    embodiment), but instead culminate in Lazarus because of its focus on an
    individual's ability to change their emotion-eliciting appraisal of the
    causal environment and/or event (i.e. stress and \ref{coping} in emotion
    processes)~\citep{lazarus1991emotion}

    \item For generating ambient agent crowd behaviour (i.e. individual agent
    behaviours (subsystem) have less impact than those of the group (system))
    that prioritizes \ref{easeAuthor} and does not need to explicitly label the
    behaviours (i.e. scoped to \textit{Emotion Effects on Behaviour} and
    \textit{Emergent Emotions}) might lead to Scherer because it views the
    emotion process as a continuously fluctuating pattern of change in several
    subsystems~\citep[p.~108]{scherer2001appraisalB}

    \item Where actions selection depends on the current social context (e.g.
    \citet{yacoubi2018teatime}) and mandates that the CME must \textit{Allow
    designers to define the association between emotions and agent response
    goals} and \textit{Allow designers to define the process(es) for choosing
    actions for goal achievement} might scope to \textit{Emotion Effects on
    Internal Cognitive Processes} and \textit{Mind-Body Interactions}, and
    subsequently lead to Roseman due to its definition of emotion-specific
    general responses that can manifest in different ways  (i.e. emotion
    strategies)~\citep{roseman2011emotional}

    \item For an empathetic agent (e.g. \citet{ochs2012formal}) that is similar
    to \progname{} in both requirements and design scope---but disregards
    \textit{Fast, Primary Emotions}---might choose OCC because of its
    ``Fortunes-of-others'' emotions~\citep[p.~109]{occ2022}

    \item Which has multiple appraisal processes with distinct mechanisms that
    collectively produce a single emotion response (e.g.
    \citet{broekens2004scalable}) and stresses the importance of
    \ref{flexCustom}, \ref{flexNew}, \ref{flexComplex}, and \ref{flexScale}
    might culminate in Smith \& Kirby due to its description of appraisal
    detectors that monitor for, collect, and combine information from disparate
    sources to determine the overall emotion response~\citep{smith2001toward}

    \item That learns about and responds to stimuli based on an underlying
    emotion elicitation mechanism (e.g. \citet{becker2008wasabi}) might examine
    the neurophysiological perspective more closely and find that one or more
    of \citet{damasio2005descartes}, \citet{ledoux1996emotional}, and/or
    \citet{sloman2005architectural} satisfies their needs
\end{itemize}

The nature of this methodology makes different outcomes possible because it
provides focus and structure for the theory-choosing process rather than
imposing constraints on it. Again, it depends on one's understanding of the
high-level requirements and design scope that one applies to their
interpretations of static affective theory/model academic texts to choose a
CME's foundation. However, following this process justifies \textit{why} a CME
uses a theory/model in a reproducible way. This lends credibility to the
choices and to the CME built on those choices.

\section{Summary}
The proposed theory-selection methodology has many potential outcomes due to
differences in one's understanding of the high-level requirements, their
relative priorities, and one's understanding of the \ref{as} literature. For
game development, and domain-specific CMEs in general, this variation is of
little concern because the goal is \textit{interesting} behaviours rather than
strictly \textit{realistic} ones. This encourages variation because the process
leads to different designs which might be better suited for some games and not
others. Ultimately, this gives game developers more tools to choose from so
that they can continue to create a cornucopia of games to play.

\clearpage
\vspace*{\fill}
\begin{keypoints}
    \begin{itemize}

        \item The proposed methodology for choosing affective theories and/or
        models has many potential outcomes and is sensitive to changes in
        high-level requirements and design scope

    \end{itemize}
\end{keypoints}

\parasep
\vspace*{\fill}

%% file: appraisalequations.tex
\chapter{Spec Your \progname{}: Defining the Pieces}\label{chapter:equations}
\def\epigraphflush{center}
\setlength{\epigraphwidth}{0.85\textwidth}
\def\textflush{center}
\epigraph{Isn't it strange, to create something that hates you?}{Ava,
\textit{Ex Machina}}

At this point, \progname{} has only specified its high-level design goals
(Chapter~\ref{sec:userReqs}) and systematically chosen emotion theories and
models that appear to best support those goals. On their own, these
theories/models are inadequate specifications and cannot be directly translated
into software. Therefore, specifications for \progname{}'s models develop
progressively from representations of the environment
(Section~\ref{sec:envDataTypes}) and emotions (Section~\ref{sec:emotionRepr})
to task-specific models that depend on them
(Sections~\ref{sec:emotiongenerationfunctions}, \ref{sec:intensity},
\ref{sec:emDecay}, and \ref{sec:padspace}). Although this process looks
straightforward, there  were many instances of backtracking to and concurrent
work between the models.

Other CMEs have documented their models (e.g. \citet{reilly1996believable,
el2000flame, becker2008wasabi, jain2019modeling, alfonso2017toward,
yacoubi2018teatime}, and \citet{bourgais2017enhancing}\footnote{Em/Oz
(\ref{em}), FLAME (\ref{flame}), WASABI (\ref{wasabi}), HybridC
(\ref{hybridc}), GenIA\textsuperscript{3} (\ref{genia3}), TEATIME
(\ref{teatime}), and GAMA-E (\ref{gamae}), included in the CME survey
(Chapter~\ref{chapter:cmeOverview})}), although they do not appear to do so
with the same depth or thoroughness as \progname{}. There is also a set-based
formalism for representing appraisal theories~\citep{broekens2008formal}
whereas \progname{} relies on formal types instead.

\section{Interfacing \progname{}'s Underlying Affective Theories}
An inevitable challenge that occurs when combining emotion theories and/or
models into a formal model is bridging the gap in their underlying assumptions.
For example, a common conflict is how many---if any---emotions are
``basic''~\citep[p.~57]{ortony2021all}. Luckily, these issues have little
effect on \progname{} because each of the chosen theories interact somewhat
superficially:
\begin{itemize}

    \item Plutchik models the internal structure of emotion, its storage and
    organization,

    \item Oatley \& Johnson-Laird generate emotions for storage in the internal
    structure, and

    \item PAD Space provides an alternate representation of the internal emotion
    structure.

\end{itemize}

This implies that it is sufficient for \progname{}, a Computational Model of
Emotion (CME) designed for entertainment rather than research, to find
comparable definitions of the emotions represented by each theory. While the
resulting mappings between definitions might not be strictly correct or
precise, they do rely on an English-speaking North American understanding of
emotion terms in everyday language (\citepg{oatley1992best}{75};
\citepg{robert1980emotion}{93, 152}; \citepg{mehrabian1980basic}{39--40})
suggesting that the same word represents an identical or comparable concept.
Since Plutchik's emotion kind definitions serve as the baseline
(Section~\ref{sec:emotionRepr}), guiding \progname{}'s development and
subsequent testing~\citep[p.~367]{broekens2021emotion}, it acts like an
``interface'' between Oatley \& Johnson-Laird
(Section~\ref{sec:emotiongenerationfunctions}) and PAD Space
(Section~\ref{sec:padspace}).

\section{Data Types for Communicating with the
Environment}\label{sec:envDataTypes}
Some non-affective data types, chiefly focused on the entity's environment, are
necessary if \progname{} is to generate any emotions at all. \progname{}
assumes that variables represent the environment and that events are changes in
those variables.

\subsection{User-Implemented Data Types}
\progname{} defines some data types as Application Programming Interface (API)
specifications because it expects their implementation to change with each
game/application. Users must provide the minimum functionality described in the
APIs, expanding it with any additional functionality they need. This improves
\progname{}'s portability and minimizes authorial effort by only requiring
users to provide implementations for predesigned modules. Since this is
\progname{}'s point-of-contact with an external system, defining these APIs
allows \progname{} to assume that certain behaviours are available for its use
without constraining users to specific implementations of them.

\subsubsection{Time and Delta Time Data Types} Emotion is time-dependent
(Chapter~\ref{sec:affectiveDefs}), so \progname{} requires representations of
time---assumed to be linearly ordered---and a difference or ``change'' between
two points in time (i.e. delta $\Delta$):
\begin{equation}\label{eq:timetype}
    \timetype, \deltatimetype
\end{equation}
These are essential for defining the Attention Data Type
(Equation~\ref{eq:attentiontype}), defining emotion as a set of states over
time (Equation~\ref{eq:emotiontype}), and Emotion Decay
(Equations~\ref{eq:decayIntensity}, \ref{eq:decayastate}, and
\ref{eq:autodecay}).

\subsubsection{World State View (WSV) Data Type} \progname{} requires a
representation of the world that an entity exists in because emotion is a
response to it. In games, ``the world'' $\worldtype$ is an imaginary universe
where game events occur and are typically two or three dimensional spaces
containing characters and objects~\citep[p.~640]{adams2014fundamentals}.
\progname{} needs to know about the configuration of characters, objects, and
variables in ``the world'' in order to evaluate an entity's relation to it.
However, it does not necessarily need to  know \textit{everything} about ``the
world''---only those aspects that are relevant to the entity. For example, if
$\worldtype$ contained variables \texttt{health} and \texttt{wealth} but the
entity only cares about \texttt{health}, then \progname{} does not care about
the \texttt{wealth} variable. \progname{} requires a type that describes this
world state ``view'':
\begin{equation}\label{eq:worldstatetype}
    \worldstatetype \subseteq \worldtype
\end{equation}
This is essential for: defining the Goal and Plan Data Types
(Equations~\ref{eq:goaltype} and \ref{eq:plantype}); eliciting \textit{Joy}
(Equation~\ref{eq:generatejoy}), \textit{Sadness}
(Equation~\ref{eq:generatesadness}), \textit{Fear}
(Equation~\ref{eq:generatefear}), \textit{Anger}
(Equation~\ref{eq:generateanger}), \textit{Disgust}
(Equation~\ref{eq:generatedisgust}), \textit{Acceptance}
(Equation~\ref{eq:generateemotionAcceptance}), and \textit{Surprise}
(Equation~\ref{eq:generateemotionSurprise}); and calculating the intensity of
\textit{Sadness} (Equation~\ref{eq:evalIntensitySadness}).

\subsubsection{World Event Data Type} \progname{} also requires a
representation of changes (i.e. events) in an entity's ``world'' to know when
to trigger emotion processes. Games can understand an event as a game action
that changes a game world's configuration of characters, objects, and
variables. \progname{} takes this to mean that it is a ``change'' (i.e. delta
$\Delta$) in the game world. As with WSV, \progname{} only needs to know about
``changing'' aspects relevant to the entity. Therefore, the World Event Data
Type represents a ``change'' in a WSV $\worldstatetype$:
\begin{equation}\label{eq:worldeventtype}
    \worldstatechangetype \subseteq \worldstatetype
\end{equation}
Applying the event to the current WSV gives the next WSV (i.e. $s \oplus
s_{\Delta} : \worldstatetype \times \worldstatechangetype \rightarrow
\worldstatetype$).

This is essential for: defining the Goal and Plan Data Types
(Equations~\ref{eq:goaltype} and \ref{eq:plantype}); and eliciting \textit{Joy}
(Equation~\ref{eq:generatejoy}), \textit{Sadness}
(Equation~\ref{eq:generatesadness}), \textit{Fear}
(Equation~\ref{eq:generatefear}), \textit{Anger}
(Equation~\ref{eq:generateanger}), \textit{Disgust}
(Equation~\ref{eq:generatedisgust}), \textit{Acceptance}
(Equation~\ref{eq:generateemotionAcceptance}), and \textit{Surprise}
(Equation \ref{eq:generateemotionSurprise}).

\subsubsection{Distance Between WSVs Data Type} \progname{} needs a way to
compare two WSVs $\worldstatetype_1, \worldstatetype_2$ to evaluate their
relative desirability for some target (e.g. goals). It takes this as a
``distance'' between them, describing the differences between each element in
the compared WSVs:
\begin{equation}\label{eq:statedisttype}
    \statedistancetype
\end{equation}
This is essential for: defining the Goal Data Type
(Equation~\ref{eq:goaltype}); eliciting \textit{Joy}
(Equation~\ref{eq:generatejoy}), \textit{Sadness}
(Equation~\ref{eq:generatesadness}), \textit{Fear}
(Equation~\ref{eq:generatefear}), and \textit{Disgust}
(Equation~\ref{eq:generatedisgust}); and calculating the intensity of
\textit{Sadness} (Equation~\ref{eq:evalIntensitySadness}) and \textit{Disgust}
(Equation~\ref{eq:evalIntensityDisgust}).

\subsubsection{Change in Distance Between WSVs Data Type} \progname{} also needs
some way to measure how much a game event $\worldstatechangetype$ changes a WSV
$\worldstatetype$. \progname{} assumes this to be equivalent to the magnitude
of ``change'' (i.e. delta $\Delta$) that the event causes in the  WSV
configuration, describing the differences between each ``changed'' element in
the WSV:
\begin{equation}\label{eq:statedistchgtype}
    \statedistancechangetype
\end{equation}
This is essential for defining the Goal Data Type (Equation~\ref{eq:goaltype});
eliciting \textit{Joy} (Equation~\ref{eq:generatejoy}), \textit{Fear}
(Equation~\ref{eq:generatefear}), \textit{Disgust}
(Equation~\ref{eq:generatedisgust}), and \textit{Acceptance}
(Equation~\ref{eq:generateemotionAcceptance}); and calculating the intensity of
\textit{Joy} (Equation~\ref{eq:evalIntensityJoy}), \textit{Fear}
(Equation~\ref{eq:evalIntensityFear}), and \textit{Acceptance}
(Equation~\ref{eq:evalIntensityAcceptance}).

\subsection{\progname{}-Implemented Data Types}
When given with information about the ``world'', \progname{} forms structures
about an entity's relation to it: goals, plans, what they pay attention to, and
their social attachments to other entities. Users can opt to provide
information for goals alone, which is enough for \progname{} to generate five
of eight emotions---\textit{Joy}, \textit{Fear}, \textit{Disgust},
\textit{Surprise}, and a limited evaluation of \textit{Sadness}. If users
require the other three emotion kinds, they must provide information for one or
more of the other data types to enable their evaluation.

\subsubsection{Goal Data Type} Unless they are lucky, an entity will not always
exist in a game world state that satisfies them. They will need a way to
represent the WSV that they want so that they know when events impact them and
by how much~\citep[p.~208]{ortony2002making}. They need goals, which are
essential to emotion elicitation. Typically, all Non-Player Character (NPC)
entities already have some~\citep[p.~223]{broekens2016emotional}. \progname{}
represents a goal as a record containing a predicate on a WSV
$\mathtt{goalState}$, two functions $\mathtt{goal}$ and $\mathtt{goal'}$, a
non-negative, real-valued $\mathtt{importance}$ value, and a $\mathtt{type}$
set:
\begin{equation}\label{eq:goaltype}
    \begin{gathered}
        \goaltype : \{ \mathtt{goalState} : \worldstatetype \rightarrow
        \mathbb{B}, \mathtt{goal} : \worldstatetype \rightarrow
        \statedistancetype, \mathtt{goal'} : \worldstatetype \times
        \worldstatechangetype \rightarrow \statedistancechangetype, \\
        \mathtt{importance} : \mathbb{R}_{\geq0}, \mathtt{type} \subseteq \{
        \mathtt{SelfPreservation}, \mathtt{Gustatory} \} \}
    \end{gathered}
\end{equation}
\begin{itemize}

    \item The $\mathtt{goalState}$ predicate represents the entity's desired
    WSV $\worldstatetype$ (Equation~\ref{eq:worldstatetype}). If they are not
    already in $\mathtt{goalState}$ and striving to maintain it, they are in
    another state and want to move towards or away from $\mathtt{goalState}$.

    \item The function $\mathtt{goal}$ maps a WSV $\worldstatetype$ to a
    distance $\statedistancetype$ (Equation~\ref{eq:statedisttype}) between it
    and $\mathtt{goalState}$ to measure the difference between the current WSV
    and the desired one.

    \item The function $\mathtt{goal'}$ is the derivative of $\mathtt{goal}$,
    measuring a change in the distance $\statedistancechangetype$
    (Equation~\ref{eq:statedistchgtype}) to $\mathtt{goalState}$ when a game
    event $s_{\Delta} : \worldstatechangetype$ changes a WSV $s:
    \worldstatetype$ (Equation~\ref{eq:worldeventtype}). \progname{}
    evaluates $\worldstatetype \times \worldstatechangetype$ as $s \oplus
    s_{\Delta}$, shorthand for $\mathtt{apply}(x, y)$, which is a function that
    changes $x$ by $y$.

    \item The goal's perceived relative $\mathtt{importance}$ to the entity
    such that higher values reflect a higher importance, mimicking the tendency
    for higher importance goals to motivate an individual more than lower
    importance ones~\citep[p.~204]{izard1977human}. If this is set to zero,
    \progname{} assumes that the goal has no importance to the entity and does
    not trigger emotion processes when affected by world events.

    \item Oatley \& Johnson-Laird's descriptions of emotion-elicitation
    conditions imply that goals can have the types of
    \textit{Self-Preservation} and/or \textit{Gustatory}
    (Table~\ref{tab:cte_patterns}). Goal $\mathtt{type}$ stores this
    information, allowing a goal to have none, one, or both of these types.

\end{itemize}

\subsubsection{Plan Data Type} According to Oatley \& Johnson-Laird, a plan is
necessary to elicit \textit{Anger} and offers another way to elicit
\textit{Sadness}~\citep[p.~36]{oatley1987towards}. \progname{} represents a
plan as:
\begin{equation}\label{eq:plantype}
    \begin{gathered}
        \plantype : \{ \mathtt{actions} : ({\worldstatechangetype}_1, ...,
        {\worldstatechangetype}_n), \mathtt{toProgress} : ((\worldstatetype
        \rightarrow \mathbb{B})_0, ..., (\worldstatetype \rightarrow
        \mathbb{B})_n), \\
        \mathtt{nextStep} : \worldstatetype \times \mathbb{N} \rightarrow
        \worldstatetype, \mathtt{isFeasible} : \worldstatetype \rightarrow
        \mathbb{B} \} \\
        \text{where } n : \mathbb{N}_{>0}, \\
        \mathtt{nextStep}(s : \worldstatetype, i : \mathbb{N}) = \begin{cases}
            s, & i = 0 \\
            \mathtt{nextStep}(s, i) \oplus \mathtt{planActions}(i), &
            \mathit{Otherwise}
        \end{cases}, \\
        \text{and } \mathtt{isFeasible}(s : \worldstatetype) =
        \bigwedge_{i=0}^{n} \mathtt{toProgress}(i, \mathtt{nextStep}(s, i))
    \end{gathered}
\end{equation}
\begin{itemize}

    \item A sequence of $\mathtt{actions}$ such that applying them to an
    initial WSV generates a series of ``good'' WSVs that satisfy a sequence of
    predicates on them representing plan progression ($\mathtt{toProgress}$),
    where each element in $\mathtt{actions}$ is something the entity can do
    (i.e. there are no elements in $\mathtt{actions}$ that the entity believes
    are impossible)

    \item At some step $i : \mathbb{N}$, the function $\mathtt{nextStep}$
    evaluates the next WSV in the plan by applying the $ith$ plan action to the
    $ith$ $\mathtt{nextStep}$ where $\mathtt{nextStep}(s, 0)$ is the initial
    state $s : \worldstatetype$

    \item A constant $\mathtt{isFeasible}$ generated by checking that, for each
    step $i : \mathbb{N}$ starting from $i = 0$ and a WSV $s :
    \worldstatetype$, each evaluation of $\mathtt{nextStep}(s, i)$ satisfies
    the $ith$ condition in $\mathtt{toProgress}$

\end{itemize}

The Plan Data Type is not explicitly connected to the Goal type so that users
can apply it to entity plans that do not target an \progname{}-specific goal
$\goaltype$.

\subsubsection{Social Attachment\protect\footnote{\normalfont \progname{} uses
the term ``Social Attachment'' instead of ``Social Relationship'' because it
appears to represent a simpler concept~\citep{rempel1985trust}.} Data Type}
Modelling \textit{Acceptance} as a social emotion
(Section~\ref{sec:evalAcceptance}) implies the need for relations between
entities~\citep[p.~360]{broekens2021emotion}. Therefore, \progname{} provides a
Social Attachment Data Type, defined as a discrete ``degree'' or ``level'' of
attachment to some entity where ``degree'' can be negative to represent
disliking that entity:
\begin{equation}\label{eq:socialtype}
    \socialattachmenttype : \mathbb{Z}
\end{equation}
The Social Attachment Data Type is linearly ordered such that higher
``degrees'' or ``levels'' represents a stronger attachment to another entity
that can reflect a ``history'' with it. Users can extend this type with
additional information about an entity's attachment to another as needed. Since
Social Attachment is just an association between two
entities~\citep[p.~359--360]{broekens2021emotion}, users do not have to limit
the ``other'' entity to characters---it could refer to objects, actions, or
other game elements.

\subsubsection{Attention Data Type} Oatley \& Johnson-Laird hypothesize that
attention is foundational to
\textit{Interest}~\citep[p.~33]{oatley1987towards}. Researchers view attention
as a set of mechanisms that allow a limited-capacity system to select salient or
goal-relevant information~\citep[p.~54]{oxfordAttention}. This leads
\progname{} to define the Attention Data Type as the consecutive elapsed time
$\deltatimetype$ (Equation~\ref{eq:timetype}) spent focusing on some $x$:
\begin{equation}\label{eq:attentiontype}
    \attentiontype : \deltatimetype
\end{equation}
Users can extend this type with additional information about an entity's focus
on $x$ and the resources available for attention as needed.

\section{Emotion Representation}\label{sec:emotionRepr}
\progname{} represents emotions with data types. The most basic type is Emotion
Intensity, a non-negative real value:
\begin{equation}\label{eq:intensitytype}
    \emotionintensitytype : \mathbb{R}_{\geq0}
\end{equation}
The domain of real numbers allows \progname{} to represent intensity as a
continuous value while affording users the ability to define emotion as a
discrete value if they wish. Emotion Intensity is strictly non-negative because
\progname{} assumes that Plutchik's concept of a ``deep sleep'' state refers to
an absence of emotion. ``Deep sleep'' implies that the entity is not
experiencing emotion at all because this is when it loses
consciousness~\citep[p.~1--2]{mondino2021definitions}. However, this is
unlikely to be unique to Plutchik due to the responsive nature  of emotion
(Chapter~\ref{sec:affectiveDefs}) and a common sense understanding assumes that
an entity is either experiencing emotion or not (i.e. feeling some intensity or
none).

\progname{} combines an Emotion Intensity with an Emotion Intensity Change
($\responsestrength$, see Section \ref{sec:intensity}) using a logarithmic
function:
\begin{equation}\label{eq:combineintensity}
    \begin{gathered}
        \mathtt{UpdateIntensity}(i : \emotionintensitytype, i_{\Delta} :
        \responsestrength) : \emotionintensitytype \defEq \begin{cases}
            0.1 \cdot \log_2 \left(2^{10 \cdot i} + 2^{10 \cdot
                i_{\Delta}}\right), & i_{\Delta} > 0 \\
            0.1 \cdot \log_2 \left(2^{10 \cdot i} - 2^{10 \cdot | i_{\Delta}
                |}\right), & i_{\Delta} < 0 \\
            i, & \mathit{Otherwise} \\
        \end{cases}
    \end{gathered}
\end{equation}
\progname{} bases this function on one from Em/Oz~\citep{reilly2006modelling}
and a reinforcement learning agent \citep[p.~370]{broekens2021emotion}. It
relies on a logarithm so that it is not strictly additive and both values
contribute to the output such that its magnitude is at least as much as the
highest input. Although not experimentally verified, it emulates these desired
behaviours and reportedly works well.

\subsubsection{Emotion Kinds}
The Emotion Kinds enumeration encodes Plutchik's eight emotion types as labels,
ensuring that the ``primary'' emotions that \progname{} supports are finite and
consistently ordered:
\begin{equation}\label{eq:kindstype}
    \emotionkindstype : \left< \mFear, \mAnger, \mSadness, \mJoy, \mInterest,
    \mSurprise, \mDisgust, \mTrust \right>
\end{equation}
An invariant on $\emotionkindstype$ requires that labels be uniquely ``paired''
with exactly one other emotion type label. While there are no data types or
functions that need this information, it embeds a fundamental characteristic of
Plutchik's emotion \ref{circumplex} into the type. \progname{} assumes that
functionally coupling ``paired'' emotion kinds such that the experience of one
reduces the experience of the other would make a state of ``deep sleep''
impossible, so it does not do so.

Defining Emotion Kinds independently of Emotion Intensity
(Equation~\ref{eq:intensitytype}) allows them to be theory-agnostic (i.e. not
strictly tied to Plutchik). This means that \progname{} can still be functional
if there is \textit{some} definition for Emotion Kinds, which could have
differing types and number of labels than Plutchik.

\subsubsection{Emotion State} Emotion State Data Type composes the Emotion
Intensity and Kinds Data Types into a new structure. \progname{} represents an
emotion state as a record containing functions $\mathtt{intensities}$ and
$\mathtt{max}$:
\begin{equation}\label{eq:emotionstatetype}
    \emotionstatetype : \left\{ \mathtt{intensities} : \emotionkindstype
    \rightarrow \emotionintensitytype, \mathtt{max} : \emotionkindstype
    \rightarrow \emotionintensitytype \right\}
\end{equation}
\begin{itemize}

    \item The function $\mathtt{intensities}$ maps Emotion Kinds to Emotion
    Intensities ($\emotionkindstype \rightarrow \emotionintensitytype$) to
    represent the current intensity of each emotion kind in the state. This
    is similar to a vector of intensity values, which CMEs commonly use to
    represent affective states~\citep[p.~358]{broekens2021emotion}. The
    function must satisfy the invariant:
    $$\forall k : \emotionkindstype \rightarrow \mathtt{intensities}(k) \leq
    \mathtt{max}(k)$$

    \item \progname{} assumes that emotion intensity is a finite
    quantity, so the function $\mathtt{max}$ maps Emotion Kinds to Emotion
    Intensities ($\emotionkindstype \rightarrow \emotionintensitytype$). This
    encodes a maximum intensity for each emotion kind individually, allowing
    users to vary this value between emotion kinds in a state. Storing maximum
    intensities in the Emotion State Data Type localizes these constraints to a
    specific state, allowing the maximum intensities of other states to vary.
    This also makes it easy to ensure that updates to $\mathtt{intensities}$
    satisfies its constraints. The $\mathtt{max}$ function must satisfy
    the invariant:
    $$\exists k : \emotionkindstype \rightarrow \mathtt{max}(k) > 0$$
    This prevents situations where every value in $\mathtt{max}$ is zero (i.e.
    constantly zero). Due to the association between zero and ``deep sleep'' in
    Emotion Intensity (Equation~\ref{eq:intensitytype}), at least one emotion
    type in the state must be non-zero for the entity to be ``awake''.

\end{itemize}

Users might need to simultaneously update multiple emotion intensities in one
state. For this task, \progname{} provides a helper function that uses the
function for combining an emotion intensity with an emotion intensity change
(Equation~\ref{eq:combineintensity}):
\begin{equation}\label{eq:updatestate}
    \begin{gathered}
        \mathtt{NewStateByIntensityChanges}(es : \emotionstatetype, i_\Delta :
        \emotionkindstype \rightarrow \responsestrength) : \emotionstatetype \\
        \defEq es' \text{ with } (\forall k : \emotionkindstype \rightarrow
        es'.\mathtt{intensities}\left(k\right) = \mathtt{clamp}\left(I_k, 0,
        es.\mathtt{max}(k) \right), \\
        es'.\mathtt{max}(k) = es.\mathtt{max}(k)) \\
        \text{where } I_k =
        \mathtt{UpdateIntensity}(es.\mathtt{intensities}\left(k\right),
        i_{\Delta}\left(k\right))
    \end{gathered}
\end{equation}
A logarithm is an unbounded function and \progname{} assumes that emotion
intensities are finite (i.e. have a maximum value), so it is necessary to clamp
the updated intensity.

\subsubsection{User-Defined Emotion Types}\label{sec:userEmotions} An advantage
of defining emotion kinds and intensities as data types is the ability to
combine them into new structures. Effectively, this allows users to define
custom emotion types (\ref{flexEm}). User-defined emotion types do not have to
be scientifically correct as long as it has the desired behaviour.

Both Plutchik and Oatley \& Johnson-Laird acknowledge the concept of ``complex''
emotions, which \progname{} uses as guidelines for the data types a user could
create:
\begin{itemize}

    \item Combinations of Emotion Kinds, e.g. \textit{Fear} $+$
    \textit{Surprise} is \textit{Awe}~\citep[p.~162--165]{robert1980emotion}

    \item A partition of an Emotion Kind's maximum intensity in an Emotion
    State, e.g. an entity experiences \textit{Rage} if the intensity of
    \textit{Anger} is $ \geq 90\%$ of the maximum
    intensity~\citep[p.~159--160]{robert1980emotion}

    \item An Emotion Kind with additional propositional meaning, e.g.
    \textit{Disgust} directed at a person is
    \textit{Contempt}~\citep[p.~60]{oatley1992best} (see
    Equation~\ref{eq:generateemotionAcceptance} for an example)

\end{itemize}

Rather than encoding these in models, \progname{} simply recognizes these as
functional requirements so that users can apply these guidelines as they see
fit.

\subsubsection{Emotion as States Over Time} \progname{} represents the temporal
nature of emotion by assigning Emotion States to instances in time
(Equation~\ref{eq:timetype}):
\begin{equation}\label{eq:emotiontype}
    \emotiontype : \timetype \rightarrow \emotionstatetype
\end{equation}
As time progresses and users create new emotion states, they can update the
contents of and retrieve emotion states from Emotion at a time $t : \timetype$:
\begin{equation}\label{eq:updateemotion}
    \mathtt{UpdateEmotion}(e : \emotiontype, t : \timetype, es :
    \emotionstatetype) : \emotiontype \defEq \{ e \text{ with } e\left(t\right)
    = es \}
\end{equation}
\begin{equation}\label{eq:getemotion}
    \mathtt{GetStateFromEmotion}(e : \emotiontype, t : \timetype) :
    \emotionstatetype \defEq e(t)
\end{equation}

\section{Emotion Elicitation}\label{sec:emotiongenerationfunctions}
\progname{} draws chiefly from Oatley \& Johnson-Laird to define its emotion
elicitation models. Several assumptions about emotion elicitation narrow the
models' scope to afford more flexibility to choose when and where \progname{}'s
tasks operate (\ref{flexTasks}):
\begin{itemize}

    \item Emotion elicitation and emotion intensity evaluation does not have to
    be done simultaneously, so \progname{} models emotion intensity separately
    (Section~\ref{sec:intensity})

    \item An entity can experience multiple emotions simultaneously but can
    only be in one state at any given time, so users can select and combine the
    results into a single emotion state with Equation~\ref{eq:updatestate}
    after intensity evaluations

\end{itemize}

Before developing a model for emotion elicitation from Oatley \& Johnson-Laird's
work, \progname{} must determine how to map its outputs to the Plutchik-based
Emotion State. This is a non-trivial issue because Plutchik accounts for more
emotions than Oatley \& Johnson-Laird (Equation~\ref{eq:emotionstatetype}).
Ignoring this will render part of the state unusable. \progname{} examines both
theories for commonalities that it can exploit to address this issue.

A core hypothesis in Oatley \& Johnson-Laird---that emotions signal a change
has happened which changes a goal-oriented plan's likely outcome---follows
Plutchik~\citep[p.~55--56, 422 Note 27]{oatley1992best}. Although they differ
significantly in other aspects, this comparison is enough for \progname{} to
assume that an emotion kind shared by both theories refers to the same or
nearly the same concept. Therefore, \progname{} maps Oatley \& Johnson-Laird's
five emotions directly to those on Plutchik's \ref{circumplex}: \textit{Joy}
(\textit{Happiness}), \textit{Sadness}, \textit{Fear} (\textit{Anxiety}),
\textit{Anger}, and \textit{Disgust}. \progname{} must connect the remaining
Plutchik emotions---\textit{Acceptance}, \textit{Surprise}, and
\textit{Interest}---to Oatley \& Johnson-Laird indirectly. Oatley \&
Johnson-Laird do not dismiss the potential for these three to be ``basic''
emotions\footnote{Their confidence in the inclusion of \textit{Disgust} as a
``basic'' emotion is also shaky.}~\citep[p.~59--61]{oatley1992best}. However,
\progname{} does not need Oatley \& Johnson-Laird and Plutchik to agree if an
emotion is ``basic'' or not---it only needs them to agree on their definition
for modelling. Therefore, the models for \textit{Acceptance},
\textit{Surprise}, and \textit{Interest} draw from additional research in their
specification.

\subsection{Eliciting \textit{Joy}, \textit{Sadness}, \textit{Fear},
\textit{Anger}, and \textit{Disgust}}\label{sec:evalOJL}
Oatley \& Johnson-Laird's theory connects five ``basic'' emotions to plan
junctures where the likelihood of a plan's success might change
(Table~\ref{tab:cte_patterns}). \progname{} defines separate models for each of
these emotions so that users can choose which ones to use (\ref{flexEm}). This
also allows entities to experience multiple emotions simultaneously depending
on how the user calls and provides information to the models.
\begin{table}[!tb]
    \centering
    \caption[Oatley \& Johnson-Laird's Connection of Goals and Plans to Five
    Emotions]{Oatley \& Johnson-Laird's Connection of Goals and Plans to Five
    Emotions (Adapted from~\citet[p.~55]{oatley1992best})}
    \label{tab:cte_patterns}
    \renewcommand{\arraystretch}{1.2}
    \small
    \begin{tabular}{lll}
        \toprule
        \textbf{Emotion} & \textbf{Juncture of Current Plan} & \textbf{Next
            State} \\ \midrule

        \colourRow\textit{Happiness} & Sub-goals being achieved & Continue with
        plan, modifying if needed \\

        \textit{Sadness} & Failure of a major plan or loss of an active goal &
        Do nothing/Search for a new plan \\

        \colourRow\textit{Anxiety} & Self-preservation goal threatened or goal
        conflict & Stop, Attend to Environment/Escape \\

        \textit{Anger} & Active plan frustrated & Try harder/Aggress \\

        \colourRow\textit{Disgust} & Gustatory goal violated & Reject
        substance/Withdraw \\

        \bottomrule
    \end{tabular}
\end{table}

At first glance, the descriptions appear to have a clear translation to formal
models with respect to goals and plans. A closer inspection reveals aspects
that people could interpret differently (e.g. what is a ``sub-goal''?).
Therefore, \progname{} rephrases Oatley \& Johnson-Laird's descriptions to
remove potential ambiguities before translating them into mathematical models.

Each model for \textit{Joy}, \textit{Sadness}, \textit{Fear}, \textit{Anger},
and \textit{Disgust} outputs an Option Type ($A^?$) containing a tuple of
processed goal and/or plan data as WSV $\worldstatetype$, event
$\worldstatechangetype$, distance $\statedistancetype$
(Equation~\ref{eq:statedisttype}), and/or change in distance
$\statedistancechangetype$ (Equation~\ref{eq:statedistchgtype}) information. If
the entity is not experiencing an emotion, the corresponding Option type is
empty. Using Option Types improves \progname{}'s efficiency (\ref{flexScale}),
because it avoids reevaluations of potentially expensive functions that users
might want to use as inputs to others such as emotion intensity
(Section~\ref{sec:intensity}).

\subsubsection{Global Functions on Goals}
These functions simplify the models of \textit{Joy}, \textit{Sadness},
\textit{Fear}, \textit{Anger}, and \textit{Disgust} by collecting common
evaluations and assigning them meaningful names:
\begin{itemize}

    \item An event progresses an entity \textit{towards} goal achievement iff
    the distance to the goal state is larger in the WSV unchanged by the event
    compared to the WSV changed by the event:
    $$\mathtt{IsCloserAfterEvent}(g : \goaltype, s : \worldstatetype, s_\Delta
    : \worldstatechangetype) : \mathbb{B} \defEq g.\mathtt{goal}(s) >
    g.\mathtt{goal}\left(s \oplus s_\Delta\right)$$

    \item An event moves an entity from a WSV into another where a goal is
    unachievable iff the distance to the goal state is infinitely large in WSV
    changed by the event:
    $$\mathtt{IsUnachievableAfterEvent}(g : \goaltype, s : \worldstatetype,
    s_\Delta : \worldstatechangetype) : \mathbb{B} \defEq | \,
    g.\mathtt{goal}\left(s \oplus s_\Delta\right) | \, = +\infty$$

    \item An event causes a noticeable change in distance to a goal from a WSV
    iff its magnitude exceeds a minimum ``threshold'':
    $$\mathtt{IsNoticeable}(g : \goaltype, s : \worldstatetype, s_\Delta :
    \worldstatechangetype, \epsilon : \statedistancechangetype) : \mathbb{B}
    \defEq | \, g.\mathtt{goal'}\left(s, s_\Delta\right)| > \epsilon$$

\end{itemize}

\subsubsection{\textit{Joy} (\textit{Happiness})\protect\footnote{\normalfont
See Chapter~\ref{sec:atc_Joy} for a partial test of
Equation~\ref{eq:generatejoy}}}
This emotion occurs when an action moves an entity closer to a goal's
achievement, assuming that each intermittent WSV is a goal state itself (i.e. a
``sub-goal'' of the goal). \progname{} conceives this as an evaluation where:
\begin{quote}
    \textit{An event transitions the previous WSV to the current WSV such that
    there is a change in the distance to a goal state where there is less
    distance between it and the current WSV compared to the distance between it
    and the previous WSV}
\end{quote}

\noindent For its model, \progname{} uses an entity goal
(Equation~\ref{eq:goaltype}), a WSV (Equation~\ref{eq:worldstatetype}), an
event (Equation~\ref{eq:worldeventtype}), and a ``tolerance'' threshold for
distance changes between WSVs:
\begin{equation}\label{eq:generatejoy}
    \begin{gathered}
        J(g : \goaltype, s_{prev} : \worldstatetype, s_\Delta :
        \worldstatechangetype, \epsilon_{J} : \statedistancechangetype) :
        (\mathit{dist}_{prev} : \statedistancetype, \mathit{dist}_{now} :
        \statedistancetype, \mathit{dist}_\Delta : \statedistancechangetype)^?
        \\[5pt]
        \defEq \begin{cases}

            \parbox{0.2\linewidth}{$(g.\mathtt{goal}(s_{prev}), \\
                g.\mathtt{goal}(s_{prev} \oplus s_\Delta), \\
                g.\mathtt{goal'}(s_{prev}, s_\Delta))$, } &
            \parbox{0.35\linewidth}{$\mathtt{IsCloserAfterEvent}(g, s_{prev},
            s_\Delta) \\
                \land \mathtt{IsNoticeable}(g, s_{prev}, s_\Delta,
                \epsilon_{J})$} \\[20pt]

            \text{None}, & Otherwise \\

        \end{cases}
    \end{gathered}
\end{equation}
If the distance to the goal state was larger in \textit{the previous WSV}
compared to the current WSV (i.e. \textit{decreases} the distance,
$\mathtt{IsCloserAfterEvent}(g, s_{prev}, s_\Delta)$) and the change in
distance exceeds a minimum threshold ($\mathtt{IsNoticeable}(g, s_{prev},
s_\Delta, \epsilon_{J})$), then the entity is achieving its ``sub-goals'' and
it experiences \textit{Joy}. The threshold $\epsilon_{J}$ controls the entity's
``sensitivity'' to changes such it experiences \textit{Joy} more easily with
lower threshold values compared to high ones.

\subsubsection{\textit{Sadness}\protect\footnote{\normalfont See
Chapter~\ref{sec:atc_Sadness} for a partial test of
Equation~\ref{eq:generatesadness}}}\label{sec:evalSadness}
\textit{Sadness} occurs when an entity has a plan that becomes impossible to
carry out (i.e. ``failure of a plan'') or a goal becomes impossible to achieve
(i.e. ``loss of a goal''). \progname{} conceives this as an evaluation where:
\begin{quote}
    \textit{An event transitions the previous WSV to the current WSV such that
    there is an unreachable plan state, implying that the plan is no longer
    viable, or when the distance from the current WSV to a goal state is
    insurmountably large, implying that it is not possible to reach that goal
    state (i.e. ``lost'')}
\end{quote}

\noindent For its model, \progname{} uses one or both of an entity goal and plan
(Equation~\ref{eq:plantype}), a WSV, and an event:
\begin{equation}\label{eq:generatesadness}
    \begin{gathered}
        S(g : \goaltype^?, p : \plantype^?, s_{prev} : \worldstatetype,
        s_\Delta : \worldstatechangetype) : ( g_{sadness} : \goaltype^?,
        p_{sadness} : \plantype^?, s_{now} : \worldstatetype,
        \mathit{dist}_{now} : \statedistancetype^? )^? \\[5pt]
        \defEq \begin{cases}
            (\text{None}, p, s_{prev} \oplus s_\Delta, \text{None}), &
            \parbox{0.54\linewidth}{$p \neq \text{None} \land
            p.\mathtt{isFeasible}(s_{prev}) \\
            \land \neg p.\mathtt{isFeasible}(s_{prev} \oplus s_\Delta)$}
            \\[10pt]

            \parbox{0.22\linewidth}{$(g, \text{None}, s_{prev} \oplus s_\Delta,
            \\
                g.\mathtt{goal}\left(s_{prev} \oplus s_\Delta\right)),$} &
            g \neq \text{None} \land \mathtt{IsUnachievableAfterEvent}(g,
            s_{prev}, s_\Delta) \\[10pt]

            \text{None}, & Otherwise \\
        \end{cases}
    \end{gathered}
\end{equation}
It is not necessary to provide both a goal and plan because the predicates on
them are mutually exclusive. If the plan was feasible in \textit{the previous
WSV} ($p.\mathtt{isFeasible}(s_{prev})$) and the world event transitions to a
WSV where the plan is no longer feasible ($\neg p.\mathtt{isFeasible}(s_{prev}
\oplus s_\Delta)$), or the distance to $g$ becomes impossible to travel after
the event ($\mathtt{IsUnachievableAfterEvent}(g, s_{prev}, s_\Delta)$), then
the entity experiences \textit{Sadness}. Including $g_{sadness} : \goaltype^?$
and $p_{sadness} : \plantype^?$ in the output makes it easy to determine which
of an entity's goal or plan elicited \textit{Sadness} at $s_{prev}$.

\subsubsection{\textit{Fear} (\textit{Anxiety})\protect\footnote{\normalfont
Equation~\ref{eq:generatefear} not yet tested}}
This emotion occurs when there is a threat to self-preservation (i.e.
``self\-/preservation goal threaten-ed''), which requires a prediction about a
future WSV based on the current world state and an action that could change it,
or there are at least two goals that are mutually exclusive (i.e. ``goal
conflict'', the entity cannot satisfy both). \progname{} conceives this as an
evaluation where:
\begin{quote}
    \textit{A potential event transitions the current WSV to a future WSV where
    the distance between the future WSV and a goal state is larger than the
    distance from the current WSV and a goal state for a goal of type
    ``Self-Preservation'' OR it is impossible to satisfy the desired states of
    two different goals}
\end{quote}

\noindent For its model, \progname{} uses a WSV, event, two entity goals where
one of them might be empty, and a ``tolerance'' threshold for distance changes
between WSVs:
\begin{equation}\label{eq:generatefear}
    \begin{gathered}
        \mathit{F}(g : \goaltype, g' : \goaltype^?, s_{now} : \worldstatetype,
        s_\Delta : \worldstatechangetype, \epsilon_{F} :
        \statedistancechangetype) \\
        : ( g_{fear} : \goaltype, \mathit{dist}_{now} : \statedistancetype,
        \mathit{dist}_{next} : \statedistancetype, \mathit{dist}_\Delta :
        \statedistancechangetype, g_{lost} : \goaltype^? )^? \\[5pt]
        \defEq \begin{cases}
            \parbox{0.22\linewidth}{$(g, g.\mathtt{goal}(s_{now}), \\
                g.\mathtt{goal}(s_{now} \oplus s_\Delta), \\
                g.\mathtt{goal'}(s_{now}, s_\Delta), \\
                \text{None})$,} &
            \parbox{0.46\linewidth}{$\mathtt{SelfPreservation} \in
            g.\mathtt{type} \\
                \land \neg \mathtt{IsCloserAfterEvent}(g, s_{now}, s_\Delta) \\
                \land \mathtt{IsNoticeable}(g, s_{now}, s_\Delta,
                \epsilon_{F})$} \\[25pt]

            \parbox{0.22\linewidth}{$(g, g.\mathtt{goal}(s_{now}), \\
                g.\mathtt{goal}(s_{now} \oplus s_\Delta), g')$,} &
            \parbox{0.46\linewidth}{$g' \neq \text{None} \land
            \mathtt{WillConflict}(g, g', s_{now}, s_\Delta)$} \\[15pt]

            \parbox{0.22\linewidth}{$(g', g'.\mathtt{goal}(s_{now}), \\
                g'.\mathtt{goal}(s_{now} \oplus s_\Delta), g)$,} &
            \parbox{0.46\linewidth}{$g' \neq \text{None} \land
                \mathtt{WillConflict}(g', g, s_{now}, s_\Delta)$} \\[10pt]

            \text{None}, & Otherwise \\

        \end{cases} \\
    \text{where } \mathtt{WillConflict}(g_1, g_2, s_{now}, s_\Delta) \\
    = \mathtt{IsCloserAfterEvent}(g_1, s_{now}, s_\Delta) \land
    \mathtt{IsUnachievableAfterEvent}(g_2, s_{now}, s_\Delta)
    \end{gathered}
\end{equation}
If the goal concerns self-preservation ($\mathtt{SelfPreservation} \in
g.\mathtt{type}$) and there is a potential event that would make the distance
to the goal state larger in the next WSV compared to \textit{the current WSV}
(i.e. would \textit{increase} the distance, $\neg
\mathtt{IsCloserAfterEvent}(g, s_{now}, s_\Delta)$), and the change in distance
exceeds a minimum threshold ($\mathtt{IsNoticeable}(g, s_{now}, s_\Delta,
\epsilon_{F})$), then the entity perceives a threat to $g$ and it experiences
\textit{Fear}. The threshold $\epsilon_{F}$ controls the entity's
``sensitivity'' to changes such it experiences \textit{Fear} more easily with
lower threshold values compared to high ones.

Alternatively, if $g' : \goaltype^?$ contains a goal ($g' \neq \text{None}$)
and there is a potential event that would reduce the distance to the goal state
of either $g$ or $g'$ ($\mathtt{IsCloserAfterEvent}(g, s_{now}, s_\Delta)$ or
$\mathtt{IsCloserAfterEvent}$ $(g', s_{now}, s_\Delta)$) that also makes it
impossible to reach the other ($\mathtt{IsUnachie-}$
$\mathtt{vableAfterEvent}(g', s_{now}, s_\Delta)$ and
$\mathtt{IsUnachievableAfterEvent}(g, s_{now}, s_\Delta)$, respectively), then
the goals $\mathtt{WillConflict}$ and the entity experiences \textit{Fear}. In
this case, neither goal $g$ or $g'$ have to concern self-preservation.

Including $g_{fear} : \goaltype$ and $g_{lost} : \goaltype^?$ in the output
makes it easy to determine if two conflicting goals elicited \textit{Fear} and
which one will become lost if the event occurs.

\subsubsection{\textit{Anger}\protect\footnote{\normalfont
Equation~\ref{eq:generateanger} not yet tested}}\label{sec:evalAnger}
\textit{Anger} occurs when an entity has a plan where the next step cannot be
reached after an intentional action to achieve it, but there are one or more
other actions that can (i.e. the plan is ``frustrated'' because it is still
feasible, but it had to change to remain so). \progname{} conceives this as an
evaluation where:
\begin{quote}
    \textit{An event transitions the previous WSV into the current WSV that is
    not part of the entity's plan, but there is a series of events that
    transitions the current WSV to another state that makes progress in the
    entity's plan}
\end{quote}

\noindent For its model, \progname{} uses a WSV, event, and a set of plans
targeting the same end-state:
\begin{equation}\label{eq:generateanger}
    \begin{gathered}
        \mathit{A}(s_{prev} : \worldstatetype, s_\Delta :
        \worldstatechangetype, ps : \{\plantype\} ) : ( s_{now} :
        \worldstatetype, p_{fail} : \plantype, ps_{alt} : \{\plantype\} )^?
        \\[5pt]
        \defEq \begin{cases}

            \parbox{0.3\linewidth}{$(s_{prev} \oplus s_\Delta, p_\alpha,
            \forall p \in \{\plantype\} \\ \rightarrow
            p.\mathtt{isFeasible}(s_{prev} \oplus s_\Delta) $,} &
            \parbox{0.4\linewidth}{$\exists p_\alpha \in ps \rightarrow
            (\forall p \in ps \\
            \text{   }\rightarrow p \neq p_\alpha \land \mathtt{Cost}(p_\alpha)
            \leq \mathtt{Cost}(p)) \\
            \land \neg p_\alpha.\mathtt{isFeasible}(s_{prev} \oplus s_\Delta) \\
            \land \exists p \in ps \rightarrow p.\mathtt{isFeasible}(s_{prev}
            \oplus s_\Delta)$} \\[25pt]

            \text{None}, & Otherwise \\

        \end{cases}
    \end{gathered}
\end{equation}
The world event transitioned \textit{the previous WSV} into the current WSV
which makes the entity's lowest effort plan ($\exists p_\alpha \in ps
\rightarrow (\forall p \in ps \rightarrow p \neq p_\alpha \land
\mathtt{Cost}(p_\alpha) \leq \mathtt{Cost}(p)$) impossible to progress ($\neg
p_\alpha.\mathtt{isFeasible}(s_{prev} \oplus s_\Delta)$), but there is at least
one other plan for achieving the same end-state ($\exists p \in ps \rightarrow
p.\mathtt{isFeasible}(s_{prev} \oplus s_\Delta)$). Therefore, the entity can
continue working towards a desired end-state but must use a plan that requires
more effort (``frustrated'') and the entity experiences \textit{Anger}. The
function $\mathtt{Cost} : \plantype \rightarrow \mathbb{R}$ evaluates the
``cost'' of the plan such that low costs are desirable.

Including $p_{fail} : \plantype$ in the output makes it easy to determine which
plan ``failed'' and caused the elicitation of \textit{Anger}. Note that the set
of plans that the model returns is a strict subset of the provided set of plans
$ps_f \subset ps$ because it has at least one plan fewer due to the
infeasibility of $p_\alpha$.

\subsubsection{\textit{Disgust}\protect\footnote{\normalfont
Equation~\ref{eq:generatedisgust} not yet tested}}
\textit{Disgust} occurs when an entity has been ``contaminated'' or encounters
``contaminated'' substances that it wants to avoid (i.e. ``gustatory goal
violated''). \progname{} conceives this as an evaluation where:
\begin{quote}
    \textit{An event transitions the previous WSV, where the entity's gustatory
    goal was satisfied, to the current WSV that dissatisfies the goal such that
    the distance between the goal state and the current WSV is larger than the
    distance to the previous WSV}
\end{quote}

\noindent For its model, \progname{} uses an entity goal, WSV, event, and two
``tolerance'' thresholds for distance changes between WSVs:
\begin{equation}\label{eq:generatedisgust}
    \begin{gathered}
        \mathit{D}(g : \goaltype, s_{prev} : \worldstatetype, s_\Delta :
        \worldstatechangetype, \epsilon_{DS} : \statedistancechangetype,
        \epsilon_{DN} : \statedistancechangetype) : (\mathit{dist}_{prev} :
        \statedistancetype, \mathit{dist}_{now} : \statedistancetype,
        \mathit{dist}_\Delta : \statedistancechangetype)^? \\[5pt]
        \defEq \begin{cases}

            \parbox{0.2\linewidth}{$(g.\mathtt{goal}(s_{prev}), \\
                g.\mathtt{goal}(s_{prev} \oplus s_\Delta), \\
                g.\mathtt{goal'}(s_{prev}, s_\Delta))$,} &
            \parbox{0.4\linewidth}{$\mathtt{Gustatory} \in
                g.\mathtt{type} \\
                \land g.\mathtt{goal}(s_{prev}) \leq \epsilon_{DS} \\
                \land g.\mathtt{goal}(s_{prev} \oplus s_\Delta) > \epsilon_{DS}
                \\
                \land \mathtt{IsNoticeable}(g, s_{prev}, s_\Delta,
                \epsilon_{DN}) $} \\[25pt]

            \text{None}, & Otherwise \\

        \end{cases}
    \end{gathered}
\end{equation}
If the goal is gustatory-related ($\mathtt{Gustatory} \in g.\mathtt{type}$),
\textit{the previous WSV} satisfied that goal within some ``satisfaction
threshold'' ($g.\mathtt{goal}(s_{prev}) \leq \epsilon_{DS}$), but the event
transitioned into the current WSV where the goal is unsatisfied
($g.\mathtt{goal}(s_{prev} \oplus s_\Delta) > \epsilon_{DS}$) and the
difference is noticeable ($\mathtt{IsNoticeable}(g, s_{prev}, s_\Delta,
\epsilon_{DN})$), then the entity experiences \textit{Disgust}.

The threshold $\epsilon_{DS}$ defines an entity's ``tolerance'' for goal
dissatisfaction such that higher values means that the entity allows larger
distances between the current WSV and its goal state before experiencing
\textit{Disgust}. The threshold $\epsilon_{DN}$ controls the entity's
``sensitivity'' to changes such it experiences \textit{Disgust} more easily
with lower threshold values compared to high ones.

\subsection{Eliciting \textit{Acceptance}\protect\footnote{\normalfont
See Chapter~\ref{sec:atc_Acceptance} for a partial test of
Equation~\ref{eq:generateemotionAcceptance}}}\label{sec:evalAcceptance}
There is no explicit reference to an emotion like \textit{Acceptance} in Oatley
\& Johnson-Laird. It is difficult to find information about it in the
literature, so \progname{} uses \textit{Trust} as its conceptual reference
because Plutchik's \ref{circumplex} associates it with \textit{Acceptance} via
the intensity dimension.

``Affective trust'' builds on past experiences with, feelings of security,
confidence, and satisfaction towards, and the perceived level of selfless
concern demonstrated by a partner regardless of what the future
holds~\citep[p.~96]{rempel1985trust}. The idea of a ``partner'' seems to align
with the concept of joint planning in Oatley \& Johnson-Laird, where joint
plans are only possible if each member believes that they can rely on all other
members~\citep[p.~178--179, 192]{oatley1992best}. From Oatley \&
Johnson-Laird's description of ``complex'' emotions as ``basic'' emotions with
additional propositional meaning, \textit{Trust} might be a ``basic'' emotion
elaborated with relationship-focused information.

The role of oxytocin in social attachments and affiliation is one biological
basis of affective trust~\citep[p.~393]{oxfordTrust}, supporting
\textit{Trust}'s reliance on relationships. This is another potential
connection between \textit{Trust} and Oatley \& Johnson-Laird: they use
``emotions of attachment'' as an example of infant-level social emotions,
linking them to the ``basic'' emotion
\textit{Happiness}~\citep[p.~192]{oatley1992best}. Therefore, \progname{} takes
the proposal that \textit{Trust} is \textit{Happiness} elaborated with
information about social attachment as the basis for the \textit{Acceptance}
evaluation model.

Due to Plutchik's differentiation of \textit{Trust} and \textit{Acceptance}
through emotion intensity and Oatley \& Johnson-Laird's concept of ``complex
emotions'', \progname{} conceives \textit{Acceptance} as an evaluation of
\textit{Joy} elaborated with social attachments such that:
\begin{quote}
    \textit{An event transitions the previous WSV to the current WSV such that
    there is a change in the distance to a goal state where there is less
    distance between it and the current WSV compared to the distance between it
    and the previous WSV AND a socially-relevant entity caused the event}
\end{quote}

\noindent For its model, \progname{} uses an Option type that might contain a
social attachment (Equation~\ref{eq:socialtype}), entity goal, WSV, event, and
two ``tolerance'' thresholds for distance changes between WSVs, and the model
for \textit{Joy} elicitation (Equation~\ref{eq:generatejoy}):
\begin{equation}\label{eq:generateemotionAcceptance}
    \begin{gathered}
        \mathit{Acc}(r_A : {\socialattachmenttype}^?, g : \goaltype, s_{prev} :
        \worldstatetype, s_\Delta : \worldstatechangetype, \epsilon_{A1} :
        \worldstatechangetype, \epsilon_{A2} : \worldstatechangetype) :
        (r_A : \socialattachmenttype, \mathit{distAttribToA}_\Delta :
        \statedistancechangetype)^? \\
        \defEq \begin{cases}

            (r_A, dist_\Delta - \epsilon_{A2}), &
            \parbox{0.46\linewidth}{$r_A \neq \text{None} \, \land \, | J(g,
            s_{prev}, s_\Delta, \epsilon_{A1}).dist_\Delta | > \epsilon_{A2} \\
                \land \mathtt{CausedBy}(s_\Delta, A)$} \\[10pt]

            \text{None}, & Otherwise \\

        \end{cases}
    \end{gathered}
\end{equation}
If an event elicits \textit{Joy} ($J(g, s_{prev}, s_\Delta, \epsilon_{A1})$),
the change in distance between \textit{the previous WSV} and the current WSV
exceeds a minimum threshold ($dist_\Delta > \epsilon_{A2}$), and the entity
attributes the event to another entity $A$ ($\mathtt{CausedBy}(s_\Delta, A)$)
that it has a social attachment to ($r_A \neq \text{None}$), then the entity
experiences \textit{Acceptance} towards the other. The threshold
$\epsilon_{A2}$ controls the entity's ``sensitivity'' to changes such it
experiences \textit{Acceptance} more easily with lower threshold values
compared to high ones. This threshold also moderates the elicitation
``magnitude'' by returning the change in distance between WSVs that exceeds it
($dist_\Delta - \epsilon_{A2}$) so that the ``magnitude'' is relative to how
easily ``impressed'' the entity is. The function $\mathtt{CausedBy} :
\worldstatechangetype \times A \rightarrow \mathbb{B}$ evaluates event
causality, returning $\True$ if the entity believes that $A$ is responsible for
causing an event.

If an entity has no social attachment yet, users could use this as a mechanism
for establishing one:
\begin{enumerate}
    \item Create a new social attachment $r_A' : \socialattachmenttype$
    \item Use it to evaluate the presence of \textit{Acceptance}:
    \begin{enumerate}
        \item If \textit{Acceptance} is present, store $r_A'$
        \item Otherwise discard it
    \end{enumerate}
\end{enumerate}

\subsection{Eliciting \textit{Interest}\protect\footnote{\normalfont
Equation~\ref{eq:generateemotionInterest} not yet tested}}
Oatley \& Johnson-Laird state that \textit{Interest} ``...implies sustained
attention to certain external events''~\citep[p.~33]{oatley1987towards}. This
seems to align with the ``starting'' behaviour tendencies that Plutchik's
theory associates with \textit{Interest}~\citep[p.~202]{plutchik1984emotions}.
From this connection, and ignoring the implied limitation to events,
\progname{} conceives \textit{Interest} as an evaluation where:
\begin{quote}
    \centering
    \textit{A significant amount of attention is paid to something}
\end{quote}

\noindent For its model, \progname{} uses the amount of attention spent on some
$x$ (Equation~\ref{eq:attentiontype}) and a ``tolerance'' threshold on it such
that:
\begin{equation}\label{eq:generateemotionInterest}
    \begin{gathered}
        \mathit{Inr}(\mathit{at}_x : \attentiontype, \epsilon_{Inr} :
        \attentiontype) : \attentiontype^? \\
        \defEq \begin{cases}

            \mathit{at}_x - \epsilon_{Inr}, & at > \epsilon_{Inr} \\

            \text{None}, & Otherwise \\

        \end{cases}
    \end{gathered}
\end{equation}
If the amount of attention paid to $x$ exceeds the threshold ($at >
\epsilon_{Inr}$), then the entity experiences \textit{Interest} towards $x$.
This threshold also moderates the elicitation ``magnitude'' by returning the
amount of attention that exceeds it ($\mathit{at}_x - \epsilon_{Inr}$) so that
the ``magnitude'' is relative to how much $x$ ``fascinates'' the entity.

\subsection{Eliciting \textit{Surprise}\protect\footnote{\normalfont
Equation~\ref{eq:generateemotionSurprise} not yet tested}}
Oatley \& Johnson-Laird state that \textit{Surprise} ``...is elicited by a
sudden unexpected event...''~\citep[p.~33]{oatley1987towards}. This seems to
align with the ``stopping'' behaviour tendencies that Plutchik's theory
associates with \textit{Surprise}~\citep[p.~202]{plutchik1984emotions}.
However, what is meant by a ``sudden unexpected'' event needs clarification.

Researchers have proposed that events appraised to be a contradiction of
explicitly or implicitly held expectations and beliefs elicit
\textit{Surprise}, which lab-based experiments found convincing supporting
evidence~\citep{reisenzein2019cognitive}. Quantitative models of
\textit{Surprise} intensity rely on event probabilities such that an
``unexpected'' event is an improbable one, and assume that intensity increases
monotonically with the degree of unexpectedness. This has no obvious conflicts
with Oatley \& Johnson-Laird's concept of emotions as system-wide
non-propositional communication signals, so \progname{} conceives
\textit{Surprise} as an evaluation where:
\begin{quote}
    \centering
    \textit{A significantly-improbable event happens}
\end{quote}

\noindent For its model, \progname{} uses a WSV, an event, and a ``tolerance''
threshold such that:
\begin{equation}\label{eq:generateemotionSurprise}
    \begin{gathered}
        \mathit{Sur}(s_{prev} : \worldstatetype, s_\Delta :
        \worldstatechangetype, \epsilon_P : [0,1]) : [0,1]^? \\
        \defEq \begin{cases}
            \epsilon_P - P(s_\Delta | s_{prev}), & P(s_\Delta | s_{prev}) <
            \epsilon_P \\

            \text{None}, & Otherwise \\

        \end{cases}
    \end{gathered}
\end{equation}
If the improbability of the event in \textit{the previous WSV} is below the
threshold ($P(s_\Delta | s_{prev}) < \epsilon_P$), then the entity experiences
\textit{Surprise}. This threshold also moderates the elicitation ``magnitude''
by returning how much the event's improbability falls below it ($\epsilon_P -
P(s_\Delta | s_{prev})$) so that the ``magnitude'' is relative to how
``impossible'' the entity believes the event is. The probability of the event
\textit{depends on} the previous WSV because what is ``expected'' and
``unexpected'' depends on its preconditions (e.g. water falling on someone is
unexpected on a sunny day bt not on a rainy one). This model assumes that there
are exactly two outcomes for any given event---either it happens or it does
not. This is for simplicity~\citep[p.~56]{reisenzein2019cognitive} and an
assumption that users will want more control over entity reactions in complex
scenarios.

\subsection{A Word About NPCs' World Knowledge and
Perception}\label{sec:worldKnowledge}
Definitions for the functions $\mathtt{Cost} : \plantype \rightarrow
\mathbb{R}$ in Equations~\ref{eq:generateanger} and
\ref{eq:evalIntensityAnger}, $\mathtt{CausedBy} : \worldstatechangetype \times
A \rightarrow \mathbb{B}$ in Equation~\ref{eq:generateemotionAcceptance},
$P(s_\Delta | s)$ in Equation~\ref{eq:generateemotionSurprise}, and
$\mathtt{Dist} : \worldstatetype \times \worldstatetype \rightarrow
\statedistancetype$ in Equation~\ref{eq:evalIntensitySadness} are not part of
\progname{}'s models by design because they are not emotion-specific
evaluations---they are evaluations about an NPC's perception of their ``world''
and their relation to elements in it. This implies that there is additional
``world knowledge'' that NPCs need to access to evaluate how desirable the
``world'' is with respect to their internal goals and plans. The dependence of
\progname{}'s elicitation models on ``external'' functions is unsurprising
given the role that cognition plays in the emotion system
(Chapter~\ref{sec:affectiveDefs}) and suggests that linking \progname{} with
other NPC systems such as decision-making and planning would improve the
cohesiveness and consistency of NPC behaviours overall.

\section{Emotion Intensity}\label{sec:intensity}
Compared to other aspects of emotion, intensity is an understudied
topic~\citep[p.~60]{frijda1992complexity} so \progname{} must draw from
informal accounts and a general understanding of it to define models of it.
\progname{}'s definition of Emotion Intensity (Equation~\ref{eq:intensitytype})
implicitly assumes that this value can change such that an entity can
experience more or less of any given emotion kind---implying that changes can
be both ``positive'' and ``negative''. Therefore, \progname{} defines Emotion
Intensity Change separately from, and unconstrained by, Emotion Intensity while
maintaining the conceptual relation between them:
\begin{equation}\label{eq:intensitychangetype}
    \responsestrength : \mathbb{R}
\end{equation}
When evaluating the value of an intensity change, there appear to be four
determinant categories~\citep[p.~71]{frijda1992complexity}:
\begin{itemize}

    \item How much the entity ``values'' affected internal conditions (e.g.
    goals),

    \item The ``seriousness'' or ``value'' of the event that affected those
    internal conditions,

    \item Contextual considerations of elements such as coping, support, and
    unexpectedness, and

    \item The entity's personality attributes that affect factors such as
    emotion response thresholds and dispositions towards different emotions.

\end{itemize}

Of these factors, only the value of the internal condition and the event are
within \progname{}'s scope. Users can extend \progname{}'s intensity
evaluations by integrating contextual considerations and/or personality
attributes to the evaluation after getting the initial Emotion Intensity Change.

\subsection{Evaluating the Intensity of \textit{Joy}, \textit{Sadness},
\textit{Fear}, \textit{Anger}, and \textit{Disgust}}\label{sec:CTE-Intensity}
Oatley \& Johnson-Laird propose that an emotion's intensity is proportional to
the force causing an emotion (``entrained in an emotion mode'') and how fixed
or non-adjustable that force is (``degree it is locked into that
mode'')~\citep[p.~34]{oatley1987towards}. From this, \progname{} assumes that
emotion intensity directly relates to the degree that something impacts a goal
or plan such that an entity would want to maintain the momentum caused by an
emotion ``mode'' as long as that ``something'' affects that goal and/or plan.
This aligns with the concept of an affected ``internal condition''.

\subsubsection{\textit{Joy}\protect\footnote{\normalfont See
Chapter~\ref{sec:atc_Joy} for a partial test of
Equation~\ref{eq:evalIntensityJoy}}}
From the model of \textit{Joy} elicitation (Equation~\ref{eq:generatejoy}), the
affected ``internal condition'' is the entity goal (Equation~\ref{eq:goaltype})
causing the elicitation and its ``value'' is its $\mathtt{importance}$. The
magnitude of change in distance caused by the event is its ``seriousness'' or
``value'' because it measures how much the event moved the entity towards the
desired goal state. To evaluate the intensity of \textit{Joy}, \progname{}
treats the event's ``value'' as an objective measure that it scales with the
entity's goal's subjective (i.e. personal) ``value'':
\begin{equation}\label{eq:evalIntensityJoy}
    J_\Delta(g : \goaltype, d_\Delta : \statedistancechangetype) :
    \responsestrength \defEq |d_\Delta| \cdot g.\mathtt{importance}
\end{equation}
Using goal $\mathtt{importance}$ as a scaling factor moderates intensity
changes such that its magnitude varies for entities observing the same
$d_\Delta$ whose goals only differ in their $\mathtt{importance}$. Note that
the \textit{Joy} elicitation model outputs a tuple with element
$\mathit{dist}_\Delta$, which users can provide as the input $d_\Delta$.

\subsubsection{\textit{Sadness}\protect\footnote{\normalfont
See Chapter~\ref{sec:atc_SadnessI} for a partial test of
Equation~\ref{eq:evalIntensitySadness}}}
From the model of \textit{Sadness} elicitation
(Equation~\ref{eq:generatesadness}), there are two possible ``internal
conditions'' affected by the event that determines the eliciting event's
``value'' or ``seriousness'':
\begin{itemize}

    \item An entity plan (Equation~\ref{eq:plantype}) with a ``value'' equal to
    the distance to the desired end-state before it became infeasible, such
    that the event's ``value'' or ``seriousness'' is inversely proportional to
    the distance between the plan's end-state and the previous WSV where the
    plan was feasible. This means that plans the entity was close to completing
    elicit more intense \textit{Sadness} compared to ones that were farther
    from completion.

    \item An entity goal with a ``value'' equal to its $\mathtt{importance}$,
    but the event's ``value'' or ``seriousness'' is not necessarily tied to the
    event---an entity can experience intense \textit{Sadness} if they were
    significantly far from the goal state (e.g. if there is a goal to see a
    loved one before they pass, losing them feels equally painful if one just
    began saving money for a plane ticket or if they have already spent a week
    with them). This means that goals with higher $\mathtt{importance}$ elicit
    more intense \textit{Sadness} compared to less important ones relative to
    some maximum \textit{Sadness} an entity can experience.

\end{itemize}

To evaluate the intensity of \textit{Sadness}, \progname{} treats the event's
or maximum intensity's ``value'' as an objective measure that it scales with
the entity's goal's or plan's subjective (i.e. personal) ``value'':
\begin{equation}\label{eq:evalIntensitySadness}
    \begin{gathered}
        S_\Delta(g : \goaltype^?, p : \plantype^?, s_{prev} : \worldstatetype,
        i_{\mathit{max}\Delta} : \responsestrength)
        : \responsestrength \defEq \begin{cases}

            \dfrac{1}{|\mathit{dist}_p|}, & p \neq \text{None} \\[15pt]

            \dfrac{g.\mathtt{importance}}{m_G} \cdot i_{\mathit{max}\Delta}, & g
            \neq \text{None} \\

        \end{cases} \\
        \text{where } \mathit{dist}_p : \statedistancetype =
        \mathtt{Dist}(s_{prev}, p.\mathtt{nextStep}(s_{prev},
        |p.\mathtt{actions}|))
    \end{gathered}
\end{equation}
For evaluating the ``seriousness'' of a plan becoming infeasible, the function
generates the plan's end-state from the previous WSV by applying every plan
action to it ($p.\mathtt{nextSteps}(s_{prev}, |p.\mathtt{actio-}$
$\mathtt{ns}|)$), then calculates the distance between the generated plan
end-state and the previous WSV ($s_{prev}$) using the function $\mathtt{Dist} :
\worldstatetype \times \worldstatetype \rightarrow \statedistancetype$
(similar to the function $\mathtt{goal}$ in $\goaltype$, see
Section~\ref{sec:worldKnowledge}). This emulates an evaluation of ``how close''
the entity was to plan completion.

For evaluating the ``seriousness'' of a goal being ``lost'', the goal's
$\mathtt{importance}$ relative to $m_G$ is a scaling factor that moderates a
maximum intensity change $i_{\mathit{max}\Delta}$ such that its magnitude
varies for entities with different $\mathtt{importance}$ valuations in
otherwise identical goals. The value $m_G$ is a user-defined maximum for goal
$\mathtt{importance}$, effectively normalizing it to $[0,1]$. The function can
access $m_G$ itself so that the user does not have to provide it. Users also
provide the value of $i_{\mathit{max}\Delta}$ so that they have more control
over how much an entity experiences emotion changes (i.e. the model does not
have to be relative to the maximum possible \textit{Sadness} intensity).

Note that the \textit{Sadness} elicitation model outputs a tuple with elements
$g_{sadness} : \goaltype^?$ and $p_{sadness} : \plantype^?$, which users can
supply as the inputs $g$ and $p$.

\subsubsection{\textit{Fear}\protect\footnote{\normalfont
Equation~\ref{eq:evalIntensityFear} not yet tested}}
From the model of \textit{Fear} elicitation (Equation~\ref{eq:generatefear}),
the affected ``internal conditions'' are entity goals causing the elicitation
and their ``value'' is their $\mathtt{importance}$. The change in distance
caused by the event is its ``seriousness'' or ``value'' because it measures how
much the event will move the entity away from the desired goal state or, when
there are two goals, move the entity towards one while making the other
unachievable. To evaluate the intensity of \textit{Fear}, \progname{} treats
the event's ``value'' as an objective measure that it scales with the entity's
goal's subjective (i.e. personal) ``value'':
\begin{equation}\label{eq:evalIntensityFear}
    F_\Delta(g : \goaltype, g_{lost} : \goaltype^?, d_\Delta :
    \statedistancechangetype) : \responsestrength \defEq \begin{cases}

        d_\Delta \cdot g.\mathtt{importance}, & g_{lost} = \text{None} \\[10pt]

        d_\Delta \cdot
        \dfrac{g_{lost}.\mathtt{importance}}{g.\mathtt{importance}}, & g_{lost}
        \neq \text{None} \\

    \end{cases}
\end{equation}
Using goal $\mathtt{importance}$ as a scaling factor moderates intensity
changes such that its magnitude varies for entities with different
$\mathtt{importance}$ valuations in otherwise identical goals that observe the
same $d_\Delta$. In the case where conflicting goals elicit \textit{Fear}, the
scaling factor is a ratio between their $\mathtt{importance}$ values such that
the ``value'' of the progressed goal tempers that of the ``lost'' goal---the
intensity of \textit{Fear} is higher when the $\mathtt{importance}$ of the
``lost'' goal is larger than the $\mathtt{importance}$ of the other goal.

Note that the \textit{Fear} elicitation model outputs a tuple with elements
$g_{fear} : \goaltype$, $g_{lost} : \goaltype^?$, and $\mathit{dist}_\Delta$,
which users can use as the inputs $g$, $g_{lost}$, and $d_\Delta$.

\subsubsection{\textit{Anger}\protect\footnote{\normalfont
        Equation~\ref{eq:evalIntensityAnger} not yet tested}}
From the model of \textit{Anger} elicitation (Equation~\ref{eq:generateanger}),
the affected ``internal condition'' is the change in entity plan availability
and their ``value'' is the amount of effort the entity needs to execute them
(i.e. plan ``cost''). The difference in ``cost'' between the infeasible plan
and the next lowest ``cost'' plan is its ``seriousness'' or ``value'' because
it measures how much additional effort the entity needs to achieve the same
result. Both of these values are subjective because the entity assigns them.
Therefore, \progname{} evaluates \textit{Anger} intensity as the difference
between the subjective ``cost'' of the ``frustrated'' plan and the next lowest
``cost'' plan:
\begin{equation}\label{eq:evalIntensityAnger}
    \begin{gathered}
        A_\Delta(p : \plantype, ps : \{\plantype\}) : \responsestrength \\
        \defEq \exists p_\beta \in ps \rightarrow (\forall p \in ps \rightarrow
        p \neq p_\beta \land \mathtt{Cost}(p_\beta) \leq \mathtt{Cost}(p))
        \rightarrow \mathtt{Cost}(p_\beta) - \mathtt{Cost}(p)
    \end{gathered}
\end{equation}
Taking the difference between plan ``costs'' ensures that \textit{Anger} is
more intense if the ``cost'' of the next most desirable plan increases compared
to the original one. The model calculates plan ``cost'' using the function
$\mathtt{Cost} : \plantype \rightarrow \mathbb{R}$ such that low ``costs'' are
desirable (\textit{Anger} elicitation also uses this function, see
Section~\ref{sec:worldKnowledge}). If these plans are for achieving a goal,
users can choose to scale the resulting \textit{Anger} intensity with the
goal's importance manually.

Note that the \textit{Anger} elicitation model outputs a tuple with elements
$p_{fail} : \plantype$ and $ps_{alt} : \{\plantype\}$, which users can use as
the inputs $p$ and $ps$.

\subsubsection{\textit{Disgust}\protect\footnote{\normalfont
Equation~\ref{eq:evalIntensityDisgust} not yet tested}}
From the model of \textit{Disgust} elicitation
(Equation~\ref{eq:generatedisgust}), the affected ``internal condition'' is the
entity goal causing the elicitation and its ``value'' is its
$\mathtt{importance}$. The distance between the current state and the desired
goal state is the event's ``seriousness'' or ``value'' because it measures how
much the event moved the entity out of it. To evaluate the intensity of
\textit{Disgust}, \progname{} treats the distance's ``value'' as an objective
measure that it scales with the entity's goal's subjective (i.e. personal)
``value'':
\begin{equation}\label{eq:evalIntensityDisgust}
    D_\Delta(g : \goaltype, d : \statedistancetype) : \responsestrength \defEq
    d \cdot g.\mathtt{importance}
\end{equation}
Using goal $\mathtt{importance}$ as a scaling factor moderates intensity
changes such that its magnitude varies for entities observing the same
$d_\Delta$ whose goals only differ in their $\mathtt{importance}$. Note that
the \textit{Disgust} elicitation model outputs a tuple with element
$\mathit{dist}_{now}$, which users can supply as the input $d$.

Although this model looks similar to the one for \textit{Joy} intensity
(Equation~\ref{eq:evalIntensityJoy}), there is a key difference between them.
\textit{Joy} is a response to events that move an entity \textit{towards} an
unsatisfied goal state, whereas \textit{Disgust} is a response to events that
move an entity \textit{away} from a satisfied goal state. This means that in
\textit{Joy}, intensity is inversely proportional to the distance from a goal
state ($|d|$), but in \textit{Disgust} the intensity grows proportionally with
the distance from a goal state ($d$).

\subsection{Evaluating the Intensity of
\textit{Acceptance}\protect\footnote{\normalfont See
Chapter~\ref{sec:atc_Acceptance} for a partial test of
Equation~\ref{eq:evalIntensityAcceptance}}}\label{sec:Acceptance-Intensity}
From the model of \textit{Acceptance} elicitation
(Equation~\ref{eq:generateemotionAcceptance}), the affected ``internal
condition'' is an entity goal. However, \progname{} assumes that the entity's
relation to $A$ (Equation~\ref{eq:socialtype}) is the relevant ``internal
condition'' that an event's ``value'' relates to in \textit{Acceptance}. The
magnitude of change in distance caused by the event \textit{attributed} to $A$
is its ``seriousness'' or ``value'' because it measures how much $A$ helped
moved the entity towards the desired goal state. To evaluate the intensity of
\textit{Acceptance}, \progname{} treats the event's ``value'' as an objective
measure and it scales with the entity's attachment to $A$ as the subjective
(i.e. personal) ``value'':
\begin{equation}\label{eq:evalIntensityAcceptance}
    \mathit{Acc}_\Delta(r_A : \socialattachmenttype, r_\mathit{min} :
    \socialattachmenttype, d_\Delta : \statedistancechangetype) :
    \responsestrength \defEq \begin{cases}
        |d_\Delta| \cdot \dfrac{r_A}{r_\mathit{min}}, & r_A < r_\mathit{min} \\
        |d_\Delta|, & \mathit{Otherwise}
    \end{cases}
\end{equation}
The value $r_\mathit{min}$ represents the minimum social attachment that an
entity must have with $A$ to ``fully'' experience \textit{Acceptance} towards
it. Using $r_\mathit{min}$ to ``normalize'' $r_A$ creates a scaling factor that
moderates intensity changes such that its magnitude varies with social
attachment level. Tuning the minimum ``level'' changes the entity's resistance
to the experience of \textit{Acceptance} so that they appear more trustful or
distrustful of other entities. Consequently, the entity's Emotion Intensity
Change value depends on the entity's relationship to the other relative to
their minimum ``trust level''.

Note that the \textit{Acceptance} elicitation model outputs a tuple with
element $\mathit{distAttribToA}_\Delta$, which users can use as the input
$d_\Delta$.

\subsection{Evaluating the Intensity of
\textit{Interest}\protect\footnote{\normalfont
Equation~\ref{eq:evalIntensityInterest} not yet
tested}}\label{sec:Interest-Intensity} From the model of \textit{Interest}
elicitation (Equation~\ref{eq:generateemotionInterest}), the affected ``internal
condition'' is an entity's attention (Equation~\ref{eq:attentiontype}). The
``event'' driving \textit{Interest} elicitation is not necessarily the same as
``world events'', implying that the ``value'' or ``seriousness'' of increased
attention is entity-specific. Therefore, to evaluate the intensity of
\textit{Interest}, \progname{} uses a subjective attention ``value'' that it
scales with the entity's attention paid to $x$:
\begin{equation}\label{eq:evalIntensityInterest}
    \begin{gathered}
        \mathit{Inr}_\Delta(at : \attentiontype, at_{min} : \attentiontype,
        i_{\delta_x} : \responsestrength) : \responsestrength \defEq
        \begin{cases}
           i_{\delta_x} \cdot \dfrac{at}{at_{min}} & at < at_{min} \\
           i_{\delta_x}, & \mathit{Otherwise}
        \end{cases}
    \end{gathered}
\end{equation}
The subjective attention ``value'' $i_{\delta_x}$ represents the entity's
``fascination'' with $x$ such that higher values elicit more intense
\textit{Interest} with smaller changes in attention. Users can specify
different values for $i_{\delta_x}$ so that entities are more ``intrigued'' by
some $x$ than others.

The value $at_\mathit{min}$ represents the minimum attention that an entity
must spend on $x$ to ``fully'' experience \textit{Interest} towards it. Using
$at_\mathit{min}$ to ``normalize'' $at$ creates a scaling factor that moderates
intensity changes such that its magnitude varies with uninterrupted, invested
attention. Tuning the minimum ``level'' changes the entity's resistance to the
experience of \textit{Interest} so that they appear to be more or less
``captivated'' by $x$.

Note that the \textit{Interest} elicitation model outputs a value with type
$\attentiontype$, which users can supply as the input $at$.

\subsection{Evaluating the Intensity of
\textit{Surprise}\protect\footnote{\normalfont
Equation~\ref{eq:evalIntensitySurprise} not yet
tested}}\label{sec:Surprise-Intensity}
From the model of \textit{Surprise} elicitation
(Equation~\ref{eq:generateemotionSurprise}), the affected ``internal
condition'' is an entity's prediction about an event's probability. Although
the ``event'' driving \textit{Surprise} is a ``world event'', its elicitation
is driven by an entity's internal prediction about it rather than some event
``valuation'' or ``seriousness''. Therefore, \progname{} evaluates the
intensity of \textit{Surprise} using a subjective ``unexpectedness value'' that
it scales with the ``discrepancy'' between the event and its improbability
based on a common-sense hypotheses about the monotonically increasing relation
between \textit{Surprise} intensity and event unexpectedness~\citep[p.~54,
56]{reisenzein2019cognitive}:
\begin{equation}\label{eq:evalIntensitySurprise}
    \mathit{Sur}_\Delta( \mathit{discr}_{s_\Delta} : [0,1],
    i_{\mathit{max}\Delta} : \responsestrength) : \responsestrength
    \defEq i_{\mathit{max}} \cdot \mathit{discr}_{s_\Delta}
\end{equation}
The subjective ``unexpectedness value'' $i_{\mathit{max}\Delta}$ measures how
easily an entity is ``startled'' such that higher values elicit more intense
\textit{Surprise} with smaller event probability discrepancies. Users can
specify different values for $i_{\mathit{max}\Delta}$ so that entities are more
``startled'' by some events than others.

Note that the \textit{Surprise} elicitation model outputs a value with type
$[0,1]^?$, which users can supply as the input $\mathit{discr}_{s_\Delta}$ if it
is not None.

\section{Emotion Decay}\label{sec:emDecay}
Emotions characteristically ``fade'' over time, implying that their dynamics
include a decay function (\citepg{hudlicka2014computational}{320};
\citepg{hudlicka2014habits}{16}). The surveyed theories/models
(Chapter~\ref{chapter:cmeOverview}, Table~\ref{tab:theoryOverview}) do not
discuss emotion decay. Like emotion intensity, emotion decay is relatively
understudied~\citep[p.~236]{el2000flame}. Therefore, \progname{} approaches its
modelling with instinctual knowledge of emotion dynamics. Other CMEs appear to
have done the same, using a variety of solutions such as:
\begin{itemize}

    \item Number of time steps (\citepg{ojha2017emotional}{4};
    \citepg{loyall1997believable}{94}; \citepg{reilly1996believable}{79}),

    \item Linear functions of time (\citepg{bourgais2017enhancing}{98};
    \citepg{durupinar2016psychological}{2149};
    \citepg{duy2004creating}{125--126}; \citepg{breazeal2003emotion}{136};
    \citepg{gratch2000emile}{330}),

    \item Exponential functions of time with
    (\citepg{kazemifard2011design}{2646};
    \citepg{dang2009experimentation}{138}) and without a probability
    distribution (\citepg{qi2019building}{213};
    \citepg{shvo2019interdependent}{68}; \citepg{zhang2016modeling}{226};
    \citepg{moshkina2011tame}{212}; \citepg{kasap2009making}{26};
    \citepg{park2009robot}{260}; \citepg{parunak2006model}{995};
    \citepg{dias2005feeling}{131}; \citepg{el2000flame}{236}),

    \item The ``half-life'' of emotions~\citep[p.~75]{aydt2011computational},
    and

    \item Multidimensional spring systems~\citep[p.~90]{becker2008wasabi}.

\end{itemize}

A commonality between these solutions is time-dependence and baseline values.
CMEs that do not combine emotion intensity and decay into a single function
only decay values in the absence of emotion-eliciting events.

\subsection{Base Model: Damped Harmonic Oscillator}
\progname{} assumes that entities have ``equilibrium'' intensities that
emotion-eliciting events disrupt. Normalizing forces restore disrupted
intensities to their ``equilibrium'' intensities over time. These normalizing
forces are proportional to the ``distance'' that an intensity is from the
``equilibrium'' intensity. \progname{} models this behaviour with a damped
harmonic oscillator mass-spring system of the form:
$$x''\left(t\right) + c \cdot x'\left(t\right) + k_s \cdot x\left(t\right) = 0$$
where $x''\left(t\right)$, $x'\left(t\right)$, and $x\left(t\right)$ are the
acceleration, speed, and position of a mass $m$ at time $t$, $c$ is a strictly
positive and real-valued damping coefficient, and $k_s$ is a spring constant.
Emotion intensity is equivalent to the ``position'' of the mass in the system
and the system's oscillation behaviour is a way for users to define an entity's
``emotional stability''.

The damping ratio $\zeta$ and natural angular frequency of the system
$\omega_n$ govern its behaviour:
$$\zeta = \dfrac{c}{2 \cdot \sqrt{m \cdot k_s}}, \; \omega_n =
\sqrt{\dfrac{k_s}{m}}$$
Closed-forms\footnote{Closed forms from \citet{alexiou2013solution}.} of each
case for $\zeta$ are necessary because the general solution allows for
imaginary numbers which are not computationally tractable:
\begin{itemize}
    \item If $0 < \zeta < 1$, the system is \textit{underdamped} such that it
    oscillates as it returns to equilibrium. As $\zeta$ approaches 1, the
    oscillations decrease more quickly. This could represent an entity that is
    ``emotionally unstable'', alternating between high and low emotion
    intensities before returning to their ``normal'' state. Position $x$ at
    time $t$ is given by:
    \begin{equation}\label{eq:underdamped}
        \begin{gathered}
            x(t) = e^{-r \cdot t} \cdot \left( A \cdot \cos(\omega \cdot t) + B
            \cdot \sin(\omega \cdot t) \right) \\
            \text{where } A = x_0, \; B = \dfrac{v_0 + x_0 \cdot \omega_n \cdot
            \zeta}{\omega}, \; r = \omega_n \cdot \zeta \\
            \text{and } \omega = \omega_n \cdot \sqrt{1 - \zeta^2}
        \end{gathered}
    \end{equation}

    \item If $\zeta = 1$, the system is \textit{critically damped} such that
    it returns to equilibrium as quickly as possible without overshooting it.
    This could represent an entity that is the most ``emotionally stable'',
    recovering more quickly and directly than entities with other $\zeta$
    values. Position $x$ at time $t$ is given by:
    \begin{equation}\label{eq:criticallydamped}
        \begin{gathered}
            x(t) = e^{-\omega_n \cdot t} \cdot \left(x_0 + \left(v_0 + x_0
            \cdot \omega_n\right) \cdot t \right)
        \end{gathered}
    \end{equation}

    \item If $\zeta > 1$, the system is \textit{overdamped} such that it does
    not oscillate as it returns to equilibrium. As $\zeta$ increases, the
    system reaches equilibrium more slowly. This could represent an entity that
    experiences their emotions longer than others. Position $x$ at time $t$ is
    given by:
    \begin{equation}\label{eq:overdamped}
        \begin{gathered}
            x(t) = C \cdot e^{-r_1 \cdot t} + D \cdot e^{-r_2 \cdot t} \\
            \text{where } C = \dfrac{1}{2} \cdot \left( x_0 + \dfrac{v_0 + x_0
                \cdot \omega_n \cdot \zeta}{\omega}\right), \; D = \dfrac{1}{2}
                \cdot \left( x_0 - \dfrac{v_0 + x_0 \cdot \omega_n \cdot
                \zeta}{\omega}\right), \\
            r_1 = \omega_n(\zeta - \sqrt{\zeta^2 - 1}), \; r_2 = \omega_n(\zeta
            + \sqrt{\zeta^2 - 1}) \\
            \text{and } \omega = \omega_n \cdot \sqrt{\zeta^2 - 1}
        \end{gathered}
    \end{equation}
\end{itemize}

\subsection{Decaying Emotion Intensity}
\progname{} encodes the spring constant $k_s$ in a damped harmonic oscillator,
a strictly positive and real-valued constant, in the Emotion Intensity Decay
Rate Data Type:
\begin{equation}\label{eq:decayratetype}
    \emotiondecaytype : \mathbb{R}_{>0}
\end{equation}
Note that this tightly couples the emotion decay model with the data type such
that changing the underlying model of one likely means changing the other as
well.

\begin{figure}[!tbh]
    \centering
    \includegraphics[width=\linewidth]{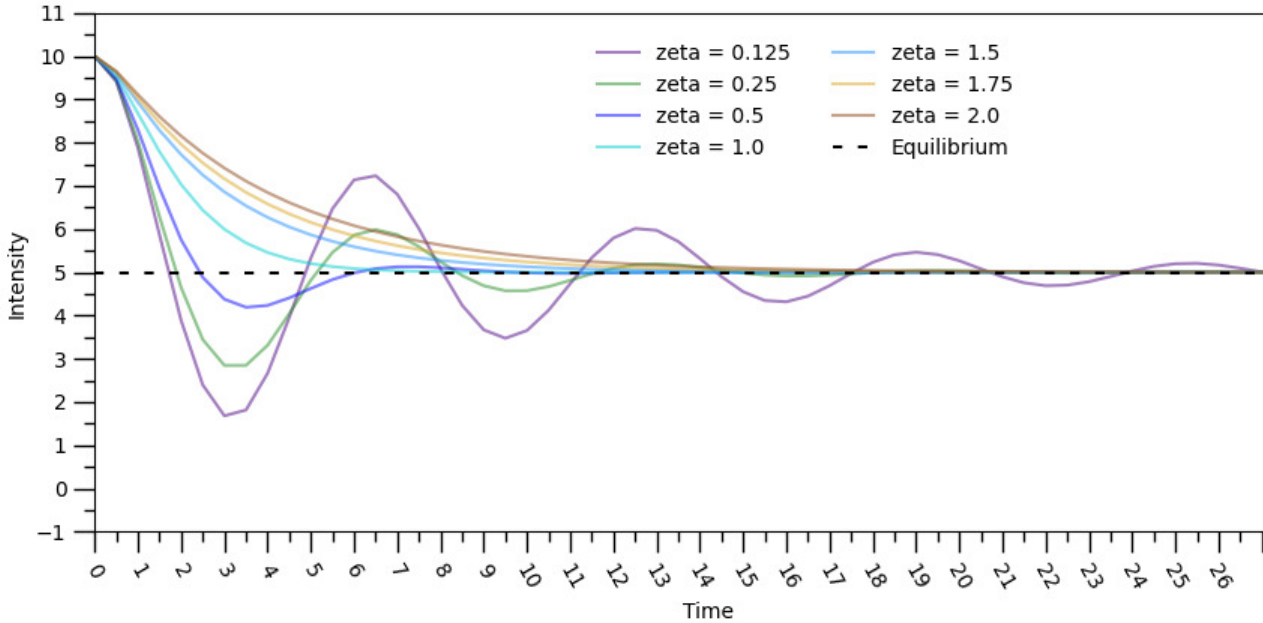}

    \caption[Examples of Emotion Decay Profiles]{Examples of Emotion Decay
    Profiles with $i_0 = 10.0$, $i_{Eq} = 5.0$, $i_\lambda = 1.0$, and $\zeta =
    $ \texttt{zeta}}
    \label{fig:dampingProfiles}
\end{figure}

\progname{} uses this data type to evaluate an intensity's ``decayed'' value.
It substitutes $k_s = i_\lambda$, $m = 1$ such that $\omega_n =
\sqrt{i_\lambda}$, $x_0 = x\left(t_0\right) = i_0 - i_{eq}$, $v_0 = 0$, and
$t_0 = 0$ such that $t = 0 + \Delta t = \Delta t$ into each model of position,
where $i_0 : \emotionintensitytype$ is the ``initial'' intensity, $i_{eq} :
\emotionintensitytype$ is the ``equilibrium'' intensity, $\Delta t :
\deltatimetype$ is the elapsed time since $t : \timetype = 0$, and $i_\lambda :
\emotiondecaytype$ is the intensity's decay rate. After simplifying,
Equations~\ref{eq:underdamped}, \ref{eq:criticallydamped}, and
\ref{eq:overdamped} become:
\begin{equation}\label{eq:decayIntensity}
    \begin{gathered}
        \mathtt{Decay}(i_0 : \emotionintensitytype, i_{Eq} :
        \emotionintensitytype, i_\lambda : \emotiondecaytype, \Delta t :
        \deltatimetype, \zeta : \mathbb{R}_{>0}) : \emotionintensitytype \\
        \defEq I_\lambda + i_{Eq} \\
        \text{where } I_\lambda = \begin{cases}
            \begin{gathered}
                e^{\text{\mbox{\footnotesize $-\sqrt{i_\lambda} \cdot \zeta
                \cdot \Delta t$}}} \cdot (i_0 - i_{Eq}) \cdot \Bigg(\cos(\omega
                \cdot \Delta t) \\
                + \left(\dfrac{\sqrt{i_\lambda} \cdot \zeta}{\omega}\right)
                \cdot \sin(\omega \cdot \Delta t) \Bigg) \\
                \text{where } \omega = \sqrt{i_\lambda} \cdot \sqrt{1 - \zeta^2}
            \end{gathered}, & 0 < \zeta < 1 \\
        & \\
            \begin{gathered}
                e^{\text{\mbox{\footnotesize $-\sqrt{i_\lambda} \cdot \Delta
                t$}}} \cdot (i_0 - i_{Eq}) \cdot \left( 1 + \sqrt{i_\lambda}
                \cdot \Delta t\right)
            \end{gathered}, & \zeta = 1 \\
        & \\
           \begin{gathered}
               \dfrac{i_0 - i_{Eq}}{2} \cdot \Bigg( \left(1 +
               \dfrac{\zeta}{Q}\right) \cdot e^{\text{\mbox{\footnotesize
               $-\sqrt{i_\lambda} \cdot \left(\zeta - Q\right) \cdot \Delta
               t$}}} \\
               + \left(1 - \dfrac{\zeta}{Q}\right) \cdot
               e^{\text{\mbox{\footnotesize $-\sqrt{i_\lambda} \cdot
               \left(\zeta + Q\right) \cdot \Delta t$}}}\Bigg) \\
               \text{where } Q = \sqrt{\zeta^2 - 1}
           \end{gathered} , & \zeta > 1
        \end{cases}
    \end{gathered}
\end{equation}
Adding $i_{Eq}$ to $I_\lambda$ shifts the position from 0 to the equilibrium
point. The damping ratio $\zeta : \mathbb{R}_{>0}$ determines how much emotion
intensity oscillates as an returns to equilibrium
(Figure~\ref{fig:dampingProfiles}). Most entities will have a profile with
$\zeta \geq 1$, but there might be a few whose personality is ``emotionally
unstable'' that warrants the oscillating behaviour that $0 < \zeta < 1$ offers.

\subsection{Decaying an Emotion State}
\progname{} assumes that each Emotion
Kind $k : \emotionkindstype$ (Equation~\ref{eq:kindstype}) can have a decay
rate and ``equilibrium'' value defined independently of others in an Emotion
State $es: \emotionstatetype$ (Equation~\ref{eq:emotionstatetype}). This allows
entities to vary how long it takes for them to return to ``normal'' for each
emotion kind (e.g. if they experience \textit{Joy} they might extend that
state by prolonging the decay to ``equilibrium''). To this end, \progname{}
defines the Emotion Decay State Data Type as a record tied to Emotion State via
$\emotionkindstype$ that has two functions $\mathtt{equilibrium}$ and
$\mathtt{decayRates}$:
\begin{equation}\label{eq:emotiondecaystatetype}
    \emotionstatedecaytype : \left\{ \mathtt{equilibrium} : \emotionkindstype
    \rightarrow \emotionintensitytype, \mathtt{decayRates} : \emotionkindstype
    \rightarrow \emotiondecaytype \right\}
\end{equation}
\begin{itemize}

    \item The function $\mathtt{equilibrium}$ maps Emotion Kinds to Emotion
    Intensities ($\emotionkindstype \rightarrow \emotionintensitytype$),
    encoding the ``equilibrium'' intensity for each $k : \emotionkindstype$. It
    must satisfy the invariant:
    $$\exists k \in\emotionkindstype \rightarrow \mathtt{equilibrium}(k) > 0$$
    This prevents situations where every value in $\mathtt{equilibrium}$ is
    zero (i.e. constantly zero). As with Emotion State
    (Equation~\ref{eq:emotionstatetype}), at least one emotion type in the
    equilibrium state must be non-zero for the entity to be ``awake''.

    \item The function $\mathtt{decayRates}$ maps Emotion Kinds to Emotion
    Intensity Decay Rates ($\emotionkindstype \rightarrow \emotiondecaytype$),
    encoding the decay rate for each emotion kind. This allows users to vary
    decay rates between kinds.

\end{itemize}

Collecting this information into Emotion Decay State makes it easier to
maintain and access for decay process automation (\ref{easeAuto}). While users
might want to decay individual emotion types in a state, allowing them to decay
some emotions while simultaneously exciting others, there might be situations
where users want to decay \textit{every} emotion type simultaneously.
\progname{} provides a helper function for this:
\begin{equation}\label{eq:decayastate}
    \begin{gathered}
        \mathtt{DecayState}(es_0 : \emotionstatetype, es_\lambda :
        \emotionstatedecaytype, \Delta t : \deltatimetype, \zeta :
        \mathbb{R}_{>0}) : \emotionstatetype \\
        \defEq \left\{ \, \forall k \rightarrow es \text{ with }
        es.\mathtt{intensities}\left(k\right) = D \, \right\} \\
        \text{where } D = \mathtt{Decay}(\,
        es_0.\mathtt{intensities}\left(k\right), \Delta t, \zeta, \\
        es_\lambda.\mathtt{decayRates}\left(k\right),
        es_\lambda.\mathtt{equilibrium}\left(k\right) \, )
    \end{gathered}
\end{equation}
Users specify an Emotion State $es_0 : \emotionstatetype$, a time difference
$\Delta t : \deltatimetype$ (Equation~\ref{eq:timetype}), an Emotion Decay
State $es_{\lambda} : \emotionstatedecaytype$, and a single damping ratio
$\zeta : \mathbb{R}_{>0}$.  From this, \progname{} decays the intensity of each
Emotion Kind $k : \emotionkindstype$ in $es_0$ using the Emotion Intensity
$\mathtt{Decay}$ function (Equation~\ref{eq:decayIntensity}). The function
returns a new Emotion State $es : \emotionstatetype$ such that $es_0$ is
unmodified.

If the user declared an Emotion $e : \emotiontype$
(Equation~\ref{eq:emotiontype}), \progname{} can extract $es_0$ from $e$ at a
user-specified time $t$ to generate a new state for some time $\{t' : \timetype
\rightarrow t < t'\}$ using the $\mathtt{DecayState}$ function
(Equation~\ref{eq:decayastate}):
\begin{equation}\label{eq:autodecay}
    \begin{gathered}
        \mathtt{NextStateByDecay}(e : \emotiontype, es_\lambda :
        \emotionstatedecaytype, \left\{ t : \timetype, t' : \timetype
        \rightarrow t < t' \right\}, \zeta : \mathbb{R}_{>0}) : \emotiontype \\
        \defEq \left\{ \, e \text{ with } e\left(t'\right) =
        \mathtt{DecayState}\left(\, e\left(t\right), t' -
        t, \zeta, es_\lambda \, \right) \, \right\}
    \end{gathered}
\end{equation}

\section{Emotion States in PAD Space}\label{sec:padspace}
\progname{} distinguishes a PAD Space point from points in other dimensional
spaces with a 3-tuple where each field's label is a PAD dimension:
\begin{equation}\label{eq:padpoint}
    \padpoint : \left( \mathtt{pleasure} : \left[-1,1\right] \subset
    \mathbb{R}, \mathtt{arousal} : \left[-1,1\right] \subset \mathbb{R},
    \mathtt{dominance} : \left[-1,1\right] \subset \mathbb{R} \right)
\end{equation}
The dimension ranges are mean ratings for emotion terms derived empirically by
\citet[p.~40--41]{mehrabian1980basic}. There were $16$ to $31$ subjects per
term whose ratings were linearly transformed to the $-1$ to $1$ scale.
Although not captured in \progname{}'s data type, mean ratings of each term
differ in statistical significance---measured from a mean of $0$ with $(p >
0.01)$---and standard deviation as shown in Table~\ref{tab:pes-pad-terms}.

\subsection{Mapping Emotion States to PAD Space}
\progname{} assumes that an entity can only be in one PAD Space location at any
given time, so it must transform an eight-element Emotion State
(Equation~\ref{eq:emotionstatetype}) into a three\-/dimensional coordinate.
However, there is no direct mapping between these spaces. \progname{} must
define reference points in PAD Space for each emotion kind
(Equation~\ref{eq:kindstype}), effectively ``mapping'' Plutchik emotion kinds
to PAD points. \progname{}'s approach for this heavily relies on the comparison
of emotion terms in each model for equal or comparable semantic meaning, which
is both subjective and error-prone. Unfortunately, there does not appear to be
a more reliable alternative.

\subsubsection{Assumptions Based on Plutchik's and Mehrabian's Empirical
Studies}
\progname{} assumes that laypeople would judge the natural language meaning of
the terms in Plutchik and PAD Space identically or nearly identically. It
derives this assumption from published, independently run empirical studies by
Plutchik and Mehrabian where they were evaluating their own affective models.
For his study on emotion language, Plutchik created a list of 145 emotion terms
and asked participants to judge how ``similar'' the terms are~\citep[p.~159,
168--170]{robert1980emotion}. They then assigned angular placements on a
\ref{circumplex} to terms based on their relative ``similarity'' to each other.
Mehrabian asked participants to judge the contribution of each
dimension---\textit{pleasure}, \textit{arousal}, and \textit{dominance}---in
the experiences described by 151 terms from which it derived statistical mean
and standard deviation values~\citep[p.~39--45]{mehrabian1980basic}. Reports
about the participants in these studies suggest that they were:
\begin{itemize}

    \item Likely in the same age group (university undergraduates, college and
    graduate students), and

    \item Likely had an North American cultural perspective (studies done in the
    United States).

\end{itemize}

The publication dates (1980) further suggest that Plutchik and Mehrabian likely
conducted their studies around the same time. Taken together, this implies that
the laypeople in these studies shared common temporal and cultural experiences
that would have influenced their interpretation of natural language terms.

\subsubsection{Selection Process for PAD Space Reference Points of Plutchik
Emotion Kinds}
From Plutchik's list of emotion terms and angular placements, \progname{}
defined eight ``boundaries'' around \ref{circumplex} areas for each of its
emotion kinds. ``Boundary'' terms are those where the perceived qualitative
meaning changes between it and the next listed term (Figure~\ref{fig:PESAreas},
Table~\ref{tab:pes-area}). For example, the change from ``Attentive'' to
``Joyful'' distinguishes a ``boundary'' between the \textit{Interest} and
\textit{Joy} areas because they do not ``feel'' like they have the same
qualitative meaning (i.e. ``Joyful'' implies a higher degree of pleasantness
than ``Attentive'', which does not imply either pleasantness or
unpleasantness). This mimics the idea of discrete emotion ``families'' such
that \progname{} can use one affective term to specify a PAD Space reference
point to serve as the emotion ``family'''s dimensional representation
(Figure~\ref{fig:theoryResolution}). Gaps between ``boundary'' terms are
inevitable due to the discrete nature of angular placements. However, they are
relatively small (between 0.3\textdegree{} and 7.3\textdegree{}), so
\progname{} ignored them instead of trying to compensate for them to avoid
introducing additional ``translation errors''.

\afterpage{\clearpage
    \begin{figure}[!t]
        \centering
        \includegraphics[width=0.65\linewidth]{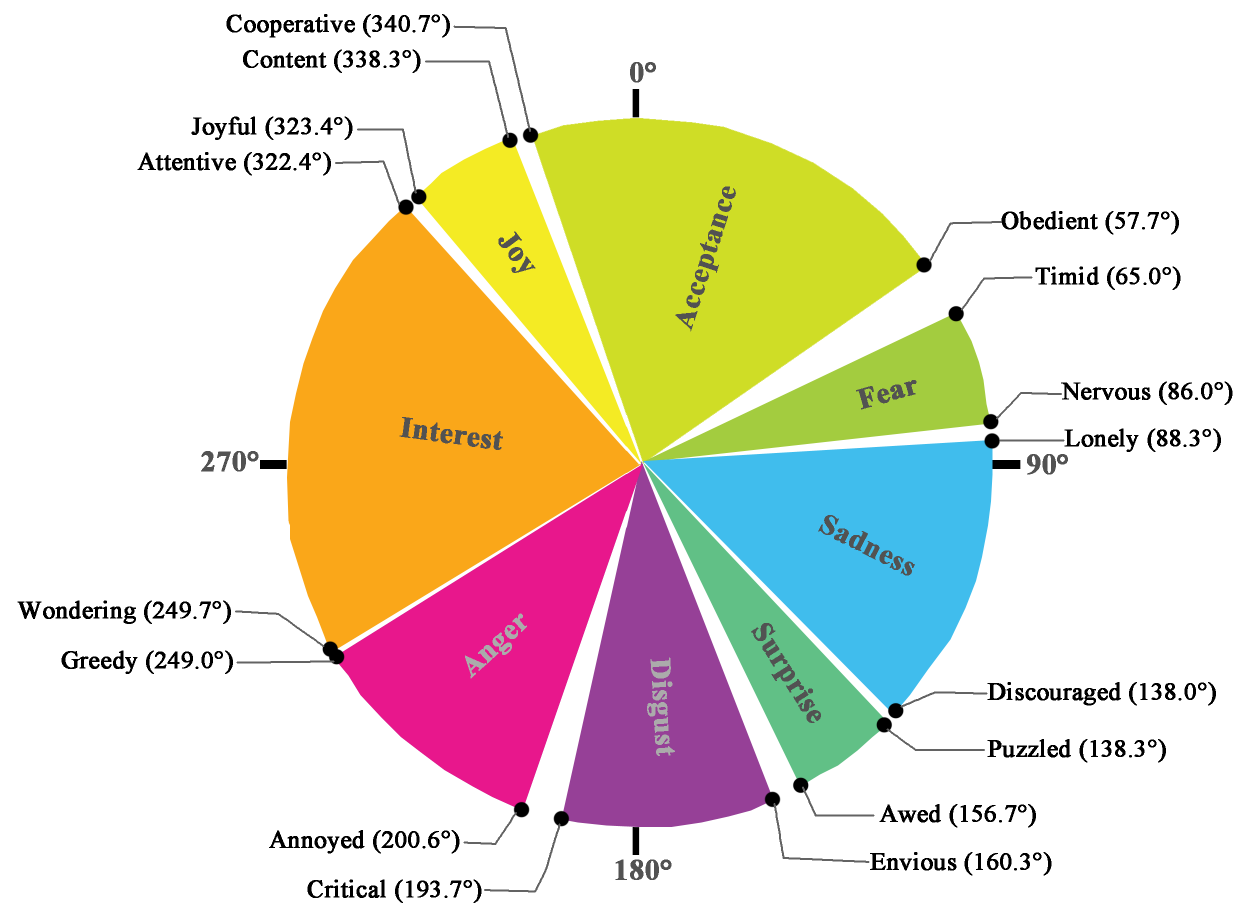}

        \caption[\progname{}-Specific Emotion Kind Boundaries
        Circumplex]{\progname{}-Specific Emotion Kind Boundaries on Plutchik's
        \ref{circumplex} based on Plutchik's Empirical Data}
        \label{fig:PESAreas}
    \end{figure}

    \begin{table}[!ht]
        \renewcommand{\arraystretch}{1.2}
        \centering
        \caption{Summary of \progname{}-defined Areas on the Plutchik
        Circumplex based on Plutchik's Empirical Data}
        \label{tab:pes-area}
        \small
        \begin{tabular}{C{0.17\linewidth} C{0.18\linewidth} C{0.15\linewidth}}
            \toprule
            \textbf{Label} & \textbf{Range (\textdegree)} & \textbf{Midpoint
                (\textdegree)} \\ \midrule

            \colourRow\textit{Acceptance} & [340.7, 57.7] & 19.20 \\

            \textit{Fear} & [65.0, 86.0] & 75.50 \\

            \colourRow \textit{Sadness} & [88.3, 138.0] & 113.15 \\

            \textit{Surprise} & [138.3, 156.7] & 147.50 \\

            \colourRow \textit{Disgust} & [160.3, 193.7] & 177.00 \\

            \textit{Anger} & [200.6, 262.0] & 231.30 \\

            \colourRow \textit{Interest} & [249.7, 322.4] & 286.05 \\

            \textit{Joy} & [323.4, 338.3] & 330.85 \\

            \midrule\bottomrule
        \end{tabular}
    \end{table}

    \begin{figure}[!b]
        \centering
        \includegraphics[width=0.6\linewidth]{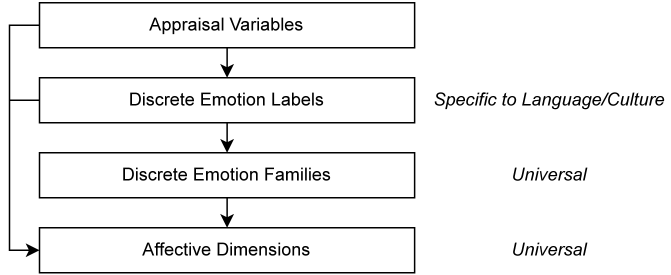}

        \caption[Mapping Data Between Perspectives]{Mapping Data Between
            Perspectives, Adapted from \citet[p.~15]{scherer2010emotion}}
        \label{fig:theoryResolution}
    \end{figure}
}

\progname{} compiled eight lists---one for each derived \ref{circumplex}
``area''---of exact or nearly exact matches of emotion terms in Plutchik and
PAD Space (Table~\ref{tab:pes-pad-terms}). \progname{} takes a single term from
each list as its PAD Space reference point for that Plutchik emotion, giving
preference to PAD terms that have statistically significant means for each
dimension, then to semantically equivalent terms. If there was no term
equivalence, \progname{} took the term closest to the midpoint of the Plutchik
\ref{circumplex} ``area''. \progname{} ignores number of ratings and standard
deviation of PAD terms for simplicity.

\afterpage{\clearpage
    {\topskip0pt
        \vspace*{\fill}
        \begin{table}[!ht]
            \renewcommand{\arraystretch}{1.2}
            \centering
            \caption{Emotion Terms in Both Plutchik and PAD Space with their
            Associated Empirical Data from Each}
            \label{tab:pes-pad-terms}
            \footnotesize
            \begin{tabular}{llc|lcccccccc}
                \toprule
                \multicolumn{3}{c|}{\textbf{Plutchik}} &
                \multicolumn{9}{c}{\textbf{PAD Space}} \\

                &  &  &  &  & & \multicolumn{2}{c}{\textbf{P}} &
                \multicolumn{2}{c}{\textbf{A}} & \multicolumn{2}{c}{\textbf{D}}
                \\

                \textbf{Area} & \textbf{Term} & \textbf{Angle} & \textbf{Term}
                & \textbf{\#} & \textbf{N} & \textbf{Mean} & \textbf{SD} &
                \textbf{Mean} & \textbf{SD} & \textbf{Mean} & \textbf{SD} \\
                \midrule

                \multirow{2}{*}{\begin{tabular}[x]{@{}l@{}}\textit{Accept-} \\
                \textit{ance} \end{tabular}} & \colourCell
                {\footnotesize\textpmhg{\Hi}}Affectionate &
                \colourCell 52.3\textdegree & \colourCell Affectionate &
                \colourCell 34 & \colourCell 29 & \colourCell 0.64* &
                \colourCell 0.26 & \colourCell 0.35* & \colourCell 0.34 &
                \colourCell 0.24* & \colourCell 0.40 \\

                & Cooperative & 340.7\textdegree & Cooperative & 43 & 31 &
                0.39* & 0.32 & 0.13* & 0.27 & 0.03 & 0.34 \\

                \midrule

                \multirow{7}{*}{\textit{Fear}} & \colourCell Anxious &
                \colourCell 78.3\textdegree & \colourCell Anxious &
                \colourCell 50 & \colourCell 28 & \colourCell 0.01 &
                \colourCell 0.45 & \colourCell 0.59* & \colourCell 0.31 &
                \colourCell -0.15 & \colourCell 0.32 \\

                & Humiliated & 84.0\textdegree & Humiliated & 99 & 27 & -0.63*
                & 0.18 & 0.43* & 0.34 & -0.38* & 0.30 \\

                & \colourCell {\footnotesize\textpmhg{\Hi}}Terrified &
                \colourCell 75.7\textdegree & \colourCell Terrified &
                \colourCell 102 & \colourCell 29 & \colourCell -0.62* &
                \colourCell 0.20 & \colourCell 0.82* & \colourCell 0.25 &
                \colourCell -0.43* & \colourCell 0.34 \\

                & Helpless & 80.0\textdegree & Helpless & 104 & 29 & -0.71* &
                0.18 & 0.42* & 0.45 & -0.51* & 0.32 \\

                & \colourCell Embarrassed & \colourCell 75.3\textdegree &
                \colourCell Embarrassed & \colourCell 110 & \colourCell 29 &
                \colourCell -0.46* & \colourCell 0.30 & \colourCell 0.54* &
                \colourCell 0.26 & \colourCell -0.24* & \colourCell 0.40 \\

                & Shy & 72.0\textdegree & Shy & 117 & 29 & -0.15 & 0.33 & 0.06
                & 0.30 & -0.34* & 0.28 \\

                & \colourCell Timid & \colourCell 65.0\textdegree &
                \colourCell Timid & \colourCell 131 & \colourCell 28 &
                \colourCell -0.15 & \colourCell 0.41 & \colourCell -0.12 &
                \colourCell 0.37 & \colourCell -0.47* & \colourCell 0.31 \\

                \midrule

                \multirow{12}{*}{\textit{Sadness}} & Gloomy & 132.7\textdegree
                & Solemn & 72 & 29 & 0.03 & 0.39 & -0.32* & 0.26 & -0.11 & 0.33
                \\

                & \colourCell \begin{tabular}[x]{@{}l@{}} Grief- \\ Stricken
                \end{tabular} & \colourCell 127.3\textdegree &
                \colourCell Anguished & \colourCell 107 &
                \colourCell 29 & \colourCell -0.50* & \colourCell 0.30 &
                \colourCell 0.08 & \colourCell 0.46 & \colourCell -0.20* &
                \colourCell 0.34 \\

                & Guilty & 102.3\textdegree & Guilty & 118 & 29 & -0.57* & 0.19
                & 0.28* & 0.38 & -0.34* & 0.28 \\

                & \colourCell Remorseful & \colourCell 123.3\textdegree &
                \colourCell Regretful & \colourCell 123 & \colourCell 30 &
                \colourCell -0.52* & \colourCell 0.24 & \colourCell 0.02 &
                \colourCell 0.32 & \colourCell -0.21* & \colourCell 0.28 \\

                & Depressed & 125.3\textdegree & Depressed & 126 & 27 & -0.72*
                & 0.21 & -0.29* & 0.44 & -0.41* & 0.28 \\

                & \colourCell Despairing & \colourCell 133.0\textdegree &
                \colourCell Despairing & \colourCell 127 & \colourCell 27 &
                \colourCell -0.72* & \colourCell 0.21 & \colourCell -0.16 &
                \colourCell 0.34 & \colourCell -0.38* & \colourCell 0.25 \\

                & Lonely & 88.3\textdegree & Lonely & 128 & 29 & -0.66* & 0.35
                & -0.43* & 0.36 & -0.32* & 0.30 \\

                & \colourCell Meek & \colourCell 91.0\textdegree &
                \colourCell Meek & \colourCell 129 & \colourCell 29 &
                \colourCell -0.19 & \colourCell 0.58 & \colourCell -0.25* &
                \colourCell 0.32 & \colourCell -0.41* & \colourCell 0.42 \\

                & Bored & 136.0\textdegree & Bored & 132 & 28 & -0.65* & 0.19 &
                -0.62* & 0.24 & -0.33* & 0.21 \\

                & \colourCell Rejected & \colourCell 136.0\textdegree &
                \colourCell Rejected & \colourCell 137 & \colourCell 29 &
                \colourCell -0.62* & \colourCell 0.24 & \colourCell -0.01 &
                \colourCell 0.38 & \colourCell -0.33* & \colourCell 0.27 \\

                & Discouraged & 138.0\textdegree & Discouraged & 150 & 30 &
                -0.61* & 0.25 & -0.15 & 0.32 & -0.29* & 0.32 \\

                & \colourCell {\footnotesize\textpmhg{\Hi}}Sad &
                \colourCell 108.5\textdegree & \colourCell Sad &
                \colourCell 151 & \colourCell 30 & \colourCell -0.63* &
                \colourCell 0.23 & \colourCell -0.27* & \colourCell 0.34 &
                \colourCell -0.33* & \colourCell 0.22 \\

                \midrule

                \multirow{4}{*}{\textit{Surprise}} & Surprised &
                146.7\textdegree & Surprised & 52 & 29 & 0.40* & 0.30 & 0.67* &
                0.27 & -0.13 & 0.38 \\

                & \colourCell Awed & \colourCell 156.7\textdegree &
                \colourCell Awed & \colourCell 56 & \colourCell 30 &
                \colourCell 0.18* & \colourCell 0.34 & \colourCell 0.40* &
                \colourCell 0.30 & \colourCell -0.38* & \colourCell 0.21 \\

                & {\footnotesize\textpmhg{\Hi}}Astonished & 148.0\textdegree &
                Astonished & 74 & 30 & 0.16* & 0.26 & 0.88* & 0.19 & -0.15* &
                0.26 \\

                & \colourCell Confused & \colourCell 141.3\textdegree &
                \colourCell Confused & \colourCell 121 & \colourCell 30 &
                \colourCell -0.53* & \colourCell 0.20 & \colourCell 0.27* &
                \colourCell 0.29 & \colourCell -0.32* & \colourCell 0.28 \\

                \midrule

                \multirow{7}{*}{\textit{Disgust}} &
                {\footnotesize\textpmhg{\Hi}}Disgusted & 161.3\textdegree &
                Disgusted & 75 & 29 & -0.60* & 0.20 & 0.35* & 0.41 & 0.11 &
                0.34 \\

                & \colourCell \begin{tabular}[x]{@{}l@{}} Contempt- \\ uous
                \end{tabular} & \colourCell 192.0\textdegree &
                \colourCell Contempt & \colourCell 85 & \colourCell 29 &
                \colourCell -0.23* & \colourCell 0.39 & \colourCell 0.31* &
                \colourCell 0.33 & \colourCell 0.18* & \colourCell 0.29 \\

                & Suspicious & 182.7\textdegree & Suspicious & 90 & 29 & -0.25*
                & 0.23 & 0.42* & 0.21 & 0.11 & 0.32 \\

                & \colourCell Distrustful & \colourCell 185.0\textdegree &
                \colourCell Skeptical & \colourCell 91 & \colourCell 29 &
                \colourCell -0.22* & \colourCell 0.28 & \colourCell 0.21* &
                \colourCell 0.25 & \colourCell 0.03 & \colourCell 0.33 \\

                & Displeased & 181.5\textdegree & Displeased & 109 & 29 &
                -0.55* & 0.21 & 0.16 & 0.34 & -0.05 & 0.41 \\

                & \colourCell Indignant & \colourCell 175.0\textdegree &
                \colourCell \begin{tabular}[x]{@{}l@{}}Quietly \\ Indignant
                \end{tabular} & \colourCell 114 & \colourCell 26 &
                \colourCell -0.28* & \colourCell 0.35 & \colourCell 0.04 &
                \colourCell 0.36 & \colourCell -0.16 & \colourCell 0.40 \\

                & Dissatisfied & 183.0\textdegree & Dissatisfied & 122 & 30 &
                -0.50* & 0.22 & 0.05 & 0.28 & 0.13 & 0.32 \\

                \midrule\bottomrule
            \end{tabular}
        \end{table}
        \vspace*{\fill}}

    \clearpage

    {\topskip0pt
        \addtocounter{table}{-1}
        \captionsetup{list=no}
        \begin{table}[!hb]
            \renewcommand{\arraystretch}{1.2}
            \centering
            \caption{\textit{(Continued)} Emotion Terms in Both Plutchik and
            PAD Space with their Associated Empirical Data from Each}
            \footnotesize
            \begin{threeparttable}
                \begin{tabular}{llc|lcccccccc}
                    \toprule
                    \multicolumn{3}{c|}{\textbf{Plutchik}} &
                    \multicolumn{9}{c}{\textbf{PAD Space}} \\

                    &  &  & & &  & \multicolumn{2}{c}{\textbf{P}} &
                    \multicolumn{2}{c}{\textbf{A}} &
                    \multicolumn{2}{c}{\textbf{D}} \\

                    \textbf{Area} & \textbf{Term} & \textbf{Angle} &
                    \textbf{Term} & \textbf{\#} & \textbf{N} & \textbf{Mean} &
                    \textbf{SD} & \textbf{Mean} & \textbf{SD} & \textbf{Mean} &
                    \textbf{SD} \\
                    \midrule

                    \multirow{8}{*}{\textit{Anger}} & \colourCell Aggressive &
                    \colourCell 232.0\textdegree & \colourCell Aggressive &
                    \colourCell 13 & \colourCell 28 & \colourCell 0.41* &
                    \colourCell 0.30 & \colourCell 0.63* & \colourCell 0.25 &
                    \colourCell 0.62* & \colourCell 0.24 \\

                    & Irritated & 202.3\textdegree & Irritated & 78 & 29 &
                    -0.58* & 0.16 & 0.40* & 0.37 & 0.01 & 0.40 \\

                    & \colourCell Defiant & \colourCell 230.7\textdegree &
                    \colourCell Defiant & \colourCell 79 & \colourCell 28 &
                    \colourCell -0.16* & \colourCell 0.30 & \colourCell 0.54* &
                    \colourCell 0.37 & \colourCell 0.32* & \colourCell 0.42 \\

                    & Hostile & 222.0\textdegree & Hostile & 81 & 29 & -0.42* &
                    0.31 & 0.53* & 0.36 & 0.30* & 0.32 \\

                    & \colourCell {\footnotesize\textpmhg{\Hi}}Angry &
                    \colourCell 212.0\textdegree & \colourCell Angry &
                    \colourCell 82 & \colourCell 29 & \colourCell -0.51* &
                    \colourCell 0.20 & \colourCell 0.59* & \colourCell 0.33 &
                    \colourCell 0.25* & \colourCell 0.39 \\

                    & Annoyed & 200.6\textdegree & Mildly Annoyed & 83 & 29 &
                    -0.28* & 0.16 & 0.17* & 0.28 & 0.04 & 0.31 \\

                    & \colourCell Furious & \colourCell 221.3\textdegree &
                    \colourCell Enraged & \colourCell 84 & \colourCell 29 &
                    \colourCell -0.44* & \colourCell 0.25 & \colourCell 0.72* &
                    \colourCell 0.29 & \colourCell 0.32* & \colourCell 0.44 \\

                    & Scornful & 227.0\textdegree & Scornful & 89 & 28 & -0.35*
                    & 0.21 & 0.35* & 0.27 & 0.29* & 0.32 \\

                    \midrule

                    \multirow{7}{*}{\textit{Interest}} &
                    \colourCell Adventurous & \colourCell 270.7\textdegree &
                    \colourCell Bold & \colourCell 1 & \colourCell 27 &
                    \colourCell 0.44* & \colourCell 0.32 & \colourCell 0.61* &
                    \colourCell 0.24 & \colourCell 0.66* & \colourCell 0.30 \\

                    & Proud & 262.0\textdegree & Proud & 7 & 29 & 0.77* & 0.21
                    & 0.38* & 0.34 & 0.65* & 0.33 \\

                    & \colourCell {\footnotesize\textpmhg{\Hi}}Interested &
                    \colourCell 315.7\textdegree & \colourCell Interested &
                    \colourCell 8 & \colourCell 29 & \colourCell 0.64* &
                    \colourCell 0.20 & \colourCell 0.51* & \colourCell 0.21 &
                    \colourCell 0.17 & \colourCell 0.40 \\

                    & Elated & 311.0\textdegree & Elated & 17 & 28 & 0.50* &
                    0.47 & 0.42* & 0.14 & 0.23* & 0.36 \\

                    & \colourCell Hopeful & \colourCell 298.0\textdegree &
                    \colourCell Hopeful & \colourCell 18 & \colourCell 29 &
                    \colourCell 0.51* & \colourCell 0.30 & \colourCell 0.23* &
                    \colourCell 0.33 & \colourCell 0.14 & \colourCell 0.41 \\

                    & Wondering & 249.7\textdegree & Wonder & 54 & 30 & 0.27* &
                    0.37 & 0.24* & 0.35 & -0.17* & 0.26 \\

                    & \colourCell Curious & \colourCell 261.0\textdegree &
                    \colourCell Curious & \colourCell 58 & \colourCell 28 &
                    \colourCell 0.22* & \colourCell 0.30 & \colourCell 0.62* &
                    \colourCell 0.20 & \colourCell -0.01 & \colourCell 0.34 \\

                    \midrule

                    \multirow{2}{*}{\textit{Joy}}&
                    {\footnotesize\textpmhg{\Hi}}Joyful & 323.4\textdegree &
                    Joyful & 20 & 29 & 0.76* & 0.22 & 0.48* & 0.26 & 0.35* &
                    0.31 \\

                    & \colourCell Happy & \colourCell 323.7\textdegree &
                    \colourCell Happy & \colourCell 31 & \colourCell 29 &
                    \colourCell 0.81* & \colourCell 0.21 & \colourCell 0.51* &
                    \colourCell 0.26 & \colourCell 0.46* & \colourCell 0.38 \\

                    \hline\bottomrule
                \end{tabular}
                \begin{tablenotes}

                    \footnotesize
                    \vspace*{2mm}

                    \item N \textit{Number of Ratings}

                    \item SD \textit{Standard Deviation}

                    \item {\normalsize*} \textit{Statistically Significant ($p
                    > 0.01$)}

                    \item {\small\textpmhg{\Hi}} \textit{Chosen term}

                \end{tablenotes}
            \end{threeparttable}%
        \end{table}
    }
    \captionsetup{list=yes}
}

\subsection{Converting \progname{} Emotion States into PAD Space Coordinates}
\progname{} uses the chosen reference points to translate an emotion state
(Equation~\ref{eq:emotionstatetype}) into a PAD point
(Equation~\ref{eq:padpoint}) by finding the PAD Space point for each individual
emotion kind (Equation~\ref{eq:kindstype}) in the state. The model evaluates
these points such that an emotion kind with zero intensity has the coordinates
$(0, 0, 0)$---the neutral PAD value~\citep[p.~40]{mehrabian1980basic}---and one
at maximum intensity has the same value as the corresponding reference point.
It then sums the individual points into an overall PAD point, clamping it to
$[-1, 1]$,  because \progname{} assumes that an entity can only occupy one
point in PAD Space at any given time:
\begin{equation}\label{eq:convert2PAD}
    \begin{gathered}
        \mathtt{ConvertStateToPADPnt}(es : \emotionstatetype): \padpoint \\
        \defEq \mathtt{clamp}\left(0.1 \cdot \log_2 \left( {\mathlarger\sum_{k
        \in \emotionkindstype}} 2^{10 \cdot v\left(k\right) \cdot I_k }
        \right), -1, 1 \right) \\
        \text{where } I_k =
        \dfrac{es.\mathtt{intensities}(k)}{es.\mathtt{max}(k)} \\
         \text{and } v\left(k : \emotionkindstype\right) : \padpoint =
         \begin{cases}
            \left( -0.62, +0.82, -0.43 \right), & k = \mFear \\
            \left( -0.51, +0.59, +0.25 \right), & k = \mAnger \\
            \left( -0.63, -0.27, -0.33 \right), & k = \mSadness \\
            \left( +0.76, +0.48, +0.35 \right), & k = \mJoy \\
            \left( +0.64, +0.51, +0.17 \right), & k = \mInterest \\
            \left( +0.16, +0.88, -0.15 \right), & k = \mSurprise \\
            \left( -0.60, +0.35, +0.11 \right), & k = \mDisgust \\
            \left( +0.64, +0.35, +0.24 \right), & k = \mTrust \\
        \end{cases}
    \end{gathered}
\end{equation}
\progname{} bases the function inside $\mathtt{clamp}$ the one from
Em/Oz~\citep{reilly2006modelling} and GAMYGDALA
\citep[p.~38]{popescu2014gamygdala}. It relies on a logarithm so that it is not
strictly additive and all values contribute to the output such that its
magnitude is at least as much as the highest input. Although not experimentally
verified, it emulates these desired behaviours and reportedly works well.

This model does lose information about the converted emotion state because it
is combining information from eight discrete categories into one point in a
three-dimensional space (\citepg{schaap2008towards}{172};
\citepg{broekens2021emotion}{353}). After conversion, it is nearly impossible
to determine which emotion kind-intensity combinations contributed to the
point's generation. Should a user need this information, they must associate
the state with the point manually.

\section{Documenting \progname{}'s Models}\label{sec:docSRS}
\progname{} adheres to Document Driven Design (DDD), progressively documenting
requirements, design, implementation, and
testing~\citep[p.~41--42]{smith2016advantages}. While it is an adaptation of the
waterfall development model, this does not mean that the execution of the
process follows it~\citep[p.~1157]{smith2009document}. It depends on
traceability between the requirements and all other products of the design and
testing process to account for changes at any point in the process, ensuring
that the final documentation ``will be rational and
accurate''~\citep[p.~256]{parnas1986rational}.

A Software Requirements Specification (SRS) documents the context of
\progname{}'s design and acts as a benchmark to compare the final product
against~\citep[p.~1157]{smith2009document}. \progname{} uses a variation of the
SRS template proposed by \citet{SmithAndLai2005} and
\citet{SmithEtAl2007}---demonstrated in \citet{smith2009document}---because it
accounts for formal models that designers can realize differently based on what
assumptions they make. This is relevant to \progname{}, which must translate
informal emotion theories into formal models which necessarily requires
assumptions. This template also encourages references to other CMEs, so that
\progname{} can use their solutions and avoid past
mistakes~\citep[p.~59]{parnas1996why}. \progname{}'s SRS\footnote{See
\progname{}'s SRS at
\href{https://github.com/GenevaS/EMgine/blob/main/docs/SRS/EMgine_SRS.pdf}{https://github.com/GenevaS/EMgine/blob/main/docs/SRS/EMgine\_SRS.pdf}.}
 documents four kinds of models:
\begin{enumerate}
    \item \textit{Conceptual Models}, which \progname{} adds, describing the
    emotion theories as they are in the literature and how this designer
    understands them using natural language, offering transparency and
    maintaining a direct connection to the primary literature;

    \item \textit{Theoretical Models} begin to refine the Conceptual Models
    using natural language and explicit assumptions to improve their precision,
    reducing ambiguities about the connection between primary sources and the
    formal Data Types and Instance Models;

    \item \textit{Data Types} to define the formal structures that \progname{}
    needs to realize Instance Models; and

    \item \textit{Instance Models} describe the formalization of Theoretical
    Models using type definitions and additional assumptions as needed.
\end{enumerate}

The SRS also documents critical contextual information including constraints,
functional and nonfunctional requirements, and anticipated changes.

\section{Summary}
\progname{}'s design relies heavily on data types, allowing it to represent
abstract concepts such as ``the world'' and ``emotion intensity''. While
\progname{} defines most data types, it leaves some as API specifications so
that it does not constrain users to implementations that might not be ideal for
their game. These data types allow the definition of ``interfaces'' between
models of different theories (i.e. Plutchik, Oatley \& Johnson-Laird, and PAD
Space) and models derived from other domains to represent underdeveloped
concepts such as emotion decay. The contents of each model and \progname{}'s
high-level requirements, documented in an SRS, then guide its design and
implementation stages.

\clearpage
\vspace*{\fill}
\begin{keypoints}
    \begin{itemize}

        \item Plutchik's emotion kind definitions serve as the baseline for
        finding ``interfaces'' between Oatley \& Johnson-Laird and PAD Space
        when specifying \progname{}'s models

        \item \progname{} primarily represents conceptual information embedded
        in affective theories, such as goals and emotion intensity, with data
        types

        \item The translation of affective theory and model concepts into
        \progname{}'s models move through an intermediary stage where the
        natural language descriptions are ``rewritten'' with more formal
        natural language to clarify their relationship and necessary assumptions

        \item \progname{} has separate elicitation and intensity evaluations
        for each emotion kind to afford more control over when and where
        emotion generation happens

        \item \progname{} bases its model of emotion decay on the behaviour of
        damped harmonic oscillators from an intuitive understanding of how
        emotions ``fade over time'', which affords the creation of different
        decay ``behaviours'' by varying the damping ration $\zeta$

        \item To ``map'' discrete Plutchik emotions into the PAD dimensional
        space, \progname{} defined reference points by comparing lists of
        affective terms used in empirical studies of that theory and model to
        select representative terms from each that have comparable semantic
        meanings

        \item \progname{} derives the PAD point ``equivalent'' of an emotion
        state by summing the ``equivalent'' PAD points for each emotion kind in
        the state, where emotion intensity in the state ``scales'' the
        associated reference PAD point

        \item \progname{} bases its software requirements documentation on
        templates proposed by \citet{SmithAndLai2005} and \citet{SmithEtAl2007}
        due to its focus on documenting assumptions that progressively refine
        mathematical models

    \end{itemize}
\end{keypoints}

\parasep
\vspace*{\fill}

%% file: designAndImplement.tex
\chapter{Build Your \progname{}: Some Assembly
    Required}\label{chapter:designImplement}
\def\epigraphflush{center}
\setlength{\epigraphwidth}{0.85\textwidth}
\def\textflush{center}
\epigraph{Hmm. Art?}{Giant, \textit{The Iron Giant}}

It is easier to integrate \progname{}'s high-level design goals into its
architecture design (Section~\ref{sec:emgineArch}) and subsequent
implementation efforts (Section~\ref{sec:emgineImplement}) after specifying its
models (Chapter~\ref{chapter:equations}) because they describe how many and
what types of components \progname{} might need. Although its development looks
like it followed the waterfall method, \progname{} treated this development
model more like guidelines than strict rules---there were many instances of
backtracking to and concurrent work between stages.

\section{Specifying \progname{}'s Architecture}\label{sec:emgineArch}
\progname{} cannot just provide a list of data types and functions---it must
also describe how it groups them into units and how they communicate. A
software's \textit{architecture} describes these aspects. Generally,
nonfunctional requirements guide the architecture style\footnote{They are also
called software architecture patterns.} selection because each style promotes
different sets of competing software qualities~\citep[p.~9]{qian2010software}.
\progname{} included the user needs (Chapter~\ref{sec:userReqs}) in its
nonfunctional requirements, which it uses to inform its architecture style. In
turn, the style aids module decomposition and the definition of their
relationships. This process lead \progname{} to a \textit{library of
components} design that comes packaged with a default \textit{``engine''} that
is itself a system of components. A library-based approach also alleviates some
of \progname{}'s design pressure, as it no longer has to ``...be generic enough
to encompass all possible forms of a perception-action cycle in an
agent''~\citep[p.~8:12]{mascarenhas2022fatima}.

\subsection{Requirements to Architecture Style}
The \textit{Independence from an agent architecture} (\ref{flexArch}),
\textit{Hiding the complexity of emotion generation} (\ref{easeHide}), and
\textit{Providing a clear API} (\ref{easeAPI}) requirements suggest that
\progname{}'s processes should be a black-box. However, it would be difficult
to support \textit{Allowing \progname{}'s outputs to be traceable and
understandable} (\ref{easeTrace}) if one could not see how \progname{}'s parts
passed information to each other so that its overall behaviour can be
explained~\citep[p.~20]{guimaraes2022fatima}. Users do not need to know
\textit{how} \progname{} decays emotion but they do need to know where the
inputs are from and where the outputs are going. Therefore, \progname{}'s
components should be black-boxes, but not their connections. This suggests that
a component-based software architecture~\citep[p.~248--261]{qian2010software}
is best, where each of \progname{}'s ``tasks'' is a discrete component that
users can include, exclude, and change as needed. A component-based design
approach is common in CME development~\citep[p.~141]{osuna2021toward}. For
\progname{}, this approach would:
\begin{itemize}

    \item Increase its portability by specifying what each component is
    guaranteed to provide so that designers can use existing validated systems
    and integrate new components more easily and
    quickly~\citep[p.~443]{rodriguez2015computational}, supporting
    \ref{flexArch} for both agent architectures and game engines

    \item Mandate well-defined interfaces to show what each component requires
    and provides, including its configuration parameters, supporting
    \textit{Allowing the customization or redefinition of \progname{}'s
        preexisting configuration parameters} (\ref{flexCustom}),
        \textit{Allowing
        developers to choose which kinds of emotion \progname{}
        produces} (\ref{flexEm}), \textit{Allowing developers to specify how to
        use
        \progname{}'s outputs} (\ref{flexOut}), \textit{Providing a clear API}
    (\ref{easeAPI}), \textit{Allowing \progname{}'s outputs to be traceable and
        understandable} (\ref{easeTrace}), and \textit{Allowing designers the
        option to automate the storing and decaying of \progname{}'s emotion
        state}
    (\ref{easeAuto})

    \item Allow designers to call \progname{}'s components as needed,
    supporting \textit{Allowing the game designer to choose which of
        \progname{}'s tasks to use} (\ref{flexTasks})
    and---potentially---\textit{Showing that \progname{} improves the player
        experience} (\ref{easePX}) and \textit{Providing examples as to how
        \progname{} can create novel game experiences} (\ref{easeNovel}) if
        done in
    a unique way

    \item Allow designers to add or swap \progname{}'s components, supporting
    \textit{Allow the integration of new components} (\ref{flexNew}),
    \textit{Allowing developers to choose which kinds of emotion \progname{}
        produces} (\ref{flexEm}), \textit{Ability to operate on different
        levels of
        NPC complexity} (\ref{flexComplex}), and \textit{Be efficient and
        scalable}
    (\ref{flexScale})~\citep[p.~466]{carbone2020radically}

    \item Have the potential for developing authoring tools that interface with
    and manage \progname{} components, supporting \textit{Minimizing authorial
        burden} (\ref{easeAuthor})

\end{itemize}

Ongoing work on the FAtiMA architecture and its descendants, the FAtiMA Modular
framework and FAtiMA Toolkit, found this approach
successful~\citep[p.~8:2, 8:12--8:13]{mascarenhas2022fatima}. After gaining
feedback from members of the games industry\footnote{As part of the ``Realising
    an Applied Gaming Eco-system'' (RAGE) project
    (\href{https://cordis.europa.eu/project/id/644187}{https://cordis.europa.eu/projec
        t/id/644187}).}, the FAtiMA Toolkit's designers realized it as a
        library of
components so that its parts can work autonomously. This also increased the
Toolkit's chances of adoption, as it allows game designers to use it in their
existing systems and/or frameworks while avoiding the complexity and
accessibility issues of other agent
architectures~\citep[p.~3]{guimaraes2022fatima}. This also lets users focus on
what emotion processing sequence works for their needs rather than constraining
them to ``\progname{}'s process'' since no one really knows what order emotion
processes truly run in~\citep[p.~142]{moffat1997personality}.

While each of \progname{}'s components are
black-boxes~\citep[p.~253]{qian2010software}, a game designer might not know
when they should use a component or if it is necessary. This could jeopardize
\textit{Hiding the Complexity of Emotion Generation} (\ref{easeHide}). A
component-based software architecture supports this need too by allowing the
specification of a prebuilt component that is itself a ``system'' of
components~\citep[p.~249]{qian2010software} that minimizes the necessary
decisions and inputs needed for emotion processing. It would accept data and
return an emotion state without the designer knowing how it
works~\citep[p.~443]{rodriguez2015computational}.

This prebuilt component or ``engine'' is similar to GAMYGDALA, which compares
itself to a physics engine~\citep[p.~32]{popescu2014gamygdala}. Due to its
plugin nature, designers have successfully applied GAMYGDALA to: arcade and
puzzle games~\citep{broekens2015emotion}; a narrative generation framework to
drive character emotions~\citep{kaptein2015affective}; and implement affective
decision-making in fighting game characters~\citep{yuda2019creating}. A
developer has also successfully integrated the fuzzy logic, classical
conditioning, and learning from another CME\footnote{FLAME (\ref{flame}),
    included in the CME survey (Chapter~\ref{chapter:cmeOverview})} with
GAMYGDALA~\citep[p.~4]{code2015learning}. The breadth of games that developers
have applied GAMYGDALA to and the potential for extensibility suggests that
\progname{}'s inclusion of a large, prefabricated component could further
increase its chances for success.

\subsection{Turning Models into Modules}\label{sec:moduleDecomp}
\progname{} creates the building blocks for its architecture by decomposing its
requirements specification into software elements or \textit{modules} using the
\textit{information hiding} principle~\citep{Parnas1972a}. The goal is to
isolate each of \progname{}'s parts that are likely to change into one module,
supporting design for change, incremental development, and the separation of
concerns. Defining a module's ``likely change'' also promotes highly cohesive
internal elements because it often indicates the module's goal or concern that
they work towards. Due to the \textit{Allowing the game designer to choose
which of \progname{}'s tasks to use} (\ref{flexTasks}) and \textit{Allowing the
customization or redefinition of \progname{}'s preexisting configuration
parameters} (\ref{flexCustom}) requirements, \progname{} conceptualizes a
potential user task as a module's goal to work towards the seamless addition
and removal of those tasks. The hidden ``likely changes'' describe the internal
representations of data types, functions, and the relationships between them,
which also moves \progname{} towards satisfying \textit{Hiding the complexity
    of emotion generation} (\ref{easeHide}). To this end, \progname{} identifies
seven logical units:
\begin{itemize}

    \item Emotion Intensity, containing the $\emotionintensitytype$ and
    $\responsestrength$ data types (Equations~\ref{eq:intensitytype} and
    \ref{eq:intensitychangetype}), a function for ``combining'' an emotion
    intensity with an intensity change (Equation~\ref{eq:combineintensity}), and
    functions for evaluating emotion intensity
    (Chapters~\ref{sec:CTE-Intensity}, \ref{sec:Acceptance-Intensity},
    \ref{sec:Interest-Intensity}, and \ref{sec:Surprise-Intensity}).
    \progname{} subdivides the unit into Emotion Intensity Type and Emotion
    Intensity Function modules because at least five other units rely on
    $\emotionintensitytype$ and the functions for evaluating emotion intensity
    are very likely to change.

    \item Emotion State, containing the $\emotionkindstype$ and
    $\emotionstatetype$ data types (Equations~\ref{eq:kindstype} and
    \ref{eq:emotionstatetype}), functions for manipulating an Emotion State
    (Equation~\ref{eq:updatestate}), and emotion generation
    (Chapter~\ref{sec:emotiongenerationfunctions}). \progname{} subdivides this
    unit into Emotion State Type and Emotion Generation modules to isolate the
    Plutchik-specified state from the Oatley \& Johnson-Laird-specific
    generation functions.

    \item Emotion Decay, containing the $\emotiondecaytype$ and
    $\emotionstatedecaytype$ data types (Equations~\ref{eq:decayratetype} and
    \ref{eq:emotiondecaystatetype}) and functions for evaluating decayed
    emotion intensity and state (Equations~\ref{eq:decayIntensity} and
    \ref{eq:decayastate}). \progname{} subdivides this unit into Emotion
    Intensity Decay and Emotion State Decay to separate the tasks of decaying a
    single emotion intensity and decaying all intensities in an emotion state.

    \item PAD, containing the $\padpoint$ data type
    (Equation~\ref{eq:padpoint}) and function for converting and emotion state
    to a PAD point (Equation~\ref{eq:convert2PAD}). \progname{} subdivides the
    unit into PAD Type and PAD Function modules because the definition of
    $\padpoint$ is highly unlikely to change whereas the conversion function is
    likely to change.

    \item Emotion, containing the $\emotiontype$ data type
    (Equation~\ref{eq:emotiontype}), functions for manipulating it
    (Equations~\ref{eq:updateemotion} and \ref{eq:getemotion}), and the
    function for generating a decayed emotion state from a previous one stored
    in $\emotiontype$ (Equation~\ref{eq:autodecay}). \progname{} subdivides the
    unit into Emotion Type and Emotion Function modules because the functions
    are for ease-of-use, which a user might not need because they defined their
    own or they are not relevant to the task.

    \item Entity, containing the $\goaltype$, $\plantype$,
    $\socialattachmenttype$, and $\attentiontype$ data types
    (Equations~\ref{eq:goaltype}, \ref{eq:plantype}, \ref{eq:socialtype}, and
    \ref{eq:attentiontype}). \progname{} subdivides these into Goal, Plan,
    Social Attachment, and Attention modules. Only the Goal module is mandatory
    and there are no dependencies between the Plan, Attention and Social
    Attachment modules, so separating them into dedicated modules allows users
    to add and remove them as needed.

    \item World, containing the function signatures that users must implement
    for $\timetype$, $\deltatimetype$, $\worldstatetype$,
    $\worldstatechangetype$, $\statedistancetype$, and
    $\statedistancechangetype$ (Equations~\ref{eq:timetype},
    \ref{eq:worldstatetype}, \ref{eq:worldeventtype}, \ref{eq:statedisttype},
    and \ref{eq:statedistchgtype}). \progname{} subdivides these into Time and
    World State modules because the World State is only necessary for emotion
    generation and evaluating emotion intensity, whereas Time is essential to
    modules that are independent of those components (e.g. Emotion Intensity
    Decay).

\end{itemize}

\begin{figure}[!tb]
    \centering
    \includegraphics[width=\linewidth]{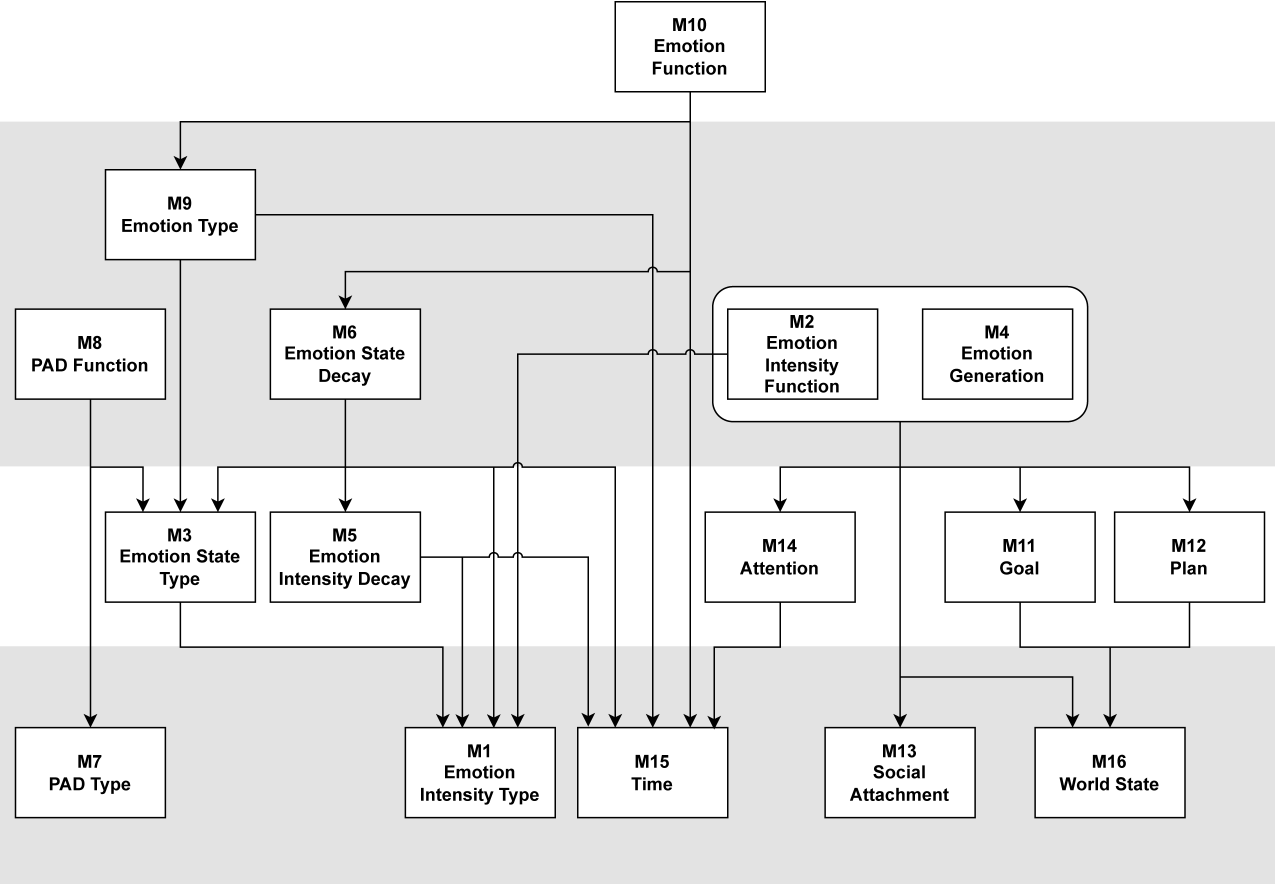}
    \caption{Module Dependency Hierarchy}
    \label{fig:moduleDependencies}
\end{figure}

\progname{} further organizes these units into two groups: behaviour-hiding
modules, which \progname{} specifies in its requirements and is responsible for
implementing; and software decision modules, which \progname{} does not specify
in its requirements. \progname{} recognizes the World logical unit and its
modules---Time and World State---as software decision modules because, although
it associates them with requirements, ultimately the user specifies these
modules. This gives \progname{} a total of three levels of decomposition and 16
modules to specify interfaces for and implement
(Table~\ref{tab:modulehierarchy}) of which only Time, World State, Emotion
Intensity Type, and Emotion State Type are highly coupled
(Figure~\ref{fig:moduleDependencies}).

\begin{table}[!tb]
    \centering
    \small
    \renewcommand{\arraystretch}{1.2}
    \begin{tabular}{P{0.16\linewidth}P{0.31\linewidth}P{0.45\linewidth}}
        \toprule
        \textbf{Level 1} & \textbf{Level 2} & \textbf{Level 3} \\
        \midrule

        \multirow{14}{\linewidth}{Behaviour\-/Hiding Module} & \colourCell &
        \colourCell \textbf{M1} Emotion Intensity Type Module \\
        & \colourCell \multirow{-2}{\linewidth}{Emotion Intensity Module} &
        \colourCell \textbf{M2} Emotion Intensity Function Module \\

        &  & \textbf{M3} Emotion State Type Module \\
        & \multirow{-2}{\linewidth}{Emotion State Module} & \textbf{M4} Emotion
        Generation Module \\

        & \colourCell & \colourCell \textbf{M5} Emotion Intensity Decay Module
        \\
        & \colourCell \multirow{-2}{\linewidth}{Emotion Decay Module} &
        \colourCell \textbf{M6} Emotion State Decay Module \\

        &  &  \textbf{M7} PAD Type Module \\
        & \multirow{-2}{\linewidth}{PAD Module} & \textbf{M8} PAD Function
        Module \\

        & \colourCell & \colourCell \textbf{M9} Emotion Type Module \\
        & \colourCell \multirow{-2}{\linewidth}{Emotion Module} & \colourCell
        \textbf{M10} Emotion Function Module \\

        &  & \textbf{M11} Goal Module \\

        &  & \textbf{M12} Plan Module \\

        &  & \textbf{M13} Attention Module \\

        & \multirow{-4}{\linewidth}{Entity Module} & \textbf{M14} Social
        Attachment Module \\

        \midrule

        \multirow{2}{\linewidth}{Software Decision Module} & \colourCell &
        \colourCell \textbf{M15} Time Module \\

        & \colourCell \multirow{-2}{\linewidth}{World Module} & \colourCell
        \textbf{M16} World State Module \\

        \bottomrule
    \end{tabular}
    \caption[\progname{}'s Module Hierarchy]{\progname{}'s Module Hierarchy
        (Implements numbered modules)}
    \label{tab:modulehierarchy}
\end{table}

\subsubsection{Module Decomposition Known Issues}
Although this modularization makes it easy to add and remove \progname{}'s
tasks, it does have points where it ``leaks'' information that should be
hidden. These are a byproduct of modularization and it is worth noting that it
is extremely difficult to remove them all. There are some situations where the
``leak'' is necessary because it supports other attributes of the
modularization. Therefore, the goal is to reduce ``leaks'' as much as possible.

A non-exhaustive list of known information ``leaks'' highlights opportunities
for future re-modularization efforts. Notably, there are no ``leaks'' in the
World Module (\textbf{M15} and \textbf{M16}) because they lack implementation
details by design.

\paragraph{Emotion Intensity Module} The most concerning information leak is in
the Emotion Intensity Type Module (\textbf{M1}), where it is apparent that it
manipulates its internal values arithmetically. While other modules need to
access those values to perform arithmetic operations, it implies that such
operations are meaningful outside of this context (e.g. adding two
\textit{values} from the emotion intensity module is not equivalent to adding
two \textit{intensities}). Due to the number of modules that depend on
$\emotionintensitytype$ and $\responsestrength$, there is no obvious module
decomposition that seals the leak while retaining the ease of adding/removing
\progname{} tasks.

There are no significant information ``leaks'' in the Emotion Intensity
Function Module (\textbf{M2}). While it is apparent that some functions depend
on particular information (e.g. \textit{Acceptance} requires information about
$\socialattachmenttype$), these follow the emotion kind's definitions. Since
this matches the conceptual model of these kinds (e.g. an attachment to
something must exist to experience \textit{Acceptance} towards it), this is not
an information ``leak''.

\paragraph{Emotion State Module} There is potential for information ``leakage''
in the Emotion Generation Module (\textbf{M4}) because it uses its outputs in
the internal evaluation. However, users cannot simply ``generate'' an emotion
if they manually calculate each output because \progname{} hides \textit{how}
it interprets them. Therefore, it is not a true information ``leak''.

There is a small ``leak'' in the Emotion State Type Module (\textbf{M3}) such
that the PAD Function Module (\textbf{M8}) knows the structure of
$\emotionkindstype$. An obvious solution is to include the PAD functions in
\textbf{M3}, but this would no longer be a clear separation of tasks. After
this, the riskiest operation in \textbf{M3} is combining an $\responsestrength$
with an $\emotionintensitytype \in \emotionstatetype$. However,
\progname{}-defined data types handle this such that users cannot manipulate
the state directly. This prevents external, ``illegal'' operations on an
emotion state and only allows internal data manipulations.

\paragraph{Emotion Decay Module} Revealing that the Emotion Intensity Decay
Module (\textbf{M5}) models a damped harmonic oscillator is a potential
information ``leak''. However, giving this information to users improves
usability because they can leverage existing tools to create their desired
emotion decay profile. Since this module isolates this information, it is not a
true ``leak''.

There are no known information ``leaks'' in the Emotion State Decay Module
(\textbf{M6}) because it is simply a container for decay objects so that users
have a convenient way to simultaneously decay all intensities in an emotion
state.

\paragraph{PAD Module} An information ``leak'' in the PAD Type Module
(\textbf{M7}) is the scaling function, which takes a $\padpoint$ and scales it
by a real value $v \in \mathbb{R}$. The intended use is to ``scale'' a
$\padpoint$ with normalized values so that it stays in PAD Space. However, the
``scaling'' function implies that users can use it for other PAD Point scaling
needs, such as animation parameters. As with the Emotion Intensity Type Module
(\textbf{M1}), this operation is not meaningful outside of this context but its
definition implies that it is. Sealing this ``leak'' might be a simple matter
of renaming the function, but must be mentioned should it cascade into further
issues.

There are no known information ``leaks'' in the PAD Function Module
(\textbf{M8}).

\paragraph{Emotion Module} There are no known information ``leaks'' in either
the Emotion Type (\textbf{M9}) or Emotion Function (\textbf{M10}) modules,
likely because they are mainly for tracking and automatically decaying emotion
state data.

\paragraph{Entity Module} Social Attachment Module (\textbf{M14}) has a
potential information ``leak'' because it exposes its underlying representation
of attachment levels. Since the representation is numeric, it implies that
arithmetic operations on them are meaningful outside the module. As with the
Emotion Intensity Type (\textbf{M1}) and PAD Type (\textbf{M7}) Modules, this
is not the case and users might do so without realizing it. Sealing this
``leak'' might simply require a reconsideration of its interface, but
alternative decomposition approaches should not discount it.

There are no known ``leaks'' in either the Goal (\textbf{M11}) or Plan
(\textbf{M12}) modules, successfully hiding how they query and manipulate World
Module data. There are no known ``leaks'' in the Attention Module
(\textbf{M13}), which hides how it tracks discrete time steps.

\subsection{Turning Modules into Components}
Module decomposition reduced \progname{}'s models into atomic units contained
in unimplemented ``virtual'' modules (Levels 1 and 2 in
Table~\ref{tab:modulehierarchy}) but this is not necessarily how a user would
visualize their groupings. For example, the hierarchy does not group the
Emotion Generation and Emotion Intensity Function modules together despite
their common dependencies (Figure~\ref{fig:moduleDependencies}). A user might
view them as a single unit because emotion intensity is only relevant if the
associated emotion is present. Therefore, \progname{}'s assignment of the 16
modules to eight components aims to collect highly-related functionality into a
comprehensive unit (Figure~\ref{fig:components}), exchangeable for another with
comparable abilities while reducing inter-component connections
(Figure~\ref{fig:componentConnections}):
\afterpage{
    \begin{figure}[!tb]
        \centering
        \includegraphics[width=\linewidth]{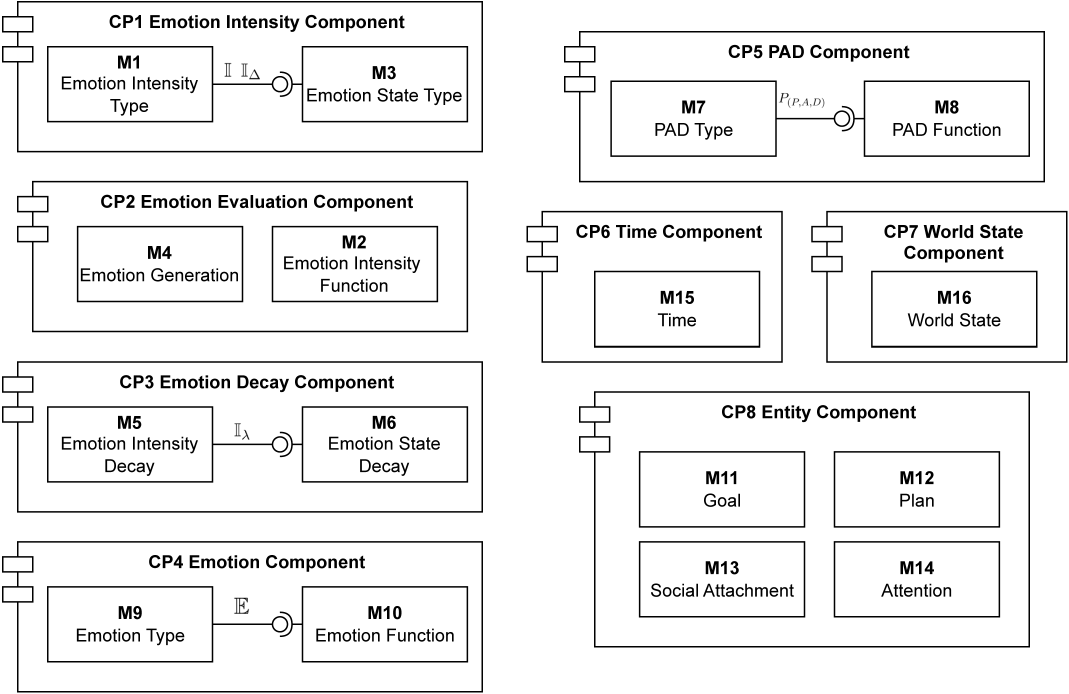}
        \caption[Organization of Modules in Components]{Organization of Modules
        in Components (``Ball'' provides functionality that a ``Cup'' needs)}
        \label{fig:components}
    \end{figure}
}

\afterpage{
    \begin{figure}[!tb]
        \centering
        \includegraphics[width=0.7\linewidth]{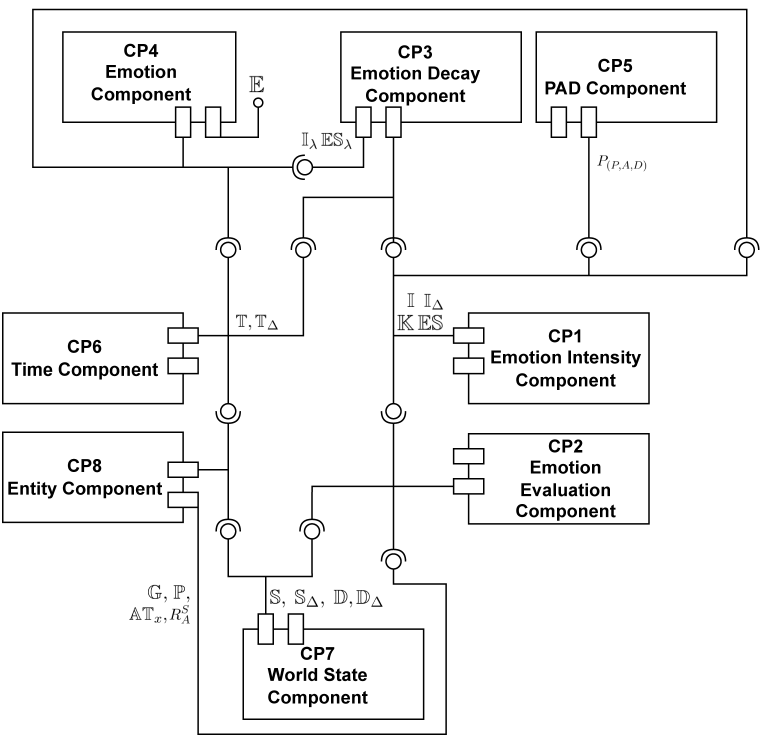}
        \caption[Relationships Between Modules]{Relationships Between Modules
        (``Ball'' provides functionality that a ``Cup'' needs)}
        \label{fig:componentConnections}
    \end{figure}
}
\begin{enumerate}[label=\textbf{C\arabic*}]

    \item Emotion Intensity collects the Emotion Intensity Type (\textbf{M1})
    and dependant Emotion State Type (\textbf{M3}) modules because they
    represent \progname{}'s core emotion types. This also collects all
    \progname{}-specific models necessary for users to define custom emotion
    kinds (Chapter~\ref{sec:userEmotions}). Since Emotion Intensity Type does
    not depend on other modules, this only reduces visible dependencies by one
    as the component hides Emotion State Type's dependency on Emotion Intensity
    Type.\vspace*{2mm}

    \item Emotion Evaluation collects the Emotion Generation (\textbf{M4}) and
    Emotion Intensity Function (\textbf{M2}) modules due to the previously
    described interdependence of emotion generation and intensity evaluations.
    Since these two modules require all the same inputs, grouping them as one
    unit also halves the visible dependencies on other modules.

    \item Emotion Decay collects the Emotion Intensity Decay (\textbf{M5}) and
    Emotion State Decay (\textbf{M6}) modules because they capture the ``decay
    emotion'' task. This hides Emotion State Decay's dependency on Emotion
    Intensity Decay while also halving the visible dependencies on other
    modules.

    \item Emotion collects the Emotion Type (\textbf{M9}) and Emotion Function
    (\textbf{M10}) modules because \progname{} intends the latter to be helper
    functions that make Emotion Type easier to use, effectively making Emotion
    Function wholly dependant on Emotion Type. This hides Emotion Function's
    dependency on Emotion Type while also halving the visible dependencies on
    other modules.\vspace*{1mm}

    \item PAD collects the PAD Type (\textbf{M7}) and PAD Function
    (\textbf{M8}) modules because \progname{} only requires a PAD data type to
    contain an emotion state that the PAD Function module has transformed into
    a PAD point. Since PAD Type does not depend on other modules, this only
    reduces visible dependencies by one as the component hides PAD Function's
    dependency on PAD Type.

    \item Time contains only the Time (\textbf{M15}) module. \progname{}
    separated it from the World State module (\textbf{M16}) because there are
    significantly more dependencies on Time than World State, including ones
    that are otherwise independent of the world such as Emotion Decay
    (\textbf{M5} and \textbf{M6}). Time does not depend on other modules, so
    there are no visible dependency reductions.

    \item World State contains only the World State (\textbf{M16}) module due
    to the need for Time to be its own component. World State does not depend
    on other modules, so there are no visible dependency reductions.

    \item Entity contains the Goal (\textbf{M11}), Plan (\textbf{M12}),
    Attention (\textbf{M13}), and Social Attachment (\textbf{M14}) modules
    because of their common ``entity representation'' task. Unlike Time
    (\textbf{M15}) and World State (\textbf{M16}), there is no need to separate
    them into different components due to a higher number of dependencies on
    one module and not the others. The Emotion Evaluation component
    (\textbf{C2}) relies on all these modules, so collecting the Entity modules
    together is convenient. There are no dependencies between Goal, Plan,
    Attention, or Social Attachment, so the component does not hide visible
    dependencies within itself. However, Goal and Plan's mutual dependency on
    World State (\textbf{M16}) is visible as a single connection rather than
    two separate ones.

\end{enumerate}

\section[Implementing \progname{}]{Implementing
\progname{}\protect\normalfont\footnote{See \progname{}'s implementation at
\href{https://github.com/GenevaS/EMgine/tree/main/src}{https://github.com/GenevaS/EMgine/tree/main/src}.}}\label{sec:emgineImplement}
\progname{}'s implementation must use a non-proprietary programming language
and environment is common in game development so that it is not limited to a
specific development ``style'' and  minimal work is necessary to prepare it for
user studies. Therefore, it uses the C\# programming language because it is one
of the languages supported in Unity, a well-known game development
platform~\citep{unity3Dcsharp}. Unity's default script editor for C\# is the
Microsoft Visual Studio (MVS) Integrated Development Environment (IDE), so
\progname{}'s implementation relies on it (Table~\ref{tab:implementEnv_Summary}
summarizes relevant components/packages and their versions). This IDE also
affords access to the following testing tools, which forwards verification and
validation efforts:
\begin{itemize}

    \item \textbf{NUnit Unit Testing Framework} \\
    This supports the bulk of the automated testing approach for unit,
    integration, system, and regression testing. The IDE configuration allows
    existing unit tests to run automatically when compiling the code base.
    Unity Testing Framework uses custom integration of NUnit
    3.5~\citep{unity3Dtestingfw}.

    \item \textbf{Moq Library for .NET} \\
    This supports tests that rely on components that lack a concrete
    implementation, such as the user-implemented data types
    (Chapter~\ref{sec:envDataTypes}). It allows the definition of type-safe,
    mocked interface calls in unit tests~\citep{moq}.

    \item \textbf{Performance Analysis} \\
    \progname{} uses the performance tools built into MVS 2022, which includes
    CPU, memory, and time usage tools~\citep{vs2022perf}.

    \item \textbf{Code Style and Quality Analyzers} \\
    \progname{}'s development uses the official .NET Compiler Platform
    (Roslyn)~\citep{roslyn} and the third-party Roslynator~\citep{roslynator}
    analyzers to help adhere to good code quality and style practices. The
    Unity documentation also references Roslyn analyzers for code style and
    quality~\citep{unity3Droslyn}.

\end{itemize}

\begin{table}[!tb]
    \centering
    \renewcommand{\arraystretch}{1.2}
    \caption{Summary of \progname{}'s Implementation Environment}
    \label{tab:implementEnv_Summary}
    \begin{tabular}{P{0.55\linewidth}P{0.3 \linewidth}}
        \toprule
        \textbf{Name} & \textbf{Version} \\
        \midrule
        \colourRow\multicolumn{2}{l}{Operating System} \\
        Windows 10 Pro & 10.0.19044 Build 19044 \\
        \colourRow\multicolumn{2}{l}{IDE (Development), Project Target
            Framework: .NET Standard 2.0} \\
        Microsoft Visual Studio 2022 Community Edition (64-bit) &
        17.4.3+33205.214 \\
        C\# Tools & 4.4.0-6.22580.4+d7a61210a88 b584ca0827585ec6e871c6b1c5 a14
        \\
        Microsoft .NET Framework & 4.8.04084 \\
        Microsoft .NET Standard Library & 2.0.3 \\
        Microsoft .NET Core Platforms & 1.1.0 \\
        NuGet Package Manager & 6.4.0 \\
        \midrule
        \colourRow\multicolumn{2}{l}{IDE (Testing), Project Target Framework:
            .NET 6.0} \\
        Microsoft .NET Test SDK & 17.4.1 \\
        NUnit 3 Framework (Targeting .NET Standard 2.0 Framework, Microsoft
        .NET Core Platforms 1.1.0) & 3.13.3 \\
        NUnit 3 Analyzers & 3.5.0 \\
        NUnit 3 Test Adapter & 4.3.1 \\
        Moq & 4.18.3 \\
        \midrule
        \colourRow\multicolumn{2}{l}{Coding Support and Analysis} \\
        MVS IntelliCode & 2.2 \\
        Roslynator Analyzers & 4.1.2 \\
        Coverlet & 3.2.0 \\
        \bottomrule
    \end{tabular}
\end{table}

\section{Documenting \progname{}'s Design}\label{sec:designDoc}
\progname{}'s design focuses on modularization to make it easier to change or
swap \progname{}'s components without interfering with otherwise independent
units. This depends on the principle of \textit{information hiding} such that
as few modules as possible contain likely changes and modules do not know how
others work internally. Continuing with a Document Driven Design (DDD) approach
(Chapter~\ref{sec:docSRS}), \progname{} uses two complementary templates from
\citet{smith2009document} to document these designs: a Module Guide (MG),
summarizing \progname{}'s modules' intended tasks and their relationships (i.e.
high-level architecture)\footnote{See \progname{}'s MG at
\href{https://github.com/GenevaS/EMgine/blob/main/docs/Design/MG/EMgine_MG.pdf}{https://github.com/GenevaS/EMgine/blob/main/docs/Design/MG/EMgine\_MG.pdf}.};
and a Module Interface Specification (MIS), describing the syntax and semantics
of each module's public interface (i.e. low-level module design)\footnote{See
\progname{}'s MIS at
\href{https://github.com/GenevaS/EMgine/blob/main/docs/Design/MIS/EMgine_MIS.pdf}{https://github.com/GenevaS/EMgine/blob/main/docs/Design/MIS/EMgine\_MIS.pdf}.}.
 They also support DDD by documenting traceability information back to the
Software Requirement Specification (SRS, Chapter~\ref{sec:docSRS}), likely
changes, and between modules; and help plan systematic testing of \progname{}'s
implementation (Chapter~\ref{sec:docTestPlan}) because modules can have
independent evaluations and have clearly defined input and output interfaces.

\section{Summary}
\progname{} moves towards a component-based software architecture because it
can behave like a software library. This affords users maximum flexibility over
how and when to use \progname{} elements, aligning with many of its high-level
requirements. However, this could be overwhelming for users if they are unsure
when or if they need an \progname{} component. To alleviate this issue,
\progname{} could include a prebuilt ``component of components'' that behaves
like a software engine. This would alleviate user stress because they would not
need to know how \progname{} works---only what inputs to give it. \progname{}'s
module decomposition supports this decision, but has several points where it
``leaks'' hidden information. Although removing all ``leaks'' is difficult,
identifying them might help future modularization efforts to reduce them.
\progname{}'s implementation relies on C\# because it is a non-proprietary
programming language supported by Unity, a well-known game development
platform. This positions \progname{} well for verification and validation
efforts both in the target domain and for reuse elsewhere.

\clearpage
\vspace*{\fill}
\begin{keypoints}
    \begin{itemize}

        \item \progname{}'s architecture supports its decision to be a
        \textit{library of components} that comes packaged with a default
        \textit{``engine''} that is itself a system of components to maximize
        its overall flexibility and ease-of-use

        \item The module decomposition focuses on the ability to add and remove
        task components from \progname{}, but other decomposition approaches
        might further reduce information ``leaks''

        \item \progname{} development used C\# because it is a non-proprietary
        programming language that Unity, a well-known game development
        platform, supports

        \item \progname{} uses the module guide and interface specification
        documentation templates proposed by \citet{smith2009document} because
        it encourages modularization to accommodate predictable changes to
        different aspects of a design

    \end{itemize}
\end{keypoints}

\parasep
\vspace*{\fill}

%% file: validation.tex
\chapter{Gather Your Tools: Defining Acceptance Test Case
Templates}\label{chapter:testcasedefinition}
\def\epigraphflush{center}
\setlength{\epigraphwidth}{0.85\textwidth}
\def\textflush{center}
\epigraph{I taught the children what fear is. I felt they had to know so that
they wouldn't run heedlessly into danger.}{Pascal, \textit{Neir: Automata}}

\progname{} is successful if game developers can create Non-Player Characters
(NPCs) that players find believable. Evaluating this requires at least two user
studies: one to evaluate the usability of \progname{} and another to evaluate
the player experience (PX) with its NPCs. User studies are ideal for evaluating
subjective judgments of believability. However, they can be expensive to plan,
execute, and analyze. It is preferable to run some preliminary tests that do
not include player agency to evaluate a Computational Model of Emotion (CME) on
``obvious'' scenarios with an expected output. Once these tests pass, then it
might be time to run user studies because there are fewer erroneous results to
confound them.

\progname{}'s models are psychologically valid because they are based on
theories from affective science (i.e. hypotheses about human affective
processes and behaviour) and tests can show that their implementation works as
designed. It is unknown if \progname{}'s implementation creates plausible
emotions that make sense to players~\citep[p.~216--217]{broekens2016emotional}.
\progname{} relies on acceptance test cases extended from templates built with
the proposed development process (Chapter~\ref{sec:atcProcess}). It starts with
generic templates that use ``fuzzy'' values to clearly separate them from
\progname{}'s models to demonstrate that the process is not
\progname{}-specific and the templates could serve as a common nexus for
building test cases to compare CMEs with similar functionality. Example
acceptance test case templates based on Elsa from Disney's
\textit{Frozen}~\citep{frozen}---an in-depth one for \textit{Grief} and a
sketch for \textit{Admiration}---illustrates this process, followed by their
extension into implementable test cases specific to \progname{}
(Chapter~\ref{chapter:testcaseEMgine}).

\section{Choosing a Test Case Source Medium}\label{sec:testsource}
Choosing a test case source medium similar to video games, \progname{}'s
intended domain, helps reduce the time, effort, and potential mistakes
associated with translating a storyteller's tales into test cases. The obvious
choice would be video games themselves, but they are not ideal for replicability
due to their interactive nature. Since a player's role cannot be entirely
scripted and their actions vary between sessions, their influence on the game
state varies. This makes it more difficult to reproduce the scenario and,
consequently, could make test case synthesis less reproducible. The only game
aspects that do not change are cutscenes---non-interactive sequences that only
make up a small fraction of most games' runtime. Character analysis can be
intensive, so it is prudent to pick a data-rich source. Taking video game
cutscenes as short films, film is the next medium to consider.

Broadly, films are great because they too are an audiovisual medium. A
character's emotional responses are most evident because there are more and
clearer cues to signal it than words alone (e.g. body language, facial
expressions, vocal tone). Animated films, in particular, are likely best for
believable emotion-focused CMEs because reality grounds them without limiting
them:
\begin{aquote}{\citet[p.~34]{williams2001animator}}
     ``In order to depart from reality, [animation] has to be based \textit{on}
     reality.''
\end{aquote}
\afterpage{
    \vspace*{\fill}
    \begin{figure}[!ht]
        \begin{center}
            \subfloat[Marge Belcher as Snow White~\citep{snowwhite}]{
                \includegraphics[width=0.63\linewidth]{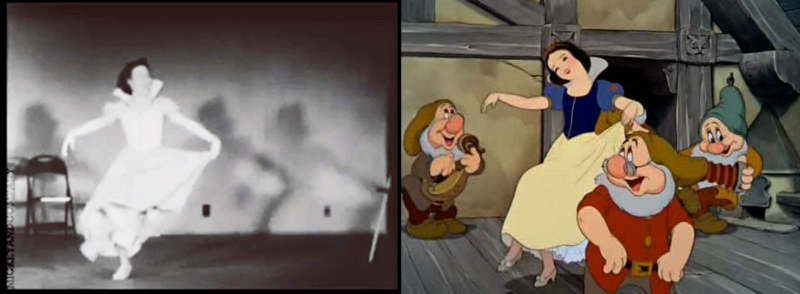}
            }

            \subfloat[Kathyrn Beaumont as Alice~\citep{alice}]{
                \includegraphics[width=0.63\linewidth]{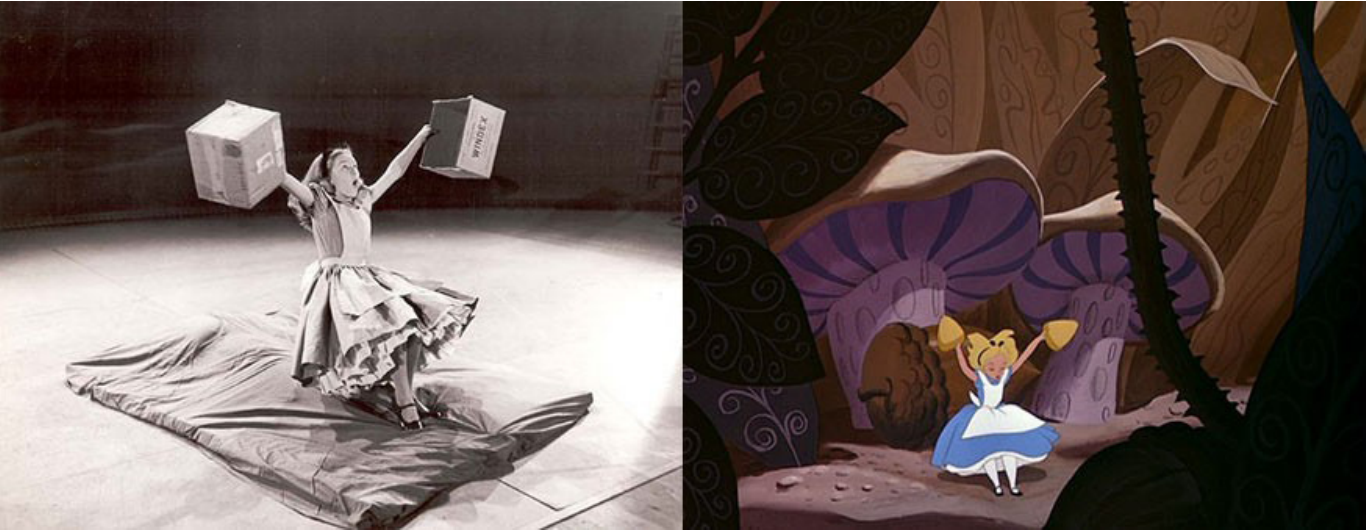}
            }

            \subfloat[Sherri Stoner as Ariel~\citep{littlemermaid}]{
                \includegraphics[width=0.63\linewidth]{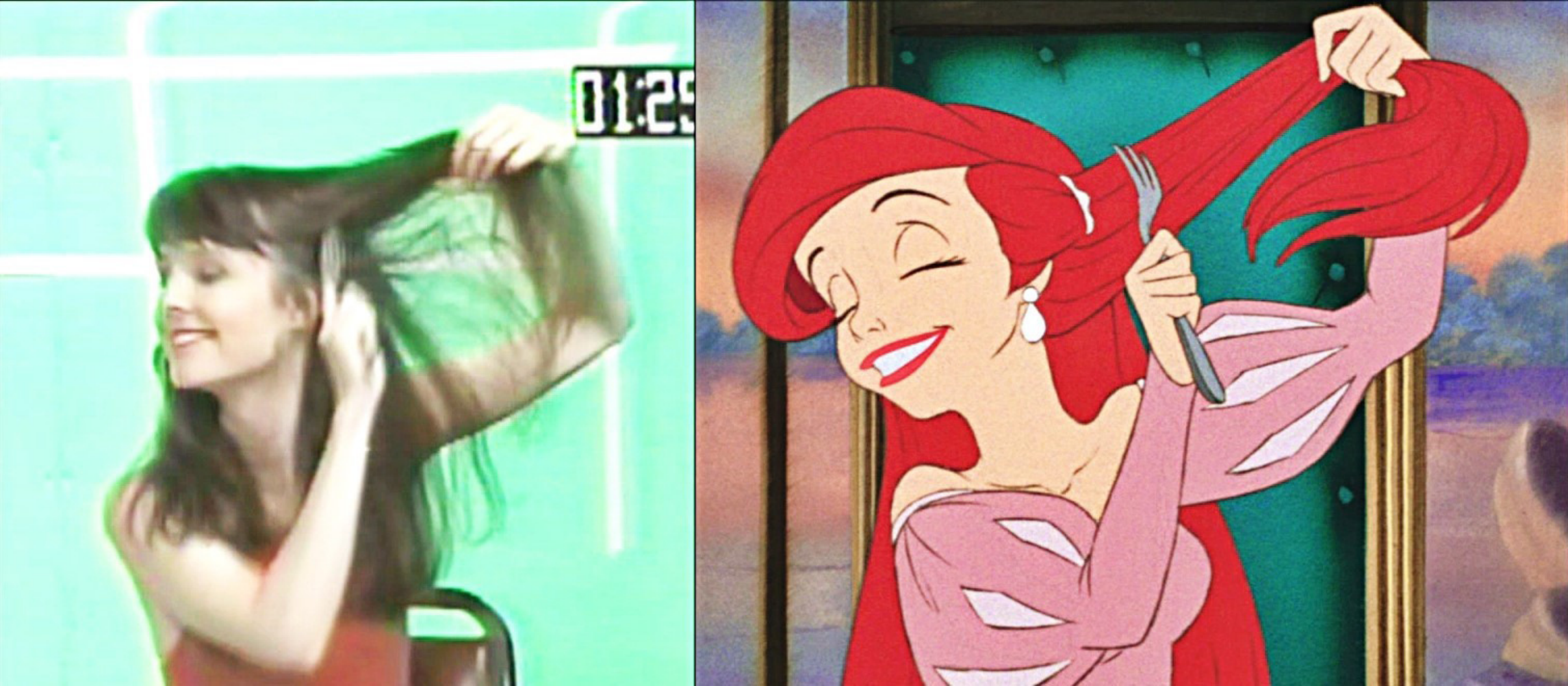}
            }

            \subfloat[Kyausha Simpson as Calliope~\citep{hercules}]{
                \includegraphics[width=0.63\linewidth]{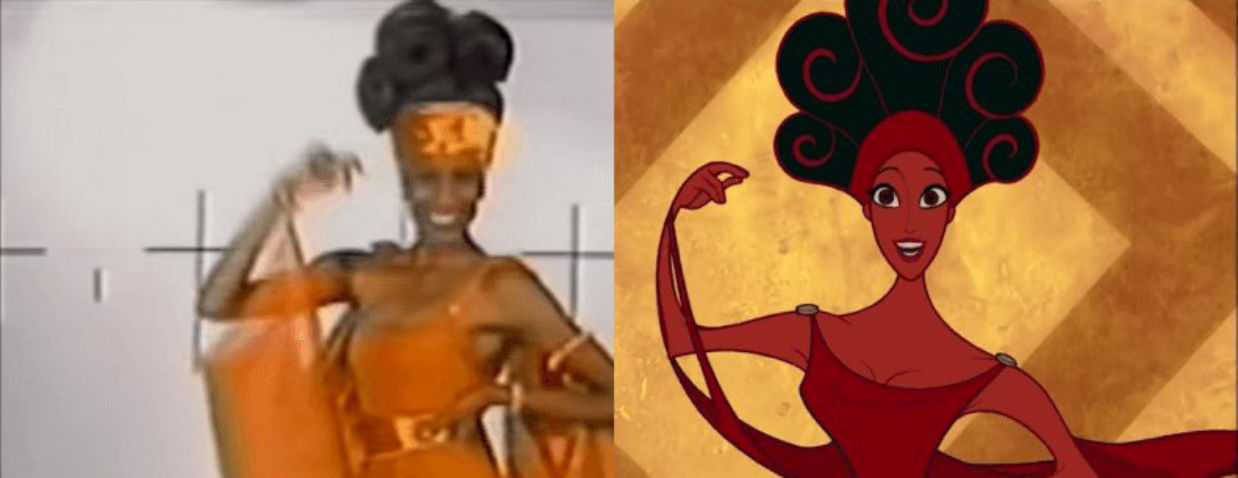}
            }

            \caption[Comparison of live action reference and final
            animation]{Comparison of live action reference and final animation
                in Disney's \textit{Snow White and the Seven Dwarfs} (1937),
                \textit{Alice in Wonderland} (1951), \textit{The Little Mermaid}
                (1989) and \textit{Hercules} (1997)}
            \label{fig:animationRefs}
        \end{center}
    \end{figure}
    \vspace*{\fill}\clearpage
}

Animators often use live action film as inspiration and reference for their
work (\citepg{thomas1981illusion}{71--72, 319--320},
Figure~\ref{fig:animationRefs}). Walt Disney famously brought performers and
animals to the studio for his animators ``...to try to capture a more realistic
believable figure''~\citep{korkis2022live}. When casting for live action
references, staff were careful to ``...select an actor whose natural voice and
mannerisms are caricatures of a normal
person's.''~\cite[p.~550]{thomas1981illusion} likely because caricatures are
the most unambiguous depictions of real behaviours (see
Chapter~\ref{sec:discreteperspective} for a brief discussion). Animators then
``...accentuate and suppress aspects of the model's character to make it more
vivid''~\citep[p.~34]{williams2001animator} using their own knowledge and
observations~\citep[p.~201, 217]{johnson1992basic}. These caricatures include
emotion, making it easier to identify what a character is experiencing and
deduce the eliciting factors. Film scene reenactment has proven useful for
evaluating the influence of CME parameters on viewer perceptions of animated
agents~\citep[p.~343--344]{bidarra2010growing}\footnote{The Soul (\ref{soul}),
included in the CME survey (Chapter~\ref{chapter:cmeOverview})}, so it is a
reasonable hypothesis that they would also be good resources for building test
cases. The audiovisual nature of film also eliminates the need to express
emotion with specific words, making it easier to overcome language and cultural
barriers. This is evident in successful translations to other
languages~\citep[p.~315]{thomas1981illusion} and improves the potential to
reproduce and/or localize test cases built on animated characters by itemizing
observable ``symptoms'' of emotions (see Chapter~\ref{sec:language} for a
discussion).
%

\section{Emotion Profiles for Believable Characters}\label{sec:profiles}
Building emotion profiles means describing the characteristics and observable
signs of emotion that others can reference to recreate test cases. A core
feature of the discrete perspective on emotion is distinct emotion kinds
distinguishable with sets of observable features (see
Chapter~\ref{sec:discreteperspective} for discussion). Many of these features
are non-verbal, which often carry significantly more information than words
alone (\citeg{mehrabian2008communication}; \citepg{isbister2006sensual}{1164}).
Therefore, it is the primary resource for building emotion profiles for
believable characters. Each profile describes:
\begin{enumerate}[label=(\alph*)]

    \item The emotion's purpose, cognitive impact, and how it changes at
    different intensities. This provides a reference for deducing ``transient''
    and ``persistent'' knowledge about a character.

    \item Action tendencies, physiological changes, and verbal and nonverbal
    signals. Together with facial expressions, this serves as a guide for
    identifying what emotion a character is experiencing.

    \item Facial expressions associated with the emotion. Together with action
    tendencies, physiological changes, and verbal and nonverbal signals, facial
    expressions are a guide for identifying what emotion a character is
    experiencing. This is especially useful for identifying animated character
    emotions due to their caricaturisation.

    \item Examples from the source medium to demonstrate how different parts of
    the profile appear in the source medium.

\end{enumerate}

The profile separates facial expressions from other observable signals like
physiological changes and nonverbal utterances because they appear to convey
the most information about an individual's affective state than any other
channel~\citep{vesterinen2001affective} and people tend to notice them more
frequently~\citep[p.~322]{scherer1994evidence}. Although facial expressions are
an imperfect indicator of emotion~\citep[p.~440]{roseman2011emotional}, the
Facial Action Coding System (FACS)~\citep{facs} is encouraging for emotion
specificity in facial movement~\citep[p.~39]{barrett2006emotions}. Previous
work in \ref{ac} successfully used combinations of individual FACS actions to
recognize facial expressions in human users and videos
(\citeg{torre2011facial}; \citepg{pantic2003toward}{1385--1386}) and CMEs that
express emotion in a face (e.g. \citet{duy2004creating},
\citet{breazeal2003emotion}, and \citet{bidarra2010growing}\footnote{ParleE
(\ref{parlee}), Kismet (\ref{kismet}), and The Soul (\ref{soul}), included in
the CME survey (Chapter~\ref{chapter:cmeOverview})}). Therefore, this system
guides facial expression descriptions\footnote{Although sufficient for these
profiles, an individual untrained in FACS made these code assignments.}. When
an emotion is not explicitly represented as a complete facial expression in
FACS, knowledge about individual facial changes, such as eyebrows, informs the
creation of one~\citep[p.~241]{smith_scott_mandler_1997}. A character's
sleeping face is the baseline to compare changes in facial features to
determine what, if any, facial expression they make.

The following profile for \textit{Sadness} shows what it might look like. The
remaining profiles for \textit{Joy}, \textit{Fear}, \textit{Anger},
\textit{Disgust}, \textit{Acceptance}, \textit{Interest}, and \textit{Surprise}
are in Appendix~\ref{appendixEmotionClassification}.

\begin{notes}[frametitle=\textit{Sadness} Profile]
    Defined by loss, \textit{Sadness} is a relatively intense and long-lasting
    emotion (\citepg{scherer1994evidence}{324};
    \citepg{ekman2007emotions}{84}). It is a social emotion for signalling to
    the self and others that something is not
    well~\citep[p.~291]{izard1977human}\footnote{Izard refers to this as
    \textit{Distress}, which this profile takes as a synonym for
    \textit{Sadness}~\citep[p.~285]{izard1977human}.} or that a situation seems
    uncontrollable (\citepg{smith1985patterns}{834}; \citeg{oxfordSadness}),
    acting as a cry for help and comfort. \textit{Sadness} encourages problem
    solving by providing tolerable or less tense motivation to change the
    situation, while simultaneously supplying little energy to do anything
    about it~\citep[p.~326]{izard1977human}. The tendency to withdraw into
    oneself supports this by conserving energy while redirecting cognitive
    resources towards finding potential compensations or
    readjustments~\citep[p.~251]{lazarus1991emotion}. To exit the state of
    \textit{Sadness} quickly, people often act on coping strategies without a
    reasonable evaluation of their future effects. This implies reduction in
    self-control, increasing the likelihood of accepting immediate over delayed
    gratification or discounting the safety of the self or bystanders.
    Subjectively, people tend to describe \textit{Sadness} as an unpleasant
    feeling.

    As the intensity of \textit{Sadness} increases, the individual becomes less
    active, withdrawing into themselves and away from their surroundings.
    \textit{Sadness} naturally cycles through periods of high\footnote{Ekman
    refers to this as \textit{Agony}, which this profile takes a synonym for
    \textit{Grief}.} and low intensity~\citep[p.~84--85]{ekman2007emotions},
    likely a result of the individual's continued appraisals of their ability
    to cope and the appraised value of compensations and adjustments made. This
    might also be why it is common for other emotions to manifest during
    periods of \textit{Sadness} such as \textit{Fear}, \textit{Anger}, and
    \textit{Joy}.

    \paragraph{Signs of \textit{Sadness}} The action tendencies in
    \textit{Sadness} are passive: withdrawal by the individual while
    unintentionally signalling for help. Others can perceive this as
    inaction~\citep[p.~252]{lazarus1991emotion} with the assumed expectation
    that the individual is waiting for others to do something for them.
    Therefore, an individual experiencing \textit{Sadness} might not do
    anything. A potential exception is to approach what has been lost to
    further evaluate its status or to approach something else that the
    individual perceives as comforting, such as a loved one. Although it is
    usually accompanied by strong nonverbal expressions to signal for
    help---notably crying---there are few vocal, verbal, or nonverbal
    expressions~\citep[p.~322--324]{scherer1994evidence}.

    Vocally, the individual's voice becomes softer and
    lower~\citep[p.~56]{ekman2007emotions}. Physically, the body shuts down for
    energy conservation, signalled by a lowered body
    temperature~\citep[p.~321--322, 326]{scherer1994evidence}. However, there is
    also an increase in heart rate and muscle tension, in addition to a
    perceived constriction in the throat. The increased tension might serve as
    an additional social signal to convey the individual's stressful state and
    elicit soothing contact-based responses from others.

    \paragraph{Characteristic Facial Expression} \textit{Sadness} registers on
    all facial areas (Tables~\ref{tab:pensivenessFACS}, \ref{tab:sadnessFACS},
    and \ref{tab:griefFACS}) and is difficult to mimic due to the inner eyebrow
    movement (\citepg{ekman2003unmasking}{117, 121--122, 126};
    \citepg{izard1977human}{287--288}). Other facial movements are not reliable
    indicators of \textit{Sadness} by themselves, so those expressions can be
    ambiguous.

    The most reliable facial feature for detecting \textit{Sadness} is the inner
    eyebrows because many people have difficulty moving them voluntarily. The
    inner corners of the eyebrows draw together and upwards in
    \textit{Sadness}, which can cause creases to appear between them and on the
    forehead. As the eyebrows rise, these creases become exaggerated. Eyebrow
    movement also causes the inner corner of the upper eyelid to rise. The
    upper eyelid itself might lower, particularly if the eyes are cast
    downwards to avoid eye contact.

    In the lower face, the outer corners of the mouth draw down and become more
    exaggerated as the intensity of the emotion increases. Tension in the cheek
    muscles increases with the intensity of \textit{Sadness}, causing them to
    rise. This pushes the lower eyelids up, making the eyes look like they are
    closing.

    \paragraph{Examples} Elsa from Disney's \textit{Frozen}~\citep{frozen} has
    built her life around her powers despite the unhappiness it brings her
    because she believes them to be dangerous and uncontrollable.

    Elsa's facial expressions clearly convey \textit{Sadness} via her raised
    inner eyebrows and subsequent rising of the inner eyelids
    (Figure~\ref{fig:elsaSadness}). In all cases, Elsa is making eye contact
    with her conversation partner so her eyes are not downcast. As her
    \textit{Sadness} intensifies, the tension in her cheeks increases and
    pushes her lower eyelids higher. The corners of her mouth also pull down as
    emotion intensity increases, eventually causing her mouth to open. Elsa
    only has the suggestion of creases between her eyebrows and running from
    her nose to mouth, likely due to artistic choice.

    \textbf{Sad01}: Elsa wakes in the dungeons to find that her kingdom has
    entered a state of permanent winter. Wishing to avoid further harm, she is
    \textit{pensive} because:
    \begin{itemize}
        \item Her powers have compromised the safety of her kingdom
        \item She believes that the damage is irreversible because she has
        little control over her powers
        \item Elsa believes her home is lost to her and needs to distance
        herself from it for its safety

        \begin{center}
            \includegraphics[width=0.9\linewidth]{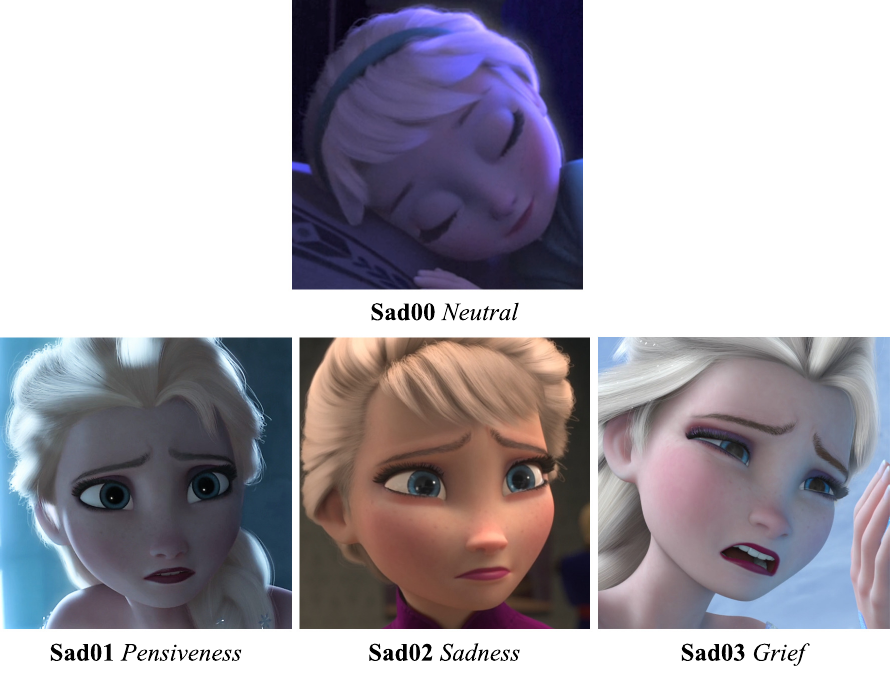}
            \captionsetup{format=hang, justification=raggedright}
            \captionof{figure}{Examples of \textit{Sadness} in Elsa's Facial
                Expressions}\label{fig:elsaSadness}
        \end{center}

        \item Her sister forewarned of the situation, giving Elsa time to
        adjust to the idea that she had caused significant damage
    \end{itemize}

    Elsa's trembling voice and soft pleading to be released so that she can put
    distance between herself and the kingdom show her \textit{pensiveness}. Her
    body language also implies it because she does not move and appears to be
    curling into herself.

    \textbf{Sad02}: Elsa and her sister are arguing, which ends with her sister
    proclaiming that she cannot continue to live isolated from the outside. Elsa
    becomes \textit{sad} because:
    \begin{itemize}
        \item She loves her sister dearly
        \item She has damaged her relationship with her sister by refusing her
        request and telling her to leave
        \item Elsa knows that distancing herself from her sister will not repair
        their relationship, but does not know how else to protect her
    \end{itemize}

    Elsa's unwillingness to speak to her sister further, apparent by a heavy
    sigh and turning away, as well as her crossed arms, drooping head, and
    curling shoulders, communicate her \textit{Sadness}.

    \textbf{Sad03}: Elsa's sister has been frozen solid, a fate equivalent to
    death, causing her \textit{Grief}:
    \begin{itemize}
        \item Death cannot be reversed, so Elsa has no way to regain her sister
        and her safety
        \item The cause was Elsa's uncontrolled ice powers
        \item It has been a few minutes, at most, since her sister froze, so
        there has been no time to compensate or adjust
    \end{itemize}

    Elsa's \textit{Grief} is apparent via her bodily collapse, hanging onto her
    sister's body, loud sobbing, and vocal denial of the situation.
\end{notes}

\vspace*{\fill}
\begin{table}[!ht]
    \centering
    \caption{Facial Sketches of \textit{Pensiveness} with Suggested FACS Codes}
    \label{tab:pensivenessFACS}
    \renewcommand{\arraystretch}{1.2}
    \begin{threeparttable}
        \begin{tabular}{P{0.6\linewidth}P{0.3\linewidth}}
            \toprule
            \begin{center}
                \includegraphics[width=0.3\linewidth]{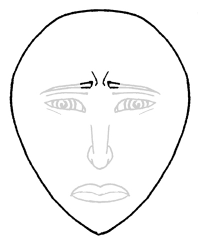}
            \end{center} & \begin{tabular}{ll}
                \textbf{Label} & Pensiveness \\
                \textbf{Intensity} & Low \\
            \end{tabular} \\ \midrule
            \textbf{Description} & \textbf{FACS Codes} \\ \midrule
            \colourRow Inner eyebrow corners raised and brought together & AU
            1+4 \\
            A faint vertical wrinkle might appear between the inner eyebrows &
            AU 1* \\
            \colourRow Inner corner of the upper eyelid is pulled upwards by
            the raised eyebrows & AU 1* \\
            Upper eyelids might be lowered & AU 43 \\
            \colourRow Gaze might be cast downwards & 64 \\
            Slightly raised cheeks & AU 6 \\
            \colourRow Lower eyelids pushed up due to the raised cheek muscles
            which make the eyes look like they are closing & AU 6* \\
            Corner of the lips are neutral OR pulled down slightly & -- OR AU
            15 \\
            \bottomrule
        \end{tabular}
        \begin{tablenotes}

            \footnotesize
            \vspace*{2mm}

            \item {*} \textit{Causes change indirectly}

        \end{tablenotes}
    \end{threeparttable}
\end{table}
\vspace*{\fill}

\vspace*{\fill}
\begin{table}[!ht]
    \centering
    \caption{Facial Sketches of \textit{Sadness} with Suggested FACS Codes}
    \label{tab:sadnessFACS}
    \renewcommand{\arraystretch}{1.2}
    \begin{threeparttable}
        \begin{tabular}{P{0.6\linewidth}P{0.3\linewidth}}
            \toprule
            \begin{center}
                \includegraphics[width=0.6\linewidth]{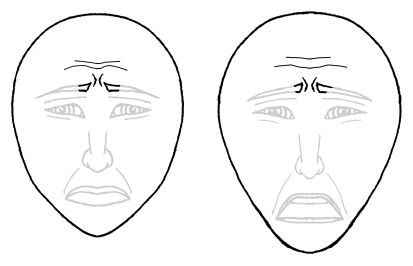}
            \end{center} & \begin{tabular}{ll}
                \textbf{Label} & Sadness \\
                \textbf{Intensity} & Medium \\
            \end{tabular} \\ \midrule
            \textbf{Description} & \textbf{FACS Codes} \\ \midrule
            \colourRow Inner eyebrow corners raised and brought together & AU
            1+4 \\
            A vertical wrinkle might appear between the inner eyebrows & AU 1*
            \\
            \colourRow Inner corner of the upper eyelid is pulled upwards by
            the raised eyebrows & AU 1* \\
            Upper eyelids might be lowered & AU 43 \\
            \colourRow Gaze might be cast downwards & 64 \\
            Raised cheeks & AU 6 \\
            \colourRow Lower eyelids pushed up due to the raised cheek muscles
            which make the eyes look like they are closing & AU 6* \\
            Corner of the lips pulled down which might cause furrows on the
            cheeks & AU 15 \\
            \colourRow Lips might be trembling & -- \\
            Lower lip might be pushed up & AU 17 \\
            \bottomrule
        \end{tabular}
        \begin{tablenotes}

            \footnotesize
            \vspace*{2mm}

            \item {*} \textit{Causes change indirectly}

        \end{tablenotes}
    \end{threeparttable}
\end{table}
\vspace*{\fill}

\vspace*{\fill}
\begin{table}[!ht]
    \centering
    \caption{Facial Sketches of \textit{Grief} with Suggested FACS Codes}
    \label{tab:griefFACS}
    \renewcommand{\arraystretch}{1.2}
    \begin{threeparttable}
        \begin{tabular}{P{0.6\linewidth}P{0.3\linewidth}}
            \toprule
            \begin{center}
                \includegraphics[width=0.3\linewidth]{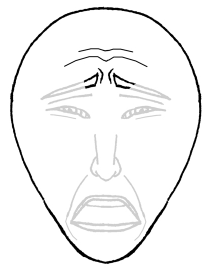}
            \end{center} & \begin{tabular}{ll}
                \textbf{Label} & Grief \\
                \textbf{Intensity} & High \\
            \end{tabular} \\ \midrule
            \textbf{Description} & \textbf{FACS Codes} \\ \midrule
            \colourRow Inner eyebrow corners raised and brought together & AU
            1+4 \\
            Likely a vertical wrinkle between the inner eyebrows & AU 1* \\
            \colourRow Inner corner of the upper eyelid is pulled upwards by
            the raised eyebrows & AU 1* \\
            Upper eyelids might be lowered & AU 43 \\
            \colourRow Gaze might be cast downwards & 64 \\
            High, raised cheeks & AU 6 \\
            \colourRow Lower eyelids pushed up due to the raised cheek muscles
            which make the eyes look like they are closing & AU 6* \\
            Corner of the lips pulled down causing furrows to appear on the
            cheeks & AU 15 \\
            \colourRow Dropped jaw that might be trembling & AU 26 \\
            \bottomrule
        \end{tabular}
        \begin{tablenotes}

            \footnotesize
            \vspace*{2mm}

            \item {*} \textit{Causes change indirectly}

        \end{tablenotes}
    \end{threeparttable}
\end{table}
\vspace*{\fill}

\clearpage\section{Character Analysis: Elsa from Disney's \textit{Frozen}
(2013)}
To illustrate how the process transforms an animated film character analysis
into an extendable test case template, this example expands the scenario of
Elsa\footnote{Elements of Elsa's character analysis verified against her
character page on Disney Wiki, a free, public, and collaborative encyclopedia
on Walt Disney and the Disney corporation:
\href{https://disney.fandom.com/wiki/Elsa}{https://disney.fandom.com/wiki/Elsa}
(Last Accessed 19 February 2023).} expressing \textit{Grief} from the
\textit{Sadness} profile (Example \textbf{Sad03}).

\subsection{Collecting Local ``Transient'' Knowledge}
Extending the description of \textbf{Sad03}, Elsa is primarily expressing
\textit{Sadness} with body language (Table~\ref{tab:elsa_Local}). Anna's
physical state (frozen solid) is most likely the cause because \textit{Sadness}
is defined by loss, such as the death of loved ones. Note that Elsa was already
experiencing \textit{Sadness} before this, reacting to news that Anna was dead
because of Elsa's powers (``Your sister is dead...because of you.'').
\begin{table}[!b]
    \centering
    \renewcommand{\arraystretch}{1.2}
    \caption[Summary of ``Transient'' Knowledge About Elsa for Example]{Summary
    of ``Transient'' Knowledge About Elsa for Example \textbf{Sad03}}
    \label{tab:elsa_Local}
    \begin{tabular}{P{0.25\linewidth}P{0.65\linewidth}}
        \toprule
        \colourRow\textbf{In Scene} An Act of Love & \textbf{Approx. Time}
        1:26:24--1:27:08 \\
        \textbf{Character} Elsa & \textbf{Emotion} \textit{Grief} (Intense
        \textit{Sadness}) \\
        \multicolumn{2}{c}{
            \includegraphics[width=0.5\linewidth]{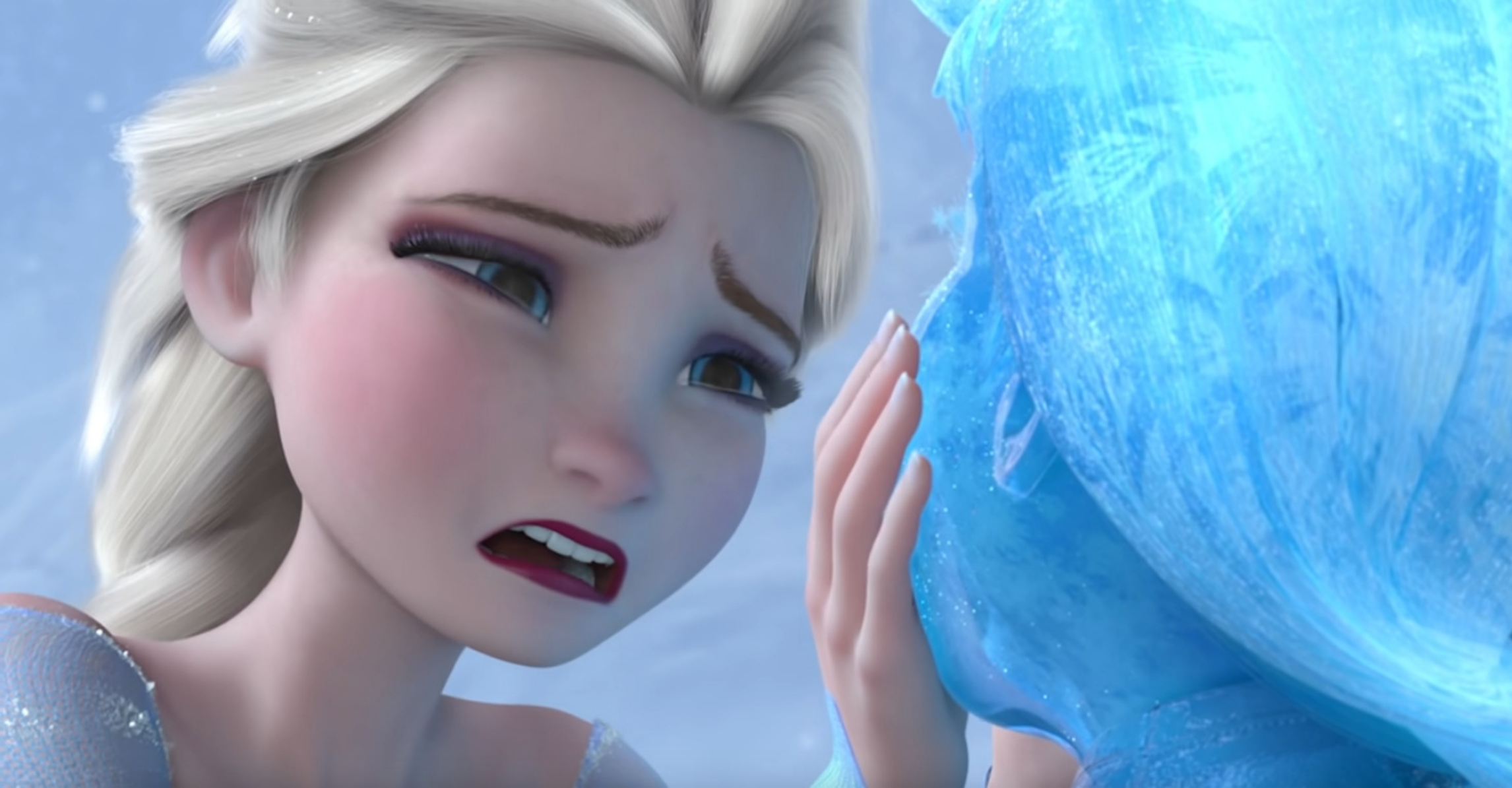}
        } \\
        \midrule
        \colourRow\textbf{Actions} & Loud sobbing; Hanging her head; Hugging
        Anna's shoulders (not supporting herself with her legs/feet) and slowly
        releasing her hold (kneeling at the end); Powers are not active
        (initially stopped when Hans told her that she killed Anna, mirrors
        their parents' funeral during ``Do You Want to Build a Snowman?'')  \\
        \textbf{Dialogue} & ``Anna! Oh, Anna...no...no, please no.'' (pleading
        tone) \\
        \colourRow\textbf{Physical State} & Uninjured; Not in danger of injury
        \\
        \midrule
        \textbf{Character} & Anna \\
        \colourRow\textbf{Relation} & Little Sister (Anna is 18 to Elsa's 21);
        Best Friend (from ``Do You Want to Build a Snowman?'', reunion at
        coronation party) \\
        \textbf{Actions} & -- \\
        \colourRow\textbf{Dialogue} & -- \\
        \textbf{Physical State} & Frozen solid (``dead'') \\
        \bottomrule
    \end{tabular}
\end{table}

\subsection{Inferring Global ``Persistent'' Knowledge}\label{sec:elsa_Global}
Animated characters often have simple goals and
personality~\citep[p.~2--3]{peng2018trip}. This example focuses on two pieces
of ``persistent'' knowledge about Elsa: her personality, to help contextualize
her responses to the world; and her goal to \texttt{Protect Anna}. All NPCs
have at least one goal of some form~\citep[p.~223]{broekens2016emotional} (e.g.
``watch the race'', ``generate income''), serving as a common nexus between the
source medium and the aim to create believable NPCs with CMEs.\vspace*{1em}

\begin{notes}[frametitle={Summary of ``Persistent'' Knowledge About Elsa's
Personality and \texttt{Protect Anna} Goal}]
    \vspace*{-1em}\paragraph{Personality} Elsa is a central character in
    Disney's 2013 film
    \textit{Frozen}~\citep{frozen}. She presents herself as a calm, reserved,
    and regal person, but also demonstrates a kind and generous nature (e.g.
    allowing young Anna to wake her during the night to play, creating a
    skating rink for the people of Arendelle in the summer). However, the
    danger posed by her powers make her insecure, depressed, and anxious.

    Elsa was born with the power of ice and snow, which allows her to conjure,
    manipulate, and create sentient (e.g. Olaf, Marshmallow) and non-sentient
    (e.g. palace, skates) constructions from them. However Elsa's powers can
    cause harm if uncontrolled. Thus Elsa believes her powers make her
    monstrous. She wears gloves, believing that they help her control her
    powers (``Conceal it, don't feel it''), but falsified when she uses her
    powers to escape her jail cell by freezing manacles that completely cover
    her hands. Instead, Elsa manifests her powers unconsciously when she is
    severely distressed and/or frightened (e.g. after injuring Anna when they
    were children, at the overwhelming coronation party, discovering that
    Arendelle is frozen, escaping execution). In contrast, Elsa appears to have
    full control of her powers when not under stress (e.g. playing as children,
    ``Let it Go'', deicing Arendelle, making a skating rink in the castle
    courtyard).

    \paragraph{Goal: \texttt{Protect Anna}} Elsa does not want to harm anyone,
    especially those close to her (Distressed when she injures Anna as
    children; ``No. Don't touch me. I don't want to hurt you.'' to her parents
    during ``Do You Want to Build a Snowman?''). She is particularly concerned
    with keeping Anna safe, evident by Elsa's self-isolation after harming Anna
    with her powers when they were playing as children and after arriving at
    the North Mountain after the coronation party (in ``Let it Go''). Elsa also
    demonstrates her desire to protect Anna by refusing to bless her engagement
    to Hans (``You can't marry a man you just met [Anna]...You asked for my
    blessing, but my answer is no.''); by forcing Anna to leave the ice palace
    without her after coming for her (``I'm just trying to protect you
    [Anna].''); and by asking Hans to take care of Anna after her execution
    (``...Just take care of my sister.''). This differs from her desire to
    protect her kingdom (experiences fear when Anna tells her Arendelle is
    frozen and distress when she sees it from her prison cell) and herself
    (asking for Anna to be cared for after Elsa's execution). Elsa also has no
    qualms with using her powers for defence (fighting thugs in her ice palace).
\end{notes}

\clearpage\section{Translating Character Analyses into Acceptance Test
Case Templates}\label{sec:formalATC}
The translation of character analyses into test cases should be
implementation\-/agnostic for reusability. Therefore, the example test cases
use ``fuzzy'' values like percentages, the set $\{ \mathtt{Low}, \mathtt{Mid},$
$\mathtt{High} \}$, and the constant $\mathtt{MIN}$ to avoid
over-specification. These test cases are small for illustrative purposes, but
extended versions for \progname{} show their usefulness. The test cases use the
following types:
\begin{itemize}
    \item World State View (WSV) $\worldstatetype$
    (Equation~\ref{eq:worldstatetype});
    \item World Event $\worldstatechangetype$
    (Equation~\ref{eq:worldeventtype});
    \item Goal $\goaltype_{ATC}$, is a predicate on a WSV ($\mathtt{goalState} :
    \worldstatetype \rightarrow \mathbb{B}$) that a character wants to satisfy,
    and its relative importance in $\{ \mathtt{Low}, \mathtt{Mid},
    \mathtt{High} \}$; and
    \item Emotion Intensity $\emotionintensitytype_{ATC}$, in $\{ \mathtt{Low},
    \mathtt{Mid}, \mathtt{High} \}$.
\end{itemize}

These are the minimum working data types necessary for defining these test
cases. Although WSV and World Event are part of \progname{}'s data types, they
are abstractions that do not assume anything about the underlying emotion
theories and/or models. Therefore, they are suitable for defining
theory\-/agnostic test cases. \progname{} does have specifications for Goal
$\goaltype$ and Emotion Intensity $\emotionintensitytype$, but their
specification is likely to change with the underlying emotion theories and/or
models. The types $\goaltype_{ATC}$ and $\emotionintensitytype_{ATC}$ represent
their simplest form, capturing only essential information necessary for test
case specification.

\subsection{Acceptance Test Case Template of Elsa's
\textit{Grief}}\label{sec:grieftemplate}
Assuming that the characters have properties $\mathtt{Health}$ and
$\mathtt{IsAlive}$ and using ``persistent'' character knowledge, the definition
of Elsa's goal to \texttt{Protect Anna} is:
\begin{align*}
    \mathtt{ProtectAnna} : \goaltype_{ATC} = \{ &\mathtt{goalState} = \{
    \mathtt{Anna.Health} \geq 75\% \wedge \mathtt{Anna.IsAlive} \}, \\
    &\mathtt{importance} = \mathtt{High} \}
\end{align*}
\noindent This model assumes that $\mathtt{Health} = 0$ is unconsciousness (a
changeable state) and $\mathtt{IsAlive} = \False$ is a permanent death state,
to reflect the ability to ``revive'' unconscious characters in video games.

Using $\mathtt{Health}$ in $\mathtt{ProtectAnna}$ reflects Elsa's fear of
hurting others with her powers, which would be physical injuries. The arbitrary
value of $75\%$ reflects Anna's fearless and impulsive actions, which often
leads to minor injuries like scrapes and bruises that Elsa would affectionately
disapprove of. Goal importance is $\mathtt{High}$ because Elsa's responses in
the story are strongest when Anna is involved.

From the ``transient'' knowledge about the scenario, Anna's health in the
\textit{current} world state $\mathtt{S}_i : \worldstatetype$ is below Elsa's
goal ($h \in ( \mathtt{MIN}\%, 25\%]$) and Elsa's current \textit{Sadness}
intensity as $\mathtt{Mid}$, reflecting that unmet, transient goal component:
\begin{equation*}
    \mathtt{S}_i : \worldstatetype = \{ \mathtt{Anna.Health} = h,
    \mathtt{Anna.IsAlive} = \True \}; \qquad
    \mathtt{Sadness}_i : \emotionintensitytype_{ATC} = \mathtt{Mid}
\end{equation*}

This WSV reflects Elsa's reaction to \textit{hearing} that Anna is dead rather
than \textit{seeing} it, which she perceives as Anna being seriously injured
rather than dead ($\mathtt{MIN}\% < $ $\mathtt{Anna.Health} \leq 25\% \wedge
\mathtt{Anna.IsAlive}$). Elsa's \textit{Sadness} is still elevated by the news
because her goal, $\mathtt{ProtectAnna}$, is currently unsatisfied.

The event of concern is Anna becoming solid ice, i.e. dying (``Anna, your life
is in danger...to solid ice will you freeze, forever.''):
$$\mathtt{AnnaFreezes_E} : \worldstatechangetype = \{ \mathtt{Anna.IsAlive} =
\False \} $$
\noindent Applying this to $\mathtt{S}_i$ produces:
$$\mathtt{S}_{i+1} : \worldstatetype = \{ \mathtt{Anna.Health} = h;
\mathtt{Anna.IsAlive} = \False \}$$

Finally, the expected output is $\mathtt{Sadness} : \emotionintensitytype_{ATC}
= \mathtt{High}$ (completed test case in Table~\ref{tab:elsa_TestCase}). If a
CME's emotion intensity function rejects $\mathtt{Sadness}_i$ as an input, a
function $\mathtt{Combine}(i_1 : \emotionintensitytype_{ATC}, i_2 :
\emotionintensitytype_{ATC})$ should produce the expected output.

Although both world states $\mathtt{S}_i$ and $\mathtt{S}_{i+1}$ fail to
satisfy $\mathtt{ProtectAnna}$, there is a subtle difference between them:
$\mathtt{Anna.Health}$ is a changeable quantity while $\mathtt{Anna.IsAlive}$
is not. This reflects \textit{world knowledge} and \textit{self knowledge} about
one's goals that a CME needs to know, but the test cases do not need to embed
that information. Nevertheless, it is the reason for the intensity of Elsa's
\textit{Sadness} (see Table~\ref{tab:elsa_TestCaseSketch} for the test case
resulting in $\mathtt{S}_{i}$).

\vspace*{\fill}
\begin{table}[!ht]
    \centering
    \renewcommand{\arraystretch}{1.2}
    \caption{Test Case of Elsa's \textit{Grief} When Anna Becomes Solid Ice}
    \label{tab:elsa_TestCase}
    \begin{tabular}{P{0.13\linewidth}P{0.77\linewidth}}
        \toprule
        \colourRow\textbf{Setup} & $\mathtt{ProtectAnna} : \goaltype_{ATC} = \{
        \mathtt{goalState} = \{ \mathtt{Anna.Health} \geq 75\%$ \newline[5pt]
        $\wedge \; \mathtt{Anna.IsAlive} \},$ $\mathtt{importance} =
        \mathtt{High} \}$, \newline[5pt]
        $\mathtt{Sadness}_i : \emotionintensitytype_{ATC} = \mathtt{Mid}$,
        \newline[5pt]
        $\mathtt{S}_i : \worldstatetype = \{ \mathtt{Anna.Health} = h,
        \mathtt{Anna.IsAlive} = \True \} \text{ where } h \in ( \mathtt{MIN}\%,
        25\%]$ \\
        \textbf{Input} & $\mathtt{AnnaFreezes_E} : \worldstatechangetype = \{
        \mathtt{Anna.IsAlive} = \False \}$ \\
        \midrule
        \colourRow\textbf{Expected Output} & $\mathtt{Sadness}_{i+1} :
        \emotionintensitytype_{ATC} = \mathtt{High}$ \\
        \bottomrule
    \end{tabular}
\end{table}
\vspace*{\fill}
\begin{table}[!ht]
    \centering
    \renewcommand{\arraystretch}{1.2}
    \caption{Test Case of Elsa's \textit{Sadness} When Told That Anna is Dead}
    \label{tab:elsa_TestCaseSketch}
    \begin{tabular}{P{0.13\linewidth}P{0.77\linewidth}}
        \toprule
        \colourRow\textbf{Setup} & $\mathtt{ProtectAnna} : \goaltype_{ATC} = \{
        \mathtt{goalState} = \{ \mathtt{Anna.Health} \geq 75\%$ \newline[5pt]
        $\wedge \; \mathtt{Anna.IsAlive} \},$ $\mathtt{importance} =
        \mathtt{High} \}$, \newline[5pt]
        $\mathtt{Fear}_i : \emotionintensitytype_{ATC} = \mathtt{Mid},
        \mathtt{Sadness}_i : \emotionintensitytype_{ATC} = \emptyset$,
        \newline[5pt]
        $\mathtt{S}_i : \worldstatetype = \{ \mathtt{Anna.Health} = h_0,
        \mathtt{Anna.IsAlive} = \True \} \text{ where } h_0 \in [75\%, 100\%]$
        \\
        \textbf{Input} & $\mathtt{AnnaHurt_E} : \worldstatechangetype = \{
        \mathtt{Anna.Health} = h_1 \} \text{ where } h_1 \in ( \mathtt{MIN}\%,
        25\%]$ \\
        \midrule
        \colourRow\textbf{Expected Output} & $\mathtt{Fear}_{i+1} :
        \emotionintensitytype_{ATC} = \mathtt{Low}, \mathtt{Sadness}_{i+1} :
        \emotionintensitytype_{ATC} = \mathtt{Mid}$ \\
        \bottomrule
    \end{tabular}
\end{table}
\vspace*{\fill}

\subsection{Sketch of an Acceptance Test Case Template of Elsa's
\textit{Admiration} of Anna}
\begin{figure}[!b]
    \centering
    \includegraphics[width=0.6\linewidth]{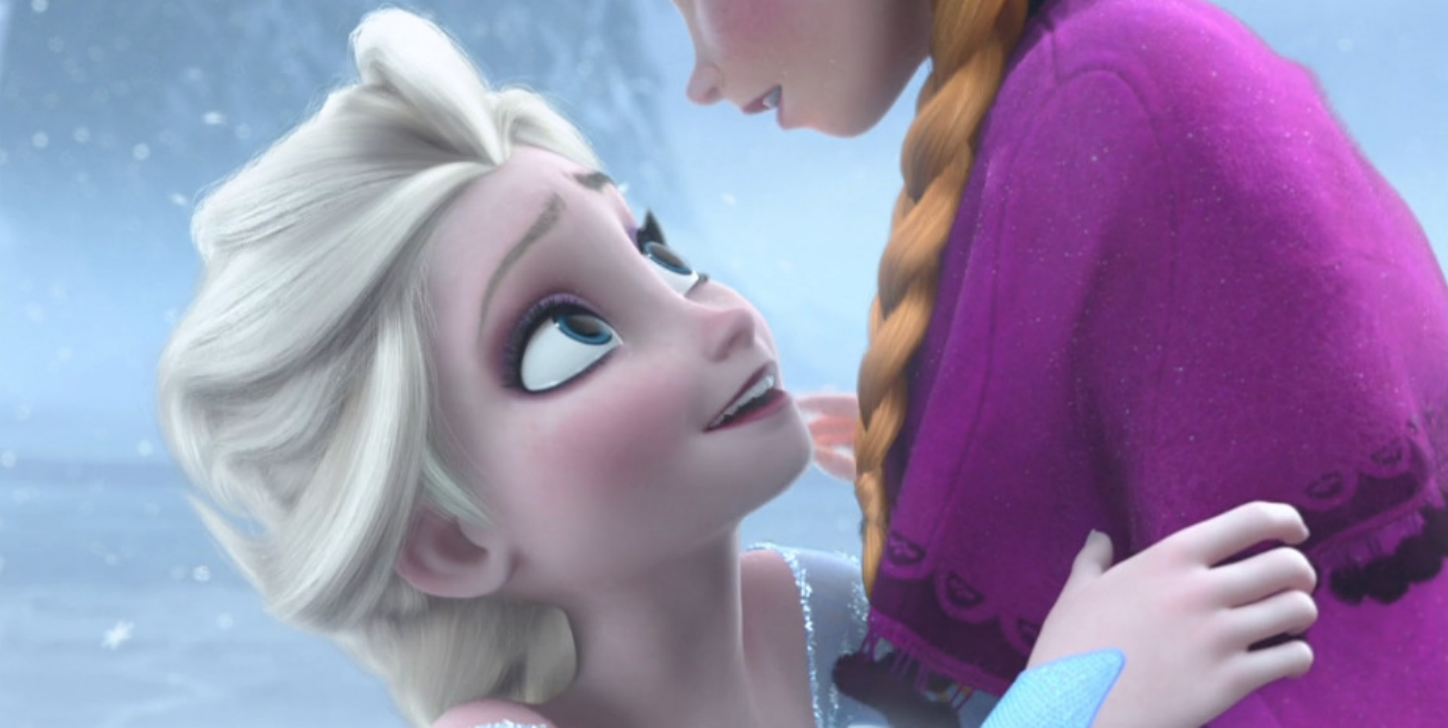}

    \caption{Elsa looks at Anna in \textit{Admiration} After She Thaws}
    \label{fig:elsaAdmiration}
\end{figure}

Immediately following the \textit{Grief} caused by Anna's transformation into
ice, Elsa experiences intense \textit{Acceptance} (i.e. \textit{Admiration})
towards Anna when she thaws and appears otherwise unharmed
(Figure~\ref{fig:elsaAdmiration}). Elsa's \textit{Admiration} is apparent via
her facial expression, tight clasp on Anna's arm and move to stand closer to
her, short utterance implying excitement (``Anna!''), and obvious concern for
Anna (see \textit{Acceptance} profile in Appendix~\ref{sec:acceptanceProfile}).

Based on the scenario, where the expected output is $\mathtt{Acceptance}_{i+2}
: \emotionintensitytype_{ATC} = \mathtt{High}$ and either no or very little
$\mathtt{Sadness}_{i+2} \in \{\emptyset, \mathtt{Low}\}$, the \textit{current}
world state is where Elsa is experiencing \textit{Grief} evaluated as
$\mathtt{S}_{i+1} = \mathtt{S}_i \oplus \mathtt{AnnaFreezes_E}$ from
Table~\ref{tab:elsa_EMgineTestCase}. The event of concern is Anna thawing,
which also seems to restore her physical health:
$$\mathtt{AnnaThaws_E} : \worldstatechangetype = \{ \mathtt{Anna.Health} = h_E,
\mathtt{Anna.IsAlive} = \True \} \text{ where } h_E \in [75\%, 100\%]$$

\noindent Applying this to $\mathtt{S}_{i+1}$ produces:
$$\mathtt{S}_{i+2} : \worldstatetype = \{ \mathtt{Anna.Health} = h_E,
\mathtt{Anna.IsAlive} = \True \} \text{ where } h_E \in [75\%, 100\%]$$

A distinguishing feature of \textit{Acceptance} is its reliance on social
attachments describing an entity's degree of liking or disliking another
entity. Representing this information requires another type, Social Attachment,
describing the relative degree of ``disliking'' and ``liking'':
$$\socialattachmenttype_{ATC} \in \{ \mathtt{Despises}, \mathtt{Dislikes},
\mathtt{Does Not Care For}, \mathtt{None}, \mathtt{Cares For}, \mathtt{Likes},
\mathtt{Loves} \}$$

From the ``transient'' knowledge about Elsa in Table~\ref{tab:elsa_Local}, her
reaction when Anna thaws, and supported by ``persistent'' knowledge about her
goal (Section~\ref{sec:elsa_Global}), it is clear that Elsa $\mathtt{Loves}$
Anna.

Adding Elsa's social attachment to Anna to the setup section of
Table~\ref{tab:elsa_TestCase} and exchanging some of its other components is
sufficient because it immediately follows that scenario in the story. The
necessary exchanges for the remaining elements of the acceptance test case
template are: $\mathtt{Sadness}_{i}$ with $\mathtt{Sadness}_{i+1}$ and
$\mathtt{S}_{i}$ with $\mathtt{S}_{i+1}$ in the setup section; the input
$\mathtt{AnnaFreezes_E}$ with $\mathtt{AnnaThaws_E}$; and the expected output
$\mathtt{Sadness}_{i+1}$ with $\mathtt{Sadness}_{i+2}$ and
$\mathtt{Acceptance}_{i+2}$. Table~\ref{tab:elsa_TestCaseAdmiration} shows the
resulting test case specification.

\begin{table}[!ht]
    \centering
    \renewcommand{\arraystretch}{1.2}
    \caption{Test Case of Elsa's \textit{Admiration} When Anna Thaws}
    \label{tab:elsa_TestCaseAdmiration}
    \begin{tabular}{P{0.13\linewidth}P{0.77\linewidth}}
        \toprule
        \colourRow\textbf{Setup} & $\mathtt{ProtectAnna} : \goaltype_{ATC} = \{
        \mathtt{goalState} = \{ \mathtt{Anna.Health} \geq 75\%$ \newline[5pt]
        $\wedge \; \mathtt{Anna.IsAlive} \},$ $\mathtt{importance} =
        \mathtt{High} \}$, \newline[5pt]
        $\mathtt{Anna} : \socialattachmenttype_{ATC} = \mathtt{Loves}$,
        \newline[5pt]
        $\mathtt{Sadness}_i : \emotionintensitytype_{ATC} =
        \mathtt{High}$,\newline[5pt]
        $\mathtt{S}_{i+1} : \worldstatetype = \{ \mathtt{Anna.Health} = h,
        \mathtt{Anna.IsAlive} = \False \} \text{ where } h \in ( \mathtt{MIN}\%,
        25\%]$ \\
        \textbf{Input} & $\mathtt{AnnaThaws_E} : \worldstatechangetype = \{
        \mathtt{Anna.Health} = h_E, \mathtt{Anna.IsAlive} = \True \}$ \newline
        where $h_E \in [75\%, 100\%]$ \\
        \midrule
        \colourRow\textbf{Expected Output} & $\mathtt{Sadness}_{i+2} :
        \emotionintensitytype_{ATC} \in \{\emptyset, \mathtt{Low}\}$,
        $\mathtt{Acceptance}_{i+2} : \emotionintensitytype_{ATC} =
        \mathtt{High}$ \\
        \bottomrule
    \end{tabular}
\end{table}

\begin{figure}[!b]
    \begin{center}
        \subfloat{
            \includegraphics[width=0.48\linewidth]{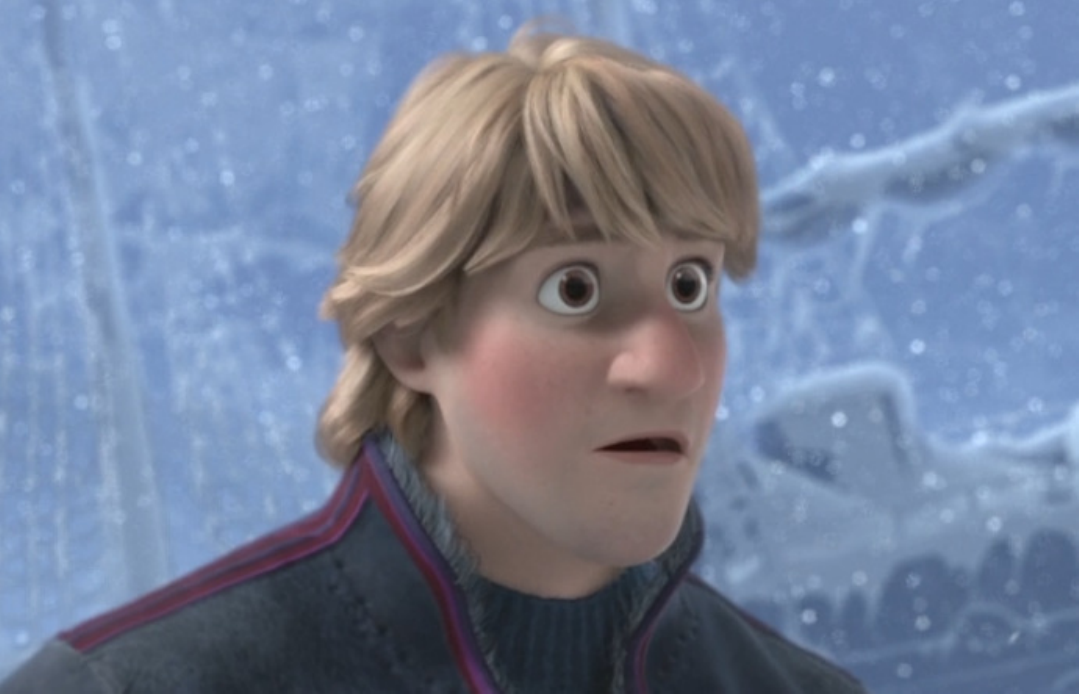}
        }
        \subfloat{
            \includegraphics[width=0.48\linewidth]{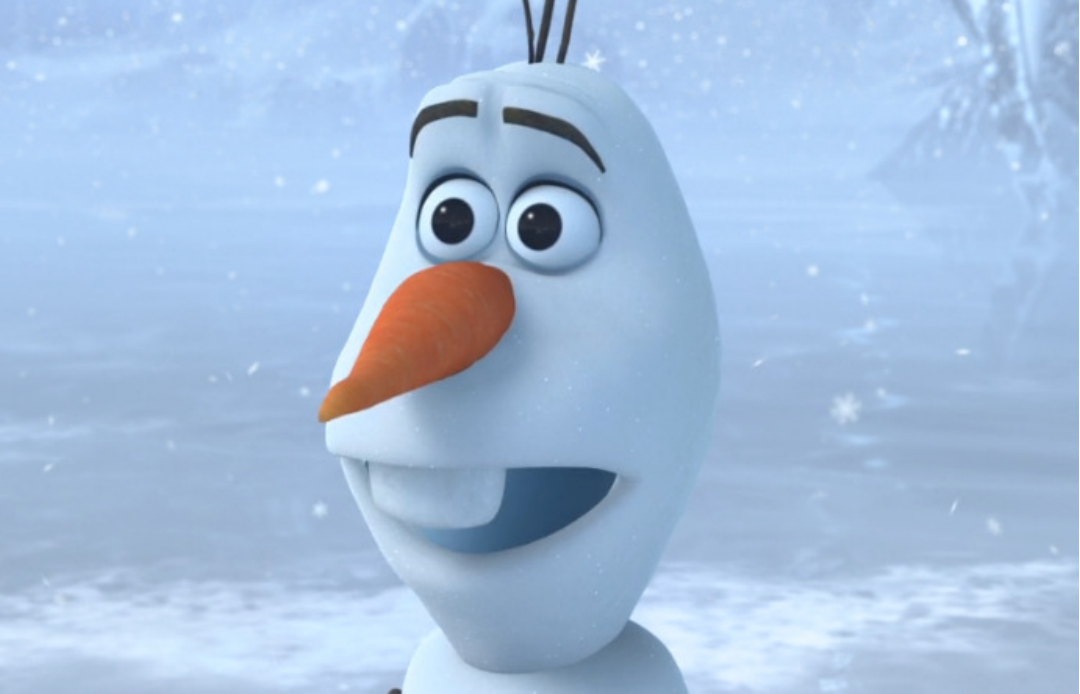}
        }

        \caption{Kristoff and Olaf look at Anna in \textit{Amazement} After She
            Thaws}
        \label{fig:kristoffAndOlaf}
    \end{center}
\end{figure}

\paragraph{A \textit{Surprising} Test Case} Although Elsa's reaction to Anna's
miraculous thawing and recovery is \textit{Admiration}, other characters---like
Kristoff and Olaf (Figure~\ref{fig:kristoffAndOlaf})---would more likely
experience an intense \textit{Surprise} because it is probability-based
(see \textit{Surprise} profile in Appendix~\ref{sec:surpriseProfile}). Elsa's
\textit{Admiration} might overwhelm any experience of \textit{Surprise} because
she feels responsible for Anna's ``death'' and her recovery absolves Elsa of
that guilt. This suggests that there could be an additional mechanism that
prioritizes emotion evaluations. This could be a simple ordering on the
available emotion kinds or something more complex such as an ordering based on
the external $\mathtt{CausedBy}$ function (Chapter~\ref{sec:worldKnowledge}).

\section{Summary}
Evaluating \progname{}'s success requires at least two user studies, which can
be both time and resource intensive to run well. Therefore, to maximize the
utility of these user studies, \progname{} must pass preliminary acceptance
test cases to ensure that it is behaving correctly with respect to its user
requirements: that the emotions it generates are plausible.

Following the proposed methodology, animated films are the source medium
because they are an audiovisual medium that are grounded in reality with
significant freedom to convey a character's thoughts and emotions in
``unrealistic'' ways. This means that they afford more cues to communicate
emotion with and makes it easier for audiences to agree on a character's
emotion state. ``Profiles'' for each emotion kind based on the discrete
perspective on emotion create a reference to aid in test case development
and recreation. They describe the emotion's purpose, its observable signs such
as action tendencies and facial expressions, and examples from animated
films---the source medium.

Extending the example of \textit{Grief} from the \textit{Sadness} profile, a
character study of Elsa from Disney's \textit{Frozen} led to an acceptance
test case template specification that uses ``fuzzy'' values like percentages to
improve its relevance to different CME designs and implementations. The
specification relies on a limited set of data types representing: a view of
relevant aspects of the world state; a world event that changes that state; a
goal describing the desired world state as a predicate, and the goal's relative
importance to Elsa; and the expected emotion's intensity. However, the
template's ``fuzziness'' means that testing efforts for specific CMEs---like
\progname{}---require extensions.

\clearpage
\vspace*{\fill}
\begin{keypoints}
    \begin{itemize}

        \item Animated films are likely the best source medium for CMEs like
        \progname{} because they are an audiovisual medium and grounded in,
        yet not limited by, reality

        \item The emotion ``profiles'' describe the emotion's purpose,
        observable signs, and examples of the emotion from the source medium
        (animated films)

        \item An example character analysis of Elsa from Disney's
        \textit{Frozen} demonstrates how both ``persistent'' and ``transient''
        character knowledge contribute to implementation-agnostic acceptance
        test case template specifications

        \item Due to its ``fuzziness'', templates require extensions for use
        with specific CMEs

    \end{itemize}
\end{keypoints}

\parasep
\vspace*{\fill}

%% file: validationExtend.tex
\chapter{Inspect Your \progname{}: Extending Acceptance Test
Case Templates}\label{chapter:testcaseEMgine}
\def\epigraphflush{center}
\setlength{\epigraphwidth}{0.85\textwidth}
\def\textflush{center}
\epigraph{Only by accepting this can one discover what they truly
want...}{Aigis, \textit{Shin Megami Tensei: Persona 3 FES}}

The acceptance test case templates aim to describe an emotion eliciting
scenario in the simplest and most generic form possible
(Chapter~\ref{sec:formalATC}): a World State View (WSV) and triggering event; a
way for the entity to know if they have achieved their goal and its relative
importance to them; and the expected emotion intensity as a relative quantity.
Comparing their specifications, they share a small subset of type definitions
with \progname{}. However, they are sufficiently different it is necessary to
extend them for \progname{} evaluations. This is ideal, despite the additional
effort, because each type of specification has its own concern. \progname{}
seeks to model particular theories with assumptions and design decisions that
someone else might not, whereas an acceptance test case wants to represent a
scenario with no expectations about the underlying processes, assumptions, or
design decisions. This separation helps build confidence in both kinds of
specification because it shows that one does not necessarily rely on the other,
implying that the acceptance test cases move towards an unbiased evaluation of
\progname{} and can serve as one measure for comparing with other CMEs.

Extensions of the Elsa-focused templates of \textit{Grief} and
\textit{Admiration} for \progname{}'s models demonstrates a way to translate
them while remaining consistent with the original test case and character
study. To build confidence in the \progname{}-specific extensions, they serve
as preliminary evaluations of some of \progname{}'s models ``on paper''. These
evaluations also start building confidence in \progname{}'s models before
making significant implementation efforts and separates modelling errors from
implementation errors.

\section[Extending the Acceptance Test Case Template of Elsa's
\textit{Grief}]{Extending the Acceptance Test Case Template of Elsa's
\textit{Gr-ief}}
\progname{} cannot use the test case describing Elsa's \textit{Grief}
(Table~\ref{tab:elsa_TestCase}) directly because it has different definitions
for $\goaltype_{ATC}$ and $\emotionintensitytype_{ATC}$. Further translations
must convert them to $\goaltype$ (Equation~\ref{eq:goaltype}) and
$\emotionintensitytype$ (Equation~\ref{eq:intensitytype}), respectively.

In both $\goaltype_{ATC}$ and $\goaltype$, the definition of
$\mathtt{goalState}$ is identical so no changes are necessary. Both types also
have $\mathtt{importance}$, but use different specifications that are not
directly compatible: $\{ \mathtt{Low}, \mathtt{Mid},$ $\mathtt{High} \}$ and
$\mathbb{R}_{\geq 0}$. To connect the specifications, a partition suitable for
\progname{} on $\mathbb{R}_{\geq 0}$ mapping to the ``subsets'' $\{
\mathtt{Low}, \mathtt{Mid},$ $ \mathtt{High} \}$ is:
$$\mathtt{importance} \in \left\{ [0], \biggl(0, \dfrac{1}{3}\biggr],
\biggl(\dfrac{1}{3}, \dfrac{2}{3} \biggr], \biggl(\dfrac{2}{3}, 1\biggr]
\right\}$$

\progname{} reserves $0$ to represent an irrelevant goal
(Equation~\ref{eq:goaltype}), necessitating a separate partition for $0 :
\mathbb{R}_{\geq 0}$. Other CMEs might decide to represent irrelevant goals
differently, so the partition $[0]$ is a \progname{}-specific element.

The partitions divide the range $(0,1]$ into three parts such that there is a
linear relation between $\mathtt{importance}$ values. For example, a goal with
$\mathtt{importance} = 1.0$ has $0.2$ more impact on an entity's evaluations
than one with $\mathtt{importance} = 0.8$. Users can define some maximum
$\mathtt{importance}$ value $m_G$, then use it to normalize a goal's
$\mathtt{importance}$ value to map it to a partition.

\progname{}'s definition of $\goaltype$ has a function $\mathtt{goal}(s)$ that
evaluates the distance between a WSV and $\mathtt{goalState}$, and its
derivative $\mathtt{goal'}(s, s_{\Delta})$ evaluates a change in a WSV's
distance to $\mathtt{goalState}$ caused by an event. The desirability a WSV
relative to $\mathtt{goalState}$ defines these functions as:
\begin{gather*}
    \mathtt{goal}(s) : \worldstatetype \rightarrow \statedistancetype =
    \begin{cases}
        +\infty, & \mathtt{s.Anna.IsAlive} = \False \\[5pt]
        \dfrac{75\% - \mathtt{s.Anna.Health}}{75\%}, & \mathtt{s.Anna.Health} <
        75\% \\
        0, & \mathit{Otherwise}
    \end{cases} \\
    \mathtt{goal'}(s, s_{\Delta}) : \worldstatetype \times \worldstatechangetype
    \rightarrow \statedistancechangetype = \mathtt{goal}(s \oplus s_{\Delta}) -
    \mathtt{goal}(s)
\end{gather*}

This model assumes that if $\mathtt{goal}(s \oplus s_{\Delta}) =
\mathtt{goal}(s) = +\infty$, then $\mathtt{goal'}(s, s_{\Delta}) = 0$.

The definition of $\goaltype$ also has a $\mathtt{type}$ so that goals can be
marked as \textit{Self-Preservation} and/or \textit{Gustatory} as needed.
Elsa's character analysis implies that Elsa would be permanently changed if
Anna died. Common sense suggests that people avoid this because it can be
mentally and emotionally draining. Therefore, \progname{} associates this goal
as self-preservational, $\mathtt{type} = \{ \mathtt{SelfPreservation} \}$,
indicating that Elsa wants to ``preserve'' her mental state. There is nothing
to suggest that $\mathtt{ProtectAnna}$ is gustatory.

Transforming emotion Intensity $\emotionintensitytype_{ATC} \in \{
\mathtt{Low}, \mathtt{Mid}, \mathtt{High} \}$ to $\emotionintensitytype \in
\mathbb{R}_{\geq 0}$ uses the same approach as goal $\mathtt{importance}$,
becoming:
$$\emotionintensitytype \in \left\{ [0], \biggl(0, \dfrac{1}{3}\biggr],
\biggl(\dfrac{1}{3}, \dfrac{2}{3} \biggr], \biggl(\dfrac{2}{3}, 1\biggr]
\right\}$$

\progname{} uses $0$ to represent the absence of an emotion
(Equation~\ref{eq:intensitytype}), necessitating a separate partition for $0 :
\emotionintensitytype$. Users can define some maximum $\emotionintensitytype$
value $m_I$, then use it to normalize an intensity value to map it to a
partition. Table~\ref{tab:elsa_EMgineTestCase} shows the final
\progname{}-specific test case.

\begin{table}[!tb]
    \centering
    \renewcommand{\arraystretch}{1.2}
    \caption{Test Case of Elsa's \textit{Grief} Extended for \progname{} Type
    Definitions}
    \label{tab:elsa_EMgineTestCase}
    \begin{tabular}{P{0.13\linewidth}P{0.77\linewidth}}
        \toprule
        \colourRow\textbf{Setup} & $\mathtt{ProtectAnna} : \goaltype = \{
        \mathtt{goalState} = \{ \mathtt{Anna.Health} \geq 75\% \wedge
        \mathtt{Anna.IsAlive} \}$, \newline[10pt]
        $\mathtt{goal}(s) : \worldstatetype \rightarrow \statedistancetype =
        \begin{cases}
            +\infty, & \mathtt{s.Anna.IsAlive} = \False \\[5pt]
            \dfrac{75\% - \mathtt{s.Anna.Health}}{75\%}, &
            \mathtt{s.Anna.Health} < 75\% \\
            0, & \mathit{Otherwise}
        \end{cases}$, \newline[10pt]
        $\mathtt{goal'}(s, s_{\Delta}) : \worldstatetype \times
        \worldstatechangetype \rightarrow \statedistancechangetype =
        \mathtt{goal}(s \oplus s_{\Delta}) - \mathtt{goal}(s)$, \newline[10pt]
        $\mathtt{importance} = \mathit{im} \in \biggl(\dfrac{2}{3}, 1\biggr],
        \mathtt{type} = \{ \mathtt{SelfPreservation} \} \}$, \newline[10pt]
        $\mathtt{Sadness}_i : \emotionintensitytype = I_i \in
        \biggl(\dfrac{1}{3}, \dfrac{2}{3} \biggr]$, \newline[10pt]
        $\mathtt{S}_i : \worldstatetype = \{ \mathtt{Anna.Health} = h,
        \mathtt{Anna.IsAlive} = \True \} \text{ where } h \in ( \mathtt{MIN}\%,
        25\%]$ \\
        \textbf{Input} & $\mathtt{AnnaFreezes_E} : \worldstatechangetype =
        \left\{ \mathtt{Anna.IsAlive} = \False \right\}$ \\
        \midrule
        \colourRow\textbf{Expected Output} & $\mathtt{Sadness}_{i+1} :
        \emotionintensitytype = I_{i + 1} \in \biggl(\dfrac{2}{3}, 1\biggr]$ \\
        \bottomrule
    \end{tabular}
\end{table}

\subsection{Trying the Extended Test Case Specification on \progname{}'s
\textit{Sadness} Elicitation Model}\label{sec:atc_Sadness}
The \textit{Sadness} elicitation model (Equation~\ref{eq:generatesadness})
requires a goal or a plan (Equation~\ref{eq:plantype}), a WSV, and an event as
inputs. The test case defines all of these except a plan for achieving
$\mathtt{ProtectAnna}$, so its definition is None. Recalling the
\textit{Sadness} elicitation model:
\begin{gather*}
    S(g : \goaltype^?, p : \plantype^?, s_{prev} : \worldstatetype,
    s_\Delta : \worldstatechangetype) : ( g_{sadness} : \goaltype^?,
    p_{sadness} : \plantype^?, s_{now} : \worldstatetype,
    \mathit{dist}_{now} : \statedistancetype^? )^? \\[5pt]
    \defEq \begin{cases}
        (\text{None}, p, s_{prev} \oplus s_\Delta, \text{None}), &
        \parbox{0.54\linewidth}{$p \neq \text{None} \land
            p.\mathtt{isFeasible}(s_{prev}) \\
            \land \neg p.\mathtt{isFeasible}(s_{prev} \oplus s_\Delta)$}
            \\[10pt]
        \parbox{0.22\linewidth}{$(g, \text{None}, s_{prev} \oplus s_\Delta, \\
            g.\mathtt{goal}\left(s_{prev} \oplus s_\Delta\right)),$} &
        g \neq \text{None} \land \mathtt{IsUnachievableAfterEvent}(g,
        s_{prev}, s_\Delta) \\[10pt]
        \emptyset, & Otherwise \\
    \end{cases}
\end{gather*}

\noindent Substituting (``$\rightsquigarrow$'') data from
Table~\ref{tab:elsa_EMgineTestCase} and $p = $ None into the \textit{Sadness}
elicitation model $S$ then starting to solve leads to its goal-focused branch:
\begin{gather*}
    S(g = \mathtt{ProtectAnna}, p = \text{None}, s_{prev} = \mathtt{S}_i,
    s_{\Delta} = \mathtt{AnnaFreezes_E}) \\
    \rightsquigarrow \begin{cases}
        \parbox{0.24\linewidth}{$(\mathtt{ProtectAnna}, \text{None}, \\
            \mathtt{S}_i \oplus \mathtt{AnnaFreezes_E}, \\
            \mathtt{ProtectAnna}.\mathtt{goal}\\
            \left(\mathtt{S}_i \oplus \mathtt{AnnaFreezes_E}\right)),$} &
        \parbox{0.5\linewidth}{$\mathtt{IsUnachievableAfterEvent}(\mathtt{ProtectAnna},
            \mathtt{S}_i, \\
            \mathtt{AnnaFreezes_E})$} \\[10pt]
        \emptyset, & Otherwise \\
    \end{cases}
\end{gather*}

\noindent Now, \textit{Sadness} elicitation depends on the evaluation of
$\mathtt{IsUnachievableAfterEvent}$ (Chapter~\ref{sec:evalOJL}). Focusing on
that piece gives:
\begin{gather*}
    \mathtt{IsUnachievableAfterEvent}(\mathtt{ProtectAnna}, \mathtt{S}_i,
    \mathtt{AnnaFreezes_E}) \\
    \rightsquigarrow | \, \mathtt{ProtectAnna}.\mathtt{goal}\left(\mathtt{S}_i
    \oplus \mathtt{AnnaFreezes_E}\right) | \, = +\infty \\
    \rightsquigarrow | \, \mathtt{ProtectAnna}.\mathtt{goal}(\{
    \mathtt{Anna.IsAlive} = \True \} \oplus \left\{ \mathtt{Anna.IsAlive} =
    \False \right\}) | \, = +\infty \\
    \rightsquigarrow | \, \mathtt{ProtectAnna}.\mathtt{goal}(\{
    \mathtt{Anna.IsAlive} = \False \}) | \, = +\infty
\end{gather*}

\noindent Looking up the value of $\mathtt{ProtectAnna}.\mathtt{goal}(\{
\mathtt{Anna.IsAlive} = \False \})$ in the \progname{}-specific test case
specification (Table~\ref{tab:elsa_EMgineTestCase}) gives:
\begin{gather*}
    \rightsquigarrow | +\infty | = +\infty \\
    \rightsquigarrow \True
\end{gather*}

The WSV $\mathtt{S}_i$ also has the field $\mathtt{Anna.Health}$, but this
example omits it for simplicity because it is not relevant to this evaluation.

\noindent Therefore, the model outputs a tuple indicating that the event
$\mathtt{AnnaFreezes_E}$ elicits \textit{Sadness} from Elsa:
\begin{gather*}
    ( g_{sadness} = \mathtt{ProtectAnna},
    \text{None}, s_{now} = \{ \mathtt{Anna.Health} = h, \mathtt{Anna.IsAlive} =
    \False \}, \\
    \mathit{dist}_{now} = +\infty) \text{ where } h \in (\mathtt{MIN}\%, 25\%]
\end{gather*}

\subsection{Trying the Extended Test Case Specification on \progname{}'s
\textit{Sadness} Intensity Model}\label{sec:atc_SadnessI}
Recalling the \textit{Sadness} intensity model
(Equation~\ref{eq:evalIntensitySadness}):
\begin{gather*}
    S_\Delta(g : \goaltype^?, p : \plantype^?, s_{prev} : \worldstatetype,
    i_{\mathit{max}\Delta} : \responsestrength)
    : \responsestrength \defEq \begin{cases}
        \dfrac{1}{|\mathit{dist}_p|}, & p \neq \text{None} \\[15pt]
        \dfrac{g.\mathtt{importance}}{m_G} \cdot i_{\mathit{max}\Delta}, & g
        \neq \text{None} \\
    \end{cases} \\
    \text{where } \mathit{dist}_p : \statedistancetype =
    \mathtt{Dist}(s_{prev}, p.\mathtt{nextStep}(s_{prev},
    |p.\mathtt{actions}|))
\end{gather*}

\noindent Substituting (``$\rightsquigarrow$'') data from the output of the
\textit{Sadness} elicitation evaluation into $S_\Delta$ and setting $m_G = 1$
and $i_{\mathit{max}\Delta} = m_I = 1$ to reflect the maximum values
\progname{} defines for goal $\mathtt{importance}$ and emotion intensity
$\emotionintensitytype$, then starting to solve leads to its goal-focused
branch:
\begin{gather*}
    S_\Delta(g = \mathtt{ProtectAnna}, p = \text{None},  s_{prev} =
    \mathtt{S}_i, i_{\mathit{max}\Delta} = 1) \\
    \rightsquigarrow \dfrac{\mathtt{ProtectAnna}.\mathtt{importance}}{1} \cdot
        1
\end{gather*}

\noindent Looking up the required values in Table~\ref{tab:elsa_EMgineTestCase}
gives:
\begin{gather*}
    \rightsquigarrow \mathit{im} \text{ where } \mathit{im} \in
    \biggl(\dfrac{2}{3}, 1\biggr]
\end{gather*}

\noindent The intensity of Elsa's \textit{Sadness} is directly proportional to
the importance of $\mathtt{ProtectAnna}$ ($\emotionintensitytype \propto
\mathit{im}$), suggesting that it would be relatively high. This aligns with the
expected outcome $\mathtt{Sadness}_{i+1}$ (Table~\ref{tab:elsa_EMgineTestCase}).

\section{Extending the Acceptance Test Case Template of Elsa's
\textit{Admiration} of Anna}
For extending the \textit{Admiration} test case template
(Table~\ref{tab:elsa_TestCaseAdmiration}), it is sufficient to use the
\progname{}-specific test case in Table~\ref{tab:elsa_EMgineTestCase} and
replace some data with that of the \textit{Admiration} test case. This first
requires a mapping between $\socialattachmenttype_{ATC}$ and \progname{}'s
social attachment type $\socialattachmenttype$ (Equation~\ref{eq:socialtype}).
To connect the specifications, a partition suitable for \progname{} on
$\mathbb{Z}$ mapping to the ``subsets'' $\{ \mathtt{Despises},
\mathtt{Dislikes}, \mathtt{Does Not Care For}, \mathtt{None},$ $\mathtt{Cares
For}, \mathtt{Likes}, \mathtt{Loves} \}$ is:
\begin{gather*}
    \socialattachmenttype \in \{ [-m_{\mathit{SA}}, -a_{Mid}), [-a_{Mid},
    -a_{Low}), [-a_{Low}, 0), [0], (0, a_{Low}], (a_{Low}, a_{Mid}], (a_{Mid},
    m_{\mathit{SA}}] \} \\
    \text{ where } a_{Low} = \biggl\lceil
    \dfrac{m_{\mathit{SA}}}{3}\biggr\rceil, a_{Mid} = \biggl\lceil\dfrac{2
    \cdot m_{\mathit{SA}}}{3}\biggr\rceil \text{ and } m_{\mathit{SA}} \in
    \mathbb{N}_{\geq 3}
\end{gather*}

The partitions divide the range $[-m_{\mathit{SA}}, m_{\mathit{SA}}]$ into
seven parts such that there is a linear relation between social attachment
``degrees''. A ``degree'' of zero represents no attachment ($0 =
\mathtt{None}$). The maximum attachment ``degree'' $m_{\mathit{SA}}$ must be at
least three to distinguish the three ``degrees'' of liking ($\mathtt{Cares
For}, \mathtt{Likes}, \mathtt{Loves}$) and disliking ($\mathtt{Does Not Care
For}, \mathtt{Dislikes}, \mathtt{Despises}$). The ceiling function ensures that
the partition boundaries are integers to satisfy the constraints on
$\socialattachmenttype$. Table~\ref{tab:elsa_EMgineTestCase2} shows the
resulting test case specification.

\begin{table}[!b]
    \centering
    \renewcommand{\arraystretch}{1.2}
    \caption{Test Case of Elsa's \textit{Admiration} of Anna for \progname{}
    Type Definitions}
    \label{tab:elsa_EMgineTestCase2}
    \begin{tabular}{P{0.13\linewidth}P{0.77\linewidth}}
        \toprule
        \colourRow\textbf{Setup} & $\mathtt{ProtectAnna} : \goaltype = \{
        \mathtt{goalState} = \{ \mathtt{Anna.Health} \geq 75\% \wedge
        \mathtt{Anna.IsAlive} \}$, \newline[10pt]
        $\mathtt{goal}(s) : \worldstatetype \rightarrow \statedistancetype =
        \begin{cases}
            +\infty, & \mathtt{s.Anna.IsAlive} = \False \\[5pt]
            \dfrac{75\% - \mathtt{s.Anna.Health}}{75\%}, &
            \mathtt{s.Anna.Health} < 75\% \\
            0, & \mathit{Otherwise}
        \end{cases}$, \newline[10pt]
        $\mathtt{goal'}(s, s_{\Delta}) : \worldstatetype \times
        \worldstatechangetype \rightarrow \statedistancechangetype =
        \mathtt{goal}(s \oplus s_{\Delta}) - \mathtt{goal}(s)$, \newline[10pt]
        $\mathtt{importance} = \mathit{im} \in \biggl(\dfrac{2}{3}, 1\biggr],
        \mathtt{type} = \{ \mathtt{SelfPreservation} \} \}$, \newline[10pt]
        $\mathtt{Anna} : \socialattachmenttype = a \in
        \biggl(\biggl\lceil\dfrac{2 \cdot m_{\mathit{SA}}}{3}\biggr\rceil,
        m_{\mathit{SA}}\biggr]$, \newline[10pt]
        $\mathtt{Sadness}_{i+1} : \emotionintensitytype = I_{i + 1} \in
        \biggl(\dfrac{2}{3}, 1\biggr]$, \newline[10pt]
        $S_{i+1} = \{ \mathtt{Anna.Health} = h_{i+1}, \mathtt{Anna.IsAlive} =
        \False \}$ \newline
        where $h_{i+1} \in ( \mathtt{MIN}\%, 25\%]$ \\
        \textbf{Input} & $\mathtt{AnnaThaws_E} : \worldstatechangetype = \{
        \mathtt{Anna.Health} = h_E, \mathtt{Anna.IsAlive} = \True \}$ \newline
        where $h_E \in [75\%, 100\%]$ \\
        \midrule
        \colourRow\textbf{Expected Output} & $\mathtt{Sadness}_{i+2} :
        \emotionintensitytype = Is_{i + 2} \in \left[0, \dfrac{1}{3}\right]$,
        $\mathtt{Acceptance}_{i+2} : \emotionintensitytype = Ia_{i + 2} \in
        \biggl(\dfrac{2}{3}, 1\biggr]$ \\
        \bottomrule
    \end{tabular}
\end{table}

\subsection{Trying the Extended Test Case Specification with \progname{}'s
\textit{Joy} and \textit{Acceptance} Models}
\progname{} treats \textit{Acceptance} as a complex emotion based on
\textit{Joy}.  Therefore, the acceptance test case in
Table~\ref{tab:elsa_EMgineTestCase2} is sufficient to evaluate \progname{}'s
models of \textit{Joy} and \textit{Acceptance} elicitation and intensity
(Equations~\ref{eq:generatejoy}, \ref{eq:generateemotionAcceptance},
\ref{eq:evalIntensityJoy}, and \ref{eq:evalIntensityAcceptance}) ``on paper''.
This dependency means that the first evaluation must be of \textit{Joy}.

\subsubsection{Evaluating \progname{}'s \textit{Joy} Elicitation and Intensity
Models}\label{sec:atc_Joy}
The \textit{Joy} elicitation model (Equation~\ref{eq:generatejoy}) requires a
goal, WSV, event, and a ``tolerance'' value for distance changes between WSVs.
The test case defines all of these except for a ``tolerance'' value, which
relies on Elsa. Recalling the \textit{Joy} elicitation model:
\begin{gather*}
    J(g : \goaltype, s_{prev} : \worldstatetype, s_\Delta :
    \worldstatechangetype, \epsilon_J : \statedistancechangetype) :
    (\mathit{dist}_{prev} : \statedistancetype, \mathit{dist}_{now} :
    \statedistancetype, \mathit{dist}_\Delta : \statedistancechangetype)^?
    \\[5pt]
    \defEq \begin{cases}
        \parbox{0.2\linewidth}{$(g.\mathtt{goal}(s_{prev}), \\
            g.\mathtt{goal}(s_{prev} \oplus s_\Delta), \\
            g.\mathtt{goal'}(s_{prev}, s_\Delta))$, } &
        \parbox{0.35\linewidth}{$\mathtt{IsCloserAfterEvent}(g, s_{prev},
            s_\Delta) \\
            \land \mathtt{IsNoticeable}(g, s_{prev}, s_\Delta,
            \epsilon_J)$} \\[20pt]
        \emptyset, & Otherwise \\
    \end{cases}
\end{gather*}

\noindent Substituting (``$\rightsquigarrow$'') data from
Table~\ref{tab:elsa_EMgineTestCase2} into the \textit{Joy} elicitation model
$J$ gives:
\begin{gather*}
    J(g = \mathtt{ProtectAnna}, s_{prev} = \mathtt{S}_{i+1}, s_{\Delta} =
    \mathtt{AnnaThaws_E}, \epsilon_J) \\
    \rightsquigarrow \begin{cases}
        \parbox{0.43\linewidth}{$(\mathtt{ProtectAnna}.\mathtt{goal}(\mathtt{S}_{i+1}),
         \\
            \mathtt{ProtectAnna}.\mathtt{goal}(\mathtt{S}_{i+1} \oplus
            \mathtt{AnnaThaws_E}), \\
            \mathtt{ProtectAnna}.\mathtt{goal'}(\mathtt{S}_{i+1},
            \mathtt{AnnaThaws_E}))$, } &
        \parbox{0.43\linewidth}{$\mathtt{IsCloserAfterEvent}(\mathtt{ProtectAnna},
         \mathtt{S}_{i+1}, \\
         \mathtt{AnnaThaws_E}) \\
            \land \mathtt{IsNoticeable}(\mathtt{ProtectAnna}, \mathtt{S}_{i+1},
            \\
            \mathtt{AnnaThaws_E}, \epsilon_J)$} \\[25pt]
        \emptyset, & Otherwise \\
    \end{cases}
\end{gather*}

\noindent \textit{Joy} elicitation depends on the evaluation of
$\mathtt{IsCloserAfterEvent}$ and $\mathtt{IsNoticeable}$ (Chapter
\ref{sec:evalOJL}). Focusing on those pieces and using the definition of
$\mathtt{S}_{i+1}$ from Table~\ref{tab:elsa_EMgineTestCase2} gives:
\begin{gather*}
    \mathtt{IsCloserAfterEvent}(\mathtt{ProtectAnna}, \mathtt{S}_{i+1},
    \mathtt{AnnaThaws_E}) \\
    \land \, \mathtt{IsNoticeable}(\mathtt{ProtectAnna}, \mathtt{S}_{i+1},
    \mathtt{AnnaThaws_E}, \epsilon_J) \\
    \rightsquigarrow \mathtt{ProtectAnna}.\mathtt{goal}(\mathtt{S}_{i+1}) >
    \mathtt{ProtectAnna}.\mathtt{goal}\left(\mathtt{S}_{i+2}\right) \\
    \land \, | \, \mathtt{ProtectAnna}.\mathtt{goal'}\left(\mathtt{S}_{i+1},
    \mathtt{AnnaThaws_E}\right)| > \epsilon_J \\
    \text{ where } \mathtt{S}_{i+2} = \{ \mathtt{S}_{i+1} \oplus
    \mathtt{AnnaThaws_E} \} \rightsquigarrow \{ \mathtt{Anna.Health} = h_E,
    \mathtt{Anna.IsAlive} = \True \}, \\
    h_{i+1} \in ( \mathtt{MIN}\%, 25\%],  \text{ and } h_E \in [75\%, 100\%]
\end{gather*}

\noindent Looking up the values of
$\mathtt{ProtectAnna}.\mathtt{goal}(\mathtt{S}_{i+1})$,
$\mathtt{ProtectAnna}.\mathtt{goal}(\mathtt{S}_{i+2})$, and
$\mathtt{ProtectAnna}.\mathtt{go-}$ $\mathtt{al'}(\mathtt{S}_{i+1},
\mathtt{AnnaThaws_E})$ in the \progname{}-specific test case specification
(Table~\ref{tab:elsa_EMgineTestCase2}) gives:

\begin{gather*}
    \rightsquigarrow \mathtt{ProtectAnna}.\mathtt{goal}(\mathtt{S}_{i+1}) >
    \mathtt{ProtectAnna}.\mathtt{goal}\left(\mathtt{S}_{i+2}\right) \\
    \land \, | \, \mathtt{ProtectAnna}.\mathtt{goal}(\mathtt{S}_{i+2}) -
    \mathtt{ProtectAnna}.\mathtt{goal}(\mathtt{S}_{i+1})| > \epsilon_J \\
    \rightsquigarrow \left(\dfrac{75\% - h_{i+1}}{75\%} + \infty \right) > 0
    \land \biggl| \, 0 - \left(\dfrac{75\% - h_{i+1}}{75\%} + \infty\right)
    \biggr| > \epsilon_J \\
    \rightsquigarrow \infty > 0 \land | \, 0 - \infty | > \epsilon_J \\
    \rightsquigarrow \True \land |-\infty| > \epsilon_J \\
    \rightsquigarrow \True \\
    \text{ where } \mathtt{S}_{i+2} = \{ \mathtt{S}_{i+1} \oplus
    \mathtt{AnnaThaws_E} \} \rightsquigarrow \{ \mathtt{Anna.Health} = h_E,
    \mathtt{Anna.IsAlive} = \True \}, \\
    h_{i+1} \in ( \mathtt{MIN}\%, 25\%],  \text{ and } h_E \in [75\%, 100\%]
\end{gather*}

\noindent Therefore, the model outputs a tuple indicating that the event
$\mathtt{AnnaThaws_E}$ elicits \textit{Joy} from Elsa:
\begin{gather*}
    \left( \mathit{dist}_{prev} = \infty, \mathit{dist}_{now} = 0,
    \mathit{dist}_\Delta = -\infty \right)
\end{gather*}

\noindent Recalling the \textit{Joy} intensity model
(Equation~\ref{eq:evalIntensityJoy}):
\begin{gather*}
    J_\Delta(g : \goaltype, d_\Delta : \statedistancechangetype) :
    \responsestrength \defEq |d_\Delta| \cdot g.\mathtt{importance}
\end{gather*}

\noindent Substituting (``$\rightsquigarrow$'') data from the output of the
\textit{Joy} elicitation evaluation into $J_\Delta$ gives:
\begin{gather*}
    J_\Delta(g = \mathtt{ProtectAnna}, d_\Delta = -\infty) \\
    \rightsquigarrow |-\infty| \cdot \mathtt{ProtectAnna}.\mathtt{importance} \\
    \rightsquigarrow |-\infty| \\
    \rightsquigarrow \infty
\end{gather*}

The intensity of Elsa's \textit{Joy} is maximized because she believed that her
goal to $\mathtt{ProtectAnna}$ was unachievable and the event
$\mathtt{AnnaThaws_E}$ made it achievable again, which \textit{should} elicit a
very strong reaction. Assuming that the intensity change of \textit{Joy} is
proportional to that of \textit{Acceptance} due to their interconnection
(${\responsestrength}_{\mathit{Joy}} \propto
{\responsestrength}_{\mathit{Acceptance}}$), this aligns with the expected
outcome $\mathtt{Joy}_{i+2} \propto \mathtt{Acceptance}_{i+2}$
(Table~\ref{tab:elsa_EMgineTestCase2}).

\subsubsection{Evaluating \progname{}'s \textit{Acceptance} Elicitation
and Intensity Models}\label{sec:atc_Acceptance}
The \textit{Acceptance} elicitation model
(Equation~\ref{eq:generateemotionAcceptance}) requires a social attachment,
goal, WSV, event, and two ``tolerance'' values for distance changes between
WSVs. The test case defines all of these except for the ``tolerance'' values,
which rely on Elsa. Recalling the \textit{Acceptance} elicitation model:
\begin{gather*}
    \mathit{Acc}(r_A : {\socialattachmenttype}^?, g : \goaltype, s_{prev} :
    \worldstatetype, s_\Delta : \worldstatechangetype, \epsilon_{A1} :
    \worldstatechangetype, \epsilon_{A2} : \worldstatechangetype) :
    (r_A : \socialattachmenttype, \mathit{distAttribToA}_\Delta :
    \statedistancechangetype)^? \\
    \defEq \begin{cases}
        (r_A, dist_\Delta - \epsilon_{A2}), &
        \parbox{0.46\linewidth}{$r_A \neq \text{None} \, \land \, | J(g,
            s_{prev}, s_\Delta, \epsilon_{A1}).dist_\Delta | > \epsilon_{A2} \\
            \land \mathtt{CausedBy}(s_\Delta, A)$} \\[10pt]
        \emptyset, & Otherwise \\
    \end{cases}
\end{gather*}

\noindent Substituting (``$\rightsquigarrow$'') data from
Table~\ref{tab:elsa_EMgineTestCase2} into the \textit{Acceptance} elicitation
model $\mathit{Acc}$ gives:
\begin{gather*}
    \mathit{Acc}(r_A = \mathtt{Anna}, g = \mathtt{ProtectAnna}, s_{prev} =
    \mathtt{S}_{i+1}, s_\Delta = \mathtt{AnnaThaws_E}, \epsilon_{A1} =
    \epsilon_J, \epsilon_{A2}) \\
    \rightsquigarrow \begin{cases}
        (\mathtt{Anna}, dist_\Delta - \epsilon_{A2}), &
        \parbox{0.55\linewidth}{$\mathtt{Anna} \neq \text{None} \\
            \land \, | J(\mathtt{ProtectAnna}, \mathtt{S}_{i+1},
            \mathtt{AnnaThaws_E}, \epsilon_J).dist_\Delta | > \epsilon_{A2} \\
            \land \mathtt{CausedBy}(\mathtt{AnnaThaws_E}, \text{Anna})$}
            \\[10pt]
        \emptyset, & Otherwise \\
    \end{cases}
\end{gather*}

\noindent \textit{Acceptance} elicitation depends on the evaluation of
\textit{Joy} elicitation and some externally defined $\mathtt{CausedBy}$
function. Focusing on those pieces and using the output of the \textit{Joy}
elicitation acceptance test case example gives:
\begin{gather*}
    \mathtt{Anna} \neq \text{None} \land | -\infty | > \epsilon_{A2} \land
    \mathtt{CausedBy}(\mathtt{AnnaThaws_E}, \text{Anna}) \\
    \mathtt{Anna} \neq \text{None} \land \infty > \epsilon_{A2} \land
    \mathtt{CausedBy}(\mathtt{AnnaThaws_E}, \text{Anna}) \\
    \True \land \True \land \mathtt{CausedBy}(\mathtt{AnnaThaws_E},
    \text{Anna}) \\
    \mathtt{CausedBy}(\mathtt{AnnaThaws_E}, \text{Anna})
\end{gather*}

\noindent The components that \progname{} evaluates---$\mathtt{Anna} \neq
\text{None}$ and $| -\infty | > \epsilon_{A2}$---resolve to the expected value
$\True$, so the output is $(\mathtt{Anna}, -\infty)$ if
$\mathtt{CausedBy}(\mathtt{AnnaThaws_E}, \text{Anna}) = \True$. This external
function depends on how Elsa attributes causality to entities, which likely
relies on some reasoning process unless users hard-code values
(Chapter~\ref{sec:worldKnowledge}). Therefore, the output depends on the
assumption that Elsa attributes Anna's thawing to Anna herself. Anticipating
that Elsa attributes the event $\mathtt{AnnaThaws_E}$ seems reasonable because
Elsa knows that she is not responsible, no one else is capable of reviving
Anna, and Elsa does not appear to attribute the event to a spirit or higher
being. However, the output of $\mathtt{CausedBy}$ is not known for certain and
is beyond the model's scope.

Recalling the \textit{Acceptance} intensity model
(Equation~\ref{eq:evalIntensityAcceptance}):
\begin{gather*}
    \mathit{Acc}_\Delta(r_A : \socialattachmenttype, r_\mathit{min} :
    \socialattachmenttype, d_\Delta : \statedistancechangetype) :
    \responsestrength \defEq \begin{cases}
        |d_\Delta| \cdot \dfrac{r_A}{r_\mathit{min}}, & r_A < r_\mathit{min} \\
        |d_\Delta|, & \mathit{Otherwise}
    \end{cases}
\end{gather*}

\noindent Substituting (``$\rightsquigarrow$'') data from the output of the
\textit{Acceptance} elicitation evaluation into $\mathit{Acc}_\Delta$ gives:
\begin{gather*}
    \mathit{Acc}_\Delta(r_A = \mathtt{Anna}, r_\mathit{min}, d_\Delta =
    -\infty) \\
    \rightsquigarrow \begin{cases}
        |-\infty| \cdot \dfrac{\mathtt{Anna}}{r_\mathit{min}}, & \mathtt{Anna} <
        r_\mathit{min} \\
        |-\infty|, & \mathit{Otherwise}
    \end{cases} \\
    \rightsquigarrow \begin{cases}
    \infty \cdot \dfrac{\mathtt{Anna}}{r_\mathit{min}}, & \mathtt{Anna} <
    r_\mathit{min} \\
    \infty, & \mathit{Otherwise}
    \end{cases}
\end{gather*}

In this case, the value of $r_\mathit{min}$ does not matter because anything
multiplied by $\infty$ is $\infty$. However, as with \textit{Joy} intensity, it
makes sense that Elsa's \textit{Acceptance} is maximized because she believed
that her goal to $\mathtt{ProtectAnna}$ was unachievable and the event
$\mathtt{AnnaThaws_E}$ that was \textit{caused by Anna} made it achievable
again, which \textit{should} elicit a very strong reaction. This aligns with
the expected outcome $\mathtt{Acceptance}_{i+2}$
(Table~\ref{tab:elsa_EMgineTestCase2}).

\section{Documenting \progname{}'s Verification and
Validation}\label{sec:docTestPlan}
Since \progname{}'s chosen Software Requirements Specification (SRS) template
draws from an IEEE standard~\citep[p.~94]{SmithEtAl2007}, it bases its test
plan and report documentation\footnote{See \progname{}'s Master Test Plan at
\href{https://github.com/GenevaS/EMgine/blob/main/docs/TestPlans/MTP/EMgine_MTP.pdf}{https://github.com/GenevaS/EMgine/blob/main/docs/TestPlans/MTP/EM-
 gine\_MTP.pdf}.} on IEEE Standard 829-2008 for Software and System Test
Documentation~\citep{vvDocIEEE} and modifies it to match the order and
structure of \progname{}'s other documentation.

\section{Summary}
Extending the acceptance test case templates of Elsa's \textit{Grief} and
\textit{Admiration} demonstrates a way to translate them into
\progname{}-specific data types while remaining consistent with the original
test case and associated character study of Elsa. This suggests that it is
possible to transform implementation-agnostic acceptance test case templates
into other comparable forms, serving as a common point to begin comparing CMEs.
Preliminary evaluations of \progname{}'s models for \textit{Sadness},
\textit{Acceptance}, and---by extension---\textit{Joy} elicitation and
intensity ``on paper'' using these test cases starts to build confidence in
\progname{}'s models for those emotions, justifying subsequent implementation
and testing efforts.

\clearpage
\vspace*{\fill}
\begin{keypoints}
    \begin{itemize}

        \item Although they share some data types, the acceptance test case
        template specifications and \progname{}'s  models are intentionally
        different because they have different concerns---one models a scenario
        without knowing about the underlying processes and the other models the
        underlying processes with no conception of specific scenarios

        \item Extensions of the implementation-agnostic test case templates of
        \textit{Grief} and \textit{Admiration} show how the translation into
        \progname{}'s data types remains consistent with the original test case
        and associated character study

        \item Preliminary evaluations of some \progname{} models ``on paper''
        demonstrate that the models and test case extensions are consistent,
        building confidence in their correctness

        \item \progname{} bases its test plan and report documentation on IEEE
        Standard 829-2008 for Software and System Test Documentation, modified
        to match the structure of its other documents

    \end{itemize}
\end{keypoints}

\parasep
\vspace*{\fill}

%% file: conclusion.tex
\chapter{Looking Up at the Sky From Down the Rabbit
Hole}\label{chapter:conclusion}
\def\epigraphflush{center}
\setlength{\epigraphwidth}{0.85\textwidth}
\def\textflush{center}
\epigraph{Well, certainly no one could have been unaware of the very strange
stories floating around before we left.}{HAL 9000, \textit{2001: A Space
Odyssey}}

Player engagement is part of ``good'' player experiences (PX) which believable
Non-Player Characters (NPCs) can help build (Chapter~\ref{chapter:believable}).
This work focuses on NPC emotion (Chapter~\ref{chapter:primer}) because it can
help players empathize with an NPC and build stronger attachments to them.
However, creating NPCs that have ``correct'' emotional reactions to an
unpredictable, dynamic world is challenging. A designer's predictions of
possible scenarios limit rule and logic-based approaches such as scripts and
state machines, and would quickly multiply to unmanageable amounts as the game
grows. Another approach is tools that generate NPC emotions and/or emotional
behaviours by modelling game world information. This led to \progname{}: a
Computational Model of Emotion (CME) for generating emotion in game entities.
\progname{} is a response to the research questions from the start of this
journey (Chapter~\ref{chapter:intro}):

\begin{quote}
    \centering
    \textbf{RQ0} \textit{What software engineering-based methods and/or
        techniques can aid the creation of game development tools for believable
        characters with emotion?}
\end{quote}

The focus on NPC emotions led to CMEs---software systems influenced by emotion
research---as the basis for building game development tools. This work assumes
that game developers find processes for creating tools more useful than any
single tool itself so that they can tailor make them for their needs. Delving
into accounts of common CME development practices revealed that there is little
information about some areas such that they appear \textit{ad hoc}, making it
difficult to achieve desirable software qualities like reusability and
replicability (Chapter~\ref{chapter:se-ee-design}). For example, a CME's
desired qualities, reasons for choosing them, and how they translate to a CME's
underlying theories is often vague. This leads to low reusability and
replicability in those CMEs. The systematic approach and tools afforded by
software engineering help address this challenge by encouraging developers to
explain their design decisions and ensure traceability between concept and
realization (Chapters~\ref{chapter:equations} and
\ref{chapter:designImplement}). There are also opportunities for systematic
methods in specific software development stages, such as requirements analysis
and validation.

\clearpage\begin{quote}
    \centering
    \textbf{RQ1} \textit{How can user-oriented software requirements and domain
        knowledge inform the selection of emotion theories and/or models for
        CMEs built as game development tools?}
\end{quote}

The survey of existing CMEs (Chapter~\ref{chapter:cmeOverview}) suggests that
designers use some qualities that they want the CME to have in a target domain
to choose its theories. This suggests that there might be a systematic way to
choose a CME's underlying theories based on its requirements. One of the
challenges is the nature of the affective science literature itself.
Descriptions of theories and models in the affective science literature often
use natural language which means that they are informal and unsystematic. This
makes it difficult to see how they could support a CME's needs without making
subjective assumptions about unspecified behaviours. Document analysis, a
qualitative research method, is one way to minimize subjectivity because it
offers a structured approach for justifying decisions by encouraging
documentation of assumptions and design decisions. Using the CME's software
requirements---user-oriented or otherwise---and domain knowledge to form the
analysis context for document analysis deeply embeds them in the theory
selection process to improve its replicability and inform decisions about reuse
in other designs (Chapters~\ref{chapter:reqsAndScope},
\ref{chapter:theoryAnalysis}, and \ref{chapter:choosingExamples}).

\begin{quote}
    \centering
    \textbf{RQ2} \textit{How can existing narratives inform the development of
        test cases for evaluating CMEs built as game development tools?}
\end{quote}

Acceptance testing evaluates systems to see if it meets its end-users
expectations without knowing how the system works internally. Before conducting
expensive acceptance tests with user studies, which directly involve end-users
in the process, creating acceptance test cases from stable, known scenarios
would help build confidence in the ``correctness'' of a CME's behaviours
without additional conditions that could impact them. This works best when
acceptance test development happens separately from system development to avoid
biasing the tests in the systems favour. Accepting storytellers as domain
experts of believable characters, the narratives they create are a rich source
of test data. With methods and techniques from literary studies, testers can
extract character data from those narratives to form the preconditions, inputs,
and expected outputs of acceptance test cases to perform preliminary
evaluations of believable NPCs (Chapters~\ref{chapter:testcasedefinition} and
\ref{chapter:testcaseEMgine}).

\begin{quote}
    \centering
    \textbf{RQ3} \textit{What steps can be taken during the development of
        domain-specific CMEs to improve their reusability and replicability?}
\end{quote}

Reusability and replicability are possible when there is sufficient information
available to: trace a CME's implementation through its design and requirements;
see how its verification and validation efforts build confidence in its
``correctness''; and allow others to use a CME's components independently of
the larger system. A CME's code and documentation is usually the primary way to
relay the thought processes behind its design choices because there are few
opportunities to communicate with the designers directly. Therefore,
significant effort went into documenting \progname{}'s development process
using highly organized templates that support traceability between components
in the same document, across different documents, and in the code base
(Chapters~\ref{sec:docSRS}, \ref{sec:designDoc}, and \ref{sec:docTestPlan}). To
further encourage reusability and replicability, \progname{}'s code and
documentation is in a public GitHub repository:
\begin{center}
    \href{https://github.com/GenevaS/EMgine}{https://github.com/GenevaS/EMgine}
\end{center}

\section{A Journey of Many Disciplines}
This work began by asking what seemed like a simple question: would video games
be more entertaining if NPCs were not oblivious of the player? There are many
examples of strange NPC behaviours: becoming angry with the player, but
immediately forgetting that when they choose a different conversation option or
enter another area; continuing to interact cheerfully with their surroundings
while the player repeatedly hits them with a stick or damages their home; not
reacting at all while the player does an odd series of little jumps and
crouches; and many others. The only certainty at the beginning of this work was
that NPCs usually act like robots---it is difficult to take them seriously and
it can detract from PX. This led to a curious rabbit hole of vastly different
research and creative disciplines using qualitative and quantitative techniques.

Although the target domain suggests otherwise, game design proper does
\textit{not} have a starring role in this work. The goal is to help game
developers by providing tools and methods for them to realize their own designs
rather than proposing designs for them, which would have a narrow use scope.
Instead, the initial focus was Human-Computer Interaction (HCI) because
creating the best possible PX is a common game design goal
(Chapter~\ref{chapter:believable}). Knowing how games engross players would
help explain why ``broken'' NPC behaviour significantly interrupts their
experience to make ``fixing'' them worthwhile. The impact of game
narratives---and the NPCs in them---on a player's emotional engagement
suggest that it is worthwhile. The key is something that artists have long
since solved: \textit{believability} and the illusion of personality and
self-awareness when there is none.

One of the elements that make a believable character is emotion, which led to
\ref{as} to define what an ``emotion'' is (Chapter~\ref{chapter:primer}), then
\ref{ac} for Computational Models of Emotion (CMEs,
Chapter~\ref{chapter:se-ee-design}) because the plausibility of character
behaviours depends on their psychological validity---they must be grounded in
affective science. This idea of grounding behaviours in real-world observations
is also true in the arts, where artists observe the real world and draw from
their life experience to create their work. Artists do not need to formally
study affective science to make their work believable, suggesting that it is
not knowledge of the exact workings of emotions that drives believability---it
is the effects that they have on behaviour. This matches the purpose of a
domain-specific CME, whose first concern is the qualities and behaviours it
should have rather than accurately simulating affective processes and
structures. This is also ideal for game development tools because it means that
they do not require their users to have a formal understanding of emotion
either. It should be enough to have a layperson's understanding of emotion to
use the tool, which hides the underlying structures and processes.

Some common issues in CME development include difficulties in reusing,
replicating, and verifying them, which also makes it difficult to compare
different CMEs. It was here that software engineering could help because it has
a repertoire of processes and techniques to address those issues. The
difference between domain-specific CMEs and research-oriented ones appear to
involve two areas: requirements analysis and validation.

\subsection{Domain Versus Research: Requirements Analysis}
The purpose of research-oriented systems is to test hypotheses about affect and
its elicitation, so it already knows which affective theories, models,
structures, and/or mechanisms to build. In contrast, domain-specific CMEs want
to emulate aspects of affect without a care for how it happens. During
\textit{requirements analysis}, this affords both the freedom and stress of
choosing the CME's underlying theories, models, structures, and/or mechanisms
while ensuring support for the CME's high-level design goals. This also comes
with additional work because documenting the decision is critical to that CME's
reusability, replicability, and the ability to compare it to other systems by
describing assumptions and early design decisions that impact its design. A
survey of existing CMEs (Chapter~\ref{chapter:cmeOverview}) revealed that it is
uncommon to report why a CME uses a theory/model and how it supports its
high-level design goals. However, the survey also revealed trends between what
a CME needs to do and the theories/models it uses. This suggested that there
might be a systematic way to decide which ones to use.

Developing the proposed methodology for choosing a domain-specific CME's
theories/models (Chapter~\ref{sec:reqProcess}) hinged on two key elements:
\begin{enumerate}

    \item The criteria for evaluating and comparing theories/models, and

    \item Controlling and minimizing the inherent subjectivity in readings of
    the affective literature because of its natural language descriptions.

\end{enumerate}

The chosen criteria are the CME's high-level requirements and design scope
because they can critically influence the CME's usefulness to the intended
users. Introducing them into theory selection helps integrate them into the
CME's design early and has a trickle down effect such that there is some
implicit support for those requirements in each development stage.

It is impossible to completely eliminate subjectivity from the process because
it largely exists as non-numerical data and simply taking careful notes of the
literature is insufficient for replicability. Qualitative research methods are
the only viable option where subjectivity is expected so that it can be
controlled and minimized rather than eliminated outright. This led to document
analysis, which formed the foundation of the proposed CME theory/model
selection methodology. By embedding CME-specific concerns into this
process---requirements and scope as the context, groups of theories/models as
the ``themes'', and organizing information by level of requirement
``satisfaction''---the methodology emerged as a way to improve a CME's software
qualities by encouraging systematic documentation of the CME developer's
thought process so that others can see and trace decisions from the literature,
through the CME's design goals, out to the selected theories and/or models.

\subsection{Domain Versus Research: Validation}
A research-oriented CME is valid if it proves the hypothesis it is built to
test because a system that disproves it implies that the underlying structures
and mechanisms are not responsible for the phenomena under study. In contrast,
domain-specific CMEs are valid if they meet some developer and/or user
acceptance criteria. This implies that the methods and data for validating
domain-specific CMEs must come from its intended domain. For a CME that is to
produce believable characters through emotion, this means looking to believable
characters and extracting test case specifications from them.

Believable characters primarily exist in fictional stories where there are no
true quantitative measures, so qualitative methods are necessary. Therefore,
the proposed methodology for building acceptance test cases for believable
character-focused CMEs draws from literary art and analysis and identifies
character analysis/studies as a useful method of qualitative data collection
(Chapter~\ref{sec:atcProcess}). Rather than proposing character studies
broadly, creating a methodology provides a direction for test case designers so
that they have a guide for identifying and extracting salient information about
a character. This forwards the test case's verifiability and replicability by
providing a trace from the narrative source to the specification while further
separating test case design from the models it must evaluate.

\subsection{Putting Theory into Practice: \progname{}}
Stopping at proposals for methodologies and good development practices has
limited value because there is no practical demonstration. This is where
\progname{} comes in, putting the proposed methodologies to work while also
aiming to emulate good documentation practices.

The methodology for choosing affective theories/models based on a CME's
requirements led to \progname{}'s complementary set of theories---Oatley \&
Johnson-Laird, Plutchik, and PAD Space (Chapters~\ref{chapter:reqsAndScope} and
\ref{chapter:theoryAnalysis}). This process also inspired a series of short
examples showing how changes in the methodologies ``inputs'', the CME's
high-level requirements and design scope, lead to reasonable and different
selections for other kinds of CME (Chapter~\ref{chapter:choosingExamples}).
This provides additional examples to study and builds confidence in the
methodology's capabilities.

The role of animated films in the example acceptance test case specifications
(Chapters~\ref{chapter:testcasedefinition} and \ref{chapter:testcaseEMgine})
draws on visual art and analysis to identify their emotion from their visual
elements as well as their audio cues and narrative context. In doing so, this
highlights the undeniable medium-agnostic interconnection of believable
characters and art which ultimately does not prevent the creation of formal
specifications describing them.

Evaluating the viability of the acceptance test case building methodology
required models to test. Therefore, creating formal specifications of emotion
elicitation and intensity for \progname{} (Chapter~\ref{chapter:equations})
became a larger concern than anticipated. However, this also presented an
opportunity to propose an approach for transforming a natural language
description of an affective process or structure into a formal specification.
Consequently, a three-stage process emerged. The ``intermediary'' step helps
clarify implicit assumptions and design decisions by describing connections
between the source description and the CME developer's interpretation of it.

Although there is room for improvement, realizing the architecture design,
module specification, and implementation of \progname{}'s models
(Chapter~\ref{chapter:designImplement}) builds further confidence in the choice
of \progname{}'s theories and model and, consequently, provides additional
support for the viability of the proposed theory/model selection methodology.

\section{Avenues of Future Work}
This work began with what seemed like a simple question: how can the
development of a CME for believable NPCs ``with emotion'' leverage software
engineering? What at first looked like a regular piece of glass turned out to
be a prism, refracting into many distinct and overlapping components. Just like
light, not all questions are visible from here. These are only a few questions,
and it is not obvious how many---if any---others they might rouse and how many
are yet to be found.

Although \progname{}'s design emphasizes user's needs
(Chapter~\ref{sec:userReqs}), there was no opportunity to solicit their
feedback on it nor elicit additional requirements. Consequently, \progname{}'s
usefulness to them is not known. Investigating this requires direct interaction
with \progname{}'s intended users, which can answer questions such as:
\begin{quote}
    \centering
    \textbf{FRQ1} \textit{To what degree does \progname{} help or hinder game
        developers in the creation of NPCs ``with emotion''?}

    \textbf{FRQ2} \textit{What features would make \progname{} attractive as a
        game development tool (e.g. visual representations of emotion kinds,
        analysis tools)?}

    \textbf{FRQ3} \textit{What types of agent architectures that game developers
        want to use can \progname{} integrate with?}

    \textbf{FRQ4} \textit{How can \progname{}'s performance be improved so that
        it is feasible to use in commercial games?}

    \textbf{FRQ5} \textit{To what degree do players attribute their engagement
        with a game to its NPCs ``with emotions'' that are created with
        \progname{}?}

    \textbf{FRQ6} \textit{How much testing is necessary for game developers to
        consider a CME for game development sufficiently validated?}
\end{quote}

\progname{}'s component-based software architecture seems like the optimal
choice to satisfy the known user requirements, but there are other potential
the module decompositions that might be better suited for \progname{}
(Chapter~\ref{sec:moduleDecomp}). Drawing from knowledge about software design
patterns and architectures, some of which are
CME-specific~\citep{osuna2021toward, osuna2022interoperable, osuna2023towards},
the obvious question that follows is:
\begin{quote}
    \centering
    \textbf{FRQ7} \textit{What kinds of software design would improve
    \progname{}'s module's information hiding while maintaining the flexibility
    of \progname{}'s component-based architecture?}
\end{quote}

An element integral to test case specification but not \progname{} itself is
the many ways to \textit{express} emotion (Chapter~\ref{sec:profiles}). This is
a necessary component of believable NPCs because it conveys their invisible,
internal state and/or processes to external observers
(Chapter~\ref{sec:believable}). Therefore, it might play an equal---or even
greater---role in the believability of NPCs ``with emotion''. Some questions
that follow are:
\begin{quote}
    \centering
    \textbf{FRQ8} \textit{What expression modalities (e.g. facial expressions,
        gestures) should NPCs have to maximize a player's engagement with them?}

    \textbf{FRQ9} \textit{How complex do the processes driving emotion
    expression need to be to effectively communicate an NPC's internal emotion
    state?}
\end{quote}

While surveying other CMEs (Chapter~\ref{chapter:cmeOverview}), it was common
to find designs that integrated more types of affect alongside emotion such as
personality---a critical component of believable characters
(Chapter~\ref{sec:believable})---and mood, which might prevent unnaturally fast
fluctuations in emotion states~\citep[p.~88]{becker2008wasabi}. Since other CMEs
have already shown that it is possible to computationally model other types of
affect, the next question to ask with respect to \progname{} is:
\begin{quote}
    \centering
    \textbf{FRQ10} \textit{What additional models are necessary to extend
    \progname{} with other types of affect (e.g. personality, mood, attitudes)?}
\end{quote}

The development of acceptance test cases and \progname{}'s models showed their
dependence on knowledge that exists independently of emotion processes. This
``world'' knowledge and an NPC's perception of it
(Chapter~\ref{sec:worldKnowledge}) proved indispensable for describing
information ``outside'' of the NPC. This is unsurprising because of the
connection between emotion and an individual's environment
(Chapter~\ref{sec:affectiveDefs}). However, it raises questions about this
information's availability that impacts \progname{}'s viability as a game
development tool such as:
\begin{quote}
    \centering
    \textbf{FRQ11} \textit{How can CMEs built as game development tools use
    existing information in games to support their tasks?}

    \textbf{FRQ12} \textit{How should ``world'' knowledge and ``self'' knowledge
    be represented so that CMEs built as game development tools can use them in
    their processes?}
\end{quote}

The acceptance test case for \textit{Joy} and \textit{Acceptance}
(Chapter~\ref{sec:atc_Acceptance}) also raised questions about emotion
elicitation and intensity models:
\begin{quote}
    \centering
    \textbf{FRQ13} \textit{What are some best practices for tuning
    \progname{}'s models that require inputs such as ``threshold'' values?}

    \textbf{FRQ14} \textit{How should a CME decide which emotion kind to
    prioritize for scenarios that satisfy more than one elicitation model and
    have comparably strong intensities?}
\end{quote}

With respect to emotion representation models, \progname{} uses a function that
maps Plutchik's emotion kinds to intensities whose underlying representation is
a real value (Equations~\ref{eq:emotionstatetype} and \ref{eq:intensitytype}
respectively). One of the attractive features of Plutchik is its account of
emotion ``mixtures'', which a different type of emotion representation might
better support. Due to the similarity-based ordering of Plutchik's
\ref{circumplex}, modelling emotion kinds as fuzzy sets is another potential way
to represent them~\cite[p.~209, 215]{russell1997how}. Creating emotion
``mixtures'' could then be a matter of defining the membership of each
component set, which could be organized as a fuzzy hierarchy to mimic an
inheritance-based one without needing strict class inclusion. Alternatively,
polar coordinates use the native language of a circle and could make emotion
``mixing'' a matter of defining \ref{circumplex} axes and using them for
factorial composition on those axes~\citep[p.~89, 91--92]{gurtman1997studying}.
These alternatives for representing Plutchik's \ref{circumplex} begs the
question:
\begin{quote}
    \centering
    \textbf{FRQ15} \textit{How do other formal representations of emotion
    compare to \progname{}'s type-based ones?}
\end{quote}

Continuing the development of \progname{} itself is another way forward,
improving it so that it can stand as an example of reusable and maintainable
CME development. Addressing this, the proposed questions, or any other
questions they inspire would be an asset---big or small---to the \ref{ac}
community.

\parasep

%% file: appendix_CMESurveySupp.tex
\chapter[Supplementary Material for CME Survey]{Supplementary
Material for CME Survey\footnote{\normalfont{\footnotesize\textcopyright{}}
2022 IEEE. Reprinted with permission from
\citet{smith2021what}.}}\label{appendix:survey}
\def\epigraphflush{center}
\setlength{\epigraphwidth}{0.75\textwidth}
\def\textflush{center}
\epigraph{I think he suffered from mood swings, personally. I'm not a therapist
in any way, but I---you let me know when I'm rambling!}{B.E.N.,
\textit{Treasure Planet}}

This material is part of the survey of emotion theories in Computational Models
of Emotion (CMEs) (Chapter~\ref{chapter:cmeOverview}).

\section{Search Protocol}\label{sec:searchprotocol}
We created a search protocol following the PRISMA-S guidelines~\citep{prismas}
to answer the questions:

\textit{What emotion theories do designers use to build their
emotion-generating Computational Model of Emotion (CME)? Why do they use these
    theories?}

\subsection{Information Sources and Methods}\label{sec:sources}
\begin{enumerate}

    \item Databases Searched
    \begin{itemize}
        \item IEEE Xplore
        \item ACM Digital Library
        \item AAAI Digital Library
        \item SpringerLINK
        \item ScienceDirect (Elsevier)
    \end{itemize}
    \textit{Rationale}: These represent some of the major organizations that
    publish work in affective computing as conference proceedings and journals.
    They tend to be limited to English publications only.

    \item Multi-Database Searching
    \begin{itemize}
        \item Not used
    \end{itemize}

    \item Study Registries
    \begin{itemize}
        \item Not applicable
    \end{itemize}

    \item Online Resources and Browsing

    The following publications are available online, listed by the database
    where they are located.

    One author examined their table of contents (TOC) by hand for relevant
    papers. Paper relevancy was assessed by its title and abstract. The author
    examined a paper's contents if relevancy could not be determined from the
    title and abstract.
    \begin{itemize}
        \item IEEE Xplore
        \begin{itemize}
            \item IEEE Transactions on Affective Computing
            \item IEEE Transactions on Computational Intelligence and AI in
            Games
        \end{itemize}
        \item ACM Digital Library
        \begin{itemize}
            \item Proceedings of the ACM International Conference on
            Intelligent Virtual Agents
        \end{itemize}
        \item SpringerLINK
        \begin{itemize}
            \item Proceedings of the International Conference on Intelligent
            Virtual Agents
            \item Proceedings of the International Conference on Agents and
            Artificial Intelligence (Selected and Revised Papers)
        \end{itemize}
    \end{itemize}

    \textit{Rationale}: They focus on intelligent virtual agents.

    \item Citation Searching

    The references/citations listed in papers found using Databases [Section
    \ref{sec:sources}, Item 1] and Online Resources and Browsing [Section
    \ref{sec:sources}, Item 4] were manually screened for potential papers.

    \textit{Rationale}: Part of understanding why a design decision was made is
    understanding what influenced it. Here, that includes cited CMEs. This
    method also reduces the probability of missing a CME that frequently
    appears in the literature reviews of subsequent ones, as there is a high
    probability of multiple papers citing it directly or citing another paper
    that leads back to it.

    \item Contacts
    \begin{itemize}
        \item Not used
    \end{itemize}

    \item Other Methods

    Some papers were found by fellow researchers during their own searches. One
    author examined these by hand for relevancy. Paper relevancy was assessed
    by its title and abstract. The author examined a paper's contents if
    relevancy could not be determined from the title and abstract.

    \textit{Rationale}: There were few papers gathered this way, so it was
    worth examining them in case they were not found by other search strategies
    (Section \ref{sec:strategies}).
\end{enumerate}

\subsection{Search Strategies}\label{sec:strategies}

\begin{enumerate}
    \item Full Search Strategies

    The initial search followed strategy (a) to gather results from databases
    (Section \ref{sec:sources}, Item 1). Additional results were added after
    searching online resources (Section \ref{sec:sources}, Item 4) and using
    other methods (Section \ref{sec:sources}, Item 7).

    After reducing the total results using paper eligibility criteria (Section
    \ref{sec:strategies}, Item 3), strategy (b) was executed on the papers in
    the reduced list to gather results from citations (Section
    \ref{sec:sources}, Item 5).
    \begin{enumerate}
        \item\label{search:keyword} Keyword Search

        \begin{enumerate}
            \item IEEE Xplore

            Search executed from \texttt{Advanced Search > Command Search}.
            Query split into two parts due to limitation on wildcard (*)
            characters.

            Search results were downloaded as a CSV file using IEEE Xplore's
            built in \texttt{Export} function.

            \begin{itemize}
                \item ((``comput* model*'') AND ``emotion*'') NOT
                (``recognition*''
                OR ``predict*'')
                \item (``affective comput*'' OR ``comput* emotion engine*'' or
                ``emotion engine'')  NOT (``recognition*'' OR ``predict*'')
            \end{itemize}

            \item ACM Digital Library

            Search executed from \texttt{Advanced Search} using \texttt{The ACM
            Full-Text Collection} \\and \texttt{Search Within ``Anywhere''}.
            Query split into three parts to manage search terms.

            Search results were downloaded by navigating to each results page,
            checking \texttt{Select All} and using \texttt{Export Citations} to
            download a text file of the results in \texttt{ACM Ref} format.

            \begin{itemize}
                \item ``computational emotion model'' OR ``computational
                emotion models'' OR \\``computational model of emotion'' OR
                ``computational models of emotion'' \\-predict* -recognition*
                \item ``affective computing'' -predict* -recognition*
                \item ``emotion engine'' OR ``computational emotion engine''
                -predict* -recognition*
            \end{itemize}

            \item AAAI Digital Library

            Search executed from the search bar at the top right corner of
            \texttt{\url{www.aaai.org}}. There were no Advanced Search
            functions.

            Since the AAAI Search function is a ``Google Custom Search'' and
            limits the results to the first 10 pages, these searches were
            repeated in \texttt{Google Search} by pre-pending
            \texttt{site:www.aaai.org} to each query.

            There was no method for exporting search results, so each item was
            examined individually. Potentially relevant results were manually
            saved in an Excel sheet.

            \begin{itemize}
                \item computational model of emotion -recognition* -predict*
                \item emotion engine -recognition* -predict*
                \item affective comput* -recognition* -predict*
            \end{itemize}

            \item SpringerLINK

            Search executed from the main search bar. Searches omit wildcards
            because search function is designed to look for words with the same
            ``stem'' (e.g. ``computer'' also finds ``computes'' and
            ``computation'').

            Search includes ``Include Preview-Only Content''.

            Search was filtered by sub-disciplines [Section
            \ref{sec:strategies}, Item \ref{filter:springer}].

            Search results were downloaded as a CSV file using SpringerLINK's
            built in \texttt{Downlo\-a\-d} function.

            \begin{itemize}
                \item computational model of emotion NOT predict NOT recognition
                \item emotion engine NOT predict NOT recognition
                \item computational emotion engine NOT predict NOT recognition
                \item affective computing elicitation generation NOT predict
                NOT recognition
            \end{itemize}

            \item ScienceDirect (Elsevier)

            Searches executed with \texttt{Advanced Search} to search for
            individual articles rather than full journals. They were filtered
            by Publication Title [Section \ref{sec:strategies}, Item
            \ref{filter:elsevier}].

            Searches capture spelling variations (e.g. ``color'' and
            ``colour'') and plural forms of search terms (e.g. ``code'' and
            ``codes'').

            Search results were downloaded by navigating to each results page,
            checking \texttt{Select All} and using \texttt{Export} to
            download a text file of the results as citations.

            \begin{itemize}
                \item computational model of emotion -recognition -prediction
                \item affective computing generation elicitation -recognition
                -prediction
                \item emotion engine -recognition -prediction
            \end{itemize}

        \end{enumerate}

        \item One author screened the reference lists of included papers found
        from [Section \ref{sec:strategies}, Item \ref{search:keyword}] by hand
        for relevancy. Paper relevancy was assessed by its title and abstract.
        The author examined a paper's contents if relevancy could not be
        determined from the title and abstract.
    \end{enumerate}

    \item Limits and Restrictions

    Papers were limited to those written in English because it is a language
    shared by both authors. Specific database restrictions were as follows:
    \begin{enumerate}
        \item IEEE Xplore, ACM Digital Library, and AAAI Digital Library

        No additional restrictions.

        \textit{Rationale}: These databases specialize in electrical, computer,
        and software engineering, and computer science. There was little chance
        that papers from unrelated fields would be captured in the search.

        \item\label{filter:springer} SpringerLINK

        Search results are limited to the sub-discipline of ``Artificial
        Intelligence''.

        \textit{Rationale}: Of the available sub-disciplines, this most closely
        matched the kinds of CMEs that the authors wished to examine. Limiting
        results to this sub-discipline better focused the results so that there
        were fewer to examine by hand.

        \item\label{filter:elsevier} ScienceDirect (Elsevier)

        Search results restricted to the following journals:
        \begin{itemize}
            \item Biologically Inspired Cognitive Architectures
            \item Cognitive Systems Research
            \item Computers \& Education
            \item Computers \& Graphics
            \item Computers in Human Behaviour
            \item Expert Systems with Applications
            \item Information and Software Technology
            \item International Journal of Human-Computer Studies
            \item Journal of Systems and Software
            \item Knowledge-based Systems
            \item Neurocomputing
            \item Procedia Computer Science
            \item Trends in Cognitive Sciences
        \end{itemize}

        \textit{Rationale}: These journals focus on computer science and
        engineering.
    \end{enumerate}

    \item Paper Eligibility Criteria

    One author examined the results gathered from the search strategies
    (Section \ref{sec:strategies}, Item 1) for papers to include in the survey
    using the following criteria:

    \textit{Inclusion Criteria}
    \begin{itemize}
        \item Papers describing CMEs with an emotion
        generation/elicitation/appraisal component that is built on at least
        one emotion theory

        \textit{Rationale}:
        \begin{itemize}
            \item Directly relates to research question

            \item Disqualifies CMEs that use empirical data, neurology/brain
            anatomy, and psychological/sociological theories of human behaviour

            \item Disqualifies CMEs that lack an emotion generation component
        \end{itemize}

        \item Papers representing the most recent version of a CME that had
        emotion generation-related design decisions (i.e. CME is given the
        same/variation of a name, has at least one common author, has the same
        designer intent, and uses the same emotion theories as its predecessors)

        \textit{Rationale}:
        \begin{itemize}
            \item Assumes that the most recent paper reflects current
            understanding of CME requirements and available emotion theories
        \end{itemize}
    \end{itemize}

    \textit{Exclusion Criteria}
    \begin{itemize}
        \item Papers describing experiments on/with CMEs where a previously
        published paper describes the design of that CME

        \textit{Rationale}:
        \begin{itemize}
            \item Do not focus on the CME's design or why decisions were made,
            therefore they do not serve research question

            \item Citation Searching (Section \ref{sec:sources}, Item 5) would
            find paper(s) describing the design of these CMEs
        \end{itemize}

        \item Papers describing CMEs that are solely/primarily combinations of
        other CMEs

        \textit{Rationale}:
        \begin{itemize}
            \item Do not help understand why the component CMEs made their
            design decisions, therefore they do not serve research question

            \item Citation Searching (Section \ref{sec:sources}, Item 5) would
            find paper(s) describing the design of component CMEs
        \end{itemize}

        \item Papers that are surveys or describe design guidelines/frameworks

        \textit{Rationale}:
        \begin{itemize}
            \item Do not directly serve research question

            \item Citation Searching (Section \ref{sec:sources}, Item 5) would
            find additional paper(s) describing CMEs discussed
        \end{itemize}

        \item Papers that describe CMEs designed solely on brain structures
        and/or empirical data

        \textit{Rationale}:
        \begin{itemize}
            \item Do not directly serve research question
        \end{itemize}
    \end{itemize}

    \item Search Filters
    \begin{itemize}
        \item Not used
    \end{itemize}

    \item Prior Work
    \begin{itemize}
        \item Not used
    \end{itemize}

    \item Updates

    All searches rerun at worst three months before submission on June 14th,
    2022.

    \item Dates of Searches

    Table~\ref{tab:lastrun} lists the last dates that full search strategies
    were executed (Section \ref{sec:strategies}, Item 1).

\end{enumerate}

\begin{table}[!tb]
    \centering
    \caption{Date Searches Were Last Executed \textcopyright{} 2022 IEEE}
    \label{tab:lastrun}
    \small
    \begin{tabular}{@{}P{0.8\linewidth}c@{}}
        \toprule

        \textbf{Source} & \textbf{Last Run Date} \\

        \midrule

        &\\

        \multicolumn{2}{l}{\textit{IEEE Xplore}} \\

        ((``comput* model*'') AND ``emotion*'') NOT (``recognition*'' OR
        ``predict*'')
        & April 21, 2022 \\

        (``affective comput*'' OR ``comput* emotion engine*'' or ``emotion
        engine'')
        NOT (``recognition*'' OR ``predict*'') & April 21, 2022 \\

        &\\

        \multicolumn{2}{l}{\textit{ACM Digital Library}} \\

        ``computational emotion model'' OR ``computational emotion models'' OR
        ``computational model of emotion'' OR ``computational models of
        emotion''
        -predict* -recognition* & April 25, 2022 \\

        ``affective computing'' -predict* -recognition* & April 25, 2022 \\

        ``emotion engine'' OR ``computational emotion engine'' -predict*
        -recognition* & April 25, 2022 \\

        &\\

        \multicolumn{2}{l}{\textit{AAAI Digital Library}} \\

        computational model of emotion -recognition* -predict* & April 26, 2022
        \\

        emotion engine -recognition* -predict* & April 26, 2022 \\

        affective comput* -recognition* -predict* & April 26, 2022 \\

        &\\

        \multicolumn{2}{l}{\textit{SpringerLINK}} \\

        computational model of emotion NOT predict NOT recognition & April 26,
        2022 \\

        emotion engine NOT predict NOT recognition & April 26, 2022 \\

        computational emotion engine NOT predict NOT recognition & April 26,
        2022 \\

        affective computing elicitation generation NOT predict NOT recognition
        & April 26, 2022 \\

        &\\

        \multicolumn{2}{l}{\textit{ScienceDirect (Elsevier)}} \\

        computational model of emotion -recognition -prediction & April 29,
        2022 \\

        affective computing generation elicitation -recognition -prediction &
        April 29, 2022 \\

        emotion engine -recognition -prediction & April 29, 2022 \\

        &\\

        \multicolumn{2}{l}{\textit{Online Resource Browsing}} \\

        IEEE Transactions on Affective Computing TOC & April 22, 2022 \\

        IEEE Transactions on Computational Intelligence and AI in Games TOC &
        April 22, 2022 \\

        Proceedings of the International Conference on Intelligent Virtual
        Agents TOC & May 2, 2022 \\

        Proceedings of the ACM International Conference on Intelligent Virtual
        Agents TOC & May 2, 2022 \\

        Proceedings of the International Conference on Agents and Artificial
        Intelligence (Selected and Revised Papers) TOC & May 2, 2022 \\

        &\\

        \bottomrule
    \end{tabular}
\end{table}

\subsection{Peer Review}
Two research librarians, one familiar with surveys, reviewed the protocol.
Their feedback resulted in a more thorough list of databases to search and
eliminated the need for Google Scholar results.

\subsection{Managing Records}

\begin{enumerate}
    \item Total Records

    Table~\ref{tab:searchhits} lists the total records found using Keyword
    Search (Section \ref{sec:strategies}, Item \ref{search:keyword}).

    \item Deduplication

    Search results were manually combined in an Excel spreadsheet, then sorted
    by ``Title'' and ``Author(s)'' to identify potential duplicate papers.
    These were manually removed so that there was only one record per unique
    paper.

\end{enumerate}

\begin{table}[!tb]
    \centering
    \caption{Total Records Found with Keyword Search \textcopyright{} 2022 IEEE}
    \label{tab:searchhits}
    \small
    \begin{tabular}{@{}P{0.8\linewidth}c@{}}
        \toprule

        \textbf{Source} & \textbf{Total Results} \\

        \midrule

        &\\

        \multicolumn{2}{l}{\textit{IEEE Xplore}} \\

        ((``comput* model*'') AND ``emotion*'') NOT (``recognition*'' OR
        ``predict*'')
        & 565 \\

        (``affective comput*'' OR ``comput* emotion engine*'' or ``emotion
        engine'')
        NOT (``recognition*'' OR ``predict*'') & 1,069 \\

        &\\

        \multicolumn{2}{l}{\textit{ACM Digital Library}} \\

        ``computational emotion model'' OR ``computational emotion models'' OR
        ``computational model of emotion'' OR ``computational models of
        emotion'' -predict* -recognition* & 32 \\

        ``affective computing'' -predict* -recognition* & 465 \\

        ``emotion engine'' OR ``computational emotion engine'' -predict*
        -recognition* & 61 \\

        &\\

        \multicolumn{2}{l}{\textit{AAAI Digital Library}} \\

        computational model of emotion -recognition* -predict* & 754 \\

        emotion engine -recognition* -predict* & 169 \\

        affective comput* -recognition* -predict* & 58 \\

        &\\

        \multicolumn{2}{l}{\textit{SpringerLINK}} \\

        computational model of emotion NOT predict NOT recognition & 7,213 \\

        emotion engine NOT predict NOT recognition & 1,061 \\

        computational emotion engine NOT predict NOT recognition & 537 \\

        affective computing elicitation generation NOT predict NOT recognition
        & 37 \\

        &\\

        \multicolumn{2}{l}{\textit{ScienceDirect (Elsevier)}} \\

        computational model of emotion -recognition -prediction & 501 \\

        affective computing generation elicitation -recognition -prediction &
        134 \\

        emotion engine -recognition -prediction & 282 \\

        &\\

        \bottomrule
    \end{tabular}
\end{table}

\clearpage
\section{CME ``Genealogy''}\label{sec:genes}
Table~\ref{tab:ancestryOverview} highlights some key systems that contributed
to the design of a CME.

\begin{table}[!b]
    \renewcommand{\arraystretch}{1.3}
    \centering
    \caption{Overview of the Contributing Designs of CMEs \textcopyright{} 2022
    IEEE}
    \label{tab:ancestryOverview}
    \small
    \begin{tabular}{P{0.03\linewidth}P{0.2\linewidth}P{0.67\linewidth}}
        \toprule
        & \textbf{System} & \textbf{Builds On} \\
        \midrule

        \colourRow\ref{ar} & \textbf{AffectR} & -- \\

        \ref{cathexis} & \textbf{Cathexis} & -- \\

        \colourRow\ref{emMod} & \textbf{EmMod} & -- \\

        \ref{flame} & \textbf{FLAME} & -- \\

        \colourRow\ref{scream} & \textbf{SCREAM} & AffectR
        (\ref{ar}), Em/Oz (\ref{em})$^{1}$ \\

        \ref{mamid} & \textbf{MAMID} & -- \\

        \colourRow\ref{tabasco} & \textbf{TABASCO} &
        3T~\citep{bonasso1997experiences}$^{2}$ \\

        \ref{wasabi} & \textbf{WASABI} &
        MAX~\citep{becker2004simulating}$^{3}$ \\

        \colourRow\ref{maggie} & \textbf{Maggie} & -- \\

        \ref{akr} & \textbf{AKR} & Will (\ref{will})$^{2}$,
        GOLEM~\citep{castelfranchi1997social}$^{2}$ \\

        \colourRow\ref{gvh} & \textbf{GVH} & Autonomous
        Virtual Human Dialog System~\citep{magnenat2000communicating}$^{3}$ \\

        \ref{parlee} & \textbf{ParleE} & Cathexis (\ref{cathexis})$^{2}$,
        FLAME (\ref{flame}), \'Emile (\ref{emile})$^{1, 2}$, Em/Oz (\ref{em}) \\

        \colourRow\ref{impmeb} & \textbf{IM-PMEB} & ALMA
        (\ref{alma}) \\

        \ref{genia3} & \textbf{GenIA$^3$} & EMA (\ref{ema}), ALMA
        (\ref{alma}), ERDAMS (\ref{erdams}),
        O3A~\citep{alfonso2014open}$^{3}$,
        AgentSpeak~\citep{vieira2007formal}$^{2}$ \\

        \colourRow\ref{infra} & \textbf{InFra} & FLAME
        (\ref{flame}) \\

        \ref{fatimaM} & \textbf{FAtiMA-M} & FAtiMA
        (\ref{fatima})$^{3}$, ORIENT~\citep{lim2012creating}$^{2}$,
        Computational Appraisal
        Architecture~\cite[p.~31]{marsella2010computational} \\

        \colourRow\ref{hybridc} & \textbf{HybridC} &
        EMIA~\citep{jain2015emia}$^{3, 4}$ \\

        \ref{gema} & \textbf{GEmA} & FLAME (\ref{flame})$^2$ \\

        \hline

        \colourRow\ref{som} & \textbf{SOM} & -- \\

        \hline

        \ref{soar} & \textbf{Soar} & Em/Oz
        (\ref{em})$^{1}$, PEACTIDM~\citep{newell1994unified}$^{2}$ \\

        \colourRow\ref{lida} & \textbf{LIDA} & Computational
        Appraisal Architecture~\cite[p.~31]{marsella2010computational} \\

        \ref{clarion} & \textbf{CLARION} & -- \\

        \hline

        \colourRow\ref{acres} & \textbf{ACRES} & -- \\

        \ref{ema} & \textbf{EMA} & AffectR (\ref{ar}), Soar
        (\ref{soar})$^{2}$, Will (\ref{will})$^{2}$, \'Emile (\ref{emile}) \\

        \colourRow\ref{will} & \textbf{Will} & ACRES
        (\ref{acres})$^{3}$ \\

        \ref{elsa} & \textbf{ELSA} & -- \\

        \colourRow\ref{gamae} & \textbf{GAMA-E} & SocioEmo
        (\ref{socio})$^2$, OCC Logical Formalism~\citep{adam2007emotions},
        GAMA~\citep{grignard2013gama}$^{2}$ \\

        \hline

        \ref{emile} & \textbf{\'Emile} & AffectR (\ref{ar}), Cathexis
        (\ref{cathexis})$^{1}$, Em/Oz (\ref{em}),
        NML1~\citep{beaudoin1994goal}$^{2}$,
        Steve~\citep{rickel1999animated}$^{5}$, Affect
        Editor~\citep{cahn1989generating}$^{5}$ \\

        \colourRow\ref{emotion} & \textbf{EMOTION} & GVH
        (\ref{gvh}): Generic Model~\citep{egges2004generic} \\

        \midrule\bottomrule

        \multicolumn{3}{r}{\textit{Continued on next page}} \\
    \end{tabular}
\end{table}
\addtocounter{table}{-1}
\captionsetup{list=no}
\begin{table}[!t]
    \renewcommand{\arraystretch}{1.3}
    \centering
    \caption{(\textit{Continued.}) Overview of the Contributing Designs of CMEs
    \textcopyright{} 2022 IEEE}
    \small
    \begin{tabular}{P{0.03\linewidth}P{0.2\linewidth}P{0.67\linewidth}}
        \toprule
        & \textbf{System} & \textbf{Builds On} \\
        \midrule

        \colourRow\ref{humdpme} & \textbf{HumDPM-E} &
        HumDPM~\citep{luo2010modeling}$^2$ \\

        \ref{jbdiemo} & \textbf{JBdiEmo} & Jadex~\citep{pokahr2005jadex}$^{2}$
        \\

        \colourRow\ref{dett} & \textbf{DETT} &
        MANA~\citep{lauren2002map}$^{1, 2}$ \\

        \ref{epbdi} & \textbf{EP-BDI} & -- \\

        \colourRow\ref{microcrowd} & \textbf{MicroCrowd} &
        Soar (\ref{soar})$^{2}$ \\

        \hline

        \ref{puppet} & \textbf{Puppet} & S3A (\ref{s3a})$^{2}$ \\

        \colourRow\ref{cbi} & \textbf{CBI} & -- \\

        \ref{fatima} & \textbf{FAtiMA} & TABASCO (\ref{tabasco})$^{2}$, EMA
        (\ref{ema}), CBI (\ref{cbi}), S3A (\ref{s3a}),
        FearNot!\citep{aylett2005fearnot}$^{3}$ \\

        \colourRow\ref{tardis} & \textbf{TARDIS} & Greta
        (\ref{greta})$^{5}$, ALMA (\ref{alma})$^{2}$, SocioEmo
        (\ref{socio})$^{2}$ \\

        \ref{puma} & \textbf{PUMAGOTCHI} & -- \\

        \hline

        \colourRow\ref{greta} & \textbf{Greta} & -- \\

        \ref{alma} & \textbf{ALMA} & EmotionEngine~\citep{gebhard2003adding},
        \citep{gebhard2004coloring}$^{3}$ \\

        \colourRow\ref{eva} & \textbf{Eva} & ALMA
        (\ref{alma})$^{2}$ \\

        \ref{ppad} & \textbf{PPAD-Algo} & ALMA (\ref{alma}), Eva
        (\ref{eva})$^{2}$ \\

        \colourRow\ref{peedy} & \textbf{Peedy} & -- \\

        \ref{erdams} & \textbf{ERDAMS} & AffectR (\ref{ar}), ParleE
        (\ref{parlee})$^{1}$, DER~\citep{tanguy2005dynamic}, \'Emile
        (\ref{emile})/``Jack and Steve''$^{1}$, Em/Oz (\ref{em}), Corpora
        Coding~\citep{ochs2007emotion}$^{3}$ \\

        \colourRow\ref{teatime} & \textbf{TEATIME} & -- \\

        \ref{mmt} & \textbf{MMT} & -- \\

        \colourRow\ref{presence} & \textbf{Presence} &
        PPP~\citep{andre1999employing}$^{2, 3}$ \\

        \hline

        \ref{pomdp} & \textbf{POMDP-CA} & -- \\

        \colourRow\ref{iphonoid} & \textbf{iPhonoid} &
        Interactive Robot System with
        Memory~\citep{masuyama2017application}$^3$,
        AEIS~\citep{han2012robotic}$^2$ \\

        \ref{eegs} & \textbf{EEGS} & Computational Appraisal
        Architecture~\cite[p.~31]{marsella2010computational} \\

        \colourRow\ref{pwe} & \textbf{PWE-I} &
        HED~\citep{steephen2013hed}$^{2}$, Mood
        Prediction~\citep{katsimerou2015predicting}$^{2}$ \\

        \ref{kismet} & \textbf{Kismet} & Cathexis (\ref{cathexis})$^3$ \\

        \colourRow\ref{robo} & \textbf{R-Cept} &
        Vickia~\citep{bruce2002role}$^{2, 5}$ \\

        \ref{grace} & \textbf{GRACE} & EmotiRob~\citep{saint2007emotirob}$^{3}$
        \\

        \colourRow\ref{tame} & \textbf{TAME} & -- \\

        \ref{aee} & \textbf{AEE} & -- \\

        \colourRow\ref{feelme} & \textbf{FeelMe} & -- \\

        \ref{socio} & \textbf{SocioEmo} & ParleE (\ref{parlee})$^{2}$, \'Emile
        (\ref{emile}), ALMA (\ref{alma})$^{2}$, Em/Oz (\ref{em}), E/P
        Model~\citep{sehaba2007emotional}$^{3}$ \\

        \colourRow\ref{soul} & \textbf{The Soul} & ALMA
        (\ref{alma})$^{2}$, Animating
        Expressions~\citep{schaap2008towards}$^{3}$ \\

        \ref{gamygdala} & \textbf{GAMYGDALA} & Em/Oz (\ref{em})$^{1}$ \\

        \colourRow\ref{mobsim} & \textbf{MobSim} & ALMA
        (\ref{alma}) \\

        \ref{apf} & \textbf{APF} & SocioEmo (\ref{socio}) \\

        \midrule\bottomrule

        \multicolumn{3}{r}{\textit{Continued on next page}} \\
    \end{tabular}
\end{table}

\clearpage
\makeatletter
\setlength{\@fptop}{0pt}
\setlength{\@fpbot}{0pt plus 1fil}
\makeatother

\addtocounter{table}{-1}
\captionsetup{list=no}
\begin{table}[!th]
    \renewcommand{\arraystretch}{1.3}
    \centering
    \caption{(\textit{Continued.}) Overview of the Contributing Designs of CMEs
    \textcopyright{} 2022 IEEE}
    \small
    \begin{threeparttable}
        \begin{tabular}{P{0.03\linewidth}P{0.2\linewidth}P{0.67\linewidth}}
            \toprule

            & \textbf{System} & \textbf{Builds On} \\
            \midrule

            \colourRow\ref{mexica} & \textbf{MEXICA} & -- \\

            \ref{npe} & \textbf{NPE} & Emotional
            Planner~\citep{gratch1999you}, Possible Worlds
            Model\citep{shirvani2017possible}$^{2, 3}$ \\

            \colourRow\ref{em} & \textbf{Em/Oz} &
            Tok~\citep{bates1992architecture}$^{2}$,
            Hap~\citep{loyall1997believable}$^{2}$ \\

            \ref{s3a} & \textbf{S3A} & Will (\ref{will})$^{4}$, Em/Oz
            (\ref{em}) \\

            \hline\bottomrule
        \end{tabular}%
        \begin{tablenotes}
            \footnotesize
            \item [1] \textit{For domain specific agent capabilities that are
                affective in nature, but have unclear theoretical roots.}

            \item [2] \textit{For domain specific agent capabilities that do
                not explicitly model agent emotion, influence emotion via other
                factors, map emotion to another affective type, or are
                implementation-specific.}

            \item [3] \textit{Direct or close descendant of this system.}

            \item [4] \textit{The relationship is inferred from chosen
                affective theories and model
                definitions~\cite[p.~61]{jain2019modeling}, \cite[p.~37,
                48]{martinho2000emotions}.}

            \item [5] \textit{For agent embodiment only.}

        \end{tablenotes}
    \end{threeparttable}
\end{table}

\clearpage
\makeatletter
\setlength{\@fptop}{0pt plus 1fil}
\setlength{\@fpbot}{0pt plus 1fil}
\makeatother
\captionsetup{list=yes}

%% file: affect_and_emotion.tex
\chapter{Affect and Emotion (In Theory)}
\label{chapter:affect}
\def\epigraphflush{center}
\setlength{\epigraphwidth}{0.75\textwidth}
\def\textflush{center}
\epigraph{Could you please continue the petty bickering? I find it most
intriguing.}{DATA, \textit{Star Trek: The Next Generation}}

One way that researchers categorize affective theories/models for analysis is
in four broad perspectives\footnote{There are finer-grained
distinctions~\citep[p.~280]{scherer2021towards}.}: \textit{discrete},
\textit{dimensional}, \textit{appraisal} or \textit{componential}, and
\textit{neurophysiologic}~\citep[p.~95--99]{lisetti2015and}. Designers have
used theories from each perspective to generate and express emotions in
computer-driven agents.

The discrete, dimensional, and appraisal perspectives use states, dimensions,
and processes to explain emotion generation. This makes them good candidates
for a Computational Model of Emotion (CME) for believable Non-Player Characters
(NPC), since they are not dependent on specific component implementations. The
neurophysiologic perspective explains emotion generation via neural circuitry
and physiology. Simulating parts of the brain and body is likely to be too
complex for entertainment-focused CMEs~\citep[p.~234]{ojha2016ethically} but
examining these theories is useful for finding other ideas.

\section{Discrete Theories}
The discrete, or categorical, perspective has roots in Darwin's theory of
evolution which assumes that emotions have an adaptive functions and must have
developed via natural selection due to their complexity~\citep{darwin18721965}.
It emphasizes a small set of fundamental emotions, often called \textit{basic}
or \textit{primary} emotions, that have evolved via natural
selection~\citep[p.~305]{hudlicka2014computational}. These theories commonly
assume that emotions are innate, hard-wired features with dedicated neural
circuitry. This circumvents cognitive processing
entirely~\citep[p.~250]{reisenzein2013computational}. However, this also means
that discrete theories have been unable to define a general framework for
evaluating a variety of emotions using the same
terms~\citep[p.~816--817]{smith1985patterns}. Theories differ in their
definitions of ``primary'' emotions but these are not mutually
exclusive~\citep[p.~2]{ortony2021all}. However, they do change which emotions
a theory considers ``basic''. Ekman \& Friesen and Izard are part of the
``biologically basic'' view which tend to focus on facial expressions as
indicators of primality, whereas Plutchik is part of the ``elemental'' view
that seeks ``atomic'' emotions that others cannot define (i.e. ``mixtures'' of
other emotions)~\citep[p.~2--3]{ortony2021all}.

The discrete perspective has the most empirical data available for building
emotion signatures (Chapter~\ref{sec:affectiveDefs}) because specific effects
link to each emotion category~\citep[p.~10]{hudlicka2014habits}. The data
mainly focuses on facial expressions and language~\citep[p.~2]{ortony2021all},
likely due to the difficulty, complexity, and time necessary to collect
it~\citep[p.~219]{johnson1992basic}.

Discrete theories are appealing for CMEs because they have a clear mapping
between a small set of universal \ref{antecedent}s to emotion kinds and related
action tendencies~\citep[p.~100]{lisetti2015and}. CMEs using this approach
``sense'' a set of triggers and respond with reflexes using action-reaction
rules. This suggests that CMEs using discrete theories script the emotion
generation process.

\subsection{Ekman \& Friesen}\label{adx:ef}
This theory assumes that universal, innate triggers and learned variations of
them cause emotion~\citep{ekman2007emotions}. The brain stores these triggers
in a mental ``database''. Individuals experience an emotion when there is a
match between stimuli and this ``database'' of triggers. Emotion intensity is a
function of the degree of ``mismatch'' such that a closer match produces a more
intense emotion. Ekman \& Friesen identify seven ``universal'' emotions:
\textit{Happiness}, \textit{Sadness}, \textit{Fear}, \textit{Anger},
\textit{Surprise}, \textit{Disgust}, and \textit{Contempt}. Facial expressions
play a critical role in Ekman \& Friesen's theory. The theory accounts for
cultural differences in facial expressions with display rules---socially
learned rules about when and to whom an emotional expression can be shown.

\begin{figure}[!b]
    \centering
    \begin{subfigure}[b]{0.25\textwidth}
        \centering
        \includegraphics[width=\textwidth]{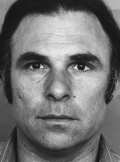}
        \caption{Neutral}
        \label{fig:au1Neutral}
    \end{subfigure}
    \hspace{0.1\textwidth}
    \begin{subfigure}[b]{0.25\textwidth}
        \centering
        \includegraphics[width=\textwidth]{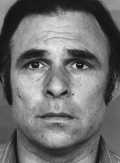}
        \caption{Active}
        \label{fig:au1Active}
    \end{subfigure}
    \caption[Movement of FACS Action Unit 1 ``Inner Brow Raiser'']{Movement of
    FACS Action Unit 1 ``Inner Brow Raiser''~\citep[p.~20]{facs}}
    \label{fig:facsExample}
\end{figure}

This theory has had appeal for CMEs because it created the Facial Action Coding
System (FACS)~\citep{facs}, which maps facial muscles and movements to
individual changes in facial expressions (Figure~\ref{fig:facsExample}). CMEs
have used it to identify configurations for the seven named ``universal''
emotions.

\subsection{Izard's Differential Emotions Theory (DET)}\label{adx:det}
\begin{figure}[!t]
    \centering
    \begin{subfigure}[b]{0.25\textwidth}
        \centering
        \includegraphics[width=\textwidth]{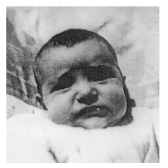}
        \caption{Sadness}
        \label{fig:maxSad}
    \end{subfigure}
    \hspace{0.1\textwidth}
    \begin{subfigure}[b]{0.25\textwidth}
        \centering
        \includegraphics[width=\textwidth]{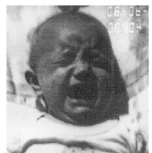}
        \caption{Anger}
        \label{fig:maxAnger}
    \end{subfigure}
    \caption[Examples of MAX-coded Expressions]{Examples of MAX-coded
    Expressions~\citep{izard1979maximally}}
    \label{fig:maxExample}
\end{figure}

DET is a personality-focused theory, where emotions contribute to personality
development and individual differences~\citep{izard1993stability,
izard2000motivational}. It suggests that the perceptional, cognitive, and
behavioural patterns of emotion manifest in stable ways and that emotion kinds
interact in individual-specific relationships. These patterns develop in
infancy and remain stable, producing personality and individual differences.
DET proposes that there are twelve emotion kinds: \textit{Interest},
\textit{Enjoyment}, \textit{Surprise}, \textit{Sadness}, \textit{Anger},
\textit{Disgust}, \textit{Contempt}, \textit{Fear}, \textit{Guilt},
\textit{Shame}, \textit{Shyness}, and
\textit{Self-Hostility}~\citep{izard1993stability}. DET suggests that the
discrete and dimensional emotion perspectives are complementary but discrete
emotion kinds provide more information. DET suggests that facial expressions
are innate to emotion kinds, developing the lesser used Maximally
Discriminative Facial Movement Coding System (MAX)~\citep{izard1979maximally}.
Researchers generally use the MAX system in infants and young children research
(Figure~\ref{fig:maxExample}).

\subsection{Plutchik's Psycho-evolutionary Synthesis (PES)}\label{adx:pes}
\begin{figure}[!ht]
    \centering
    \begin{subfigure}[b]{0.35\textwidth}
        \centering
        \includegraphics[width=\textwidth]{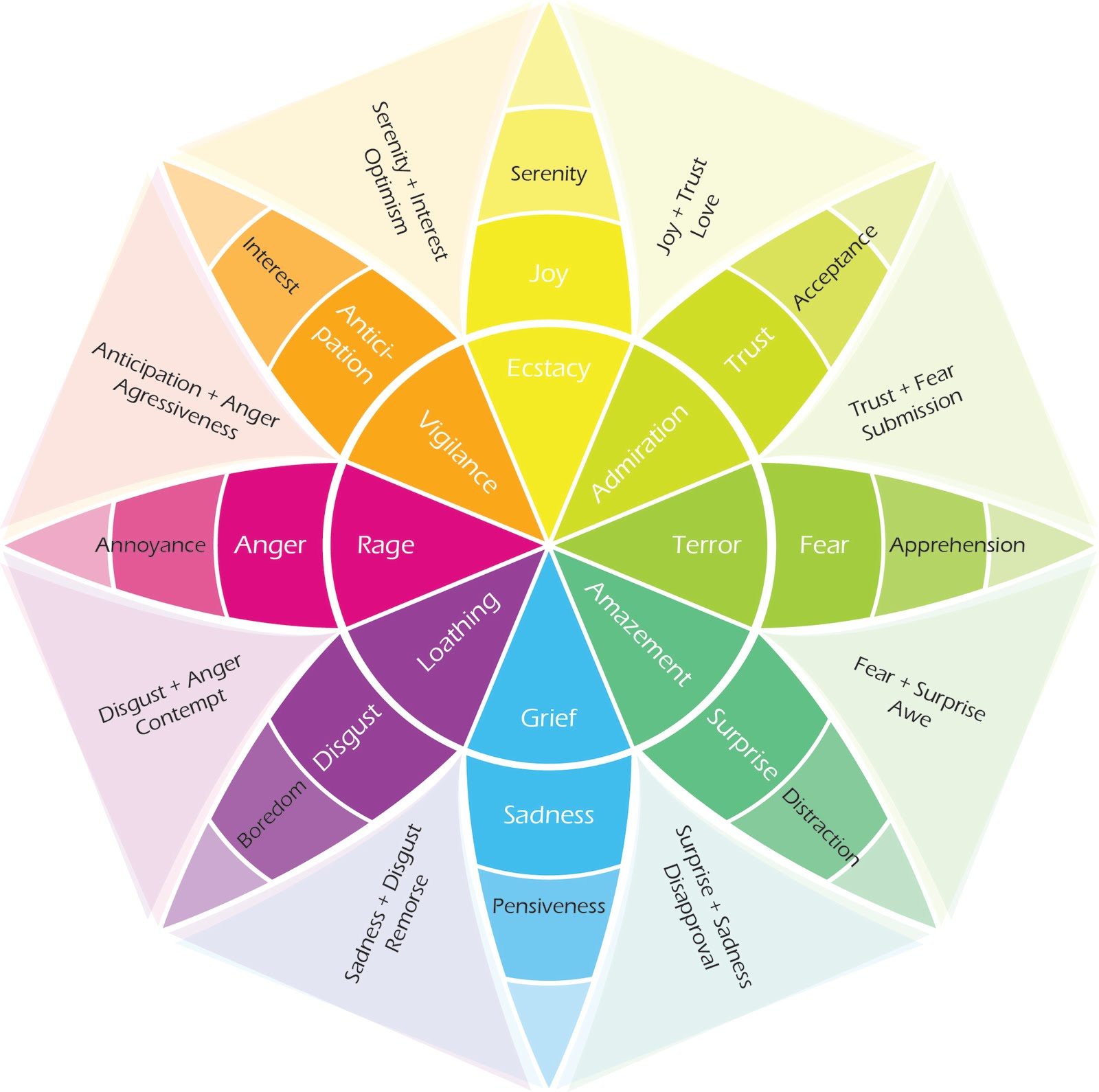}
        \caption{2D \ref{circumplex}}
        \label{fig:PESCircumplex}
    \end{subfigure}
    \hspace{0.1\textwidth}
    \begin{subfigure}[b]{0.35\textwidth}
        \centering
        \includegraphics[width=\textwidth]{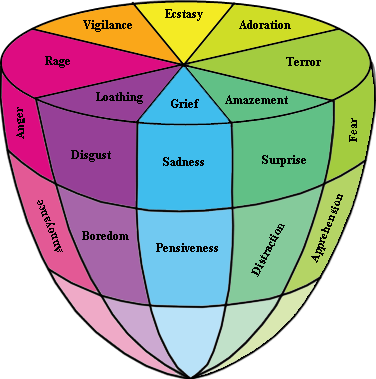}
        \caption{\ref{circumplex} with Intensity}
        \label{fig:PESCone}
    \end{subfigure}
    \caption[PES Structural View]{PES Structural View~\citep[p.~203,
    205]{plutchik1984emotions}}
    \label{fig:PESStructuralView}
\end{figure}

PES aims to synthesize four major, but traditionally separated, fields of
study: evolution, psychophysiology, neurology, and
psychoanalysis~\citep{robert1980emotion, plutchik1984emotions,
plutchik1997circumplex}. PES proposes that emotions are responses to survival
issues common to all organisms. It derived one of its criteria for claiming
a state is an ``emotion'' from this proposal---association with goal-directed
behaviour. PES's eight primary emotions therefore follow from the
identification of four adaptational issues that have two, opposing response
behaviours (Table~\ref{tab:cognitionstoemotions}). The theory organizes these
emotions---\textit{Joy}, \textit{Sadness}, \textit{Fear}, \textit{Anger},
\textit{Anticipation}, \textit{Surprise}, \textit{Disgust}, and
\textit{Trust}---into a \ref{circumplex} structure by their relative
similarities based on layman evaluations of everyday
language~\citep{robert1980emotion, block1957studies, conte1976circumplex}. The
resulting arrangement places ``opposing'' emotions opposite each other as
predicted (Figure~\ref{fig:PESCircumplex}). PES treats emotion intensity as a
third dimension, extending from most intense to a state of ``deep sleep'' where
organisms experience no emotion at all (Figure~\ref{fig:PESCone}). The
\ref{circumplex} nature of PES's structure implies that the primary emotions
need not have equidistant spacing between them because there are no fundamental
axes. There is also the implication that there is conflict between opposing
elements on the circle, representing polarized behaviours (e.g. approach versus
retreat). The mathematical concept of the \ref{circumplex} suggests other
implications which also implies that PES has more depth than explicitly
described.
\begin{table}[!ht]
    \centering
    \caption{Connection Between Behaviours and Emotions in PES}
    \label{tab:cognitionstoemotions}
    \resizebox{\linewidth}{!}{%
    \begin{tabular}{P{0.3\linewidth}P{0.24\linewidth}P{0.2\linewidth}P{0.15\linewidth}}
        \toprule
        \textbf{Event} & \textbf{Cognition} &  \textbf{Behaviour} &
        \textbf{Emotion}
        \\ \midrule
        \colourRow Threat & ``Danger'' & Protection &
        \textit{Fear} \\

        Obstacle & ``Enemy'' & Destruction & \textit{Anger} \\

        \colourRow Loss of a valued individual &
        ``Abandonment'' & Reintegration & \textit{Sadness} \\

        Potential Mate & ``Possess'' & Reproduction & \textit{Joy} \\

        \colourRow New Territory & ``What's out there?'' &
        Exploration & \textit{Interest} \\

        Unexpected Object & ``What is it?'' & Orientation & \textit{Surprise} \\

        \colourRow Gruesome Object & ``Poison'' & Rejection &
        \textit{Disgust} \\

        Group Member & ``Friend'' & Incorporation & \textit{Trust} \\
        \bottomrule
    \end{tabular}%
    }
\end{table}

PES compares its \ref{circumplex} structure to the colour wheel. This has some
interesting implications such as the ability to ``mix'' the emotions like
primary colours. This creates ``complex'' emotions that have the prototypical
behaviours of each component emotion (Table~\ref{tab:emotionMix}). For example,
mixing the primary emotions \textit{Disgust} and \textit{Anger} creates the
secondary emotion \textit{Contempt}, which has the prototypical behaviour
tendencies of both component emotions---rejection and destruction. The theory
does provide evidence for some mixtures collected from studies that asked
participants to name the underlying primary emotions. PES also says that mixing
could produce a valid emotion that does not yet have a name.
\begin{table}[!ht]
    \centering
    \caption{Examples of Emotion ``Mixtures'' in PES}
    \label{tab:emotionMix}
    \begin{tabular}{P{0.2\linewidth}P{0.25\linewidth}}
        \toprule
        \textbf{Emotion Label} & \textbf{``Mixture'' Of} \\ \midrule

        \colourRow \textit{Pride} & \textit{Joy} +
        \textit{Anger} \\

        \textit{Outrage} & \textit{Anger} + \textit{Surprise}   \\

        \colourRow \textit{Panic} & \textit{Terror} +
        \textit{Interest}  \\

        \textit{Hope} & \textit{Trust} + \textit{Interest}   \\

        \colourRow \textit{Hatred} & \textit{Loathing} +
        \textit{Anger} \\ \bottomrule
    \end{tabular}
\end{table}

PES proposes that the necessary sequence of events between a stimulating event
and emotion is cognitive appraisal, subjective reaction, and behavioural
reaction. It theorizes that cognition developed to predict future needs and
events based on prior experience, a mental model of the environment, and
available information. Sensory input and expanded memory gives organisms the
capacity to name objects, create concepts, and identify relationships between
them. Therefore, cognitive processes evolved with the brain to serve emotions
and biological needs.

\section{Dimensional Theories}
Dimensional theories focus on what kind of mental states emotions are, how to
construct them, and how they fit into a general taxonomy of mental
states~\citep[p.~250]{reisenzein2013computational}. They define a coordinate
space for ``core'' affect (Chapter~\ref{sec:affectiveDefs}) using two or three
affective dimensions, such as \textit{\ref{valence}}, \textit{\ref{arousal}},
and \textit{Dominance} or \textit{Stance}~\citep[p.~97]{lisetti2015and}.
Psychologists have located emotions in this
space~\citep[p.~9]{hudlicka2014habits}, suggesting that affective states are
systematically related. However, dimensional representations can lose
information about an emotional state if its resolution exceeds the named
dimensions~\citep[p.~172]{schaap2008towards}. This results in a lower
resolution of information than other perspectives. Defining an emotion with
dimensions can also be unreliable, as their meaning changes with the
context~\citep[p.~61]{lazarus1991emotion}, and the chosen statistical analysis
technique could bias the dimensions found in the
data~\citep[p.~825]{smith1985patterns}.

Although appealing in their simplicity and relative ease to
implement~\citep[p.~440]{rodriguez2015computational}, dimensional theories say
relatively little about how to generate emotions and what their effects
are~\citep[p.~10]{hudlicka2014habits} making them unsuitable for defining a
complete CME \citep[p.~250]{reisenzein2013computational}.

\subsection{\ref{valence}-\ref{arousal} (V-A)}\label{adx:va}
Perhaps the simplest way to approximate affect, V-A describes emotion with two
dimensions---\textit{\ref{valence}} and \textit{\ref{arousal}}---which appear
more consistently across affective dimension studies than any other
dimensions~\citep[p.~814, 816]{smith1985patterns}. Psychologists have treated
\textit{\ref{valence}} and \textit{\ref{arousal}} as independent dimensions,
although affect fits more consistently in a \ref{circumplex}
structure~\citep[p.~12]{barrett1999structure}.

The idea of the V-A dimensions forming part of affective space goes at least as
far back as \cite{wundt1912introduction} (\citepg{izard1971face}{83};
\citepg{lang1995emotion}{373}; \citepg{barrett1999structure}{10}). Other
iterations of these dimensions appear in more recent affective models such as
\cite{russell1980circumplex}.

\subsection{Pleasure-\ref{arousal}-Dominance (PAD) Space}\label{adx:pad}
This theory views emotion as a mediator between stimuli and
behaviours~\citep{mehrabian1980basic, mehrabian1996pleasure}. With the goal of
quantifying different types of affective phenomena and drawing from studies in
a variety of related fields, PAD Space describes three nearly orthogonal
dimensions for analyzing emotional states and behaviours while relating them to
other affect types and experiences:
\begin{itemize}

    \item \textit{Pleasure} measures the positive-negative aspects of the
    emotion state,

    \item \textit{\ref{arousal}} is how alert and active the individual is in
    that state, and

    \item \textit{Dominance} is how much control the individual feels they have
    in that state.

\end{itemize}

These dimensions are present in all affective reactions operative in a
situation. Three dimensions are optimal for general characterizations and
measurements of emotional states because two dimensions cannot distinguish
between clusters of affect and more than three does not further improve cluster
distinctions.

\section{Appraisal Theories}
Appraisal theories emphasize distinct emotion components such as appraisal
dimensions or variables~\citep[p.~97]{lisetti2015and}. Appraisal variables
represent particular emotions by mapping onto an $n$-dimensional
space~\citep[p.~306]{hudlicka2014computational} which might appear identical in
a lower-dimensional space~\citep[p.~489]{lerner2000beyond}. Analyzing stimuli
for meaning and consequences with respect to an individual generates values for
these variables (\citepg{reisenzein2013computational}{250};
\citepg{smith1985patterns}{819}), regardless of process
sophistication~\citep[p.~273]{gratch2004domain} and independent of biological
processes~\citep[p.~559]{arbib2004emotions}. Appraisals are continuous and
change with the situation, the individual's behaviours, and their attempts to
appraise the situation differently~\citep[p.~28]{siemer2007appraisals}.

Individuals appraise situations differently, accounting for different reactions
to the same situation (\citepg{siemer2007same}{598};
\citepg{smith2009putting}{1353}) due to their personality and biases. This
makes appraisal theories of particular interest for decision-making, action
selection, facial animations, and personality-based
CMEs~\citep[p.~274]{gratch2004domain}. Models of social intelligence could also
use appraisals because they also seem important for mediating social
relationships. However, there is little empirical data associating individual
appraisal variables to expressive behaviours or behavioural
choices~\citep[p.~10]{hudlicka2014habits}.

\subsection{Frijda's Concern Realization (CR)}
CR views emotions as ``changes in action readiness'' that prepare individuals
to change their behaviours in response to or to change something about their
current environment~\citep{frijda1986emotions}. Action readiness can relate to
features in the environment or to a general ``activation mode''.

Concerns, including goals, are  dispositions towards internal conditions.
Cognitive processes use concerns to assign emotional significance to events and
stimuli. CR provides empirical evidence for several appraisal dimensions
relating events and stimuli to concerns, including~\citep{frijda1987emotion}:
\begin{itemize}
    \item \textit{\ref{valence}}, describing the pleasantness or unpleasantness
    of an event,

    \item The \textit{impact} of the event, or if it is an event at all,

    \item \textit{Interestingness}, similar to novelty, which naturally
    accompanies the emotion of \textit{Interest},

    \item \textit{Globality}, or if the individual can locate the event in
    space,

    \item The \textit{uncertainty} of the event's outcome,

    \item If the \textit{responsibility} of the event's occurrence was due to
    the self or another, and

    \item If the event was \textit{relevant} to the individual.
\end{itemize}

Studies also found weak evidence for the dimensions of \textit{certainty},
\textit{controllability}, \textit{accessibility}, \textit{self-esteem},
\textit{modifiability}, \textit{manageability}, and \textit{time reference}. CR
then connected the appraisal dimensions to proposed action tendencies and some
named emotions and moods (Table~\ref{tab:crAppraisalRelations}).
\begin{table}[!tb]
    \centering
    \caption{Relationship Between Action Tendencies and Some Emotions and Moods
        in CR}
    \label{tab:crAppraisalRelations}
    \renewcommand{\arraystretch}{1.2}
    \begin{threeparttable}
        \begin{tabular}{ll}
            \toprule
            \textbf{Action Tendency} & \textbf{Emotion/Mood} \\ \hline

            \colourRow Approach & Energetic Mood \\

            Avoidance & Fear \\

            \colourRow Being-With & Happiness, Energetic Mood \\

            Attending & Happiness, Energetic Mood \\

            \colourRow Rejection & -- \\

            Indifference & -- \\

            \colourRow Antagonism & Anger, Irritable Mood, Distrust \\

            Interruption & -- \\

            \colourRow Dominance & -- \\

            Submission & Sadness\textpmhg{\Hibp} \\

            \colourRow Apathy & Sadness, Tired, Uninterested \\

            Excitement & Anger\textpmhg{\Hibp}, Energetic Mood,
            Distrust\textpmhg{\Hibp} \\

            \colourRow Exuberance & Happiness, Energetic Mood \\

            Passivity & Tired, Uninterested \\

            \colourRow Inhibition & Fear \\

            Helplessness & Nervousness \\

            \colourRow Blushing & -- \\

            Rest & -- \\ \hline

            \bottomrule
        \end{tabular}

        \begin{tablenotes}

            \vspace*{2mm}

            \item []{\normalsize\textpmhg{\Hibp}} \textit{Only somewhat
            related.}

            \vspace*{-2mm}
        \end{tablenotes}
    \end{threeparttable}
\end{table}

\begin{figure}[!tb]
    \centering
    \includegraphics[width=0.75\linewidth]{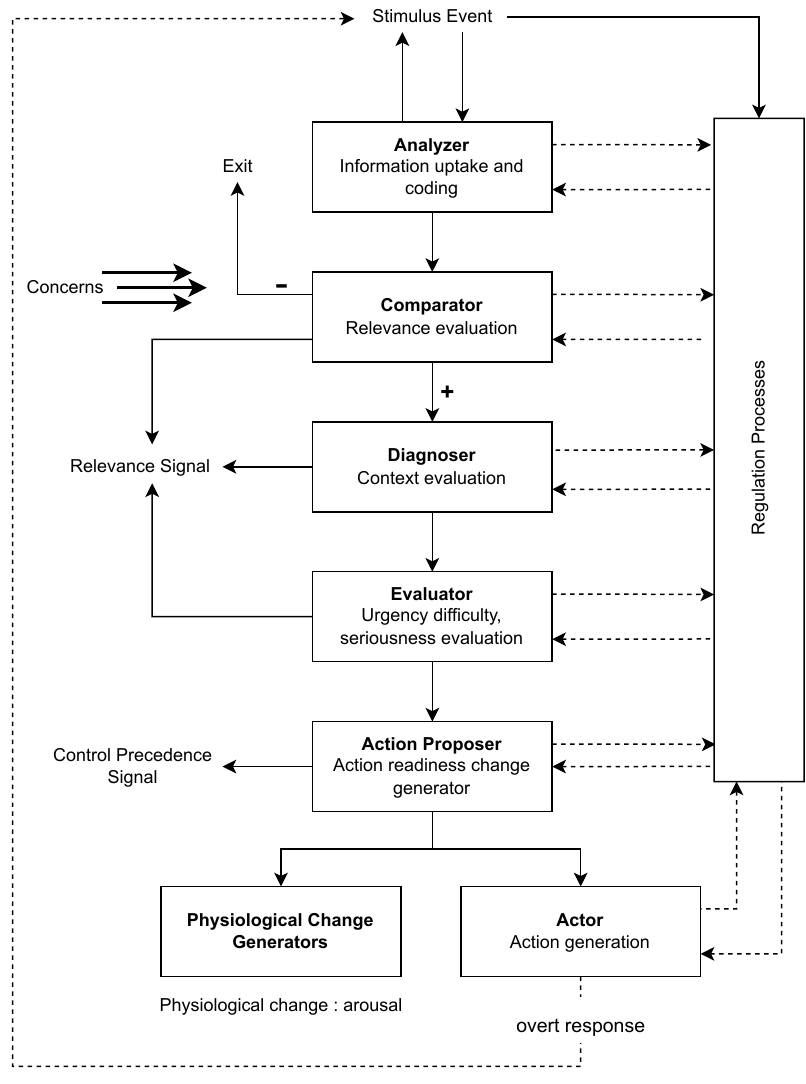}

    \caption[System View of CR]{System View of CR (Adapted from
    \citeauthor{frijda1986emotions}, \citeyear{frijda1986emotions}, p.~454)}
    \label{fig:FrijdaSystemModel}
\end{figure}

CR compares the emotion process to a continuous information processing system
(Figure~\ref{fig:FrijdaSystemModel}). The emotion system constantly takes in
information, analyzing its relevance to an individual's concerns. The system
then determines what the individual can do to cope and if it should interrupt
the individual's current behaviours to address this information. Finally, the
system proposes an action readiness change---an action plan, tendency, or
activation mode---which tries to take control precedence and instigates
physiological and behavioural changes. Any of these processes can take
additional inputs or feedback from subsequent processes. The system can skip,
interrupt, or evaluate processes in different orders. In this view, what
psychologists see as primary emotions are discrete events that emerge from a
continuous system that can produce a much broader range of affect.

\subsection{Lazarus's Cognitive Appraisal (CA)}
CA is a system theory focusing on both the transactional nature of emotions and
recurrent, stable patterns in the environment and individual that can explain
stable emotion patterns~\citep{lazarus1991emotion}. It relies on the concept of
the person-environment relationship---a series of adaptational encounters
between the individual and their environment, influenced by the individual's
personality. An individual evaluates the person-environment relationship's
relevance to themselves with \textit{primary} and \textit{secondary} appraisals
to assign personal significance to their knowledge (e.g. goals, beliefs).
Their relative importance orders appraisals---if the results of primary
appraisal determine the event to be irrelevant, then secondary appraisal has
limited, if any, value.

Primary appraisal asks if the current person-environment relationship is:
relevant to the individual's well-being, establishing benefits and harms; and
if the benefits/harms are a certain, currently observable outcome, or a
potential future outcome. It has three components:
\begin{itemize}
    \item \textit{Goal Relevance} establishes if there are personal goals or
    stakes affected by the current person-environment relationship,

    \item \textit{Goal Congruence} is the extent to which an encounter benefits
    or harms the affected goal, and

    \item \textit{Type of Ego-involvement} establishes what aspect of the self
    is affected---self or social esteem, moral values, ego ideals, meanings and
    ideas, other persons and their well-being, and life goals. This
    differentiates similar emotions such as \textit{Joy} (no involvement) and
    \textit{Affection} (identity).
\end{itemize}

Secondary appraisal asks what resources and actions are available to the
individual for coping with the person-environment relationship, dictating the
experienced emotion. Expectations and beliefs about potential actions and the
individual's ability to act on them can influence it. It establishes:
\begin{itemize}
    \item \textit{Accountability}, assigning blame and credit for the
    person-environment relationship. It requires knowledge of who or what is
    responsible for the current situation and if they had control over the
    action that caused it.

    \item \textit{Coping Potential}, evaluating if and how the individual can
    manage the current individual\-/environment relationship's demands or
    actualized personal commitments. At this point, the individual does not
    act. How much energy the individual believes they have is likely a factor,
    as the pursuit of goals must require it.

    \item \textit{Future Expectations} is the predicted likelihood of
    psychological change in the future (i.e. a change in goal congruence).
\end{itemize}

Answers to appraisal questions filter potential emotions in a structure
reminiscent of a decision tree, from least to most specific emotion, starting
with its \textit{goal congruence}. CA calls a configuration of these variables
a \textit{core relational theme}. It uniquely associates each theme with an
emotion kind (Table~\ref{tab:ca_relationalthemes}), which is itself partially
defined by an action tendency~\citep[p.~59]{lazarus1991emotion}. How the
individual interprets these themes depends on their personality and learned
meanings.
\begin{table}[!b]
    \centering
    \caption[The Core-Relational Themes of Emotions in CA]{The Core-Relational
    Themes of Emotions in CA~\citep[p.~122]{lazarus1991emotion}}
    \label{tab:ca_relationalthemes}
    \renewcommand{\arraystretch}{1.2}
    \resizebox{\linewidth}{!}{%
    \begin{threeparttable}
        \begin{tabular}{P{0.25\linewidth}P{0.68\linewidth}}
            \toprule
            \textbf{Emotion} & \textbf{Core Relational Theme} \\ \hline

            \colourRow Anger & Demeaning offence to me and mine \\

            Fright & Imminent physical harm \\

            \colourRow Anxiety & Uncertain, existential threat \\

            Guilt & Having transgressed a moral imperative \\

            \colourRow Shame & Failure to live up to an ego-ideal \\

            Sadness & Irrevocable loss \\

            \colourRow Envy & Wanting what someone else has \\

            Jealousy & Resenting a third party for loss \textit{OR} threat to
            another's affection \\

            \colourRow Disgust & Taking in or being too close to an
            indigestible object or idea (metaphorically) \\

            Happiness/Joy & Making reasonable progress towards our goals \\

            \colourRow Pride & Enhancement of one's ego-identity by taking
            credit for a valued object or achievement, either our own or that
            of someone or group with whom we identify \\

            Love/Affection & Desiring or participating in affection, usually but
            not necessarily reciprocated \\

            \colourRow Relief & A distressing goal incongruent condition has
            changed for the better or gone away \\

            Hope~{\Large\textpmhg{\Hl}} & Fearing the worst but yearning for
            better \\

            \colourRow Compassion~{\Large\textpmhg{\Hl}} & Being moved by
            another's suffering and wanting to help \\

            Aesthetic Emotions~{\Large\textpmhg{\Hl}} & \textpmhg{\HKi} \\
            \hline
            \bottomrule
        \end{tabular}

        \begin{tablenotes}

        \vspace*{2mm}

        \item []{\Large\textpmhg{\Hl}} \textit{``Problematic'' emotions that
        have uncertain status as proper emotions in CA.}

        \item []{\normalsize\textpmhg{\HKi}} \textit{Does not involve new
        emotions; generated emotion determines core relational theme.}

        \end{tablenotes}
    \end{threeparttable}%
    }
\end{table}

\begin{figure}[!tb]
    \centering
    \includegraphics[width=0.75\linewidth]{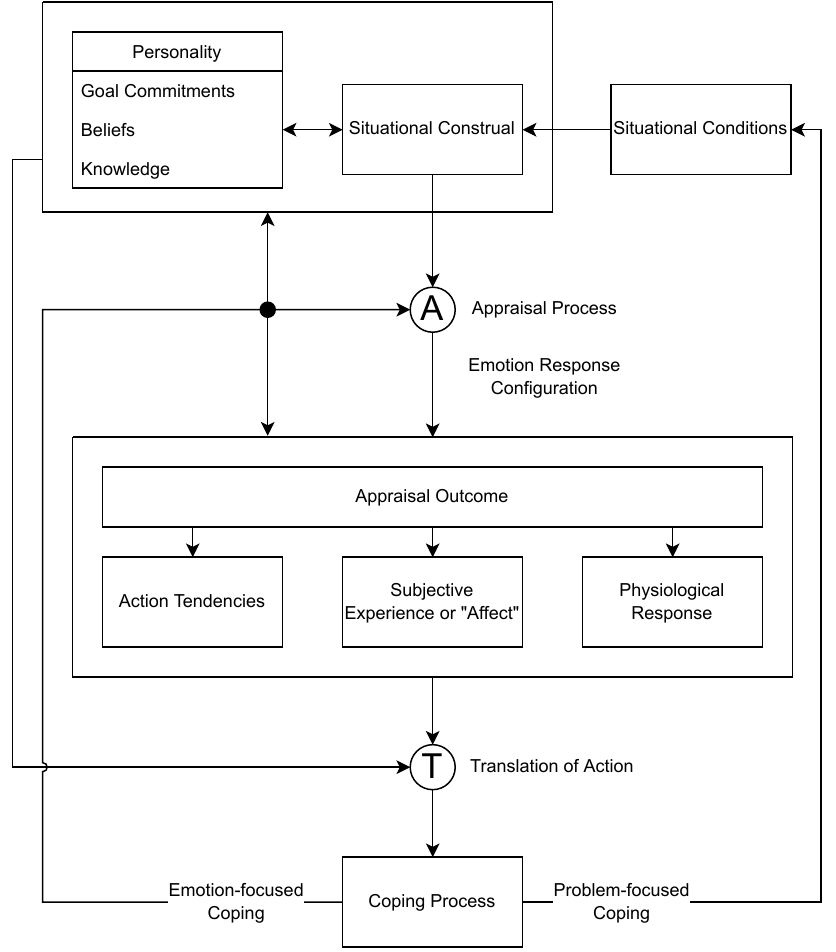}

    \caption[System View of CA]{System View of
    CA~\citep[p.~623]{smith1990emotion}}
    \label{fig:CASystemModel}
\end{figure}

CA's system view, when taken as a series of time slices, shows how emotions
develop and change with an individual's evaluation of their situation
(Figure~\ref{fig:CASystemModel}). This view also emphasizes how system
variables impact each other, and the integral role of \ref{coping}---an
individual's ability to change their relationship with the environment---in the
process. Coping is a more deliberate process than action
tendencies~\citep[p.~39, 197]{lazarus1991emotion} and typically causes a
\textit{reappraisal} to consider how it changed the person-environment
relationship. In contrast to general appraisals, reappraisals are usually a
conscious and deliberate decision.

\subsection{Scherer's Component Process Model of Emotion (CPE) and Sequential
Check Theory of Emotion Differentiation (SCT)}
This theory views emotion as a continuous mechanism that developed through
evolution to afford flexible adaptation to a changing environment by decoupling
stimuli from responses~\citep{scherer2001appraisalB}. This creates latency time
for optimizing responses that affect subsystems such as the central nervous
system. An emotion, then, is an episode of interrelated, synchronized changes
across cognition, physiology, motivation, expression, and subjective feelings
in response to an evaluation of events that are relevant to the organism.

SCT is a component of CPE, describing how to evaluate events and how they lead
to specific emotions. It defines Stimulus Evaluation Checks (SECs) as a minimal
set of criteria necessary to differentiate emotion families
(Table~\ref{tab:schererSECs}). SCT organizes SECs, which can be scalar or
multidimensional, by appraisal objective:
\begin{itemize}
    \item \textit{Relevance} to determine if the event impacts the organism and
    warrants further processing;

    \item \textit{Implication} to evaluate the extent of the impact on the
    organism's survival, adaptation, and goal satisfaction;

    \item \textit{Coping potential} to see what responses are available and the
    consequences each one will have on the organism; and

    \item For social species, the \textit{normative significance} of an event
    is an evaluation of how the organism expects its social group members and
    its own self-concept to perceive its response.
\end{itemize}
\begin{table}[!b]
    \caption{CPE/SCT Stimulus Evaluation Checks (SECs)}
    \label{tab:schererSECs}
    \renewcommand{\arraystretch}{1.2}
    \resizebox{\linewidth}{!}{%
    \begin{tabular}{P{0.15\textwidth}P{0.28\textwidth}P{0.49\textwidth}}
        \toprule
        \textbf{Appraisal Objective} & \textbf{Component} &
        \textbf{Evaluations} \\ \midrule

        \multirow{3}{*}{Relevance}& \colourCell Novelty & \colourCell
        Suddenness, Familiarity, Predictability \\

        & Intrinsic Pleasantness & Feature of the stimulus \\

        & \colourCell Goal Relevance & \colourCell Affected Goals \\\hline

        \multirow{7}{*}{Implication} & Causal Attribution & Responsibility
        Assignment, Motive/Intention Inference \\

        & \colourCell Outcome Probability & \colourCell Likelihood/Certainty of
        Event Outcomes \\

        & Discrepancy from Expectation & Number of features in current
        situation matching original expectation of it \\

        & \colourCell Goal/Need Conduciveness & \colourCell Conduciveness of
        the event to goal achievement \\

        & Urgency & Event significance to  goal, temporal contingencies \\\hline

        \multirow{6}{*}{\begin{tabular}[l]{@{}l@{}}Coping \\
                Potential\end{tabular}}&
        \colourCell Control & \colourCell Extent that people or animals can
        influence the event \\

        & Power & (If Control is possible) Evaluation of resources to change
        contingencies and outcomes according to individual interests \\

        & \colourCell Adjustment & \colourCell Ability to adapt to the outcomes
        of the event, especially if Control and Power are poor \\\hline

        \multirow{4}{*}{\begin{tabular}[l]{@{}l@{}}Normative \\
                Significance\end{tabular}} & Internal Standards & Comparison of
                an action to internal standards (e.g. self-ideal) \\

        & \colourCell External Standards & \colourCell Comparison of action to
        the social norms, values, and standards of a reference group (e.g.
        culture) \\\hline

        \bottomrule
    \end{tabular}%
    }
\end{table}

CPE assumes that a SEC sequence can predict emotions. Proposed SEC sequences
for predicting modal emotions (``kinds'') include:
\textit{Enjoyment/Happiness}, \textit{Elation/Joy},
\textit{Displeasure/Disgust}, \textit{Contempt/Scorn},
\textit{Sadness/Dejection}, \textit{Despair}, \textit{Anxiety/Worry},
\textit{Fear}, \textit{Irritation/Cold Anger}, \textit{Rage/Hot Anger},
\textit{Boredom/Indifference}, \textit{Shame}, \textit{Guilt}, and
\textit{Pride}.

\begin{figure}[!t]
    \centering
    \includegraphics[width=0.75\linewidth]{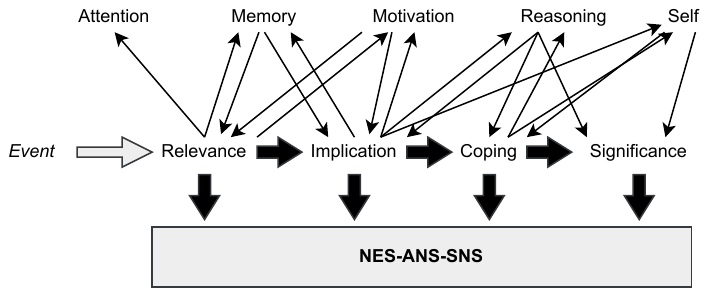}

    \caption[SCT Appraisal Process]{SCT Appraisal
    Process~\citep[p.~100]{scherer2001appraisalB}}
    \label{fig:schererProcess}
\end{figure}

SCT evaluates SECs sequentially in a fixed order so that they act as conditions
for further processing (Figure~\ref{fig:schererProcess}). For example,
evaluating the nature and consequences of an event---its \textit{implications}
for an organism---is not worth the processing expense if it does not pass at
least one of the \textit{relevance} checks. Due to the continuous nature of
detection, events can trigger reappraisal cycles. These cycles update SECs
until the monitoring subsystem signals termination or an adjustment to the
original stimulating event. CPE can process SECs in parallel, but does not
finalize a result until prerequisite SECs are complete. Realizing SECs as
registers makes this possible where each SEC stores its own results, holding
the best estimate of the current evaluation for reference at any time. The
process combines values in SEC registers using weighted functions.

The evaluation speed of different criteria within a SEC can differ because they
can involve mechanisms of varying complexity at the sensory-motor, schematic,
and conceptual levels. These levels account for the variation of an emotional
response over time, representing feature detection and reflexes, learned
responses, and deliberate reasoning.

SEC evaluations cause underlying, highly interdependent subsystems to change
what they do such that it records a ``historical account'' of the original
stimulus. An outcome profile collects the changes that each SEC makes to its
predecessor's changes, creating patterns that can distinguish between emotion
types~\citep[p.~114--115]{scherer2001appraisalB} and evaluate their intensity.

\subsection{Roseman's Emotion System Model (ESM)}
The ESM states that, given the number of appraisal theories available, moving
towards a unified theory requires a consideration of the empirical evidence of
emotion appraisals~\citep{roseman1996appraisal}. To this end, the ESM continues
to undergo incremental revisions as new evidence emerges
(\citepg{roseman2011emotional}{436}; \citepg{roseman2018functions}{149}). The
ESM also proposes mediating answers to debated topics such as the discrete
versus dimensional nature of emotions (\citepg{roseman2011emotional}{435};
\citepg{roseman2013appraisal}{146--148}). For example, it proposes that the
discrete and dimensional perspectives are complementary and a single system can
represent them~\citep[p.~441]{roseman2011emotional}. Appraisal dimensions are
thought to be continuous, but their combinations result in categories of coping
strategies that are characteristic of discrete
emotions~\citep[p.~147]{roseman2013appraisal}.

Improving on previous work~\citep{roseman1996appraisal}, a study presented
subjects with an experience involving either two negative or two non-negative
discrete emotions. Evaluators asked subjects to describe the experience, what
they felt caused their reaction, and rate the extent to which proposed
appraisals from competing appraisal theories caused the emotion. A model of the
17 discrete emotions from the study incorporated statistically significant
appraisal dimensions (Figure~\ref{fig:roseman}). The dimensions in the revised
ESM are:
\begin{figure}[!t]
    \centering
    \includegraphics[width=0.8\textwidth]{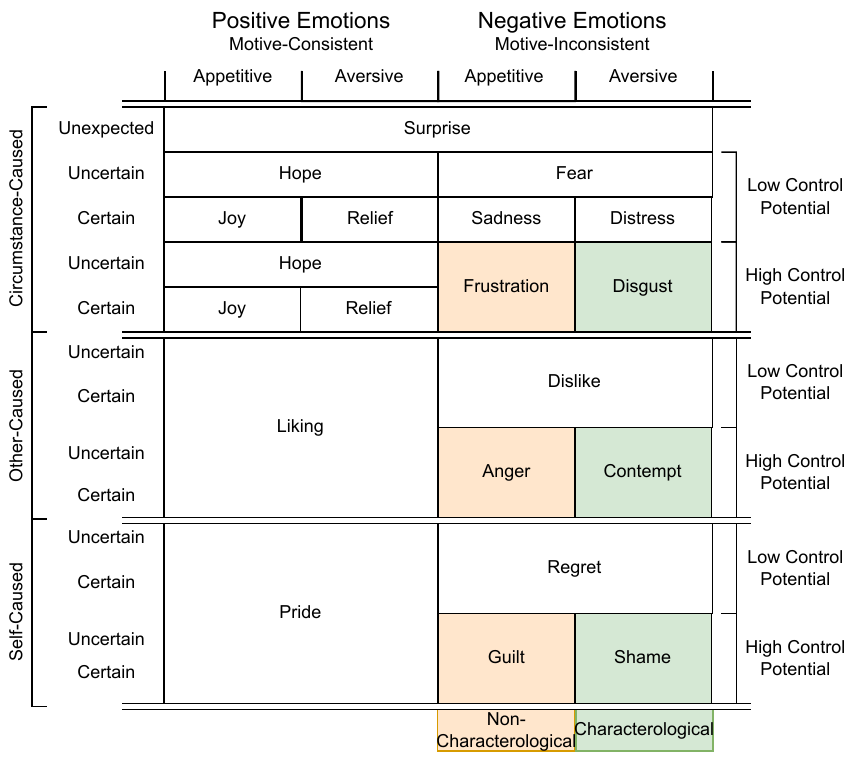}

    \caption[ESM Structure]{ESM Structure (\citepg{roseman1996appraisal}{269};
        \citepg{roseman2013appraisal}{143})}
    \label{fig:roseman}
\end{figure}
\begin{itemize}
    \item \textit{Situational state}, or if the event aligns with the
    individual's motives (motive\-/consistent) or not (motive-inconsistent);

    \item \textit{Motivational state}, describing if the individual sees a
    potential reward (appetitive) or punishment (aversive) in the event;

    \item The \textit{unexpectedness} of the situation, exclusively for the
    appraisal of \textit{Surprise};

    \item \textit{Agency}, describing if the situation was circumstantial,
    caused by others, or caused by the individual
    (self)\footnote{\citet[p.~825]{smith1985patterns} found a similar
    multi-part representation of \textit{agency}.};

    \item \textit{Probability}, or the uncertainty of the event's outcome;

    \item \textit{Control potential}, a combined evaluation of the individual's
    ability to do something about the situation and their \ref{coping}
    potential; and

    \item The \textit{problem source}, or if the problem is intrinsic to the
    source of the event (characterological) or not (non-characterological),
    necessary to distinguish similar negative emotions such as \textit{Anger}
    and \textit{Contempt}.
\end{itemize}

\subsection{The Ortony, Clore, and Collins (OCC) Model}
The OCC model focuses on the system of cognitive representations (i.e. value
system) underlying emotions~\citep{occ2022}. It proposes structures for the
overall organization and individual structure of emotions, as well as how to
evaluate their intensity with respect to personal and interpersonal situation
descriptions. These rely on the structure, content, and organization of
knowledge and the processes that work on them. An individual's construal of the
eliciting conditions determines what kind of emotion the scenario elicits. This
accommodates individual and cultural differences without changing the structure
or definitions of the emotions themselves.

The OCC model defines emotions as valenced reactions to three types of world
aspects associated with knowledge structures:
\begin{itemize}
    \item \textit{Events} and their consequences for an individual's Goals
    (i.e. representations of desired world states)

    \item \textit{Agents}---which contribute to \textit{Events} either
    instrumentally or by the attribution of agency---whose actions an
    individual compares to their Standards (i.e. representations of points of
    reference or criteria, often moral in nature)

    \item \textit{Objects} and their qualities as they relate to an
    individual's tastes (i.e. representations of dispositional likes and
    dislikes)
\end{itemize}

Each emotion definition is modular, differentiated by their cognitive origins.
The OCC model organizes the global emotion structure by types
(Figure~\ref{fig:occStructure}) recognizable with many linguistic ``tokens''
such as \textit{Content} and \textit{Elated} for \textit{Joy}. The structure
groups emotion types by their structurally-related eliciting conditions. Each
group contains two emotion types representing the positive and negative
emotions arising from one of three central appraisal variable evaluations:
\textit{desirability} and \textit{undesirability} for Events relative to Goals;
\textit{praiseworthiness} and \textit{blameworthiness} for Agents relative to
Standards; and \textit{appealing} and \textit{unappealing} for Objects relative
to Tastes.
\afterpage{
    \vspace*{\fill}
    \begin{figure}[!ht]
        \centering
        \includegraphics[width=0.9\linewidth]{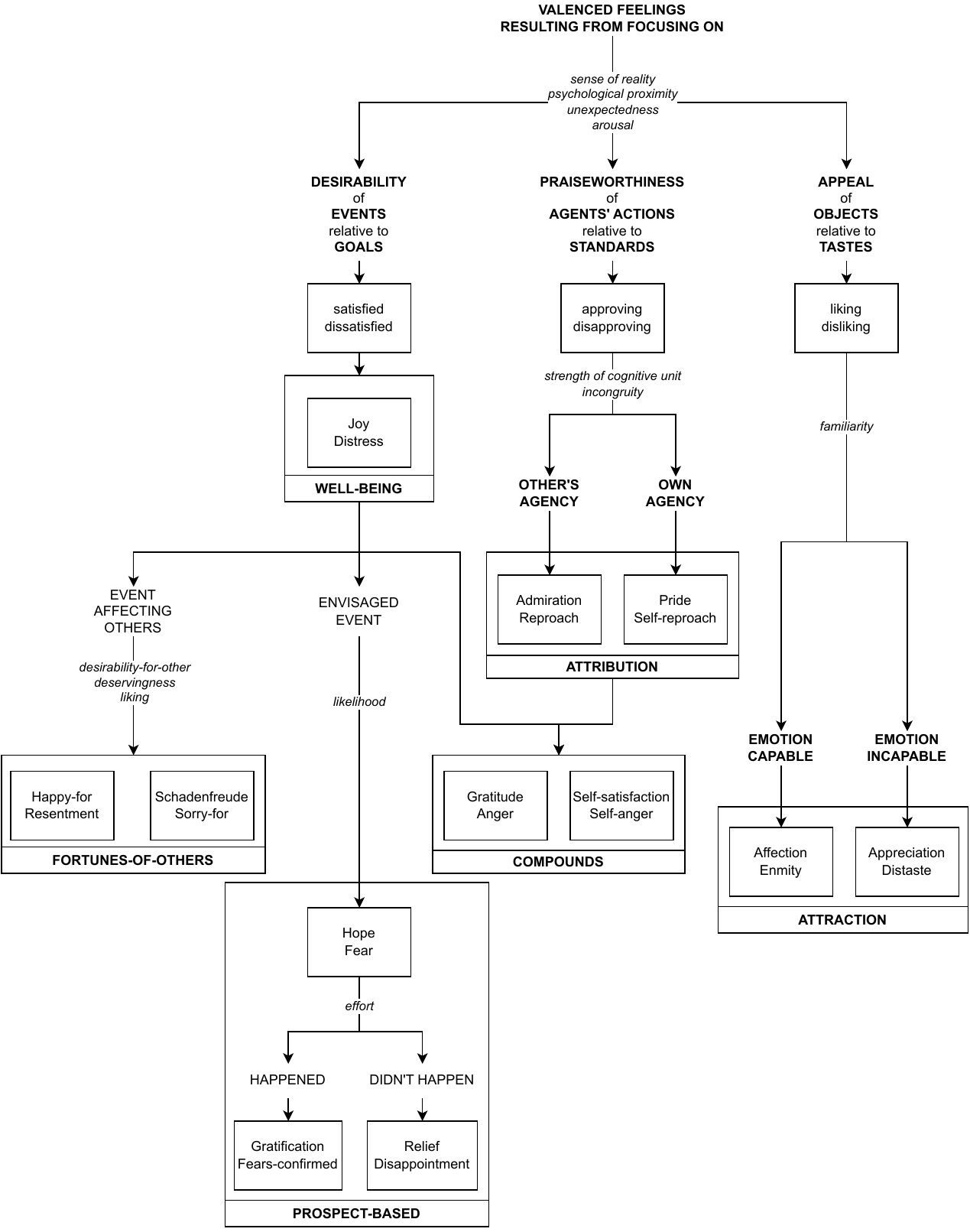}

        \caption[Global, Logical Structure of Emotion Types in OCC]{Global,
        Logical Structure of Emotion Types in OCC (Appraisal Variables are
        \textit{italicized})~\citep[p.~29, 84]{occ2022}}
        \label{fig:occStructure}
    \end{figure}
    \vspace*{\fill}
}

The OCC theory also proposes a method for evaluating emotion intensity based on
its appraisal variables (Table~\ref{tab:occVariables}). The
\textit{desirability}, \textit{praiseworthiness}, and \textit{appeal} central
variables drive all emotion intensity evaluations, which other global and local
variables influence. Each emotion type specification includes a list of such
variables impacting its intensity. These variables follow from the emotion type
structure which appraisal collects as it progresses through the structure.
\afterpage{
\begin{landscape}
    \begin{table}[!ht]
        \centering
        \renewcommand{\arraystretch}{1.2}
        \caption{OCC Appraisal Variables}
        \label{tab:occVariables}
        \resizebox{\linewidth}{!}{%
        \begin{threeparttable}
            \begin{tabular}{P{0.07\linewidth}P{0.2\linewidth}P{0.79\linewidth}}
                \toprule
                \textbf{Location} & \textbf{Variable} & \textbf{Definition}
                \\ \midrule

                \multirow{7}{*}{Global} &
                \colourCell Sense of Reality & \colourCell How ``real'' the
                Event/Agent/Object is to the individual (e.g. probability
                of success); based on cognitive interpretations of the
                appraisal target (e.g. imagined scenarios) \\

                & Psychological Proximity & How psychologically ``close''
                the appraisal target is (e.g. spatially, temporally);
                connected to, but not dependent on, ``Sense of Reality'' \\

                & \colourCell Unexpectedness & \colourCell Degree of
                mismatch between the appraisal target and what was
                predicted (i.e. mispredicted or not predicted at all,
                retrospectively) \\

                & Arousal & A state of enhanced mental
                activation/excitement; slow to decay; non-emotional effects
                can impact and carry the value forward; must be attributed
                to the appraisal target or previous reactions to the same
                situation \\
                \hline

                \multirow{7}{*}{Events} &
                \colourCell Desirability{\Large\textpmhg{\Hl}}/
                Undesirability{\Large\textpmhg{\Hl}} &
                \colourCell Evaluated on Goals; the total values of Goals
                that achieving this currently impacted Goal helps
                (desirability) and hinders (undesireability)
                \\

                & Desirability-for-Others & An evaluation of the presumed
                desirability of an Event with respect to another
                individual's Goal(s) \\

                & \colourCell (Dispositional) Liking & \colourCell The
                degree to which this individual likes another individual \\

                & Deservingness & The degree of belief that the outcome of
                an Event is fair to another individual \\

                & \colourCell Likelihood & \colourCell Perceived chances of
                an Event's outcome (prospectively); comparative and
                qualitative, not constant over time \\

                & Effort & The investment an individual has made towards an
                Event's desirable outcome; includes mental, physical, and
                materialistic investments \\

                \hline

                \multirow{5}{*}{Agents} &
                \colourCell Praiseworthiness{\Large\textpmhg{\Hl}}/
                Blameworthiness{\Large\textpmhg{\Hl}} & \colourCell
                Evaluated on Standards; the degree to which an Agent
                appears to uphold (praiseworthiness) and violate
                (blameworthiness) a Standard with their  action(s) \\

                & Strength of Cognitive Unit & The degree that this
                individual associates with the Agent; accommodates cases
                where the individual experiencing the emotion is not the
                target Agent, but is affiliated with them \\

                & \colourCell Incongruity & \colourCell The degree that an
                Agent deviates from what the individual expects from them;
                can be a mismatch between the Agent's social role and their
                actions \\

                \hline

                \multirow{2}{*}{Objects} & Appealing{\Large\textpmhg{\Hl}}/
                Unappealing{\Large\textpmhg{\Hl}} & Evaluated on Tastes;
                the individual's disposition towards liking (appealing) and
                disliking (unappealing) something; does not require an
                evaluation of significance \\

                & \colourCell Familiarity & \colourCell The level of
                exposure the individual has experienced with the target
                Object \\

                \midrule\bottomrule
            \end{tabular}

            \begin{tablenotes}

                \vspace*{2mm}

                \item []{\Large\textpmhg{\Hl}} Central Variable

            \end{tablenotes}
        \end{threeparttable}%
    }
    \end{table}
\end{landscape}
}

\afterpage{
    \vspace*{\fill}
    \begin{figure}[!tbh]
        \centering
        \begin{subfigure}[b]{\linewidth}
            \centering
            \includegraphics[width=0.77\linewidth]{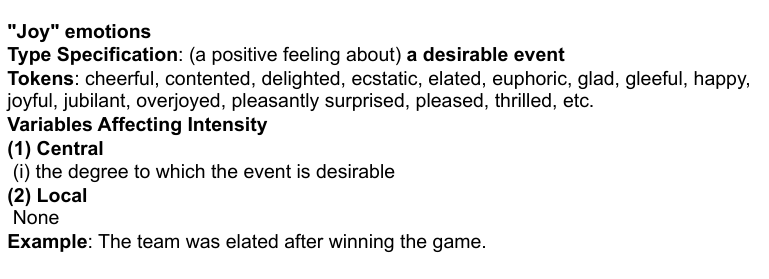}
            \caption{Description}
            \label{fig:joyDescription}
        \end{subfigure}

        \begin{subfigure}[b]{\linewidth}
            \centering
            \includegraphics[width=0.77\linewidth]{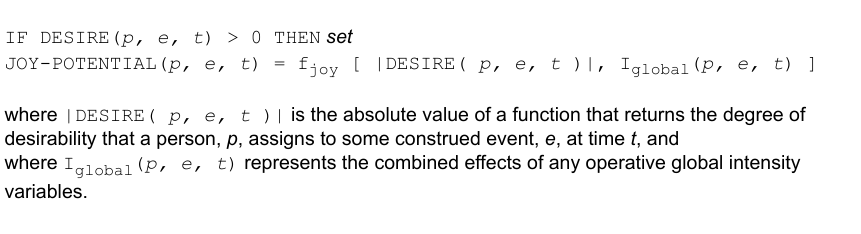}
            \caption{Computational Rule}
            \label{fig:joyRule}
        \end{subfigure}
        \caption[Example of OCC \textit{Joy}]{Example of OCC
            \textit{Joy}~\citep[p.~102, 220]{occ2022}}
        \label{fig:occJoyExample}
    \end{figure}
    \vspace*{\fill}
    \begin{figure}[!tbh]
        \centering
        \includegraphics[width=0.55\linewidth]{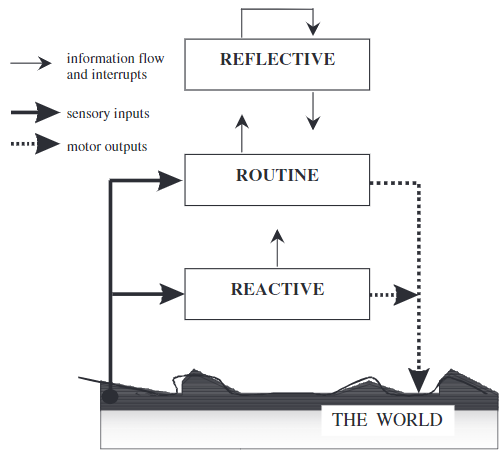}
        \caption[Schematic of OCC Processing Levels and their Main
        Interactions]{Schematic of OCC Processing Levels and their Main
            Interactions~\citep[p.~175]{ortony2005affect}}
        \label{fig:occProcess}
    \end{figure}
    \vspace*{\fill}
    \clearpage
}

Variables have a value and an assigned weight. Appraisal mechanisms produce
values from information available to the individual. These do not have to be
precise---if a mechanism cannot evaluate a variable, it retains its default
value. Default values tend to skew towards the positive ones. A variable's
weight is its ``degree'' of influence over intensity, which can vary between
emotion types and members in the same type. Finally, the process compares the
raw intensity value (i.e. emotion potential) with an emotion-specific threshold
value. Individuals only experience an emotion if its potential exceeds this
threshold and the experienced intensity is the difference between them.

The OCC model is clear about its goals and limitations: it focuses on the
cognitive structure of the emotion system and its emotion types without relying
on language and cultural meanings of emotion words; it does not theorize about
the the physiological, behavioural, or expressive components of emotion; it
does not propose a way to conceptualize how to appraise or combine variables;
it does not associate action tendencies with emotion kinds because OCC
hypothesizes that they are not characteristic of all emotions; it is meant to
be sufficiently specific for empirical testing; and the design aims to be
computationally tractable (Figure~\ref{fig:occJoyExample}).

Other work puts OCC's conception of emotion into a three-layered
information processing structure (Figure~\ref{fig:occProcess}). Guided by
neurophysiological findings, each layer represents different process types that
get progressively more complex as one moves through
them~\citep{ortony2005affect}:
\begin{itemize}

    \item At the reactive level, the system assigns value to a stimuli---called
    ``proto-affect''---which has multiple potential meanings. It represents
    simple environmental interpretations that have hard-wired responses. The
    routine level encompasses automatic, learned responses and cognitive
    processes that do not need to be consciously controlled.

    \item The routine level has enough information to identify four
    ``primitive'' emotion states: \textit{Happiness/Joy}, \textit{Distress},
    \textit{Excitement}, and \textit{Fear}. They can exist separately from
    cognition, so they are not yet ``full'' emotions.

    \item ``Full'' emotions are only possible at the reflective level, where
    complex reasoning, representation, and planning processes live. This allows
    it to interpret ``proto-affect'' and ``primitive'' emotion states from the
    lower levels and generate a discrete emotion label. This means that
    organisms cannot experience emotion without conscious reflection.

\end{itemize}

\subsection{Smith \& Kirby}
This theory's goal is to improve the validity of appraisal theories through
theoretical and empirical efforts~\citep{smith2001toward, smith2000structure,
smith1985patterns}. Smith \& Kirby focus on answering if and to what degree the
appraisal construct can explain the \ref{antecedent}s and organization of
physiological activity in emotion. They constructed the model using a
functionalist perspective and take Lazarus as a starting
point\footnote{Specifically the work of \cite{smith1990emotion}.},
``...recast[ing] the theory in more computational
terms''~\citep[p.~58]{marsella2015appraisal}. The theory describes appraisal in
two distinct, but complementary, ways:
\begin{itemize}
    \item \textit{Dimensional appraisal components}---which closely correspond
    with appraisal dimensions---representing questions that an individual
    evaluates, and

    \item \textit{Categorical relational themes}\footnote{Although they have the
    same origin, these themes are \textit{different} from Lazarus's core
    relational themes~\citep[p.~138]{smith2001toward}. The themes proposed by
    Smith \& Kirby are both more limited in scope and tightly linked.}
    representing significant answers to appraisal components and associated
    with distinct emotion kinds.
\end{itemize}

Smith \& Kirby put significant effort into matching the functions an emotion
serve  with a theme and appraisal component pattern. They identify seven
appraisal components:
\begin{enumerate}
    \item \textit{Motivational relevance}: How important is the situation to
    the individual?

    \item \textit{Motivational congruence}: To what extent is the situation
    consistent or inconsistent with the individual's current goals?

    \item \textit{Problem-focused coping potential}: To what extent can the
    individual act on the situation to increase or maintain its desirability?

    \item \textit{Emotion-focused coping potential}: To what extent can the
    individual psychologically adjust to the situation if it does not go
    favourably?

    \item \textit{Self-accountability}: To what degree is the individual
    responsible for the situation?

    \item \textit{Other-accountability}: To what degree is someone or something
    else responsible for the situation?

    \item \textit{Future expectancy}: For any reason, how much does the
    individual expect the situation to become more or less desirable?
\end{enumerate}

Smith \& Kirby assume that appraisals are relational, representing a stimulus
evaluation relative to an individual's needs, goals, beliefs, and values. This
allows appraisal to address the adaptivity of emotion and individual, temporal,
and contextual differences. They partially explored the relational nature of
emotions by creating test-specific relational \ref{antecedent} models for the
\textit{motivational relevance} and \textit{problem-focused \ref{coping}
potential} appraisal components. Smith \& Kirby made similar efforts to
empirically link physiological activity with individual appraisal components.
Separate studies found appraisal patterns for \textit{Affection},
\textit{Amusement}, \textit{Anger}, \textit{Anxiety}, \textit{Awe},
\textit{Compassion}, \textit{Determination}, \textit{Disgust},
\textit{Embarrassment}, \textit{Fear}, \textit{Gratitude}, \textit{Guilt},
\textit{Hope}, \textit{Interest}, \textit{Joy}, \textit{Pride},
\textit{Relief}, \textit{Sadness}, \textit{Shame}, and \textit{Tranquility}
(\citeg{yih2016distinct}; \citepg{yih2020profiles}{489}). These studies also
identified the appraisal dimensions of \textit{acceptability}, \textit{goal
attainment}, \textit{vastness}, \textit{involvement of others},
\textit{involvement of unknown}, \textit{likeability}, \textit{negative
evaluation by others}, \textit{positive evaluation by others}, and
\textit{urgency}.

\begin{figure}[!b]
    \centering
    \includegraphics[width=0.71\linewidth]{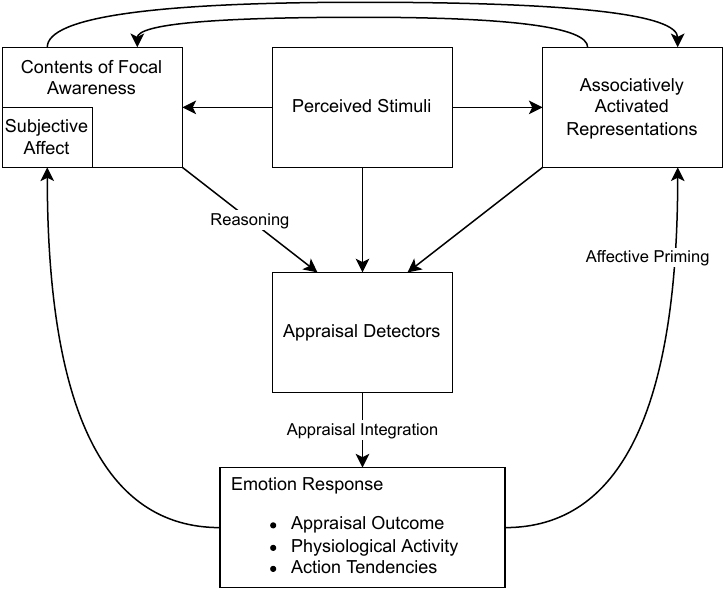}

    \caption[Appraisal Process Model Sketch]{Appraisal Process Model
    Sketch~\citep[p.~130]{smith2001toward}}
    \label{fig:skAppraisal}
\end{figure}

Appraisal theory critics believe that the process is too slow and deliberate to
account for the quick, automatic, and unconscious nature of emotion. To address
this, Smith \& Kirby propose a process model with multiple, parallel appraisal
processes that use distinct cognitive mechanisms
(Figure~\ref{fig:skAppraisal}). This model represents a process that produces
information rich signals---emotions---while limiting attentional resource
use~\citep[p.~91]{smith2000consequences}.

Smith \& Kirby link quick, reactive cognition to associative processing,
including priming and memory activation~\citep{smith2001toward}. This happens
automatically and continuously such that an individual is always monitoring the
environment with minimal attentional cost. The theory ties deliberative
cognition to reasoning, which is more controlled, deliberate, and flexible.
Associative processing can use any type of information but reasoning can only
access semantically-encoded information. A third appraisal source is inherent
to stimuli, which might carry appraisal meanings an individual can detect
directly.

Appraisal detectors gather information from each appraisal process by
continuously monitoring and responding to incoming appraisal information. The
detectors combine and feed the information into appraisal evaluation whose
outcome initiates other processes that generate emotional response components.
This includes physiological activity, subjective feelings, and action
tendencies. The resulting emotion also influences associative processing and
deliberative cognition~\citep[p.~96]{smith2000consequences}.

\subsection{Oatley \& Johnson-Laird's Communicative Theory of Emotions (CTE)}
CTE's aim is to be a formally testable emotion system representation for
simulation and other computationally-based theories of language and
perception~\citep{oatley1987towards, johnson1992basic, oatley1992best}. CTE
proposes that emotions are functional, preparing an individual for action and
helping them construct plans and new parts of their mental models and cognitive
system. Emotions are central to cognitive processing, coordinating multiple
goals, plans, and agents under time, knowledge, and resource constraints while
operating in uncertain environments and imperfect rationality. A key assumption
is that the cognitive system is modular and asynchronous. A top level module
holds a model of the whole system and can reorganize:
\begin{itemize}
    \item \textit{Goals}, symbolic representations of the environment, and
    \item \textit{Plans}, transformations from environment representations
    to goals that the system can act on.
\end{itemize}

Each goal and plan has its own monitoring mechanism evaluating success
probabilities~\citep[p.~50--51, 54, 62--63, 101--102]{oatley1992best}. This
decoupling of concerns affords the ability to satisfy multiple goals in an
unpredictable environment. CTE proposes a schema-driven emotion generation
process that follows a set of steps:
\begin{enumerate}
    \item One or more cognitive modules---acting as monitors---evaluate an
    event to determine if there is a change in the probability of achieving an
    active or latent goal (usually at plan junctions)

    \item If the modules detect a change, they emit a distinctive emotion
    control signal or ``alarm''\footnote{Emotions lack symbolic
    representations.} that sets up an emotion ``mode'',

    \item The emotion ``mode'' becomes input to other cognitive modules and
    ``colours'' the resulting emotional state. This allows for rapid module
    calls to create a unified response and maintain focus on the affected goal
    or plan. Semantic signals typically influence these states, generated
    parallel to the control signal. They carry information about the cause of
    the change and/or plan components if it is available.

    \item The emotion state brings the affected goal or plan into focus,
    shifting cognitive resources to the current situation and the possibility of
    doing something about it, reminiscent of problem and emotion-focused
    coping. This can create additional semantic information and change the type
    of emotion experienced
\end{enumerate}

CTE treats emotions as products of both nature and nurture---biological and
involuntary events whose interpretation changes with culture and social
rules~\citep[p.~216]{oatley1992best}. Instinctual actions, which act at
recurring plan junctions, are default plans that evolution has hard-wired. With
the evolution of cognition, the system built additional processes on these
existing structures to bridge the gap between fallible, time-consuming reasoning
processes and fixed action patterns.

CTE considers an emotion to be ``universal'' if it has some biological basis
that is recognizable across languages and cultures~\citep[p.~113--115,
119]{oatley1992best}. It emphasizes that the labels assigned to the basic
emotions only represent the closest descriptor in the English
language---different terms might be more appropriate in other languages. CTE
identifies five basic emotion ``modes'' that can be elicited unconsciously and
by instinct---\textit{Happiness}, \textit{Sadness}, \textit{Fear},
\textit{Anger}, and \textit{Disgust} (see Table~\ref{tab:cte_patterns} in
Chapter~\ref{sec:evalOJL})---heavily influenced by the work of Ekman \& Friesen
(\citepg{oatley1992best}{55, 91, 103, 212}; \citepg{johnson1992basic}{209,
217})\footnote{\citet[p.~209]{johnson1992basic} list \textit{Desire/Interest}
as a primary emotion. However, they need more evidence to confirm its
status~\citep[p.~61]{oatley1992best}. Other emotions in the same situation are
\textit{Surprise}, \textit{Disgust}, \textit{Hatred}, and \textit{Contempt}}.
``Modes'' are heuristic in nature to afford flexible responses to an
unpredictable environment~\citep[p.~87]{oatley2000sentiments}. These responses
are effectively the opening phases of new plans~\citep[p.~75]{oatley1992best}.
CTE hypothesizes that moods and personality are directly related to
emotion---moods are either temporary dispositions or self-sustaining low
intensity emotion ``modes'' that are divorced from their initial cause, and
personality traits are enduring predispositions to ``modes''.

CTE emphasizes the social role of emotions, which become important as early as
infancy. Emotion signals like facial expressions and vocal tone relay
information to the social group to coordinate their behaviour when accompanied
by semantic information understood by all parties~\citep[p.~66,
212]{oatley1992best}. CTE argues that many human actions are joint ventures to
take advantage of collective resources, the distribution of cognitive labour
and agency, and the ability to criticize assumptions and biases. This requires
cognition and a model of the self, developed by culture and
language~\citep[p.~195]{oatley1992best}. CTE also describes the concept of a
group plan---a mutual plan with others whose implicit parameters vary with
culture and between individuals~\citep[p.~178--179, 192, 196, 204--205,
212]{oatley1992best}. This augmentation of individual plans can account for
interpersonal emotions and those based on evaluations of the self and the
self-in-relation-to-others. The semantic content afforded by social
interactions varies by cultural and social context, changing interpretations of
emotion ``modes''.

CTE suggests that emotion descriptions can exist in ways recognizable in
everyday language~\citep[p.~74--75, 125, 414--417]{oatley1992best}. This is
just as important as empirical data and scientific inference because an
understanding of emotion is also based in life experiences and interpretations
of them by both actors and observers. CTE argues that part of emotional
development is the ability to simulate mental states and imagine the emotions of
others~\citep[p.~122]{oatley1992best}. This is also crucial for understanding
and enjoying stories. CTE proposes that narrative literature is a good source
for collecting information about emotion because, by viewing a story plot as a
plan, it can combine the experience of an emotion with an understanding of
it~\citep[p.~6--7, 127, 220]{oatley1992best}. CTE uses notable pieces of
literature such as Tolstoy's \textit{Anna Karenina} as case studies of emotion
episodes.

\section{Neurophysiologic Theories}\label{sec:neuro}
Biological neural circuitry and brain structures inspire the
\textit{neurophysiologic} theories of affect, which offer a grounded view of
potential organizations of emotion systems and their connection to the
body~\citep[p.~98--99]{lisetti2015and}. Though uncommon in
\ref{ac}~\citep[p.~451]{rodriguez2015computational} and fundamentally different
from psychologically-based theories~\cite[p.~7]{hudlicka2014habits}, they often
influence biologically-inspired CMEs.

\subsection{Sloman}
This theory frames emotion as a consequence of intelligence rather than a
separate subsystem~\citep{sloman1987motives, sloman2005architectural}.
Emotions---whose type depend on the underlying architecture and the host's
needs---are states of powerful motivations that respond to relevant beliefs,
triggering mechanisms in a system with limited resources. In this view, emotion
serves a functional purpose and is inseparable from cognition, but it is
possible to suppress or override affective states. This also means that
searching for discrete emotions is not as productive as finding a set of core
processes that create different affective states. Interestingly, Sloman does
not believe that physiological changes are necessary to define emotion.

Goals are essential to Sloman's theory, representing motivations. Their ability
to interrupt other processes is directly proportional to their urgency, whose
intensity determines how aggressively the system pursues it. Planning generates
new goals in service to a higher level one, possibly under constraints. Unlike
the parent goal, a system can easily replace or abandon these goals if it
cannot satisfy them. This tightly links planning to emotion, leading to the
claim that no taxonomy of emotion terms can sufficiently capture all the
potential behaviours of such a system.

Sloman also differentiates between different types of affect: moods are
dispositions of the system without a clear focus; attitudes are a collection of
beliefs, motivations, and mechanisms focused on one thing expressed as
tendencies to certain actions in a particular situation; and personality is a
collection of long-term attitudes.

Sloman's theory is computationally-friendly, describing itself as a
computational theory of mind, harmonious with the concepts of ``beliefs'',
``intentions'', and ``desires'' in a Belief-Desire-Intention (BDI) system.
Beliefs have the same structure as goals. Sloman lists design constraints that
are consistent with computational ones, including: multiple, often
inconsistent, internal and external motivations; speed limitations; missing and
erroneous beliefs about the environment; and motivations with different degrees
of urgency. This necessitates a fast and ``stupid'' system to react to
immediate changes, a specialized central decision-making mechanism, a
meta-management process, and the ability to learn new information.

\subsection{Damasio}
This theory posits that ``emotion'' is a product of body state changes that
trigger neural activity causing dispositional responses that affect the body
and mind~\citep{damasio2005descartes}. It hypothesizes that there are two types
of emotion---primary and secondary---that share expression channels but serve
different purposes:
\begin{itemize}

    \item Primary emotions are innate, powerful manifestations of drives and
    instincts relevant to survival, and

    \item Secondary emotions build on these regulation structures by
    incorporating social, cultural, and environmental influences via learning.
    They are the product of systematic connections between object and situation
    categories and primary emotions.

\end{itemize}

Secondary emotions begin with cognition, which create signals that the brain
unconsciously picks up, triggering the primary emotion system and causing the
associated changes. The signals are a product of the association between
emotion and experience. Some emotion-causing cognitive evaluations can bypass
the body via symbolic states. These are another product of learning and can aid
decision-making.

``Feelings'' are cognitive evaluations of body state changes---a monitor---that
juxtapose those changes with some other mental image. Feeling can be present in
the absence of emotion, in a way similar, but not equivalent, to mood. This
led to the creation of the Somatic Marker Hypothesis (SMH). Somatic markers are
a special instance of feeling generated by secondary emotions causally linked
to its trigger. Learning connects them in service of predictive functions,
which can account for individual differences. Somatic markers act as a biasing
device, raising alarms or indicating incentives for decision-making. They
depend on attention and memory, driven by biological regulation processes,
``tagging'' information with bodily states. Damasio hypothesizes that intuition
(i.e. ``gut feelings'') is the work of unconscious somatic states. In the full
SMH, somatic states can also boost attention and memory, energizing the
cognitive system, a consequence of knowing that something is being evaluated
for an individual's preferences and goals.

\subsection{LeDoux}
This theory views emotions as biological functions with different neural
systems that evolution has maintained across
species~\citep{ledoux1996emotional}. It refers to what one would call an
emotion as the subjective feeling of it, which  requires the conscious brain's
awareness of unconscious processes to realize. This requires working memory, the
amygdala and arousal systems to be active, and bodily feedback. LeDoux's
research focuses on the defensive behaviour system, which is consciously
recognized as \textit{Fear}.

At the neural level, each emotion system has a set of inputs and outputs and an
appraisal mechanism. The system programs the mechanism to detect innate stimuli
relevant to its function, but it can also learn about other stimuli that it
tends to associate with, or predict the occurrence of, innate ones. This ability
afforded by cognition allows the systems to learn new triggers while retaining
the same behaviour as innate ones. These systems bypass cognition, improving
their speed, but cannot distinguish similar stimuli. This results in ``quick
and dirty'' responses that could be correct and life-saving, drawing attention
to and buying time for cognition to form plans about those stimuli. A separate
type of memory associates stimuli with others present at the time the system
learned the trigger, which can elicit a response even if the core trigger is
missing. The ability to unconsciously and quickly perceive and form persistent
emotional memories is one of the brain's most efficient learning and memory
functions~\cite[p.~266]{ledoux1996emotional}.

\section{A Brief Analysis of the Theories by Perspective}
While each perspective has its strengths and weaknesses, the theories within
them differ in their approach towards the perspective's goals. This implies
that different theories from the same perspective might be better or worse
suited to an application than the others. A brief analysis and comparison of
the theories in each perspective highlights some of these differences and
proposes the type of CME that might benefit from them.

\subsection{Discrete Theories}
Ekman \& Friesen (Section~\ref{adx:ef}), DET (Section~\ref{adx:det}), and PES
(Section~\ref{adx:pes}) generally agree on six primary emotions:
a variant of \textit{Joy}, \textit{Sadness}, \textit{Fear}, \textit{Anger},
\textit{Surprise}, and \textit{Disgust}. What other emotions a theory considers
primary depends on its focus.

Ekman \& Friesen's theory focuses on facial expressions. The associated set of
primary emotions reflect seven ``universal'' facial expressions found in all
humans, which they tested in a number of
cultures~\citep[p.~1]{ekman2007emotions}. The theory also defines display rules
that inform expression mechanisms. However, Ekman said that the presence of a
unique facial expression is not enough to identify all emotions because some
emotions might lack one~\citep[p.~48]{ekman1999basic}\footnote{There is also
evidence that a person's visual and social perception influences how they
evaluate facial expressions~\citep[p.~42]{brooks2019neuroimaging}.}. Izard
connects emotions to personality development and includes additional emotion
categories thought to correlate with personality
traits~\citep[p.~850]{izard1993stability}. Due to this, a significant portion
of DET's validation is from data collected from infants and young children.
Izard also forwards the idea that facial expressions unique to an emotion. He
mostly agrees with Ekman \& Friesen's findings, but adds some additional facial
expressions~\citep[p.~236--237]{izard1971face}. PES focuses on ``universal''
adaptation problems, expanding the list of emotion kinds with additional
behaviours. Plutchik developed the theory's structure as a \ref{circumplex} and
organizes emotion kinds on it based on how people evaluated the similarity and
dissimilarity of affective terms~\citep[p.~24]{plutchik1997circumplex}. Like
Izard, Plutchik also connected emotion with personality
traits~\citep[p.~27]{plutchik1997circumplex}. There is also a shared concept of
interacting emotions: Ekman describes ``blends'' of facial
expressions~\citep[p.~69]{ekman2007emotions}; Izard suggests interacting
emotion patterns connected to
personality~\citep[p.~254]{izard2000motivational}; and Plutchik proposes that
complex emotions are combinations of primary ones in the same vein as colour
mixing~\citep[p.~204--205]{plutchik1984emotions}.

For applications where emotion expression is critical, Ekman \& Friesen's
theory of facial expressions is the strongest candidate. CMEs can use it alone,
or extend it with the facial expressions from DET. Ekman \& Friesen's display
rules have potential for defining mechanisms to control when and how to express
an emotion. If personality is a focus, DET is likely the best choice because it
proposes ways that personality develops from ongoing emotional experiences. As
a behaviour-oriented theory, PES seems to be a good general choice for
modelling emotion. It focuses on classes of behaviours, which can connect to
facial expressions and other tendencies associated with an emotion. There is
also potential to develop a personality model due to the connection between its
emotion and personality circumplexes. Perhaps most useful is the ability to
create complex emotions by blending the given primary ones. While both Ekman \&
Friesen's theory and DET appear to have a fixed set of defined emotions, PES
affords the inclusion of additional emotions that a user might need including
those from other discrete theories and ones that it does not consider to be
primary.

\subsection{Dimensional Theories}
There is a clear increase in complexity between V-A (Section~\ref{adx:va}) and
PAD Space (Section~\ref{adx:pad}). The V-A structure is general, describing
affective space with two bipolar dimensions. Given that the surveyed CMEs
(Chapter~\ref{chapter:cmeOverview}) tend to use the \ref{valence} and
\ref{arousal} as independent dimensions (\citepg{andre2000integrating}{162};
\citepg{breazeal2003emotion}{133--134}; \citepg{hudlicka2019modeling}{136}),
there is little to analyze. Therefore, the analysis uses the Russell
\ref{circumplex}~\citep{russell1980circumplex} to compare V-A and PAD space.
The Russell \ref{circumplex} is a reasonable stand-in for V-A because the space
represented by the dimensions tend to fit a \ref{circumplex} structure better
than a simple structure~\citep[p.~12]{barrett1999structure} and it emerges
across several European and Asian languages~\citep[p.~82, 90]{yik2002relating}.
There is also evidence that several \ref{circumplex} representations of affect
might be variations on the same space (\citepg{yik1999structure}{615--617};
\citepg{yik2002relating}{80--82}).

The Russell \ref{circumplex} describes a two-dimensional \ref{circumplex} using
\textit{\ref{valence}} and \textit{\ref{arousal}}. Researchers validated this
structure with empirical data gathered from laymen's self-reports and judgment
tasks on affective terms~\citep[p.~1174--1176]{russell1980circumplex}. The
resulting \ref{circumplex} can account for a significant portion of the data
variance. Including a \textit{dominance-submission} dimension is one hypothesis
to explain the remaining variance, which accounts for a small, but significant,
portion of it. PAD Space adds this third dimension to the affective space to
differentiate data clusters like \textit{Fear} and
\textit{Anger}~\citep[p.~264]{mehrabian1996pleasure}. Mehrabian validated the
PAD dimensions with self-report data~\citep[p.~21--24]{mehrabian1980basic}.
Independent studies found comparable dimensions
(\citepg{elefant2005affect}{30--32}; \citepg{li2005reliability}{517}) and links
to other psychological theories~\citep[p.~12--13]{bakker2014pleasure}. PAD
Space can also define personality
traits~\citep[p.~266--267]{mehrabian1996pleasure}.

PAD Space seems to be well-suited for CMEs because its computations are simple,
yet better explains data variance than V-A. It captures and expands the
information in both V-A and the Russell \ref{circumplex}. The addition of a
third dimension, \textit{dominance}, distinguishes affective states that many
would consider emotions, such as \textit{Fear} and \textit{Anger}. It also has
the potential to expand a CME's capabilities with personality traits using its
common space. If a CME is modelling general affect or needs a very simple
model, V-A or the Russell \ref{circumplex} is likely the best choice. They have
equivalent dimensions, but the Russell \ref{circumplex} is better defined and
validated than the general V-A dimensions.

\subsection{Appraisal Theories}
\begin{figure}[!b]
    \centering
    \includegraphics[width=0.75\linewidth]{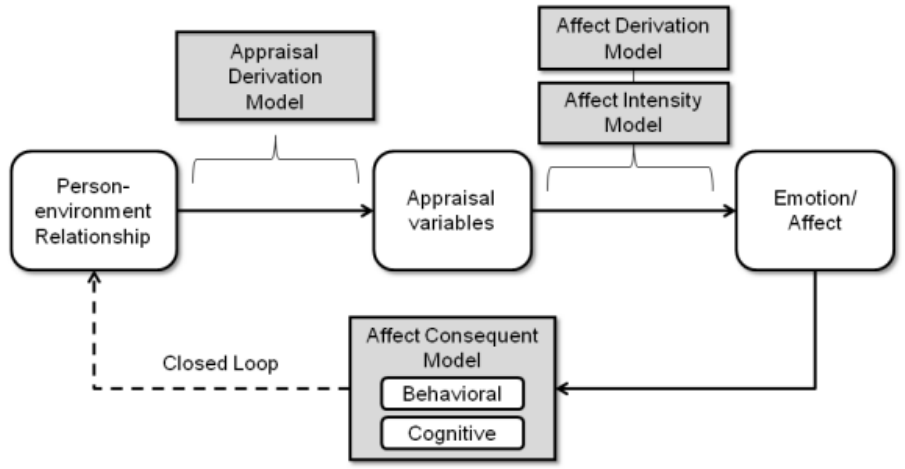}

    \caption[Idealized Component Model of Computational Appraisal
    Models]{Idealized Component Model of Computational Appraisal
        Models~\citep[p.~31]{marsella2010computational}}
    \label{fig:generalAppraisalProcess}
\end{figure}

The appraisal theories tend to follow the same process steps, modifying or
expanding them according to their focus and goals
(Figure~\ref{fig:generalAppraisalProcess}\footnote{This work  focuses on CME
components, but appear to work just as well for decomposing the theories
themselves.}). The description and form of the steps vary between theories
based on the affective phenomena that they want to explain. The final outcome
of this process can impact the person-environment relationship due to the
closed-loop nature of the system.

\subsubsection{Appraisal Derivation Model}
The appraisal derivation model translates an individual\-/environment
relationship representation into appraisal variables\footnote{They could also
be viewed as features for feature analysis where \textit{emotion} is the object
to be recognized~\citep[p.~101]{oatley1992best}.} or
dimensions~\citep[p.~32]{marsella2010computational}. There are many
commonalities in the theories' appraisal dimensions
(Table~\ref{tab:appraisalDimCompare}).

The presence of some variation of a \textit{goal congruence} appraisal
dimension\footnote{With the exception of CTE, which lacks explicitly defined
appraisal dimensions.} implies that all examined theories operate on an
individual's goals or motivations (\citepg{frijda1987emotion}{116};
\citepg{smith2000consequences}{87}). CR proposes that goals and
motivations\footnote{Frijda collectively calls them \textit{Concerns}.} are
intimately linked to the emotion system~\citep[p.~467]{frijda1986emotions}. CA
echoes this, including an individual's goal hierarchy in their
personality~\citep[p.~94]{lazarus1991emotion}, as well as OCC whose goal
hierarchy is critically linked to some appraisal mechanisms
(\citepg{occ2022}{59--60}). A \textit{goal relevance} dimension acts as a
gatekeeper to the appraisal process. It could perform a similar function in CMEs
to prevent resource use when environmental changes do not affect the agent. A
relevance dimensions is not present in OCC and ESM, but one could easily add it
due to their reliance on goals. CTE lacks \textit{goal relevance} and
\textit{goal congruence}. However, it does tie changes in goal achievement
probability to the emotion's quality~\citep[p.~49--50, 98]{oatley1992best}
which could implicitly define these dimensions. This conception also decouples
\ref{valence} from the ``pleasantness'' label, making it possible to talk about
events like watching horror films for fun. Instead, \ref{valence} simply
indicates if the system should continue the current plan or change it.

\afterpage{
    \begin{landscape}
        \begin{table}[!t]
            \renewcommand{\arraystretch}{1.2}
            \centering
            \caption[Approximately Comparable Appraisal Dimensions Across
            Appraisal Theories]{Approximately Comparable Appraisal Dimensions
                Across Appraisal
                Theories\textsuperscript{\normalsize\textpmhg{\Hi}}}
            \label{tab:appraisalDimCompare}
            \resizebox{\linewidth}{!}{%
            \begin{tabular}{P{0.15\linewidth}P{0.15\linewidth}P{0.11\linewidth}
                    P{0.16\linewidth}P{0.14\linewidth}P{0.17\linewidth}P{0.13\linewidth}}
                \toprule
                {\begin{tabular}[c]{@{}c@{}}\textbf{CR} \\
                        \citep{frijda1987emotion}\end{tabular}} &
                {\begin{tabular}[c]{@{}c@{}}\textbf{CA} \\
                        \citep{lazarus1991emotion}\end{tabular}} &
                {\begin{tabular}[c]{@{}c@{}}\textbf{CPE/SCT} \\
                        \citep{scherer2001appraisalB}\end{tabular}} &
                {\begin{tabular}[c]{@{}c@{}}\textbf{ESM} \\
                        \citep{roseman1996appraisal}\end{tabular}} &
                {\begin{tabular}[c]{@{}c@{}}\textbf{OCC} \\
                        \citep{occ2022}\end{tabular}} &
                {\begin{tabular}[c]{@{}c@{}}\textbf{S \&
                            K}\textsuperscript{\normalsize\textpmhg{\HF}} \\
                        (\citeauthor{yih2016distinct},
                        \citeyear{yih2016distinct}, \\
                        \citeyear{yih2020profiles})\end{tabular}} &
                {\begin{tabular}[c]{@{}c@{}}{\textbf{CTE}\textsuperscript{\normalsize\Moon}}
                        \\ \citep{oatley1992best}\end{tabular}} \\
                \hline

                \colourRow  Relevance & Goal Congruence & Relevance &
                Motivational State & Desirability, Desirability-for-Others &
                Motivational Congruence/Incongruence &
                {\normalsize\textpmhg{\Hibp}} \\

                Responsibility & Accountability & Causal Attribution & Agency &
                Praiseworthiness/ Blameworthiness & Self/Other Accountability &
                {\normalsize\textpmhg{\Hibp}\textpmhg{\HR}} \\

                \colourRow {\normalsize\textpmhg{\Hibp}} & Goal Relevance &
                {\normalsize\textpmhg{\Hibp}} & {\normalsize\textpmhg{\Hibp}} &
                {\normalsize\textpmhg{\Hibp}} & Motivational Relevance &
                {\normalsize\textpmhg{\Hibp}} \\

                Controllability + Modifiability + Manageability & Coping
                Potential & Control + Power + Adjustment & Control Potential &
                \textminus & Problem-focused/Emotion-focused Coping Potential &
                {\normalsize\textpmhg{\Hibp}} \\

                \colourRow Uncertainty of Outcome & \textminus & Outcome
                Probability & Probability & {\normalsize\textpmhg{\Hibp}} &
                \textminus & {\normalsize\textpmhg{\Hibp}} \\

                Certainty of Outcome & \textminus & Expectation &
                {\normalsize\textpmhg{\Hibp}} & Likelihood & \textminus &
                {\normalsize\textpmhg{\Hibp}} \\

                \colourRow Self-Esteem & Type of Ego-Involvement & Internal
                Standards & \textminus & \textminus & Acceptability &
                {\normalsize\textpmhg{\Hibp}\textpmhg{\HR}} \\

                Event Impact & \textminus & Urgency & \textminus & \textminus &
                Urgency & \textminus \\

                \colourRow Interestingness & \textminus & Novelty & \textminus
                & \textminus & Vastness & \textminus \\

                Globality + Time Reference & \textminus & \textminus &
                Situational State & Psychological Proximity & \textminus &
                \textminus \\

                \colourRow Valence & \textminus & Intrinsic Pleasantness &
                \textminus & Appealing/ Unappealing & \textminus & \textminus \\

                \textminus & \textminus &
                Suddenness\textsuperscript{\Large\Jupiter} &
                Unexpectedness & Unexpectedness & \textminus & \textminus \\

                \colourRow \textminus & \textminus & External Standards &
                \textminus & \textminus & Positive/Negative Evaluation by
                Others & {\normalsize\textpmhg{\Hibp}\textpmhg{\HR}} (p.~114)\\

                \textminus & Future Expectancy & \textminus & \textminus &
                \textminus & Future Expectancy & \textminus \\

                \hline\bottomrule
                \multicolumn{7}{r}{\textit{Continued on next page}}
            \end{tabular}%
        }
    \end{table}

    {\addtocounter{table}{-1}
        \captionsetup{list=no}
        \begin{table}[!t]
            \renewcommand{\arraystretch}{1.2}
            \centering
            \caption{Approximately Comparable Emotion Kinds Across
                Appraisal Theories\textsuperscript{\normalsize\textpmhg{\Hi}}
                (\textit{Cont'd})}
            \resizebox{\linewidth}{!}{%
            \begin{threeparttable}
                    \begin{tabular}{P{0.15\linewidth}P{0.15\linewidth}P{0.11\linewidth}
                            P{0.16\linewidth}P{0.14\linewidth}P{0.17\linewidth}P{0.13\linewidth}}
                    \hline
                    {\begin{tabular}[c]{@{}c@{}}\textbf{CR} \\
                            \citep{frijda1987emotion}\end{tabular}} &
                    {\begin{tabular}[c]{@{}c@{}}\textbf{CA} \\
                            \citep{lazarus1991emotion}\end{tabular}} &
                    {\begin{tabular}[c]{@{}c@{}}\textbf{CPE/SCT} \\
                            \citep{scherer2001appraisalB}\end{tabular}} &
                    {\begin{tabular}[c]{@{}c@{}}\textbf{ESM} \\
                            \citep{roseman1996appraisal}\end{tabular}} &
                    {\begin{tabular}[c]{@{}c@{}}\textbf{OCC} \\
                            \citep{occ2022}\end{tabular}} &
                    {\begin{tabular}[c]{@{}c@{}}\textbf{S \&
                                K}\textsuperscript{\normalsize\textpmhg{\HF}} \\
                            (\citeauthor{yih2016distinct},
                            \citeyear{yih2016distinct},
                            \\ \citeyear{yih2020profiles})\end{tabular}} &
                    {\begin{tabular}[c]{@{}c@{}}{\textbf{CTE}\textsuperscript{\normalsize\Moon}}
                            \\
                            \citep{oatley1992best}\end{tabular}} \\
                    \midrule

                    \colourRow \textminus & \textminus & Discrepancy from
                    Expectation & \textminus & Incongruity & \textminus &
                    \textminus \\

                    \textminus & \textminus &
                    Familiarity\textsuperscript{\Large\Jupiter} &
                    \textminus & Familiarity & \textminus & \textminus \\

                    \colourRow \textminus & \textminus & \textminus &
                    \textminus & Liking & Likeability & \textminus \\

                    Accessibility & \textminus & Goal/Need Conduciveness &
                    \textminus & \textminus & \textminus & \textminus \\

                    \colourRow \textminus & \textminus &
                    Predictability\textsuperscript{\Large\Jupiter} &
                    \textminus & \textminus & \textminus & \textminus \\

                    \textminus & \textminus & \textminus & Problem Source &
                    \textminus & \textminus & \textminus \\

                    \colourRow \textminus & \textminus & \textminus &
                    \textminus & Sense of Reality & \textminus & \textminus
                    \\

                    \textminus & \textminus & \textminus & \textminus &
                    Arousal & \textminus & \textminus \\

                    \colourRow \textminus & \textminus & \textminus &
                    \textminus & Deservingness & \textminus & \textminus \\

                    \textminus & \textminus & \textminus & \textminus &
                    Effort & \textminus & \textminus \\

                    \colourRow \textminus & \textminus & \textminus &
                    \textminus & Strength of Cognitive Unit & \textminus &
                    \textminus \\

                    \textminus & \textminus & \textminus & \textminus &
                    \textminus & Goal Attainment & \textminus \\

                    \colourRow \textminus & \textminus & \textminus &
                    \textminus & \textminus & Involvement of Others &
                    \textminus \\

                    \textminus & \textminus & \textminus & \textminus &
                    \textminus & Involvement of Unknown & \textminus \\

                    \midrule\bottomrule
                \end{tabular}
                \begin{tablenotes}

                    \footnotesize
                    \vspace*{2mm}

                    \item []{\normalsize\textpmhg{\Hi}} \textit{Inferred
                        from definitions given by each theory and guided by
                        \citet[p.~128--131]{frijda1987emotion} and
                        \citet[p.~56]{marsella2015appraisal}}

                    \item []{\normalsize\textpmhg{\HF}} \textit{Smith \&
                        Kirby}

                    \item []{\normalsize\Moon} \textit{Due to the nature of
                        CTE, all dimensions are implied/inferred from
                        descriptions}

                    \item []{\normalsize\textpmhg{\Hibp}}
                    \textit{Implied/inferred dimension}

                    \item []{\normalsize\textpmhg{\HR}}
                    \textit{Non-critical dimension because it requires
                        semantic content which is not needed for emotion
                        signals~\citep[p.~76]{oatley1992best}}

                    \item []{\Large\Jupiter} \textit{Component of
                        \textit{Novelty}}

                    \item []{\normalsize(+)} \textit{Combined effect}

                \end{tablenotes}
            \end{threeparttable}%
        }
        \end{table}
        \captionsetup{list=yes}
    }
\end{landscape}
}

A \textit{responsibility} dimension is also present in all the examined
theories (\citepg{frijda1987emotion}{116}; \citepg{smith2000consequences}{87}),
although it is implicit and
non-critical\footnote{\label{foot:cte_noncritical}This would require semantic
content, which is not necessary for triggering an emotion
signal~\citep[p.~76]{oatley1992best}.} in CTE~\citep[p.~62]{oatley1992best}.
This suggests that there is a mechanism for assigning causality to
person-environment relationship changes. Some variant of a
\textit{controllability} or \textit{coping} dimension is also common, enforcing
the transactional nature of emotions and an individual's ability to act on
appraisal inputs and processes. CTE appears to have a comparable dimension
implicitly because it describes what should happen in response to changes in
plan junctures~\citep[p.~55, 102]{oatley1992best}. The ways that an individual
can respond can be construed as a measure of their coping potential. This
dimension is not present at all in OCC, possibly because it is for
reasoning about emotions instead of generating them~\citep[p.~219]{occ2022}.

Some variation of a \textit{certainty} and/or \textit{uncertainty} dimension
appears in most of the examined theories~\citep[p.~116]{frijda1987emotion}. CTE
appears to incorporate a version of this dimension based on its evaluation of
changes in goal and/or plan outcomes~\citep[p.~48, 98]{oatley1992best}. CA and
Smith \& Kirby lack this dimension, instead defining a \textit{future
expectancy} dimension. One could view this as a specialized certainty
evaluation focusing on changes in \textit{goal congruence}. This implies a
mechanism for predicting future events or the outcomes of a proposed action.

Less common is a \textit{self-esteem} related dimension, also implicit but
non-critical\footnote{See footnote \ref{foot:cte_noncritical}.} in
CTE~\citep[p.~44--45, 114]{oatley1992best}, demonstrating the role of
self-perception in emotion processes. Evaluating this dimension requires a
mechanism for reasoning about the self and what the person-environment
relationship means for it. ESM lacks a \textit{self-esteem} dimension, which
the associated study appears to omit~\citep[p.~275--277]{roseman1996appraisal},
and neither does the OCC model---potentially due to its focus on reasoning and
understanding emotions. The remaining dimensions only appear in half of the
examined appraisal theories. The focus of each theory accounts for this wider
variation and a CME will need additional mechanisms depending on the dimensions
it models.

While the theories are somewhat clear on \textit{what} the appraisal derivation
model produces, they are relatively silent on \textit{how} it does this in
terms of inputs and transformations. This is unsurprising because most theories
are structural and more concerned with the contents of cognition than the
processes that produce it~\citep[p.~88]{smith2000consequences}. ESM says
nothing about \textit{how} derivation works, but seems to agree with the
general implication that an individual combines what they take in from the
environment with internal knowledge structures such as goals and beliefs. This
creates an interpretation of the situation to examine for affective patterns.

Smith \& Kirby appears to be the only theory that has begun to derive the
inputs necessary for evaluating explicitly defined dimensions. So far, it has
derived inputs for \textit{problem-focused coping potential}
(\citepg{smith2009putting}{1361}; \citepg{smith2009relational}{483, 97--499};
\citepg{smith2001toward}{127--128}), \textit{emotion-focused coping potential}
(\citepg{smith2009putting}{1365}; \citepg{smith2009relational}{483--485, 495}),
and \textit{motivational relevance} (\citepg{smith2001toward}{125--127};
\citepg{smith2009putting}{1358, 1361}). It has not yet begun testing the
remaining dimensions~\citep[p.~1357, 1369]{smith2009putting}. However, this
might be transferable knowledge to theories that share these dimensions. The
OCC model provides another piece of the puzzle, stating that the appraisal
derivation model does not need to be precise because of the assumed imprecision
of psychophysics\footnote{The study of the relationship between the objective
characteristics of stimuli and the subjective perception of
it.}~\citep[p.~96]{occ2022}. It also does not need to produce values for all
dimensions, using default values where necessary.

The theories of CPE/SCT, Smith \& Kirby, and CTE describe some mechanisms for
appraisal derivation that transform inputs into appraisal dimension values.
CPE/SCT proposes a four stage process following a fixed sequence which controls
information processing by predetermining if evaluations are necessary and
activating them in increasing levels of
cost~\citep[p.~99--100]{scherer2001appraisalB}. The multi-stage appraisal also
accounts for different processing levels such that appraisals can access quick,
simple information as well as deliberative and slow
information~\citep[p.~102--103]{scherer2001appraisalB}. Smith \& Kirby propose
a general mechanism for gathering and combining information from multiple
sources~\citep[p.~129--130]{smith2001toward}. An appraisal detector
consistently monitors for changes in different, interacting information
sources. When it detects a change, the detector gathers and combines
information into a single unit and triggers its appraisal. CTE proposes that
each goal and plan has its own monitoring mechanism that triggers a global
emotion signal when there is a change in success
probability~\citep[p.~98]{oatley1992best}.

The affect derivation model appears to need some common elements in the
examined theories: goal representation; a causality assignment mechanism; the
ability to predict the outcome of actions and their chances of causing change;
and possibly a representation of the self. Unfortunately, the theories
emphasize the \textit{what}, not the \textit{how}, of appraisal derivation.
This makes them insufficient for modelling. However, the theories offer pieces
that form the begin of an appraisal derivation model. OCC seems to have the
best developed structure for goals, accounting for the intimate link described
by CR and further developing the hierarchical structure from CA. It also
provides a theoretically grounded reason for imprecision and missing values in
the model's outputs. CMEs that can access multiple information sources could
use CPE/SCT to design variable cost cognitive processes and incremental
appraisals to control their execution. Smith \& Kirby offer a mechanism for
combining multiple information sources into a single unit for appraisal, while
CTE offers an option for plan-focused designs. It is likely that any given CME
will take one of these theories as their foundation and take parts from other
theories as needed.

\subsubsection{Affect Derivation Model}
The affect derivation model maps appraisal variables to emotions or affect,
specifying an individual's reaction-based appraisal value
patterns~\citep[p.~33]{marsella2010computational}. All examined appraisal
theories have patterns for \textit{Fear}, \textit{Joy}, \textit{Anger}, and
\textit{Sadness} (Table~\ref{tab:appraisalEmotions}). After this, they begin to
diverge based on the definition and kinds of emotion under study
(\citepg{frijda1987emotion}{117}; \citepg{oatley1992best}{104}).

There does not appear to be a correlation between the number of appraisal
variables and emotions that a theory offers
(Table~\ref{tab:appraisalDims2Kinds}). Assuming that a CME wants to use the
fewest dimensions for the most emotion kinds, CA looks to be the best choice at
first glance. However, its mappings between dimensions and emotions is somewhat
ambiguous. For example, the patterns for \textit{Anxiety} and \textit{Disgust}
only differ in how they describe \textit{type of
ego-involvement}~\citep[p.~237, 261]{lazarus1991emotion}. What distinguishes
``protection'' from ``risk of contamination''? This question leaves room for
unintentional assumptions that might conflicts with CA. Smith \& Kirby have a
similar problem---\textit{Affection} and \textit{Compassion} are one instance
of emotions sharing an appraisal pattern~\citep[p.~489]{yih2020profiles}. There
is a clear difference in the statistical significance of the elements in the
pattern, but only an assumption can make the difference explicit. CTE, too, has
this issue because it lacks explicit appraisal dimensions. However, its
descriptions of changes at plan junctures seems to remove some
ambiguity~\citep[p.~55, 104--106]{oatley1992best}. It names evaluations of goal
and plan types that precede emotion ``modes'' which do not appear to overlap.
CTE also seems to account for simultaneous goal or plan evaluation, proposing
that more than one emotion ``mode'' can be active with each containing
different semantic contents from different interpretations of the same
situation. As with CA and Smith \& Kirby, assumptions are still necessary to
make terms like ``major plan'' and ``gustatory goal'' unambiguous.

\afterpage{
    \begin{landscape}
        \begin{table}[!ht]
            \renewcommand{\arraystretch}{1.2}
            \centering
            \caption{Approximately Comparable Emotion Kinds Across Appraisal
                Theories}
            \label{tab:appraisalEmotions}
            \resizebox{\linewidth}{!}{%
            \begin{tabular}{lccccccc}
                \toprule
                \textbf{Emotion} &
                {\begin{tabular}[c]{@{}c@{}}\textbf{CR}\textsuperscript{\normalsize$\mathrightbat$}
                        \\ \citep{frijda1987emotion}\end{tabular}} &
                {\begin{tabular}[c]{@{}c@{}}\textbf{CA}\textsuperscript{\normalsize\textpmhg{\Hi}}
                        \\ \citep{lazarus1991emotion}\end{tabular}} &
                {\begin{tabular}[c]{@{}c@{}}\textbf{CPE/SCT} \\
                        \citep{scherer2001appraisalB}\end{tabular}} &
                {\begin{tabular}[c]{@{}c@{}}\textbf{ESM}
                        \\ \citep{roseman1996appraisal}\end{tabular}} &
                {\begin{tabular}[c]{@{}c@{}}\textbf{OCC} \\
                        \citep{occ2022}\end{tabular}} &
                {\begin{tabular}[c]{@{}c@{}}\textbf{S \&
                K}\textsuperscript{\normalsize\textpmhg{\HF}} \\
                (\citeauthor{yih2016distinct}, \citeyear{yih2016distinct}, \\
                \citeyear{yih2020profiles})\end{tabular}} &
                {\begin{tabular}[c]{@{}c@{}}\textbf{CTE} \\
                \citep{oatley1992best}\end{tabular}} \\
                \midrule

                \colourRow Fear & X & (Fright) & X & X & X &
                X & X \\

                Happiness/Joy & X & X & X & X & X & X & X \\

                \colourRow Anger & X & X & X & X & X & X & X\\

                Sadness & X & X & X & X &
                (Distress)\textsuperscript{\Large\textpmhg{\Hl}} & X & X \\

                \colourRow Disgust &  & X & X & X &
                \begin{tabular}[c]{@{}c@{}}(Reproach/ \\
                    Distaste)\textsuperscript{\Large\textpmhg{\Hl}}\end{tabular}
                     & X & X \\

                Anxiety & (Nervousness) & X & X &  &
                (Fear)\textsuperscript{\Large\textpmhg{\Hl}} & X & \\

                \colourRow Guilt &  & X & X & X &
                (Self-Reproach)\textsuperscript{\Large\textpmhg{\Hl}} & X & \\

                Shame &  & X & X & X &
                (Self-Reproach)\textsuperscript{\Large\textpmhg{\Hl}} & X & \\

                \colourRow Pride &  & X & X & X & X & X & \\

                Love/Affection &  & X &  & (Liking) &
                \begin{tabular}[c]{@{}c@{}}X, (Liking/ \\
                Appreciation)\textsuperscript{\Large\textpmhg{\Hl}}\end{tabular}
                 & X & \\

                \colourRow Relief &  & X &  & X & X & X & \\

                Hope &  & X &  & X & X & X & \\

                \colourRow Distress &  &  & (Despair) & X & X &  & \\

                Contempt &  &  & X & X &
                (Reproach)\textsuperscript{\Large\textpmhg{\Hl}} & & \\

                \colourRow Irritability &  &  & X &
                (Frustration) & (Anger)\textsuperscript{\Large\textpmhg{\Hl}} &
                & \\

                Compassion &  & X &  &  &
                (Sorry-For)\textsuperscript{\Large\textpmhg{\Hl}} & X & \\

                \colourRow Uninterested & X &  &
                {\begin{tabular}[c]{@{}l@{}}(Boredom/ \\
                        Indifference)\end{tabular}} &  &  & & \\

                Interest & {\Large\Uranus} &  &  &  &  & X & \\

                \colourRow Envy &  & X &  &  &
                (Resentment)\textsuperscript{\Large\textpmhg{\Hl}} & & \\

                Jealousy &  & X &  &  &
                (Resentment)\textsuperscript{\Large\textpmhg{\Hl}}
                & & \\

                \colourRow Dislike &  &  &  & X & \begin{tabular}[c]{@{}c@{}}
                X, (Enmity/ \\
                Distaste)\textsuperscript{\Large\textpmhg{\Hl}}\end{tabular} & &
                \\

                \hline\bottomrule
                \multicolumn{8}{r}{\textit{Continued on next page}}
            \end{tabular}%
        }
        \end{table}
    \end{landscape}
    \clearpage
    \begin{landscape}
        {\addtocounter{table}{-1}
            \captionsetup{list=no}
            \begin{table}[!ht]
                \renewcommand{\arraystretch}{1.2}
                \centering
                \caption{Approximately Comparable Emotion Kinds Across
                    Appraisal Theories (\textit{Cont'd})}
                \resizebox{\linewidth}{!}{%
                \begin{threeparttable}
                    \begin{tabular}{lccccccc}
                        \toprule
                        \textbf{Emotion} &
                        {\begin{tabular}[c]{@{}c@{}}\textbf{CR}\textsuperscript{\normalsize$\mathrightbat$}\\
                                \citep{frijda1987emotion}\end{tabular}} &
                        {\begin{tabular}[c]{@{}c@{}}\textbf{CA}\textsuperscript{\normalsize\textpmhg{\Hi}}\\
                                \citep{lazarus1991emotion}\end{tabular}} &
                        {\begin{tabular}[c]{@{}c@{}}\textbf{CPE/SCT} \\
                                \citep{scherer2001appraisalB}\end{tabular}} &
                        {\begin{tabular}[c]{@{}c@{}}\textbf{ESM} \\
                                \citep{roseman1996appraisal}\end{tabular}} &
                        {\begin{tabular}[c]{@{}c@{}}\textbf{OCC} \\
                                \citep{occ2022}\end{tabular}} &
                        {\begin{tabular}[c]{@{}c@{}}\textbf{S \&
                        K}\textsuperscript{\normalsize\textpmhg{\HF}} \\
                        (\citeauthor{yih2016distinct},
                        \citeyear{yih2016distinct}, \\
                        \citeyear{yih2020profiles})\end{tabular}}
                        & {\begin{tabular}[c]{@{}c@{}}\textbf{CTE} \\
                            \citep{oatley1992best}\end{tabular}}
                        \\
                        \midrule

                        \colourRow Regret &  &  &  & X &
                        (Self-anger)\textsuperscript{\Large\textpmhg{\Hl}} & &
                        \\

                        Surprise &  &  &  & X & {\Large\Pluto} & &  \\

                        \colourRow \begin{tabular}[c]{@{}l@{}}
                        Gratitude\end{tabular} &  &  &  &  & X & X & \\

                        \begin{tabular}[c]{@{}l@{}}Appreciation\end{tabular} &
                        &  &  &  & X &  & \\

                        \colourRow Awe &  &  &  &  &
                        \begin{tabular}[c]{@{}c@{}} (Admiration/ \\
                        Appreciation)\textsuperscript{\Large\textpmhg{\Hl}}\end{tabular}
                         & X & \\

                        Embarrassment &  &  &  & &
                        (Self-Reproach)\textsuperscript{\Large\textpmhg{\Hl}}
                         & X & \\

                        \colourRow Distrust & X &  &  &  &  & &  \\

                        Tired & X &  &  &  &  & & \\

                        \colourRow Happy-For &  &  &  &  & X & &  \\

                        Schadenfreude &  &  &  &  & X & & \\

                        \colourRow Satisfied &  &  &  &  & X & &  \\

                        Dissatisfied &  &  &  &  & X & &  \\

                        \colourRow Gratification &  &  &  &  & X & & \\

                        Fears-Confirmed &  &  &  &  & X & & \\

                        \colourRow Disappointment &  &  & &  & X & & \\

                        Self-satisfaction &  &  &  &  & X & & \\

                        \colourRow Approving &  &  &  &  & X & & \\

                        Disapproving &  &  & &  & X & & \\

                        \colourRow Amusement &  &  &  &  &  & X & \\

                        Determination &  &  &  &  &  & X & \\

                        \colourRow Tranquility &  &  &  & &  & X & \\

                        \midrule\bottomrule
                    \end{tabular}
                    \begin{tablenotes}

                        \small
                        \vspace*{2mm}

                        \item []{\normalsize$\mathrightbat$} \textit{Excludes
                        moods.}

                        \item []{\normalsize\textpmhg{\Hi}} \textit{Excludes
                        ``Aesthetic Emotions'' because it lacks an explicit
                        appraisal profile.}

                        \item []{\normalsize\textpmhg{\HF}} \textit{Smith \&
                            Kirby}

                        \item []{\Large\textpmhg{\Hl}} \textit{OCC emotion
                        token identifying that this emotion word is part of an
                        emotion ``family''.}

                        \item []{\Large\Uranus} \textit{Mentioned as an
                        explanation for the \texttt{Interestingness} dimension.}

                        \item []{\Large\Pluto} \textit{Not an emotion, but
                        elicited by \textit{Unexpectedness} appraisal
                        dimension.}

                    \end{tablenotes}
                \end{threeparttable}%
            }
            \end{table}
            \captionsetup{list=yes}
        }
    \end{landscape}
}

\begin{table}[!b]
    \renewcommand{\arraystretch}{1.2}
    \begin{center}
        \begin{threeparttable}
            \caption{Comparison of the Number of Appraisal Dimensions and
            Emotion Kinds in Appraisal Theories\label{tab:appraisalDims2Kinds}}
            \begin{tabular}{ccc}
                \toprule
                \textbf{Theory} & \textbf{\# of Dimensions} & \textbf{\# of
                    Emotions} \\ \hline

                \colourRow CR & 14 & 9
                (18~{\large\textpmhg{\Hl}}) \\

                CA & 6 & 15 \\

                \colourRow CPE/SCT & 15 & 13 \\

                ESM & 7 & 17 \\

                \colourRow OCC & 19 & 29
                (22~{\normalsize\textpmhg{\HR}}) \\

                Smith \& Kirby & 17 & 20 \\

                \colourRow CTE & 5~{\Large\Pluto} & 5 \\

                \hline\bottomrule

            \end{tabular}

            \begin{tablenotes}
                \footnotesize
                \vspace*{2mm}

                \item []{\Large\textpmhg{\Hl}} \textit{Count of action
                    tendencies}

                \item []{\large\textpmhg{\HR}} \textit{Count of unique
                    emotion ``families''}

                \item []{\Large\Pluto} \textit{Inferred/implied necessary
                dimensions}

                \vspace*{-7mm}
            \end{tablenotes}
        \end{threeparttable}
    \end{center}
\end{table}

ESM, OCC, and CPE/SCT created their affect derivation models systematically,
assigning values to all appraisal dimensions for each emotion. ESM visually
delineates the boundaries between emotion kinds based on empirical studies of
the appraisal-emotion relationship~\citep[p.~269]{roseman1996appraisal}. The
model includes elements of what they would look like, leaving little room for
assumptions~\citep[p.~143]{roseman2013appraisal}. The OCC model organizes its
emotion ``families'' in a hierarchy based on their logical description such
that a derived emotion becomes more specific as it considers more
dimensions~\citep[p.~36--38]{occ2022}. While it appears to be clear, a group of
computer scientists and logicians proposed an alternative hierarchy to remove
some lingering ambiguities~\citep{steunebrink2009occ}. The assignment of
``tokens'' to the OCC emotion families is not empirically
supported~\citep[p.~205]{occ2022}, so using them to compare to other emotions
within and without the same family is dubious and leaves room for implicit
assumptions. CPE/SCT views emotion ``families'' as labels for common responses,
so affect derivation is not necessarily limited to
them~\citep[p.~113]{scherer2001appraisalB}. It clearly notes which dimensions
have multiple valid appraisal values, claiming them as potential points of
within-``family'' variation. Different intensity levels in a dimension could
also indicate differences between similar emotion families. However, like OCC,
this is only speculation and defining new emotion kinds this way might not be
theoretically sound. CTE also considers its emotion ``modes'' as related
families of emotion sharing similar antecedents and responses but differing in
semantic contents~\citep[p.~76--78, 105]{oatley1992best}. The connection to
folk understanding of the contents of different emotions indicate what
additional cognitive processes their identification might need. However, like
OCC and CPE/SCT, this might not be theoretically sound. Depending on the
application, this might still be a viable option for systematically adding more
emotions to a CME.

CR differs from the other theories in this regard, mapping appraisal values to
action tendencies instead of emotion kinds~\citep[p.~455,
479]{frijda1986emotions}. This affords a continuous system where emotion kinds
are discrete events. CR assumes that appraisals and action tendencies are
correlated and emotion kinds are labels that relate different types of
affective data~\citep[p.~141]{frijda1987emotion}.

\textit{Surprise} and \textit{Interest} appear to be special cases of affect
derivation. ESM and OCC state that the \textit{unexpectedness}
dimension alone indicates \textit{Surprise}
(\citepg{roseman1996appraisal}{257}; \citepg{occ2022}{45--46}). However, OCC
does not consider \textit{Surprise} an emotion because it is not an inherently
positive or negative state. CR and Smith \& Kirby recognize \textit{Interest},
tying it to the comparable dimensions of \textit{interestingness} and
\textit{vastness} respectively (\citepg{frijda1987emotion}{130};
\citepg{yih2020profiles}{493}). Smith \& Kirby use additional dimensions to
distinguish \textit{Interest} from other emotions, with preliminary results
suggesting that it is a positive emotion.

In general, the emotion kinds that a CME needs should determine which model to
use. Given the number of ambiguities, one should avoid CA for affect
derivation. Smith \& Kirby could be an alternative due to its notes on
statistical significance, but assumptions are unavoidable since no mention is
made about how they make a difference. Assumptions are also unavoidable in CTE,
although there seem to be fewer to make. ESM, OCC, and CPE/SCT appear to be
computationally-friendly, ensuring that each emotion has appraisal values for
all dimensions. However, one should avoid defining individual members of the
emotion families given in CPE/SCT and OCC as there are no guidelines for doing
so. CTE also has no explicit guidelines for defining different members of an
emotion family, but its ties to everyday understanding of emotions might make
it viable for some applications. CR offers an alternative model, mapping
appraisal values to action tendencies. This could be useful for systems that
are more interested in how emotion affects behaviour rather than producing
specific emotion states. One could use a secondary mapping to match action
tendencies with an emotion label. ESM is ideal for modelling \textit{Surprise},
as it is the only appraisal theory in the survey that considers it a proper
emotion. Either CR or Smith \& Kirby could act as a guideline for defining an
appraisal mapping for \textit{Interest} as neither appear to have an obvious
advantage.

\subsubsection{Affect Intensity Model}
The affect intensity model specifies the strength of the emotion response given
the appraisal values~\citep[p.~33]{marsella2010computational}. Unfortunately,
there is precious little research on it (\citepg{frijda1992complexity}{60};
\citepg{marsella2010computational}{33}) and the examined theories do not offer
much insight. ESM and Smith \& Kirby hypothesize that their dimensions of
\textit{motivational state} and \textit{motivational relevance} are factors of
emotion intensity (\citepg{roseman2013appraisal}{147};
\citepg{smith2001toward}{125}), but do not state how they work. CTE has a
similar proposal, claiming that the importance of an affected goal and/or plan
is one factor of emotion intensity, likely in relation to the change in success
probabilities at plan junctions~\citep[p.~23, 98]{oatley1992best}. CTE does not
list other factors, their evaluation, or their combination. This does, however,
imply that goal/plan importance is scalar so that intensity can vary with it.
CR proposes that appraisal dimensions use thresholds that determine if the
system should trigger a response~\citep[p.~300]{frijda1986emotions}. The
individual's personality, physical and mental state, and event history
influence threshold values. While CR is clear that intensity changes over time,
becoming more or less sensitive to changes, it is less clear on how factors
combine in a single time step. OCC provides a more developed intensity model:
\begin{itemize}

    \item It assumes that all global and local variables exert their influence
    through the three central variables: \textit{desirability},
    \textit{praiseworthiness}, and \textit{appealing}~\citep[p.~99]{occ2022}.

    \item It calculates an emotion potential such that emotion intensity is the
    difference between the potential and a context-sensitive, emotion-specific
    threshold~\citep[p.~220--222]{occ2022}. Therefore, an emotion only
    manifests if its potential exceeds the threshold.

    \item It also assigns weights to appraisal values, which can vary between
    and within emotion families~\citep[p.~96--97]{occ2022}. This affords
    control over how changes in those factors change the overall emotion
    experienced.

\end{itemize}

Unfortunately, OCC does not specify what the emotion thresholds, variable
weights, or default values should be due to a lack of empirical
data~\citep[p.~132]{occ2022}.

Ultimately, none of the examined theories are sufficient for designing the
affect intensity model. At best, CR or OCC could act as a guideline for the
affect intensity model but it is equally helpful to look at other \ref{as}
research for this component.

\subsubsection{Affect Consequent Model}
The affect consequent model maps appraisal outputs to behavioural and cognitive
changes~\citep[p.~34]{marsella2010computational}. In most cases, this means
mapping affect to those changes. Notable exceptions are CR---which maps
directly from appraisals to action tendencies---and OCC, which is silent on the
matter by design\footnote{One author started to change this, but the work
appears unfinished (\citeg{gilboa1991structure}) or does not cover all emotions
from the original theory (\citepg{ortony2002making}{198,
201}).}~\citep[p.6]{occ2022}.

There is general agreement that the affect consequent model includes
physiological and behavioural changes. CR calls this an \textit{action
readiness change} which is a signal that presses for control of the
individual~\citep[p.~455]{frijda1986emotions}. CTE specifies that this
``readiness for action'' is a call for a predetermined plan suite that can
prompt a quick and resource-friendly response to manage transitions between
individual and social plans at plan junctures~\citep[p.~19--21,
176--179]{oatley1992best}. A distinctive feeling, differentiating them from
other mental and body states, pairs with the transition and might also be
accompanied by conscious preoccupation, bodily disturbances, and outward
expressions. ESM also cites expressive changes, as well as emotion-specific
goals and general strategies for achieving
them~\citep[p.~143--144]{roseman2013appraisal}. It describes profiles for each
emotion kind with entries for each consequent type. Both CA and Smith \& Kirby
call the behavioural changes \textit{action tendencies} and include a
subjective feeling aspect. CA groups the subjective feeling and appraisal
outcome, but Smith \& Kirby include it as part of deliberative cognition
(\citepg{lazarus1991emotion}{210}; \citepg{smith2001toward}{130}). These
theories group the changes together into a single appraisal outcome. In
contrast, CPE/SCT distributes changes between SECs such that each group affects
physiological systems, cognition, and other SECs
independently~\citep[p.~99--100]{scherer2001appraisalB}. These become
interrelated and synchronized changes that CPE/SCT calls emotion as the system
evaluates each SEC group~\citep[p.~93]{scherer2001appraisalB}. This creates
tightly linked physiological changes, behaviours, and appraisal outcomes.

The examined theories also generally agree that the appraisal process is a
closed loop system. An individual's actions might change their relationship
with the environment, triggering a new appraisal cycle. Both CR and CA state
that the changes can be direct actions on the environment or cognitive
changes~\citep[p.~456]{frijda1986emotions}. However, CA differentiates the two
because of its intentional inclusion of \ref{coping} which the individual's
personality modulates~\citep[p.~112, 134]{lazarus1991emotion}. Problem-focused
\ref{coping} is about directly acting on the environment, changing the external
information coming into the appraisal process. Emotion-focused \ref{coping}
involves internal, cognitive actions that can change the individual's goals,
beliefs, or knowledge, which changes their interpretation of the situation, the
appraisal process itself, or the appraisal outcome. CA calls this initiation of
another appraisal process \textit{reappraisal}. Smith \& Kirby also split the
emotional response into two paths~\citep[p.~130]{smith2001toward}: a priming
signal for associative processing, which can direct memory activation; and
deliberative cognition as the subjective feeling component of the response. The
differences in CA and Smith \& Kirby might not be mutually exclusive in this
area. One can imagine the outcome of emotion-focused \ref{coping} from CA as
input to the two paths from Smith \& Kirby. CTE heavily implies that emotion
generation is a closed loop system~\citep[p.~102]{oatley1992best}. Shifting
cognitive resources and attention can change a situation's interpretation and
any actions taken to modify it can also result in shifting goal evaluations.
This can cause changes in emotion intensity and quality. The distinction
between control and semantic messages also implies the presence of two
pathways---a quick, reactionary path and another that is slower and
deliberate~\citep[p.~51]{oatley1992best}---comparable to the pathways in Smith
\& Kirby.

Excluding OCC, all the examined theories imply---if not explicitly describe---a
closed-loop system. This means that any changes elicited by the affect
consequence model feeds back into the emotion process and might initiate
another cycle. ESM is ideal for systems that produce specific emotion
categories because it organizes information into profiles reminiscent of
discrete theories. Both CR and CPE/SCT describe an emergent system such that
recognizable ``emotion'' is the product of overlapping system changes. These
theories are ideal for CMEs that are more interested in how affective
processing influences behaviours and internal processes than producing discrete
emotion kinds. CR is the more lightweight option because CPE/SCT's described
changes are tightly linked to external systems. CR also offers a control signal
for affective processing to gain control of behavioural and cognitive
resources. Due to its focus on goals and plans, CTE is likely ideal for
planning applications, including narrative planning and multi-agent
coordination.

Both CR and CA separate appraisal outcome effects into two pathways for
independently impacting the external environment and internal processes. CA is
likely the better choice for specifying this because it associates the pathways
with problem-focused and emotion-focused \ref{coping} respectively. It also
indicates that the affected internal processes are the appraisal process and
outcome, as well as the individual's goals, beliefs, and knowledge. CR does not
appear to make comparable distinctions between the paths. Smith \& Kirby
further divide the emotion-focused \ref{coping} path by sending an affective
priming signal to associative processing and memory, and a signal to
deliberative cognition that manifests as subjective affect. This makes it ideal
for CMEs with different internal information gathering mechanisms. Smith \&
Kirby do not, however, say how the process affects the external environment.
CTE could be a candidate for this, as it seems to mirror the pathways in Smith
\& Kirby, but allows for the potential to connect to both the internal and
external environment like CA can.

\subsection{Neurophysiologic Theories}
Of the explored perspectives, the neurophysiologic theories agree the most.
They all propose that emotion: is part of a functional mind; aids
decision-making; and operates under time constraints. They also agree that
learning can mediate ``stupid'' emotional behaviours. The differences between
theories is a matter of their focus.

The conceptual, schematic, and sensorimotor processing levels described by
\citet[p.~17]{leventhal1987relationship} can relate the neurophysiologic
theories:
\begin{itemize}

    \item Sloman operates on the highest level of the mind---the conceptual
    level---where cognition is agnostic of the body, evident in the central
    role it assigns to goals and planning;

    \item Damasio operates on the schematic level, as the SMH describes how
    emotions incorporate new information to regulate survival-based processes;
    and

    \item LeDoux operates on the sensorimotor level, describing how emotions
    are fundamental biological functions that evolution has tuned to react
    quickly and unconsciously to survival issues.

\end{itemize}

Damasio and LeDoux have more in common than Sloman does with either. Both
Damasio and LeDoux assign bodily feedback a critical role in emotion processes,
claiming that the layman's notion of ``emotion'' is a conscious, cognitively
driven assignment of meaning to the collective effect of emotion
processes~\citep[p.~329]{ledoux1996emotional}. The processes themselves have
dedicated neural pathways that react automatically and unconsciously to innate
stimuli and mediating structures can ``tune'' the affective response through
learning and ``created'' stimuli.

The constraints that these theories imply as part of their implementation could
limit the type of CMEs that could use them to their full capacity. Therefore,
the decision to use these theories is highly dependent on the CME's
requirements. Damasio and LeDoux could complement each other in a CME because
they have nearly identical assumptions. The theories' basis in brain structures
and replicable studies suggests that this combination could create a more
plausible emotion processing model than other theories. Damasio's claim that
LeDoux has the most comprehensive study on primary
emotions~\citep[p.~133]{damasio2005descartes} and LeDoux's that Damasio's work
is notable~\citep[p.~36, 250, 293]{ledoux1996emotional} further supports this
idea. However, emotion processes' dependency on bodily feedback could impose
constraints that cannot be easily addressed in virtual environments. A virtual
environment itself does not contain the information richness that the real
world does innately. However, since they operate in the real world, robots are
a good platform for CMEs using neurophysiologic theories. For example,
researchers use Damasio to justify the inclusion of emotions in
robots~\cite[p.~806]{malfaz2004new}. In contrast, Sloman designed this theory
for computers using symbolic representations and transformations. However, it
depends on an underlying system that operates with symbolic goals and plans.
Sloman's emphasis on beliefs, desires, and
intentions~\citep[p.~230]{sloman1987motives} implies that it works best with a
BDI-based system. Overall, this strengthens the proposal that Sloman is best
suited to the domain of agent planning.

\afterpage{
    \begin{landscape}
        \begin{table}[!ht]
            \renewcommand{\arraystretch}{1.2}
            \centering
            \caption{Summary of Potential Uses for Examined Theories}
            \label{tab:theoryUsesSummary}
            \resizebox{0.885\linewidth}{!}{%
            \begin{tabular}{P{0.11\linewidth}P{0.16\linewidth}P{0.69\linewidth}}
                \toprule
                \textbf{Perspective} & \textbf{Theory} & \textbf{Potential
                Uses} \\
                \midrule

                \multirow{9}{*}{Discrete} & \colourCell Ekman \&
                Friesen~\textpmhg{\HF} & \colourCell
                \begin{itemize}\setlength{\itemsep}{0pt}
                    \vspace*{-1.3em}
                    \item Defining facial expressions
                    \item Defining display rules for emotion expression
                    \vspace*{-0.8em}\end{itemize} \\

                & DET\textpmhg{\HF} &
                \begin{itemize}\setlength{\itemsep}{0pt}\vspace*{-1.3em}
                    \item Personality developed by ongoing emotional
                    experiences
                    \item Supplementing facial expressions from Ekman \& Friesen
                    \vspace*{-0.8em}\end{itemize} \\

                & \colourCell PES & \colourCell
                \begin{itemize}\setlength{\itemsep}{0pt}\vspace*{-1.3em}
                    \item General emotion model~\textpmhg{\Hi}
                    \item Can connect to facial expression and action
                    tendencies~\textpmhg{\HF}
                    \item Potential for personality model
                    development~\textpmhg{\HF}
                    \item Supplement existing emotions by mixing them to create
                    new ones~\textpmhg{\Hi}
                    \vspace*{-0.8em}\end{itemize} \\

                \midrule

                \multirow{5}{*}{Dimensional} & V-A~\textpmhg{\Hi}{\Large\Pluto}
                & \begin{itemize}\vspace*{-1.3em}
                    \item General model of affect
                    \vspace*{-0.8em}\end{itemize} \\

                & \colourCell PAD Space~\textpmhg{\Hi} & \colourCell
                \begin{itemize}\setlength{\itemsep}{0pt}\vspace*{-1.3em}
                    \item Computation-friendly
                    \item Expands on V-A
                    \item Potential to integrate personality in a common space
                    \vspace*{-0.8em}\end{itemize} \\

                \midrule

                \multirow{4}{*}{Appraisal} & CR &
                \begin{itemize}\setlength{\itemsep}{0pt}\vspace*{-1.3em}
                    \item Computation-friendly~\citep{frijda1987can}
                    \item Defining an alternative affect derivation model by
                    mapping appraisal values to action
                    tendencies~{\large\textpmhg{\HR}}
                    \item Guideline for defining an affect derivation model for
                    \textit{Interest}
                    \item Guideline for defining an affect intensity model
                    \item Specifying an affect consequent model that produces
                    continuous, emergent changes
                    \item Specifying a control precedence signal for the affect
                    consequent model
                    \vspace*{-0.8em}\end{itemize}  \\

                & \colourCell CA & \colourCell
                \begin{itemize}\setlength{\itemsep}{0pt}\vspace*{-1.3em}
                    \item Defining an affect consequent model with distinct
                    pathways for influencing the environment and internal
                    cognitive processes (i.e. \ref{coping} behaviour)
                    \vspace*{-0.8em}\end{itemize} \\

                \midrule\bottomrule
                \multicolumn{3}{r}{\textit{Continued on next page}}
            \end{tabular}%
        }
        \end{table}

        {\addtocounter{table}{-1}
            \captionsetup{list=no}
            \begin{table}[!ht]
                \renewcommand{\arraystretch}{1.2}
                \centering
                \caption{Summary of Potential Uses for Examined Theories
                (\textit{Cont'd})}
                \resizebox{0.885\linewidth}{!}{%
                    \begin{tabular}{P{0.11\linewidth}P{0.16\linewidth}P{0.69\linewidth}}
                        \toprule
                        \textbf{Perspective} & \textbf{Theory} &
                        \textbf{Potential Uses} \\
                        \midrule

                        \multirow{20}{*}{{\begin{tabular}[c]{@{}c@{}}Appraisal
                        \\ \textit{Cont'd}\end{tabular}}} & CPE/SCT &
                        \begin{itemize}\setlength{\itemsep}{0pt}\vspace*{-1.3em}
                            \item
                            Computation-friendly~\citep[p.~103--106]{scherer2001appraisalB,
                             scherer2010component}
                            \item Defining an appraisal derivation model with
                            incremental processing stages based on cost and
                            accounts for multiple processing levels
                            \item Defining an affect derivation
                            model~{\Large\textpmhg{\Hl}}
                            \item Specifying an affect consequent model that
                            produces continuous, emergent changes where there
                            are cognitive processing systems
                            \vspace*{-0.8em}\end{itemize} \\

                        & \colourCell ESM & \colourCell
                        \begin{itemize}\setlength{\itemsep}{0pt}\vspace*{-1.3em}
                            \item Defining an affect derivation model that
                            accounts for all appraisal dimensions for each
                            emotion kind
                            \item Defining an affect derivation model for
                            \textit{Surprise}
                            \item Defining an affect consequent model that
                            treats emotion kinds as discrete entities
                            \vspace*{-0.8em}\end{itemize} \\

                        & OCC &
                        \begin{itemize}\setlength{\itemsep}{0pt}\vspace*{-1.3em}
                            \item Computation-friendly
                            \item Defining a goal hierarchy for the appraisal
                            derivation model
                            \item Justifying imprecision and missing values in
                            the appraisal derivation model's outputs
                            \item Defining an affect derivation
                            model~{\Large\textpmhg{\Hl}}
                            \item Guideline for defining an affect intensity
                            model
                            \vspace*{-0.8em}\end{itemize} \\

                        & \colourCell Smith \& Kirby & \colourCell
                        \begin{itemize}\setlength{\itemsep}{0pt}\vspace*{-1.3em}
                            \item Defining a mechanism for the appraisal
                            derivation model that combines multiple input
                            sources into one unit for evaluation
                            \item Guideline for defining an affect derivation
                            model
                            \item Guideline for defining an affect derivation
                            model for \textit{Interest}
                            \item Potential guideline for defining an affect
                            derivation model in place of CA
                            \item Defining an affect consequent model with
                            distinct pathways for independently influencing
                            fast, reactive associative processing and slow,
                            deliberative cognition (emotion-focused
                            \ref{coping})
                            \vspace*{-0.8em}\end{itemize} \\

                        \midrule\bottomrule
                        \multicolumn{3}{r}{\textit{Continued on next page}}
                    \end{tabular}%
                }
            \end{table}

        {\addtocounter{table}{-1}
            \captionsetup{list=no}
            \begin{table}[!ht]
                \renewcommand{\arraystretch}{1.2}
                \centering
                \caption{Summary of Potential Uses for Examined Theories
                    (\textit{Cont'd})}
                \resizebox{0.885\linewidth}{!}{%
                \begin{threeparttable}
                    \begin{tabular}{P{0.11\linewidth}P{0.16\linewidth}P{0.69\linewidth}}
                        \toprule
                        \textbf{Perspective} & \textbf{Theory} &
                        \textbf{Potential Uses} \\
                        \midrule

                        \multirow{1}{*}{{\begin{tabular}[c]{@{}c@{}}Appraisal
                                    \\ \textit{Cont'd}\end{tabular}}} &
                                    \colourCell CTE & \colourCell
                        \begin{itemize}\setlength{\itemsep}{0pt}\vspace*{-1.3em}
                            \item Computation-friendly

                            \item Defining a plan-focused appraisal derivation
                            model with implicit and variable dimensions

                            \item Guideline for defining an affect derivation
                            model

                            \item Defining an affect consequent model with
                            distinct pathways for independently influencing
                            fast, reactive action and slow, deliberative
                            cognition

                            \item Defining an affect consequent model that
                            makes and alters computational plans (e.g.
                            narrative planning, agent coordination)

                            \item Tied to narratives and character
                            intentionality

                            \item Can rely on an ``everyday'' understanding of
                            emotion to define some components

                            \vspace*{-0.8em}\end{itemize} \\

                        \midrule

                        \multirow{7}{*}{{Neurophysiologic}} & Sloman &
                        \begin{itemize}\setlength{\itemsep}{0pt}\vspace*{-1.3em}
                            \item Computation-friendly

                            \item When built on a system with goals and
                            planning, does not require a separate emotion system

                            \vspace*{-0.8em}\end{itemize} \\

                        & \colourCell Damasio & \colourCell
                        \begin{itemize}\setlength{\itemsep}{0pt}\vspace*{-1.3em}
                            \item Defines learned/conditioned emotional
                            reactions as an associative network

                            \item Connected to a CME's ``body''
                            \vspace*{-0.8em}\end{itemize} \\

                        & LeDoux &
                        \begin{itemize}\setlength{\itemsep}{0pt}\vspace*{-1.3em}
                            \item Defines core emotion systems as discrete,
                            dedicated circuits

                            \item ``Quick and dirty'' model of emotion
                            elicitation

                            \item Connected to a CME's ``body''
                            \vspace*{-0.8em}\end{itemize} \\

                        \midrule\bottomrule
                    \end{tabular}
                    \begin{tablenotes}

                        \footnotesize
                        \vspace*{2mm}

                        \item []{\normalsize\textpmhg{\HF}} \textit{Can link to
                        the affect consequent model in appraisal theories.}

                        \item []{\normalsize\textpmhg{\Hi}} \textit{Can link to
                        the affect derivation model in appraisal theories.}

                        \item []{\Large\Pluto} \textit{Could use the Russell
                        \ref{circumplex} instead.}

                        \item []{\large\textpmhg{\HR}} \textit{A secondary
                        mapping can connect groups of behaviours to emotion
                        kinds.}

                        \item []{\Large\textpmhg{\Hl}} \textit{Strictly for the
                        emotion families as there is no information for
                        defining individual members.}

                    \end{tablenotes}
                \end{threeparttable}%
            }
            \end{table}
            \captionsetup{list=yes}
        }
        }
    \end{landscape}
}

\clearpage
\vspace*{\fill}
\begin{keypoints}
    \begin{itemize}

        \item Discrete theories of emotion emphasize a small set of fundamental
        emotions that are innate, hard-wired features with dedicated neural
        circuitry

        \item The discrete theory of: Ekman \& Friesen has excellent support
        for describing the connection between emotion and facial expressions;
        DET has elements describing connections between emotion and
        personality; and PES has mechanisms describing a structural
        representation of emotion and emotion ``mixtures''

        \item Dimensional theories define a coordinate space for affect using
        two or three dimensions and focus on connecting emotions to mental
        states in a general taxonomy and to their construction

        \item V-A is useful for describing a general affective space, whereas
        PAD Space can integrate multiple types of affect in the same coordinate
        space

        \item Appraisal theories tend to combine discrete and dimensional
        emotion characteristics, proposing different methods of emotion
        production and manifestation

        \item CR directly maps appraisals to action tendencies; CA describes
        how \ref{coping} behaviours impact appraisal inputs; CPE/SCT is useful
        for defining complex systems with multi-stage appraisals that access
        different types of cognitive processing; ESM supports the emotion
        definitions using both appraisal patterns and discrete entities; OCC is
        useful for defining a goal hierarchy for appraisal derivation and
        mapping between appraisals and emotion kinds; Smith \& Kirby is good
        for defining a mechanism that combines multiple appraisal inputs and
        for specifying how appraisal outcomes affect some types of cognition;
        and CTE is especially useful for plan and/or language-focused
        applications

        \item Neurophysiologic theories could produce more plausible behaviours
        than the other perspectives, but appear to have more constraints

        \item Sloman supports goal and plan-based emotion processes; Damasio
        describes a learning mechanism for learned or conditioned stimuli; and
        LeDoux describes emotion with dedicated neural processes
    \end{itemize}
\end{keypoints}

\parasep
\vspace*{\fill}

%% file: appendix_theory2reqsNotes.tex
\chapter{(Notes) Of High-Level Requirements and Emotion Theories}
\label{chapter:reqsTheoryNotes}
\def\epigraphflush{center}
\setlength{\epigraphwidth}{0.75\textwidth}
\def\textflush{center}
\epigraph{If we all reacted the same way, we'd be predictable, and there's
always more than one way to view a situation.}{Major Motoko Kusanagi,
\textit{Ghost in the Shell}}

These are the notes about emotion theories with respect to high-level
requirements made during analysis (Chapter~\ref{chapter:theoryAnalysis}). Notes
for the dimensional theories and \ref{flexEm} do not appear here because they
serve as an example in Chapter~\ref{chapter:theoryAnalysis}.
Tables~\ref{tab:theory-req-sys-summary-flexibility},
\ref{tab:theory-req-sys-summary-easeofuse},
\ref{tab:theory-req-comp-summary-flexibility}, and
\ref{tab:theory-req-comp-summary-easeofuse} summarize the resulting scores.

A reminder that the scores are somewhat subjective and depend on one's
understanding of the requirements and current state of \ref{as} literature.

\section{Discrete Theories}
The discrete theories are the most likely candidates for satisfying high-level
requirements that depend on understanding what emotion an NPC has, as this is
their core focus. This includes the flexibility requirements for
\textit{Allowing the Integration of New Components} (\ref{flexNew}) and
\textit{Choosing Which NPC Emotions to Use} (\ref{flexEm}), and the ease-of-use
requirements for \textit{Having a Clear API (Output)} (\ref{easeAPI}) and
\textit{Showing That Emotions Improve the Player Experience} (\ref{easePX}).

\subsection{Flexibility: Allowing the Integration of New Components
    (\ref{flexNew})}
All three discrete theories provide variable levels of native support for
personality but only Plutchik does not touch on mood. None of them touch on
core affect.

\begin{itemize}
    \item \textbf{Ekman \& Friesen} (\weak)
    \begin{itemize}
        \item Moods and personality are inferred from emotion
        signals (e.g. many \textit{Joy}-related signals could suggest a
        cheerful mood)~\citep[p.~48, 55--56]{ekman1999basic}

        \item Little information beyond these definitions $\rightarrow$
        developers would need to create patterns of emotions for each mood and
        trait, could become too time consuming and error-prone

        \item No coverage of Core Affect, Personality and Mood are error-prone
        and time consuming
    \end{itemize}

    \item \textbf{Izard} (\weak)
    \begin{itemize}
        \item Natively accounts for personality
        (\citepg{izard2000motivational}{253--254}; \citepg{izard1977human}{44})
        \begin{itemize}
            \item Emergent phenomena that begins at birth and develops as the
            individual interacts with their environment

            \item Treated as a product of emotions associated with patterns in
            perception, cognition, and behaviour

            \item [$\rightarrow$] Requires developers to create patterns for
            each personality trait which is likely to be too time consuming
            and error-prone
        \end{itemize}

        \item Seems to acknowledge two definitions of mood
        \begin{itemize}
            \item Defined as a ``continuing total life condition'' similar to
            what he calls an emotion trait, or tendencies towards certain
            emotion experiences~\citep[p.~17, 171]{izard1991psychology}
            $\rightarrow$ in the view of stable traits and fluid states,
            conceptualization appears to be closer to personality than mood

            \item As a state, defined as an enduring emotion state that is too
            mild to enter consciousness but can influence mental health and
            bodily systems such as the immune
            system~\citep[p.~21]{izard1991psychology} $\rightarrow$ closer to
            working definition of mood in \progname{}'s context

            \item [$\rightarrow$] could be realized as a timed function that
            monitors an NPC's emotion state and acts on those that have not
            surpassed a given threshold (minimal effort to implement)
        \end{itemize}

        \item No coverage of Core Affect, Mood requires minimal effort,
        Personality is error-prone and time consuming
    \end{itemize}

    \item \textbf{Plutchik} (\good)
    \begin{itemize}
        \item Natively accounts for personality
        \begin{itemize}
            \item Connects its emotion \ref{circumplex} directly to a
            \ref{circumplex} of personality traits\footnote{The assumption that
                the circumplexes can be connected this way might be naive. They
                might be unique to the modelled domain rather than showing
                similarities across them~\citep[p.~815]{feldman1995variations}.}
            \citep[p.~27--28]{plutchik1997circumplex} $\rightarrow$ could
            mechanize with a simple weighting mechanism such that emotions are
            easier or harder to elicit

            \item Built around a layperson's understanding of personality
            $\rightarrow$ upholds the \textit{Hiding the Complexity of Emotion
                Generation} (\ref{easeHide}) requirement
        \end{itemize}

        \item \ref{circumplex} is a way to incorporate a model of mood
        \begin{itemize}
            \item Agreement that it can be represented by an elliptical
            \ref{circumplex} with \ref{arousal} as the shorter
            dimension~\citep[p.~806, 812, 814]{feldman1995variations}

            \item Could add an additional element such that the length of the
            \ref{arousal} dimension changes with context $\rightarrow$ afford
            more creative freedom than a fixed model

            \item Can uphold the \textit{Hiding the Complexity of Emotion
                Generation} requirement (\ref{easeHide}) with a well-designed
            interface with details available for advanced users
        \end{itemize}

        \item Dimensional nature of theory could aid in a non-native
        representing core affect $\rightarrow$ could map the intensity
        dimension to \ref{arousal} and the relative positions of an emotion to
        some anchor points as \ref{valence}
        \begin{itemize}
            \item Mapping might not be understandable due to debates about the
            valence of \textit{Surprise} and its relation to
            \textit{Anticipation}~\citep[p.~98]{susanto2020hourglass}
            $\rightarrow$ concessions could be made, such as listing some
            categories as zero \ref{valence}, since \progname{} is unconcerned
            with realism

            \item Can uphold the \textit{Hiding the Complexity of Emotion
                Generation} requirement (\ref{easeHide}) with a well-designed
            interface with details available for advanced users
        \end{itemize}

        \item Ability to build \textit{some} type of Core Affect and Mood
        representation on top of existing theory, Personality native to theory
        and built on a layperson's perspective
    \end{itemize}
\end{itemize}

\subsection{Flexibility: Choosing What Emotions the NPC can Have
(\ref{flexEm})}\label{sec:notes-discrete-choose-emotions}
Within each discrete theory's set of emotions, it would be easy to exclude any
unneeded ones. However, \textit{adding} more emotions to a set is less clear
cut. There does not seem to be any convincing empirically validated or
verifiable rules for creating ``non-basic'' emotions in discrete
theories~\citep[p.~6]{ortony2021all}. This does not impact \progname{} because
it does not need to replicate true affective phenomena---it need only produce
convincing results.

\begin{itemize}
    \item \textbf{Ekman \& Friesen} (\weak)
    \begin{itemize}
        \item Do not believe that there are ``non-basic''
        emotions~\citep[p.~55, 57]{ekman1999basic}
        \begin{itemize}
            \item Each emotion represents a family of related states that share
            a theme and variations between members are the result of learning

            \item [$\rightarrow$] Requires developers to either associate
            different situations for the desired variations manually or create
            a learning mechanism to create them as the NPC interacts with the
            game environment

            \item Unideal $\rightarrow$ Could be difficult to adapt this kind
            of system to simple games, violating the \textit{Ability to Operate
                on Different Levels of NPC Complexity} requirement
            (\ref{flexComplex}); requires some knowledge of psychology and
            neuroscience, violating the \textit{Hiding the Complexity of
                Emotion Generation} (\ref{easeHide}) requirement
        \end{itemize}

        \item Facial expressions can be blends of prototypical primary
        ones~\citep[p.~69]{ekman2007emotions}
        \begin{itemize}
            \item A step removed from the emotion generation process, and would
            likely happen in an emotion expression component

            \item Would need to translate the blended expression into an
            emotion to use it for emotion generation $\rightarrow$ feasible,
            but error-prone, method as facial expression interpretations can be
            subjective
        \end{itemize}

        \item No ``non-basic'' emotions, translating from facial expressions is
        error-prone
    \end{itemize}

    \item \textbf{Izard} (\disqualified)
    \begin{itemize}
        \item ``New emotions'' are the product of affective-cognitive
        structures~\citep[p.~564--565]{izard1992basic}
        \begin{itemize}
            \item Association of primary emotion patterns or clusters with
            images, thoughts, and memories

            \item [$\rightarrow$] Requires developers to either create these
            structures manually or create a learning mechanism to create them
            as the NPC interacts with the game environment

            \item Unideal $\rightarrow$ Same reasoning as Ekman \& Friesen,
            violating both the \textit{Ability to Operate on Different Levels
                of NPC Complexity} and \textit{Hiding the Complexity of Emotion
                Generation} requirements (\ref{flexComplex}, \ref{easeHide})
        \end{itemize}

        \item Difficult to adapt to different NPC complexities, requires
        knowledge for connecting emotions to cognitive patterns
    \end{itemize}

    \item \textbf{Plutchik} (\strong)
    \begin{itemize}
        \item One of the better developed theories of emotion mixes
        (\citepg{ledoux1996emotional}{113}; \citepg{ortony2021all}{3, 5})

        \item Might be the only discrete theory to focus on this aspect of the
        ``primary'' emotions~\citep[p.~47]{ekman1999basic}

        \item Colour wheel analogy uses concepts and terms that are generally
        understood by laypeople $\rightarrow$ does not require any knowledge of
        the theory to use
        \begin{itemize}
            \item Lacks clarity about technical rules for combining emotions
            (\citepg{johnson1992basic}{208--209}; \citepg{ortony2021all}{5})
            \textbf{BUT} laypeople tend to attribute the same underlying
            primary emotions to named emotions outside the primary
            set~\citep[p.~204--205]{plutchik1984emotions}

            \item [$\rightarrow$] Implies that a game developer can apply their
            own experiences when deciding how to represent a new emotion with
            the Plutchik \ref{circumplex}
        \end{itemize}

        \item Ability to build additional emotions from existing set based on
        a layperson's understanding of emotions and their combinations
    \end{itemize}
\end{itemize}

\subsection{Ease-of-Use: Having a Clear API (Output)
(\ref{easeAPI})}\label{sec:notes-discrete-api-output}
The discrete theories are generally easy for laypeople to understand. All three
theories connect their emotions to distinctive behaviours applicable to
situations of variable complexity. This makes for a clean output API, providing
an emotion category that developers can attach to ``buckets'' of related
behaviours and expressions that are ``familiar''.

\begin{itemize}
    \item \textbf{Ekman \& Friesen} (\strong)
    \begin{itemize}
        \item Have publications that are meant for the general public (e.g.
        \cite{ekman2007emotions}) $\rightarrow$ accessible to laypeople

        \item Use of facial expressions is a helpful tool for conveying meaning
        about their primary emotions
    \end{itemize}

    \item \textbf{Izard} (\weak)
    \begin{itemize}
        \item Gains understandability by connecting its emotions to facial
        expressions, although some emotions are not connected to one

        \item [$\rightarrow$] Weakens its usability, as some developers might
        actively avoid the emotions that cannot be readily represented on the
        face
    \end{itemize}

    \item \textbf{Plutchik} (\good)
    \begin{itemize}
        \item Construction based on similarities and differences between
        affective terms as they are understood in (English) language
        $\rightarrow$ can help developers understand each emotion based on
        their understanding of the word's meaning and its relative position to
        other emotion words on the \ref{circumplex}

        \item Each primary emotion is also connected to an intended behaviour
        pattern, like rejection and
        exploration~\citep[p.~202]{plutchik1984emotions}
        \begin{itemize}
            \item  Addresses problems of ``missing'' facial expressions with
            characteristic or typical behaviours
            (\citepg{julle2020there}{20--21};
            \citepg{schindler2013admiration}{101})

            \item [$\rightarrow$] Could help developers conceptualize what each
            emotion could look and act like, can include facial expressions
        \end{itemize}

        \item Does not directly benefit from assigned facial expressions but
        some connections could be made with Ekman \& Friesen and Izard
        (Table~\ref{tab:discreteEmotions}) $\rightarrow$ understanding of
        emotion terms and associated behaviours is not as ``clear cut''
    \end{itemize}
\end{itemize}

\begin{table}[!tb]
    \begin{center}
        \renewcommand{\arraystretch}{1.2}
        \begin{threeparttable}
            \caption{Primary Emotions in Discrete Theories}
            \label{tab:discreteEmotions}
            \small
            \begin{tabular}{lccc}
                \toprule
                \textbf{Emotion} & {\begin{tabular}[c]{@{}c@{}}\textbf{Ekman \&
                            Friesen} \\
                        \textbf{\citep{ekman2007emotions}}\end{tabular}} &
                \textbf{\citet{izard1993stability}} &
                \textbf{\citet{plutchik1997circumplex}} \\ \hline

                \colourRow
                Happiness/Enjoyment/Joy\textsuperscript{\large\Jupiter\Pluto}
                & \checkmark & \checkmark & \checkmark \\

                Sadness\textsuperscript{\large\Jupiter\Pluto} & \checkmark &
                \checkmark & \checkmark \\

                \colourRow Fear\textsuperscript{\large\Jupiter\Pluto}
                & \checkmark & \checkmark & \checkmark \\

                Anger\textsuperscript{\large\Jupiter\Pluto} & \checkmark &
                \checkmark & \checkmark \\

                \colourRow Surprise\textsuperscript{\large\Jupiter\Pluto}
                & \checkmark & \checkmark & \checkmark \\

                Disgust\textsuperscript{\large\Jupiter\Pluto} & \checkmark &
                \checkmark &
                \checkmark \\

                \colourRow
                Contempt\textsuperscript{{\normalsize\Moon}{\large\Pluto}}
                & \checkmark & \checkmark &
                {\small\textpmhg{\Hibp}} \\

                Interest\textsuperscript{\large\Pluto} &  & \checkmark &
                \checkmark \\

                \colourRow Guilt\textsuperscript{\large\textpmhg{\Hl}}
                 & & \checkmark & {\small\textpmhg{\Hibp}} \\

                Shame\textsuperscript{\large\Pluto} &  & \checkmark &
                {\small\textpmhg{\Hibp}} \\

                \colourRow Shyness\textsuperscript{\large\Pluto}
                &  & \checkmark & \\

                Acceptance\textsuperscript{{\large\textpmhg{\Hl}}\textpmhg{\Hi}}
                &  & & \checkmark \\
                \hline

                \bottomrule

            \end{tabular}

            \begin{tablenotes}
                \footnotesize
                \vspace*{2mm}

                \item []{\Large\Jupiter} \textit{Associated with a facial
                    expression by \citet{ekman2003unmasking}.}

                \item []{\Large\Pluto} \textit{Associated with a facial
                    expression by \citet[p.~236--237]{izard1971face},
                    \citet[p.~85--91]{izard1977human}.}

                \item []{\normalsize\Moon} \textit{Associated with a facial
                    expression by \citet[p.~184--186]{ekman2007emotions}.}

                \item []{\normalsize\textpmhg{\Hibp}} \textit{As a mixture of
                    the primary emotions.}

                \item []{\normalsize\textpmhg{\Hl}} \textit{Might lack a
                characteristic expression
                    (\citepg{keltner1996evidence}{155};
                    \citepg{schindler2013admiration}{106}), but artistic
                    renditions of facial expressions exist (e.g.
                    \citet{lebrun1760admiration}).}

                \item []{\normalsize\textpmhg{\Hi}} \textit{Plutchik is noted
                as the only researcher to consider \textit{Adoration}---the
                highest intensity of \textit{Acceptance}---a primary
                emotion~\citep[p.~87--88]{schindler2013admiration}.}

                \vspace*{-7mm}
            \end{tablenotes}
        \end{threeparttable}%
    \end{center}
\end{table}

\subsection{Ease-of-Use: Showing that Emotions Improve the Player Experience
    (\ref{easePX})}
Like the \textit{Having a Clear API (Output)} requirement  (\ref{easeAPI}),
discrete theories are generally understandable by laypeople. This helps
identify ways to design studies to evaluate and ways to build the player
experience.

\begin{itemize}
    \item \textbf{Ekman \& Friesen} (\good)
    \begin{itemize}
        \item Emotions could be directly connected to an expression module
        built on the Facial Action Coding System (FACS), which is part of the
        theory itself~\citep{facs}

        \item Players could report on their experiences based on NPC
        expressions $\rightarrow$ relatively easy to test how emotions impact a
        player but limited to facial expressions alone
    \end{itemize}

    \item \textbf{Izard} (\good)
    \begin{itemize}
        \item Considerable overlap between Ekman \& Friesen and Izard regarding
        facial expressions \citep[p.~3]{ekman2007emotions} $\rightarrow$ could
        connect to an emotion expression component built on FACS, probably with
        minimal effort

        \item Players could report on their experiences based on NPC
        expressions $\rightarrow$ relatively easy to test how emotions impact a
        player but limited to facial expressions alone
    \end{itemize}

    \item \textbf{Plutchik} (\good)
    \begin{itemize}
        \item Could be connected to facial expressions, but there is no obvious
        match for the \textit{Acceptance} emotion type

        \item Associates each emotion with a behaviour that could be applied to
        many actions and expressions that an NPC could need $\rightarrow$
        design studies around these behaviour classes

        \item Players could report on their experiences based on NPC
        expressions $\rightarrow$ relatively easy to test how emotions impact a
        player, but likely an element of subjectivity in matching behaviours to
        meaning
    \end{itemize}
\end{itemize}

\subsection{Examining the Remaining Requirements}
The absence of a defined emotion elicitation tasks in the discrete theories is
a double-edged sword---some requirements are trivial to satisfy, while others
are impossible. The lack of elicitation processes makes it impossible for
discrete theories to satisfy most task-related requirements. This means that
they \textit{cannot} be categorized for the component-level requirements
(Tables~\ref{tab:theory-req-comp-summary-flexibility} and
\ref{tab:theory-req-comp-summary-easeofuse}). For the remaining system-level
requirements (Tables~\ref{tab:theory-req-sys-summary-flexibility} and
\ref{tab:theory-req-sys-summary-easeofuse}), the theories satisfy the
requirements in similar ways, so they are examined as a single unit.
\begin{itemize}
    \item \textit{Flexibility: Independence from an Agent Architecture
        (\ref{flexArch})} (\good)
    \begin{itemize}
        \item No specific tasks $\rightarrow$ effectively architecture-agnostic

        \item Only require processes that satisfy input and output requirements
    \end{itemize}

    \item \textit{Flexibility: Allowing Developers to Specify How to Use
        Outputs (\ref{flexOut})} (\good)
    \begin{itemize}
        \item No specific tasks $\rightarrow$ affords flexibility for
        specifying how to use \progname{}'s outputs

        \item Theoretically could hook up any process to \progname{} using the
        emotions as ``buckets'' for collecting related behaviours
    \end{itemize}

    \item \textit{Flexibility: Ability to Operate on Different Levels of NPC
        Complexity (\ref{flexComplex})} (\good)
    \begin{itemize}
        \item Could add processes and parameters as needed $\rightarrow$ does
        not affect core \progname{} processes
    \end{itemize}

    \item \textit{Flexibility: Be Efficient and Scalable (\ref{flexScale})}
    (\good)
    \begin{itemize}
        \item Could add processes and parameters as needed $\rightarrow$ does
        not affect core \progname{} processes
    \end{itemize}

    \item \textit{Ease-of-Use: Providing Examples of Novel Game Experiences
        (\ref{easeNovel})} (\weak)
    \begin{itemize}
        \item No immediately obvious features for novel game mechanics,
        challenges, or other elements
    \end{itemize}
\end{itemize}

\section{Dimensional Theories}
The dimensional theories are the most likely candidates for satisfying
requirements related to CME expansion, as they aim to discover the structure of
emotion and how they relate to other mental
states (\citepg{reisenzein2013computational}{250};
\citepg{broekens2021emotion}{353}). This mainly concerns the flexibility
requirements for \textit{Allowing the Integration of New Components}
(\ref{flexNew}) and \textit{Choosing What Emotions the NPC can Have}
(\ref{flexEm}).

This analysis treats the \ref{valence}-\ref{arousal} model as a \ref{circumplex}
because it is a reasonable representation of affective
states~\citep[p.~296]{remington2000reexamining} and is more consistent with
affective structure~\citep[p.~12]{barrett1999structure}. The \ref{circumplex}
also tends to emerge regardless of the data collected, research domain, and
analysis~\citep[p.~211]{russell1997how}. This representation still has issues,
such as inclusion/exclusion of terms, self-report weaknesses, and the effect of
context on state positions~\citep[p.~298]{remington2000reexamining}. However,
it also provides more structure to an otherwise two-dimensional and nebulous
space.

\subsection{Flexibility: Allowing the Integration of New Components
    (\ref{flexNew})}
Both V-A and PAD can trivially represent core affect because they both natively
include the dimensions of \ref{valence} and \ref{arousal}. Like Plutchik, both
dimensional theories could also model mood as an elliptical \ref{circumplex}
with relative ease. Unlike Plutchik, this mapping is native due to the presence
of both a \ref{valence}/\textit{pleasantness} and \ref{arousal} dimension. This
only leaves an evaluation of the ability to represent personality in V-A and
PAD.

The dimensional theories seem to have variable levels of built-in support for
representing personality. This analysis focuses on support for the Five-Factor
Model OCEAN\footnote{Defined as ``psychological entities with causal
    force''~\citep{ffmdef}. Although they have the same dimensions, this differs
    from the Big Five Model which ``views the five personality dimensions as
    descriptions of behaviour and treats the five-dimensional structure as a
    taxonomy of individual differences''.} personality
traits~\citep{costa1992normal}. With research ongoing in personality
psychology, there is currently a good consensus on the usefulness of OCEAN as a
descriptive model (\citepg{yik2002relating}{100--101};
\citepg{raad2002big}{3}). OCEAN has, arguably, also become known among the
general populace as a personality profile tool due to its accessible
language~\citep[p.~1]{raad2002big} and use in career counselling
(\citepg{costa1995persons}{135}; \citeg{howard1995big};
\citepg{hurtado2019five}{528}). This familiarity makes it ideal for
\progname{} which cannot assume that a user will have an academic understanding
of psychology. For game design, the OCEAN model will also likely prove
convenient for defining NPC personalities, as \citet{costa1992normal} provide a
questionnaire consisting of five-point Likert scales representing statement
agreement to measure how each factor contributes to personality. It has also
been translated to several languages~\citep[p.~84]{yik2002relating}. This
implies that a simple tool presenting game designers with the questionnaire is
sufficient for defining a new NPC personality in \progname{} with OCEAN traits.

\begin{itemize}
    \item \textbf{V-A} (\strong)
    \begin{itemize}
        \item Relating OCEAN personality traits with \ref{circumplex}
        structures is more consistent than simple structures $\rightarrow$ two
        structures have close-fitting probability plots supporting an ideal
        \ref{circumplex} structure and the third has a convincing and
        serviceable, but less satisfactory, probability plot~\citep[p.~84--87,
        90]{gurtman1997studying}
        \begin{itemize}
            \item Some evidence that the interpersonal traits of Extroversion
            and Agreeableness are best described with a \ref{circumplex}
            $\rightarrow$ Extroversion can be related to the \ref{valence}
            dimension~\citep[p.~590, 593]{mccrae1989structure}

            \item Extroversion/Neuroticism $\rightarrow$ represents the
            affective plane~\citep[p.~84--87, 90]{gurtman1997studying}

            \item Unnamed or ``mixed'' Agreeableness/Neuroticism
            plane~\citep[p.~84--87, 90]{gurtman1997studying}

            \item All three planes can be layered in the polar coordinate
            system~\citep[p.~84--87, 90]{gurtman1997studying}

            \item Does not appear to be support for Openness or
            Conscientiousness~\citep[p.~84--87, 90]{gurtman1997studying}
        \end{itemize}

        \item Alternate hypothesis puts OCEAN traits as points on the
        \ref{circumplex} $\rightarrow$ a high value in a trait implies a higher
        tendency to experience the type of affect represented in
        the same space~\citep[p.~94--96]{yik2002relating}
        \begin{itemize}
            \item Locates the angles for each trait in five
            languages---English, Spanish, Korean, Chine-se, and Japanese

            \item Configuration option $\rightarrow$ prebuild some cultural
            differences into \progname{}
        \end{itemize}
    \end{itemize}

    \item \textbf{PAD} (\strong)
    \begin{itemize}
        \item Personality\footnote{Mehrabian refers to \textit{emotional
                traits} or \textit{temperament}. Since he defines them as
                ``...stable
            over periods of years or even a
            lifetime''~\citep[p.~262]{mehrabian1996pleasure} and temperament is
            a
            biologically-based bias in personality
            development~\citep{oxfordTemperament}, they are assumed to be
            equivalent to personality traits in \progname{}.} can be inferred by
        averaging an individual's emotional states across a representative
        sample of day-to-day situations~\citep[p.~262]{mehrabian1996pleasure}

        \item As traits, the PAD dimensions were found to be a good base
        description of personality~\citep[p.~64]{mehrabian1980basic}

        \item  Other personality scales are represented as linear combinations
        of the three dimensions~\citep[p.~267]{mehrabian1996pleasure}, forming
        a line through the space
        \begin{itemize}
            \item Provides lines estimates for the OCEAN personality
            traits\footnote{\textit{Trait Sophistication} is assumed to be
                equivalent to \textit{Trait
                    Openness}~\citep[p.~826--827]{mccrae1997conceptions}.} using
            \textit{pleasure}, \ref{arousal}, and
            \textit{dominance}~\citep[p.~91
            Eq.~11C--13C]{mehrabian1996analysis}, and from the dimensions to
            PAD space~\citep[p.~90 Eq.~1D--5D]{mehrabian1996analysis}

            \item Gender agnostic~\citep[p.~89]{mehrabian1996analysis}
            $\rightarrow$ removes a layer of complexity that one might consider
            when adding the OCEAN model of personality to \progname{}
        \end{itemize}
    \end{itemize}
\end{itemize}

\subsection{Examining the Remaining Requirements}
Dimensional theories are  similar to the discrete theories in that they have no
defined emotion elicitation tasks, so they also cannot satisfy the
component-level requirements
(Tables~\ref{tab:theory-req-comp-summary-flexibility} and
\ref{tab:theory-req-comp-summary-easeofuse}). However, for the remaining
system-level requirements (Tables~\ref{tab:theory-req-sys-summary-flexibility}
and \ref{tab:theory-req-sys-summary-easeofuse}), the dimensional theories do
not necessarily satisfy the same requirements as discrete theories. Again, the
dimensional theories satisfy some requirements in similar ways, so they are
examined as a single unit.
\begin{itemize}
    \item \textit{Flexibility: Independence from an Agent Architecture
        (\ref{flexArch})} (\good)
    \begin{itemize}
        \item Coordinate space that does not depend on its surrounding
        environment $\rightarrow$ effectively architecture-agnostic

        \item Only require processes that satisfy input and output requirements
    \end{itemize}

    \clearpage
    \item \textit{Flexibility: Allowing Developers to Specify How to Use CME
        Outputs (\ref{flexOut})} (\strong)
    \begin{itemize}
        \item Numerical representation $\rightarrow$ easy to pipe them to other
        computational processes such as facial expression generation and
        decision-making

        \item Potential to violate \textit{Having a Clear API (Output)}
        (\ref{easeAPI}) $\rightarrow$ resolve by providing alternate
        definitions of the dimensions that are easier to understand for
        non-experts
    \end{itemize}

    \item \textit{Flexibility: Ability to Operate on Different Levels of NPC
        Complexity} (\ref{flexComplex}) (\weak) AND \textit{Flexibility: Be
        Efficient and Scalable} (\ref{flexScale})
    (\weak)
    \begin{itemize}
        \item Numerical representation could satisfy these requirements
        $\rightarrow$ requires one of:
        \begin{itemize}
            \item Developers to have some understanding of what the dimensions
            mean and how different factors impact them $\rightarrow$ violates
            \textit{Hiding the Complexity of Emotion Generation}
            (\ref{easeHide})

            \item Providing alternate definitions of the dimensions that are
            easy to understand as with \textit{Allowing Developers to Specify
                How to Use CME Outputs} (\ref{easeAPI}) $\rightarrow$ potential
                to
            violate \textit{Hiding the Complexity of Emotion Generation}
            (\ref{easeHide})
        \end{itemize}
    \end{itemize}

    \item \textit{Ease-of-Use: Having a Clear API (Output) (\ref{easeAPI})}
    (\weak)
    \begin{itemize}
        \item Output API has three numerical components $\rightarrow$ requires
        developers to know how each dimension affects NPC behaviours

        \item Inference on quantities like \textit{pleasantness}
        (\ref{valence}) and \textit{excitement} (\ref{arousal}) likely not as
        automatic as identifying \textit{Joy} and \textit{Fear}
        \begin{itemize}
            \item Could minimize problem with a \ref{circumplex} structure

            \item Disagreements between different models as to where certain
            data points should be~\citep[p.~287]{remington2000reexamining}
            $\rightarrow$ could reduce the psychological validity of \progname{}
        \end{itemize}
    \end{itemize}

    \item \textit{Ease-of-Use: Showing that Emotions Improve the Player
        Experience (\ref{easePX})} (\good)
    \begin{itemize}
        \item Numerical representation with limited variables $\rightarrow$
        easy to manipulate in  experimental settings

        \item Might be difficult for future user study participants to answer
        questions about combinations of values $\rightarrow$ could use a proxy
        mapping values to affective labels
    \end{itemize}

    \item \textit{Ease-of-Use: Providing Examples of Novel Game Experiences
        (\ref{easeNovel})} (\good)
    \begin{itemize}
        \item Numerical representation $\rightarrow$ leverage as a game
        mechanic where players manipulate affective variables as they would
        other resources like character and item statistics~\citep[p.~292, 466,
        559--560, 578]{adams2014fundamentals}

        \item Can be implemented alongside similar mechanics (e.g. status
        attributes in Computer Role-Playing Games (CRPGs), character-related
        puzzles in adventure and social simulation games)
    \end{itemize}
\end{itemize}
\clearpage
\section{Appraisal Theories}
Due to their nature, the appraisal theories are the only ones that can satisfy
the \textit{component-level} requirements in addition to \textit{system-level}
ones.

\subsection{Flexibility: Independence From an Agent Architecture
    (\ref{flexArch})}
Appraisal theories assume that cognition is essential to emotion processing
(\citepg{marsella2015appraisal}{55}; \citepg{broekens2021emotion}{354}) and
that emotions are \textit{about} something that has been intentionally
evaluated~\citep[p.~11]{ortony2021all}. This prevents complete separation from
agent architectures because of the information required for the appraisal
process. Therefore, the goal is \textit{not} to identify theories that can
exist independently of an external system---it is to identify which theories
are agnostic about what that architecture is.

\begin{itemize}
    \item \textbf{Frijda} (\strong)
    \begin{itemize}
        \item Core process is an information processing
        system~\citep[p.~453--456]{frijda1986emotions} that begins with an
        encoding stage that tries to match incoming events with known types and
        their implications for causes and consequences
        \begin{itemize}
            \item Also need to encode actions to evaluate coping potential

            \item Matching process requires users to define event and action
            types, then tag relevant game elements with them
            \begin{itemize}
                \item Event types are tailored to the external architecture
                $\rightarrow$ affords maximal architecture independence
            \end{itemize}
        \end{itemize}

        \item Concerns are dispositions towards the achievement or
        non-achievement of situations that remain dormant as long as its
        satisfaction conditions are met~\citep[p.~335--336,
        466--467]{frijda1986emotions}
        \begin{itemize}
            \item Do not have to generate emotion from ``active'' pursuits
            alone (e.g. goals and motivations), can also be driven by events
            that just \textit{happen} that change a satisfaction condition

            \item [$\rightarrow$] Can account for a much wider range of events,
            supports independence from specific architectures and information
            structures
        \end{itemize}

        \item Action tendencies only specify \textit{what} type of action should
        happen, not \textit{how}~\citep[p.~70]{frijda1986emotions}
        \begin{itemize}
            \item Freedom to connect the actions represented in the
            architecture to any type of action readiness $\rightarrow$ separate
            process can decide which action to execute

            \item [$\rightarrow$] Can account for a much wider range of
            behaviours, supports independence from specific architectures and
            information structures
        \end{itemize}
    \end{itemize}

    \item \textbf{Lazarus} (\good)
    \begin{itemize}
        \item Relational themes described in context of goal achievement,
        requires preexisting knowledge to drive appraisal~\citep[p.~81,
        145]{lazarus1991emotion} $\rightarrow$ goal-based architecture or system

        \item Multiple references to goals, beliefs, and knowledge
        requirements~\citep[p.~39, 151, 177, 210]{lazarus1991emotion}
        $\rightarrow$ implies a Belief-Desire-Intention (BDI) architecture,
        coping coded as intentions
        \begin{itemize}
            \item Has been used to model
            players~\citep[p.~208--209]{yannakakis2018artificial}, unsure of
            use for creating NPCs
        \end{itemize}
    \end{itemize}

    \item \textbf{Scherer} (\disqualified)
    \begin{itemize}
        \item Conceptualizes theory as an information processing
        system~\citep[p.~103--104]{scherer2001appraisalB}
        \begin{itemize}
            \item Structure based on \citet{cowan1988evolving}
            \begin{itemize}
                \item Requires components for: attention, memory,
                goal/need/motivation, reasoning, and a self-model to evaluate
                appraisal dimensions~\citep[p.~100]{scherer2001appraisalB}

                \item Goals/needs/motivations do not have to be
                conscious~\citep[p.~96, 119]{scherer2001appraisalB}

                \item Potential to violate \textit{Ability to Operate on
                    Different Levels of NPC Complexity} (\ref{flexComplex}) if
                    some
                parts cannot be excluded
            \end{itemize}

            \item Assumes multiple processing levels of varying
            complexity~\citep[p.~103]{scherer2001appraisalB}
            \begin{itemize}
                \item Faster, less sophisticated levels call ``higher'' levels
                when they cannot resolve an evaluation

                \item Add more processing layers as needed $\rightarrow$
                potential to support \textit{Ability to Operate on Different
                    Levels of NPC Complexity} (\ref{flexComplex})
            \end{itemize}

            \item Parts of the system are represented with a neural
            network~\citep[p.~105]{scherer2001appraisalB}
            \begin{itemize}
                \item An implementation of Scherer this way was found to be at
                least partially
                black-box~\citep[p.~143--144]{meuleman2015computational}
                $\rightarrow$ violation of \textit{Traceable CME Outputs}
                (\ref{easeTrace})
            \end{itemize}

            \item [$\rightarrow$] Emotion generation is \textit{not}
            independent of the surrounding processes
        \end{itemize}
    \end{itemize}

    \item \textbf{Roseman} (\good)
    \begin{itemize}
        \item Focus on the relationship between appraisal values and emotions,
        how those emotions impact different systems in response, and the
        structure of emotions~\citep[p.~68, 81]{roseman2001model} $\rightarrow$
        does not touch on the emotion process itself, effectively
        architecture-agnostic

        \item Some appraisal dimensions have cognitive
        contents~\citep[p.~265]{roseman1996appraisal} $\rightarrow$ requires
        some type of architecture to provide appraisal inputs
    \end{itemize}

    \item \textbf{OCC} (\strong)
    \begin{itemize}
        \item Requires modelling, planning, reasoning, and predictive processes
        (\citepg{ortony2005affect}{185--186}) $\rightarrow$ not unique to
        emotion~\citep[p.~36]{clore2000cognition}, do not require a separate
        architecture to support \progname{}
        \begin{itemize}
            \item Precursors to expectations about outcomes and world states,
            and self-reflection~\citep[p.~195]{ortony2005affect}

            \item Inputs include memory and
            knowledge~\citep[p.~101]{smith2000consequences}

            \item Assumes that significance detection is
            cognitive~\citep[p.~42]{clore2000cognition}
        \end{itemize}

        \item Requires representations of goals/wants, standards/beliefs, and
        tastes/attitudes (\citepg{occ2022}{54--59})
        \begin{itemize}
            \item Evaluate different input types~\citep[p.~59--60]{occ2022}

            \item Can interact to help/hinder each other~\citep[p.~59]{occ2022}

            \item Must be coherent and relatively stable internal structure,
            like a goal hierarchy, to evaluate the environment by to produce
            consistent results in both kind and
            intensity~\citep[p.~194--195]{ortony2002making} $\rightarrow$
            coherence depends on how the user defines these structures, not
            directly dependent on \progname{}
        \end{itemize}

        \item Acknowledges that there are different potential action
        outcomes~\citep[p.41]{clore2000cognition} $\rightarrow$ potential to
        create architecture-agnostic outputs

        \item Later ties emotion to changes in the body similar to
        neurophysiological theories (\citepg{ortony2005affect}{174, 177, 188,
            195}; \citepg{clore2000cognition}{24--25, 28--29}) $\rightarrow$ at
        least partially architecture dependent because of dependence on
        embodiment
    \end{itemize}

    \item \textbf{Smith \& Kirby} (\strong)
    \begin{itemize}
        \item Conceptualized as a process model, built from previously gathered
        findings on the effects of emotion and mood on
        cognition~\citep[p.~85]{smith2000consequences}
        \begin{itemize}
            \item Builds from the framework described by
            \cite{smith1990emotion}~\citep[p.~122]{smith2001toward}
            \begin{itemize}
                \item Does not appear to have the same dependencies on goals,
                beliefs, and intentions $\rightarrow$ more likely to be
                architecture-agnostic
            \end{itemize}

            \item Views emotion as a well-being monitor or guidance system for
            attentional and motivational
            functions~\citep[p.~90--91]{smith2000consequences} $\rightarrow$
            idea of a ``guidance system'' does not belong to any single
            architecture, potential to apply to many

            \item Not empirically tested $\rightarrow$ \progname{} not
            concerned with ``correct'' results, just interesting ones
        \end{itemize}

        \item Accounts for more than one appraisal process, processes work in
        parallel (\citepg{smith2000consequences}{91--92};
        \citepg{smith2001toward}{129})
        \begin{itemize}
            \item Specifies two appraisal types for automatic reactions (i.e.
            priming and activation of memories) and deliberative analysis (i.e.
            reasoning) $\rightarrow$ notes that concept appears in previous
            proposals (e.g. \cite{leventhal1987relationship},
            \cite{sloman2005architectural})

            \item Proposes that memory is a network~\citep[p.~94,
            102]{smith2000consequences}
            \begin{itemize}
                \item Allows priming and spreading activation $\rightarrow$
                appraisal is continuous, activated quickly and automatically,
                and does not require much attention

                \item Knowledge in memory does not have to be organized in
                schemas
            \end{itemize}

            \item Proposes that reasoning uses highly developed and abstract
            thinking processes (\citepg{smith2001toward}{130};
            \citepg{smith2000consequences}{95--96})
            \begin{itemize}
                \item Requires that memory items be associated with semantic
                meaning $\rightarrow$ resulting appraisals can be integrated
                back into memory for associative processing (i.e.
                learning)
            \end{itemize}

            \item Users are not required to have these processes $\rightarrow$
            core idea of appraisal unaffected because it does not rely on these
            two specific appraisal types or definitions
        \end{itemize}
    \end{itemize}

    \item \textbf{Oatley \& Johnson-Laird} (\strong)
    \begin{itemize}
        \item Assumes that the cognitive system is modular and asynchronous,
        similar to
        \citet{minsky1988society}~\citep[p.~31--32]{oatley1987towards},
        model-driven rather than
        rule-driven~\citep[p.~205--206]{johnson1992basic} $\rightarrow$ aligns
        with the idea of architecture independence
        \begin{itemize}
            \item Top-level module organizes whole system, can reorganize
            system goals and plans \citep[p.~50--51]{oatley1992best}
            $\rightarrow$ top-level control module in software architecture
        \end{itemize}

        \item Implicitly assumes that individuals have beliefs, desires, and
        needs that they make goals about and plans to
        achieve~\citep[p.~213]{johnson1992basic}
        \begin{itemize}
            \item Defines ``cognitive'' as psychological explanations with
            knowledge representations and transformations that might not be
            conscious~\citep[p.~30]{oatley1987towards} $\rightarrow$ acts on
            transformations on data, could be defined for a generalized data
            representation

            \item Core elements are goals and
            plans~\citep[p.~30]{oatley1987towards}
            \begin{itemize}
                \item Goals $\rightarrow$ symbolic representations of possible
                environment states to achieve

                \item Plans $\rightarrow$ sequences from the current
                environment state to a goal, can include instinctive and highly
                practised ones (i.e. automatic)
            \end{itemize}

            \item Emotions as a mechanism for managing cognitive resources and
            goal priorities \citep[p.~207--208]{johnson1992basic}, and
            responding to models---including social ones for cooperation and
            competitive planning---that are proven invalid in the
            moment~\citep[p.~205--206]{johnson1992basic}
            \begin{itemize}
                \item Triggered when smoothly flowing action is interrupted,
                detects significant change in goal or plan outcomes, typically
                at plan junctures (\citepg{oatley1992best}{46, 48};
                \citepg{oatley1987towards}{35--36})

                \item Cause the system to enter an ``emotion mode'' that
                inhibits other ``emotion modes'' or oscillates between multiple
                ``modes''~\citep[p.~34]{oatley1987towards} $\rightarrow$
                comparable to other system state changes

                \item ``Modes'' associated with different goal priorities,
                possible actions, and skills \citep[p.~37]{oatley1987towards}
            \end{itemize}

            \item [$\rightarrow$] Does not necessarily imply a
            Belief-Desire-Intention (BDI) architecture
        \end{itemize}

        \item Assumes a two-pathway system~\citep[p.~32--34]{oatley1987towards}
        \begin{itemize}
            \item Reactive $\rightarrow$ propagates a global ``signal" to setup
            an emotion ``mode"

            \item Deliberative $\rightarrow$ invoke individual functions,
            reason about system state for planning

            \item [$\rightarrow$] Does not depend on specific architecture
            features, assume that ``planning" does not have to be formal

            \item Can naturally cause temporal shifts in emotion quality as
            different processes add meaning (influenced by individual and
            cultural factors) to a goal/plan
            change~\citep[p.~47]{oatley1987towards}
        \end{itemize}
    \end{itemize}
\end{itemize}

\subsection{Flexibility: Choosing Which CME Tasks to Use
(\ref{flexTasks})}\label{sec:notes-appraisal-choose-tasks}
Appraisal theories are assumed to need some minimum number of processes for
emotion generation. Therefore, they are evaluated on the ability to call them
individually as needed. It is assumed that a game designer can choose when to
call the emotion generation as a complete process.

\begin{itemize}
    \item \textbf{Frijda} (\weak)
    \begin{itemize}
        \item Core emotion process is
        interdependent~\citep[p.~454]{frijda1986emotions} $\rightarrow$
        unrealistic to allow its components to be called out of turn

        \item Possible to skip and/or interrupt
        processes~\citep[p.~461--463]{frijda1986emotions}
        \begin{itemize}
            \item Direct implementation would require theory knowledge
            $\rightarrow$ violates \textit{Hiding the Complexity of Emotion
                Generation} requirement (\ref{easeHide})

            \item Could build interrupts over the emotion process, temporarily
            bypassing it (i.e. automatic responses) $\rightarrow$ emotion
            process continues at its current pace and updates emotion state
            when it finishes
        \end{itemize}

        \item Task choice difficult to realize within the process, can
        implement interrupts that bypass the system and act like automatic
        responses
    \end{itemize}

    \item \textbf{Lazarus} (\weak)
    \begin{itemize}
        \item Emotion process is
        interdependent~\citep[p.~39, 208--211]{lazarus1991emotion}
        $\rightarrow$ unrealistic to allow its components to be called out of
        turn

        \item No obvious mention of ways to skip or interrupt tasks

        \item Define separate processing levels for societal, psychological,
        and physiological tasks~\citep[p.~211]{lazarus1991emotion}
        $\rightarrow$ could turn whole levels on/off as needed

        \item Create switches/input points for designers to allow internal
        processes (i.e. emotion-based coping) to influence the appraisal
        process and outcomes~\citep[p.~210]{lazarus1991emotion}
        \begin{itemize}
            \item Potential to violate \textit{Hiding the Complexity of Emotion
                Generation} (\ref{easeHide}) $\rightarrow$ make available to
            advanced users
        \end{itemize}
    \end{itemize}

    \item \textbf{Scherer} (\strong)
    \begin{itemize}
        \item Monitoring system triggers appraisal cycles based on
        relevance~\citep[p.~99]{scherer2001appraisalB} $\rightarrow$ choose
        when to start and stop reappraisals and/or update appraisal registers

        \item Check individual appraisal units (SEC) to update systems and when
        to see what the current action tendency is~\citep[p.~104,
        106]{scherer2001appraisalB} $\rightarrow$ requires caution because it
        could cause cascading changes in interdependent modules, which also
        changes the current appraisal

        \item Define separate processing levels for different types of
        information (i.e. sensory-motor, schematic,
        conceptual)~\citep[p.~102--103]{scherer2001appraisalB} $\rightarrow$
        could turn whole levels on/off as needed
    \end{itemize}

    \item \textbf{Roseman} (\disqualified)
    \begin{itemize}
        \item Focus on the relationship between appraisal values and emotions,
        how those emotions impact different systems in response, and the
        structure of emotions~\citep[p.~68, 81]{roseman2001model} $\rightarrow$
        does not touch on the emotion process itself
    \end{itemize}

    \item \textbf{OCC} (\good)
    \begin{itemize}
        \item Emotion structure built with three distinct branches
        $\rightarrow$ could choose a subset of branches
        \begin{itemize}
            \item Some emotions only possible if multiple branches active
            (e.g. \textit{Anger} requires event and attribution branches)
            (\citepg{occ2022}{29}; \citepg{ortony2002making}{195};
            \citepg{steunebrink2009occ}{7})

            \item Each branch requires at least one evaluated variable to
            proceed, additional variables can retain neutral
            values~\citep[p.~71, 95, 98]{occ2022} $\rightarrow$ could choose
            which tasks to run based on required values

            \item Insufficient information could mean that the process will not
            produce a result
        \end{itemize}
    \end{itemize}

    \item \textbf{Smith \& Kirby} (\good)
    \begin{itemize}
        \item Builds on \cite{smith1990emotion}~\citep[p.~122]{smith2001toward}
        \wasytherefore{} assume that its core emotion process is also
        interdependent and it is unrealistic to allow its components to be
        called out of turn

        \item Control over sources of appraisal inputs
        (\citepg{smith2000consequences}{93--94, 100};
        \citepg{smith2001toward}{129--130})
        \begin{itemize}
            \item Sources interact and their disparate information integrated
            before appraisal

            \item Can control when sources provide information, when to
            integrate, and how to integrate them $\rightarrow$ control emotion
            generation at the triggering stage

            \item [$\rightarrow$] Potential to choose tasks that provide and
            integrate inputs, controlling the emotion generation process

            \item No obvious information about how to integrate information
            sources
        \end{itemize}
    \end{itemize}

    \item \textbf{Oatley \& Johnson-Laird} (\good)
    \begin{itemize}
        \item Base elements are goals and plans $\rightarrow$ can decide which
        plan junctures to call emotion generation at

        \item Emotion ``modes'' have a basic meaning that deliberative
        processes can build on~\citep[p.~35, 43]{oatley1987towards}, definition
        of two pathways that can propagate to the whole system (i.e. reactive)
        or invoke individual functions (i.e.
        deliberative)~\citep[p.~32--34]{oatley1987towards}
        \begin{itemize}
            \item Freedom to choose which tasks to call when additional
            information is needed to add nuance to emotion states

            \item Game developer would need to provide all additional tasks
            $\rightarrow$ does \textit{not} violate \textit{Hiding the
                Complexity of Emotion Generation} (\ref{easeHide}) because of
                its
            partial basis on an intuitive, ``folk'' understanding of emotion
            embedded in language~\citep[p.~74--75, 86--87]{oatley1992best}
        \end{itemize}
    \end{itemize}
\end{itemize}

\subsection{Flexibility: Customizing Existing Task Parameters
    (\ref{flexCustom})}
Differing from when game designers call emotion generation tasks is the ability
to control their functionality, such as variable sensitivity and activation
thresholds. Ideally, \progname{} should allow game designers to manipulate as
many system parameters as possible to maximize customizability, effectively
creating ``individual differences'' with each change.

\begin{itemize}
    \item \textbf{Frijda} (\strong)
    \begin{itemize}
        \item Notes many potential elements that can be parameterized, one
        hypothesized source of individual
        differences~\citep[p.~456--458]{frijda1986emotions}
        \begin{itemize}
            \item Each phase in the core emotion process can be influenced
            individually by both internal and external inputs

            \item Different and variable sensitivity levels/thresholds/concern
            priorities for matching inputs with satisfaction conditions

            \item Variable acceptance conditions for connecting a generated
            meaning structure with action readiness modes/emotions

            \item Open ended parameters $\rightarrow$ allow designers to
            customize additional parameters to influence emotion generation
        \end{itemize}

        \item Potential to implement some parameters implicitly from system
        states
    \end{itemize}

    \clearpage
    \item \textbf{Lazarus} (\disqualified)
    \begin{itemize}
        \item Discussion of appraisal styles implies that emotion process
        dispositions are part of an encoding process, not the appraisal
        itself~\citep[p.~138]{lazarus1991emotion}
        \begin{itemize}
            \item Some individual differences contained in the structure and
            organization of goals~\citep[p.~99]{lazarus1991emotion}
            $\rightarrow$ outside \progname{}'s scope

            \item Personality defined as goal commitments, beliefs, and
            knowledge as an input to emotion
            generation~\citep[p.~209]{lazarus1991emotion} $\rightarrow$ outside
            \progname{}'s scope

            \item [$\rightarrow$] No explicit mention of ``tuning'' the emotion
            generation process directly
        \end{itemize}
    \end{itemize}

    \item \textbf{Scherer} (\good)
    \begin{itemize}
        \item Parameters associated with appraisal
        registers~\citep[p.~105--106]{scherer2001appraisalB}
        \begin{itemize}
            \item Individual variables combined with weighted functions that
            change with the ``confidence'' in the data $\rightarrow$ mechanize
            as a user-defined task parameter

            \item Action tendency activation ``strength'' tied to appraisal
            profile and degree of ``definiteness'' of individual checks
            $\rightarrow$ potential for parameterized activation thresholds
            based on strength and confidence in appraisal check accuracy
        \end{itemize}
    \end{itemize}

    \item \textbf{Roseman} (\disqualified)
    \begin{itemize}
        \item Focus on the relationship between appraisal values and emotions,
        how those emotions impact different systems in response, the
        structure of emotions~\citep[p.~68, 81]{roseman2001model}, and empirical
        validation of appraisal dimension influence on resulting
        emotion~\citep[p.~242, 244]{roseman1996appraisal} $\rightarrow$ does
        not touch on the emotion process itself
    \end{itemize}

    \item \textbf{OCC} (\good)
    \begin{itemize}
        \item Parameterization of emotion intensity and activation thresholds
        $\rightarrow$ change how easily and intensely emotions are
        produced~\citep[p.~220--221]{occ2022}
        \begin{itemize}
            \item Variable weights on emotion intensity function

            \item Modulation of emotion thresholds $\rightarrow$ changes how
            strong the emotion is before it manifests
        \end{itemize}

        \item Elicitation rule conflict resolution not
        addressed~\citep[p.~228]{occ2022} $\rightarrow$ allow customization of
        rule priority

        \item Handling ``mixed emotions'', coexisting positive and negative
        emotions from the same appraisal~\citep[p.~63--64]{occ2022}
        $\rightarrow$ implement customizable mechanism to determine which to
        express at any given moment

        \item Suggest varying parameters on emotion generation
        mechanisms~\citep[p.~203]{ortony2002making} $\rightarrow$ process not
        well defined, limits ability to implement it

        \item If multiple processing levels are implemented, can parameterize
        the thresholds for control and interrupt thresholds from each
        one~\citep[p.~185]{ortony2005affect}

        \item Few guidelines about how these work $\rightarrow$ risk of
        reducing psychological validity
    \end{itemize}

    \item \textbf{Smith \& Kirby} (\good)
    \begin{itemize}
        \item Builds on \cite{smith1990emotion}~\citep[p.~122]{smith2001toward}
        \wasytherefore{} assume that its core emotion process prevents direct
        ``tuning'' too

        \item Control over appraisal input sources~\citep[p.~93--94,
        100]{smith2000consequences}
        \begin{itemize}
            \item Sources interact and their separate information integrated
            before appraisal $\rightarrow$ control degrees of interaction and
            weights during information integration

            \item Can control when sources provide information $\rightarrow$
            ``sensitivity'' or activation thresho-lds

            \item No obvious information about how to integrate information
            sources
        \end{itemize}
    \end{itemize}

    \item \textbf{Oatley \& Johnson-Laird} (\good)
    \begin{itemize}
        \item Emotions elicited by relative changes in success probabilities at
        plan junctions~\citep[p.~98]{oatley1992best} $\rightarrow$ candidate
        for implementing sensitivity thresholds

        \item Mentions temporal differences in emotion intensity, variable
        emotion decay rates of emotion, replacement with other emotions
        elicited by the same scenario~\citep[p.~22--23]{oatley1992best}
        $\rightarrow$ candidates for customizing how emotion quality and
        intensity varies over time and context

        \item Few guidelines about how to define parameters $\rightarrow$ low
        risk of violating psychological validity due to its partial basis on an
        intuitive, ``folk'' understanding of emotion embedded in
        language~\citep[p.~74--75, 86--87]{oatley1992best}
    \end{itemize}
\end{itemize}

\subsection{Flexibility: Allowing the Integration of New Components
    (\ref{flexNew})}
When included, the appraisal theories tend to define other affective types
relative to emotion. This implies that integrating them requires no additional
structures in favour of building on top of existing features. This makes them
ideally suited for \textit{Allowing the Integration of New Components}
(\ref{flexNew}) in this respect.

Although integrating non-affective components should be theory-agnostic, how
easily this can be done varies between appraisal theories. Therefore, the
analysis examines their ability to integrate both affective and non-affective
components.

\begin{itemize}
    \item \textbf{Frijda} (\strong)
    \begin{itemize}
        \item Proposes that adding, removing, and/or modifying the components of
        emotion creates different types of
        affect~\citep[p.~253]{frijda1986emotions} $\rightarrow$ does not
        require changes to emotion generation, definitions built on top of
        existing emotion definitions and functions
        \begin{itemize}
            \item Later refinement for an implemented version of the theory
            related mood, personality, and sentiments to emotion by their focus
            and duration (Figure~\ref{fig:CRAffectTypes})
            \begin{figure}[!tb]
                \centering
                \includegraphics[width=0.5\linewidth]{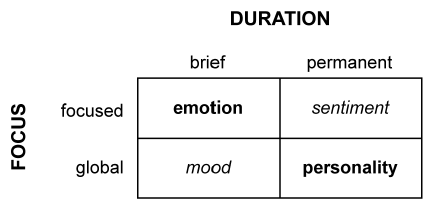}

                \caption[Proposed Relation Between Emotion and Other Affective
                Types based on Frijda]{Proposed Relation Between Emotion and
                Other Affective Types based on Frijda (Adapted from
                \citet[p.~136]{moffat1997personality})}
                \label{fig:CRAffectTypes}
            \end{figure}

            \item Could derive core affect from emotion process via the
            \textit{valence} and \textit{demand character} appraisal
            dimensions and \ref{arousal} value~\citep[p.~207,
            454]{frijda1986emotions}
        \end{itemize}

        \item Inclusion of a ``Regulation Processes" block that can affect
        nearly all parts of the emotion
        process~\citep[p.~545]{frijda1986emotions}
        \begin{itemize}
            \item Multiple points to introduce new components and processes
            $\rightarrow$ easy to add non-affective components

            \item Potential to violate \textit{Hiding the Complexity of Emotion
                Generation} (\ref{easeHide}) if users can access points directly
            $\rightarrow$ create an interface to hide entry points, make it
            easier to use and understand
        \end{itemize}
    \end{itemize}

    \newpage
    \item \textbf{Lazarus} (\weak)
    \begin{itemize}
        \item Personality not seen as a set of innate traits that manifest in
        appraisal and coping~\citep[p.~316]{lazarus1991emotion}, defined as a
        collection of goals, needs, commitments, knowledge, attitudes, and
        beliefs that influence how an event is perceived and how the individual
        acts on the resulting action tendency~\citep[p.~623--624,
        628]{smith1990emotion} $\rightarrow$ implicitly defined
        \begin{itemize}
            \item  Affords flexibility (i.e. not limited to a set of values)
            $\rightarrow$ supports creative freedom, definitions of individual
            characters based on their goals and knowledge rather than numerical
            values

            \item More difficult to define personality quickly (e.g. have to
            decide what beliefs a character with a desired personality would
            have)
        \end{itemize}

        \item Mood is ``an existential state or condition of life'' that is
        appraisal-dependent, related to subjective
        well-being~\citep[p.~266--267]{lazarus1991emotion}
        \begin{itemize}
            \item Could define as a state that aggregates appraisal results
            into a ``satisfaction/dissatisfaction'' value
        \end{itemize}

        \item Equates affect to subjective
        experience~\citep[p.~57]{lazarus1991emotion}
        \begin{itemize}
            \item Could define core affect using \textit{goal congruence} as
            \ref{valence}

            \item \ref{arousal} is part of an action tendency, tied to the
            emotion's core relational theme \citep[p.~58--59,
            150]{lazarus1991emotion} $\rightarrow$ not explicitly defined
        \end{itemize}

        \item Potential interface points for external processes part of input
        generation/output manipulation~\citep[p.~210]{lazarus1991emotion}
        $\rightarrow$ does not have to integrate with emotion generation
        processes, trivial to add non-affective components
    \end{itemize}

    \item \textbf{Scherer} (\weak)
    \begin{itemize}
        \item Proposes definitions for mood and
        personality~\citep[p.~140--141]{scherer2000psychological}, but are not
        accounted for in the working theory~\citep[p.~93,
        119]{scherer2001appraisalB}
        \begin{itemize}
            \item Personality could be defined as sensitivities in appraisal
            dimension and register functions, mood as temporary sensitivities
            caused by previous appraisals $\rightarrow$ potential to violate
            psychological validity
        \end{itemize}

        \item No clear connection to core affect

        \item Potential to integrate non-affective components during the
        information processing and appraisal objective
        steps~\citep[p.~104]{scherer2001appraisalB} $\rightarrow$ might require
        knowledge of how those components work, violating \textit{Hiding the
            Complexity of Emotion Generation} (\ref{easeHide})
    \end{itemize}

    \item \textbf{Roseman} (\weak)
    \begin{itemize}
        \item Suggests that mood and personality are tied to emotion
        generation~\citep[p.~81--83]{roseman2001model}, ways to describe
        appraisal styles for individual or families of
        emotion~\citep[p.~88--89]{roseman2001model} $\rightarrow$ implicitly
        defined
        \begin{itemize}
            \item  Affords flexibility (i.e. not limited to a set of values)
            $\rightarrow$ supports creative freedom, definitions of individual
            characters based on their goals and knowledge rather than numerical
            values

            \item More difficult to define quickly (e.g. have to decide what
            appraisal dispositions a character with a desired personality would
            have)

            \item Could be extended to represent cultural influences on emotion
            generation
        \end{itemize}

        \item No clear connection to core affect
        \begin{itemize}
            \item Could define core affect using \textit{situational state} and
            \textit{motivational state} as \ref{valence}

            \item No obvious component for \ref{arousal}
        \end{itemize}

        \item Focus on the relationship between appraisal values and
        emotions~\citep[p.~81]{roseman2001model} $\rightarrow$ does not focus
        on other parts of the generation process, no obvious place to integrate
        non-affective processes
    \end{itemize}

    \item \textbf{OCC} (\strong)
    \begin{itemize}
        \item Proposes that personality is a unique parameter profile defining
        how emotion generation behaves within and between process
        levels~\citep[p.~189--190]{ortony2005affect}
        \begin{itemize}
            \item Tuning emotion generation for each NPC $\rightarrow$
            personality implicitly supported by \textit{Customizing Existing
                CME Task Parameters} (\ref{flexCustom})

            \item Could implement personality inventories as parameter
            profiles~\citep[p.~191--192]{ortony2005affect} $\rightarrow$ does
            not provide explicit definitions, potential to violate psychological
            validity if done incorrectly, might not matter if the profiles do
            what the developer expects
        \end{itemize}

        \item Moods described as free-floating, object-less affective states
        that can influence emotion but can also arise from sources
        independently of emotion~\citep[p.~27]{clore2000cognition}
        $\rightarrow$ could be linked to personality ``parameter profiles'' by
        treating it as an initial condition
        \begin{itemize}
            \item Defined as temporally-driven parameter
            changes~\citep[p.~221--222, 228]{occ2022} $\rightarrow$ implicitly
            supported by \textit{Customizing Existing CME Task Parameters}
            (\ref{flexCustom})
        \end{itemize}

        \item Potential to represent core affect
        \begin{itemize}
            \item \ref{arousal} is a global intensity variable, roughly
            proportional to base emotion intensity or perhaps even only parts
            of this (i.e. subjective importance of the
            situation)~\citep[p.~80--83]{occ2022} $\rightarrow$ other factors
            can influence it and has a slow rate of decay, supports
            \textit{Customizing Existing CME Task Parameters} (\ref{flexCustom})

            \item It follows that \ref{valence} might be approximated as sum of
            the absolute signed values of the same variables (i.e. is the
            overall feeling positive or negative?)
        \end{itemize}

        \item Two potential ways to integrate non-affective components
        \begin{itemize}
            \item As part of the input generation process $\rightarrow$
            designer-driven, supports \textit{Ability to Operate on Different
                Levels of NPC Complexity} (\ref{flexComplex})

            \item As a method for controlling task parameters $\rightarrow$
            implicitly supported by \textit{Customizing Existing CME Task
                Parameters} (\ref{flexCustom})
        \end{itemize}
    \end{itemize}

    \item \textbf{Smith \& Kirby} (\weak)
    \begin{itemize}
        \item No clear definitions for personality, mood, or core affect
        \begin{itemize}
            \item Potential correlation between \textit{emotion-focused coping
                potential} and some personality traits~\citep[p.~1366--1368,
            1369]{smith2009putting} $\rightarrow$ suggests that personality
            traits are parameters on the emotion generation process

            \item Core affect could be constructed from \textit{motivational
                congruence} (as \ref{valence}) and \textit{motivational
                relevance}
            (as \ref{arousal}) $\rightarrow$ not necessarily empirically
            supported, potential to violate psychological validity
        \end{itemize}

        \item Appraisal registers synthesize information from multiple sources
        and levels of processing \citep[p.~130]{smith2001toward}
        \begin{itemize}
            \item Multiple points to introduce new components and processes
            $\rightarrow$ easy to add non-affective components

            \item Integrating non-affective components would require
            manipulating the detector mechanisms, potential to violate
            \textit{Hiding the Complexity of Emotion Generation}
            (\ref{easeHide}) if users can access points directly $\rightarrow$
            create an interface to hide entry points, make it easier to use and
            understand
        \end{itemize}
    \end{itemize}

    \item \textbf{Oatley \& Johnson-Laird} (\strong)
    \begin{itemize}
        \item Emotions often have moods and sentiments associated with
        them~\citep[p.~87]{oatley2000sentiments} $\rightarrow$ implies that
        adding these affective types would be an extension of existing emotion
        structures

        \item Temperaments (i.e. personality traits) hypothesized to be enduring
        predispositions towards emotion ``modes''
        (\citepg{oatley1987towards}{34}; \citepg{oatley1992best}{61})
        \begin{itemize}
            \item Also defines sentiments---enduring emotional dispositions
            about something, typically other
            individuals~\citep[p.~81]{oatley2000sentiments} $\rightarrow$
            potential to define two sets of personality traits (general and
            target-specific), affords more creative freedom
        \end{itemize}

        \item Moods defined directly in the theory as control signals that
        persist after the cause of an emotion passes/no longer associated with
        semantic content and keeps the system in a particular state
        (\citepg{oatley1987towards}{32}; \citepg{oatley1992best}{64})
        \begin{itemize}
            \item Could be realized as temporary predispositions towards
            emotion ``modes'' or a longer lasting, low intensity emotion
            state~\citep[p.~34--35]{oatley1987towards} $\rightarrow$ potential
            to allow both, give user the choice of which to use, affording more
            creative freedom
        \end{itemize}

        \item Potential to define core affect based on how a goal is affected
        (positive or negative) for \ref{valence}, emotion intensity as
        \ref{arousal} $\rightarrow$ no explicit definitions given, potential to
        violate psychological validity if done incorrectly, might not matter if
        the profiles do what the developer expects

        \item Assume that the cognitive system is modular and
        asynchronous~\citep[p.~31]{oatley1987towards} $\rightarrow$ implies
        that adding non-affective processes is feasible, should not require
        knowledge of the inner workings of emotion generation
    \end{itemize}
\end{itemize}

\subsection{Flexibility: Choosing NPC Emotions
(\ref{flexEm})}\label{sec:notes-appraisal-choose-emotions}
Like the discrete theories, the appraisal theories tend to define emotions as
categories to group different aspects of a response together. In this sense,
excluding predefined emotions is trivial. Once again, \textit{adding} new ones
is unclear and there are no obvious ``rules'' to follow. \progname{} can still
take advantage of them because new emotions need only make sense to the
developer so that they can use them. Therefore, the appraisal theories are
evaluated for what a developer would need to do and how easy it is to realize.

\begin{itemize}
    \item \textbf{Frijda} (\weak)
    \begin{itemize}
        \item Emotions as descriptions of action readiness in response to
        different combinations of events, or by the nature of the emotional
        object~\cite[p.~72--74]{frijda1986emotions} $\rightarrow$ defining new
        emotions requires defining new action tendencies
        \begin{itemize}
            \item Requires modifications to the emotion generation process (i.e.
            defining appraisal patterns) $\rightarrow$ violates \textit{Hiding
                the Complexity of Emotion Generation} (\ref{easeHide})

            \item Some ``non-basic'' emotions are blends, can define a limited
            set of additional emotions $\rightarrow$ limited flexibility

            \item Define emotions by pairing existing ones with an event or
            object type and assigning it a new name $\rightarrow$ similar to
            scripting, event-coding
        \end{itemize}
    \end{itemize}

    \item \textbf{Lazarus} (\weak)
    \begin{itemize}
        \item Emotions associated with themes that can
        coexist~\citep[p.~229]{lazarus1991emotion}
        \begin{itemize}
            \item Potential to allow developers to create named combinations
            representing ``new'' emotions $\rightarrow$ necessarily create more
            complicated emotions

            \item Defining emotions that are not combinations would require new
            appraisal pattern definitions (i.e. modify the emotion generation
            process) $\rightarrow$ violates \textit{Hiding the Complexity of
                Emotion Generation} (\ref{easeHide})
        \end{itemize}
    \end{itemize}

    \item \textbf{Scherer} (\good)
    \begin{itemize}
        \item ``Emotions'' defined as the net effect of continuous,
        fluctuating changes in subsystems \citep[p.~106,
        108]{scherer2001appraisalB}
        \begin{itemize}
            \item Adding new emotions requires identification and naming of a
            set of subsystem changes (i.e. requires an understanding of how
            emotions are generated) $\rightarrow$ violates \textit{Hiding the
                Complexity of Emotion Generation} (\ref{easeHide})
        \end{itemize}

        \item ``Innate'' emotions\footnote{Scherer calls them \textit{modal}
            emotions.} (e.g. \textit{Joy}) attributed to common adaptational
            issues
        that produce consistent system effects~\citep[p.~108,
        113]{scherer2001appraisalB} $\rightarrow$ range of known emotions are
        products of mixtures and/or blends of ``innate'' ones
        \begin{itemize}
            \item Some emotion profiles have ``open'' entries that can accept
            any value for that dimension $\rightarrow$ potential to define
            emotion family ``members'' by providing specific values for
            ``open'' entries, potential to violate \textit{Hiding the
                Complexity of Emotion Generation} (\ref{easeHide})

            \item Intensity differences can also differentiate otherwise
            identical emotions $\rightarrow$ define new emotions by intensity
            class, external to the emotion generation process so no violation
            of \textit{Hiding the Complexity of Emotion Generation}
            (\ref{easeHide})
        \end{itemize}
    \end{itemize}

    \item \textbf{Roseman} (\good)
    \begin{itemize}
        \item Proposes that more than one emotion can be experienced
        simultaneously due to different
        evaluations~\citep[p.~81]{roseman2001model}
        \begin{itemize}
            \item Definition of new emotions as mixtures $\rightarrow$ external
            to the generation process, so no clear violation of \textit{Hiding
            the Complexity of Emotion Generation} (\ref{easeHide})

            \item Emotions logically grouped by response strategy (e.g.
            \textit{attack}, \textit{exclude}) $\rightarrow$ potential for a
            design tool to guide the process of defining new emotions?
        \end{itemize}
    \end{itemize}

    \item \textbf{OCC} (\good)
    \begin{itemize}
        \item Emotions necessarily tied to cognitive abilities that build on
        four basic affective states \citep[p.~183--184]{ortony2005affect}
        implies that adding ``new'' emotions is about adding meaning
        $\rightarrow$ dependent on what is available for inputs,
        designer-driven, supports \textit{Ability to Operate on Different
            Levels of NPC Complexity} (\ref{flexComplex})

        \item Proposed a smaller emotion structure for believable agents,
        collapsing 22 emotions into five positive and five negative
        ones~\citep[p.~193--194]{ortony2002making}
        \begin{itemize}
            \item Potential to add ``new'' emotions as more cognitive processes
            are added $\rightarrow$ dependent on what is available for inputs,
            designer-driven, supports \textit{Ability to Operate on Different
                Levels of NPC Complexity} (\ref{flexComplex})
        \end{itemize}

        \item \textit{Surprise} as a special case, can be added with the
        \textit{unexpectedness} appraisal variable~\citep[p.~13--14,
        16--17]{ortony2021all}

        \item ``New'' emotions can be defined as differences in
        intensity/elicitation thresholds (e.g. \textit{Pleased} for low
        intensity and \textit{Ecstatic} for high)~\citep[p.~220--221]{occ2022}
        \begin{itemize}
            \item External to the emotion generation process $\rightarrow$ no
            violation of \textit{Hiding the Complexity of Emotion Generation}
            (\ref{easeHide})
        \end{itemize}
    \end{itemize}

    \item \textbf{Smith \& Kirby} (\weak)
    \begin{itemize}
        \item Unclear how to define additional emotions

        \item Hypothesizes that emotion categories are likely dense clusters in
        dimensional space \citep[p.~245--246]{smith_scott_mandler_1997}
        \begin{itemize}
            \item Potential to combine with V-A or PAD Space, define new
            emotion categories from existing data clusters $\rightarrow$ would
            require some understanding of source material, potential violation
            of \textit{Hiding the Complexity of Emotion Generation}
            (\ref{easeHide})

            \item Developers could collect their own data to find affective
            ``clusters'' in a dimensional space $\rightarrow$ time consuming,
            error-prone, potential violation of \textit{Hiding the Complexity
                of Emotion Generation} (\ref{easeHide})
        \end{itemize}
    \end{itemize}

    \item \textbf{Oatley \& Johnson-Laird} (\strong)
    \begin{itemize}
        \item How people describe emotions in everyday language indicates
        underlying cognitive meanings~\citep[p.~210]{johnson1992basic}
        $\rightarrow$ can build on a layperson's understanding of emotions,
        supports \textit{Hiding the Complexity of Emotion Generation}
        (\ref{easeHide})

        \item Emotion ``modes'' are not absolute definitions, only have
        heuristic properties that capture general classes of
        events~\citep[p.~87]{oatley2000sentiments} $\rightarrow$ potential to
        create more refined emotions by constraining the classes
        \begin{itemize}
            \item ``New''/''adult'' emotions are based on emotion ``modes'',
            deliberative processes attach more meaning to them
            (\citepg{oatley1987towards}{35, 43};
            \citepg{oatley1992best}{76--78})

            \item Can also be defined at junctions of mutual plans with one or
            more other agents, requires a self-model~\citep[p.~44, 46,
            48]{oatley1987towards} $\rightarrow$ way to integrate cultural
            differences due to the impact on models and reasoning
            processes~\citep[p.~47]{oatley1987towards}
        \end{itemize}

        \item Emotions that have different semantic contents can exist
        simultaneously~\citep[p.~104]{oatley1992best} $\rightarrow$ potential
        to define emotion ``mixtures''
        \begin{itemize}
            \item External to the emotion generation process $\rightarrow$ no
            violation of \textit{Hiding the Complexity of Emotion Generation}
            (\ref{easeHide})
        \end{itemize}
    \end{itemize}
\end{itemize}

\subsection{Flexibility: Allowing Developers to Specify How to Use CME Outputs
    (\ref{flexOut})}
There is a clear difference between reflexes and emotions: one is very
difficult to control how one reacts and the other has a range of them to pick
from~\citep[p.~45]{fellous2004human}. The idea of emotion \textit{components}
also suggests that there is flexibility in which system aspects emotion affects
and how (\citepg{hudlicka2019modeling}{133};
\citepg{scherer2001appraisalB}{108}; \citepg{roseman2011emotional}{436}). This
implies that \textit{any} appraisal theory should be able to strongly support
this requirement. The question is now how well each theory defines what this
means.

\begin{itemize}
    \item \textbf{Frijda} (\strong)
    \begin{itemize}
        \item Set the output boundary of \progname{} at the action proposer
        (i.e. action tendency), relevance and control precedence signals, and
        physiological change generator (i.e. \ref{arousal})
        points~\citep[p.~455]{frijda1986emotions}
        \begin{itemize}
            \item Would also allow for another layer to group these into
            emotion categories~\citep[p.~72]{frijda1986emotions}

            \item Can feed the outputs back into \progname{} $\rightarrow$
            provide an interface so that this is a matter of ``flipping a
            switch'', supported by \textit{Customizing Existing CME Task
                Parameters} (\ref{flexCustom})

            \item [$\rightarrow$] Allows maximum flexibility for defining what
            to do with outputs that is agnostic to how ``action'' is defined
        \end{itemize}
    \end{itemize}

    \item \textbf{Lazarus} (\strong)
    \begin{itemize}
        \item Set the output boundary of \progname{} at the appraisal outcome
        (i.e. action tendency, subjective experience, physiological response),
        would also allow for labelling the output with emotion
        categories~\citep[p.~209--210]{lazarus1991emotion}
        \begin{itemize}
            \item Excludes \ref{coping} process integral to the theory
            $\rightarrow$ could be added as an external component, supported by
            \textit{Allowing the Integration of New CME Components}
            (\ref{flexNew})

            \item Resulting NPC actions would impact their interpretation of
            the environment $\rightarrow$ implicitly supports the reappraisal
            process~\citep[p.~134]{lazarus1991emotion}

            \item Can feed the outputs back into \progname{} $\rightarrow$
            provide an interface so that this is a matter of ``flipping a
            switch'', supported by \textit{Customizing Existing CME Task
                Parameters} (\ref{flexCustom})

            \item [$\rightarrow$] Allows maximum flexibility for defining what
            to do with outputs that is agnostic to how ``action'' is defined
        \end{itemize}
    \end{itemize}

    \item \textbf{Scherer} (\strong)
    \begin{itemize}
        \item Set the output boundary of \progname{} at the action tendency
        level, would also allow for labelling the output with emotion
        categories~\citep[p.~107, 113]{scherer2001appraisalB} $\rightarrow$
        allows maximum flexibility for defining what to do with outputs that is
        agnostic to how ``action'' is defined

        \item Allow users to access individual appraisal
        registers~\citep[p.~104]{scherer2001appraisalB} $\rightarrow$ affords
        more flexibility
        \begin{itemize}
            \item Each part of the appraisal process makes changes to different
            subsystems, creates continuously changing
            outputs~\citep[p.~107]{scherer2001appraisalB} $\rightarrow$
            appraisal ``history" encoded in unique pattern caused by subsystem
            changes, values update frequently

            \item Potential to violate \textit{Hiding the Complexity of Emotion
                Generation} (\ref{easeHide}) $\rightarrow$ make available for
            advanced users

            \item Potential to violate psychological validity $\rightarrow$ how
            a user uses the values should be external to \progname{}, so its
            internal psychological validity would remain intact
        \end{itemize}
    \end{itemize}

    \item \textbf{Roseman} (\strong)
    \begin{itemize}
        \item Set the output boundary of \progname{} at the ``response
        strategy'', the typical physiological, phenomenological, expressive,
        behavioural, and motivational emotion contents
        (\citepg{roseman2013appraisal}{141}; \citepg{roseman2001model}{75};
        \citepg{roseman2018functions}{146--148, 151--152})
        \begin{itemize}
            \item Potential to connect components directly to existing game
            modules (e.g. ``expressive'' as an input to NPC animation,
            ``motivation'' to planning) $\rightarrow$ user chooses which ones
            to use, supported by \textit{Choosing Which CME Tasks to Use}
            (\ref{flexTasks})

            \item Differentiates between ``action tendency'' and what action is
            actually taken, informed by emotion
            intensity~\citep[p.~436]{roseman2011emotional} $\rightarrow$ input
            to the behaviour/expression selection process, explicitly built-in
            support for \textit{Allowing Developers to Specify How to Use CME
                Outputs} (\ref{flexOut})

            \item Also notes that there is consistency in what types of
            responses that each emotion elicits~\citep{roseman2011emotional}
            $\rightarrow$ creates consistent behaviour necessary for
            believability~\citep[p.~200]{ortony2002making}

            \item [$\rightarrow$] Allows maximum flexibility for defining what
            to do with outputs that is agnostic to the system at large
        \end{itemize}
    \end{itemize}

    \item \textbf{OCC} (\strong)
    \begin{itemize}
        \item Set the output boundary of \progname{} at the evaluation of
        emotion categories and intensities~\citep[p.~29, 72, 84]{occ2022}
        \begin{itemize}
            \item Propose that action tendencies are not necessary or
            sufficient to define emotion because some emotions might lack a
            ``characteristic" action tendency as defined as voluntary actions
            following emotion~\citep[p.~14]{occ2022} $\rightarrow$ allows the
            generation of emotions that lack one, removes constraint from user

            \item Claims ``action tendency'' is a set of components that
            emotions constrain themselves to but might not use all
            components~\citep[p.~198, 201]{ortony2002making} $\rightarrow$
            creates consistent behaviour necessary for
            believability~\citep[p.~200]{ortony2002making}

            \item [$\rightarrow$] Allows maximum flexibility for defining what
            to do with outputs that is agnostic to the system at large
        \end{itemize}
    \end{itemize}

    \item \textbf{Smith \& Kirby} (\strong)
    \begin{itemize}
        \item Set the output boundary of \progname{} at ``emotional response'',
        contains appraisal outcome (e.g. emotion category and dimensions,
        intensity), physiological activity (e.g. \ref{arousal}, facial
        expressions), and action tendencies~\citep[p.~123, 130]{smith2001toward}
        \begin{itemize}
            \item Users can separate components and send them to different
            system processes

            \item Can also send components back into appraisal processes via
            appraisal sources $\rightarrow$ potential for different
            interpretations due to processing in appraisal
            source~\citep[p.~99]{smith2000consequences}

            \item [$\rightarrow$] Allows maximum flexibility for defining what
            to do with outputs that is agnostic to the system at large
        \end{itemize}
    \end{itemize}

    \item \textbf{Oatley \& Johnson-Laird} (\strong)
    \begin{itemize}
        \item Set the output boundary of \progname{} at control signals,
        labelled with an emotion category~\citep[p.~50, 54]{oatley1992best}

        \item Control signals are global entities or ``alarms'' that change the
        system ``mode'' when a goal is impacted by an active plan, bringing it
        into focus~\citep[p.~62--63]{oatley1992best}
        \begin{itemize}
            \item Users can choose which system components are receptive to the
            signal and to what degree $\rightarrow$ supported by
            \textit{Choosing Which CME Tasks to Use} (\ref{flexTasks}) and
            \textit{Customizing Existing CME Task Parameters} (\ref{flexCustom})

            \item [$\rightarrow$] Allows maximum flexibility for defining what
            to do with outputs that is agnostic to the system at large
        \end{itemize}
    \end{itemize}
\end{itemize}

\subsection{Flexibility: Ability to Operate on Different Levels of NPC
    Complexity (\ref{flexComplex})}
Due to the assumed role of cognition in emotion processes
(\citepg{marsella2015appraisal}{55}; \citepg{broekens2021emotion}{354}), there
is also an assumed level of NPC complexity needed for an appraisal-based CME to
properly function. Therefore, the appraisal theories evaluation is concerned
with how much cognitive processing it requires to produce results.

\begin{itemize}
    \item \textbf{Frijda} (\weak)
    \begin{itemize}
        \item Requires encoded categories for events and rules for inputs,
        action structures for evaluating coping, process to evaluate
        event/action implications, definition of concern
        structures~\citep[p.~457]{frijda1986emotions} $\rightarrow$ effort to
        create likely to linearly increase with respect to game complexity
        \begin{itemize}
            \item Some can be implicitly evaluated in \progname{} $\rightarrow$
            relieves some of the authorial burden from game designers
        \end{itemize}

        \item Can add regulation processes as needed, not mandated in the core
        emotion process \citep[p.~454, 456]{frijda1986emotions} $\rightarrow$
        can potentially adapt to increases in NPC/game complexity
        \begin{itemize}
            \item Path in process to add planning if needed, but not critical to
            function~\citep[p.~462]{frijda1986emotions}
        \end{itemize}

        \item Requires a monitoring process (i.e. blackboard structure) to
        continuously update situational
        meaning~\citep[p.~459]{frijda1986emotions} $\rightarrow$ space
        requirements increases with information sources

    \end{itemize}

    \item \textbf{Lazarus} (\strong)
    \begin{itemize}
        \item Purpose of appraisal is to ``integrate the two [personal
        interests with environmental realities] as effectively as
        possible''~\citep[p.~135]{lazarus1991emotion} $\rightarrow$ complexity
        controlled by \progname{}, can be made relatively simple

        \item Complexity in individual factors and environmental condition
        evaluations that get passed to
        appraisal~\citep[p.~209--210]{lazarus1991emotion} $\rightarrow$
        designer controlled, can define to match game complexity
        \begin{itemize}
            \item Minimally requires definitions for goals, ``ego type", event
            causes, predictions about the impact of actions and future
            events~\citep[p.~149--150]{lazarus1991emotion}

            \item [$\rightarrow$] Could connect to hard-coded data/processes
            for low-complexity games (e.g. \textit{Pac-Man}~\citep{pacman}),
            increase complexity with game
        \end{itemize}

        \item Implication of a central data structure to store appraisal values
        as they become available~\citep[p.~134, 151, 189,
        210--211]{lazarus1991emotion}
        \begin{itemize}
            \item Stores outputs of appraisal evaluations, not the inputs

            \item [$\rightarrow$] Complexity likely to be constant or linearly
            increase with respect to the number and complexity of inputs
        \end{itemize}
    \end{itemize}

    \item \textbf{Scherer} (\strong)
    \begin{itemize}
        \item Built-in support for variable NPC complexity
        \begin{itemize}
            \item Appraisal dimension groupings (SECs) can be as complex as the
            information processing system allows, often a continuous or graded
            scalar or multidimensional
            evaluation~\citep[p.~94]{scherer2001appraisalB} $\rightarrow$
            numerical values easier to manipulate

            \item Assumes three levels of processing (sensory-motor, schematic,
            conceptual) that
            interact~\citep[p.~102--103]{scherer2001appraisalB} $\rightarrow$
            potential to derive some information in \progname{} implicitly from
            inputs
            \begin{itemize}
                \item Provide option to turn these tasks off or configure them
                $\rightarrow$ support for \textit{Choosing Which CME Tasks to
                    Use} (\ref{flexTasks}) and \textit{Customizing Existing
                    Task
                    Parameters} (\ref{flexCustom})
            \end{itemize}
        \end{itemize}

        \item Reappraisals run until a monitoring system signals termination or
        adjustment, appraisal components updated by
        reappraisals~\citep[p.~99]{scherer2001appraisalB} $\rightarrow$ game
        designer can decide when to terminate appraisal cycles based on game
        needs, support of \textit{Customizing Existing Task Parameters}
        (\ref{flexCustom}) and \textit{Be Efficient and Scalable}
        (\ref{flexScale})

        \item Appraisal registers updated as new information becomes available,
        central structure, can control relative importance of each
        value using a weighted function to represent ``goodness'' of
        data~\citep[p.~105]{scherer2001appraisalB};
        \begin{itemize}
            \item Implies a temporal and confidence value for each
            register~\citep[p.~106]{scherer2001appraisalB} $\rightarrow$ game
            designer can decide when to evaluate emotion state based on these
            values, support of \textit{Customizing Existing Task Parameters}
            (\ref{flexCustom})
        \end{itemize}
    \end{itemize}\clearpage

    \item \textbf{Roseman} (\good)
    \begin{itemize}
        \item Suggestion that there are different versions of appraisal
        mechanisms of variable complexity triggered as time
        allows~\citep[p.~77]{roseman2001model}, influenced by emotion
        intensity~\citep[p.~440]{roseman2011emotional}
        \begin{itemize}
            \item Can specify different appraisal mechanisms $\rightarrow$ game
            designers can build on top of input API, choose how information is
            synthesized into \progname{} inputs
        \end{itemize}

        \item Still requires empirical data to determine the minimum cognitive
        requirements for appraisals~\citep[p.~87--88]{roseman2001model}
        \begin{itemize}
            \item Lowest level involves fixed action
            patterns~\citep[p.~32]{clore2000cognition} $\rightarrow$
            correspondence with core \progname{} tasks, direct match between
            generated emotion and game designer-assigned behaviours

            \item With no defined process, do not know how to integrate
            cognitive processes into \progname{}
        \end{itemize}
    \end{itemize}

    \item \textbf{OCC} (\good)
    \begin{itemize}
        \item Has three levels of processing (reactive, routine,
        reflective)~\citep[p.~175--177, 179]{ortony2005affect}
        \begin{itemize}
            \item OCC proper part of the highest processing level,
            ``reflective'', does not interact with external environment
            $\rightarrow$ requires cognitive/high-level
            processes, inputs from other two levels

            \item Number of potential emotions restricted in reactive (no
            emotions) and routine (four emotions) levels $\rightarrow$ violates
            \textit{Choosing NPC Emotions} (\ref{flexEm})

            \item Assumptions make it unlikely to be applicable to
            architectures that are simpler than adult
            humans~\citep[p.~220]{sloman2005architectural}
        \end{itemize}

        \item Produce a simpler, less rigorous architecture with one processing
        level
        \begin{itemize}
            \item Complexity might lie in variable evaluations and
            representations of goals/standards/attitudes, coding of rules
            appears relatively simple~\citep[p.~220--227]{occ2022}

            \item Could allow for user-defined evaluations and representations
            $\rightarrow$ support as a Domain-Specific Language (DSL) to avoid
            violating \textit{Hiding the Complexity of Emotion Generation}
            (\ref{easeHide})
        \end{itemize}
    \end{itemize}

    \item \textbf{Smith \& Kirby} (\strong)
    \begin{itemize}
        \item Potential to design multiple appraisal mechanisms that rely
        on different functions (e.g. planning, expectation evaluation)
        $\rightarrow$ clear distinction between available functions and
        \progname{}'s abilities
        \begin{itemize}
            \item Number of appraisal variables determines types of emotion
            available (\citeg{yih2016distinct}; \citeg{yih2016patterns};
            \citepg{yih2020profiles}{488--492})

            \item [$\rightarrow$] Number of potential emotion categories tied
            to NPC complexity

            \item Potential to violate \textit{Choosing NPC Emotions}
            (\ref{flexEm}) $\rightarrow$ assuming that NPC complexity and what
            emotions the game designer wants them to have are directly
            proportional, this is unlikely to be a concern
        \end{itemize}
    \end{itemize}\clearpage

    \item \textbf{Oatley \& Johnson-Laird} (\good)
    \begin{itemize}
        \item Distinguishing emotion types changes based on cognitive
        abilities~\citep[p.~40--41]{oatley1987towards}, possible goal and plan
        representations~\citep[p.~57--58]{oatley1992best}
        \begin{itemize}
            \item Emotion ``modes'' as ``base classes'' of emotion

            \item [$\rightarrow$] Number of potential emotion categories tied
            to NPC complexity

            \item Potential to violate \textit{Choosing NPC Emotions}
            (\ref{flexEm}) $\rightarrow$ assuming that NPC complexity and what
            emotions the game designer wants them to have are directly
            proportional, this is unlikely to be a concern
        \end{itemize}
    \end{itemize}
\end{itemize}

\subsection{Flexibility: Be Efficient and Scalable (\ref{flexScale})}
The main concern for the efficiency and scalability of the appraisal theories
is how they evaluate inputs. However, this complexity seems to lie in input
creation. Since this precedes their transformation into emotions, action
tendencies, and other components, it passes a lot of this burden to the game
developer. Ideally, \progname{} would handle more of these tasks. However, this
also allows developers to choose how they want to generate inputs. Ultimately
this gives them more freedom and allows \progname{} to merge more easily into
different games and underlying architectures. What \progname{} focuses on,
then, is helping game developers manage different evaluation processes.

\begin{itemize}
    \item \textbf{Frijda} (\good)
    \begin{itemize}
        \item Uses 17--24 appraisal dimensions to create unique profiles for
        emotions, not all dimensions needed for each emotion
        (\citepg{frijda1986emotions}{205--219};
        \citepg{frijda1987emotion}{121--124})
        \begin{itemize}
            \item Efficiency might be hindered if unnecessary dimensions are
            evaluated

            \item Give developers choice of appraisal dimensions $\rightarrow$
            compromises \textit{Hiding the Complexity of Emotion Generation}
            (\ref{easeHide})
        \end{itemize}

        \item Scalability mostly driven by complexity of inputs and
        externally-defined regulation processes $\rightarrow$ depends on
        complexity of evaluations to generate inputs, designer-driven
    \end{itemize}

    \item \textbf{Lazarus} (\strong)
    \begin{itemize}
        \item Dependent on knowledge~\citep[p.~145]{lazarus1991emotion},
        knowledge evaluation mechanisms $\rightarrow$ depends on designer chosen
        architecture

        \item Appraisal process appears to be of a fixed complexity once inputs
        are given~\citep[p.~210]{lazarus1991emotion}, implies that scalability
        depends on complexity of knowledge processes and inputs $\rightarrow$
        depends on complexity of evaluations to generate inputs, designer-driven
    \end{itemize}

    \item \textbf{Scherer} (\good)
    \begin{itemize}
        \item Uses 16 appraisal dimensions divided into four
        groups~\citep[p.~114--115]{scherer2001appraisalB} $\rightarrow$ minimum
        set of appraisal dimensions necessary to differentiate emotion
        families~\citep[p.~94]{scherer2001appraisalB}

        \item Evaluates groups to avoid using unneeded expensive
        processes~\citep[p.~99--100, 102--103]{scherer2001appraisalB}
        \begin{itemize}
            \item Does not exclude potential to begin getting partial results by
            running the four components in parallel

            \item Groups can have more than one associated process
            $\rightarrow$ lower level processes for each group first, higher
            level processes only used if they do not return results

            \item Can add a central controller to allow integration of
            additional processes similar to Smith \& Kirby appraisal
            detector~\citep[p.~103--105]{scherer2001appraisalB}

            \item [$\rightarrow$] Efficiency tied to the complexity of inputs,
            partially designer-driven

            \item [$\rightarrow$] Mechanisms for scalability built-in, but
            requires more overhead to manage, might conflict with
            \textit{Ability to Operate on Different Levels of NPC Complexity}
            (\ref{flexComplex})
        \end{itemize}
    \end{itemize}

    \item \textbf{Roseman} (\weak)
    \begin{itemize}
        \item Suggests that there are different versions of appraisal
        mechanisms of variable complexity triggered as time
        allows~\citep[p.~77]{roseman2001model} $\rightarrow$ potential for
        scalability, efficiency

        \item Focuses on the structure of emotions and appraisal
        dimensions~\citep[p.~68, 81]{roseman2001model} $\rightarrow$ no further
        information given
    \end{itemize}

    \item \textbf{OCC} (\good)
    \begin{itemize}
        \item Uses 3--14 variables to differentiate emotion families, not all
        variables needed for each emotion~\citep[p.~29, 72, 84]{occ2022}
        $\rightarrow$ can prevent evaluation of some variables based on the
        active branch

        \item Efficiency tied to complexity of
        inputs~\citep[p.~220--227]{occ2022} $\rightarrow$ tied to evaluation
        mechanisms, designer-driven

        \item Can introduce mechanisms such that lower-complexity processes run
        first and call higher-complexity processes if they cannot produce a
        result~\citep[p.~179]{ortony2005affect} $\rightarrow$ more control of
        efficiency and scalability

        \item Can have two parallel emotion-elicitation mechanisms
        $\rightarrow$ can produce conflicting results~\citep[p.~37--39,
        54]{clore2000cognition}
        \begin{itemize}
            \item Memory-based heuristics, which is faster and more error-prone
            $\rightarrow$ improved efficiency, could run into memory-related
            scalability issues

            \item Deliberative processing, which is slower and less error-prone
            $\rightarrow$ reduced efficiency, more scale-friendly

            \item [$\rightarrow$] Could create a more believable result, but
            might conflict with \textit{Ability to Operate on Different Levels
                of NPC Complexity} (\ref{flexComplex})
        \end{itemize}
    \end{itemize}

    \item \textbf{Smith \& Kirby} (\strong)
    \begin{itemize}
        \item Uses 7--16 variables to create unique profiles for emotions, not
        all dimensions needed for each emotion (\citepg{yih2020profiles}{489};
        \citeg{yih2016distinct})
        \begin{itemize}
            \item Efficiency might be hindered if unnecessary dimensions are
            evaluated

            \item Give developers choice of appraisal dimensions $\rightarrow$
            compromises \textit{Hiding the Complexity of Emotion Generation}
            (\ref{easeHide})
        \end{itemize}

        \item A few variables could be evaluated by \progname{} (e.g.
        \textit{motivational relevance}) $\rightarrow$ efficiency and
        scalability controlled by \progname{}

        \item Some evaluation processes produce inputs \textit{for} \progname{}
        $\rightarrow$ designer-driven, architecture dependent
        \begin{itemize}
            \item Appraisal detector continuously monitors for changes in
            variables, combines information and called appraisal
            process~\citep[p.~129--130]{smith2001toward} $\rightarrow$
            implicitly enforces scalability as developers add and remove
            processes
            \begin{itemize}
                \item Detector must only require minimal resources to function
                well~\citep[p.~90--91]{smith2000consequences} $\rightarrow$
                acknowledges that efficiency is essential
            \end{itemize}

            \item Support for multiple, parallel user-defined processes that
            could have variable complexity
            levels~\citep[p.~91--92]{smith2000consequences} $\rightarrow$
            create a mechanism for developers to define when complex processes
            activate
        \end{itemize}
    \end{itemize}

    \item \textbf{Oatley \& Johnson-Laird} (\strong)
    \begin{itemize}
        \item Assumes a system that coordinates multiple plans and goals under
        time and resource constraints (e.g. plans only work 1--2 steps
        ahead)~\citep[p.~31, 36]{oatley1987towards} $\rightarrow$ property of
        the architecture, designer-driven

        \item Goals and plans associated with emotion has their own monitoring
        mechanisms~\citep[p.~50]{oatley1992best} $\rightarrow$ can decide which
        ones to associate with \progname{}, built-in scalability
        \begin{itemize}
            \item Mechanism would work on goal and plan information
            $\rightarrow$ can be made efficient
        \end{itemize}

        \item Emotions can be given more complex meanings by evaluating more
        information via selective function
        calls~\citep[p.~32--34]{oatley1987towards} $\rightarrow$
        designer-dependent and ties scalability, efficiency scaled to the needs
        of the game, supports \textit{Ability to Operate on Different Levels of
        NPC Complexity} (\ref{flexComplex})
    \end{itemize}
\end{itemize}

\subsection{Ease-of-Use: Hiding the Complexity of Emotion Generation
    (\ref{easeHide})}
The appraisal theories generally have strong support for this requirement.
Designers do not need to know what is done with the inputs they provide, so
the processing of those inputs can be hidden from them.

\begin{itemize}
    \item \textbf{Frijda} (\strong)
    \begin{itemize}
        \item System arranged so that information can be supplied at any point
        to black box processes, tracked by an internal monitor/situational
        meaning blackboard structure~\citep[p.~455--456,
        459]{frijda1986emotions} $\rightarrow$ do not need to know how the
        process uses information
    \end{itemize}

    \item \textbf{Lazarus} (\strong)
    \begin{itemize}
        \item Appraisal assigns personal meaning to
        knowledge~\citep[p.~145]{lazarus1991emotion} $\rightarrow$ do not need
        to know how the process uses information

        \item Six appraisal dimensions, reappraisal accounts for changes to
        input values~\citep[p.~134, 149--150]{lazarus1991emotion} $\rightarrow$
        can incorporate changing information as a queue of input values
    \end{itemize}

    \item \textbf{Scherer} (\strong)
    \begin{itemize}
        \item Inputs combined into
        registers~\citep[p.~105]{scherer2001appraisalB}, patterns of register
        values matched to emotions~\citep[p.~114--115]{scherer2001appraisalB}
        $\rightarrow$ do not need to know how this is done
        \begin{itemize}
            \item Some dimensions are pure information (e.g. \textit{intrinsic
                pleasantness})~\citep[p.~146]{lazarus1991emotion} $\rightarrow$
            inherently hides process complexity

            \item Appraisal dimension groupings (SECs) can be divided into
            hard-wired and deliberative
            units~\citep[p.~102]{scherer2001appraisalB} $\rightarrow$ do not
            need to know which ones are deliberative or not
        \end{itemize}

        \item Changes in SECs can cause continuously changing
        outputs~\citep[p.~107]{scherer2001appraisalB} $\rightarrow$ need only
        query if there has been a change, do not need to know if the generation
        process was triggered
    \end{itemize}

    \item \textbf{Roseman} (\good)
    \begin{itemize}
        \item Inputs pattern-matched to emotion families~\citep[p.~70--71,
        81]{roseman2001model} $\rightarrow$ do not need to know what the
        patterns are

        \item Focus on the relationship between appraisal values and
        emotions~\citep[p.~81]{roseman2001model} $\rightarrow$ does not focus
        on other parts of the generation process
    \end{itemize}

    \item \textbf{OCC} (\good)
    \begin{itemize}
        \item Inputs pattern-matched to emotions, combined into intensity
        values and compared to threshold rules~\citep[p.~84, 227--228]{occ2022}
        $\rightarrow$ do not need to know how this is done

        \item Need to know patterns of variables to support \textit{Choosing
            NPC Emotions} (\ref{flexEm}) $\rightarrow$ potential to expose
            variable
        patterns, but not how the variables are combined or compared to
        threshold rules
    \end{itemize}

    \item \textbf{Smith \& Kirby} (\good)
    \begin{itemize}
        \item Inputs combined into single unit by appraisal
        register, triggers generation process~\citep[p.~130]{smith2001toward}
        $\rightarrow$ do not need to know how this is done

        \item Need to know patterns of variables to support \textit{Choosing
            NPC Emotions} (\ref{flexEm}) (\citepg{yih2020profiles}{489};
        \citeg{yih2016distinct}) $\rightarrow$ potential to expose variable
        patterns, but not how they are used
    \end{itemize}

    \item \textbf{Oatley \& Johnson-Laird} (\good)
    \begin{itemize}
        \item Emotions as products of interpretations of goals and
        plans~\citep[p.~30]{oatley1987towards} $\rightarrow$ do not need to
        know how they are interpreted

        \item Need to provide additional information to support
        \textit{Choosing NPC Emotions} (\ref{flexEm})
        \begin{itemize}
            \item Add contextual information to generated
            emotion~\citep[p.~76--78]{oatley1992best} $\rightarrow$ does not
            require knowledge of how the emotion was produced

            \item Tied to ``folk'' understanding of emotions, their
            consequences, and antecedents \citep[p.~214--215]{johnson1992basic}
            $\rightarrow$ minimizes potential to violate this requirement
        \end{itemize}
    \end{itemize}
\end{itemize}

\clearpage\subsection{Ease-of-Use: Having a Clear API (Input) (\ref{easeAPI})}
It is not enough for an appraisal theory to be clear in what it requires for
appraisal. It must also be clear in what it does with those inputs to produce
an unambiguous output. Therefore, the theories are analyzed for both their
necessary inputs, how those could be realized as an input interface, and how
those inputs map to appraisal outputs.

\begin{itemize}
    \item \textbf{Frijda} (\good)
    \begin{itemize}
        \item Minimally requires definition of concerns (which include goal
        definitions), environment states/events, action (tendency)
        structures~\citep[p.~454, 457]{frijda1986emotions} $\rightarrow$
        generally do not know how to define
        inputs~\citep[p.~86--87]{roseman2001model}

        \item At least 20 variables
        listed~\citep[p.~205--216]{frijda1986emotions}, smaller list of 14
        variables have preliminary empirical
        validation~\citep[p.~128--131]{frijda1987emotion} $\rightarrow$
        potential to overwhelm users with the full list
        \begin{itemize}
            \item Some variables describe knowledge (e.g.
            \textit{valence}~\citep[p.~207]{frijda1986emotions}) $\rightarrow$
            cannot remove from input variable list

            \item Some variables could be derived from knowledge (e.g.
            \textit{change} derived from previous and current state, implicit
            in definition of ``event''~\citep[p.~209--210]{frijda1986emotions})
            $\rightarrow$ exchange variables for knowledge in input list
            \begin{itemize}
                \item Might be able to reduce the required input list if there
                are overlaps in required knowledge for many variables

                \item Replace variable names with knowledge that is generally
                understood (e.g. states, goals) $\rightarrow$ supports
                \textit{Hiding the Complexity of Emotion Generation}
                (\ref{easeHide})
            \end{itemize}

            \item Some variables might be encoded implicitly in others (e.g.
            \textit{presence/absence}, \textit{urgency}~\citep[p.~208--209,
            455]{frijda1986emotions}) $\rightarrow$ do not have to be exposed
            to user, reduce required input list in API

            \item Unique profiles for some
            emotions~\citep[p.~217--219]{frijda1986emotions}, empirical
            validation of some patterns~\citep[p.~122--123]{frijda1987emotion}
            showing that each emotion uses a subset of variables $\rightarrow$
            might be able to define subsets of variables if some emotions are
            not needed

            \item [$\rightarrow$] Options error-prone, require careful design
            of \progname{}

            \item Do not know how to define some of these
            inputs~\citep[p.~86--87]{roseman2001model}
        \end{itemize}
    \end{itemize}

    \item \textbf{Lazarus} (\disqualified)
    \begin{itemize}
        \item Minimally requires local and global ``ego-identity'' goals,
        causal agents and their control over an event, coping potential,
        predictions about future prospects~\citep[p.~102,
        149--150]{lazarus1991emotion} $\rightarrow$ generally do not know how
        to define inputs~\citep[p.~86--87]{roseman2001model}
        \begin{itemize}
            \item Correlated with five appraisal dimensions

            \item Written in natural/familiar language, conceptualized as
            entities and values $\rightarrow$ supports \textit{Hiding the
                Complexity of Emotion Generation} (\ref{easeHide})
        \end{itemize}

        \item Appraisal patterns are not clearly unique (e.g. \textit{Anxiety}
        and \textit{Disgust} only differ in ego-involvement, but what is the
        difference between ``protection against existential threats'' and
        ``being at risk of a poisonous idea''?~\citep[p.~237,
        261]{lazarus1991emotion}) $\rightarrow$ prone to assumption biases,
        threatening psychological validity
    \end{itemize}

    \item \textbf{Scherer} (\good)
    \begin{itemize}
        \item Minimally requires goals, events and their properties (e.g.
        \textit{predictability}, \textit{intrinsic pleasantness}), causality,
        predictions about events, time constraints on goals, predictions about
        the controllability over potential outcomes, ability to influence
        and/or adapt to potential outcomes, and information about the agent's
        conception of self-ideal and social norms

        \item 15 appraisal variables divided into four groups internally for
        organization and flow \citep[p.~94]{scherer2001appraisalB}
        $\rightarrow$ potential to overwhelm users with the full list
        \begin{itemize}
            \item Some variables describe knowledge (e.g. \textit{intrinsic
            pleasantness}) \citep[p.~95]{scherer2001appraisalB} $\rightarrow$
            cannot remove from input variable list

            \item Some variables could be derived from knowledge (e.g.
            \textit{discrepancy from expectation} derived from current state
            prediction about current state from previous ones
            \citep[p.~96]{scherer2001appraisalB}) $\rightarrow$ exchange
            variables for knowledge in input list
            \begin{itemize}
                \item Might be able to reduce the required input list if there
                are many variables require the same knowledge

                \item Replace variable names with knowledge that is generally
                understood (e.g. probabilities, goals) $\rightarrow$ supports
                \textit{Hiding the Complexity of Emotion Generation}
                (\ref{easeHide})
            \end{itemize}

            \item Unique profiles for some emotions that have some empirical
            support~\citep[p.~114--117]{scherer2001appraisalB} showing that
            each emotion uses a subset of variables $\rightarrow$ might be able
            to define subsets of variables if some emotions are not needed

            \item [$\rightarrow$] Options error-prone, require careful design
            of \progname{}

            \item Do not know how to define some of these
            inputs~\citep[p.~86--87]{roseman2001model}
        \end{itemize}
    \end{itemize}

    \item \textbf{Roseman} (\good)
    \begin{itemize}
        \item Minimally requires goals, current environment states, causal
        agents, predictions and confidence values about future events, coping
        potential, if a problem is intrinsic or instrumental

        \item Seven appraisal variables create 17 emotion
        categories, accounts for all variable
        combinations~\citep[p.~68--69]{roseman2001model} $\rightarrow$
        empirically validated and compared with dimensions from other appraisal
        theories~\citep[p.~256, 260,267]{roseman1996appraisal}, revised as new
        data is collected~\citep[p.~72, 75]{roseman2001model}

        \item Generally do not know how to define these
        inputs~\citep[p.~86--87]{roseman2001model} $\rightarrow$ hypothesize
        that appraisal variables influence each other, some might be inputs to
        the evaluations of others~\citep[p.~271]{roseman1996appraisal}
    \end{itemize}

    \item \textbf{OCC} (\weak)
    \begin{itemize}
        \item Clear distinctions between events, agents, and
        objects~\citep[p.~69--70]{occ2022} $\rightarrow$ minimally requires
        goals, changes to goals, agents, standards, and preferences

        \item At least four global variables and ten local variables to
        distinguish emotion families~\citep[p.~98]{occ2022} $\rightarrow$
        potential to overwhelm users with the full list
        \begin{itemize}
            \item Reduce the list by using some variables to calculate others
            (e.g. \textit{arousal} is evaluated from other
            variables~\citep[p.~83]{occ2022}) $\rightarrow$ applies to few
            variables, might violate \textit{Ability to Operate on Different
                Levels of NPC Complexity} (\ref{flexComplex})

            \item Reduce the list by ignoring some variables $\rightarrow$
            unclear how to choose which or how many variables to keep

            \item Limit the variables to the six variables that distinguish
            events, agent, and object-related emotions $\rightarrow$ unable to
            differentiate emotions in the same branch

            \item Generally do not know how to define these
            inputs~\citep[p.~86--87]{roseman2001model}
        \end{itemize}
    \end{itemize}

    \item \textbf{Smith \& Kirby} (\good)
    \begin{itemize}
        \item Minimally requires goals, environment states/events, agent
        actions and their confidence in their efficacy, agent dispositional
        traits, event causality, agent responsibility, and predictions about
        the desirability of future environment states
        \begin{itemize}
            \item Inputs required for three of seven appraisal dimensions
            empirically tested~\citep[p.~1357, 1361--1362,
            1367]{smith2009putting}

            \item A fourth variable might be dependent on one of the three
            tested variables~\citep[p.~138]{smith2001toward} $\rightarrow$
            observed tendency, not examined directly

            \item Other dimensions have yet to be
            tested~\citep[p.~1369]{smith2009putting}
        \end{itemize}

        \item Sixteen appraisal variables (seven ``core''
        variables~\citep[p.~123]{smith2001toward}, approximately nine
        additional ones derived from empirical data
        (\citepg{yih2020profiles}{489}; \citeg{yih2016distinct})) make unique
        patterns for 20 emotions $\rightarrow$ not all patterns accounted for,
        potential to overwhelm users with the full list
        \begin{itemize}
            \item Some variables appear to describe knowledge (e.g.
            \textit{likeability}~\citep{yih2016distinct}) $\rightarrow$ cannot
            remove from input variable list

            \item Some variables derived from knowledge (e.g.
            \textit{motivational relevance} derived from goals and environment
            states~\citep[p.~1361]{smith2009putting}) $\rightarrow$ exchange
            variables for knowledge in input list
            \begin{itemize}
                \item Might be able to reduce the required input list if there
                are many variables require the same knowledge

                \item Replace variable names with knowledge that is generally
                understood (e.g. probabilities, goals) $\rightarrow$ supports
                \textit{Hiding the Complexity of Emotion Generation}
                (\ref{easeHide})
            \end{itemize}

            \item Emotion appraisal profiles show that each emotion uses a
            subset of variables (\citepg{yih2020profiles}{489};
            \citeg{yih2016distinct}) $\rightarrow$ might be able to define
            subsets of variables if some emotions are not needed

            \item [$\rightarrow$] Options error-prone, require careful design
            of \progname{}
        \end{itemize}
    \end{itemize}

    \item \textbf{Oatley \& Johnson-Laird} (\strong)
    \begin{itemize}
        \item Emotions elicited at plan junctions where probabilities of goal
        success changes~\citep[p.~98]{oatley1992best} $\rightarrow$ minimally
        requires information about current state of plans and goals

        \item Goals as symbolic representations of environments states, plans
        as transformations betw-een the current environment state and a
        goal~\citep[p.~30]{oatley1987towards}
        \begin{itemize}
            \item Goals $\rightarrow$ active/dormant, achievement status, type
            (e.g. self-preservation, gustatory), known conflicts with other
            goals

            \item Plans $\rightarrow$ active/dormant, priority

            \item [$\rightarrow$] Little to no additional information required
        \end{itemize}
    \end{itemize}
\end{itemize}

\clearpage\subsection{Ease-of-Use: Having a Clear API (Output) (\ref{easeAPI})}
Many appraisal theories output emotion as a series of components rather than
explicit categories. This has the potential to clutter the output API, and
potentially violating requirements like \textit{Hiding the Complexity of
    Emotion Generation} (\ref{easeHide}) and \textit{Traceable CME Outputs}
(\ref{easeTrace}). However, the componential outputs lend themselves well to
supporting \textit{Allowing Developers to Specify How to Use CME Outputs}
(\ref{flexOut}) as this would allow users to only use the parts relevant to
their design. Fortunately, most of the examined theories relate output
component groups with a named emotion. This allows \progname{} to create output
``packages'' labelled with an emotion category and intensity that can be
unpacked by advanced users. This creates a small output API while supporting
\textit{Hiding the Complexity of Emotion Generation} (\ref{easeHide}),
\textit{Traceable CME Outputs} (\ref{easeTrace}), and \textit{Allowing
    Developers to Specify How to Use CME Outputs} (\ref{flexOut}).
\begin{itemize}
    \item \textbf{Frijda} (\good)
    \begin{itemize}
        \item Outputs three data ``packages'': an \ref{arousal} value,
        relevance and control precedence signals, and action
        tendencies~\citep[p.~454--455]{frijda1986emotions} $\rightarrow$
        provides ``how strong'', ``what priority'', and an abstracted ``how to
        behave''

        \item [$\rightarrow$] Supports \textit{Hiding the Complexity of Emotion
            Generation} (\ref{easeHide})

        \item \ref{arousal} and action tendency ``packages'' might not be
        immediately understandable
        \begin{itemize}
            \item Can associate action tendencies with emotion
            words~\citep[p.~72]{frijda1986emotions} to improve understandability

            \item Give \ref{arousal} a more recognizable name and/or made into
            an optional or hidden output
        \end{itemize}
    \end{itemize}

    \item \textbf{Lazarus} (\strong)
    \begin{itemize}
        \item Outputs one data ``package'' tagged with an emotion label and
        core theme, containing two ``sub-packages'': physiological response and
        action tendencies~\citep[p.~209--210]{lazarus1991emotion}
        \begin{itemize}
            \item Also includes subjective experience components, which
            requires reasoning about the emotion process $\rightarrow$ skip in
            design

            \item Do not have to manage ``sub-packages'' directly $\rightarrow$
            supports \textit{Allowing Developers to Specify How to Use CME
                Outputs} (\ref{flexOut}) and \textit{Hiding the Complexity of
                Emotion Generation} (\ref{easeHide})
        \end{itemize}
    \end{itemize}

    \item \textbf{Scherer} (\weak)
    \begin{itemize}
        \item No direct output $\rightarrow$ emotions emergent, related to a
        series of subsystem changes~\citep[p.~113]{scherer2001appraisalB}
        \begin{itemize}
            \item Might not be immediately clear what the information means or
            what can be done with it $\rightarrow$ violates
            \textit{Traceable CME Outputs} (\ref{easeTrace}), potential
            violation of \textit{Hiding the Complexity of Emotion Generation}
            (\ref{easeHide})
        \end{itemize}

        \item Could output an emotion term and associated action tendency
        when the changes match a known
        pattern~\citep[p.~117]{scherer2001appraisalB}
        \begin{itemize}
            \item Support \textit{Hiding the Complexity of Emotion Generation}
            (\ref{easeHide}) and \textit{Traceable CME Outputs}
            (\ref{easeTrace})

            \item Advanced users could examine the changes directly
            $\rightarrow$ supports \textit{Allowing Developers to Specify
                How to Use CME Outputs} (\ref{flexOut})

            \item Potential to have subsystem changes that do not match any
            known patterns $\rightarrow$ \progname{} might appear to be
            non-functional
        \end{itemize}
    \end{itemize}

    \item \textbf{Roseman} (\strong)
    \begin{itemize}
        \item Outputs a ``package'' labelled with an emotion category
        containing ``sub-packages'' for typical physiological,
        phenomenological, expressive, behavioural, and motivational
        contents~\citep[p.~75]{roseman2001model}
        \begin{itemize}
            \item ``Package'' contents represents coordinating systems for a
            coping strategy~\citep[p.~141]{roseman2013appraisal}

            \item Designed to impose categorical distinctions on continuous
            appraisal dimensions \citep[p.~75, 80]{roseman2001model}, might
            explain why laypeople see emotions as discrete
            entities~\citep[p.~147]{roseman2013appraisal} $\rightarrow$ achieves
            understandability of discrete theories, supports \textit{Hiding the
                Complexity of Emotion Generation} (\ref{easeHide})

            \item Differentiates reward seeking and punishment
            avoidance~\citep[p.~910]{roseman1990appraisals} $\rightarrow$
            easier to distinguish emotion states

            \item Do not have to manage ``sub-packages'' directly $\rightarrow$
            supports \textit{Hiding the Complexity of Emotion Generation}
            (\ref{easeHide}) and \textit{Allowing Developers to Specify How to
                Use CME Outputs} (\ref{flexOut})
        \end{itemize}
    \end{itemize}

    \item \textbf{OCC} (\good)
    \begin{itemize}
        \item Outputs two fields: emotion category and
        intensity~\citep[p.~220--2212]{occ2022} $\rightarrow$ simple
        presentation of ``what'' and ``how strong'', achieves understandability
        of discrete theories

        \item Supports \textit{Hiding the Complexity of Emotion Generation}
        (\ref{easeHide})
    \end{itemize}

    \item \textbf{Smith \& Kirby} (\strong)
    \begin{itemize}
        \item Outputs a ``package'' labelled with an emotion category, core
        relational theme, and intensity, wrapped around appraisal dimension
        values~\citep[p.~123, 125]{smith2001toward}
        \begin{itemize}
            \item Do not have to manage appraisal dimensions directly
            $\rightarrow$ supports \textit{Hiding the Complexity of Emotion
                Generation} (\ref{easeHide}) and \textit{Allowing Developers to
                Specify How to Use CME Outputs} (\ref{flexOut})

            \item Include a ``sub-packages'' of facial muscle movements using
            FACS and physiological
            changes~\citep[p.~133--135]{smith2001toward}, motivational goals
            and coping strategies (\citeg{yih2016distinct};
            \citeg{yih2016patterns}, \citepg{yih2020profiles}{488--492})
            derived from appraisal values $\rightarrow$ additional suggestions
            for advanced use cases
        \end{itemize}
    \end{itemize}

    \item \textbf{Oatley \& Johnson-Laird} (\strong)
    \begin{itemize}
        \item Basic emotion ``modes'' are psychologically and physiologically
        distinct~\citep[p.~48]{oatley1987towards}, assigned labels are not
        strict definitions~\citep[p.~217]{johnson1992basic} and rely on
        intuitions about emotions from experience and
        language~\citep[p.~69--71, 74--75, 82, 86--87]{oatley1992best}
        $\rightarrow$ generally understandable

        \item Outputs an emotion ``signal'' rather than a concrete
        state~\citep[p.~214]{johnson1992basic} $\rightarrow$ affords
        flexibility for both state-based and stateless emotion definitions
        \begin{itemize}
            \item Potential to connect both definitions to expressive and
            action-based mechanisms \citep[p.~31]{oatley1987towards}

            \item As signals, only have a small number
            (five)~\citep[p.~33]{oatley1987towards} $\rightarrow$ requires a
            small signal recognition process to pick up signal emissions
        \end{itemize}

        \item Supports \textit{Hiding the Complexity of Emotion Generation}
        (\ref{easeHide}) and \textit{Allowing Developers to Specify How to Use
            CME Outputs} (\ref{flexOut})
    \end{itemize}
\end{itemize}

\subsection{Ease-of-Use: Traceable CME Outputs (\ref{easeTrace})}
Generally, the process view afforded by appraisal theories also allows for
traceability between inputs and outputs, supporting testing and debugging. The
appraisal theories also produce, or connect their outputs to, emotion
categories which supports \textit{Hiding the Complexity of Emotion Generation}
(\ref{easeHide}) and \textit{Having a Clear API (Output)} (\ref{easeAPI}).

\begin{itemize}
    \item \textbf{Frijda} (\strong)
    \begin{itemize}
        \item ``Blackboard'' structure as central information hub, retains
        history of evaluation~\citep[p.~459]{frijda1986emotions} $\rightarrow$
        built-in traceability tool
        \begin{itemize}
            \item Potential to violate \textit{Hiding the Complexity of Emotion
                Generation} (\ref{easeHide}) $\rightarrow$ could be designed to
            avoid psychology jargon, use everyday language
        \end{itemize}
    \end{itemize}

    \item \textbf{Lazarus} (\strong)
    \begin{itemize}
        \item Appraisal patterns shown as a decision
        tree~\citep[p.~222]{lazarus1991emotion} $\rightarrow$ visualization of
        appraisal showing where inputs are used
        \begin{itemize}
            \item Six appraisal dimensions $\rightarrow$ tree is a manageable
            size

            \item Potential to violate \textit{Hiding the Complexity of Emotion
                Generation} (\ref{easeHide}) $\rightarrow$ could design to use
                everyday language and avoid psychology jargon
        \end{itemize}

        \item Appraisals are not sequential~\citep[p.~151]{lazarus1991emotion},
        has multiple mechanisms~\citep[p.~189]{lazarus1991emotion}, is
        transactional and temporal in
        nature~\citep[p.~210--211]{lazarus1991emotion}, reappraisal as a key
        process~\citep[p.~134]{lazarus1991emotion} $\rightarrow$ implies a
        central data structure, like a blackboard, to store information as it
        becomes available
        \begin{itemize}
            \item Potential to violate \textit{Hiding the Complexity of Emotion
                Generation} (\ref{easeHide}) $\rightarrow$ could design to use
                everyday language and avoid psychology jargon
        \end{itemize}
    \end{itemize}

    \item \textbf{Scherer} (\weak)
    \begin{itemize}
        \item Suggested system representation is a neural
        network~\citep[p.~105]{scherer2001appraisalB} $\rightarrow$ might not
        produce traceable/explainable results

        \item Each appraisal unit (SEC) changes different subsystems
        $\rightarrow$ creates a continuously changing outputs, ``history'' of
        the appraisal~\citep[p.~107]{scherer2001appraisalB} that is helpful for
        debugging
        \begin{itemize}
            \item Many possible combinations $\rightarrow$ difficult to have
            consistent outputs for testing
        \end{itemize}
    \end{itemize}\clearpage

    \item \textbf{Roseman} (\strong)
    \begin{itemize}
        \item Appraisal as a selection mechanism in a system~\citep[p.~76,
        81--83]{roseman2001model} $\rightarrow$ requires decision rules,
        potential to provide trace as a decision tree
        \begin{itemize}
            \item Seven appraisal dimensions $\rightarrow$ tree is a manageable
            size

            \item Potential to violate \textit{Hiding the Complexity of Emotion
                Generation} (\ref{easeHide}) $\rightarrow$ could design to use
                everyday language and avoid psychology jargon
        \end{itemize}
    \end{itemize}

    \item \textbf{OCC} (\strong)
    \begin{itemize}
        \item Eliciting conditions written to reflect how they are talked
        about in everyday language, how people tend to experience
        them~\citep[p.~25--26]{clore2000cognition} $\rightarrow$ connection
        between inputs and outputs (emotion category/intensity pair) generally
        understandable
        \begin{itemize}
            \item Could include a trace function showing effects of input
            variables and values, changes in expression thresholds
            $\rightarrow$ make as an advanced system function to maintain
            \textit{Hiding the Complexity of Emotion Generation}
            (\ref{easeHide})
        \end{itemize}
    \end{itemize}

    \item \textbf{Smith \& Kirby} (\strong)
    \begin{itemize}
        \item Two potential points for tracing~\citep[p.~130]{smith2001toward}
        \begin{itemize}
            \item Appraisal detector output where it merges information into
            one unit before appraisal $\rightarrow$ show trace of how
            information is combined
            \begin{itemize}
                \item Converting disparate inputs that developers know of into
                aggregated values $\rightarrow$ unlikely to violate
                \textit{Hiding the Complexity of Emotion Generation}
                (\ref{easeHide})

                \item Would require a customizable detector to support
                \progname{}'s flexibility $\rightarrow$ supported by
                \textit{Customizing Existing CME Task Parameters}
                (\ref{flexCustom})
            \end{itemize}

            \item After converting appraisal unit into an emotion $\rightarrow$
            can show trace of appraisal pattern as a decision tree
            \begin{itemize}
                \item Sixteen appraisal dimensions
                (\citepg{yih2020profiles}{489}; \citeg{yih2016distinct})
                $\rightarrow$ unmanageable tree size

                \item Could reduce tree size by removing dimensions that are
                not relevant to the elicited emotion $\rightarrow$ provides
                information about unused variables

                \item Potential to violate \textit{Hiding the Complexity of
                    Emotion Generation} (\ref{easeHide}) $\rightarrow$ could be
                designed to avoid psychology jargon, use everyday language
            \end{itemize}
        \end{itemize}
    \end{itemize}

    \item \textbf{Oatley \& Johnson-Laird} (\strong)
    \begin{itemize}
        \item Connection of emotion ``mode'' triggers with plans and
        goals~\citep[p.~36]{oatley1987towards}
        \begin{itemize}
            \item Create a trace function that presents goal and plan
            information used to produce outputs

            \item Leverage computational knowledge
        \end{itemize}

        \item Adding cognitive meanings to an emotion ``mode'' could explain
        language and cultural differences~\citep[p.~218]{johnson1992basic}
        $\rightarrow$ can trace the impact of cognitive processes on the
        elicitation of new emotions
    \end{itemize}
\end{itemize}

\subsection{Ease-of-Use: Providing Examples of Novel Game Experiences
    (\ref{easeNovel})}
Each theory presents different ways to create novel game mechanics and
interactions due to their differing focuses. The role of the self and culture
is commonly discussed, suggesting new ways that NPCs can develop and interact
with players. However, in all cases, realizing these requires additional
components or relies on a particular type of NPC functionality. Consequently,
all of the theories are only good (\good) candidates for satisfying this
requirement.

\begin{itemize}
    \item \textbf{Frijda} (\good)
    \begin{itemize}
        \item Incorporate additional components as needed to make finer
        distinctions between emotion states~\citep[p.~216]{frijda1986emotions}

        \item [$\rightarrow$] Potential for social and cultural based mechanics
    \end{itemize}

    \item \textbf{Lazarus} (\good)
    \begin{itemize}
        \item Coping as part of variable NPC
        responses~\citep[p.~112--115]{lazarus1991emotion} $\rightarrow$ part of
        action generation, outside the scope of \progname{}

        \item Role of biological and social variables in the development of the
        emotion process~\citep[p.~39]{lazarus1991emotion} $\rightarrow$ varying
        \progname{} parameters based on NPC ``age'' to alter processing

        \item Core relation themes and the stages of the emotion
        process~\citep[p.~106, 121]{lazarus1991emotion} $\rightarrow$ potential
        to relate to narrative drama
    \end{itemize}

    \item \textbf{Scherer} (\good)
    \begin{itemize}
        \item Normative significance evaluation accounts for the role of
        self-esteem/self-concept, social, and cultural influences in emotion
        generation~\citep[p.~98]{scherer2001appraisalB}

        \item [$\rightarrow$] Potential for social and cultural based mechanics
    \end{itemize}

    \item \textbf{Roseman} (\good)
    \begin{itemize}
        \item Ties response strategies to appraisal
        dimensions~\citep[p.~144]{roseman2013appraisal} $\rightarrow$ emotions
        differentiated by strategy (\citepg{roseman2013appraisal}{148};
        \citepg{roseman2001model}{76})
        \begin{itemize}
            \item Subjective interpretation of emotion state $\rightarrow$
            closer to emotion recognition in real life

            \item Requires knowledge about dimensions $\rightarrow$ violates
            \textit{Hiding the Complexity of Emotion Generation} requirement
            (\ref{easeHide})

            \item [$\rightarrow$] Unideal, available for advanced users
        \end{itemize}

        \item Proposed connected between different bases of
        racism~\citep[p.~84--85]{roseman2001model} $\rightarrow$ potential for
        a social intervention mechanics
    \end{itemize}

    \item \textbf{OCC} (\good)
    \begin{itemize}
        \item Implement NPC curiosity when no other emotion processes are
        active (i.e. resting state)~\citep[p.~194]{ortony2005affect}
        $\rightarrow$ enhance believability
        \begin{itemize}
            \item Make the state slightly positive instead of zero $\rightarrow$
            engages in exploratory behaviour

            \item Requires expectations to know when things are different and
            might warrant action
        \end{itemize}

        \item Define mood as repeated emotion
        elicitations~\citep[p.~228]{occ2022} $\rightarrow$ dynamic and
        potentially player-unique game sessions
    \end{itemize}

    \item \textbf{Smith \& Kirby} (\good)
    \begin{itemize}
        \item Proposed connection between appraisal dimensions and individual
        facial movements, physiological activities~\citep[p.~237--240,
        242--243]{smith_scott_mandler_1997}
        \begin{itemize}
            \item [$\rightarrow$] Connection to non-human (e.g. alien, animal)
            and non-humanoid (e.g. computer, city) representations for emotion
            expression
        \end{itemize}
    \end{itemize}

    \item \textbf{Oatley \& Johnson-Laird} (\good)
    \begin{itemize}
        \item English names for basic emotions imply behaviours of social
        creatures~\citep[p.~209]{johnson1992basic} $\rightarrow$ social role of
        emotions

        \item Group dynamics using communication between agents at plan
        junctions via emotion, modelling ``contagious'' emotions, mutual plans
        between NPCs~\citep[p.~31, 40--44]{oatley1987towards} $\rightarrow$
        requires a model of the self (i.e. NPC models itself)
        \begin{itemize}
            \item Potential to model computationally as a system $\rightarrow$
            recursively defined, informed by language and culture

            \item [$\rightarrow$] Maintenance and propagation of cultural
            values and norms via learning and guidance

            \item [$\rightarrow$] Learning new ways to handle plan junctions
            can lead to evolving abilities to represent individual and
            cultural differences
        \end{itemize}

        \item Sentiments $\rightarrow$ model influence of social emotions on
        NPC relationships~\citep[p.~78, 80--86]{oatley2000sentiments}
        \begin{itemize}
            \item Connection to social goals: affiliation, protection,
            dominance $\rightarrow$ potential to overlay onto PAD Space?
        \end{itemize}

        \item Narrative planning~\citep[p.~6--7, 107--108, 225]{oatley1992best}
        \begin{itemize}
            \item Mental simulation of personal plans to understand
            potential reactions in advance, simulation of others' plans to
            understand their emotions $\rightarrow$ directly tied to narratives
            and can account to responses to stories and films

            \item [$\rightarrow$] Integration into narrative planning to elicit
            specific emotions from different NPCs
        \end{itemize}

        \item Reliance on plans $\rightarrow$ partial violation of
        \textit{Independence From an Agent Architecture} requirement
        (\ref{flexArch})
    \end{itemize}
\end{itemize}

\subsection{Examining the Remaining Requirements}
The appraisal theories are sufficiently different and satisfy most of the
requirements in different ways. However, they are also relatively sparse for
certain aspects of the emotion process (e.g. emotion decay, intensity) and have
comparable information for two ease-of-use high-level requirements.

\begin{itemize}

    \item \textit{Ease-Of-Use: Allowing the Automatic Storage and Decay of the
        Emotion State (\ref{easeAuto})}
    \begin{itemize}
        \item Frijda, Lazarus, Scherer, OCC, Smith \& Kirby, Oatley \&
        Johnson-Laird (\good)
        \begin{itemize}
            \item No explicit description about how to store or decay emotion
            states

            \item Time-dependent process approach
            (\citepg{frijda1986emotions}{453}; \citepg{lazarus1991emotion}{39,
            209}; \citepg{occ2022}{220--228}; \citepg{oatley1992best}{22--23};
            \citepg{smith2000consequences}{85}) $\rightarrow$ could support its
            design

            \item Could be integrated via existing processes like
            reappraisal~\citep[p.~99]{scherer2001appraisalB} and
            monitoring~\citep[p.~129--130]{smith2001toward}

            \item Supports \textit{Customizing Existing CME Task Parameters}
            (\ref{flexCustom}) $\rightarrow$ include parameters to turn
            automation off, change entire decay functions or only some of their
            variables, allow for multiple context-dependent decay functions
        \end{itemize}

        \item Roseman (\disqualified)
        \begin{itemize}
            \item Focus on the relationship between appraisal values and
            emotions, how those emotions impact different systems in response,
            and the structure of emotions~\citep[p.~68, 81]{roseman2001model},
            empirical validation of appraisal dimension influence on resulting
            emotion~\citep[p.~242, 244]{roseman1996appraisal} $\rightarrow$
            does not touch on the emotion process itself
        \end{itemize}
    \end{itemize}

    \item \textit{Ease-Of-Use: Showing that Emotions Improve the Player
        Experience (\ref{easePX})}
    \begin{itemize}
        \item Scherer (\strong)
        \begin{itemize}
            \item Associates each emotion with action
            tendencies~\citep[p.~108]{scherer2001appraisalB} $\rightarrow$
            could be applied to many actions and expressions that an NPC might
            need

            \item [$\rightarrow$] Design studies around behaviour classes that
            players evaluate with respect to their experience

            \item [$\rightarrow$] Some cross-cultural validation, connected to
            FACS-coded facial
            expressions~\citep[p.~116--118]{scherer2001appraisalB}
        \end{itemize}

        \item Frijda, Lazarus, Roseman, and Smith \& Kirby (\good)
        \begin{itemize}
            \item Associates each emotion with action tendencies
            (\citepg{frijda1986emotions}{88}; \citepg{lazarus1991emotion}{87,
                122}; \citepg{roseman2013appraisal}{143};
                \citeg{yih2016distinct};
            \citeg{yih2016patterns}; \citepg{yih2020profiles}{488--492})
            $\rightarrow$ could be applied to many actions and expressions that
            an NPC might need

            \item [$\rightarrow$] Design studies around behaviour classes that
            players evaluate with respect to their experience
        \end{itemize}

        \item Oatley \& Johnson-Laird (\good)
        \begin{itemize}
            \item Associates each emotion with action tendencies~\citep[p.~55,
            108, 192, 212]{oatley1992best} $\rightarrow$ could be applied to
            many actions and expressions that an NPC might need

            \item [$\rightarrow$] Design studies around behaviour classes that
            players evaluate with respect to their experience

            \item Focus on the connection between emotion and plans and goals
            in narrative and language~\citep[p.~70--71]{oatley1992best}
            $\rightarrow$ suggest that it is especially amenable to studies of
            NPC intentionality and believability
        \end{itemize}

        \item OCC (\weak)
        \begin{itemize}
            \item Provides an emotion and intensity as output $\rightarrow$
            gives only a few guidelines about action tendency
            associations~\citep[p.~197]{ortony2002making}.
        \end{itemize}
    \end{itemize}

\end{itemize}

\clearpage
\vspace*{\fill}
\begin{keypoints}
    \begin{itemize}

        \item These notes guided the score assignments in
        Tables~\ref{tab:theory-req-sys-summary-flexibility},
        \ref{tab:theory-req-sys-summary-easeofuse},
        \ref{tab:theory-req-comp-summary-flexibility}, and
        \ref{tab:theory-req-comp-summary-easeofuse}

        \item Scoring is, in part, subjective based on interpretations of the
        requirements and understanding of emotion literature

    \end{itemize}
\end{keypoints}

\parasep
\vspace*{\fill}

%% file: appendix_emotionClassifications.tex
\chapter{\progname{}'s Emotion Profiles for Examining Animated Characters}
\label{appendixEmotionClassification}
\def\epigraphflush{center}
\setlength{\epigraphwidth}{0.85\linewidth}
\def\textflush{center}
\epigraph{Every game has its rules. We just need to know how to break
them.}{Maeve Millay, \textit{Westworld}}

These are the emotion profiles created during the acceptance test case
specification process (Chapters~\ref{chapter:testcasedefinition} and
\ref{chapter:testcaseEMgine}). The profile of \textit{Sadness} does not appear
here because it serves as the example in Chapter~\ref{sec:profiles}. A reminder
that the FACS scores are only suggestions because an individual untrained in
FACS made the code assignments.

\section{Joy}
\textit{Joy} and its variants are enjoyable emotions, which are critically
understudied because they are not typically
problematic~\citep[p.~191, 199--200]{ekman2007emotions}. They are essential to
survival because they motivate individuals to engage in activities that: are
good for them; build confidence, social responsiveness, and connections to
things that reduce negative mental states~\citep[p.~244, 246]{izard1977human};
and achieve personal goals~\citep[p.~50, 102]{occ2022}. Cognitively,
\textit{Joy} broadens attention and thinking and builds personal skills and
resources through play (\citeg{oxfordJoy}). However, it also lowers the
experience thresholds for negative emotions as the perceptual system appears to
``see more'' and might decrease productivity by delaying uncomfortable
situations~\citep[p.~256--257]{izard1977human}. To combat this, \textit{Joy}
encourages the cognitive system to have higher tolerances for stimuli and lower
the importance of unsolved problems. To others, \textit{Joy} can communicate
the individual's evaluation of their subjective well-being---how well-off they
are and that they expect their good fortune to continue.

A low arousal state and subjective, self-aware feelings of calmness and
tranquility characterize \textit{Serenity}, a low intensity
\textit{Joy}~\citep{oxfordSerenity}. The intensity of the arousal state grows
to match the intensity of \textit{Joy} and \textit{Ecstasy} while
simultaneously reducing feelings of calmness. Although not explicitly
mentioned, the elapsed time since making the last progression towards goal
achievement and how much progress they made also appears to define the strength
of the initial \textit{Joy} appraisal. For example, a reunion with a loved one
is more intense if the last reunion was some time ago and, presumably, the
individual achieves the goal of being with them in full. The intensity of
\textit{Joy} is also often inversely proportional to the perceived chances of
achieving a goal, such as meeting with an admired person that the individual
never thought was possible.

\paragraph{Signs of \textit{Joy}}
\textit{Joy} is a relatively intense, long-lasting emotion~\citep[p.~321,
324]{scherer1994evidence}. Its expression is rarely controlled or regulated,
marked by highly-expressive verbal and non-verbal behaviours in addition to
strong approach behaviours. A socially attractive emotion, \textit{Joy}
predisposes individuals towards actions that manifest a sense of pleasure and
security in the world including sharing and approaching
behaviours~\citep[p.~268--269]{lazarus1991emotion}. This can manifest as
expansiveness, generosity, and outgoingness. In social situations, a happy
individual is more likely to initial a conversation and observers will enjoy
their company.

The voice is more likely to distinguish \textit{Joy} from other
enjoyment-related emotions than the face ~\citep[p.~204]{ekman2007emotions}.
Laughter is a unique non-verbal signature~\citep[p.~322]{scherer1994evidence}.
Energy conservation is not important as it is needed to enable approach
behaviours associated with \textit{Joy}, which can involve physical and mental
labour, indicated by an elevated body temperature.

\paragraph{Characteristic Facial Expression}
The lower face and eyelids convey \textit{Joy}
(\citepg{ekman2003unmasking}{103, 105, 107, 112}; \citepg{izard1977human}{241})
and is easy to spot (Tables~\ref{tab:serenityFACS}, \ref{tab:joyFACS}, and
\ref{tab:ecstacyFACS}). Changes in the eyebrows and forehead are not essential
for its identification.

The ``Duchenne smile'', simultaneously affecting the eyelids and lower face, is
the most reliable indicator of an ``enjoyment smile''
(\citepg{frank1993not}{12--13, 18, 21}; \citepg{frank1993behavioral}{92}).
There are different kinds of smile, a movement drawing the corners of the lips
back and up, but the ``Duchenne smile'' specifically activates:
\begin{itemize}
    \item The \textit{zygomaticus major} muscles in the cheeks, revealing the
    nasolabial folds\footnote{The creases running from the nose to the space
    beyond the lip corners on the cheeks.}, and

    \item The \textit{orbicularis oculi} muscles around the eyes causing
    ``crow's feet'' to appear in the outer corners, which become more apparent
    with age.
\end{itemize}

As the intensity of \textit{Joy} increases, the smile becomes broader and the
lips part to reveal the teeth. The cheeks rise higher, deepening the
nasolabial folds and pushing the lower eyelids up, narrowing the eyes. This
also causes more pronounced ``crow's feet'' and creasing below the lower
eyelids.

While the cheek muscles are easy to move voluntarily, it is much more difficult
to move the \textit{orbicularis oculi} muscles circling the
eyes~\citep[p.~206]{ekman2007emotions} which pulls the eye cover
fold\footnote{The skin between the eyelid and eyebrow.} down. This produces the
``eyes sparkling with happiness'' effect, which is difficult to replicate
voluntarily. This helps disambiguate a \textit{Joy} smile from other kinds.

\paragraph{Examples}
Jasmine from Disney's \textit{Aladdin}~\citep{aladdin} has been pressured by
her father to marry according to criteria defined in the law, which she feels
severely limits her agency in the decision. Thus far, Jasmine has been unable
to find someone that she connects to from the line of suitors that she has been
presented with and she has no say in who she meets.

\begin{figure}[!tb]
    \begin{center}
        \begin{subfigure}[t]{0.25\textwidth}
            \includegraphics[width=\textwidth]{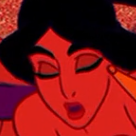}
            \caption*{\textbf{Joy00} \textit{Neutral}}
            \label{fig:jasmineSleep}
        \end{subfigure}

        \begin{subfigure}[t]{0.25\textwidth}
            \includegraphics[width=\textwidth]{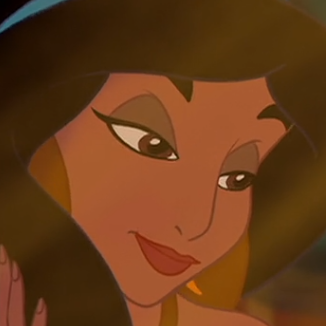}
            \caption*{\textbf{Joy01} \textit{Serenity}}
            \label{fig:serenity}
        \end{subfigure}
        \begin{subfigure}[t]{0.25\textwidth}
            \includegraphics[width=\textwidth]{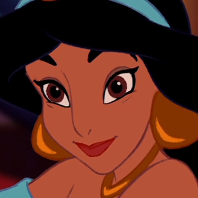}
            \caption*{\textbf{Joy02} \textit{Joy}}
            \label{fig:joy}
        \end{subfigure}
        \begin{subfigure}[t]{0.25\textwidth}
            \includegraphics[width=\textwidth]{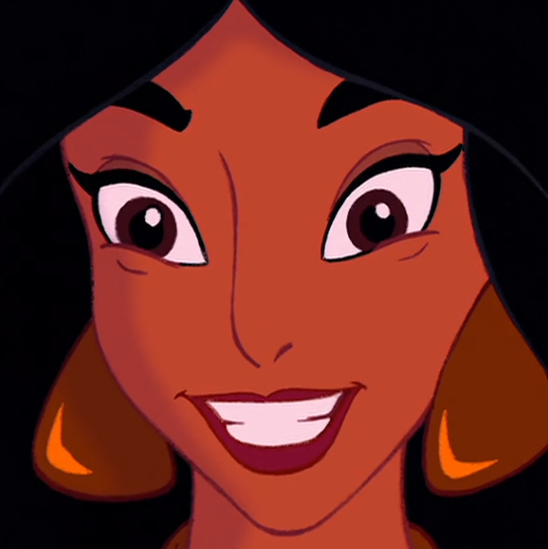}
            \caption*{\textbf{Joy03} \textit{Ecstasy}}
            \label{fig:ecstasy}
        \end{subfigure}
    \end{center}
    \caption{Examples of \textit{Joy} in Jasmine's Facial Expressions}
    \label{fig:charJoy}
\end{figure}

She clearly illustrates different intensities of \textit{Joy} through her eyes
and mouth (Figure~\ref{fig:charJoy}). In \textit{Serenity}, she has a general
lack of tension in her face, a slight smile, and lowered eyelids conveying a
sense of calm. As the intensity of the emotion increases, her smile broadens
and her cheeks rise which pushes her lower eyelids up. Animators often choose
not to illustrate the nasolabial folds or ``crow's feet'', likely to reduce the
amount of work required per frame and to create a more aesthetically pleasing
image. Instead, they rely on the broadness of the smile and changes in the lower
eyelids.

\textbf{Joy01}: After returning from an illicit date with her suitor and
choosing him as her betrothed, Jasmine experiences \textit{serenity} because:
\begin{itemize}
    \item She has chosen a prince, which will both please her father for
    following the law and herself for choosing someone she cares for
\end{itemize}

Jasmine signals her \textit{Serenity} via overall calmness and apparent
reflection as she hums the song she sang with Prince Ali to herself while
brushing her hair.

\textbf{Joy02}: Seeing her father's reaction to her choice of suitor is
\textit{joyful} for Jasmine because:
\begin{itemize}
    \item She has made the choice for herself
    \item The choice follows the law, which pleases her doting father
    \item Her father personally approves of her choice
\end{itemize}

She expresses her \textit{Joy} by maintaining physical contact with her suitor
while maintaining eye contact with her father and nodding her confirmation of
his statements. Her shoulders appear to rise towards her ears rather than
relaxing, implying a higher than normal energy expenditure.

\textbf{Joy03}: During a quiet moment with her father and love interest,
Jasmine becomes \textit{ecstatic} when:
\begin{itemize}
    \item Her father relaxes the conditions of the law, previously thought
    impossible, giving her complete control of her choice of suitor
    \item She has made the choice for herself
    \item Her father personally approves of her choice
\end{itemize}

Her excitement is shown by her hopping on the spot then running to her
betrothed to jump into his arms, both indicative of a high energy expenditure,
and her verbal confirmation that she chooses him.

\vspace*{\fill}
\begin{table}[!ht]
    \centering
    \caption{Facial Sketches of \textit{Serenity} with Suggested FACS Codes}
    \label{tab:serenityFACS}
    \renewcommand{\arraystretch}{1.2}
    \begin{threeparttable}
        \begin{tabular}{P{0.6\linewidth}P{0.3\linewidth}}
            \toprule
            \begin{center}
                \vspace*{-1em}
                \includegraphics[width=0.3\linewidth]{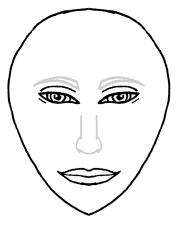}
                \vspace*{-1.5em}
            \end{center} & \begin{tabular}{ll}
                \textbf{Label} & Serenity \\
                \textbf{Intensity} & Low \\
            \end{tabular} \\ \midrule
            \textbf{Description} & \textbf{FACS Codes} \\ \midrule
            \colourRow Smooth forehead skin & -- \\
            No tension in upper eyelids & -- \\
            \colourRow No tension in the lower eyelids & -- \\
            Corners of the lips are slightly drawn back and up & AU 12 \\
            \colourRow Mouth closed & -- \\
            \bottomrule
        \end{tabular}
        \begin{tablenotes}

            \footnotesize
            \vspace*{0.5mm}

            \item \textit{AU 12 describes \textit{zygomaticus major} muscle
                movement}

        \end{tablenotes}
    \end{threeparttable}
\end{table}
\vspace*{\fill}

\begin{table}[!tb]
    \centering
    \caption{Facial Sketches of \textit{Joy} with Suggested FACS Codes}
    \label{tab:joyFACS}
    \renewcommand{\arraystretch}{1.2}
    \begin{threeparttable}
        \begin{tabular}{P{0.6\linewidth}P{0.3\linewidth}}
            \toprule
            \begin{center}
                \includegraphics[width=0.3\linewidth]{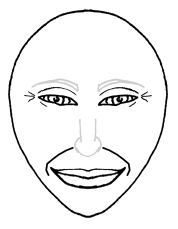}
            \end{center} & \begin{tabular}{ll}
                \textbf{Label} & Joy \\
                \textbf{Intensity} & Medium \\
            \end{tabular} \\ \midrule
            \textbf{Description} & \textbf{FACS Codes} \\ \midrule
            \colourRow Smooth forehead skin & -- \\
            Slightly lowered eyebrows/upper eyelids & AU 6* \\
            \colourRow Eyes might appear to have a
            ``sparkling'' effect & AU 6* \\
            Crow's feet in outer corners of eyes & AU 6+12* \\
            \colourRow Raised cheeks pushing the lower
            eyelids upwards & AU 6 \\
            The eyes might appear narrower due to the raised lower eyelids & AU
            6+12* \\
            \colourRow Obvious nasolabial folds & AU 6+12* \\
            Corners of the lips are drawn back and up towards the ears & AU 12
            \\
            \colourRow Mouth might open to reveal the teeth &
            AU 25 \\
            \bottomrule
        \end{tabular}
        \begin{tablenotes}

            \footnotesize
            \vspace*{0.5mm}

            \item {*} \textit{Indirect effect of AU}

            \item \textit{AU 6 describes \textit{orbicularis oculi} muscle
                movement}

            \item \textit{AU 12 describes \textit{zygomaticus major} muscle
                movement}

        \end{tablenotes}
    \end{threeparttable}
\end{table}

\begin{table}[!tb]
    \centering
    \caption{Facial Sketches of \textit{Ecstasy} with Suggested FACS Codes}
    \label{tab:ecstacyFACS}
    \renewcommand{\arraystretch}{1.2}
    \begin{threeparttable}
        \begin{tabular}{P{0.6\linewidth}P{0.3\linewidth}}
            \toprule
            \begin{center}
                \includegraphics[width=0.6\linewidth]{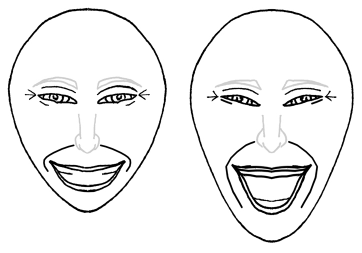}
            \end{center} & \begin{tabular}{ll}
                \textbf{Label} & Ecstasy \\
                \textbf{Intensity} & High \\
            \end{tabular} \\ \midrule
            \textbf{Description} & \textbf{FACS Codes} \\ \midrule
            \colourRow Smooth forehead skin & -- \\
            Slightly lowered eyebrows/upper eyelids & AU 6* \\
            \colourRow Eyes might appear to have a
            ``sparkling'' effect & AU 6* \\
            Crow's feet in the outer corners of eyes & AU 6+12* \\
            \colourRow Raised cheeks pushing the lower
            eyelids upwards & AU 6 \\
            The eyes might appear narrower due to the raised lower eyelids & AU
            6+12* \\
            \colourRow Obvious nasolabial folds & AU 6+12* \\
            Corners of the lips are drawn back and up towards the ears & AU 12
            \\
            \colourRow Upper lip appears tense & AU 12* \\
            Mouth open to reveal the teeth & AU 25 \\
            \colourRow Jaw might stretch, parting the teeth
            (e.g. laughing) & AU 27 \\
            \bottomrule
        \end{tabular}
        \begin{tablenotes}

            \footnotesize
            \vspace*{0.5mm}

            \item {*} \textit{Indirect effect of AU}

            \item \textit{AU 6 describes \textit{orbicularis oculi} muscle
                movement}

            \item \textit{AU 12 describes \textit{zygomaticus major} muscle
                movement}

        \end{tablenotes}
    \end{threeparttable}
\end{table}

\clearpage\section{Fear}
This is the most researched emotion primarily because it is easy to elicit in
animals (\citepg{ekman2007emotions}{152}) and there is an increase in the
number of fear-inducing scenarios in contemporary
civilizations~\citep[p.~355]{izard1977human}. The primary purpose of
\textit{Fear} is to ensure survival by either avoiding or escaping potentially
harmful situations (\citepg{lazarus1991emotion}{235--236};
\citepg{ekman2007emotions}{152}). Stimuli causing \textit{Fear} often leave
strong impressions in memory, making them easier to
recall~\citep[p.~355]{izard1977human}. It also influences perception, reducing
the individuals information intake and increasing uncertainty, and reduces the
number of behavioural alternatives. Thinking slows, becoming narrow and rigid,
paving the way for immobilization and helplessness in the most intense
scenarios. The individual also experiences the subjective feeling of
insecurity, helplessness, and looming imminent danger.

As the intensity of \textit{Fear} increases, individuals experience more
extreme physical and cognitive changes~\citep[p.~365]{izard1977human}. They
become functionally blind to all but the triggering stimulus, experience high
tension that can hinder movement, and cannot think beyond the concept of escape.
Attempting to manage \textit{Fear} can counteracted it, but becomes less
feasible as its intensity grows. If ``flight'' is not possible and the
perceived harm is great, the intensity of \textit{Fear} greatly increases and
the individual is more like to become immobilized and helpless.

\paragraph{Signs of \textit{Fear}} Compared to the other emotions,
\textit{Fear} has few outward expressions and have nearly no verbal
expressions~\citep[p.~323--324]{scherer1994evidence} which matches with the
tendency to immobilize the body. How someone behaviourally reacts to
\textit{Fear} depends on their past experiences, but follows the same
tendencies of avoidance and escape: immobilization to avoid detection followed
by either ``flight'' to escape or attributing blame to become \textit{Angry} in
an attempt to physically remove the threat
(\citepg{ekman2007emotions}{153--154}). Other behaviours aside from ``flight''
that organisms might use in the interest of protection include warily watching
the source of \textit{Fear} to increase awareness, action inhibition, trembling,
cowering, hiding, and seeking the company of others. The action tendency
associated with \textit{Fear} is avoidance or escape, which is reflected in the
common connection of ``flight''.

\textit{Fear} accumulates a high energy expenditure and body tension, but the
body temperature is low~\citep[p.~321--322, 324, 326]{scherer1994evidence}.
This matches the ``flight'' response where the body redirects blood flow to the
chest and abdomen in preparation for fast movement. This causes an increase in
muscle tension, which appears as trembling. As the intensity of \textit{Fear}
increases, trembling goes from slow and loose movements to faster and stiffer
ones. Heart acceleration also increases at a linearly proportional rate to the
intensity of \textit{Fear}. Perspiration is another physical indicator of
\textit{Fear} that prepares the body to prepare to run. \textit{Fear} is a
relatively short-lived emotion~\citep[p.~320]{scherer1994evidence}, possibly
because of the high strain on body and mind.

\paragraph{Characteristic Facial Expression} All major facial areas convey
\textit{Fear} (Tables~\ref{tab:apprehensionFACS}, \ref{tab:fearFACS}, and
\ref{tab:terrorFACS}), but how many areas it engages depends on the emotion's
intensity (\citepg{ekman2003unmasking}{50, 55, 63};
\citepg{izard1977human}{364--365}). If only one area has characteristic
changes, it implies a low intensity \textit{Fear}\footnote{Ekman refers to both
\textit{Worry} and \textit{Apprehension} as mild versions of \textit{Fear},
which this profile takes as synonymous to \textit{Apprehension}.}. Changes in
all facial areas and a high level of facial muscle tension indicates high
intensity \textit{Fear}: \textit{Terror}.

People often confuse the facial expressions of \textit{Fear} and
\textit{Surprise}, but changes in the upper face reliably distinguish them. In
\textit{Fear}, the lower eyelids tense and the upper eyelids rise, exposing the
sclera. This causes the illusion that the eyeballs are popping out of their
sockets. This is further emphasized by raised eyebrows with the inner corners
drawn together such that they appear to be straight. This can also cause
horizontal wrinkles to appear in the middle of the forehead.

Tensed lips pulled back towards the ears is another indicator of \textit{Fear}.
As the intensity grows, the mouth opens and exposes the teeth which might be
clenched.

\paragraph{Examples} Mrs. Brisby, a field mouse in Don Bluth's \textit{The
Secret of NIMH}~\citep{nimh}, is a good example of the primary purpose of
\textit{Fear}: self-preservation. She knows that she is small and is an easy
target for larger predators like cats and owls.

\begin{figure}[!b]
    \begin{center}
        \begin{subfigure}[t]{0.25\linewidth}
            \includegraphics[width=\linewidth]{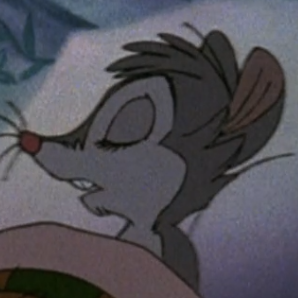}
            \caption*{\textbf{Fear00} Neutral}
            \label{fig:brisbySleep}
        \end{subfigure}

        \begin{subfigure}[t]{0.25\linewidth}
            \includegraphics[width=\linewidth]{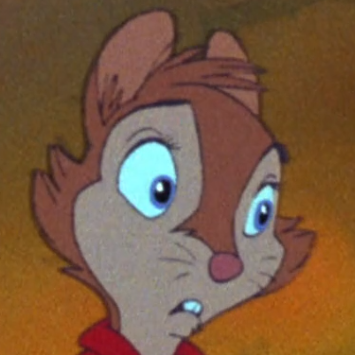}
            \caption*{\textbf{Fear01} \textit{Apprehension}}
            \label{fig:apprehension}
        \end{subfigure}
        \begin{subfigure}[t]{0.25\linewidth}
            \includegraphics[width=\linewidth]{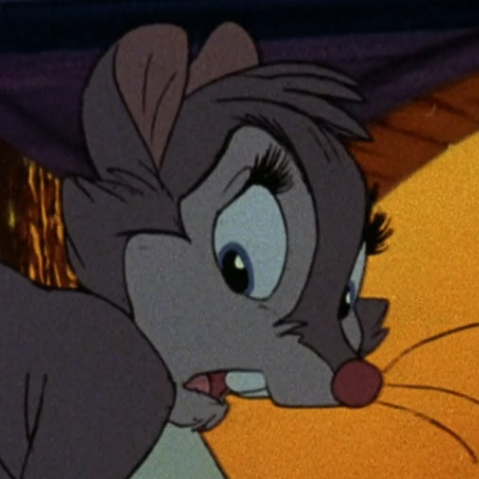}
            \caption*{\textbf{Fear02} \textit{Fear}}
            \label{fig:fear}
        \end{subfigure}
        \begin{subfigure}[t]{0.25\linewidth}
            \includegraphics[width=\linewidth]{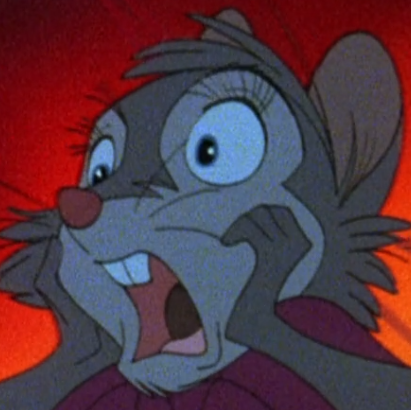}
            \caption*{\textbf{Fear03} \textit{Terror}}
            \label{fig:terror}
        \end{subfigure}
    \end{center}
    \caption{Examples of \textit{Fear} in Mrs. Brisby's Facial Expressions}
    \label{fig:mrsbrisbyFear}
\end{figure}

While the facial expressions vary in the jaw and lips, Mrs. Brisby clearly
shows \textit{Fear} through her eyelids and eyebrows
(Figure~\ref{fig:mrsbrisbyFear}). As its intensity rises, her upper eyelids
rise and her lower eyelids tense. Her shrinking iris and pupil emphasizes this,
creating the illusion that she is exposing more of her sclera. She has raised
her eyebrows and they appear to be straight, albeit angled due to the
perspective of the images. The tension in Mrs. Brisby's lips increases with the
intensity of \textit{Fear} and exposes her teeth, with her jaw dropping in
\textit{Terror}. The animators chose not to include forehead and cheek creases
in these expressions.

\textbf{Fear01}: A friend urged Mrs. Brisby to visit the Great Owl, a wise
animal of the forest who is likely to be able to help her solve a problem
threatening her family. Despite knowing that the Great Owl is no ordinary
animal, and after being told by the Owl to come inside his house or leave, Mrs.
Brisby still feels \textit{apprehensive} because:
\begin{itemize}
    \item Her physical safety is at imminent risk  because he might choose
    to eat her instead of talk (``owls EAT mice!'')
    \item She cannot defend herself against a predator and escape is
    uncertain because the Great Owl's house is a hollow high in a tree
\end{itemize}

Mrs. Brisby also displays \textit{Apprehension} by wearily watching her
surroundings and slowly proceeding into the Great Owl's hollow.

\textbf{Fear02}: To help her family, Mrs. Brisby volunteers to put a
sleeping drug in the farm cat's food dish, which is only accessible after
crossing the large, open kitchen inside the farmer's house. She is
\textit{fearful} because:
\begin{itemize}
    \item The kitchen is empty but there is a high potential for change, as
    the immediate risks---the cat and the farmer's wife---are both at the
    front door and will return to the kitchen shortly
    \item She cannot be certain how long they will be gone for
\end{itemize}

Her \textit{Fear} is apparent in her loose trembling and her reluctance to
move from her current location (i.e. freezing).

\textbf{Fear03}: In the process of helping her friend, Mrs. Brisby comes
within a few feet of the farm cat whom has seen her. She experiences
\textit{Terror} because:
\begin{itemize}
    \item There is an immediate and severe risk to her physical safety, as
    an encounter with the cat will certainly kill her
\end{itemize}

Her trembling is quick and sharp as she freezes in place, staring at the cat
who is climbing the rest of the way onto the log she is standing on.

\begin{table}[!tb]
    \centering
    \caption{Facial Sketches of \textit{Apprehension} with Suggested FACS Codes}
    \label{tab:apprehensionFACS}
    \renewcommand{\arraystretch}{1.2}
    \begin{threeparttable}
        \begin{tabular}{P{0.6\linewidth}P{0.3\linewidth}}
            \toprule
            \begin{center}
                \includegraphics[width=0.9\linewidth]{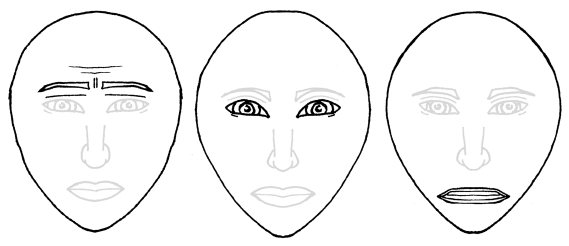}
            \end{center} & \begin{tabular}{ll}
                \textbf{Label} & Apprehension \\
                \textbf{Intensity} & Low
            \end{tabular} \\ \midrule
            \textbf{Description} & \textbf{FACS Codes} \\ \midrule
            \colourRow  \textbf{At least one of:} & \\
            Raised, straightened eyebrows pulled together; Wrinkles might
            appear in the centre of the forehead & AU 1+2; AU 1+2+5* \\
            \colourRow Raised upper eye lids exposing the
            sclera, tensed lower eye lids; Lower eyelids might be covering
            part of the iris; Eyes fixated on one point & AU 5+7; AU 7*; 69 \\
            Lips pulled back towards the ears; Open mouth exposing the teeth;
            Jaw clenched & AU 20; AU 25; AU 31 \\
            \bottomrule
        \end{tabular}
        \begin{tablenotes}

            \footnotesize
            \vspace*{0.5mm}

            \item {*} \textit{Causes change indirectly}

        \end{tablenotes}
    \end{threeparttable}
\end{table}

\begin{table}[!tb]
    \centering
    \caption{Facial Sketches of \textit{Fear} with Suggested FACS Codes}
    \label{tab:fearFACS}
    \renewcommand{\arraystretch}{1.2}
    \begin{threeparttable}
        \begin{tabular}{P{0.6\linewidth}P{0.3\linewidth}}
            \toprule
            \begin{center}
                \includegraphics[width=0.3\linewidth]{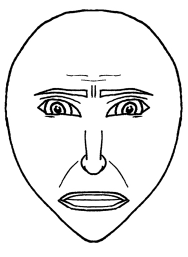}
            \end{center} & \begin{tabular}{ll}
                \textbf{Label} & Fear \\
                \textbf{Intensity} & Medium \\
            \end{tabular} \\ \midrule
            \textbf{Description} & \textbf{FACS Codes} \\ \midrule
            \colourRow Raised, straightened eyebrows pulled
            together & AU 1+2 \\
            Wrinkles might appear in the centre of the forehead & AU 1+2+5* \\
            \colourRow Raised upper eye lids exposing the
            sclera, tensed lower eye lids & AU 5+7  \\
            Lower eyelids might be covering part of the iris & AU 7* \\
            \colourRow Eyes fixated on one point & 69 \\
            Lips pulled back towards the ears & AU 20 \\
            \colourRow Open mouth exposing the teeth & AU 25
            \\
            Jaw clenched & AU 31\\
            \bottomrule
        \end{tabular}
        \begin{tablenotes}

            \footnotesize
            \vspace*{0.5mm}

            \item {*} \textit{Causes change indirectly}

        \end{tablenotes}
    \end{threeparttable}
\end{table}

\begin{table}[!tb]
    \centering
    \caption{Facial Sketches of \textit{Terror} with Suggested FACS Codes}
    \label{tab:terrorFACS}
    \renewcommand{\arraystretch}{1.2}
    \begin{threeparttable}
        \begin{tabular}{P{0.6\linewidth}P{0.3\linewidth}}
            \toprule
            \begin{center}
                \includegraphics[width=0.6\linewidth]{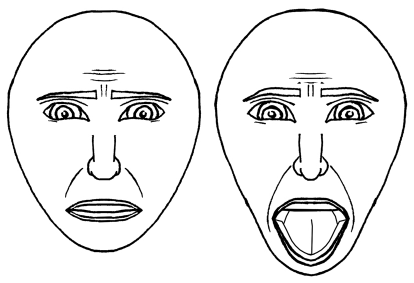}
            \end{center} & \begin{tabular}{ll}
                \textbf{Label} & Terror \\
                \textbf{Intensity} & High \\
            \end{tabular} \\ \midrule
            \textbf{Description} & \textbf{FACS Codes} \\ \midrule
            \colourRow Raised, straightened eyebrows pulled
            together & AU 1+2 \\
            Wrinkles might appear in the centre of the forehead & AU 1+2+5* \\
            \colourRow Raised upper eye lids exposing the
            sclera, tensed lower eye lids & AU 5+7 \\
            Lower eyelids might be covering part of the iris & AU 7* \\
            \colourRow Eyes fixated on one point & 69 \\
            Lips pulled back towards the ears & AU 20 \\
            \colourRow Open mouth exposing the teeth & AU 25
            \\
            Jaw clenched OR Stretched open & AU 31 OR AU 27 \\
            \colourRow Chin might appear drawn back & AU 27*
            \\
            \bottomrule
        \end{tabular}
        \begin{tablenotes}

            \footnotesize
            \vspace*{0.5mm}

            \item {*} \textit{Causes change indirectly}

        \end{tablenotes}
    \end{threeparttable}
\end{table}

\clearpage\section{Anger}
\textit{Anger} is a high energy, negative emotion that commonly arises in
response to unwanted or harmful situations, mobilizing the individual to remove
or attack the offensive person or object (\citeg{oxfordAnger}). It is rarely
felt alone for long, with emotions like \textit{Fear} and \textit{Disgust}
commonly preceding or following it~\citep[p.~113]{ekman2007emotions}.
\textit{Anger} predisposes the individual to attack a target that they believe
is preventing them from achieving a goal and/or caused offence
(\citepg{ekman2007emotions}{114}; \citepg{izard1977human}{329--330};
\citepg{lazarus1991emotion}{226}). This impulse is usually tightly controlled
due to potential consequences for social relations, cultural expectations,
long-term self-interest, and personal values. To this end, coping strategies
and the importance of other goals affected by the consequences of expressing
\textit{Anger} play a significant role in altering the individual's reaction to
experiencing it. Subjectively, people often describe \textit{Anger} as an
unpleasant feeling.

The intensity of \textit{Anger} heavily depends on personality, with some
people unable to feel anything stronger than \textit{Annoyance} and others
having short tempers, easily flying into
\textit{Rage}~\citep[p.~81]{ekman2003unmasking}. The ability to control the
action tendencies and expression of \textit{Anger} becomes more difficult as
the emotion's intensity increases. Frustration is an early warning sign that
something is preventing a goal's completion.

\paragraph{Signs of \textit{Anger}}
The tendency to attack or remove obstacles usually appears as agitation or
aggression. As intensity increases, the individual becomes more energetic and
impulsive in order to prepare them for physical
altercations~\citep[p.~331]{izard1977human}. This emotion tends to make people
more extroverted to try to sustain a high-level of focused and directed
activity to unblock progression towards goals. Although \textit{Anger} is a
highly expressive and rarely well-controlled emotion, both verbally and
non-verbally, many cultures condemn it (\citepg{ekman2007emotions}{120};
\citepg{scherer1994evidence}{320, 323--324}). Therefore, there might be signs
that an individual is trying to control their \textit{Anger} instead of acting
on it.

An increase in energy expenditure and high tension often accompany
\textit{Anger}~\citep[p.~321--322, 324, 326]{scherer1994evidence}. In
preparation for ``fight'', the body increases its heart rate to send more blood
through the body, raising its temperature while increasing its tension. The
increased energy primes individuals for some measure of violence, either
physical or verbal, presumably to remove a perceived obstacle~\citep[p.~110,
125]{ekman2007emotions} or source of blame~\citep[p.~223,
225]{lazarus1991emotion}.

\paragraph{Characteristic Facial Expression}
\textit{Anger} registers on all facial areas (\citepg{ekman2003unmasking}{82,
88, 92, 95, 97}; \citepg{izard1977human}{330}). Changes in all areas must appear
for the expression to be unambiguous (Tables~\ref{tab:annoyanceFACS},
\ref{tab:angerFACS}, and \ref{tab:rageFACS}).

In \textit{Anger}, the eyebrows draw together and lower. This can make them
look like they are pointing down towards the nose. The eyebrow movement
typically causes the upper eyelids to lower as well. Combined with tensed lower
eyelids, which appear to rise with the intensity of \textit{Anger}, the eyes
appear penetrating or as if the individual has a hard, fixed stare.

The mouth can have two forms depending on the individual's intended action. If
the individual is engaging verbally, then the lips form a squared shape around
the teeth. If they are trying not to speak or are preparing for a physical
altercation, the lips are firmly pressed together which might make them appear
thinner. Thinning of the lips is one of the earliest observable signs of
\textit{Anger} in the face. It can happen alone, likely because it is a hard
action to inhibit. This sign usually occurs before a person becomes aware of
their \textit{Anger}.

\paragraph{Examples}
Beast from Disney's \textit{Beauty and the Beast}~\citep{beautyandthebeast} is
a natural choice to demonstrate \textit{Anger} due to his angry disposition. He
is the castle's master, which comes with the ultimate level of control over its
affairs and castle staff.

\begin{figure}[!b]
    \begin{center}
        \begin{subfigure}[t]{0.25\textwidth}
            \includegraphics[width=\textwidth]{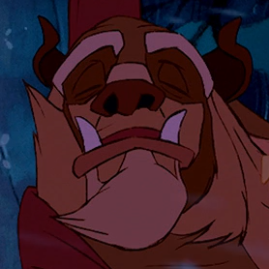}
            \caption*{\textbf{Anger00} \textit{Neutral}}
            \label{fig:beastSleep}
        \end{subfigure}

        \begin{subfigure}[t]{0.25\textwidth}
            \includegraphics[width=\textwidth]{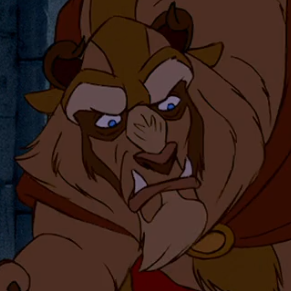}
            \caption*{\textbf{Anger01} \textit{Annoyance}}
            \label{fig:annoyance}
        \end{subfigure}
        \begin{subfigure}[t]{0.25\textwidth}
            \includegraphics[width=\textwidth]{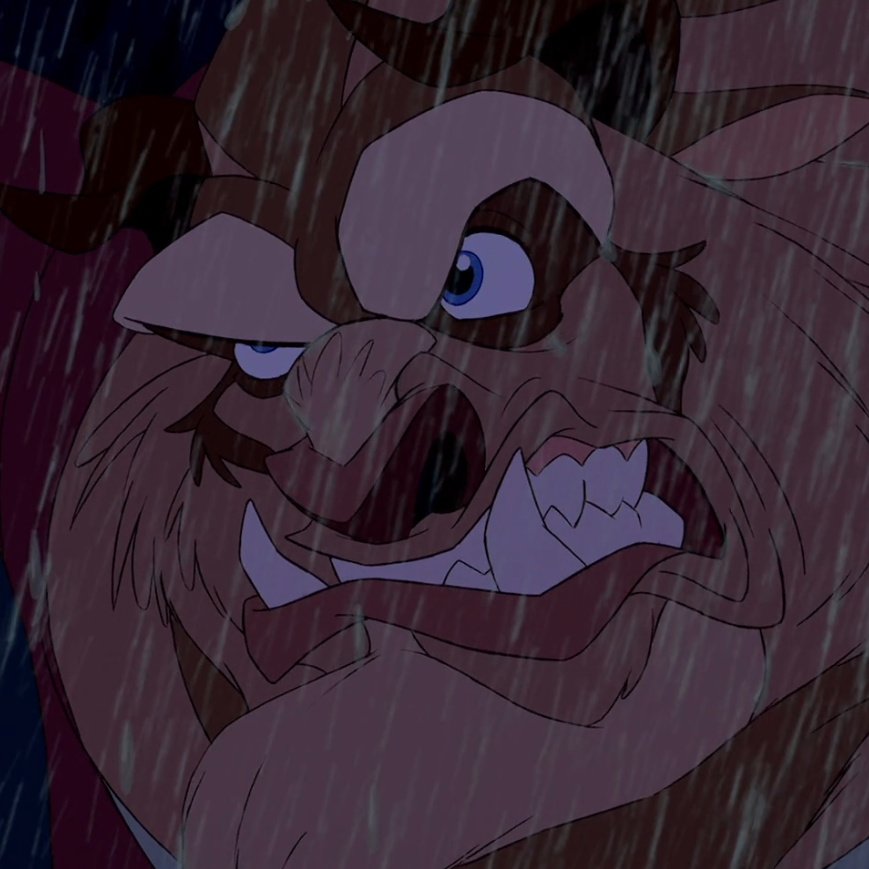}
            \caption*{\textbf{Anger02} \textit{Anger}}
            \label{fig:anger}
        \end{subfigure}
        \begin{subfigure}[t]{0.25\textwidth}
            \includegraphics[width=\textwidth]{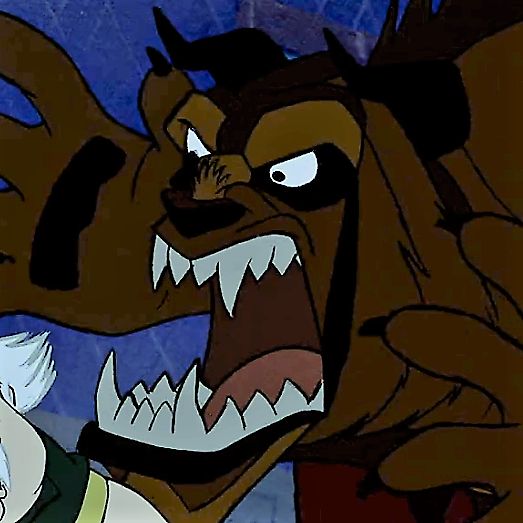}
            \caption*{\textbf{Anger03} \textit{Rage}}
            \label{fig:rage}
        \end{subfigure}
    \end{center}
    \caption{Examples of \textit{Anger} in Beast's Facial Expressions}
    \label{fig:beastAnger}
\end{figure}

Beast's angled eyebrows, lowered upper eyelids, tensed lower eyelids, and
penetrating stare become more apparent as the intensity of his \textit{Anger}
increases (Figure~\ref{fig:beastAnger}). Animators also exaggerated changes in
the pupil size and mouth, making them more reliable indicators of intensity
compared to the eyebrows. Despite its importance, thinning lips are not a
reliable sign of \textit{Anger} in animation because many animated characters
lack explicitly drawn lips. A line implies the mouth and sometimes animators
add a second line to indicate where the bottom of the lower lip is. Instead,
changes in the mouth shape imply changes in the lips. Beast demonstrates the
possible mouth shapes: the flat line of his mouth implies that his lips are
pressed together to avoid speaking in \textit{Annoyance}; a square shape is
formed by the mouth line when preparing for a fight in \textit{Anger} to show
the increased tension in the cheeks; and in \textit{Rage}, the square shape of
the mouth line is significantly larger to convey the act of yelling, which also
exposes more of Beast's teeth.

\textbf{Anger01}: Beast wants to make a good impression on Belle and expects
that moving her out of the dungeon and into a proper room will help. He becomes
\textit{annoyed} with her after she:
\begin{itemize}
    \item Questions his authority, which he perceives as harmful to his image
    as the castle's master
    \item She chose to speak from a sitting position rather than following
    silently
\end{itemize}

His sweeping gestures to the dungeon and demanding to know if the girl wants to
stay in the dungeon are other indicators of \textit{Annoyance}.

\textbf{Anger02}: The villagers are attacking the castle and Beast knows that
he is physically powerful and can match his attacker. He becomes \textit{Angry}
because:
\begin{itemize}
    \item The aggressor is attacking Beast in his home, with the intent to
    kill, putting him in immediate danger of mortal harm
    \item The attacker made it clear that he premeditated the decision
\end{itemize}

Beast's decision to physically retaliate, including intimidation via loud
roars, is evidence of his \textit{Anger}.

\textbf{Anger03}: Beast has locked himself away in his castle because he
believes himself to be hideous. When a traveller arrives at the castle, Beast
experiences \textit{Rage} because he:
\begin{itemize}
    \item Believes the traveller is mocking him because he is no longer
    ``handsome'', which is an integral aspect of his self-image
    \item Interprets the man's inability to look away as a conscious decision
    to stare rather than \textit{Fear}
\end{itemize}

Leaping to block doors and forcefully speaking to the traveller are all signs
of Beast's \textit{Rage}.

\vspace*{\fill}
\begin{table}[!ht]
    \centering
    \caption{Facial Sketches of \textit{Annoyance} with Suggested FACS Codes}
    \label{tab:annoyanceFACS}
    \renewcommand{\arraystretch}{1.2}
    \begin{tabular}{P{0.6\linewidth}P{0.3\linewidth}}
        \toprule
        \begin{center}
            \includegraphics[width=0.6\linewidth]{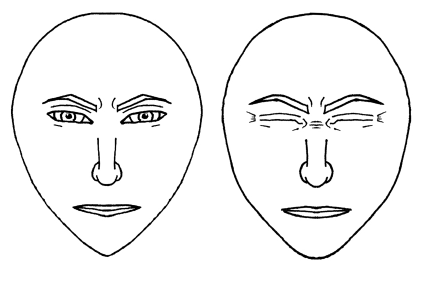}
        \end{center} & \begin{tabular}{ll}
            \textbf{Label} & Annoyance \\
            \textbf{Intensity} & Low \\
        \end{tabular} \\ \midrule
        \textbf{Description} & \textbf{FACS Codes} \\ \midrule
        \colourRow Lowered eyebrows that are close together
        and angled downwards towards the nose causing vertical wrinkles to
        appear between them & AU 4 \\
        Narrowed OR tensed, closed eyes & AU 7 OR AU 7+43 \\
        \colourRow Thinned lips that are pressed together &
        AU 17+24 \\
        \bottomrule
    \end{tabular}
\end{table}
\vspace*{\fill}

\begin{table}[!tb]
    \centering
    \caption{Facial Sketches of \textit{Anger} with Suggested FACS Codes}
    \label{tab:angerFACS}
    \renewcommand{\arraystretch}{1.2}
    \begin{tabular}{P{0.6\linewidth}P{0.3\linewidth}}
        \toprule
        \begin{center}
            \includegraphics[width=0.6\linewidth]{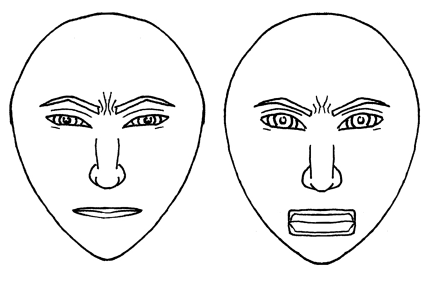}
        \end{center} & \begin{tabular}{ll}
            \textbf{Label} & Anger \\
            \textbf{Intensity} & Medium \\
        \end{tabular} \\ \midrule
        \textbf{Description} & FACS Codes \\ \midrule
        \colourRow Lowered eyebrows that are close
        together and angled downwards towards the nose causing vertical
        wrinkles to appear between them & AU 4 \\
        Narrowed OR glaring eyes (raised upper eyelids, tensed lower eyelids)
        that might appear to be bulging & AU 7 OR AU 5+7  \\
        \colourRow Might have flared nostrils & AU 38 \\
        Thinned lips that are pressed together OR Exposed teeth with tensed,
        rectangular shape mouth & AU 17+24 \linebreak OR AU 20+23+25 \\
        \colourRow Tightly clenched jaw, possibly jutting
        forward & AU 31+29 \\
        Possibly red-faced & -- \\
        \bottomrule
    \end{tabular}
\end{table}

\begin{table}[!tb]
    \centering
    \caption{Facial Sketches of \textit{Rage} with Suggested FACS Codes}
    \label{tab:rageFACS}
    \renewcommand{\arraystretch}{1.2}
    \begin{tabular}{P{0.6\linewidth}P{0.3\linewidth}}
        \toprule
        \begin{center}
            \includegraphics[width=0.3\linewidth]{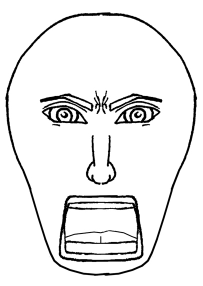}
        \end{center} & \begin{tabular}{ll}
            \textbf{Label} & Rage \\
            \textbf{Intensity} & High \\
        \end{tabular} \\ \midrule
        \textbf{Description} & \textbf{FACS Codes} \\ \midrule
        \colourRow Lowered eyebrows that are close
        together and angled downwards towards the nose causing vertical
        wrinkles to appear between them & AU 4 \\
        Glaring eyes (raised upper eyelids, tensed lower eyelids) with
        constricted pupil that might appear to be bulging OR Eyes squeezed
        shut, especially if yelling & AU 5+7 OR AU 7+43  \\
        \colourRow Flared nostrils & AU 38 \\
        Thinned lips & AU 24 \\
        \colourRow Tightly clenched jaw that might be
        jutting forward OR yelling, exposing teeth & AU 31+29 OR AU 27 \\
        Usually red-faced & -- \\
        \bottomrule
    \end{tabular}
\end{table}

\clearpage\section{Disgust}
A low-intensity emotion, \textit{Disgust} is the feeling of aversion and
rejection, usually involving some type of removal or avoidance strategy. People
experience it when they encounter something revolting, contaminated,
deteriorating, or spoiled (\citepg{izard1977human}{336};
\citepg{ekman2007emotions}{172--174}; \citepg{lazarus1991emotion}{259}).
Cognitively, \textit{Disgust} motivates individuals to redirect their attention
to avoid dealing with the object of revulsion, an impulse which can manifest as
nausea, or change it if possible~\citep[p.~336--337]{izard1977human}. If the
offensive object cannot be avoided and cannot be changed, the individual will
attempt to eject it in a manner comparable to
vomiting~\citep[p.~262]{lazarus1991emotion}. Subjectively, people describe
\textit{Boredom}---taken as low intensity \textit{Disgust}---as unpleasant,
marked by a strong desire to shut out or ignore the
situation~\citep[p.~832--833, 835]{smith1985patterns}. \textit{Disgust} is
similar, but feels like it requires more effort than \textit{Boredom}.

The intensity of \textit{Disgust} is directly proportional to the level of
revulsion and avoidance the individual feels. The weakest level of aversion,
\textit{Boredom}, is an understudied emotion. Some work conducted on
\textit{Boredom} found it to be an emotion of its
own~\citep[p.~317]{van2017boredom}. \textit{Boredom} is a desire to abandon the
current situation for a more stimulating one (\citepg{bench2013function}{468};
\citepg{vanTilburg2012}{192}). Framed as low intensity \textit{Disgust},
\textit{Boredom} is like a weak aversion to an event with an avoidance strategy
aimed at finding a more interesting one or withdrawing attention from the
current activity~\citep[p.~430]{rozin1999disgust}. Unlike \textit{Disgust},
\textit{Boredom} can be a positive experience, prompting a search for
stimulation~\citep[p.~118, 120]{vodanovich1990factor} and
interest~\citep[p.~157]{smith2009keeping}.

Although bodily function is intimately tied to \textit{Disgust}, it evolved to
include ideational and social aspects~\citep[p.~431]{rozin1999disgust}.
\textit{Disgust} towards others effectively dehumanizes them and enables the
condoning of their persecution (\citepg{ekman2007emotions}{178--179};
\citepg{miller1998anatomy}{8--9, 50, 133--134}). Tolerance for \textit{Disgust}
triggers increases when dealing with intimate relationships, such as caring for
a sick family member. This suggests that \textit{Disgust}, or having a higher
tolerance for triggers, also serves as a mark of personal commitment and
further cements its importance as a social emotion.

\paragraph{Signs of \textit{Disgust}}
\textit{Disgust} predisposes individuals to avoid contact with the offending
object. This is different from \textit{Fear} in that the object of
\textit{Disgust} might still capture the individual's attention and prompt them
to observe it from afar~\citep[p.~111--113]{rozin1999disgust}. As the intensity
increases, the experience of \textit{Disgust} is no longer positive and the
individual's aversion increases. At the most intense, \textit{Loathing},
individuals can feel physically nauseous~\citep[p.~66]{ekman2003unmasking} and
employ more extreme avoidance strategies that have higher costs than they would
normally be willing to pay, such as refusing to eat. In the most extreme case,
the individual cannot look at the object at all because the nausea is
overwhelming and potentially vomiting in response. If the object of
\textit{Disgust} is related to sanitation or hygiene, it is more likely that
the individual will change the situation rather than avoiding it altogether to
increase the probability of survival.

\textit{Disgust} uses a significantly lower amount of energy compared to
\textit{Anger}, and it consequently passes quickly and with little
expressiveness~\citep[p.~319, 323--324, 326]{scherer1994evidence}. This is
coupled with a decelerated heart rate, likely tied to parasympathetic
activation such as salivation and gastrointestinal activation which causes
heart activity to slow. Short, verbal expressions such as exclamations often
accompany \textit{Disgust}.

\paragraph{Characteristic Facial Expression}
The nose and upper lips primarily drive the facial expressions of
\textit{Disgust} (Tables~\ref{tab:boredomFACS}, \ref{tab:disgustFACS}, and
\ref{tab:loathingFACS}), which affect the surrounding facial areas
(\citepg{ekman2003unmasking}{68, 71, 76}; \citepg{izard1977human}{336}). The
eyebrows are not critical to this facial expression, but can distinguish
\textit{Disgust} from \textit{Anger} as the inner eyebrows do not draw
together.

Wrinkling the nose or raising and pushing the upper lip out signals
\textit{Disgust}. While these features can occur individually, it is more
common to observe them together. As the intensity of \textit{Disgust} increases,
these movements become more pronounced. The wrinkling of the nose causes other
wrinkles to appear on and around it, which pulls the upper lip and cheeks
upwards and the eyebrows down. In turn, this lowers the upper eyelid. These
effects can make \textit{Disgust} difficult to distinguish from \textit{Anger}
because it appears that the eyelids are tensing due to wrinkles formed by the
raised cheeks.

It is common to observe a gaping mouth, potentially with a visible or
protruding tongue, as if to physically reject the offending stimulus.

\paragraph{Examples}
Never wanting for anything in his life and being Emperor, Kuzco from Disney's
\textit{The Emperor's New Groove}~\citep{emperorsnewgroove} is self-absorbed
and has difficulty changing his habits.

\begin{figure}[!b]
    \begin{center}
        \begin{subfigure}[t]{0.25\textwidth}
            \includegraphics[width=\textwidth]{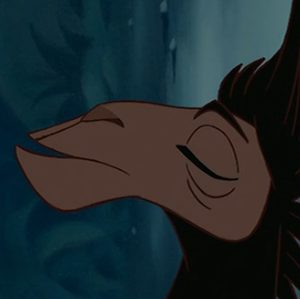}
            \caption*{\textbf{Disgust00} \textit{Neutral}}
            \label{fig:kuzcoSleep}
        \end{subfigure}

        \begin{subfigure}[t]{0.25\textwidth}
            \includegraphics[width=\textwidth]{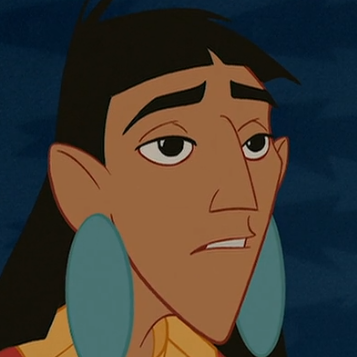}
            \caption*{\textbf{Disgust01} \textit{Boredom}}
            \label{fig:boredom}
        \end{subfigure}
        \begin{subfigure}[t]{0.25\textwidth}
            \includegraphics[width=\textwidth]{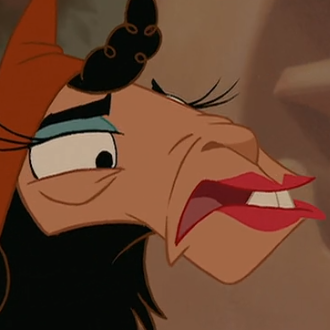}
            \caption*{\textbf{Disgust02} \textit{Disgust}}
            \label{fig:disgust}
        \end{subfigure}
        \begin{subfigure}[t]{0.25\textwidth}
            \includegraphics[width=\textwidth]{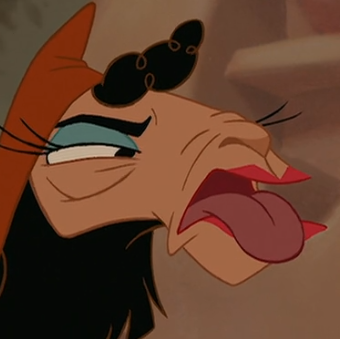}
            \caption*{\textbf{Disgust03} \textit{Loathing}}
            \label{fig:loathing}
        \end{subfigure}
    \end{center}
    \caption{Examples of \textit{Disgust} in Kuzco's Facial Expressions}
    \label{fig:charDisgust}
\end{figure}

Kuzco provides examples of the different levels of \textit{Disgust} in his
expressions (Figure~\ref{fig:charDisgust}). The changes are weak in
\textit{Boredom}, but he still displays a protruding upper lip and lowered
upper eyelids. \textit{Disgust} is easier to identify with prominent wrinkles
around his nose, a raised and jutting upper lip, lowered eyebrows, pushed up
lower eyelids due to his rising cheek muscles, and lowered upper eyelids.
\textit{Loathing} sees these aspects further exaggerated and punctuated with a
protruding tongue.

\textbf{Disgust01}: Kuzco quickly becomes \textit{bored} at a dinner because:
\begin{itemize}
    \item His hosts are not entertaining him despite his need to constantly do
    something interesting
    \item He can leave without consequence at any time
\end{itemize}

Kuzco's \textit{Boredom} is apparent from his body language---slouching forward
and leaning his elbows on the table---and his attempts to find a distraction by
playing with his fork and then by trying to start a conversation with his
hostess.

\textbf{Disgust02}: Seeing his companion eating a questionable meal fills Kuzco
with \textit{Disgust} because:
\begin{itemize}
    \item He finds the meal repulsive
    \item He does not have to eat it too
\end{itemize}

Kuzco makes his emotions clear non-verbally with several utterances of
``ewww'', shaking his head, and leaning away from the table. He is also fixated
on the event rather than averting his eyes, suggesting some level of morbid
interest.

\textbf{Disgust03}: As he continues to watch his companion eat, Kuzco's emotion
intensifies to \textit{loathing}:
\begin{itemize}
    \item He notices that his companion is enjoying his meal
    \item He cannot change his companion's behaviour or thoughts
\end{itemize}

The same signals as \textit{Disgust} mark his \textit{loathing}, but following
his facial expression and his obvious attempt to avoid vomiting further
supports this.

\vspace*{\fill}
\begin{table}[!ht]
    \centering
    \caption{Facial Sketches of \textit{Boredom} with Suggested FACS Codes}
    \label{tab:boredomFACS}
    \renewcommand{\arraystretch}{1.2}
    \begin{threeparttable}
        \begin{tabular}{P{0.6\linewidth}P{0.3\linewidth}}
            \toprule
            \begin{center}

        \includegraphics[width=0.3\linewidth]{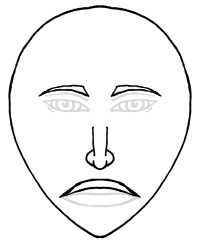}
                \end{center} & \begin{tabular}{ll}
                    \textbf{Label} & Boredom \\
                    \textbf{Intensity} & Low \\
                \end{tabular} \\ \midrule
            \textbf{Description} & \textbf{FACS Codes} \\ \midrule
            \colourRow Slightly lowered eyebrows which lowers the upper eyelids
            & AU 9* \\
            Inner eyebrows might be raised & AU 1 \\
            \colourRow No tension in the eyelids & -- \\
            No tension in the cheeks & -- \\
            \colourRow Slightly raised upper lip that juts out & AU 9 \\
            Lips might be parted & AU 25 \\
            \bottomrule
        \end{tabular}
    \end{threeparttable}
    \begin{tablenotes}

        \footnotesize
        \vspace*{0.5mm}

        \item {*} \textit{Causes change indirectly}

    \end{tablenotes}
\end{table}
\vspace*{\fill}

\begin{table}[!tb]
    \centering
    \caption{Facial Sketches of \textit{Disgust} with Suggested FACS Codes}
    \label{tab:disgustFACS}
    \renewcommand{\arraystretch}{1.2}
    \begin{threeparttable}
        \begin{tabular}{P{0.6\linewidth}P{0.3\linewidth}}
            \toprule
            \begin{center}

        \includegraphics[width=0.3\linewidth]{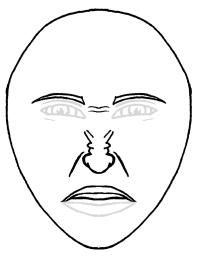}
                \end{center} & \begin{tabular}{ll}
                    \textbf{Label} & Disgust \\
                    \textbf{Intensity} & Medium \\
                \end{tabular} \\ \midrule
            \textbf{Description} & \textbf{FACS Codes} \\ \midrule
            \colourRow Eyebrows might be lowered causing the upper eyelids to
            lower & AU 9* \\
            Inner eyebrows might be raised & AU 1 \\
            \colourRow No tension in the eyelids & -- \\
            Raised cheeks causing the lower eyelids to rise & AU 9* \\
            \colourRow Nose wrinkled and drawn upwards causing wrinkles to
            appear beside and on the nose bridge & AU 9 \\
            Raised upper lip that juts out & AU 9 \\
            \colourRow Corners of the lips are drawn down and back & AU 15 \\
            \bottomrule
        \end{tabular}
    \end{threeparttable}
    \begin{tablenotes}

    \footnotesize
    \vspace*{0.5mm}

    \item {*} \textit{Causes change indirectly}

    \end{tablenotes}
\end{table}

\begin{table}[!tb]
    \centering
    \caption{Facial Sketches of \textit{Loathing} with Suggested FACS Codes}
    \label{tab:loathingFACS}
    \renewcommand{\arraystretch}{1.2}
    \begin{threeparttable}
        \begin{tabular}{P{0.6\linewidth}P{0.3\linewidth}}
            \toprule
            \begin{center}

        \includegraphics[width=0.3\linewidth]{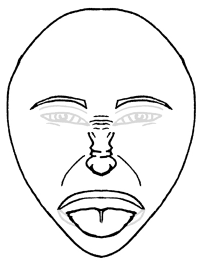}
                \end{center} & \begin{tabular}{ll}
                    \textbf{Label} & Loathing \\
                    \textbf{Intensity} & High \\
                \end{tabular} \\ \midrule
            \textbf{Description} & \textbf{FACS Codes} \\ \midrule
            \colourRow Eyebrows might be lowered causing the
            upper eyelids to lower & AU 9* \\
            Inner eyebrows might be raised & AU 1 \\
            \colourRow No tension in the eyelids & -- \\
            Crow's feet might appear beside the eyes & AU 9* \\
            \colourRow Raised cheeks causing the lower eyelids to rise & AU 9*
            \\
            Nose wrinkled and drawn upwards causing deep wrinkles to appear
            beside and on the nose bridge & AU 9 \\
            \colourRow Fully raised upper lip that juts out & AU 9 \\
            Tongue might be visible or protruding from the mouth & AD 19 \\
            \bottomrule
        \end{tabular}
    \end{threeparttable}
    \begin{tablenotes}

    \footnotesize
    \vspace*{0.5mm}

    \item {*} \textit{Causes change indirectly}

    \end{tablenotes}
\end{table}

\clearpage\section{Acceptance}\label{sec:acceptanceProfile}
\textit{Trust}, or low intensity \textit{Acceptance}, is an elementary
component in social and economic life referring to a subclass of decisions
under risk~\citep{oxfordTrust}. It is a general expectation about a subjective
probability that an individual assigns to a potential future event where they
are willing to accept risk by relying on another
(\citepg{misztal1996trust}{18--19, 24}; \citepg{rousseau1998not}{395, 399};
\citepg{nooteboom2002trust}{48}). Although there is a clear sociocultural
element, ``social emotions'' use some of the same brain functioning areas as
the other agreed-on primary
emotions~\citep[p.~8024--8025]{immordino2009neural}. The emotional element of
\textit{Trust}, or ``affective trust'', builds on past experiences with,
feelings of security, confidence, and satisfaction towards, and the perceived
level of selfless concern demonstrated by a partner regardless of what the
future holds~\citep[p.~96, 109]{rempel1985trust}. In these relationships, the
individual takes risks when disclosing information, accepting promises, and
sacrificing immediate rewards for future gains. It is necessary for coping with
the volume and complexity of information for decision-making because
\textit{Trust} reduces the number of outcomes to consider, effectively
narrowing the selection space. This makes it easier to accept risk when some
factors are unknown (\citepg{misztal1996trust}{19--20};
\citepg{nooteboom2002trust}{79--81}). Memory is a key component in this as it
allows the individual to use previous experiences to predict how likely their
partner will act in a manner sensitive to the individual's goals.

At the least intense, \textit{Acceptance}, the individual does not require a
personal relationship, likely because of the implicit agreement that others
respect and adhere to established values, norms, and behaviours or habits in
addition to internalized norms or values of ethical conduct
(\citepg{misztal1996trust}{21--23}; \citepg{nooteboom2002trust}{11, 67}). As
the relationship becomes personal and intensity increases to \textit{Trust},
the individual begins to take their partner's qualities and past actions into
account. In \textit{Admiration}, the individual weighs their partner's actions
and qualities less in favour of believing that their partner has their best
interests at heart more~\citep[p.~97]{rempel1985trust}, reducing the overall
cognitive load of risk assessment. \textit{Admiration} is also a positive
response to extraordinary displays of skill, talent, and achievement
(\citeg{oxfordAdmire}; \citepg{algoe2009witnessing}{107--108}). This might be
due to a belief that the admired will do what is best in the imagined fictional
relationship, especially in the absence of past experience with them.

\paragraph{Signs of \textit{Acceptance}}
The general action tendencies in \textit{Trust} are to approach, interact, and
sometimes touch another to convey that they are valued and secure in the
relationship~\citep[p.~278--279]{lazarus1991emotion}. This often includes
gestures of warmth, tenderness, interest, and concern for the other. As the
intensity grows to \textit{Admiration}, individuals increase the risk they are
willing to make for the other~\citep[p.~111]{rempel1985trust} and focus more
on the positive aspects that they could emulate in
themselves~\citep[p.~185--186]{smith2000assimilative}. This energizes the self
and motivates both relationship and skill building with the
other~\citep[p.~111]{algoe2009witnessing}, advancing learning via mimicry, and
the pursuit of other personal goals.

Creating a closer relationship enables individuals to learn more from the
admired while also increasing and sharing in their
prestige\footnote{Researchers consider prestige to be fundamentally different
from dominance because it evolved to improve human cultural capacity through
social learning where deference is given freely rather than under
threat~\citep[p.~167, 170]{henrich2001evolution}.} through praise. This speeds
cultural learning by allowing information transmission between individuals and
creating a common knowledge base. Some contradictory studies have found that
\textit{Admiration}, while inspiring to individuals, might not motivate them to
action~\citep[p.~790--791]{van2011envy}. Instead, individuals happily surrender
to the admired, potentially due to the difficulty of self-improvement compared
to praising someone else for completing a task that they believe they could do
themselves if they wanted to.

Physiologically, the body often feels warm and has an elevated heart rate when
experiencing \textit{Trust}~\citep[p.~115]{algoe2009witnessing}. However, at
the highest intensity \textit{Admiration} there might be chills instead.
Reminiscent of the ``flight'' \textit{Fear} response, this tends to happen when
the individual is not familiar with the admired individual and have no evidence
that there will not be an altercation that they believe themselves ill-equipped
to handle. There might also be ``tears in the eyes'' and/or a lump in the
throat at high intensities~\citep[p.~110]{algoe2009witnessing}.

\paragraph{Characteristic Facial Expression}
There are no known, empirically supported facial expressions for
\textit{Acceptance}, \textit{Trust}, or \textit{Admiration}. These sketches
(Tables~\ref{tab:acceptanceFACS}, \ref{tab:trustFACS}, and
\ref{tab:admirationFACS}) rely on the definition of each emotion and facial
movements of \textit{Joy}, which motivates individuals to engage in beneficial
activities. They also draw from evidence about the meanings of individual
facial features, such as a smile and briefly raised eyebrows encouraging social
contact (\citepg{oatley1992best}{93};
\citepg{smith_scott_mandler_1997}{248--249}).
\begin{itemize}
    \item A near complete relaxation of all muscles into a neutral or calm
    expression could be \textit{Acceptance}, conveying that the other is
    following an expectation about a known routine. It might appear to have a
    slight smile, shown by lifting the corners of the mouth.

    \item A confident, inviting expression could convey \textit{Trust}. Raised
    eyebrows and a closed-mouth smile while maintaining eye contact with the
    other could convey this intent.

    \item The face could convey \textit{Admiration} with a small, possibly
    open-mouthed, smile and widened eyes with raised, curved eyebrows angled
    upwards in the middle, conveying the child-like wonder and happiness that
    typically accompanies this emotion. A ``sparkling'' effect in the eyes
    describing excitement or happiness with another might be
    present\footnote{In addition to descriptions of \textit{Admiration}, this
        was inspired by artistic representations of the emotion such as those of
        Charles Le Brun~\citep{lebrun1760admiration}.}.
\end{itemize}

\paragraph{Examples}
Tiana from Disney's \textit{The Princess and the
    Frog}~\citep{princessandthefrog} has built her life around the belief that
working hard is the key to making your dreams come true. For Tiana, that wish
is to own a restaurant.

\begin{figure}[!tb]
    \begin{center}
        \begin{subfigure}[t]{0.25\textwidth}
            \includegraphics[width=\textwidth]{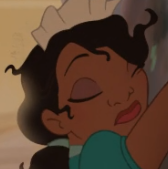}
            \caption*{\textbf{Trust00} \textit{Neutral}}
            \label{fig:tianaSleep}
        \end{subfigure}

        \begin{subfigure}[t]{0.25\textwidth}
            \includegraphics[width=\textwidth]{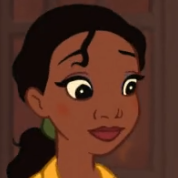}
            \caption*{\textbf{Trust01} \textit{Acceptance}}
            \label{fig:acceptance}
        \end{subfigure}
        \begin{subfigure}[t]{0.25\textwidth}
            \includegraphics[width=\textwidth]{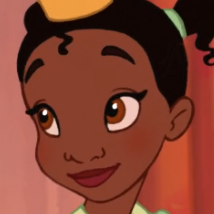}
            \caption*{\textbf{Trust02} \textit{Trust}}
            \label{fig:trust}
        \end{subfigure}
        \begin{subfigure}[t]{0.25\textwidth}
            \includegraphics[width=\textwidth]{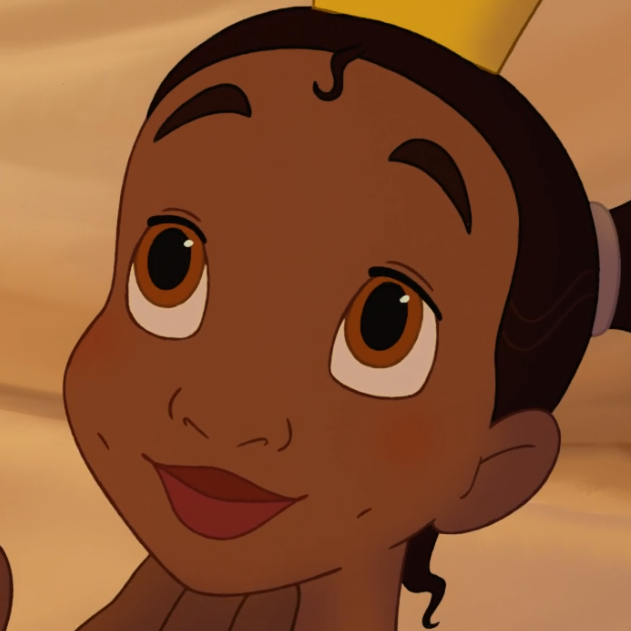}
            \caption*{\textbf{Trust03} \textit{Admiration}}
            \label{fig:admiration}
        \end{subfigure}
    \end{center}
    \caption{Examples of \textit{Trust} in Tiana's Facial Expressions}
    \label{fig:charAcceptance}
\end{figure}

Tiana has slightly upturned lip corners and an otherwise neutral expression in
\textit{Acceptance}, a happy expression with raised eyebrows in \textit{Trust},
and a more pronounced expression of \textit{Trust} with her eyes sparkling in
\textit{Admiration} (Figure~\ref{fig:charAcceptance}).

\textbf{Trust01}: Tiana has regular shifts at the diner where she serves
customers. Due to the established server-customer routine that has a
significant social element, she experiences \textit{acceptance}:
\begin{itemize}
    \item As a waitress, Tiana puts herself at social risk while attending
    customers because she has limited avenues to handle them should they behave
    poorly

    \item In return for her services, she expects her customers to treat her
    courteously, both socially and when paying their bill
\end{itemize}

This is shown through Tiana's calm and confident approach of customers and
general appearance of ease.

\textbf{Trust02}: Tiana's friend suddenly leans in close and whispers loudly to
her while Tiana's mother reads them a fairy tale, causing her to experience
\textit{trust}:
\begin{itemize}
    \item Her friend is excitable and has a deep love of fairy tales

    \item Tiana has an established, positive relationship with her friend

    \item She values her friend's interests
\end{itemize}

Tiana shows care for her friend by being physically close to them. Her head and
eye movement also indicate that she is giving her friend her complete attention.

\textbf{Trust03}: Tiana openly \textit{admires} her hard-working father when he
tells her that she can wish on stars, but she must also work hard for what she
wants:
\begin{itemize}
    \item Tiana has a strongly established, positive relationship with her
    father

    \item She feels energized to work hard with her father to open a restaurant
    together
\end{itemize}

Leaning towards her father, giving him her full attention, and her relaxed
posture all indicate Tiana's \textit{Admiration}.

\begin{table}[!tb]
    \centering
    \caption{Facial Sketches of \textit{Acceptance} with Suggested FACS
Codes}
    \label{tab:acceptanceFACS}
    \renewcommand{\arraystretch}{1.2}
    \begin{tabular}{P{0.6\linewidth}P{0.3\linewidth}}
            \toprule
            \begin{center}

        \includegraphics[width=0.3\linewidth]{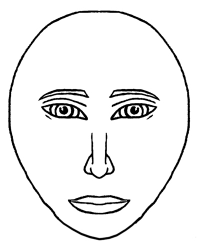}
                \end{center} & \begin{tabular}{ll}
                    \textbf{Label} & Acceptance \\
                    \textbf{Intensity} & Low \\
                \end{tabular} \\ \midrule
            \textbf{Description} & \textbf{FACS Codes} \\ \midrule
            \colourRow Eyebrows might be slightly raised & AU 1+2 \\
            No tension in the eyelids & -- \\
            \colourRow Corners of the mouth might be lifted & AU 12 \\
            \bottomrule
        \end{tabular}
\end{table}

\begin{table}[!tb]
    \centering
    \caption{Facial Sketches of \textit{Trust} with Suggested FACS Codes}
    \label{tab:trustFACS}
    \renewcommand{\arraystretch}{1.2}
    \begin{tabular}{P{0.6\linewidth}P{0.3\linewidth}}
            \toprule
            \begin{center}

        \includegraphics[width=0.3\linewidth]{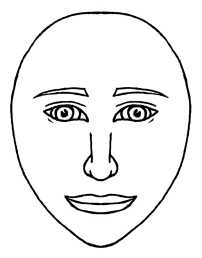}
                \end{center} & \begin{tabular}{ll}
                    \textbf{Label} & Trust \\
                    \textbf{Intensity} & Medium \\
                \end{tabular} \\ \midrule
            \textbf{Description} & \textbf{FACS Codes} \\ \midrule
            \colourRow Smooth forehead skin & -- \\
            Eyebrows raised & AU 1+2 \\
            \colourRow Neutral eyelids & -- \\
            Eye contact with subject of interest & 69 \\
            \colourRow Corners of the lips are drawn back and up & AU 12 \\
            Closed mouth & -- \\
            \bottomrule
        \end{tabular}
\end{table}

\begin{table}[!ht]
    \centering
    \caption{Facial Sketches of \textit{Admiration} with Suggested FACS
Codes}
    \label{tab:admirationFACS}
    \renewcommand{\arraystretch}{1.2}
    \begin{threeparttable}
        \begin{tabular}{P{0.6\linewidth}P{0.3\linewidth}}
            \toprule
            \begin{center}

        \includegraphics[width=0.6\linewidth]{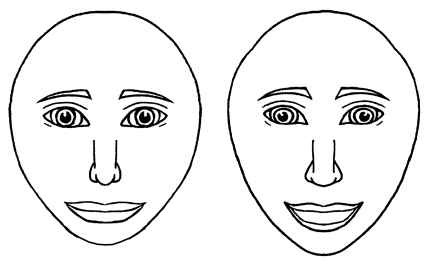}
                \end{center} & \begin{tabular}{ll}
                    \textbf{Label} & Admiration \\
                    \textbf{Intensity} & High \\
                \end{tabular} \\ \midrule
            \textbf{Description} & \textbf{FACS Codes} \\ \midrule
            \colourRow Smooth forehead skin & -- \\
            Raised, curved eyebrows with the inner corners angled upwards & AU
            1+2 \\
            \colourRow Raised upper eyelids & AU 5 \\
            Relaxed lower eyelids & -- \\
            \colourRow Eyes might appear to have a ``sparkling'' effect &
            63*+AU 5* \\
            Gaze might be cast upwards & 63 \\
            \colourRow Eye pupils might be dilated & -- \\
            Corners of the lips are drawn back and up & AU 12 \\
            \colourRow The jaw might be slack, allowing the mouth to open
            slightly & AU 25+26 \\
            \bottomrule
        \end{tabular}
    \end{threeparttable}
    \begin{tablenotes}

        \footnotesize
        \vspace*{0.5mm}

        \item {*} \textit{Causes change indirectly}

    \end{tablenotes}
\end{table}

\clearpage\section{Interest}
\textit{Interest} motivates individuals to achieve goals by enabling sustained
periods of attention on a single object or thought
(\citepg{tomkins1962affect}{186, 188--189, 191}; \citepg{izard1977human}{212,
236--238}) and explains how organisms are able to select a manageable number of
stimuli to attend to from a never-ending stream of internal and external
stimuli sources~\citep[p.~321]{izard1971face}. As the most prevalent emotion in
healthy individuals, \textit{Interest} motivates long-term commitment and
effort which are necessary in interpersonal relationships, self-development,
creativity, and the motivation to search for approaching and exploring new
experiences (\citepg{silvia2008interest}{58}). The more stimulation an object
or thought brings, the more likely that \textit{Interest} will arise. What
provides the necessary stimulation to arouse this emotion differs by
personality. \textit{Interest} critically involves attention, supported by an
increase in energy expenditure, to increase physical or symbolic interaction
with an object via fine tuning and stretching of the relevant
sensory-perceptual-cognitive mechanisms~\citep[p.~238--239]{izard1971face}.
\textit{Interest} can potentially increase predictive processes, consequently
affecting decision-making and stress levels. Subjectively, people tend to
describe \textit{Interest} as pleasant, marked by devotional
attention~\citep[p.~832]{smith1985patterns} but it becomes stressful as
intensity rises~\citep[p.~438]{warm2008vigilance}.

At the most intense, \textit{Vigilance}, an organism sustains its attention
while waiting for a stimulus to appear at an unknown time~\citep[p.~433,
435]{warm2008vigilance} to heighten their ability to protect themselves from
danger, identify self-benefits, and clarify changes in the
environment~\citep{oxfordVigilance}. Therefore, \textit{Anticipation} is
assumed to be a milder form of \textit{Vigilance} where organisms have made a
prediction about some future event and are waiting for it to occur.

Researchers contest \textit{Interest}'s status as an emotion because of its
ties to the orientation reflex, an involuntary biological survival mechanism
for rapidly directing attention towards immediate changes in the environment
that are novel or threatening before the organism evaluates
them~\citep[p.~1144]{friedman2009brain}. However, empirical, physiological
evidence collected from individuals who had a lack of interest in their daily
lives strengthen its status as an emotion. Forcing themselves to work out of
necessity, duty, or pride while \textit{Interest} was absent led to apathy and
excessive physical fatigue. Researchers suggest that the physical symptoms of
the apathetic state are due to a reduction of glucose usage by the body,
regardless of how much is present~\citep[p.~281]{rennie1942hypoglycemia}, a
type of ``vegetative state'' where the body subconsciously prepares the body for
rest even though it is working~\citep[p.~205]{alexander1944psychosomatic}. This
suggests that \textit{Interest} serves another purpose: conserving an
individual's energy when there is nothing worthwhile attending to so that there
is ample energy available when a pressing event arises.

\paragraph{Signs of \textit{Interest}}
In \textit{Interest}, an individual focuses on certain aspects of their
environment and knowledge to gather information they believe is necessary to
accurately appraise their current situation~\citep[p.~213,
225]{izard1977human}. Another emotion typically follows an
\textit{Interest}-driven appraisal once the individual has gathered the
information that they want. Head and eye tracking might accompany this emotion
to maintain the individual's focus.

Attention is usually maintained for short periods of time unless the current
activity demands a prolonged arousal state~\citep[p.~190]{tomkins1962affect}.
Maintaining attention and the continuous use of cognitive processes makes
\textit{Interest} mentally and physically draining. This results in increased
stress levels, which could trigger other negative emotions. This makes it
difficult to remain highly-attentive for long periods of time, an effect known
as the vigilance decrement~\citep[p.~434--435]{warm2008vigilance}. Training can
improve this effect but it worsens with age.

\paragraph{Characteristic Facial Expression}
Researchers have described and empirically tested a facial expression for
\textit{Interest} (\citepg{izard1977human}{215};
\citepg{tomkins1962affect}{185, 187}; \citepg{silvia2008interest}{57}), but no
descriptions were found for \textit{Anticipation} or \textit{Vigilance}.
However, the expression for \textit{Anger} is ambiguous if changes are not
observed in all three facial areas---eyebrows, eyes, and lower
face~\citep[p.~83]{ekman2003unmasking}. If only the eyebrows change, the
individual might be intently concentrating on something instead of experiencing
\textit{Anger}, which aligns with the integral role of attention in
\textit{Anticipation} and \textit{Vigilance}. The eyes have a similar
description when they alone change, as well as the mouth when the lips press
together. Therefore, the assumption is that the expressions for
\textit{Anticipation} and \textit{Vigilance} involve similar facial changes as
\textit{Anger} and the surrounding context differentiate between them when
there are changes in all three facial areas.

\textit{Interest} is shown in the face primarily by the eyes, which fixate on a
target (Tables~\ref{tab:interestFACS}, \ref{tab:anticipationFACS}, and
\ref{tab:vigilenceFACS}). The lower eyelids might rise to further sharpen
vision. If reverie or reflective problem solving caused \textit{Interest}, the
eyes tend to have a ``faraway'' look indicating that the person is not in the
present moment. There might be minor variances of eyebrow height in either
direction of movement. As the intensity increases to \textit{Anticipation},
there are similar changes in the eyes that become more exaggerated as the
eyebrows change. Directing more attention to a single source in
\textit{Vigilance}, the eyes either widen, as in \textit{Interest}, or squint
to temporarily improve the eye's focus like a pinhole camera. The eyes might
cycle between widening and squinting to reduce stress on the underlying facial
muscles. The inner corners of the eyebrows are likely further drawn together
due to the amount of effort exerted, possibly exaggerated due to stress. If
the individual is imagining a scenario, their eyes might close.

The lower face is not essential to the expression of \textit{Interest} because
individuals tend to have quirks, such as biting their lip. Generally, a
slackness in the jaw and mouth marks \textit{Interest} which can persist as
intensity increases. Alternatively, the lips might press together and the jaw
might clench in an attempt to contain energy or stress. If the jaw is not
clenched, it is likely slack as in \textit{Interest}. With a further increase
in intensity, the mouth and jaw further clench together, often accompanied by a
twisting of the mouth to one side.

\paragraph{Examples}
Elastigirl from Pixar's \textit{The Incredibles 2}~\citep{theincredibles2} is a
superhero and parent, both of which demand her attention to do well.

\begin{figure}[!tb]
    \begin{center}
        \begin{subfigure}[t]{0.25\textwidth}
            \includegraphics[width=\textwidth]{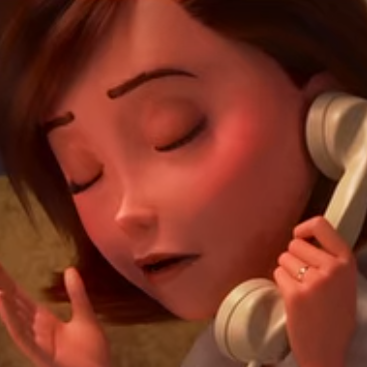}
            \caption*{\textbf{Interest00} \textit{Neutral}}
            \label{fig:elastigirlSleep}
        \end{subfigure}

        \begin{subfigure}[t]{0.25\textwidth}
            \includegraphics[width=\textwidth]{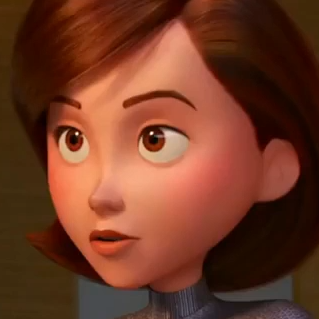}
            \caption*{\textbf{Interest01} \textit{Interest}}
            \label{fig:Interest}
        \end{subfigure}
        \begin{subfigure}[t]{0.25\textwidth}
            \includegraphics[width=\textwidth]{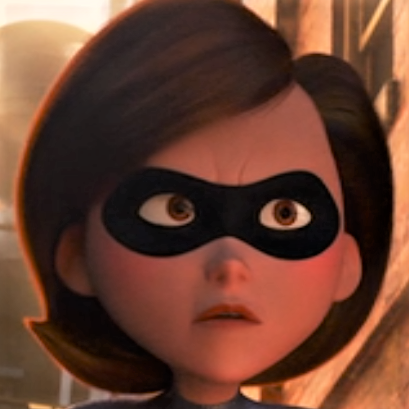}
            \caption*{\textbf{Interest02} \textit{Anticipation}}
            \label{fig:Anticipation}
        \end{subfigure}
        \begin{subfigure}[t]{0.25\textwidth}
            \includegraphics[width=\textwidth]{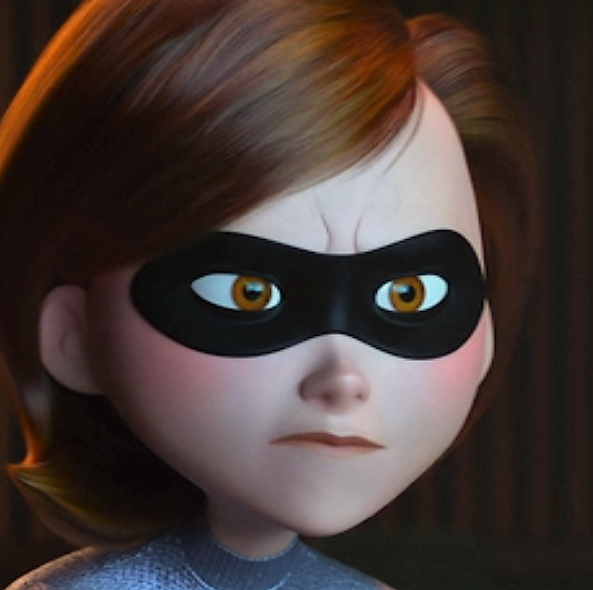}
            \caption*{\textbf{Interest03} \textit{Vigilance}}
            \label{fig:Vigilance}
        \end{subfigure}
    \end{center}
    \caption{Examples of \textit{Interest} in Elastigirl's Facial Expressions}
    \label{fig:charAnticipation}
\end{figure}

She displays \textit{Interest} through her raised eyebrows, maintenance of eye
contact, and slackness in her jaw shown by her slightly parted lips
(Figure~\ref{fig:charAnticipation}). As the intensity grows to
\textit{Anticipation}, Elastigirl's eyebrows come together and lower, forcing
her upper eyelids to lower as well, focusing her vision. Her jaw is tighter,
but her lips are still slightly parted. At the highest intensity, her furrowed
brow and narrowed eyes are more prominent. She is clenching her jaw with lips
pressed together with one side twisted up in a display of extreme concentration.

\textbf{Interest01}: Elastigirl's husband says that there is a note and package
for her, but does not tell her what it says. She focuses her attention on him
in \textit{Interest} while he hands her the note because:
\begin{itemize}
    \item The note came with a mysterious package from work, so it is likely
    related to a work goal
    \item The package excites her husband, so she tries to gather more
    information from him about it
\end{itemize}

Elastigirl's \textit{Interest} is shown by tracking her husband's face with her
head and eyes. Her posture also conveys \textit{Interest} by being relaxed and
still while she listens to him.

\textbf{Interest02}: A police scanner picks up a conversation about potential
criminal activity around the opening ceremony for a new train. Elastigirl
experiences \textit{Anticipation} because:
\begin{itemize}
    \item Her work tasked her with the city's protection and suspicious
    activity could negatively affect it
    \item Listening to the police scanner should provide critical details about
    the location and severity of the threat, helping her decide if they need her
\end{itemize}

Her \textit{Anticipation} is apparent via her increased attention to the police
scanner, evident by her stillness, and her eyes that appear alert but do not
seem to be focusing on anything.

\textbf{Interest03}: Elastigirl's instincts say that her assigned case is not
closed and she arrested the wrong person. She becomes \textit{vigilant} when an
oddity appears in her suit's camera footage because:
\begin{itemize}
    \item With the criminal still potentially free in the city, her goal to
    protect the city is at risk
    \item Reviewing the footage could reveal critical information, reducing her
    uncertainty regarding the case's status
\end{itemize}

Elastigirl's tense posture, leaning towards the screen, and eye fixation is
evidence of her \textit{Vigilance}.

\begin{table}[!tb]
    \centering
    \caption{Facial Sketches of \textit{Interest} with Suggested FACS Codes}
    \label{tab:interestFACS}
    \renewcommand{\arraystretch}{1.2}
    \begin{tabular}{P{0.6\linewidth}P{0.3\linewidth}}
        \toprule
        \begin{center}

\includegraphics[width=0.3\linewidth]{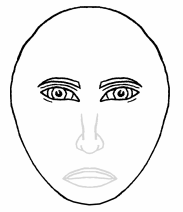}
        \end{center} & \begin{tabular}{ll}
            \textbf{Label} & Interest \\
            \textbf{Intensity} & Low \\
        \end{tabular} \\ \midrule
        \textbf{Description} & \textbf{FACS Codes} \\ \midrule
        \colourRow Eyebrows are marginally lifted or lowered
        & AU 4 \\
        Upper eyelids might be raised & AU 5 \\
        \colourRow Lower eyelids might be tensed & AU 7 \\
        Eyes are fixed on a target & 69, M69 \\
        \colourRow Jaw might be slightly slack & AU 26 \\
        Lips might be parted & AU 25 \\
        \bottomrule
    \end{tabular}
\end{table}

\begin{table}[!tb]
    \centering
    \caption{Facial Sketches of \textit{Anticipation} with Suggested FACS
Codes}
    \label{tab:anticipationFACS}
    \renewcommand{\arraystretch}{1.2}
    \begin{tabular}{P{0.6\linewidth}P{0.3\linewidth}}
        \toprule
        \begin{center}

\includegraphics[width=0.3\linewidth]{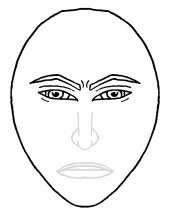}
        \end{center} & \begin{tabular}{ll}
            \textbf{Label} & Anticipation \\
            \textbf{Intensity} & Medium \\
        \end{tabular} \\ \midrule
        \textbf{Description} & \textbf{FACS Codes} \\ \midrule
        \colourRow Eyebrows lowered & AU 4 \\
        Upper eyelids are raised OR lowered & AU 5 OR AU 7 \\
        \colourRow Lower eyelids might be tensed & AU 7 \\
        Eyes are fixed on a target & 69, M69 \\
        \colourRow Jaw is slack or clenched & AU 26 OR AU 31 \\
        Lips might be parted & AU 25 \\
        \bottomrule
    \end{tabular}
\end{table}

\begin{table}[!tb]
    \centering
    \caption{Facial Sketches of \textit{Vigilance} with Suggested FACS Codes}
    \label{tab:vigilenceFACS}
    \renewcommand{\arraystretch}{1.2}
    \begin{tabular}{P{0.6\linewidth}P{0.3\linewidth}}
        \toprule
        \begin{center}

\includegraphics[width=0.6\linewidth]{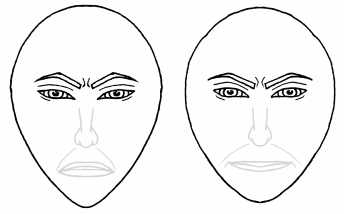}
        \end{center} & \begin{tabular}{ll}
            \textbf{Label} & Vigilance \\
            \textbf{Intensity} & High \\
        \end{tabular} \\ \midrule
        \textbf{Description} & \textbf{FACS Codes} \\ \midrule
        \colourRow Eyebrows are lowered and the inner corners
        are brought together & AU 4 \\
        Upper eyelids are lowered (squinting) or raised & AU 7 OR AU 5 \\
        \colourRow Lower eyelids be tensed & AU 7 \\
        Eyes are fixed on a target & 69, M69 \\
        \colourRow Jaw is slack or clenched & AU 26 OR AU 31 \\
        Lips might be parted & AU 25 \\
        \bottomrule
    \end{tabular}
\end{table}

\clearpage\section{Surprise}\label{sec:surpriseProfile}
\textit{Surprise} is a response to sudden and unexpected events that an
individual is ill-prepared for (\citepg{ekman2007emotions}{149};
\citepg{occ2022}{144--146}; \citepg{izard1977human}{277}). It is not possible to
feel \textit{Surprise} when the individual makes a correct prediction about
something~\citep[p.~151]{ekman2007emotions}. The main function of
\textit{Surprise} is to prepare the individual so that they can effectively
handle rapidly changing scenarios and their
consequences~\citep[p.~291]{izard1971face}. It forces the clearing ongoing
cognitive activities to make way for the immediately following, likely more
appropriate, emotion~\citep[p.~281]{izard1977human}. \textit{Surprise} itself
is not pleasant or unpleasant, but might subjectively feel like it. People
often consider \textit{Surprise} as pleasant because it typically leads to a
pleasant or interesting event~\citep[p.~832, 836]{smith1985patterns}. When
unpleasant emotions follow \textit{Surprise}, it is usually because the
individual was caught off-guard when they wanted to be prepared and they
subsequently feel overexposed.

For this profile, differing degrees of certainty with respect to what
individuals should do differentiate \textit{Surprise} intensity. At the lowest
level, \textit{Distraction}, individuals are entirely focused on one task and
are effectively blind to everything else. By the middle level,
\textit{Surprise}, the uncertainty has grown such that the individual's mind
has gone blank, but is still cognitively aware enough for other processes to
activate and produce a more appropriate emotion for the event---the brief pause
needed to effectively evaluate the situation. In \textit{Amazement}, the
uncertainty is so severe that the individual does not know what to do and they
are simply focused on taking in as much information about the unexpected person,
object, or event as possible.

\textit{Surprise} is the briefest of emotions---lasting mere seconds---and
lacks some of the characteristics of the other emotions
(\citep[p.~150--151]{ekman2007emotions}; \citepg{izard1977human}{280--281}).
This often results in its re-evaluation to determine if it is an emotion at
all. There is further evidence against its candidacy because it is often
confused with \textit{Fear} in preliterate
cultures~\citep[p.~10]{ekman2007emotions} whereas literate cultures do not,
potentially due to their exposure to expressions portrayed in the
media~\citep[p.~714, 716]{ekman1987universals} and since \textit{Fear} and
\textit{Surprise} are both caused by extreme changes in stimulation. This might
be due to the tendency for individuals to interchange \textit{Surprise} and the
startle reflex, a motor response which protects vulnerable body areas and
enables escape in sudden encounters. This reflex is part of the startle response
defence mechanism underlying the mostly unconscious response to sudden, intense
stimuli~\citep[p.~288]{davis1984mammalian}. \textit{Surprise} serves its own
functional purpose (\citepg{occ2022}{146}; \citepg{izard1977human}{281}) and its
briefness distinguishes it from other emotions and involuntary physical
reactions such as startling.

\paragraph{Signs of \textit{Surprise}}
During \textit{Surprise}, the individual's mind goes blank and they are unsure
of what to do~\citep[p.~278--279]{izard1977human}. Their muscles contract in
preparation of movement if needed with a tension level comparable to
\textit{Interest}. This might look like the freeze response of \textit{Fear},
which can account for its confusion with it. This, effectively, gives
individuals a moment to figure out what is going
on~\citep[p.~148--149]{ekman2007emotions}. \textit{Surprise} is difficult to
manage due to its unexpectedness and sudden onset, but its short duration does
not necessitate this need under typical circumstances.

\paragraph{Characteristic Facial Expression}
Ekman proposes that there are four types of \textit{Surprise} with distinct
facial expressions~\citep[p.~42--43]{ekman2003unmasking}. Plutchik's
\textit{Surprise} seems to be the type requiring changes in all three facial
areas. The other types only need two out of the three facial areas, becoming an
expression of \textit{Surprise} when the last area changes. However, Plutchik
describes \textit{Distraction} and \textit{Amazement} as different intensities
of \textit{Surprise}, implying that there are changes in all facial areas that
become more exaggerated with intensity. Expressions of \textit{Surprise} here
use this approach.

\textit{Surprise} registers on all three facial areas
(Tables~\ref{tab:distractionFACS}, \ref{tab:surpriseFACS}, and
\ref{tab:amazementFACS}) and there must be changes in all of them for the
expression to be unambiguous due to this emotion's closeness to \textit{Fear}
(\citepg{ekman2003unmasking}{37, 45}). In general, the face has less tension
when expressing \textit{Surprise} than \textit{Fear}.

The eyebrows rise and curve, causing more skin to show between the eyes and
eyebrows. This movement sometimes causes horizontal wrinkles to appear on the
forehead, especially in older adults. The eyes open wide and the upper eyelids
rise to show more of the sclera in both \textit{Surprise} and \textit{Fear}.
The relaxed lower eyelids distinguish between the expressions because there is
little or no tension in them when expressing \textit{Surprise}.

The jaw also lacks the tension that \textit{Fear} causes, simply dropping open.
This can also indicate the intensity of \textit{Surprise}, with the jaw
dropping more as the intensity increases.

\paragraph{Examples}
The title character in Disney's \textit{Mulan}~\citep{mulan} disguises herself
to enter a military training camp, which is full of people and experiences that
are completely different from what she has been exposed to her entire life.

\begin{figure}[!b]
    \begin{center}
        \begin{subfigure}[t]{0.25\textwidth}
            \includegraphics[width=\textwidth]{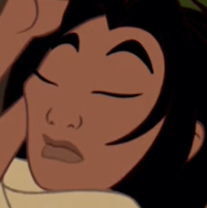}
            \caption*{\textbf{Surprise00} \textit{Neutral}}
            \label{fig:mulanSleep}
        \end{subfigure}

        \begin{subfigure}[t]{0.25\textwidth}
            \includegraphics[width=\textwidth]{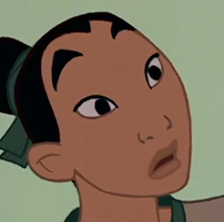}
            \caption*{\textbf{Surprise01} \textit{Distraction}}
            \label{fig:distraction}
        \end{subfigure}
        \begin{subfigure}[t]{0.25\textwidth}
            \includegraphics[width=\textwidth]{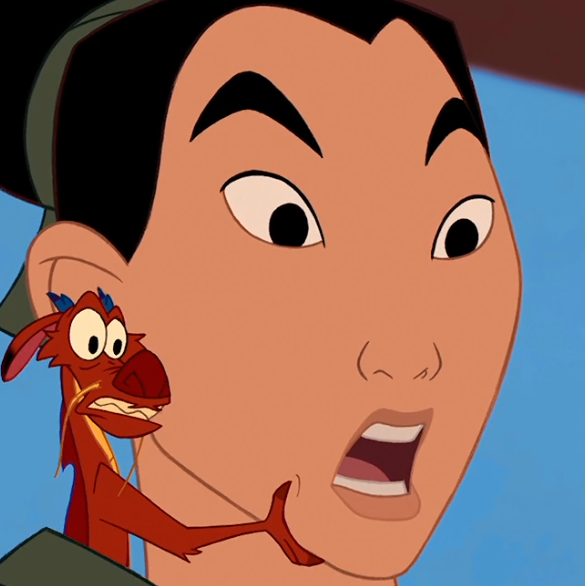}
            \caption*{\textbf{Surprise02} \textit{Surprise}}
            \label{fig:Surprise}
        \end{subfigure}
        \begin{subfigure}[t]{0.25\textwidth}
            \includegraphics[width=\textwidth]{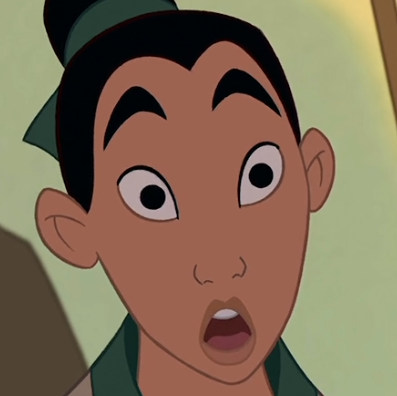}
            \caption*{\textbf{Surprise03} \textit{Amazement}}
            \label{fig:Amazement}
        \end{subfigure}
    \end{center}
    \caption{Examples of \textit{Surprise} in Mulan's Facial Expressions}
    \label{fig:charSurprise}
\end{figure}

The intensity of Mulan's \textit{Surprise} are in expressions during her first
day at a military camp (Figure~\ref{fig:charSurprise}). Her eyebrows rise and
curve making it seem that there is more skin between her eyes and eyebrows. Her
eyes are wide and expose her sclera, but show no tension in the lower lids as
they still appear to have the same shape as her neutral expression. Her jaw has
dropped, parting her lips. These changes become more exaggerated with each
intensity level.

\textbf{Surprise01}: Mulan is \textit{distracted} at the start of her first
fighting lesson:
\begin{itemize}
    \item Her captain threw several staffs to the recruits quickly without
    warning, but Mulan knew that she would be practising with one that day
    \item Mulan had a few seconds to adjust  to the situation as others caught
    their staff
\end{itemize}

She reacts by fixating her eyes and head orientation towards the staff as it
approaches and remaining still with one arm out to catch it.

\textbf{Surprise02}: Mulan was so engrossed by the conversation with her friend
that she is \textit{surprised} when she almost collides with another group of
recruits---one of which has just loudly told others to ``Look!'' at his chest
tattoo:
\begin{itemize}
    \item She was not paying attention to her surroundings and had been keeping
    her distance from others, so being that close to other people was unexpected
    \item The talking was loud and close to her, contrasting with her other
    quiet conversation
\end{itemize}

Her \textit{surprise} is apparent when she immediately stops her conversation
and movement, fixating her eyes on the people in front of her, suggesting that
Mulan is trying to understand the situation. Her straightened posture also
suggests \textit{surprise} when she reflexively draws her shoulders and head
back.

\textbf{Surprise03}: Mulan and the recruits are \textit{amazed} by their
captain's precise and graceful handling of the staff which they are only being
introduced to:
\begin{itemize}
    \item Mulan likely assumed that her captain would be showing them basic
    movements with the staff
    \item The captain had only briefly taken a resting stance before suddenly
    launching into a display of his skill
    \item Compared to the recruits, the difference in skill is unparalleled
\end{itemize}

Mulan's \textit{amazement} is shown through her body language: an inability to
move and the relaxation of her shoulders and hands as her staff slowly leans
away from her while fixating her eyes on her captain. This suggests that she is
diverting all of her attention to him.

\begin{table}[!tb]
    \centering
    \caption{Facial Sketches of \textit{Distraction} with Suggested FACS Codes}
    \label{tab:distractionFACS}
    \renewcommand{\arraystretch}{1.2}
    \begin{tabular}{P{0.6\linewidth}P{0.3\linewidth}}
        \toprule
        \begin{center}

\includegraphics[width=0.3\linewidth]{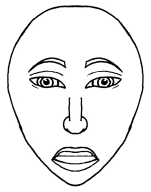}
        \end{center} & \begin{tabular}{ll}
            \textbf{Label} & Distraction \\
            \textbf{Intensity} & Low \\
        \end{tabular} \\ \midrule
        \textbf{Description} & \textbf{FACS Codes} \\ \midrule
        \colourRow Raised, curved eyebrows that might produce
        wrinkles on the forehead & AU 1+2 \\
        Upper eyelids might be raised & AU 5 \\
        \colourRow Slack jaw causing the lips and teeth to
        part & AU 25+26 \\
        No tension in the mouth & -- \\
        \bottomrule
    \end{tabular}
\end{table}

\begin{table}[!tb]
    \centering
    \caption{Facial Sketches of \textit{Surprise} with Suggested FACS Codes}
    \label{tab:surpriseFACS}
    \renewcommand{\arraystretch}{1.2}
    \begin{tabular}{P{0.6\linewidth}P{0.3\linewidth}}
        \toprule
        \begin{center}

\includegraphics[width=0.3\linewidth]{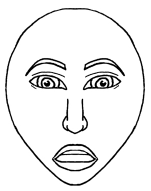}
        \end{center} & \begin{tabular}{ll}
            \textbf{Label} & Surprise \\
            \textbf{Intensity} & Medium \\
        \end{tabular} \\ \midrule
        \textbf{Description} & \textbf{FACS Codes} \\ \midrule
        \colourRow Raised, curved eyebrows that might produce
        wrinkles on the forehead & AU 1+2 \\
        Raised upper eyelids OR blinking, revealing the sclera above the iris
        & AU 5 OR AU 45 \\
        \colourRow Lower eyelids are relaxed & -- \\
        Slack jaw causing the lips and teeth to part & AU 25+26 \\
        \colourRow No tension in the mouth & -- \\
        \bottomrule
    \end{tabular}
\end{table}

\begin{table}[!tb]
    \centering
    \caption{Facial Sketches of \textit{Amazement} with Suggested FACS Codes}
    \label{tab:amazementFACS}
    \renewcommand{\arraystretch}{1.2}
    \begin{tabular}{P{0.6\linewidth}P{0.3\linewidth}}
        \toprule
        \begin{center}

\includegraphics[width=0.3\linewidth]{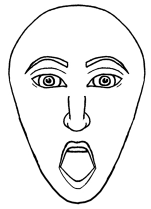}
        \end{center} & \begin{tabular}{ll}
            \textbf{Label} & Amazement \\
            \textbf{Intensity} & High \\
        \end{tabular} \\ \midrule
        \textbf{Description} & \textbf{FACS Codes} \\ \midrule
        \colourRow Raised, curved eyebrows that might produce
        wrinkles on the forehead & AU 1+2 \\
        Raised upper eyelids OR blinking, revealing the sclera above the iris &
        AU 5 OR AU 45 \\
        \colourRow Lower eyelids are relaxed & -- \\
        Slack jaw causing the lips and teeth to part & AU 25+26 \\
        \colourRow No tension in the mouth & -- \\
        \bottomrule
    \end{tabular}
    \parasep
\end{table}